\documentclass[longauth,traditabstract]{aa}

\usepackage{xspace}
\usepackage[mathscr]{eucal}
\usepackage{array}
\usepackage{morefloats}
\usepackage{multirow}
\usepackage{enumerate}

\usepackage[nonamebreak]{natbib}
\usepackage[stable]{footmisc}
\usepackage{amsmath}
\usepackage{amssymb}
\usepackage{txfonts}
\usepackage{placeins}
\usepackage{natbib}
\bibpunct{(}{)}{;}{a}{}{,} %
\usepackage{graphicx}
\usepackage{epstopdf}
\usepackage{ifthen}
\usepackage[table,usenames,dvipsnames]{xcolor}
\definecolor{linkcolor}{rgb}{0.6,0,0}
\definecolor{citecolor}{rgb}{0,0,0.75}
\definecolor{urlcolor}{rgb}{0.12,0.46,0.7}
\usepackage[breaklinks, colorlinks, urlcolor=urlcolor, linkcolor=linkcolor,citecolor=citecolor,pdfencoding=auto]{hyperref}
\hypersetup{linktocpage}
\usepackage{overpic}
\usepackage{fixltx2e}
\usepackage{rotating}
\usepackage{booktabs}
\usepackage{subcaption}
\usepackage{ulem}
\usepackage{etoolbox}
\usepackage[utf8]{inputenc} 

\def\setsymbol#1#2{\expandafter\def\csname #1\endcsname{#2}}
\def\getsymbol#1{\csname #1\endcsname}

\def\Planck{\textit{Planck}}

\newbox\tablebox    \newdimen\tablewidth
\def\leaderfil{\leaders\hbox to 5pt{\hss.\hss}\hfil}
\def\endPlancktable{\tablewidth=\columnwidth 
    $$\hss\copy\tablebox\hss$$
    \vskip-\lastskip\vskip -2pt}
\def\endPlancktablewide{\tablewidth=\textwidth 
    $$\hss\copy\tablebox\hss$$
    \vskip-\lastskip\vskip -2pt}
\def\tablenote#1 #2\par{\begingroup \parindent=0.8em
    \abovedisplayshortskip=0pt\belowdisplayshortskip=0pt
    \noindent
    $$\hss\vbox{\hsize\tablewidth \hangindent=\parindent \hangafter=1 \noindent
    \hbox to \parindent{$^#1$\hss}\strut#2\strut\par}\hss$$
    \endgroup}
\def\doubleline{\vskip 3pt\hrule \vskip 1.5pt \hrule \vskip 5pt}

\def\L2{\ifmmode L_2\else $L_2$\fi}

\def\DeltaT{\ifmmode \Delta T\else $\Delta T$\fi}
\def\deltat{\ifmmode \Delta t\else $\Delta t$\fi}
\def\fknee{\ifmmode f_{\rm knee}\else $f_{\rm knee}$\fi}
\def\Fmax{\ifmmode F_{\rm max}\else $F_{\rm max}$\fi}
\def\solar{\ifmmode{\rm M}_{\mathord\odot}\else${\rm M}_{\mathord\odot}$\fi}
\def\Msolar{\ifmmode{\rm M}_{\mathord\odot}\else${\rm M}_{\mathord\odot}$\fi}
\def\Lsolar{\ifmmode{\rm L}_{\mathord\odot}\else${\rm L}_{\mathord\odot}$\fi}
\def\inv{\ifmmode^{-1}\else$^{-1}$\fi}
\def\mo{\ifmmode^{-1}\else$^{-1}$\fi}
\def\sup#1{\ifmmode ^{\rm #1}\else $^{\rm #1}$\fi}
\def\expo#1{\ifmmode \times 10^{#1}\else $\times 10^{#1}$\fi}
\def\,{\thinspace}
\def\lsim{\mathrel{\raise .4ex\hbox{\rlap{$<$}\lower 1.2ex\hbox{$\sim$}}}}
\def\gsim{\mathrel{\raise .4ex\hbox{\rlap{$>$}\lower 1.2ex\hbox{$\sim$}}}}

\def\simprop{\mathrel{\raise .4ex\hbox{\rlap{$\propto$}\lower 1.2ex\hbox{$\sim$}}}}
\def\deg{\ifmmode^\circ\else$^\circ$\fi}
\def\pdeg{\ifmmode $\setbox0=\hbox{$^{\circ}$}\rlap{\hskip.11\wd0 .}$^{\circ}
          \else \setbox0=\hbox{$^{\circ}$}\rlap{\hskip.11\wd0 .}$^{\circ}$\fi}
\def\arcs{\ifmmode {^{\scriptstyle\prime\prime}}
          \else $^{\scriptstyle\prime\prime}$\fi}
\def\arcm{\ifmmode {^{\scriptstyle\prime}}
          \else $^{\scriptstyle\prime}$\fi}
\newdimen\sa  \newdimen\sb
\def\parcs{\sa=.07em \sb=.03em
     \ifmmode \hbox{\rlap{.}}^{\scriptstyle\prime\kern -\sb\prime}\hbox{\kern -\sa}
     \else \rlap{.}$^{\scriptstyle\prime\kern -\sb\prime}$\kern -\sa\fi}
\def\parcm{\sa=.08em \sb=.03em
     \ifmmode \hbox{\rlap{.}\kern\sa}^{\scriptstyle\prime}\hbox{\kern-\sb}
     \else \rlap{.}\kern\sa$^{\scriptstyle\prime}$\kern-\sb\fi}
\def\ra[#1 #2 #3.#4]{#1\sup{h}#2\sup{m}#3\sup{s}\llap.#4}
\def\dec[#1 #2 #3.#4]{#1\deg#2\arcm#3\arcs\llap.#4}
\def\deco[#1 #2 #3]{#1\deg#2\arcm#3\arcs}
\def\rra[#1 #2]{#1\sup{h}#2\sup{m}}

\def\dots{\relax\ifmmode \ldots\else $\ldots$\fi}
\def\WHzsr{\ifmmode $W\,Hz\mo\,sr\mo$\else W\,Hz\mo\,sr\mo\fi}
\def\mHz{\ifmmode $\,mHz$\else \,mHz\fi}
\def\GHz{\ifmmode $\,GHz$\else \,GHz\fi}
\def\mKs{\ifmmode $\,mK\,s$^{1/2}\else \,mK\,s$^{1/2}$\fi}
\def\muKs{\ifmmode \,\mu$K\,s$^{1/2}\else \,$\mu$K\,s$^{1/2}$\fi}
\def\muKRJs{\ifmmode \,\mu$K$_{\rm RJ}$\,s$^{1/2}\else \,$\mu$K$_{\rm RJ}$\,s$^{1/2}$\fi}
\def\muKHz{\ifmmode \,\mu$K\,Hz$^{-1/2}\else \,$\mu$K\,Hz$^{-1/2}$\fi}
\def\MJysr{\ifmmode \,$MJy\,sr\mo$\else \,MJy\,sr\mo\fi}
\def\MJysrmK{\ifmmode \,$MJy\,sr\mo$\,mK$_{\rm CMB}\mo\else \,MJy\,sr\mo\,mK$_{\rm CMB}\mo$\fi}
\def\microns{\ifmmode \,\mu$m$\else \,$\mu$m\fi}

\def\muK{\ifmmode \,\mu$K$\else \,$\mu$\hbox{K}\fi}
\def\microK{\ifmmode \,\mu$K$\else \,$\mu$\hbox{K}\fi}
\def\muW{\ifmmode \,\mu$W$\else \,$\mu$\hbox{W}\fi}
\def\kms{\ifmmode $\,km\,s$^{-1}\else \,km\,s$^{-1}$\fi}
\def\kmsMpc{\ifmmode $\,\kms\,Mpc\mo$\else \,\kms\,Mpc\mo\fi}
\providecommand{\sorthelp}[1]{}

\def\LCDM{$\Lambda$CDM}
\def\NHUNIT{\ifmmode {\rm \,cm^{-2}} \else $\rm \,cm^{-2}$ \fi} 

\def\wmap{\WMAP}

\def\muKcmb{\ifmmode \,\mu$K$_{\rm CMB}$\else \,$\mu$K$_{\rm CMB}$\fi}
\newcommand{\sigmaexp}{\sigma_{\mathrm{exp}}}
\newcommand{\sigmagauss}{\sigma_{\mathrm{Gauss}}}
\newcommand{\sigmadiff}{\sigma_{\mathrm{diff}}}
\newcommand{\lowl}{\mksym{lowlTT}}
\newcommand{\dustcib}{\, \mathrm{dust+CIB}}

\newcommand{\planck}{\Planck}
\newcommand{\WMAP}{WMAP}

\newcommand{\As}{A_{\rm s}}

\newcommand{\ns}{n_{\rm s}}
\newcommand{\lcdm}{$\Lambda$CDM}

\newcommand{\Alens}{A_{\rm L}}

\newcommand{\omegak}{\Omega_K}

\newcommand{\OmegaM}{\ifmmode\Omega_{\rm M}\else $\Omega_{\rm M}$\fi}

\newcommand{\nnu}{N_{\rm eff}}
\newcommand{\neff}{N_{\rm eff}}
\newcommand{\mnu}{\sum m_\nu}

\def\WMAP{{WMAP}}

\newcommand{\onesig}[1]{(68\%, \text{#1})}

\newcommand{\camspec}{{\tt CamSpec}}
\newcommand{\plik}{{\tt Plik}}
\newcommand{\pliklite}{\texttt{Plik\_lite}}

\newcommand{\commander}{{\tt Commander}}
\newcommand{\bflike}{{\tt bflike}}

\newcommand{\thetaMC}{\theta_{\rm MC}}

\providecommand{\Planck}{\textit{Planck}}
\providecommand{\planck}{\Planck}

\providecommand{\text}[1]{\rm{#1}}

\providecommand{\muK}{\mu\rm{K}}

\providecommand{\omb}{\omega_{\mathrm{b}}}
\providecommand{\omc}{\omega_{\mathrm{c}}}

\providecommand{\CAMB}{{\tt camb}}

\providecommand{\LCDM}{{$\rm{\Lambda CDM}$}}

\newcommand{\begm}{\begin{pmatrix}}
\newcommand{\enm}{\end{pmatrix}}

\def\pmb#1{\setbox0=\hbox{#1}%
    \kern-.025em\copy0\kern-\wd0
    \kern.05em\copy0\kern-\wd0
    \kern-.025em\raise.0433em\box0}

\def\p2Y{\;_2Y}
\def\m2Y{\;_{-2}Y}
\def\beglet{
  \addtocounter{equation}{1}%
  \setcounter{parentequation}{\value{equation}}%
  \setcounter{equation}{0}%
  \def\theequation{\arabic{parentequation}\alph{equation}}%
  \ignorespaces
}
\def\endlet{
  \setcounter{equation}{\value{parentequation}}%
  \def\theequation{\arabic{equation}}%
}
\providecommand{\beglet}{\begin{subequations}}
\providecommand{\endlet}{\end{subequations}}

\newcommand{\mksym}[1]{\ifmmode {\rm #1}\else #1\fi}

\newcommand{\plikTT}{\texttt{Plik}\rm TT}
\newcommand{\plikEE}{\texttt{Plik}\rm EE}
\newcommand{\plikTE}{\texttt{Plik}\rm TE}
\newcommand{\TEEE}{\mksym{{\rm TE{,}EE}}}

\newcommand{\TT}{\mksym{{\rm TT}}}
\newcommand{\EE}{\mksym{{\rm EE}}}
\newcommand{\BB}{\mksym{{\rm BB}}}
\newcommand{\TE}{\mksym{{\rm TE}}}

\newcommand{\TB}{\mksym{{\rm TB}}}
\newcommand{\EB}{\mksym{{\rm EB}}}

\newcommand{\lowE}{\mksym{lowE}}

\newcommand{\dataplus}{\allowbreak+}
\newcommand{\lensing}{\mksym{lensing}}
\newcommand{\TTTEEE}{\mksym{{\rm TT{,}TE{,}EE}}}
\newcommand{\planckTTonly}{{\it Planck\/} {\rm TT}}
\newcommand{\planckTEonly}{{\it Planck\/} {\rm TE}}
\newcommand{\planckEEonly}{{\it Planck\/} {\rm EE}}
\newcommand{\lowTEB}{\mksym{{\rm lowTEB}}}
\newcommand{\lowEB}{\mksym{{\rm lowP}}}

\newcommand{\planckTT}{{\it PlanckTT\/} {\rm {+}\lowEB}}
\newcommand{\planckall}{{\it Planck\/} {\rm TT,TE,EE{+}\lowE}}

\newcommand{\planckTTTEEE}{{\it Planck\/} {\rm TT,TE,EE}}
\newcommand{\planckalllensing}{\planckall\dataplus\lensing}

\newcommand{\lnAs}{\ln(10^{10} A_{\rm s})}

\providecommand{\text}[1]{\rm{#1}}

\providecommand{\muK}{\mu\rm{K}}

\providecommand{\omb}{\omega_{\mathrm{b}}}
\providecommand{\omc}{\omega_{\mathrm{c}}}

\newcommand{\Omch}{\Omega_{\mathrm{c}}h^2}
\newcommand{\Ombh}{\Omega_{\mathrm{b}}h^2}

\providecommand{\CAMB}{\texttt{CAMB}}

\providecommand{\PICO}{\texttt{PICO}}

\providecommand{\LCDM}{{$\rm{\Lambda CDM}$}}

\newcommand\ba{\begin{eqnarray}}
\newcommand\ea{\end{eqnarray}}
\newcommand\bea{\begin{eqnarray}}
\newcommand\eea{\end{eqnarray}}

\newcommand\be{\begin{equation}}
\newcommand\ee{\end{equation}}

\newcommand{\quickpol}{\texttt{QuickPol}\xspace}

\newcommand{\plikTTTEEE}{\texttt{Plik}\rm TT,EE,TE}

\newcommand{\degree}{\ensuremath{^\circ}}

\newcommand{\fsky}{\ensuremath{f_{\rm sky}}}
\newcommand{\calibM}{\ensuremath{y}}
\newcommand{\calibC}{\ensuremath{c}}

\newcommand{\sroll}{{\tt SRoll}}
\newcommand{\simbal}{{\tt SimBaL}}
\newcommand{\simlow}{{\tt SimLow}}
\newcommand{\simall}{{\tt SimAll}}

\newcommand{\Nside}{\ensuremath{N_{\mathrm{side}}}}
\newcommand{\cosmomc}{{\tt cosmomc}}

\defcitealias{planck2013-p08}{PPL13}
\defcitealias{planck2014-a13}{PPL15}
\defcitealias{planck2013-p11}{PCP13}
\defcitealias{planck2014-a15}{PCP15}
\defcitealias{planck2016-l06}{PCP18}
\defcitealias{planck2013-012}{PL13}
\defcitealias{planck2014-a17}{PL15}
\defcitealias{planck2016-l08}{PL18}
\defcitealias{planck2016-l03}{HFI18}
\defcitealias{planck2016-l02}{LFI18}
\defcitealias{quickpolHivon}{QPL17}

\makeatother

\begin{document}

\title{\vglue -10mm\Planck\ 2018 results. V. CMB power spectra and likelihoods}

\author{\small
Planck Collaboration: N.~Aghanim\inst{54}
\and
Y.~Akrami\inst{15, 56, 58}
\and
M.~Ashdown\inst{64, 5}
\and
J.~Aumont\inst{93}
\and
C.~Baccigalupi\inst{76}
\and
M.~Ballardini\inst{21, 41}
\and
A.~J.~Banday\inst{93, 8}
\and
R.~B.~Barreiro\inst{60}
\and
N.~Bartolo\inst{28, 61}
\and
S.~Basak\inst{83}
% ...
}
\institute{\small
AIM, CEA, CNRS, Universit\'{e} Paris-Saclay, Universit\'{e} Paris-Diderot, Sorbonne Paris Cit\'{e}, F-91191 Gif-sur-Yvette, France\goodbreak
\and
APC, AstroParticule et Cosmologie, Universit\'{e} Paris Diderot, CNRS/IN2P3, CEA/lrfu, Observatoire de Paris, Sorbonne Paris Cit\'{e}, 10, rue Alice Domon et L\'{e}onie Duquet, 75205 Paris Cedex 13, France\goodbreak
\and
African Institute for Mathematical Sciences, 6-8 Melrose Road, Muizenberg, Cape Town, South Africa\goodbreak
\and
Aix Marseille Univ, CNRS, CNES, LAM, Marseille, France\goodbreak
\and
Astrophysics Group, Cavendish Laboratory, University of Cambridge, J J Thomson Avenue, Cambridge CB3 0HE, U.K.\goodbreak
\and
Astrophysics \& Cosmology Research Unit, School of Mathematics, Statistics \& Computer Science, University of KwaZulu-Natal, Westville Campus, Private Bag X54001, Durban 4000, South Africa\goodbreak
\and
CITA, University of Toronto, 60 St. George St., Toronto, ON M5S 3H8, Canada\goodbreak
\and
CNRS, IRAP, 9 Av. colonel Roche, BP 44346, F-31028 Toulouse cedex 4, France\goodbreak
\and
Cahill Center for Astronomy and Astrophysics, California Institute of Technology, Pasadena CA,  91125, USA\goodbreak
\and
California Institute of Technology, Pasadena, California, U.S.A.\goodbreak
\and
Centre for Theoretical Cosmology, DAMTP, University of Cambridge, Wilberforce Road, Cambridge CB3 0WA, U.K.\goodbreak
\and
Computational Cosmology Center, Lawrence Berkeley National Laboratory, Berkeley, California, U.S.A.\goodbreak
\and
DTU Space, National Space Institute, Technical University of Denmark, Elektrovej 327, DK-2800 Kgs. Lyngby, Denmark\goodbreak
\and
D\'{e}partement de Physique Th\'{e}orique, Universit\'{e} de Gen\`{e}ve, 24, Quai E. Ansermet,1211 Gen\`{e}ve 4, Switzerland\goodbreak
\and
D\'{e}partement de Physique, \'{E}cole normale sup\'{e}rieure, PSL Research University, CNRS, 24 rue Lhomond, 75005 Paris, France\goodbreak
\and
Departamento de Astrof\'{i}sica, Universidad de La Laguna (ULL), E-38206 La Laguna, Tenerife, Spain\goodbreak
\and
Departamento de F\'{\i}sica, Universidad de Oviedo, C/ Federico Garc\'{\i}a Lorca, 18 , Oviedo, Spain\goodbreak
\and
Department of Astrophysics/IMAPP, Radboud University, P.O. Box 9010, 6500 GL Nijmegen, The Netherlands\goodbreak
\and
Department of Mathematics, University of Stellenbosch, Stellenbosch 7602, South Africa\goodbreak
\and
Department of Physics \& Astronomy, University of British Columbia, 6224 Agricultural Road, Vancouver, British Columbia, Canada\goodbreak
\and
Department of Physics \& Astronomy, University of the Western Cape, Cape Town 7535, South Africa\goodbreak
\and
Department of Physics and Astronomy, University College London, London WC1E 6BT, U.K.\goodbreak
\and
Department of Physics and Astronomy, University of Sussex, Brighton BN1 9QH, U.K.\goodbreak
\and
Department of Physics, Gustaf H\"{a}llstr\"{o}min katu 2a, University of Helsinki, Helsinki, Finland\goodbreak
\and
Department of Physics, Princeton University, Princeton, New Jersey, U.S.A.\goodbreak
\and
Department of Physics, University of California, One Shields Avenue, Davis, California, U.S.A.\goodbreak
\and
Department of Physics, University of California, Santa Barbara, California, U.S.A.\goodbreak
\and
Dipartimento di Fisica e Astronomia G. Galilei, Universit\`{a} degli Studi di Padova, via Marzolo 8, 35131 Padova, Italy\goodbreak
\and
Dipartimento di Fisica e Scienze della Terra, Universit\`{a} di Ferrara, Via Saragat 1, 44122 Ferrara, Italy\goodbreak
\and
Dipartimento di Fisica, Universit\`{a} La Sapienza, P. le A. Moro 2, Roma, Italy\goodbreak
\and
Dipartimento di Fisica, Universit\`{a} degli Studi di Milano, Via Celoria, 16, Milano, Italy\goodbreak
\and
Dipartimento di Fisica, Universit\`{a} degli Studi di Trieste, via A. Valerio 2, Trieste, Italy\goodbreak
\and
Dipartimento di Fisica, Universit\`{a} di Roma Tor Vergata, Via della Ricerca Scientifica, 1, Roma, Italy\goodbreak
\and
European Space Agency, ESAC, Planck Science Office, Camino bajo del Castillo, s/n, Urbanizaci\'{o}n Villafranca del Castillo, Villanueva de la Ca\~{n}ada, Madrid, Spain\goodbreak
\and
European Space Agency, ESTEC, Keplerlaan 1, 2201 AZ Noordwijk, The Netherlands\goodbreak
\and
Gran Sasso Science Institute, INFN, viale F. Crispi 7, 67100 L'Aquila, Italy\goodbreak
\and
HEP Division, Argonne National Laboratory, Lemont, IL 60439, USA\goodbreak
\and
Haverford College Astronomy Department, 370 Lancaster Avenue, Haverford, Pennsylvania, U.S.A.\goodbreak
\and
Helsinki Institute of Physics, Gustaf H\"{a}llstr\"{o}min katu 2, University of Helsinki, Helsinki, Finland\goodbreak
\and
IFPU - Institute for Fundamental Physics of the Universe, Via Beirut 2, 34014 Trieste, Italy\goodbreak
\and
INAF - OAS Bologna, Istituto Nazionale di Astrofisica - Osservatorio di Astrofisica e Scienza dello Spazio di Bologna, Area della Ricerca del CNR, Via Gobetti 101, 40129, Bologna, Italy\goodbreak
\and
INAF - Osservatorio Astronomico di Padova, Vicolo dell'Osservatorio 5, Padova, Italy\goodbreak
\and
INAF - Osservatorio Astronomico di Trieste, Via G.B. Tiepolo 11, Trieste, Italy\goodbreak
\and
INAF - Osservatorio Astronomico di Trieste, via G. B. Tiepolo 11, I-34143 Trieste, Italy\goodbreak
\and
INAF, Istituto di Radioastronomia, Via Piero Gobetti 101, I-40129 Bologna, Italy\goodbreak
\and
INAF/IASF Milano, Via E. Bassini 15, Milano, Italy\goodbreak
\and
INFN - CNAF, viale Berti Pichat 6/2, 40127 Bologna, Italy\goodbreak
\and
INFN, Sezione di Bologna, viale Berti Pichat 6/2, 40127 Bologna, Italy\goodbreak
\and
INFN, Sezione di Ferrara, Via Saragat 1, 44122 Ferrara, Italy\goodbreak
\and
INFN, Sezione di Milano, Via Celoria 16, Milano, Italy\goodbreak
\and
INFN, Sezione di Roma 1, Universit\`{a} di Roma Sapienza, Piazzale Aldo Moro 2, 00185, Roma, Italy\goodbreak
\and
INFN, Sezione di Roma 2, Universit\`{a} di Roma Tor Vergata, Via della Ricerca Scientifica, 1, Roma, Italy\goodbreak
\and
Imperial College London, Astrophysics group, Blackett Laboratory, Prince Consort Road, London, SW7 2AZ, U.K.\goodbreak
\and
Institut d'Astrophysique Spatiale, CNRS, Univ. Paris-Sud, Universit\'{e} Paris-Saclay, B\^{a}t. 121, 91405 Orsay cedex, France\goodbreak
\and
Institut d'Astrophysique de Paris, CNRS (UMR7095), 98 bis Boulevard Arago, F-75014, Paris, France\goodbreak
\and
Institute Lorentz, Leiden University, PO Box 9506, Leiden 2300 RA, The Netherlands\goodbreak
\and
Institute of Astronomy, University of Cambridge, Madingley Road, Cambridge CB3 0HA, U.K.\goodbreak
\and
Institute of Theoretical Astrophysics, University of Oslo, Blindern, Oslo, Norway\goodbreak
\and
Instituto de Astrof\'{\i}sica de Canarias, C/V\'{\i}a L\'{a}ctea s/n, La Laguna, Tenerife, Spain\goodbreak
\and
Instituto de F\'{\i}sica de Cantabria (CSIC-Universidad de Cantabria), Avda. de los Castros s/n, Santander, Spain\goodbreak
\and
Istituto Nazionale di Fisica Nucleare, Sezione di Padova, via Marzolo 8, I-35131 Padova, Italy\goodbreak
\and
Jet Propulsion Laboratory, California Institute of Technology, 4800 Oak Grove Drive, Pasadena, California, U.S.A.\goodbreak
\and
Jodrell Bank Centre for Astrophysics, Alan Turing Building, School of Physics and Astronomy, The University of Manchester, Oxford Road, Manchester, M13 9PL, U.K.\goodbreak
\and
Kavli Institute for Cosmology Cambridge, Madingley Road, Cambridge, CB3 0HA, U.K.\goodbreak
\and
Kavli Institute for the Physics and Mathematics of the Universe (Kavli IPMU, WPI), UTIAS, The University of Tokyo, Chiba, 277- 8583, Japan\goodbreak
\and
Laboratoire d'Oc{\'e}anographie Physique et Spatiale (LOPS), Univ. Brest, CNRS, Ifremer, IRD, Brest, France\goodbreak
\and
Laboratoire de Physique Subatomique et Cosmologie, Universit\'{e} Grenoble-Alpes, CNRS/IN2P3, 53, rue des Martyrs, 38026 Grenoble Cedex, France\goodbreak
\and
Laboratoire de Physique Th\'{e}orique, Universit\'{e} Paris-Sud 11 \& CNRS, B\^{a}timent 210, 91405 Orsay, France\goodbreak
\and
Lawrence Berkeley National Laboratory, Berkeley, California, U.S.A.\goodbreak
\and
Low Temperature Laboratory, Department of Applied Physics, Aalto University, Espoo, FI-00076 AALTO, Finland\goodbreak
\and
Max-Planck-Institut f\"{u}r Astrophysik, Karl-Schwarzschild-Str. 1, 85741 Garching, Germany\goodbreak
\and
Mullard Space Science Laboratory, University College London, Surrey RH5 6NT, U.K.\goodbreak
\and
NAOC-UKZN Computational Astrophysics Centre (NUCAC), University of KwaZulu-Natal, Durban 4000, South Africa\goodbreak
\and
National Centre for Nuclear Research, ul. L. Pasteura 7, 02-093 Warsaw, Poland\goodbreak
\and
Purple Mountain Observatory, No. 8 Yuan Hua Road, 210034 Nanjing, China\goodbreak
\and
SISSA, Astrophysics Sector, via Bonomea 265, 34136, Trieste, Italy\goodbreak
\and
San Diego Supercomputer Center, University of California, San Diego, 9500 Gilman Drive, La Jolla, CA 92093, USA\goodbreak
\and
School of Chemistry and Physics, University of KwaZulu-Natal, Westville Campus, Private Bag X54001, Durban, 4000, South Africa\goodbreak
\and
School of Physical Sciences, National Institute of Science Education and Research, HBNI, Jatni-752050, Odissa, India\goodbreak
\and
School of Physics and Astronomy, Cardiff University, Queens Buildings, The Parade, Cardiff, CF24 3AA, U.K.\goodbreak
\and
School of Physics and Astronomy, Sun Yat-sen University, 2 Daxue Rd, Tangjia, Zhuhai, China\goodbreak
\and
School of Physics and Astronomy, University of Nottingham, Nottingham NG7 2RD, U.K.\goodbreak
\and
School of Physics, Indian Institute of Science Education and Research Thiruvananthapuram, Maruthamala PO, Vithura, Thiruvananthapuram 695551, Kerala, India\goodbreak
\and
School of Physics, The University of New South Wales, Sydney NSW 2052, Australia\goodbreak
\and
Simon Fraser University, Department of Physics, 8888 University Drive, Burnaby BC, Canada\goodbreak
\and
Sorbonne Universit\'{e}, CNRS, UMR 7095, Institut d'Astrophysique de Paris, 98 bis bd Arago, 75014 Paris, France\goodbreak
\and
Sorbonne Universit\'{e}, Institut Lagrange de Paris (ILP), 98 bis Boulevard Arago, 75014 Paris, France\goodbreak
\and
Sorbonne Universit\'{e}, Observatoire de Paris, Universit\'{e} PSL, \'{E}cole normale sup\'{e}rieure, CNRS, LERMA, F-75005, Paris, France\goodbreak
\and
Space Research Institute (IKI), Russian Academy of Sciences, Profsoyuznaya Str, 84/32, Moscow, 117997, Russia\goodbreak
\and
Space Science Data Center - Agenzia Spaziale Italiana, Via del Politecnico snc, 00133, Roma, Italy\goodbreak
\and
Space Sciences Laboratory, University of California, Berkeley, California, U.S.A.\goodbreak
\and
The Oskar Klein Centre for Cosmoparticle Physics, Department of Physics, Stockholm University, AlbaNova, SE-106 91 Stockholm, Sweden\goodbreak
\and
Universit\'{e} de Toulouse, UPS-OMP, IRAP, F-31028 Toulouse cedex 4, France\goodbreak
% ...
}

\abstract{\vglue -3mm 
We describe the legacy \Planck\ cosmic microwave background (CMB)
likelihoods derived from the 2018 data release. The overall approach
is similar in spirit to the one retained for the 2013 and 2015 data
release, with a hybrid method using different approximations at low
($\ell<30$) and high ($\ell\ge30$) multipoles, implementing several
methodological and data-analysis refinements compared to previous
releases. With more realistic simulations, and better correction and
modelling of systematic effects, we can now make full use of the CMB
polarization observed in the High Frequency Instrument (HFI)
channels. The low-multipole $EE$ cross-spectra from the 100-GHz and
143-GHz data give a constraint on the \lcdm\ reionization
optical-depth parameter $\tau$ to better than 15\,\% (in combination
with the $TT$ low-$\ell$ data and the high-$\ell$ temperature and
polarization data), tightening constraints on all parameters with
posterior distributions correlated with $\tau$. We also update the
weaker constraint on $\tau$ from the joint TEB likelihood using the
Low Frequency Instrument (LFI) channels, which was used in 2015 as
part of our baseline analysis. At higher multipoles, the CMB
temperature spectrum and likelihood are very similar to previous
releases. A better model of the
temperature-to-polarization leakage and corrections for the effective
calibrations of the polarization channels (i.e., the polarization
efficiencies) allow us to make full use of polarization spectra,
improving the \lcdm\ constraints on the parameters $\thetaMC$, $\omc$,
$\omb$, and $H_0$ by more than 30\,\%, and $\ns$ by more than 20\,\%
compared to $TT$-only constraints.
Extensive tests on the robustness of the modelling of the polarization data demonstrate good consistency, with some residual modelling uncertainties. At high multipoles{,} we are now limited mainly by the accuracy of the polarization efficiency modelling. Using our various tests, simulations, and comparison between different high-multipole likelihood implementations, we estimate the consistency of the results to be better than the $0.5\,\sigma$ level on the \LCDM\ parameters, as well as classical single-parameter extensions for the joint likelihood (to be compared to the $0.3\,\sigma$ levels we achieved in 2015 for the temperature data alone on \LCDM\ only).
Minor curiosities already present in the previous releases remain, such as the differences between the best-fit
\LCDM\ parameters for the $\ell<800$ and $\ell>800$ ranges of the
power spectrum, or the preference
for more smoothing of the power-spectrum peaks than predicted in
\lcdm\ fits. These are shown to be driven by the temperature power spectrum and are not
significantly modified by the inclusion of the polarization data. Overall, the legacy \Planck\ CMB likelihoods provide a robust tool for constraining the cosmological model and represent a reference for future CMB observations. }


\authorrunning{Planck Collaboration}
\titlerunning{\Planck\ likelihoods}

\maketitle

\tableofcontents

\section{Introduction}

This paper presents the \Planck\footnote{\Planck\ 
(\url{http://www.esa.int/Planck}) is a project of the European Space Agency (ESA) with instruments provided by two scientific consortia funded by ESA member states and led by Principal Investigators from France and Italy, telescope reflectors provided through a collaboration between ESA and a scientific consortium led and funded by Denmark, and 
additional contributions from NASA (USA).} legacy likelihoods for the cosmic microwave background anisotropies (CMB). The data set considered for this legacy likelihood release (also known as the ``2018 release'' or ``PR3'')
is derived from full-mission \Planck\ data and consists of Stokes intensity and linear polarization maps in the frequency range 30 to 353\,GHz, complemented by intensity maps at 545 and 857\,GHz. For polarization, thanks to significant data processing improvements \citep{planck2016-l03}, the set employed is wider than what was used in the previous (2015) \Planck\ release and now includes large-angle polarization data from the High Frequency Instrument (HFI) 100- and 143-GHz channels, improving on preliminary results described in \citet{planck2014-a10}. Many methodological and analysis improvements have been carried out since the 2015 release that directly impact the \Planck\ CMB likelihoods, resulting in tighter control of the systematic and statistical error budget and more thorough validation of the final products. The use of simulations has grown significantly since 2015, along with the level of realism in the simulated data. At small scales, a better model of the temperature-to-polarization leakage, as well as better determinations of the polarization efficiencies of HFI polarimeters, allow us to use the polarization data in the baseline \Planck\ cosmological results \citep{planck2016-l06}. These improvements are extensively discussed in the remainder of this paper. 

As in 2015, we adopt a hybrid approach between so-called ``low-multipole'' and ``high-multipole'' ($\ell$) regimes, the dividing line still being at $\ell\,{=}\,30$. While in 2015 the low-$\ell$ polarization likelihood was based on Low Frequency Instrument (LFI) 70-GHz low-resolution maps, the baseline low-$\ell$ legacy polarization likelihood is based on $E$-mode ($EE$) angular cross-spectra derived from the HFI 100- and 143-GHz channels. We do nonetheless present here an improved version of the LFI 70-GHz low-$\ell$ polarization likelihood, based on the latest LFI maps discussed in \citet{planck2016-l02}. All of the different likelihood products presented in this paper, including those that are not part of the baseline, are released through the Planck Legacy Archive (PLA\footnote{\url{https://www.cosmos.esa.int/web/planck/pla}}).

In this legacy release, the CMB likelihoods are built from estimates of the angular power spectra derived from intensity and linear polarization maps, with the only exception being the LFI 70-GHz low-$\ell$ polarization likelihood, which is based on maps. Specifically, the low-$\ell$ temperature (TT) likelihood is constructed by approximating the marginal distribution of the temperature angular power spectrum derived from Gibbs sampling-based component separation. The low-$\ell$ polarization (EE) likelihood is built by comparing a cross-frequency power spectrum of two foreground-corrected maps to a set of simulations.The temperature and polarization high-$\ell$ likelihoods (TT, TE, and EE) uses multiple cross-frequency spectra estimates, assuming smooth foreground and nuisance spectra templates and a Gaussian likelihood approximation.

The information content of the CMB sky can be split into temperature, plus two polarization components, the $E$ and $B$ modes.
For a full-sky, statistically isotropic, Gaussian distributed CMB intensity and linear polarization pattern, the six CMB angular power spectra (computed between the three signals $T$, $E$, and $B$) contain all information available in the map, and thus represent an effectively lossless compression of the cosmological information. They are uniquely determined by the underlying cosmological model and its parameters. The $E$ and $B$ polarization modes are coordinate-independent quantities \citep[e.g.,][]{KamionkowskiKS1997,ZaldarriagaS1997,HuWhite1997} with different dependencies on the underlying cosmology.  Density perturbations can only source $E$-mode polarization and this is hence correlated with temperature, while $B$-mode polarization is a signature of tensor modes, i.e., primordial gravitational waves from inflation. In a standard $\Lambda$CDM scenario, only four power spectra are expected to be non-zero due to parity conservation: $TT$, $TE$, $EE$, and $BB$.

Weak gravitational lensing by the inhomogeneous mass distribution along the line of sight between the last-scattering surface and the observer distorts the intensity and polarization CMB field. This effect modifies the angular power spectrum at small scales, but also modifies the field statistics by introducing a non-Gaussian component with associated non-zero 4-point angular correlation function, or trispectrum. This information can be exploited to derive the power spectrum of the lensing potential and the lensing field map itself. A dedicated lensing likelihood is used in addition to the CMB likelihood to constrain cosmological parameters.  The \Planck\ legacy lensing likelihood is described in \citet{planck2016-l08}. The amplitude of the effect of lensing on the CMB power spectra can be used as a consistency check. Slightly surprising results of this test (i.e., values of the consistency parameter $A_{\rm L}$ above unity) are thoroughly discussed in the previous \Planck\ release papers and in particular in \citet{planck2014-a13}, \citet{planck2014-a15}, \citet{planck2016-LI}, and \citet{planck2016-l06}. This test will also be discussed in the present paper.

While \Planck\ resolution and sensitivity allow for a solid determination of the $TT$, $TE$, and $EE$ power spectra, a detection of the CMB $BB$ spectrum is still beyond the reach of the latest data analysis. At large scales, the signal is masked by 
the different sources of emission from our own Galaxy. We will show that, both in the low-frequency and in the high-frequency ranges, after a careful removal of the Galactic emission, the \Planck\ $BB$ power spectrum is compatible with zero. At smaller scales, the weak lensing effect described above dominates the signal and the resulting $B$-mode polarization can only be detected in the \Planck\ maps by cross-correlating the $B$ maps with a template built from the observed $E$ maps and a tracer of the distribution of lenses (such as the \Planck\ CMB lensing map or the \Planck\ CIB map) as shown in \citet{planck2014-a17}, \citet{planck2015-XLI}, and \citet{planck2016-l08}. We do not further investigate the $B$ polarization in the present paper.

In a real-world situation, accurate power-spectrum estimation needs to rely on models of instrumental noise and other instrumental systematic effects, as well as of residual contamination from astrophysical foregrounds. Several approximations can be built in order to facilitate this. So-called ``pseudo-power spectrum'' (also known as pseudo-$C_\ell$) estimators, \citep[e.g.,][]{Hietal02,Tristram2004,2004MNRAS.350..914C,Polenta2005} typically work well in the high-$\ell$ regime and are computationally efficient thanks to fast convolution on the sphere \citep{Muciaccia1997,gorski2005}. In the low-$\ell$ regime, methods that derive power-spectrum estimates from the likelihood function are better suited, such as the quadratic maximum likelihood (QML) approach \citep[e.g.,][]{Tegmark2001,Efstathiou:2006,Gruppuso:2009}. Consistently, the high-$\ell$ likelihood methodology presented in this paper is based on pseudo-$C_\ell$ estimators, while the baseline low-$\ell$ likelihood relies on QML methods. On the other hand, the LFI 70-GHz low-$\ell$ polarization likelihood, as in the 2015 release, is directly built from low-resolution maps without using power spectra as an intermediate step. The latter likelihood has been significantly improved since the 2015 release, although it is not used in the main parameter analysis due to its lower signal-to-noise ratio with respect to the present baseline. Currently, the LFI 70-GHz likelihood is the only low-resolution \Planck\ likelihood to contain information from the temperature-polarization correlations (i.e., the $TE$ spectrum).  Furthermore, being based on CMB maps, rather than angular power spectra, this likelihood can also be employed, pending appropriate modifications in the signal covariance matrix, to test models that do not assume statistical isotropy of the CMB field. 

Parts of this paper or earlier stages of the work make use of the \texttt{CAMB} \citep{Lewis:1999bs} and \texttt{CLASS} \citep{2011arXiv1104.2932L} Boltzmann codes and the \texttt{CosmoMC} \citep{2002PhRvD..66j3511L} and \texttt{Monte Python} \citep{2013JCAP...02..001A} Markov chain packages. The likelihood code and some of the validation work are built on the library \texttt{pmclib} from the \texttt{CosmoPMC} package \citep{2011arXiv1101.0950K}.

\paragraph{Likelihood names:}

Throughout this paper, we will follow the CMB likelihood naming convention adopted by the other \planck\ papers, defining the labels:
\begin{itemize}
 \item \planckTTonly, the likelihood formed using only the temperature data, spanning the multipole range $2\le\ell\la2500$;
 \item \lowl, the likelihood formed using only the temperature data, spanning the multipole range $2\le\ell<30$ (included in the \planckTTonly);
 \item \planckTEonly\ and \planckEEonly, the likelihood formed using exclusively the $TE$ power spectrum from $30\le\ell\la2000$ and the $EE$ power spectrum, respectively;
 \item \planckTTTEEE, the combination of \planckTTonly, \planckTEonly, and \planckEEonly, taking into account correlations between the $TT$, $TE$, and $EE$ spectra at $\ell>29$;
 \item \lowE, the likelihood formed using the $EE$ power spectrum over $2\le\ell<30$;
 \item \lowTEB, the map-based LFI likelihood, covering the range $2\le\ell<30$, which is sometimes also referred to as \bflike.
\end{itemize}

Each of these likelihoods use combinations of different approximations, which we describe in the paper. 
A summary of those approximations is given at the end of the paper (namely in Table~\ref{tab:likelihooddef}).
\planckall, the reference likelihood of this release, is given by the multiplication of 
\begin{itemize}
 \item the \commander\ likelihood in \TT\ for $2\le\ell<30$, which provides \lowl\ and contributes to \planckTTonly,
 \item the \simall\ likelihood in \EE\ for $2\le\ell<30$, which provides \lowE,
 \item the \plikTT,\TEEE\ likelihood for \TTTEEE\ over $30\le\ell\la2500$ in \TT\ and $30\le\ell\la2000$ in \TE\ and \EE, which contributes the high-$\ell$ part of \planckTTTEEE.
\end{itemize}
Two other alternative high-$\ell$ likelihood implementations are presented in the paper, namely the \camspec\ and \pliklite\ likelihoods. 
Similarly to the \plik\ case, we will denote the \TT\ \camspec\ likelihood at $\ell\ge30$ as \camspec{\rm \TT},  the \TE\ \pliklite\ likelihood at $\ell\ge30$ as \pliklite{\rm \TE}, etc.
Unless otherwise indicated, likelihoods described with the ``{\it Planck}\dots'' notation will always use the \plik\ reference implementation.

\paragraph{Citation conventions:}

In this new likelihood paper, we often refer to the likelihood papers associated
with the 2013 \Planck\
release \citep[hereafter \citetalias{planck2013-p08}]{planck2013-p08}
and that associated with the 2015 \Planck\ release
\citep[hereafter \citetalias{planck2014-a13}]{planck2014-a13}.
We also refer to results from the 2018 release cosmological parameters
\citep[hereafter \citetalias{planck2016-l06}]{planck2016-l06},
as well as the cosmological parameters paper associated with the 2015 release
\citep[hereafter \citetalias{planck2014-a15}]{planck2014-a15}.
Additionally,
whenever discussing the latest version of the \Planck\ maps, we will refer to the two papers describing the 
processing of the LFI \citep[hereafter \citetalias{planck2016-l02}]{planck2016-l02} and HFI \citep[hereafter \citetalias{planck2016-l03}]{planck2016-l03} data.

The plan of this paper is as follows. In Sect.~\ref{sec:lo-ell} we present our low-$\ell$ methodology, starting from the temperature likelihood (Sect.~\ref{sec:lo-ell:TT}). We then present the HFI low-$\ell$ $EE$ likelihood, which is used in our baseline (Sect.~\ref{sec:lo-ell:hfi}) and finally the LFI pixel-based polarization likelihood (Sect.~\ref{sec:lo-ell:lfi}). In Sect.~\ref{sec:hi-ell} we present our high-$\ell$ approach. We first describe the overall methodology, data selection, and modelling of the baseline high-$\ell$ likelihood (Sects.~\ref{sec:hi-ell:method} to~\ref{sec:hi-ell:datamodel:alltogether}), before presenting two 
alternative high-$\ell$ implementations (Sect.~\ref{sec:hi-ell:prod}). The remainder of this section (Sects.~\ref{sec:valandro} to~\ref{sec:valandro:sims}) is devoted to the validation of the main high-$\ell$ likelihood product, focusing on the issues revealed in the previous release, as well as new tests.  The joint baseline likelihood is briefly discussed in Sect.~\ref{sec:joint}, while in Sect.~\ref{sec:conclusion} we present our conclusions.

\section{Low multipoles}
\label{sec:lo-ell}

We present in this section the different low-$\ell$ likelihoods. The baseline low-$\ell$ likelihood adopted in the 2018 legacy release of \Planck\ (and thus employed for the main parameter analysis) uses the combination of a Gibbs-sampling approach in temperature (provided by the \texttt{Commander} code; \citetalias{planck2013-p08,planck2014-a13}) and a cross-spectrum-based, simulation-supported method, relying on the HFI 100- and 143-GHz channels in polarization. The latter takes into account only the $EE$ and $BB$ spectra, but not currently $TE$. 
We also update and release the pixel-based likelihood already presented in \citetalias{planck2014-a13} using the \texttt{Commander} 2018 solution in temperature and the 70-GHz LFI full maps in polarization. While less sensitive, this latter likelihood combination does sample $TE$ and, being pixel-based, can be straightforwardly adapted to handle non-rotationally invariant cosmologies.

The following three subsections describe, respectively, the \texttt{Commander} likelihood, the HFI simulation-based likelihood{,} and the LFI pixel-based likelihood. For each method{,} we present a detailed validation based on Monte Carlo simulations.


\subsection{\TT{} low-$\ell$ likelihood}
\label{sec:lo-ell:TT}

The first of these likelihood implementations is based on the Bayesian
posterior sampling framework called \texttt{Commander}, which combines
astrophysical component separation and likelihood estimation, and
employs Gibbs sampling to map out the full joint posterior
\citep{eriksen2008}. This method has been used extensively in previous
\Planck\ releases, and we therefore provide only a brief review of the
main ideas in the following, and refer interested readers to
previous papers for full details
\citepalias{planck2013-p08,planck2014-a13}.

The starting point of the \texttt{Commander} framework is an explicit
parametric data model of the form
\begin{equation}
\vec d_{\nu} = \sum_{i=1}^{N_{\mathrm{comp}}} \tens F_i(\theta_i) \vec
a_i + \vec n_{\nu},
\end{equation}
where $\vec d_{\nu}$ denotes an observed sky map at frequency $\nu$;
the sum runs over a set of distinct astrophysical emission components
(CMB, synchrotron, thermal dust emission, etc.), each parametrized by
an amplitude vector $\vec a_i$ and an effective spectral energy
density $\tens F_i$ with some set of free parameters
$\theta_i$. The instrumental noise is given by $\vec n_{\nu}$. The CMB is
assumed to be Gaussian distributed with variance given by the power
spectrum $C_{\ell} = \left<\vec |a_{\mathrm{CMB}}|^2\right>$, where
the CMB amplitude vector is now defined in terms of spherical
harmonics.

Given this model, \texttt{Commander} explores the full joint
distribution $P(\vec a_i, \theta_i, C_{\ell}|\vec d_{\nu})$ through
Gibbs sampling \citep{eriksen2008}. From this joint distribution, a
posterior mean CMB map may be extracted simply by averaging over all
individual samples, and this may be used as input in a brute-force
low-$\ell$ likelihood code, as for instance was done for the baseline
\Planck\ 2015 likelihood \citepalias{planck2014-a13} and is still done for the 2018 
LFI-based likelihood described in Sect.~\ref{sec:lo-ell:lfi} below. 
The advantage of this
approach is that it is algorithmically easy to combine different
temperature and polarization estimators. The main disadvantage,
however, is that it scales poorly with angular resolution, and is in
practice limited to very low multipoles, typically $\ell \la 30$. 
The latter problem may be solved through a so-called
Blackwell-Rao estimator \citep{chu2005}, which
approximates the marginal power spectrum distribution $P(C_{\ell}|\vec
d_{\nu})$ by the expression
\begin{equation}
\mathcal{L}(C_{\ell}) \propto \sum_{i=1}^{N} \prod_{\ell=\ell_{\mathrm{min}}}^{\ell_{\mathrm{max}}} \frac{1}{\sigma^i_{\ell}} \left(\frac{\sigma^i_{\ell}}{C_{\ell}}\right)^{\frac{2\ell+1}{2}} e^{-\frac{2\ell+1}{2} \frac{\sigma^i_{\ell}}{C_{\ell}}},
\end{equation}
where
\begin{equation}
\sigma_{\ell}^i = \frac{1}{2\ell+1}\sum_{m=-\ell}^{\ell} |a_{\ell m}^i|^2
\end{equation}
is the observed power spectrum of the $i$th Gibbs CMB sky sample, $N$
is the total number of Gibbs samples, and $\ell_{\mathrm{min}}$ and
$\ell_{\mathrm{max}}$ define the multipole range of the likelihood
estimator. As $N$ approaches infinity, this expression converges to
the exact likelihood; however, the number of samples required for
convergence scales exponentially with the multipole range, which makes
it expensive for large $\ell_{\mathrm{max}}$. To break this scaling,
the estimator may be approximated by a transformed Gaussian, as
described by \citet{rudjord2009}, which leads to linear convergence in
$\ell_{\mathrm{max}}$. As in 2015, we adopt this Gaussianized version
of the Blackwell-Rao estimator for the \Planck\ 2018 likelihood code,
with a normalization defined such that
$\log\mathcal{L}(\left<\sigma_{\ell}\right>) = 0$, corresponding to a
standard $\chi^2$-like normalization. 

Even in its Gaussianized version, this estimator requires high
signal-to-noise observations to converge quickly and hence it does
not easily support polarization analysis with the current
\Planck\ data set. We therefore use the cleaned map produced by the Gaussianized Blackwell-Rao
estimator only for the low-$\ell$ \TT{} likelihood
analysis for $\ell<30$.

The algorithms used for the low-$\ell$ \texttt{Commander}-based
analyses in the \Planck\ 2018 release are unchanged compared to 2015,
and only the data selection and model specification
differ. Specifically, while the 2015 analysis included both
\Planck\ and external data (WMAP and Haslam 408~MHz;
\citealt{bennett2012,haslam1982}) and a rich astrophysical model (CMB,
synchrotron, free-free, spinning and thermal dust, multi-line CO line
emission, etc.), the corresponding 2018 analysis employs \Planck\ data
only, and a greatly simplified foreground model (CMB, a combined
low-frequency power-law component, thermal dust, and CO line
emission). This choice is driven both by the fact that the 2018
\Planck\ data set is inherently more coarse-grained in terms of map
products (there are only full-frequency maps, no detector set or
single bolometer maps; \citetalias{planck2016-l03}), and by a desire to
make the final delivery products as conservative as possible in terms
of data selection, with minimal dependence on external data sets. The
cost of this choice is a slightly reduced effective sky fraction
compared to 2015 ($f_{\mathrm{sky}}=0.86$ in 2018 versus 0.94 in 2015;
see Sect.~\ref{ValidationTT}), reflecting the fact that our ability
to resolve the various foreground components inside the Galactic plane
is reduced without external data sets (for free-free and spinning dust
emission) or single-bolometer maps (for CO line emission). Further
details on the \texttt{Commander} 2018 temperature mask definition can be
found in appendix~A.5 of \citep{planck2016-l04}.

\subsubsection{Validation}\label{ValidationTT}

\begin{figure}
\centering
\includegraphics[width=0.5\textwidth]{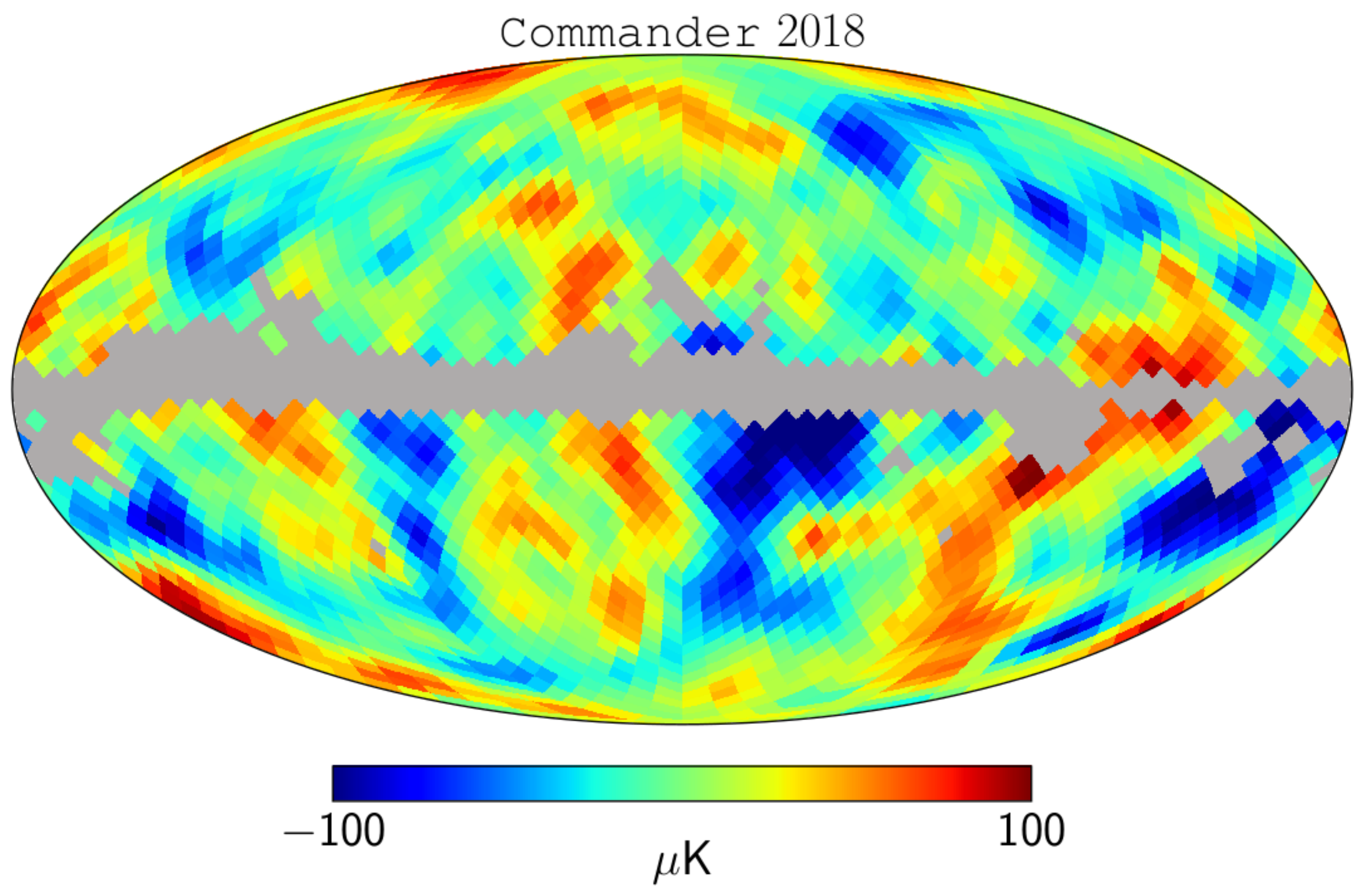}
\includegraphics[width=0.5\textwidth]{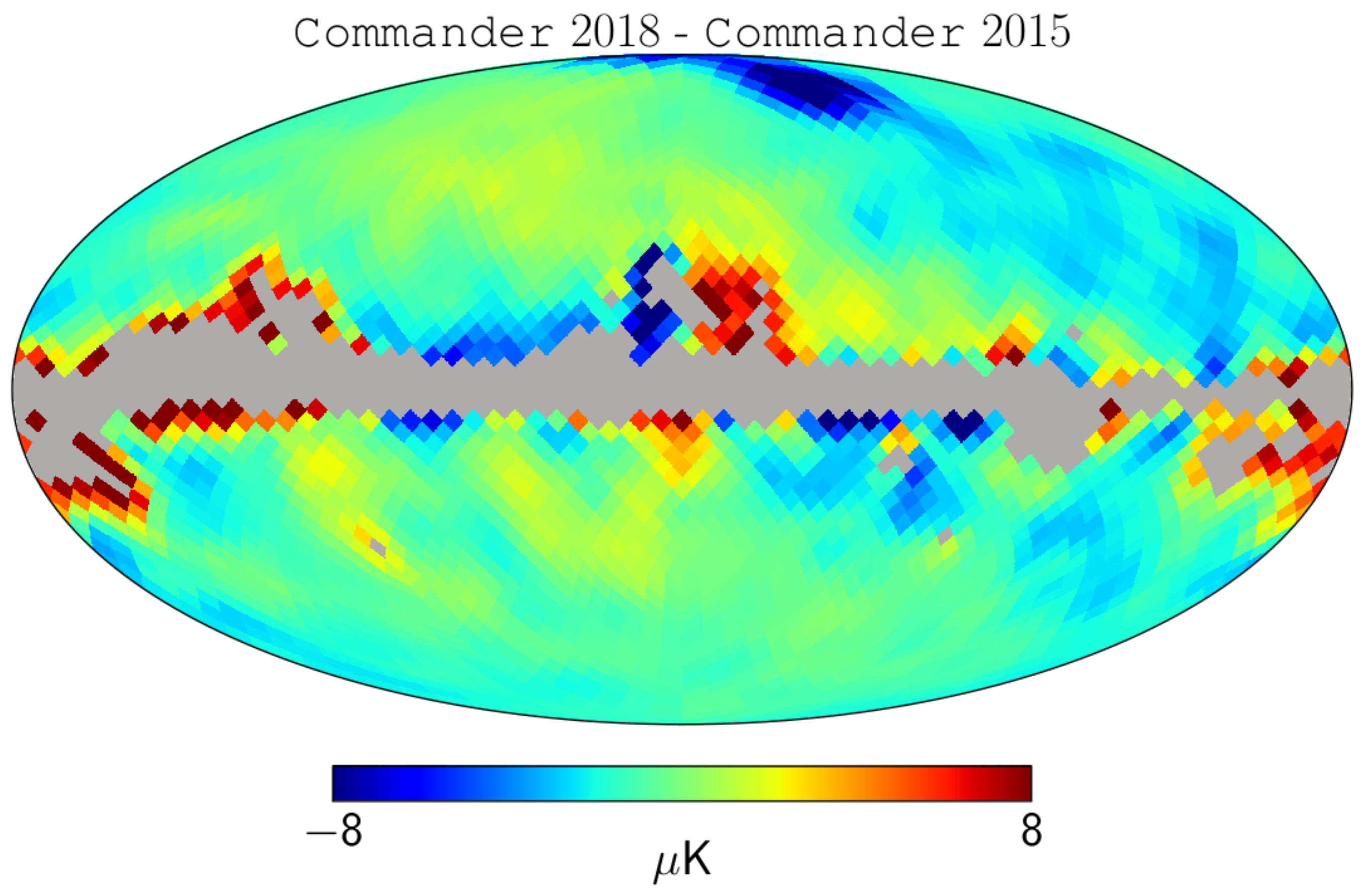}
\includegraphics[width=0.5\textwidth]{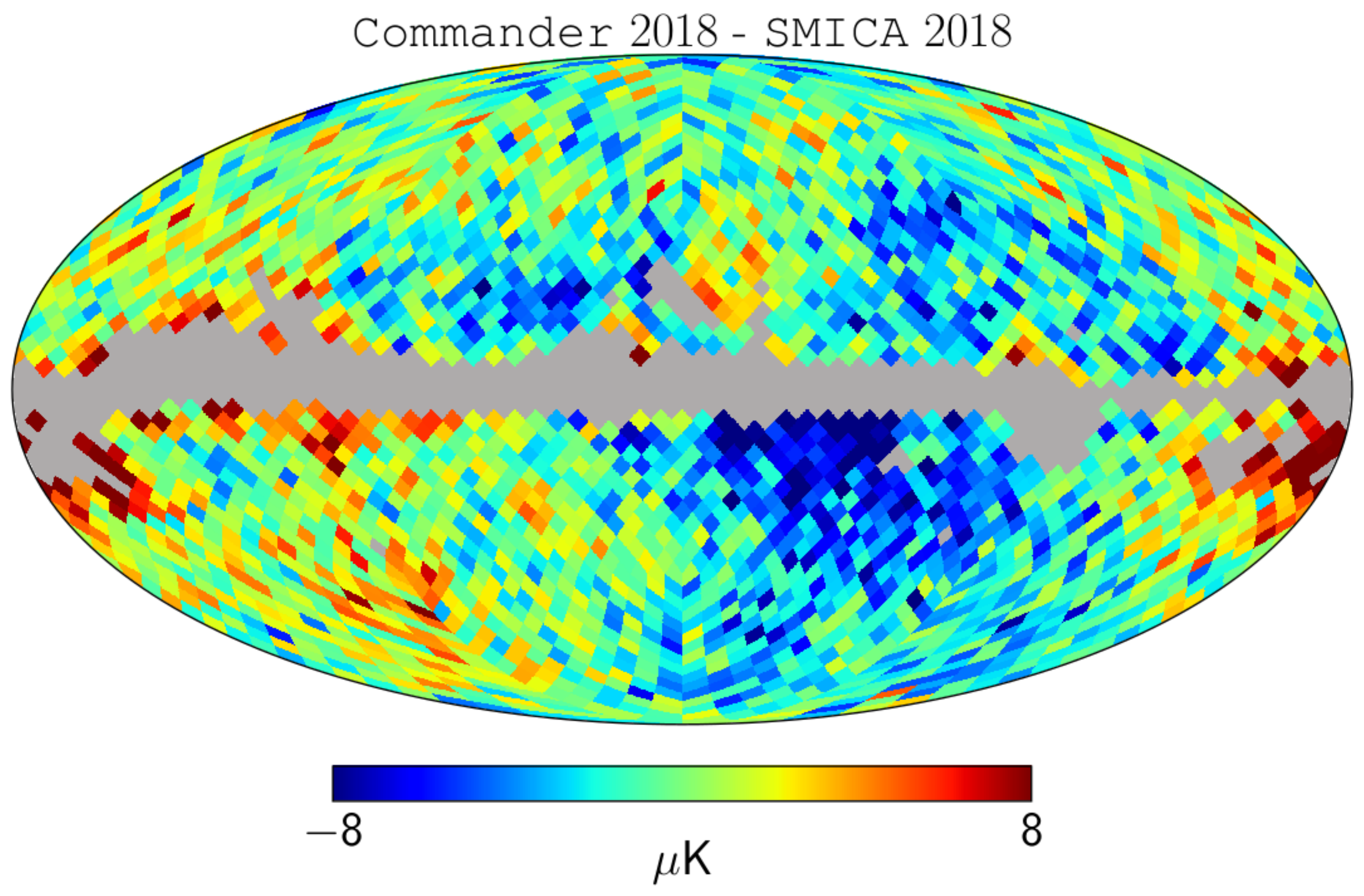}
\caption{{\it Top}: {\tt Commander} 2018 low-$\ell$ temperature map masked with the {\tt Commander} 2018 mask. {\it Middle\/} and {\it bottom}: differences between the {\tt Commander} 2018 map and the {\tt Commander} 2015 map (middle) and the {\tt SMICA}-dedicated low-$\ell$ map (bottom).
}\label{TTmaps}
\end{figure}

We performed several tests of the new \TT{} low-$\ell$ {\tt Commander} solution based on angular power spectrum extraction and parameter estimation. The validation is obtained by comparing the results with the {\tt Commander} 2015 low-$\ell$ temperature map \citepalias{planck2014-a13} and with the component-separated 2018 {\tt SMICA} map \citep{planck2016-l04} obtained starting from the same inputs as the new {\tt Commander} solution and downgraded to low resolution. Each of the maps introduced above has been delivered with a corresponding confidence mask.
In Fig.~\ref{TTmaps} we show the 2018 {\tt Commander} map masked with the 2018 {\tt Commander} mask. We also show the differences of this map with the one of 2015, and with the {\tt SMICA} map. The difference maps are masked with the 2018 {\tt Commander} mask, the most conservative mask among the ones considered in this section.

\begin{figure}
\centering
\includegraphics[width=0.5\textwidth]{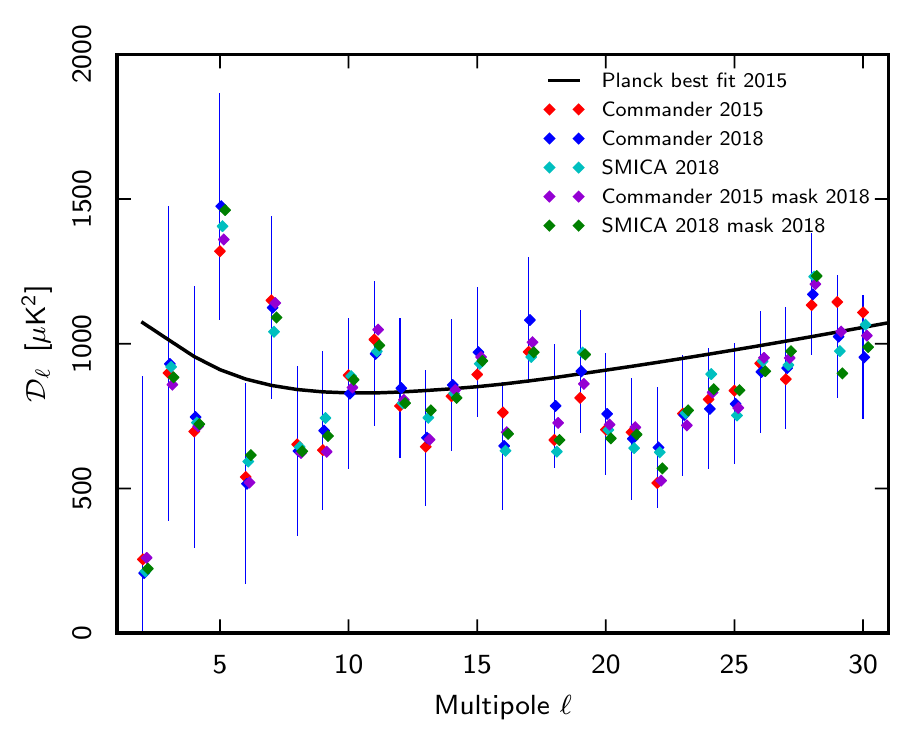}
\caption{$TT$ angular power spectra ($D_\ell\equiv\ell(\ell+1)C_\ell/2\pi$)
of the available low-$\ell$ component-separated maps: {\tt Commander} 2018 (blue points); {\tt Commander} 2015 (red points); {\tt SMICA} 2018 (cyan points); and {\tt Commander} 2015 and {\tt SMICA} 2018 masked with the 2018 {\tt Commander} mask (purple and green points, respectively).
}\label{TT_aps}
\end{figure}

In Fig.~\ref{TT_aps} we show the $TT$ angular power spectra extracted applying a quadratic maximum likelihood (QML) estimator to the maps described above, using for each its own confidence mask. Since the 2018 {\tt Commander} solution provides a more conservative mask than the 2015 one, we also show the angular power spectra of the {\tt Commander} 2015 map and {\tt SMICA} 2018 with the {\tt Commander} 2018 mask applied (purple and green points). To avoid confusion we show only the errors associated with the {\tt Commander} 2018 mask since we are cosmic-variance dominated and all the masks involved are very similar in sky coverage.

We notice good consistency between the angular power spectra, especially with respect to the low quadrupole values and the well-known dip around $\ell\approx22$. However, we notice some differences, the most striking one is at $\ell\,{=}\,5$, where {\tt Commander} 2018 is about 150$\,\mu{\rm K}^2$ higher than for the {\tt Commander} 2015 map. When masked with the 2018 {\tt Commander} mask, {\tt SMICA} also shows an amplitude similar to the {\tt Commander} 2018 spectrum at $\ell\,{=}\,5$.

\begin{figure}
\centering
\includegraphics[width=0.5\textwidth]{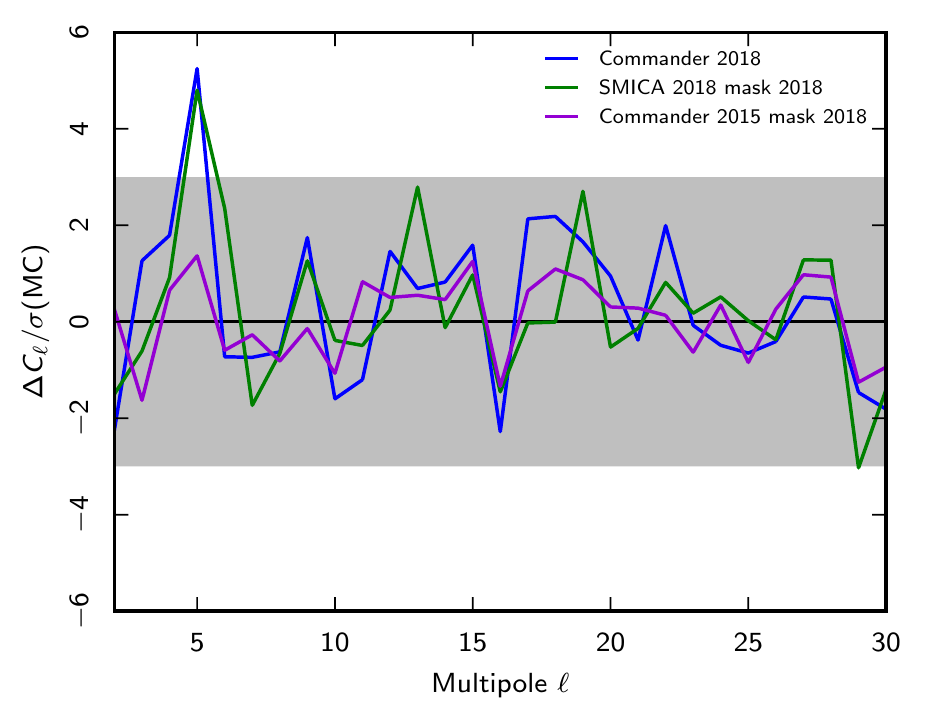}
\caption{Differences normalized to the sampling variance in the angular power spectrum with respect to the {\tt Commander} 2015 one. The grey band is the $\pm3\,\sigma$ dispersion of 10\,000 MC angular power spectra differences calculated using either the {\tt Commander} 2015 mask or the {\tt Commander} 2018 mask.
}\label{TT_aps_diff}
\end{figure}

In order to assess the statistical significance of the observed differences, we employed Monte Carlo (MC) simulations of 10\,000 pure CMB maps, based on the \Planck\ 2015 best-fit model,\footnote{We did not use FFP10 simulations for this test due to the low number of maps available; we expect that the inclusion of systematic effects and foreground residuals in the final error budget would decrease the statistical significance of these results.} We extracted the angular power spectrum from the MC maps using both the 2015 and the 2018 {\tt Commander} masks, and calculated the dispersion of the angular power spectra differences obtained map by map. In Fig.~\ref{TT_aps_diff}{,} we show as a grey band the $\pm3\,\sigma$ dispersion of the MC angular power spectrum differences. We also show the differences of the angular power spectrum shown in Fig.~\ref{TT_aps} with respect to the {\tt Commander} 2015 one.

As expected, the choice between 2015 and 2018 masks has little effect on the {\tt Commander} 2015 results (all differences are well inside the grey region).  On the other hand, the differences with respect to the other two maps are clearly larger. The most anomalous point is $\ell\,{=}\,5$ for {\tt Commander} and {\tt SMICA} 2018 maps, which is about 5.2$\,\sigma$ off the MC distribution; all the differences in the other multipoles are within 3$\,\sigma$. We tried to see if these differences are due to the unmasked regions close to the Galactic plane, but the $\ell\,{=}\,5$ multipole does not change when considering a more aggressive Galactic mask with a sky fraction of 73\,\%. Even when accounting for the look-elsewhere effect the significance is reduced only to 3$\,\sigma$. The look-elsewhere effect is obtained by counting how many maps of the 10\,000 show a shift in at least one multipole in the considered range as large as 5$\,\sigma$ from the distribution of the rest of the MC simulations. We suggest that the reason for this significant shift at $\ell\,{=}\,5$ is due to a combination of the different data and foreground models used, an improved control of systematics, and the use of a more conservative mask in 2018. 
{We note} that $\ell\,{=}\,5$ in $EE$ gives an important contribution to the $\tau$ measurement in the HFI low-$\ell$ Likelihood. As discussed below (see Sect.~\ref{subsubsec:final_considerations} for details), the effect in polarization is completely consistent with our error model, so there is no evidence of contamination from the $\ell\,{=}\,5$ temperature anomaly. This conclusion is supported by the positive outcome of the $\ell\,{=}\,5$ $TE$ null-tests.

\begin{figure}
\centering
\includegraphics[width=0.5\textwidth]{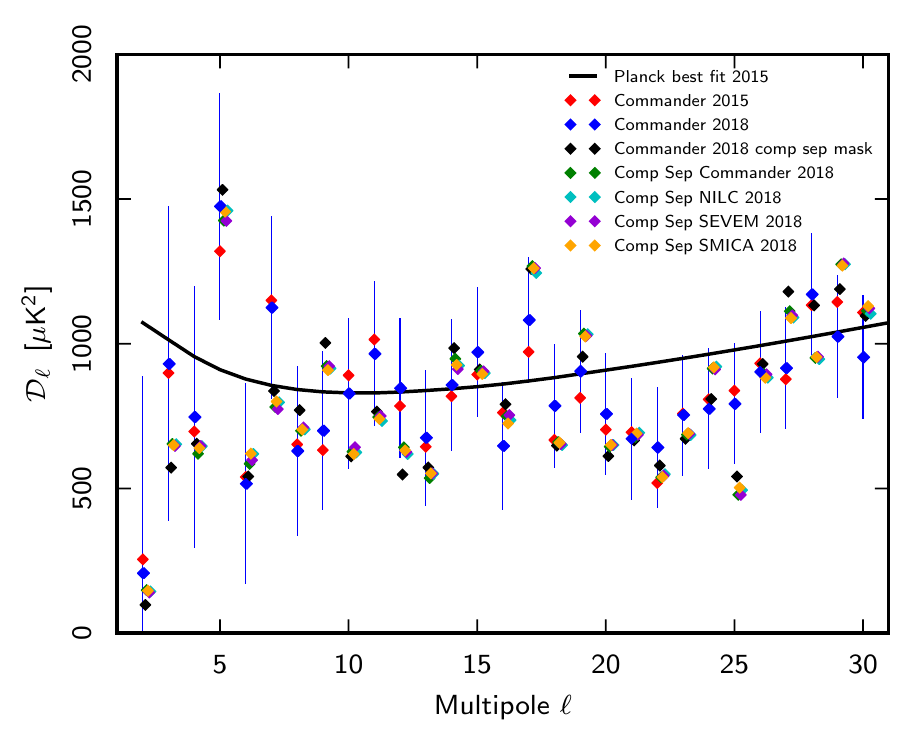}
\includegraphics[width=0.5\textwidth]{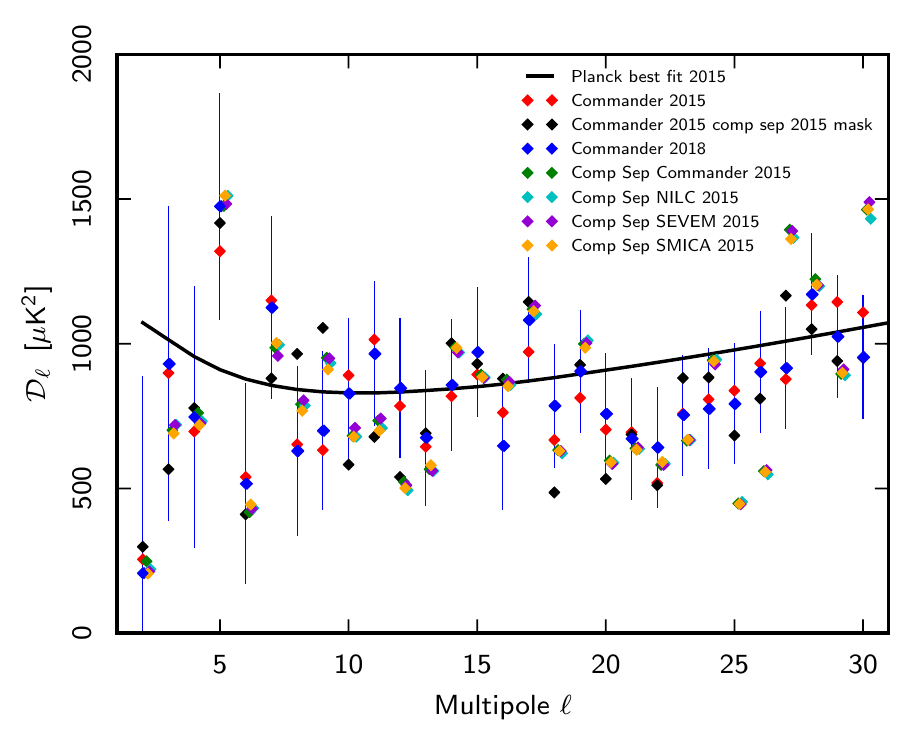}
\caption{Angular power spectrum of the low-$\ell$ {\tt Commander} maps compared to those of the other component-separated maps. {\it Top}: comparison performed with the component-separated maps of the 2018 release. Black points show the power spectrum of the 2018 low-$\ell$ {\tt Commander} map with the 2018 common mask from component separation. {\it Bottom}: comparison performed with the 2015 component-separated maps. Black points show the power spectrum of the 2015 low-$\ell$ {\tt Commander} map with the 2015 common mask from component separation. Blue and red points are, respectively, the power spectrum of the low-$\ell$ {\tt Commander} maps with the 2018 or 2015 native mask.
}\label{TT_comp_sep-2018}
\end{figure}

As a stability check, in Fig.~\ref{TT_comp_sep-2018} we compare the angular power spectrum from the low-$\ell$ maps with those of the component-separated maps generated by the {\tt Commander}, {\tt NILC}, {\tt SEVEM}, and {\tt SMICA} pipelines for the 2015 release (bottom panel) described in \cite{planck2014-a11} and for the current release (top panel) described in \cite{planck2016-l04}.  The latter paper provides a description of the differences between the low-$\ell$ \commander\ solution and the full-resolution component-separated \commander\ map. We show the angular power spectra employing the masks delivered with each cleaned map, extracted using the QML method after having degraded the component-separated maps to the same low resolution as the low-$\ell$ maps.

Generally, the $C_{\ell}$s of the angular power spectra scatter simply because a different mask is used. This is clear from the low-$\ell$ {\tt Commander} 2018 power spectrum (black points in the top panel); it is in very good agreement with the component-separated ones when extracted using the component-separated mask. If we focus on $\ell\,{=}\,5$, however, the component-separated maps from both the 2015 and the 2018 release show an amplitude of about $1450\,\mu{\rm K}^2$, compatible with the $\ell\,{=}\,5$ amplitude in the {\tt Commander} 2018 map. None of the maps considered reproduces the amplitude of the low-$\ell$ {\tt Commander} 2015 map at $\ell\,{=}\,5$. 

\begin{figure}
\centering
\includegraphics[width=0.5\textwidth]{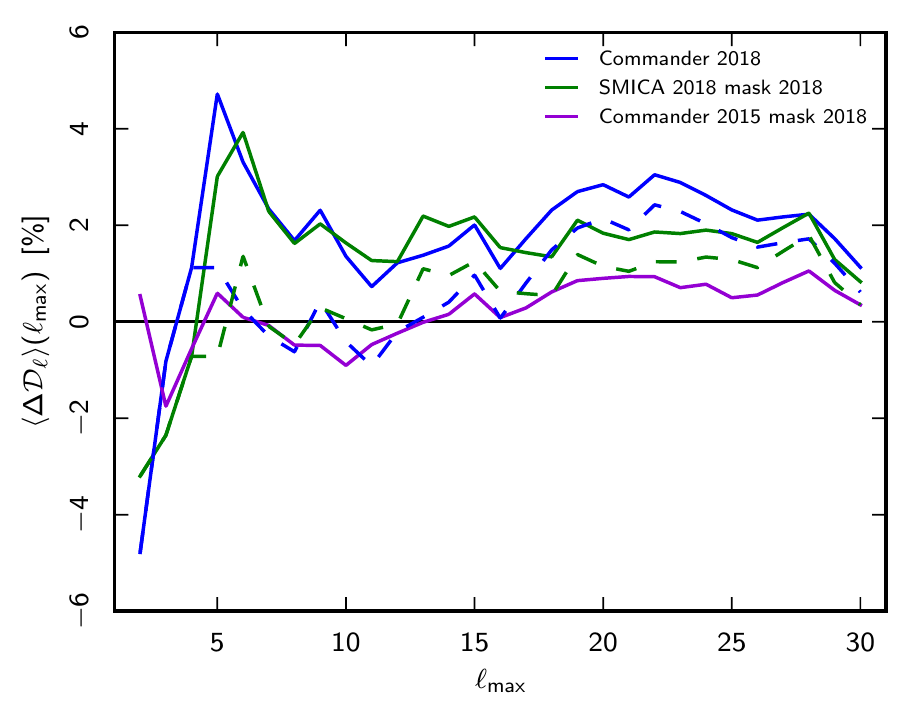}
\caption{Average of the angular power spectrum differences between $\ell\,{=}\,2$ and $\ell_{\rm max}$ as a function of $\ell_{\rm max}$, expressed as a percentage. The differences are taken with respect to {\tt Commander} 2015 results with its native mask.  The dashed lines omit $\ell\,{=}\,5$.}\label{TT_aps_lmax}
\end{figure}

We can also compare the power spectra shown in Fig.~\ref{TT_aps} by taking the angular power spectrum differences with respect to \commander\ 2015 and calculating the average from $\ell\,{=}\,2$ up to a given $\ell_{\rm max}$. This should show the average difference in the power of the maps with respect to selected angular scales. In Fig.~\ref{TT_aps_lmax} we show the average of the power spectrum differences (expressed as a percentage) as a function of $\ell_{\rm max}$ for the low-$\ell$ maps. The dashed lines are the averages when we do not consider $\ell\,{=}\,5$. As clearly shown, the larger $\ell\,{=}\,5$ difference is responsible for an average larger amplitude of about 2\,\% up to $\ell\,{=}\,15$, decreasing to about 1\,\% if we consider the total multipole range up to $\ell\,{=}\,30$. When we do not consider $\ell\,{=}\,5$ all the maps are more compatible. In addition to $\ell\,{=}\,15$, variations in the averaged amplitude are also due to differences in the angular power spectrum around $\ell\,{=}\,20$. The higher power of the 2018 maps will evidently have an impact on some large-scale anomalies, e.g., the so-called lack of power \citep[see][for more detailed analyses dedicated to the search for anomalies in the CMB maps]{planck2016-l07}.

\begin{figure}
\includegraphics[width=0.5\textwidth]{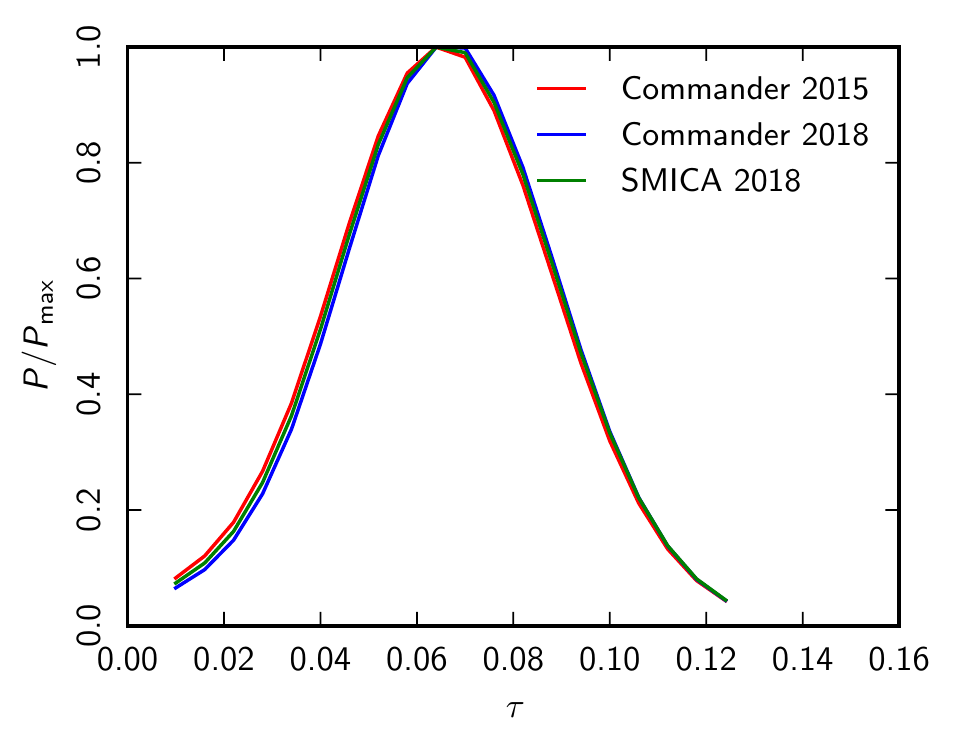}
\includegraphics[width=0.5\textwidth]{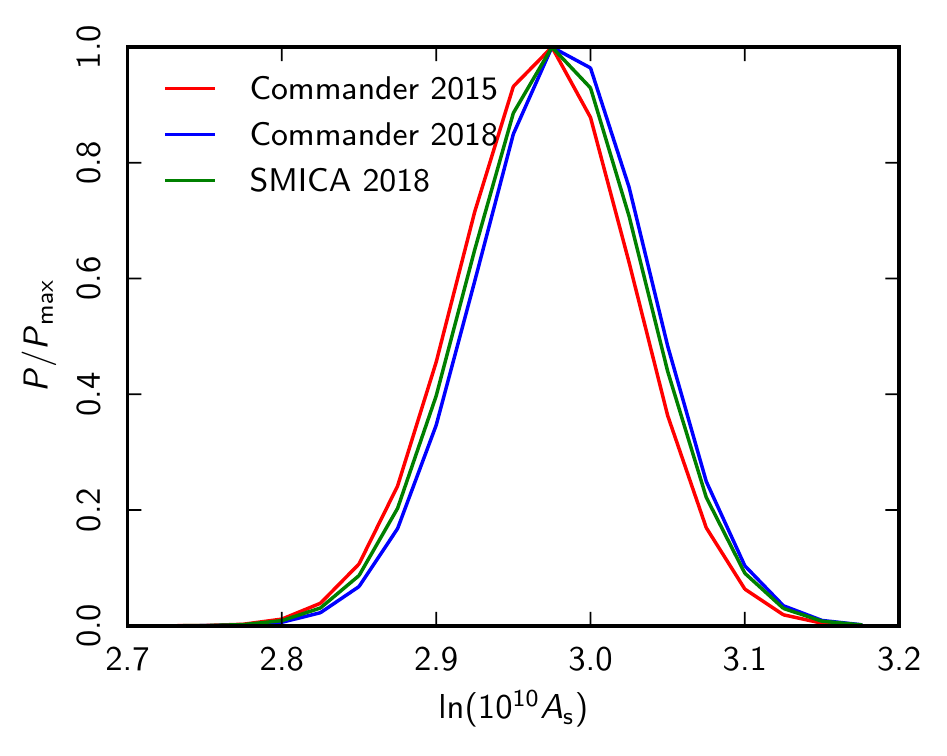}
\caption{Posteriors of the $\tau$ (upper panel) and $A_{\rm s}$ (lower panel) parameters obtained from the LFI pixel-based likelihood using the same 2015 polarized maps and considering different low-$\ell$ temperature maps.
}\label{TT_parameters}
\end{figure}

Finally{,} we would like to understand the impact of the different maps on the parameters most dependent on the largest angular scales, namely $A_{\rm s}$ and $\tau$. We thus added to the maps the 2015 low-$\ell$ likelihood polarization maps used in the LFI pixel-based likelihood and then extracted these parameters using the same pixel-based likelihood algorithm. The posterior distributions are shown in Fig.~\ref{TT_parameters} and the marginalized values are described in Table~\ref{tab:TT_parameters}. The results clearly show that in spite of the presence of a significant shift in $\ell\,{=}\,5${,} the impact on parameters is negligible.

\begin{table}[htbp!]
\begingroup
\newdimen\tblskip \tblskip=5pt
\caption{Marginalized $A_{\rm s}$ and $\tau$ parameters extracted using the LFI 2015 pixel-based likelihood considering different low-$\ell$ temperature maps.}\label{tab:TT_parameters}
\vskip -6mm
\footnotesize
\setbox\tablebox=\vbox{
\newdimen\digitwidth
\setbox0=\hbox{\rm 0}
\digitwidth=\wd0
\catcode`*=\active
\def*{\kern\digitwidth}
\newdimen\signwidth
\setbox0=\hbox{+}
\signwidth=\wd0
\catcode`!=\active
\def!{\kern\signwidth}
\newdimen\decimalwidth
\setbox0=\hbox{.}
\decimalwidth=\wd0
\catcode`@=\active
\def@{\kern\decimalwidth}
\openup 3pt
\halign{
\hbox to 1.15in{#\leaderfil}\tabskip=0.0em&
    \hfil#\hfil\tabskip=1em&
    \hfil#\hfil&
    \hfil#\hfil\tabskip=0.5em&
    \hfil#\hfil\tabskip=0pt\cr
\noalign{\doubleline}
\noalign{\vskip -2pt}
\omit\hfil Data\hfil& $\tau$& $\sigma(\tau)$& ln$(10^{10}A_{\rm s})$&
 $\sigma\big({\rm ln}(10^{10}A_{\rm s})\big)$\cr
\noalign{\vskip 3pt\hrule\vskip 5pt}
{\tt Commander} 2015& 0.06535& 0.02214& 2.96988& 0.05568\cr
{\tt Commander} 2018& 0.06663& 0.02178& 2.98154& 0.05561\cr
{\tt SMICA} 2018&     0.06608& 0.02201& 2.97676& 0.05637\cr
\noalign{\vskip 5pt\hrule\vskip 3pt}
}}
\endPlancktable
\endgroup
\end{table}


\subsection{HFI-based low-$\ell$ likelihood}
\label{sec:lo-ell:hfi}

The low-multipole HFI polarization likelihood for the 2018 release is an extension of the \simbal\ and \simlow\ algorithms for $EE$ presented in \citet{planck2014-a10}. It is based on the (cross-quasi-) QML \citep{Tegmark:1996,Bond:1998,Efstathiou:2004,Efstathiou:2006} spectrum of the 100- and 143-GHz maps cleaned using a template-fitting procedure from polarized synchrotron and polarized dust contaminations. The likelihood algorithm does not rely on an analytical shape approximation, but instead it uses the $300$ end-to-end FFP10 simulations (see Sect.~\ref{sec:valandro:sims}) in order to empirically build the probability distribution of the $EE$ and $BB$ power spectra, ignoring the off-diagonal correlations. We do not build a likelihood for $TE$, given the poor probabilities to exceed (PTEs) in the null tests of the $TE$ spectrum obtained from the {\tt Commander} temperature solution and HFI polarization maps and the difficulties of describing accurately the correlation with $EE$ and $TT$ spectra.

The LFI and HFI FFP10 simulations, used in this section, are built using two different procedures. For LFI, realistic signal-plus-noise timelines are produced, de-calibrated using the measured gains, re-calibrated with the algorithms used for data, and projected into maps. For HFI, realistic timelines are produced containing signal, all the reproducible systematics, and noise; these are then projected into maps using the official HFI mapmaking code, \sroll. The complete simulation pipeline is described in detail in \citetalias{planck2016-l02} and \citetalias{planck2016-l03}, and we also give a more complete description of the HFI FFP10 in Sect.~\ref{sec:valandro:sims}. In this section, noise-covariance matrices are used for the pixel weight in the foreground cleaning and in the cross-QML estimation. For LFI we used the \planck\ legacy products \citepalias[see][]{planck2016-l02}.

 For HFI, since the mapmaking procedure does not provide any analytical approximation to the noise covariance, we used FFP8 products \citep{planck2014-a14,keskitalo2010} which capture only the Gaussian noise, not describing the variance associated with systematics.

In the following sections{,} we describe the foreground-cleaning procedure, the power spectrum estimation, the likelihood approximation, and the consistency tests.

\subsubsection{Polarization low-resolution maps and cleaning procedure}\label{subsec:cleaning}

The low-$\ell$ polarization likelihood uses the lowest frequencies of the HFI instrument, {that is}, (typically the full-mission solutions of) the 100- and 143-GHz channels. We limit the low-$\ell$ polarization analysis to $\ell\,{<}\,\ell_{\rm max}\,{=}\,30$, adopting a {\tt HEALPix} \citep{gorski2005} pixelization corresponding to $\Nside\,{=}\,16$. As in \citet{planck2014-a10} we degrade the full-resolution maps in harmonic space using a cosine window function \citep{Benabed:2009af}:
\begin{equation}
f(\ell) = \left\{
    \begin{array} {ll}            1,  & \ell \le \Nside; \\
             {1 \over 2} \left ( 1 + \sin \left ( {\pi \over 2} {\ell \over \Nside}  \right ) \right ) & \Nside < \ell \le 3 \ \Nside \, ; \\
            0, & \ell > 3 \ \Nside.
    \end{array}
\right. \label{eq:cosdegrade}
\end{equation}
Here the suppression of the high-resolution signal limits possible aliasing.

In order to remove the foreground contamination we perform a template fitting on the $Q$ and $U$ maps using the 30-GHz full-channel map as a tracer of polarized synchrotron and the 353-Hz polarization-sensitive bolometer (hereafter PSB) only map as a tracer of polarized thermal dust. The two templates are smoothed and downgraded with the same procedure as described above. Defining $\vec{m}\equiv [Q,U]$ for each channel the foreground-cleaned polarization maps $\vec{\hat{m}}$ are
\begin{equation}
\vec{\hat{m}}_{100,143} = \frac{\vec{m}_{100,143} - \alpha \vec{m}_{30} - \beta \vec{m}_{353}}{1-\alpha-\beta},\label{eq:cleaned_dataset}
\end{equation}

\noindent where $\alpha$ and $\beta$ are the amplitudes of dust and synchrotron templates, respectively. 
The two amplitudes are estimated by minimizing the $\chi^2$ constructed from the following covariance matrices, associated with each channel:

\begin{equation}
\tens{C}_{100,143} = \tens{S}(C_\ell)+ \frac{\tens{N}_{100,143}+\alpha^2\tens{N}_{30}+\beta^2\tens{N}_{353}}{\left(1-\alpha-\beta\right)^2}.\label{eq:cov_cleaned_dataset}
\end{equation}
Here \tens{S} represents the signal covariance, assuming a theoretical $C_\ell$ from the \Planck~\TTTEEE+{\tt SIMlow} best fit of \citet{planck2014-a10} and $\tens{N}$ represents the $[Q,U]$ part of the noise-covariance matrices. At 30\GHz\ we use the 2018 release covariance matrix \citepalias[see][]{planck2016-l02}.  At 100, 143, and 353\GHz\ we use FFP8 covariance matrices \citep{planck2014-a14}. The FFP8 matrices, {which are computed from an analytical model \citep{keskitalo2010}}, capture only the Gaussian noise part of the uncertainty in the HFI channels and do not account for any systematic effects. 

\begin{figure}[htbp!]
\begin{center}
\includegraphics[width=0.475\textwidth]{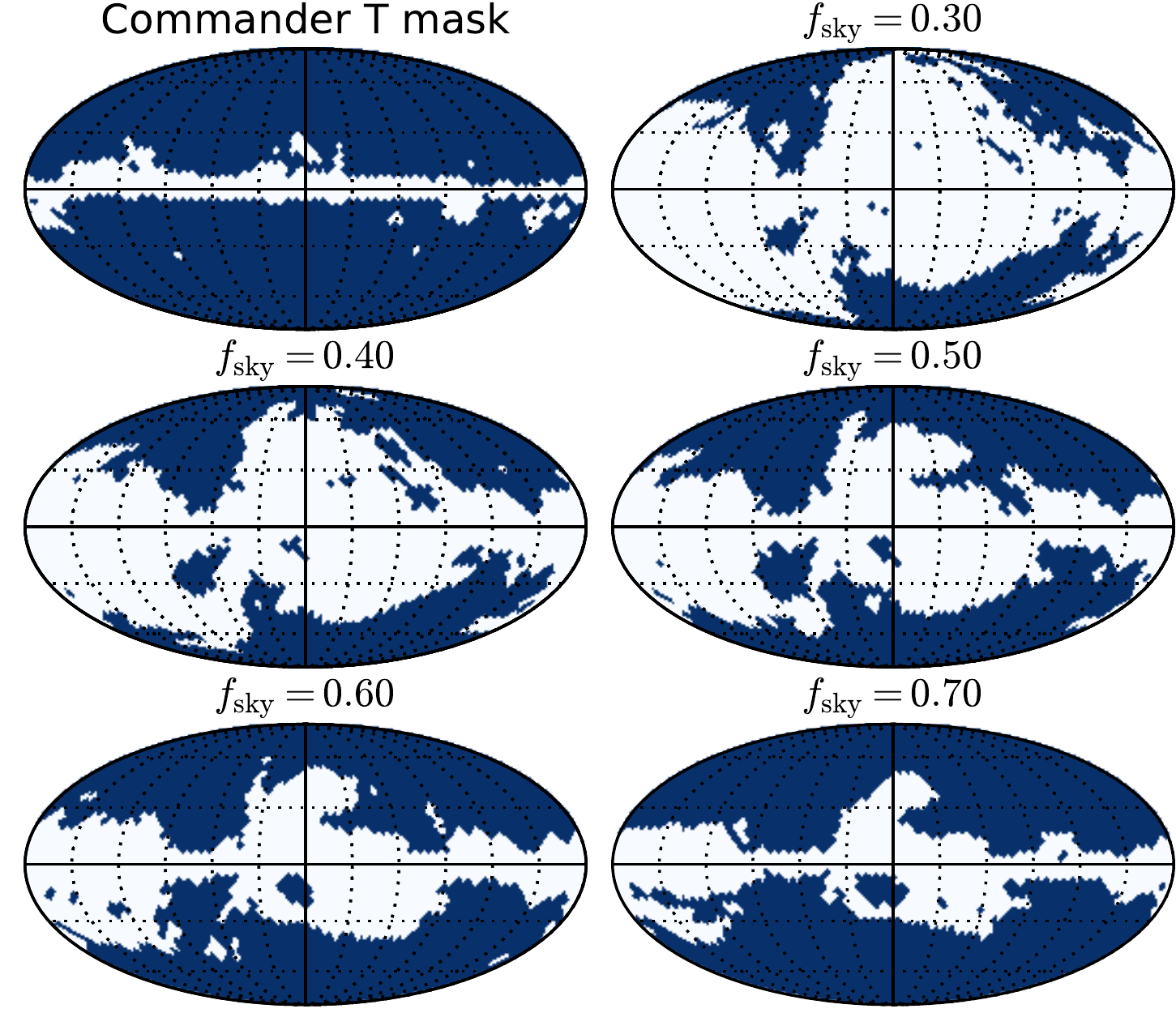}
\noindent\rule{0.475\textwidth}{0.8pt}
\includegraphics[width=0.475\textwidth]{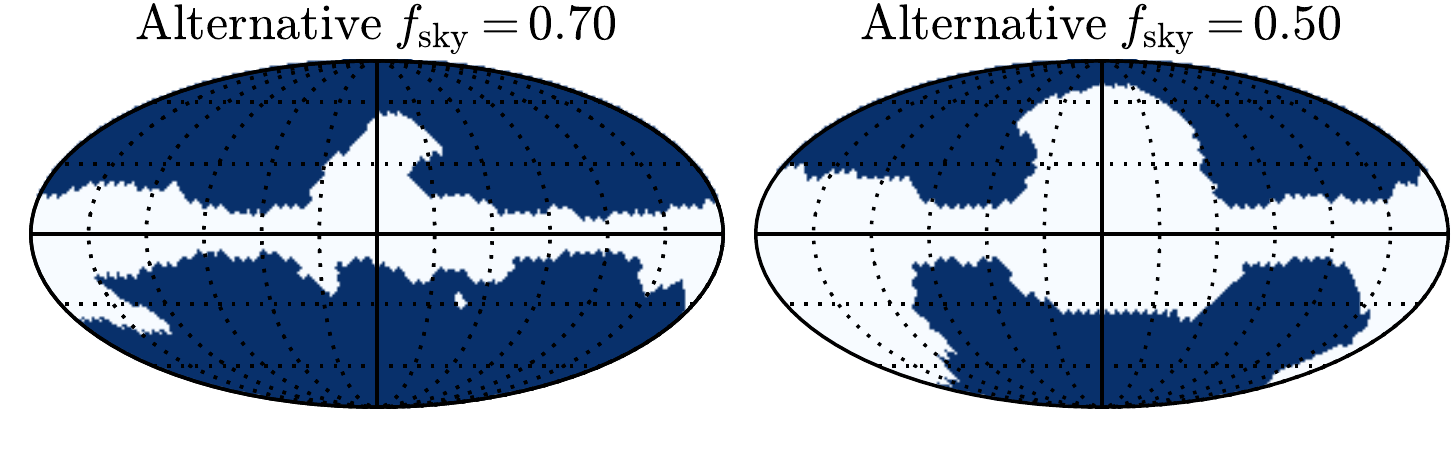}
\noindent\rule{0.475\textwidth}{0.8pt}
\includegraphics[width=0.475\textwidth]{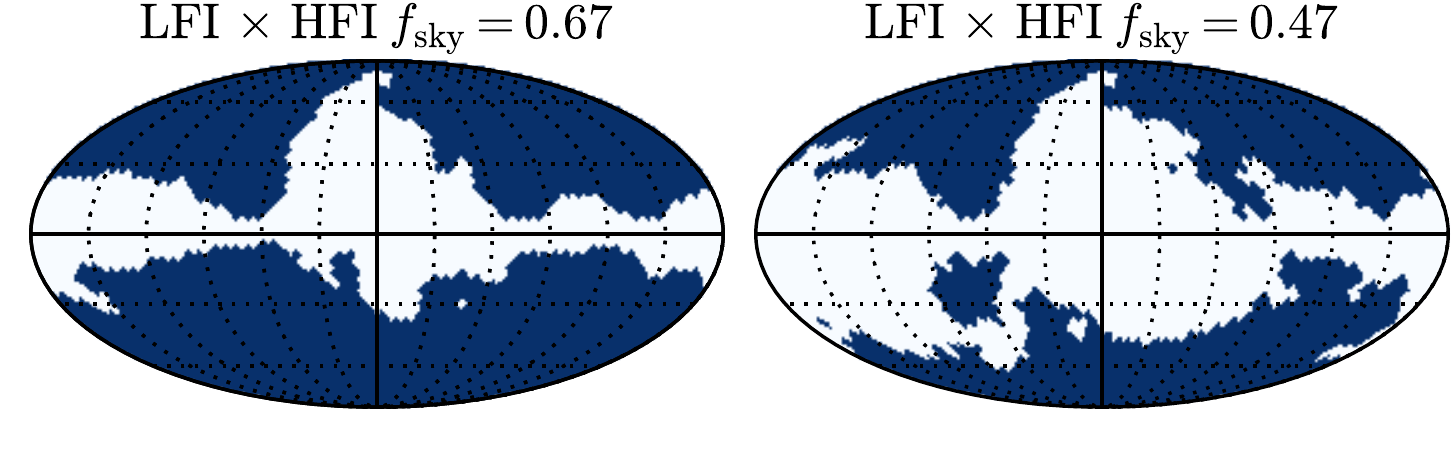}
\caption{Different masks used for the low-$\ell$ HFI likelihood. In the top panels we show the {\tt Commander} temperature confidence mask, and polarization masks obtained thresholding the 5\deg-smoothed 353-GHz polarization intensity map. In the middle panels we show the masks produced with the alternative algorithm described in Sect.~\ref{subsubsec:masks}. In the bottom panels we show the masks used for the ${\rm LFI}\times{\rm HFI}$ likelihood estimation (Sect.~\ref{subsubsec:lfi_cross_hfi}).}
\label{fig:plot_masks}
\end{center}
\end{figure}

The mask applied in the cleaning process retains a fraction $f_{\rm sky}\,{=}\,0.70$ of the entire sky and is obtained by thresholding the polarization intensity at 353\GHz, smoothed with a Gaussian beam with full-width-half-maximum of 5\deg. With the same procedure we produce other masks, used for power spectra estimation, with decreasing sky fraction, down to $f_{\rm sky}\,{=}\,0.30$. Those masks are shown in the upper panels of Fig.~\ref{fig:plot_masks}.

In the foreground-cleaning procedure presented in this section we fit both synchrotron and dust at 100\GHz, but only dust at 143\GHz, mainly to avoid auto-correlation between possible residual systematics present in the 30-GHz data \citepalias[see appendix~A of][]{planck2016-l02}. Nevertheless, we verified that the inclusion of the synchrotron cleaning at 143\GHz\ also has negligible impact on the recovered power spectra (see Sect.~\ref{subsubsec:sync_cleaning_test} for details).

The same downgrading and cleaning procedure is performed on 300 CMB + foregrounds + noise + systematics end-to-end FFP10 simulations for 100, 143, and 353\GHz, together with 300 CMB + foregrounds + noise FFP10 simulations for 30\GHz. The CMB sky used in all the FFP10 simulations is always the same CMB realization, called the ``Fiducial CMB'' (see \citetalias{planck2016-l02} and \citetalias{planck2016-l03} for details). After the cleaning procedure the Fiducial CMB is subtracted from the 300 cleaned simulations, leaving only noise + systematics + foreground residual maps. By always using the same CMB realization in cleaning simulations we implicitly neglect the accidental correlations between the CMB and the other components. In order to quantify the impact of this assumption we perform the same cleaning procedure substituting the Fiducial CMB map with a Monte Carlo suite of CMB realizations. By analysing the empirical distribution of the scalings we find consistent peak values and only a 1\,\% larger width, verifying that this assumption has a negligible impact on our results. 

Due to the non-perfect match between the foreground model used in FFP10 and the data, the scalings measured on simulations are not fully compatible with the values measured on data. We have verified that switching between the two delivered foreground models \citep{planck2016-ES} has only a marginal impact on our final results. This mismatch represents a limitation of our approach that is not correctable with the current version of data and simulations.

\begin{table}[htbp!]
\begingroup
\caption{Template scalings measured on data. The uncertainties are obtained from $\chi^2$ minimization. The errors in square brackets are computed from the 300 FFP10 simulations.}
\label{tab:scalings}
\nointerlineskip
\vskip -3mm
\setbox\tablebox=\vbox{
   \newdimen\digitwidth
   \setbox0=\hbox{\rm 0}
   \digitwidth=\wd0
   \catcode`*=\active
   \def*{\kern\digitwidth}
   \newdimen\signwidth
   \setbox0=\hbox{+}
   \signwidth=\wd0
   \catcode`!=\active
   \def!{\kern\signwidth}
\halign{\hbox to 1.0in{#\leaderfil}\tabskip=1em&
  \hfil#\hfil\tabskip=2em&
  \hfil#\hfil\tabskip=0pt\cr
\noalign{\doubleline}
\noalign{\vskip -2pt}
\omit\hfil Channel [GHz]\hfil& $\alpha\times10^{2}$& $\beta\times10^{2}$\cr
\noalign{\vskip 3pt\hrule\vskip 5pt}
100& $1.83\pm0.12\, [0.18]$& $1.950\pm0.014\,[0.015]$\cr
143& \dots& $4.078\pm0.011\,[0.013]$\cr
\noalign{\vskip 5pt\hrule\vskip 3pt}}}
\endPlancktable
\endgroup
\end{table}

In Table~\ref{tab:scalings}{,} we report the amplitudes of the templates measured on the data.  The uncertainties shown are computed in the $\chi^2$ minimization and in square parentheses we show the dispersion of the scalings of the 300 simulations. The MC-based errors are slightly larger than the ones obtained from the $\chi^2$, as one might expect given that the FFP8 covariance matrices used in the $\chi^2$ computation do not contain any variance from systematics.

Following \citetalias{planck2014-a13}, the noise-covariance matrices for the cleaned data sets defined in Eq.~\eqref{eq:cleaned_dataset} are taken to be

\begin{equation}
\tens{\tilde N}_{100,143} = \frac{\tens{N}_{100,143}+\alpha^2\tens{N}_{30}+\beta^2\tens{N}_{353}+\sigma_\alpha^2 \vec{m}_{30} \vec{m}_{30}^{\sf T}+\sigma_\beta^2 \vec{m}_{353} \vec{m}_{353}^{\sf T}}{\left(1-\alpha-\beta\right)^2},\label{eq:ncvm_final}
\end{equation}

\noindent where $\alpha$, $\beta$, $\sigma_\alpha$, and $\sigma_\beta$ are the mean values and 1\,$\sigma$ (in square parenthesis) errors shown in Table~\ref{tab:scalings}.

\subsubsection{Power spectra}\label{subsec:spectra}

In order to estimate the cross-spectra between maps we use a quasi-QML estimator as in \citet{planck2014-a10}. We used the foreground-cleaned 100- and 143-GHz polarization maps derived with the procedure described in the previous section. As a temperature map we used the \commander\ solution smoothed with a Gaussian window function of 440\arcm, degraded to $\Nside\,{=}\,16$ and with a regularization white noise with $2\,\mu{\rm K}$ rms per pixel added to the map, assuming no component-separation residuals. The signal-covariance matrix of the QML estimator is computed assuming the \Planck~\TTTEEE+{\tt SIMlow} best fit of \citet{planck2014-a10}. Unless stated otherwise, the mask used for polarization is the 50\,\% mask shown in Fig.~\ref{fig:plot_masks}, and for temperature we use the \commander\ mask (based on the $\chi^2$ of the multi-component fit), shown in the top left panel of Fig.~\ref{fig:plot_masks}.

For computing the QML spectra, we used FFP8 noise-covariance matrices as described in Eq.~\eqref{eq:ncvm_final}. This is clearly suboptimal, since the FFP8 matrices ignore relevant systematic effects; however, this will affect only the variance of the estimator, without biasing the power spectrum estimates \citep{planck2014-a10}.

\begin{figure}[htbp!]
\begin{center}
\includegraphics[width=0.475\textwidth]{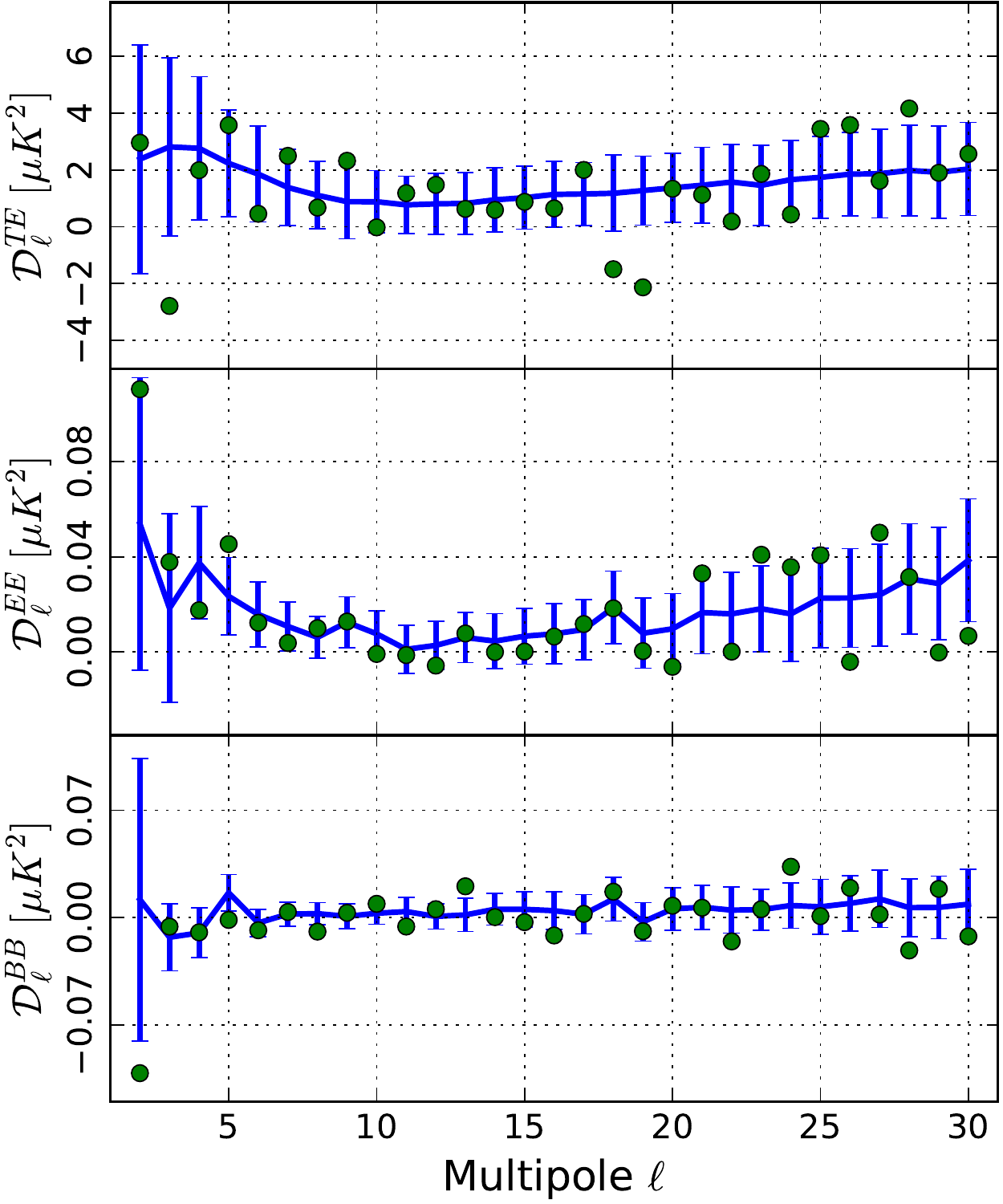}
\caption{QML power spectra for $100\times143$. The green points represent data, while the blue bars show the averages and 1\,$\sigma$ dispersions of 300 FFP10 signal + noise simulations.}
\label{fig:plot_spectra_100x143}
\end{center}
\end{figure}

In Fig.~\ref{fig:plot_spectra_100x143} we show cross-spectra between 100 and 143\GHz. The $TE$ spectrum shown is the average of $TE$ and $ET$ spectra. The green points represent the data, while the blue bars are the averages and 1\,$\sigma$ dispersions of the 300 FFP10-cleaned simulations combined with 300 CMB signal simulations.\footnote{The CMB fiducial model, used for the generation of the signal MC, assumed $\tau\,{=}\,0.05$, $\ln(10^{10}A_{\rm s})\,{=}\,3.054$, and $r\,{=}\,0$.} The distribution of simulations encompasses all data points well; however, the scatter of the data is actually smaller than that suggested by simulations. This is probably caused by the large variance of the analogue-to-digital convertor nonlinearities (ADCNL) fed into the simulations, due to an overestimation of the ADCNL effect at 100 and 143\GHz\ (see section~5.13 of \citetalias{planck2016-l03} for more details, in particular the second panel of figure~49 there comparing data with simulations). The autocorrelated noise appearing in the cross-spectrum, proportional to $\beta_{100}\times\beta_{143}\times N_{353}$, is completely negligible compared to the noise of 100 and 143\GHz.

In Table~\ref{tab:pte_100x143}, assuming the empirical distribution of the FFP10 simulations, we report, $\ell$-by-$\ell$, the percentage of simulations that have an absolute value of the difference between $\mathcal{D}_\ell$ and the mean of the distribution larger than that of the data (i.e., the PTE). Despite the presence of some outliers (such as $\ell\,{=}\,19$ in $TE$), the overall agreement between data and simulations is excellent. In Table~\ref{tab:pte_100x143_total} we report the cumulative PTEs for $\ell_{\rm max}\,{=}\,10$ and $\ell_{\rm max}\,{=}\,29$ for the cross-spectra between 100 and 143\GHz.

We perform similar analyses for the following data cuts: half-missions (HM); detector-sets (DS); and odd-even rings (OE). We independently clean HM, DS, and OE maps using 30-GHz and 353-GHz PSB-only full-mission maps and we compute all the possible cross-spectra.  We use the notation ``HM1'' to refer to the first
half of the half-mission data, etc.

\begin{table}[htbp!]
\begingroup
\caption{Percentage of simulations that have absolute difference between $\mathcal{D}_\ell$ and the mean of the empirical distribution larger than the data. }
\label{tab:pte_100x143}
\nointerlineskip
\vskip -3mm
\setbox\tablebox=\vbox{
   \newdimen\digitwidth
   \setbox0=\hbox{\rm 0}
   \digitwidth=\wd0
   \catcode`*=\active
   \def*{\kern\digitwidth}
   \newdimen\signwidth
   \setbox0=\hbox{+}
   \signwidth=\wd0
   \catcode`!=\active
   \def!{\kern\signwidth}
\halign{\hbox to 0.5in{#\leaderfil}\tabskip=1em&
  \hfil#\hfil\tabskip=1.5em&
  \hfil#\hfil\tabskip=1.5em&
  \hfil#\hfil\tabskip=1.5em&
  \hfil#\hfil\tabskip=1.5em&
  \hfil#\hfil\tabskip=0pt\cr
\noalign{\doubleline}
\noalign{\vskip -2pt}
\omit\hfil $\ell$\hfil& $EE$& $BB$& $TE$& $TB$& $EB$\cr
\noalign{\vskip 3pt\hrule\vskip 5pt}
*2&  29.6& 20.2& 87.5& 88.1& 27.3\cr
*3&  61.4& 69.6& *6.8& 79.0& 45.7\cr
*4&  37.5& 97.8& 74.9& 31.2& 42.0\cr
*5&  16.7& 13.2& 47.9& 65.3& *8.0\cr
*6&  79.4& 61.0& 39.2& 95.7& 92.2\cr
*7&  47.9& 85.0& 39.6& 34.4& 37.5\cr
*8&  65.3& 10.4& 70.7& 32.9& *3.7\cr
*9&  97.8& 76.2& 26.6& 55.4& 91.5\cr
10& 38.6& 40.1& 41.0& *7.9& 13.6\cr
11& 81.3& 29.1& 67.0& 15.6& 81.2\cr
12& 38.0& 56.4& 52.4& 95.1& 64.8\cr
13& 86.6& *8.3& 85.5& 94.3& 51.9\cr
14& 69.3& 62.1& 75.0& *7.1& 98.0\cr
15& 54.4& 42.2& 89.2& 14.7& 10.0\cr
16& 92.6& 15.3& 66.4& 74.3& 85.3\cr
17& 85.3& 98.3& 44.3& 92.0& 52.9\cr
18& 97.9& 75.3& *4.6& 12.8& 47.3\cr
19& 59.5& 61.5& *0.5& 83.7& 16.4\cr
20& 26.3& 87.2& 97.9& 57.1& 14.3\cr
21& 34.4& 98.4& 79.8& 54.6& 63.1\cr
22& 36.1& 17.2& 28.7& 82.9& 72.2\cr
23& 20.2& 98.3& 76.6& 21.4& 31.1\cr
24& 32.2& *8.3& 36.5& 31.4& 32.1\cr
25& 37.9& 71.5& 23.6& 45.3& 91.5\cr
26& 19.2& 57.0& 23.5& 92.6& 14.0\cr
27& 21.8& 58.3& 87.4& 13.4& 94.1\cr
28& 97.5& 13.8& 16.9& 65.9& 84.9\cr
29& 22.0& 54.7& 99.1& 26.7& 29.6\cr
\noalign{\vskip 5pt\hrule\vskip 3pt}}}
\endPlancktable
\endgroup
\end{table}

\begin{table}[htbp!]
\begingroup
\caption{PTEs for the cross-spectra between the 100-GHz and 143-GHz full-mission maps. ``Total'' represents the co-addition of the $TE$, $EE$, and $BB$ spectrum results.}
\label{tab:pte_100x143_total}
\nointerlineskip
\vskip -3mm
\setbox\tablebox=\vbox{
   \newdimen\digitwidth
   \setbox0=\hbox{\rm 0}
   \digitwidth=\wd0
   \catcode`*=\active
   \def*{\kern\digitwidth}
   \newdimen\signwidth
   \setbox0=\hbox{+}
   \signwidth=\wd0
   \catcode`!=\active
   \def!{\kern\signwidth}
\halign{\hbox to 0.75in{#\leaderfil}\tabskip=1em&
  \hfil#\hfil\tabskip=2em&
  \hfil#\hfil\tabskip=0pt\cr
\noalign{\doubleline}
\omit\hfil Spectrum\hfil& $\ell_{\rm max}\,{=}\,10$& $\ell_{\rm max}\,{=}\,29$\cr
\noalign{\vskip 3pt\hrule\vskip 5pt}
$TE$&  $61.3$& $50.6$\cr
$EE$&  $82.7$& $93.9$\cr
$BB$&  $56.9$& $73.7$\cr
Total& $83.2$& $90.4$\cr
\noalign{\vskip 5pt\hrule\vskip 3pt}}}
\endPlancktable
\endgroup
\end{table}

In Table~\ref{tab:pte_crosses} we report the PTEs for $\ell_{\rm max}\,{=}\,10$ and $\ell_{\rm max}\,{=}\,29$ for all the cross-spectra computed from the different data cuts. No major discrepancy is found in these tests except for a 2\,$\sigma$ inconsistency for the two cross-spectra involving 143HM1, for $\ell<10$; however, this is not seen in the full integrated channel.
Finally we analysed the null maps using HM, DS, and OE, making difference maps within frequencies and then crossing them, e.g., $(100{\rm HM}1-100{\rm HM}2) \times (143{\rm HM}1-143{\rm HM}2)$. We compute the power spectra assuming zero signal power and we compare the data points with the probability distribution of simulated noise and systematics. In Table~\ref{tab:pte_null_test} we report PTEs for those null tests. 
The nulls of the $EE$ and $BB$ spectra are in good agreement with the probability distribution of the 300 simulations, with only the $EE$ from the HM marginally problematic, exceeding 2\,$\sigma$. On the other hand we fail the $TE$ null test for $\ell < 30$, which is a possible indication of residual systematics or foregrounds in polarization correlated with residual foregrounds in temperature. The latter are not included by our simulation pipeline, as we assume perfect component separation in \commander\ for HM and OE; for DS in temperature we use 100- and 143-GHz maps. As conservative choice, in order to avoid any impact on our cosmological analysis, we thus implement a likelihood based on only the $EE$ and $BB$ spectra.

\begin{figure}[htbp!]
\begin{center}
\includegraphics[width=0.475\textwidth]{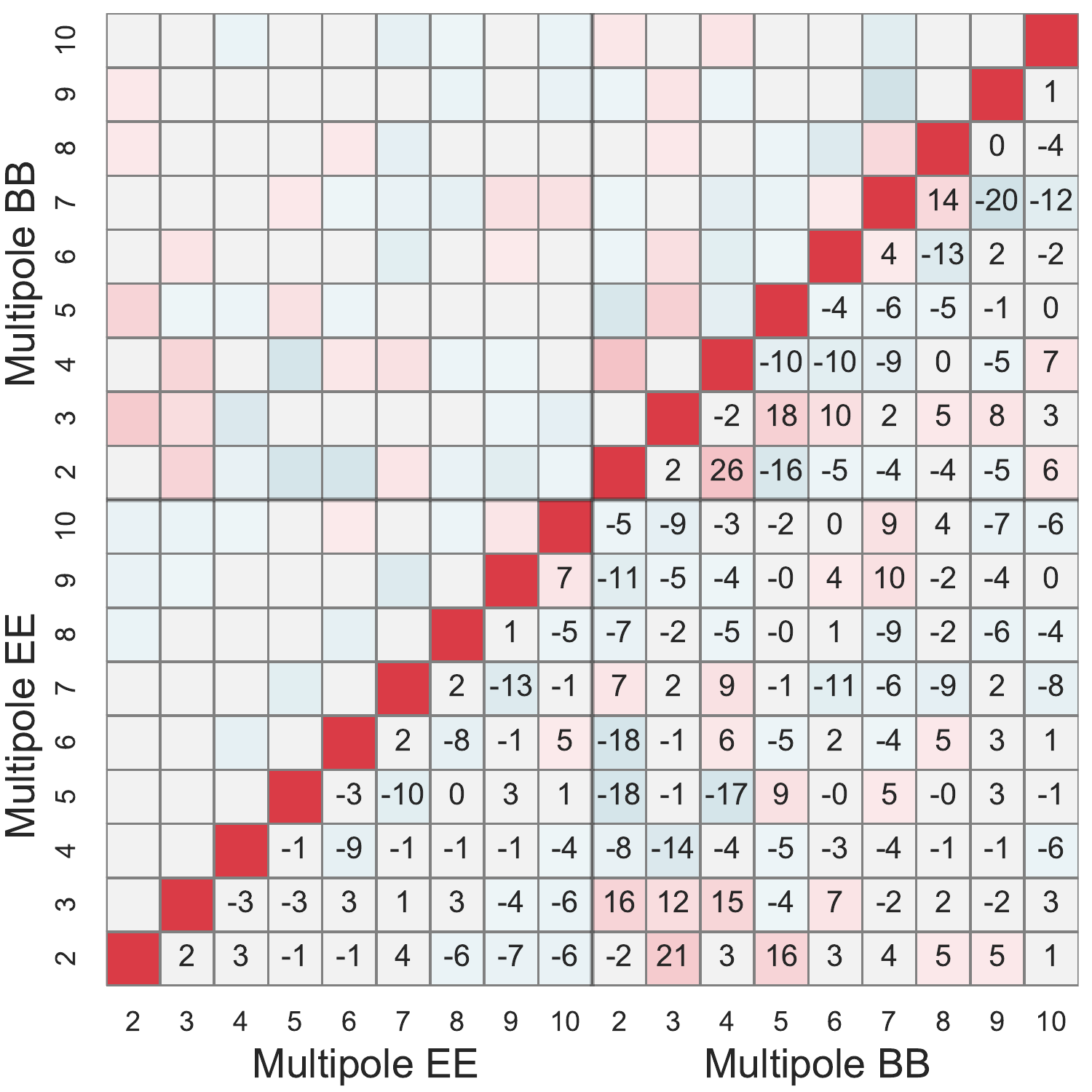}
\caption{Empirical correlation matrix (as percentages) for $EE$ and $BB$ power spectra estimated from 300 signal + noise + systematics simulations.}
\label{fig:plot_covariance_EEBB}
\end{center}
\end{figure}

Since we use a QML estimator we anticipate that the correlation between $C_\ell$ for different $\ell$ is small. We use the Monte Carlo of signal + noise + systematics described above to verify that this is the case; 
Fig.~\ref{fig:plot_covariance_EEBB} shows the $EE$-$BB$ covariance matrix for $\ell \le10$ estimated analysing the 300 simulations.

\begin{table}[htbp!]
\begingroup
\caption{PTEs for all the 100-GHz and 143-GHz
cross-spectra computed between splits for $\ell_{\rm max}\,{=}\,10$ and (shown in parentheses) for $\ell_{\rm max}\,{=}\,29$. ``Total'' represents the combined value of $TE$, $EE$, and $BB$ spectra.}
\label{tab:pte_crosses}
\nointerlineskip
\vskip -3mm
\setbox\tablebox=\vbox{
   \newdimen\digitwidth
   \setbox0=\hbox{\rm 0}
   \digitwidth=\wd0
   \catcode`*=\active
   \def*{\kern\digitwidth}
   \newdimen\signwidth
   \setbox0=\hbox{+}
   \signwidth=\wd0
   \catcode`!=\active
   \def!{\kern\signwidth}
\halign{\hbox to 0.6in{#\leaderfil}\tabskip=0.5em&
  \hfil#\hfil\tabskip=0.5em&
  \hfil#\hfil\tabskip=0.5em&
  \hfil#\hfil\tabskip=0.5em&
  \hfil#\hfil\tabskip=0pt\cr
\noalign{\doubleline}
\omit\hfil Spectrum\hfil& $\ell_{\rm max}\,{=}\,10\,(29)$& $\ell_{\rm max}\,{=}\,10\,(29)$& $\ell_{\rm max}\,{=}\,10\,(29)$& $\ell_{\rm max}\,{=}\,10\,(29)$\cr
\noalign{\vskip 3pt\hrule\vskip 5pt}
\omit& $100 \;\;\;\;\;\;   143$& $100 \;\;\;\;\;\;  143$&$100 \;\;\;\;\;\;  143$& $100 \;\;\;\;\;\;   143$\cr
\omit& ${\rm HM}1\times{\rm HM}1$& ${\rm HM}2\times{\rm HM}2$&
       ${\rm HM}1\times{\rm HM}2$& ${\rm HM}2\times{\rm HM}1$\cr
\noalign{\vskip 3pt\hrule\vskip 5pt}
$TE$&  $60.5\,(46.1)$& $22.2\,(*8.9)$& $41.7\,(26.3)$& $18.1\,(58.4)$\cr
$EE$&  $*4.9\,(32.8)$& $49.5\,(60.0)$& $95.6\,(92.7)$& $*6.2\,(19.9)$\cr
$BB$&  $75.2\,(43.3)$& $58.8\,(76.2)$& $11.8\,(23.7)$& $86.4\,(93.0)$\cr
Total& $30.5\,(37.7)$& $42.5\,(40.7)$& $48.9\,(50.4)$& $18.1\,(67.3)$\cr
\noalign{\vskip 3pt\hrule\vskip 5pt}
\omit& $100 \;\;\;\;\;   143$& $100 \;\;\;\;\;  143$&$100 \;\;\;\;\;  143$& $100 \;\;\;\;\;   143$\cr
\omit& ${\rm DS}1\times{\rm DS}1$& ${\rm DS}2\times{\rm DS}2$&
       ${\rm DS}1\times{\rm DS}2$& ${\rm DS}2\times{\rm DS}1$\cr
\noalign{\vskip 3pt\hrule\vskip 5pt}
$TE$&  $19.9\,(42.5)$& $78.6\,(35.4)$& $40.2\,(*9.7)$& $49.8\,(20.5)$\cr
$EE$&  $99.7\,(70.3)$& $69.1\,(67.5)$& $90.3\,(75.6)$& $76.2\,(98.0)$\cr
$BB$&  $90.8\,(85.3)$& $71.8\,(62.1)$& $26.6\,(51.6)$& $15.1\,(32.6)$\cr
Total& $90.9\,(80.7)$& $89.6\,(62.0)$& $59.3\,(37.1)$& $44.2\,(61.9)$\cr
\noalign{\vskip 3pt\hrule\vskip 5pt}
\omit& $100 \;\;\;\;\;   143$& $100 \;\;\;\;\;  143$&$100 \;\;\;\;\;  143$& $100 \;\;\;\;\;   143$\cr
\omit& ${\rm OE}1\times{\rm OE}1$& ${\rm OE}2\times{\rm OE}2$&
       ${\rm OE}1\times{\rm OE}2$& ${\rm OE}2\times{\rm OE}1$\cr
\noalign{\vskip 3pt\hrule\vskip 5pt}
$TE$&  $48.0\,(88.2)$& $56.1\,(32.1)$& $88.5\,(65.5)$& $26.8\,(32.4)$\cr
$EE$&  $98.2\,(98.2)$& $34.1\,(55.5)$& $98.0\,(84.6)$& $60.9\,(61.9)$\cr
$BB$&  $17.8\,(*3.6)$& $90.0\,(29.7)$& $58.2\,(85.5)$& $51.9\,(53.8)$\cr
Total& $64.8\,(79.3)$& $73.7\,(34.2)$& $97.6\,(93.5)$& $48.9\,(52.1)$\cr
\noalign{\vskip 3pt\hrule\vskip 3pt}}}
\endPlancktable
\endgroup
\end{table}

\begin{table}[htbp!]
\begingroup
\caption{PTEs for half-mission (HM), detector-set (DS), and odd-even (OE) null tests. The spectra are obtained by crossing the difference maps.}
\label{tab:pte_null_test}
\nointerlineskip
\vskip -3mm
\setbox\tablebox=\vbox{
   \newdimen\digitwidth
   \setbox0=\hbox{\rm 0}
   \digitwidth=\wd0
   \catcode`*=\active
   \def*{\kern\digitwidth}
   \newdimen\signwidth
   \setbox0=\hbox{+}
   \signwidth=\wd0
   \catcode`!=\active
   \def!{\kern\signwidth}
\halign{\hbox to 0.6in{#\leaderfil}\tabskip=0.5em&
  \hfil#\hfil\tabskip=2em&
  \hfil#\hfil\tabskip=2em&
  \hfil#\hfil\tabskip=0pt\cr
\noalign{\doubleline}
\omit\hfil Spectrum\hfil& $\ell_{\rm max}\,{=}\,10\,(29)$& $\ell_{\rm max}\,{=}\,10\,(29)$& $\ell_{\rm max}=10\,(29)$\cr
\noalign{\vskip 3pt\hrule\vskip 5pt}
\omit& HM& DS& OE\cr
\noalign{\vskip 3pt\hrule\vskip 5pt}
$TE$& $*8.0\,(*0.3)$& $73.0\,(*1.6)$& $12.9\,(*0.1)$\cr
$EE$& $*7.0\,(17.7)$& $83.1\,(69.5)$& $96.7\,(73.9)$\cr
$BB$& $71.6\,(60.0)$& $85.1\,(96.0)$& $91.1\,(83.2)$\cr
\noalign{\vskip 3pt\hrule\vskip 3pt}}}
\endPlancktable
\endgroup
\end{table}

\begin{figure}[htbp!]
\begin{center}
\includegraphics[width=0.475\textwidth]{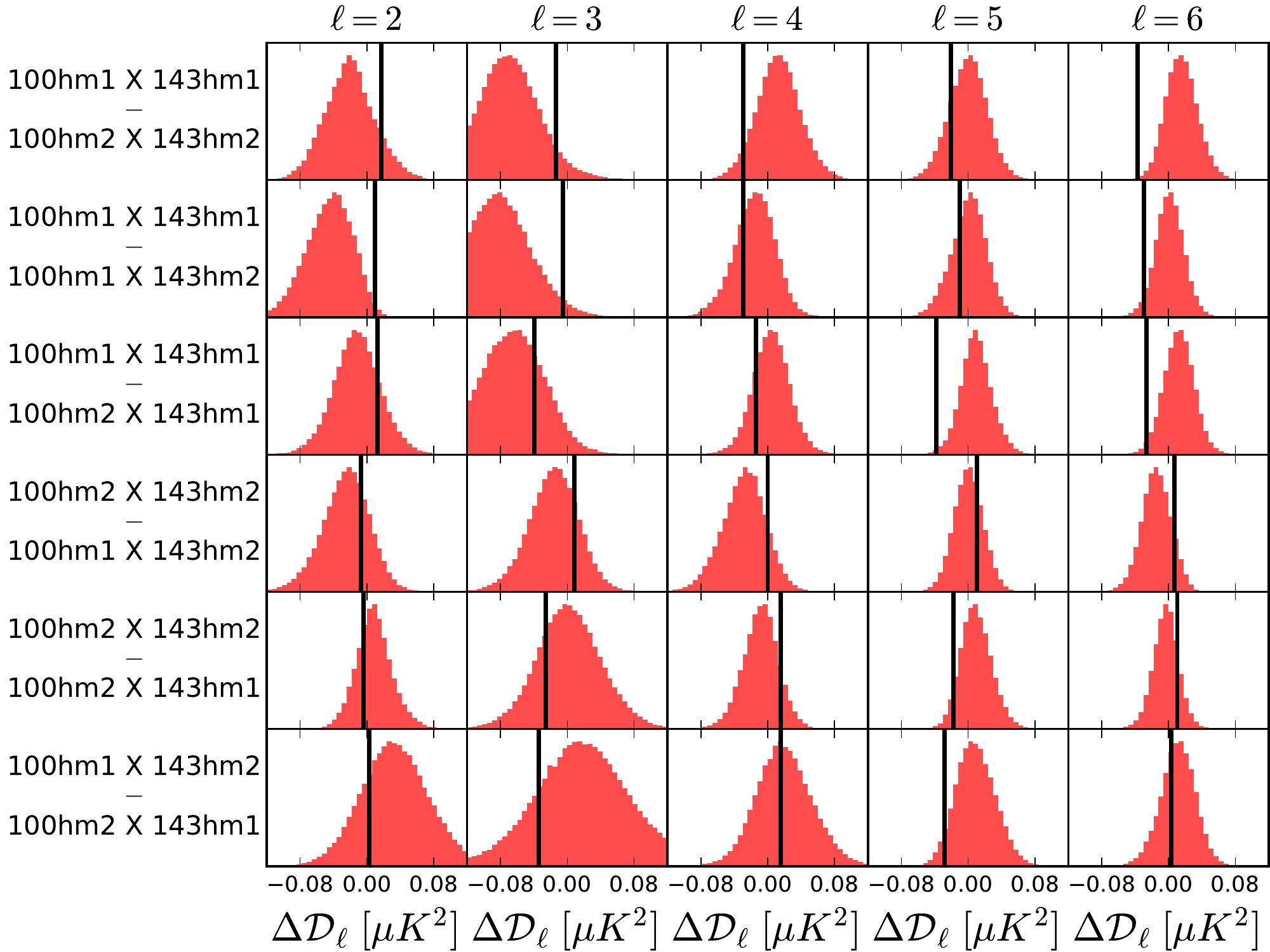}
\includegraphics[width=0.475\textwidth]{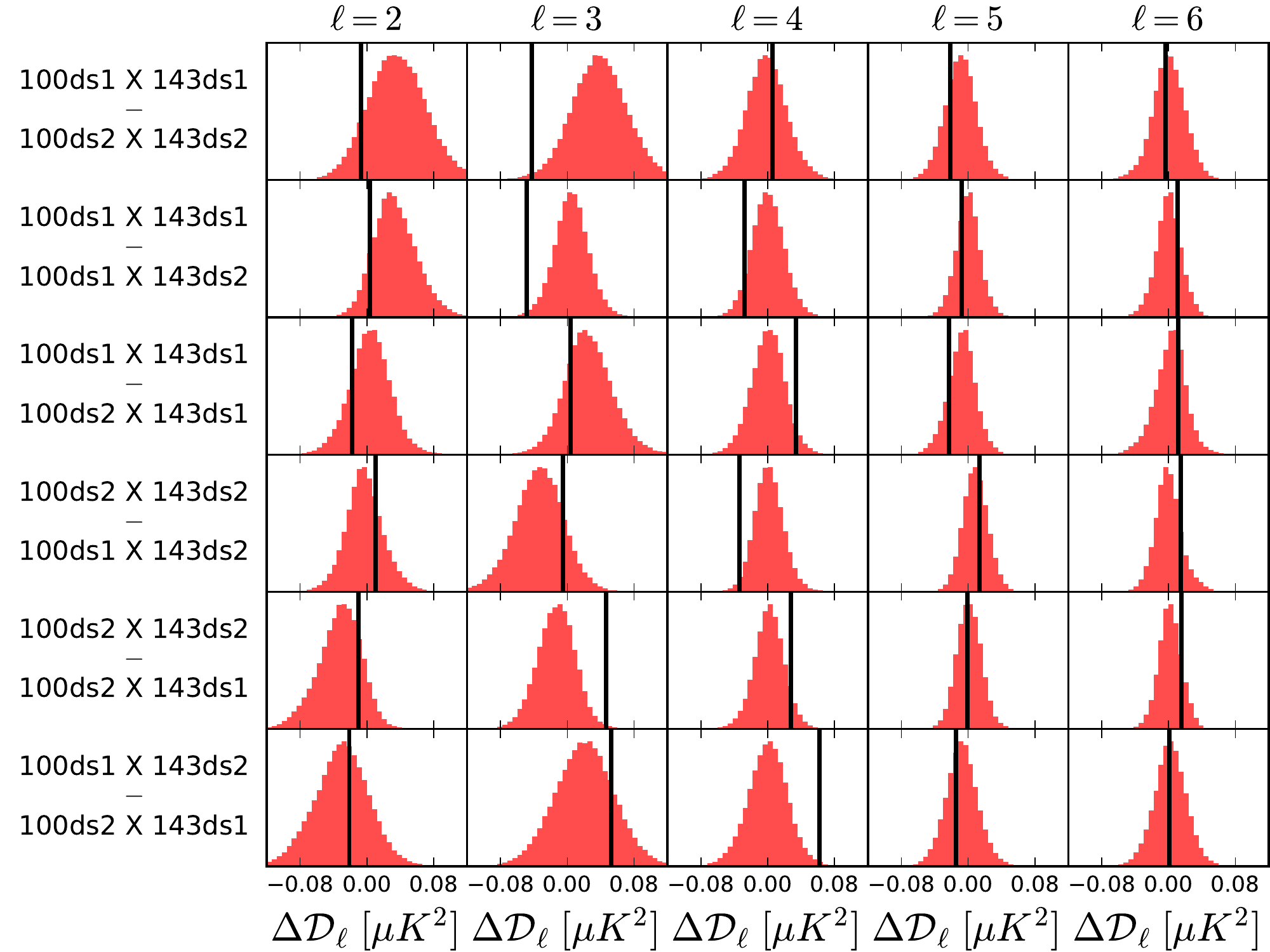}
\includegraphics[width=0.475\textwidth]{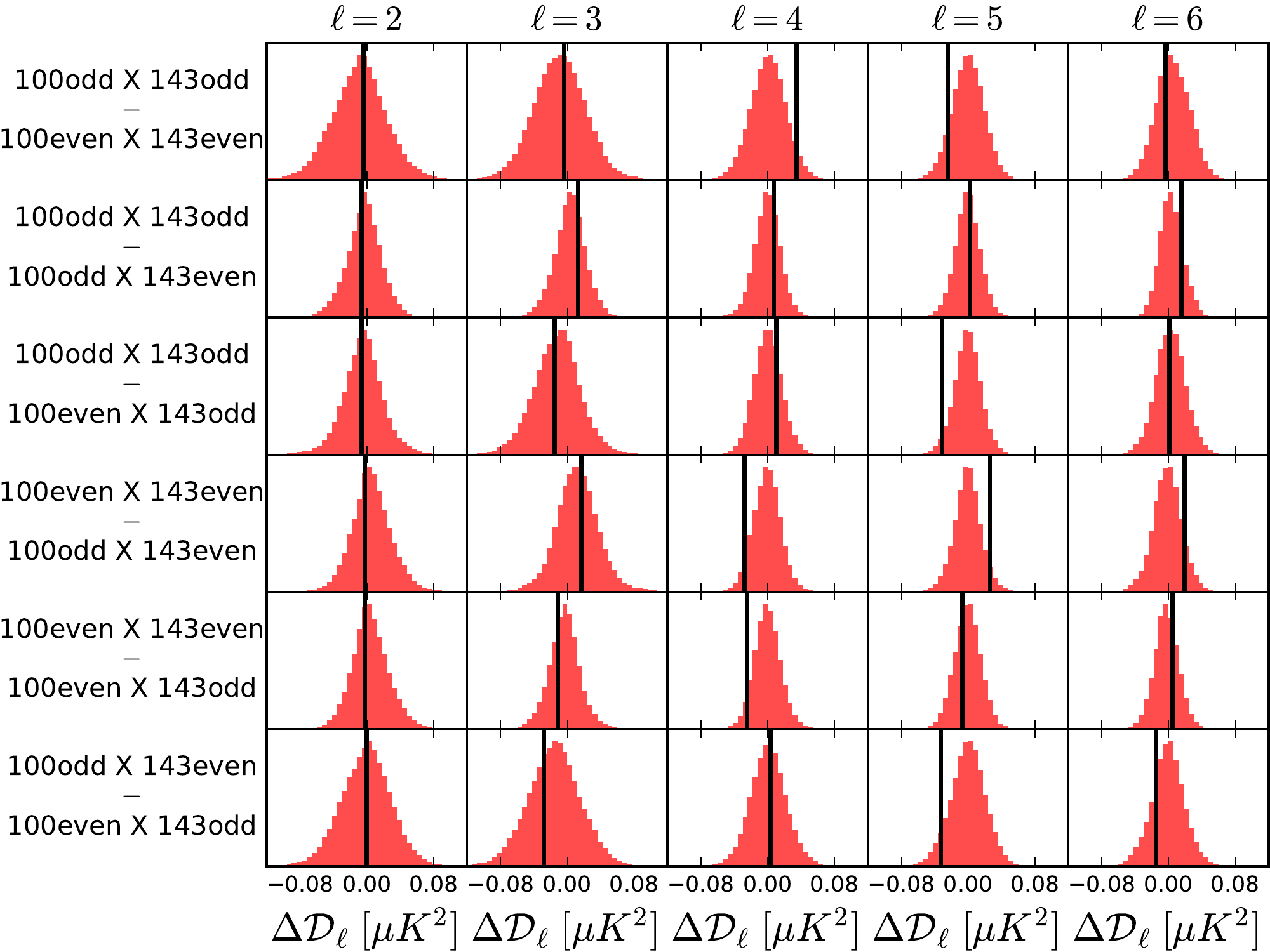}
\caption{Difference, $\Delta \mathcal{D}_\ell$, between spectra for different data splits (vertical line), compared with the distribution of the difference in FFP10 simulations (histograms).}
\label{fig:plot_diff_spectra_1d_splits}
\end{center}
\end{figure}

Figure~\ref{fig:plot_diff_spectra_1d_splits} shows the difference between the power spectra, $\Delta \mathcal{D}_\ell$, for the first five multipoles of different analysed data splits compared with the histograms of FFP10 simulations. The empirical distributions encompass the data quite well for almost all the cases shown. Some outliers are present at the 2\,$\sigma$ level, in particular for the HMs. Moreover for DSs and HMs the distributions are often not centred on $\Delta \mathcal{D}_\ell=0$ and are skewed in some cases. This behaviour is related to large signals distorted by the second-order ADCNL effect, not corrected by the gain variation in the mapmaking, which only corrects for the first-order approximation (see section~5.13 of \citetalias{planck2016-l03} for more details). The odd and even maps share the same scanning strategy, unlike the half-mission maps and use the same detectors, unlike the detector-set splits. Thus they show similar residual systematics, associated with the second-order ADCNL effect and consequently provide a more consistent null test.

\subsubsection{Likelihood function}\label{subsubsec:likelihood}

Two of the primary uses for the low-$\ell$ polarization likelihood are the investigations of $\tau$ and of $r$ using $EE$ and $BB$ spectra. As in \citet{planck2014-a10} we implement a simulation-based likelihood applied to the QML estimation of $100\times143$. 

The likelihood $\mathcal{L}(\vec{C}|\vec{C}^{\mathrm{data}})$ of a given model power spectrum vector $\vec{C}$, given the observed one $\vec{C}^{\mathrm{data}}$, is given by the conditional probability of observing the data realization of the model $\vec{C}$, $P(\vec{C}^{\mathrm{data}}|\vec{C})$. We will approximate the likelihood by counting the occurrences of our data being suitably close to a simulation produced in a Monte Carlo on a grid of models. 

To illustrate the methodology, consider estimating the variance $\sigma^2$ of an underlying Gaussian process that produces a vector $\vec{X}$ of $n$ independent samples (this will be analogous to the measurement of a single cross-power spectrum multipole from the spherical harmonic transforms of two different maps of the sky).
 Those samples are measured by two noisy, potentially biased, and weakly-correlated instruments A and B that can, however, be reasonably well simulated.  The empirical covariance between the two measurements, $\widehat{\sigma^2}_\mathrm{AB} \equiv X_\mathrm{A}^{\sf T} X_\mathrm{B}^{\vphantom{\sf T}} /n$ provides an estimate of $\sigma^2$, only biased by the noise cross-covariance between the two instruments.  Given only weak correlations between the instruments this bias should be small, certainly much smaller than that for either of the individual variances.

One may now build a simulation-based likelihood estimation for our model parameter $\sigma^2$. Assuming that one can generate realizations of the noise vectors $\vec{N}_{\rm A}$ and $\vec{N}_{\rm B}$, the entire measurement process is fully simulatable.  Given some prior on the model, one can jointly sample $\widehat{\sigma^2}_\mathrm{AB}$ and $\sigma^2$ and thus build an interpolation $\tilde{p}(\widehat{\sigma^2}_\mathrm{AB},\sigma^2)$ of the joint distribution. Setting the first variable of the interpolation $\tilde{p}$ to the measured $\widehat{\sigma^2}_\mathrm{AB}$ and dividing through by the model prior, one thus obtains an approximation to the likelihood $\mathcal{L}(\sigma^2|\widehat{\sigma^2}_\mathrm{AB}^\mathrm{data})$.  The use of only the cross-spectrum is hoped to increase the robustness of this likelihood to inaccuracies in the noise modelling. 

The problem of estimating the polarization likelihood is extremely similar; however, instead of a single variance, both theory and data become multi-dimensional. We  assume that, at the level of precision of our approximation, we can ignore correlations between all power spectrum elements and simplify the problem to the determination of $n_{\ell_\mathrm{max}}$ individual power spectrum mode likelihoods:
\begin{equation}
\mathcal{L}(\vec{C}|\vec{C}^{\mathrm{data}})=\prod_{\ell=2}^{\ell_\mathrm{max}}\mathcal{L}_\ell\left(C_\ell \big| C_\ell^{\mathrm{data}}\right), \label{eq:lowhfi:sep}
\end{equation}
for $EE$ and $BB$ independently.

The simulation procedure, however, is slightly more complicated. Even though we made the assumption that our approximated likelihood can be separated multipole by multipole, the correlations induced by the mask and the complex structure of the noise, while being in large part accounted for by the QML estimator, still prevent us from performing simulations on one multipole at a time. We must simulate full maps of the observed CMB on which to measure the QML power spectrum. Sampling the theory $C_\ell$ for each mode over a large region of the parameter space would be very inefficient, since one would typically spend too much effort exploring low probability regions. We thus only explore the region of the power spectrum around the theory $C_\ell$ of interest and produce our MCs using the \LCDM\ power spectrum from uniformly sampled cosmological parameters with uniform distributions having
\begin{equation}
10^9A_{\rm s} = 0.6\ {\rm to}\ 3.8, \;\; \tau = 0\ {\rm to}\ 0.14, \;\; r = 0\ {\rm to}\ 1, \label{eq:lowhfi:theorydist}
\end{equation}
 and all the other cosmological parameters set to their \citet{planck2014-a10} best-fit values. 
 The cosmological model distribution of Eq.~\eqref{eq:lowhfi:theorydist} translates into an exploration of the $EE$ and $BB$ spectra in the range {of}
\begin{equation}
0\,\mu{\rm K}^2 \leq \mathcal{D}_\ell^{EE} \leq 0.30\,\mu{\rm K}^2, \quad 0\,\mu{\rm K}^2 \leq \mathcal{D}_\ell^{BB} \leq 0.20\,\mu{\rm K}^2.
\end{equation}

Paralleling our methodological illustration, for each of the theory $C_\ell$, we generate CMB maps to which we add noise realizations for the 100-GHz and 143-GHz channels, allowing us to form simulated QML cross-spectra estimates. 

Generation of the noise simulations is expensive. In principle they necessitate the running of the full FFP10 pipeline on each CMB map, since the noise (including its mean) can be dependent on the sky signal (which, on the scales we are interested in, consists of CMB, dipole, and Galactic emission, with the latter two being dominant). Instead of rerunning this pipeline for each of our CMB realization, we reuse the 300 FFP10 samples, ignoring the dependence of the properties of the noise on the CMB emission. To do so, we remove the foreground and CMB from those realizations by first regressing a Galactic template from each FFP10 realization using the procedure described Sect.~\ref{subsec:cleaning} and then subtracting the known CMB input map. With this procedure, we obtain 300 FFP10 ``pure'' noise and systematics realizations that we reuse thorough the Monte Carlos. Those realizations contain realistic residuals of foreground cleaning and they are obtained not exploring the full CMB-galaxy chance correlations. Nevertheless{,} we have already shown in Sect.~\ref{subsec:cleaning} that, in the template fitting procedure, those variance terms are negligible with respect to the noise-galaxy chance correlations.

Finally, each sample of the MC is built using the following procedure. For each of the 1000 theory $C_\ell$ realizations (random parameter realizations from the distribution given in Eq.~\ref{eq:lowhfi:theorydist}), we generate 300 Gaussian CMB maps that we pair with the 300 FFP10 ``pure'' noise realizations before estimating the simulated QML cross-spectra. Even though we are reusing the noise realizations, for a given theory $C_\ell$ sample there will not be any correlation due to this limitation. This procedure provides us with 300\,000 samples of the joint distribution $p(C_\ell^{\mathrm{QML}},C_\ell^{\mathrm{theory}})$ for each multipole.

The next step of the procedure is to interpolate the joint distribution independently for each multipole. Furthermore, since we are only interested in the likelihood of a model given our specific data realization, we can avoid interpolating over the whole MC sample. In fact, we will reduce the 2D interpolation to two 1D interpolations around $C_\ell^{\mathrm{data}}$ with the following scheme. 
First, for each of sample $i$ of the 1000 $C_\ell^{\mathrm{theory}}$ realizations, we determine a low-order polynomial 1D interpolation of the log of the histogram of the $C_\ell^{\mathrm{QML}}$ over the 300 noise simulations,
\begin{equation}
f_\ell^i\left(C_\ell^{\mathrm{QML}}\right)\approx \log p\left(C_\ell^{\mathrm{QML}} \big| C_\ell^{i,\mathrm{theory}}\right).
\end{equation}
Using this approximation we can now evaluate $f_\ell^i(C_\ell^{\mathrm{data}})$, the probability of our data (i.e., the QML cross-spectrum estimated on the \Planck\ data) for each of our theory simulations. The 1000 pairs of $(C_\ell^{i,\mathrm{theory}},f_\ell^i(C_\ell^{\mathrm{data}}))$ may be viewed as a tabulated version of the log of the joint probability $\log P(C_\ell^{\mathrm{data}},C_\ell)$, up to a prior term that, if carried, would be cancelled in forming the likelihood.  This we now interpolate on $C_\ell$ with a low-order polynomial for each multipole, $g_\ell(C_\ell,C_\ell^{\mathrm{data}})$. Finally, the sum of the $g_\ell$ over $\ell$ gives us our approximate log-likelihood, up to a constant.

In summary, the likelihood approximation production can be written as the following recipe:
\begin{enumerate}
\item we clean 100\GHz\ and 143\GHz\ as described in the previous sections and we compute the QML cross-spectrum between them, $\vec{C}^{\mathrm{data}}$;
\item we clean one-by-one with the same procedure the corresponding FFP10 simulations and then remove the fiducial CMB to produce 300 ``pure'' noise FFP10 realizations;
\item we generate 1000 cosmological parameter sets $\theta^i$, under the distribution given by Eq.~\eqref{eq:lowhfi:theorydist} and compute the corresponding $C_\ell^{i,\mathrm{theory}}$;
\item we generate 300 pairs of maps for each theory realization $i$ by combining the $i$th simulated sky signal with the 300 FFP10 noise realizations at 100\GHz\ and 143\GHz, and compute the QML cross-spectrum for each of them, $C_\ell^{\mathrm{QML}}$;
\item for each theory realization $i$ and for each $\ell$, we produce $f_\ell^i(C_\ell^{\mathrm{QML}})$, a low-order polynomial interpolation of the log of the power spectrum $C_\ell^{\rm{QML}}$ histogram;
\item for each $\ell$, we use the 1000 pairs, $(C_\ell^{i,\mathrm{theory}},f_\ell^i(C_\ell^{\mathrm{data}}))$, computed by evaluating $f_\ell^i(C_\ell^{\mathrm{QML}})$ for $C_\ell^{\mathrm{QML}}=C_\ell^{\mathrm{data}}$, to produce a low-order polynomial interpolation to estimate (up to a constant) the log of the joint probability distribution, $g_\ell(C_\ell^{\mathrm{data}},C_\ell) \approx \log P(C_\ell^{\mathrm{data}},C_\ell)$;
\item we identify our approximation of $\log P(C_\ell^{\mathrm{data}},C_\ell)$ as an approximation of $\log P(C_\ell^{\mathrm{data}}|C_\ell)$ (up to a constant), and use Eq.~\eqref{eq:lowhfi:sep} to produce our approximate likelihood,
\begin{equation}
\log \mathcal{L}\left(\vec{C}^{\mathrm{theory}} \big| \vec{C}^{\mathrm{data}}\right) \approx \sum_{\ell=2}^{29} g_\ell\left(C_\ell^{\mathrm{data}},C_\ell\right) + \mathrm{constant}.
\end{equation}

\end{enumerate}

This likelihood approximation forms the basis of what is provided in the HFI low-$\ell$ component of the \Planck\ 2018 likelihood release, with a tabulated version of the $g_\ell$ functions stored in the likelihood files.

In order to validate that we can go from a uniform distribution on cosmological parameters $\theta^i$ to a $C_\ell$-based likelihood, we built a similar interpolation scheme for the likelihood $\mathcal{L}(\vec{C}^{\mathrm{data}}|\tau)$.  It was found that the constraint on $\tau$ obtained with this new likelihood perfectly overlapped the constraint obtained using our $C_\ell$-based likelihood. The corresponding test was also successfully passed for the $r$ parameter using the BB likelihood.

The low number of simulations, namely 300, can lead to inaccuracies and biases whenever the signal is small compared to the noise and systematic effects (especially for the $\tau$-based scheme mentioned above). This may occur when low values of $\tau$ are explored. Section~\ref{subsubsec:validation} describes a specific test of this effect and shows that, at the level of our validation, the $C_\ell$-based method provides an unbiased likelihood even for very low $\tau$ values.

As was done in \citet{planck2014-a10}, we combine the low-$\ell$ polarization likelihood with the low-$\ell$ temperature likelihood based on \commander\ and we estimate the cosmological parameters making use of the \cosmomc\ package \citep{2002PhRvD..66j3511L}. We sample $\ln(10^{10}A_{\rm s})$, and $\tau$ with the TT and EE likelihoods and $\ln(10^{10}A_{\rm s})$, $\tau$ and $r$ with the TT, EE, and BB likelihoods. 

We fix all parameters that are not sampled to following values: $\{\Omega_{\rm b}h^2=0.0221, \Omega_{\rm c}h^2=0.12, \theta_\ast=1.0411,n_{\rm s}=0.96\}$.\footnote{We do not fix $A_{\rm s} e^{-2 \tau}$, but leave both $A_{\rm s}$ and $\tau$ free to vary. The inflationary consistency relation is assumed for $n_{\rm t}$ \citep{planck2016-l10}.} From the low-$\ell$-only analysis we find $\tau=0.0506\pm0.0086$, consistent with previous \Planck\ large-scale analysis. For the tensor-to-scalar ratio $r$ we find a 95\,\% upper limit of $0.41$, more than a factor of 2 better than the previous \Planck\ bound from large scales \citepalias{planck2014-a13}. When the high-$\ell$ temperature likelihood is also considered, the constraint on $\tau$ is slightly dragged towards higher values by the higher $A_{\rm s}$ preferred by the smaller scales, and we obtain $\tau=0.0522\pm0.0080$. Constraints on the cosmological parameters from the low-$\ell$ only likelihoods are shown in Table~\ref{tab:results_lowell_hfi}.

\begin{table}[htbp!]
\begingroup
\caption{Constraints on $\ln(10^{10}A_{\rm s})$, $\tau$, and $r$ from the TT+EE likelihood (central column) and TT+EE+BB likelihood (right column). We show here the mean and 68\,\% confidence levels. For $r$, the 95\,\% upper limit is shown.}
\label{tab:results_lowell_hfi}
\nointerlineskip
\vskip -3mm
\setbox\tablebox=\vbox{
   \newdimen\digitwidth 
   \setbox0=\hbox{\rm 0} 
   \digitwidth=\wd0 
   \catcode`*=\active 
   \def*{\kern\digitwidth}
   \newdimen\signwidth 
   \setbox0=\hbox{+} 
   \signwidth=\wd0 
   \catcode`!=\active 
   \def!{\kern\signwidth}
\halign{\hbox to 0.9in{#\leaderfil}\tabskip=1em&
  \hfil#\hfil\tabskip=10pt& 
  \hfil#\hfil\tabskip=0pt\cr
\noalign{\doubleline}
\noalign{\vskip -2pt}
\omit\hfil Parameter\hfil& $\Lambda$CDM& $\Lambda$CDM + $r$\cr
\noalign{\vskip 4pt\hrule\vskip 4pt}
${\rm{ln}}(10^{10} A_{\rm s})$& $2.924\pm0.052$& $2.863^{+0.089}_{-0.062}$\cr
\noalign{\vskip 2pt}
$\tau$& $0.0506\pm0.0086$& $0.0503\pm0.0087$\cr
$r_{0.002}$& \dots& $\leq 0.41$\cr
\noalign{\vskip 2pt}
$10^9A_{\rm s}e^{-2\tau}$& $1.685^{+0.083}_{-0.091}$& $1.59^{+0.11}_{-0.13}$\cr
\noalign{\vskip 4pt\hrule\vskip 3pt}}}
\endPlancktable
\endgroup
\end{table}

The results on $\tau$ shown in Table~\ref{tab:results_lowell_hfi} differ by about half a $\sigma$ from the results presented in \citet{planck2014-a10}.  This is mainly driven by three factors. First of all in the 2018 release the last $1000$ anomalous rings have been removed from the HFI data \citepalias[see][for more details]{planck2016-l03}. Secondly the 30-GHz map used in \citet{planck2014-a10} was affected by a large-scale calibration leakage substantially improved in the 2018 release \citepalias[see appendix~A of][for more details]{planck2016-l02}. Thirdly, the variance of the FFP10 simulations is larger than the variance of the simulations used in \citet{planck2014-a10}, which pushes $\tau$ towards lower values. We attempt to quantify this last point by building a likelihood again from the 2018 data, but in conjunction with the 283 simulations used in \citet{planck2014-a10}, rather than with the FFP10 ones. With such a likelihood we measure $\tau = 0.053 \pm 0.008$, which is closer to the result published in \citet{planck2014-a10}.

\subsubsection{Monte Carlo validation}\label{subsubsec:validation}

In order to demonstrate that the low-$\ell$ polarization likelihood provides an unbiased estimate of $\tau$, we generate an MC data set of 300 CMB signal maps and combine them with the 300 FFP10 simulations. We then apply the likelihood construction procedure described above and proceed to estimate the $\tau$ parameter for each map of the MC set.  From these results we compute the mean and dispersion of the best-fit values. We perform this test with two different input $\tau$ values, namely 0.05 and 0.06.

For the first case (i.e., $\tau=0.05$) we find $\overline{\tau} = 0.05032\pm0.00048$, while in the latter case (i.e., $\tau=0.06$) we find $\overline{\tau} =0.05968\pm0.00044$. In both cases the likelihood method is able to recover the input value to within 0.1\,$\sigma$ of the single realization and hence has provided an essentially unbiased estimate for $\tau$.

\begin{figure}[htbp!]
\begin{center}
\includegraphics[width=0.475\textwidth]{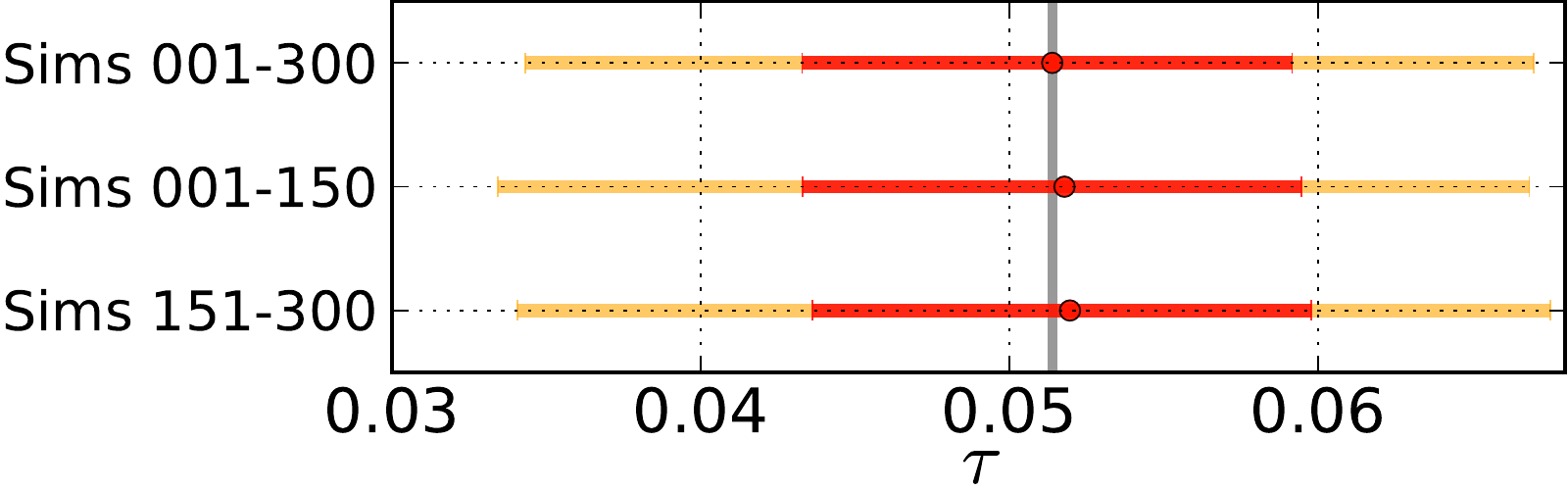}
\caption{Whisker plot, showing comparison of the best-fit values, together with the 68\,\% (red) and 95\,\% (yellow) confidence levels for $\tau$, obtained using either the complete set of 300 FFP10 simulations or the first half or the second half of them. The vertical line shows the best-fit value from the default analysis, i.e., using 300 simulations.}
\label{fig:plot_tau_half_sims}
\end{center}
\end{figure}

We also test stability with respect to the simulation database used in point 5 of the list shown in Sect.~\ref{subsubsec:likelihood}. We split the sets of 300 signal and noise simulations into two halves, and build two independent likelihoods from each of them.  When independently estimating $\tau$ with each of them, we find that the posteriors closely overlap (see Fig.~\ref{fig:plot_tau_half_sims}).

\subsubsection{Consistency analysis}\label{subsubsec:consistency_tests}

Multiple tests have been carried out to validate the low-$\ell$ polarization likelihood. In the following paragraphs we present some of those tests, focussing in particular on the estimation of the optical depth to reionization $\tau$. 

\paragraph{Dependence on masks:}\label{subsubsec:masks}

As a first consistency test, we estimate $\tau$ on all of the various masks shown in the first part of Fig.~\ref{fig:plot_masks}. We retain the cleaning procedure performed on the 70\,\% sky fraction mask and we apply the likelihood procedure described in Sect.~\ref{subsubsec:likelihood} for the different masks.

\begin{figure}[htbp!]
\begin{center}
\includegraphics[width=0.475\textwidth]{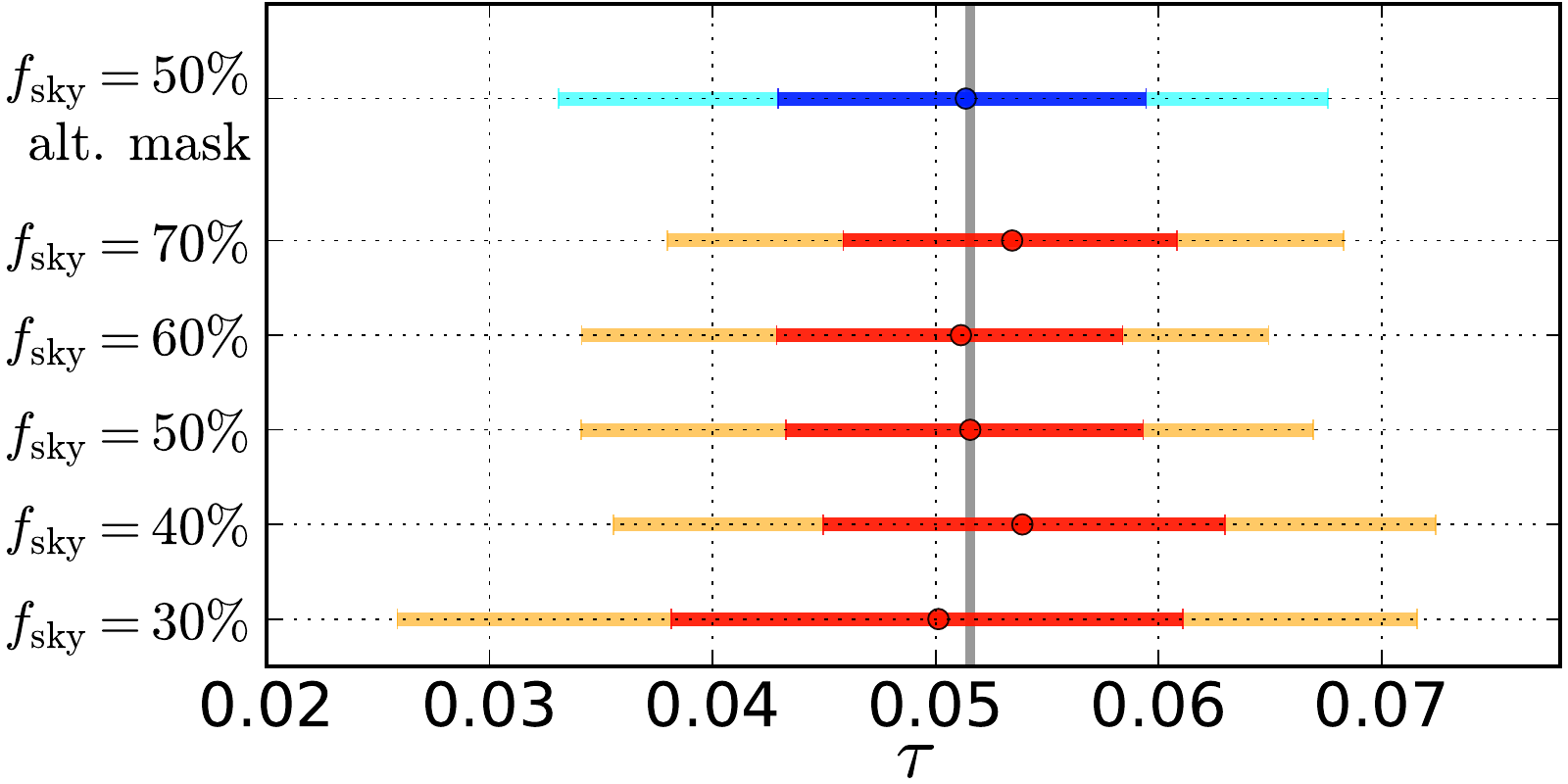}
\caption{Values of $\tau$ measured on data using different sky masks. The red points show the median values of the $\tau$ distributions measured on the masks shown in Fig.~\ref{fig:plot_masks}. The blue point represents the value of $\tau$ obtained using the 50\,\% alternative mask shown in the middle part of Fig.~\ref{fig:plot_masks}. Intense and fainter bars show 68\,\% and 95\,\% confidence levels, respectively. The vertical line shows the best-fit value from the default analysis, i.e., the 50\,\% mask.}
\label{fig:plot_tau_masks}
\end{center}
\end{figure}

\begin{figure}[htbp!]
\begin{center}
\includegraphics[width=0.475\textwidth]{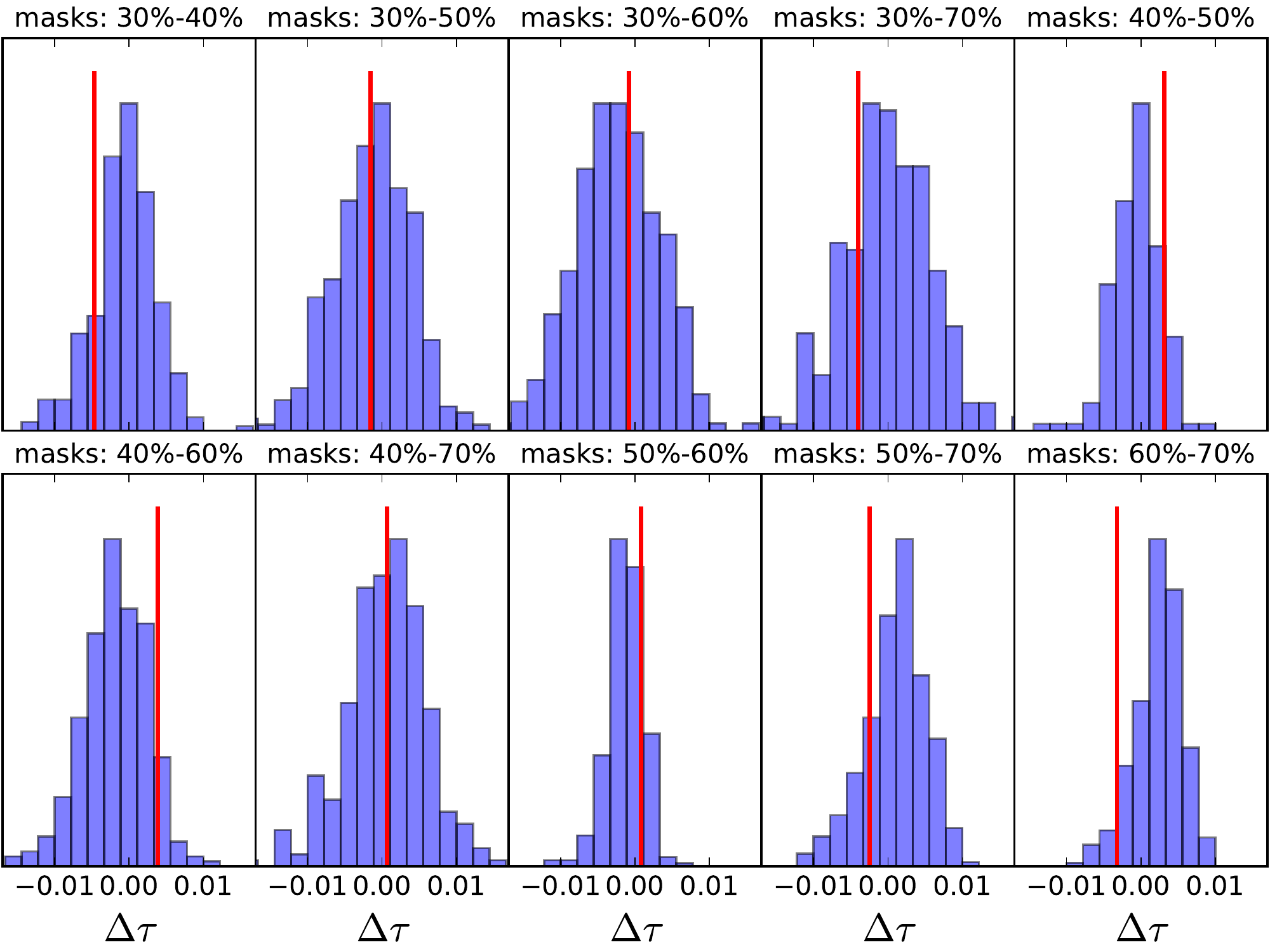}
\caption{Shifts in $\tau$ values between different masks (red vertical lines) compared with histograms of the same quantities measured in simulations.}\label{fig:plot_validation_masks}
\end{center}
\end{figure}

Figure~\ref{fig:plot_tau_masks} shows the $\tau$ values obtained with $f_{\rm sky}$ ranging from 30\,\% to 70\,\%.  In addition to the good visual consistency, we go on to estimate $\tau$, with all masks, on the validation MC simulations using a fiducial $\tau\,{=}\,0.05$, as used in Sect.~\ref{subsubsec:validation}.  We compare the differences between $\tau$ values obtained using different masks with corresponding histograms of differences for the 300 simulations. Results of this test are presented in Fig.~\ref{fig:plot_validation_masks}, showing that the discrepancy we see between different masks is indeed compatible with the empirical distribution of signal, noise, and systematics.

This test also shows that foreground cleaning is not a limiting factor for the low-$\ell$ polarization likelihood; even when highly contaminated regions are included (e.g., with the 70\,\% mask), the estimated values of $\tau$ are consistent (within scatter) with the values obtained using cleaner regions of the sky. The probability distribution of the residuals is well characterized by applying the same analysis to the FFP10 simulations. For the main likelihood analysis and for the delivered products we adopt an intermediate sky fraction, using the 50\,\% mask.

In order to further asses the stability of our results with respect to the choice of mask we consider the effect of employing an alternative algorithm for mask production.  In this scheme we smooth the polarized intensity of the 353-GHz channel with a beam of ${\rm FWHM}\,{=}\,10\deg$, instead of the 5\deg\ used for the masks shown in the top part of Fig.~\ref{fig:plot_masks}. Then we perform the same smoothing on the 30-GHz polarization map, scaling it down to 100\GHz.  Finally we threshold on the sum of the two polarization maps, obtaining masks with a different shape and with much smoother edges relative to the standard ones; see the middle part of Fig.~\ref{fig:plot_masks}.
We continue to perform the foreground cleaning on the 100- and 143-GHz maps with the 70\,\% mask as in the procedure described above.  Then we compute spectra and and the likelihood on the alternative 50\,\% mask. In Fig.~\ref{fig:plot_tau_masks} the blue point shows the $\tau$ value obtained with this alternative mask; this constraint is very consistent with those obtained using the standard series of masks.

\paragraph{Stability under the removal of a single multipole:}

\begin{figure}[htbp!]
\begin{center}
\includegraphics[width=0.475\textwidth]{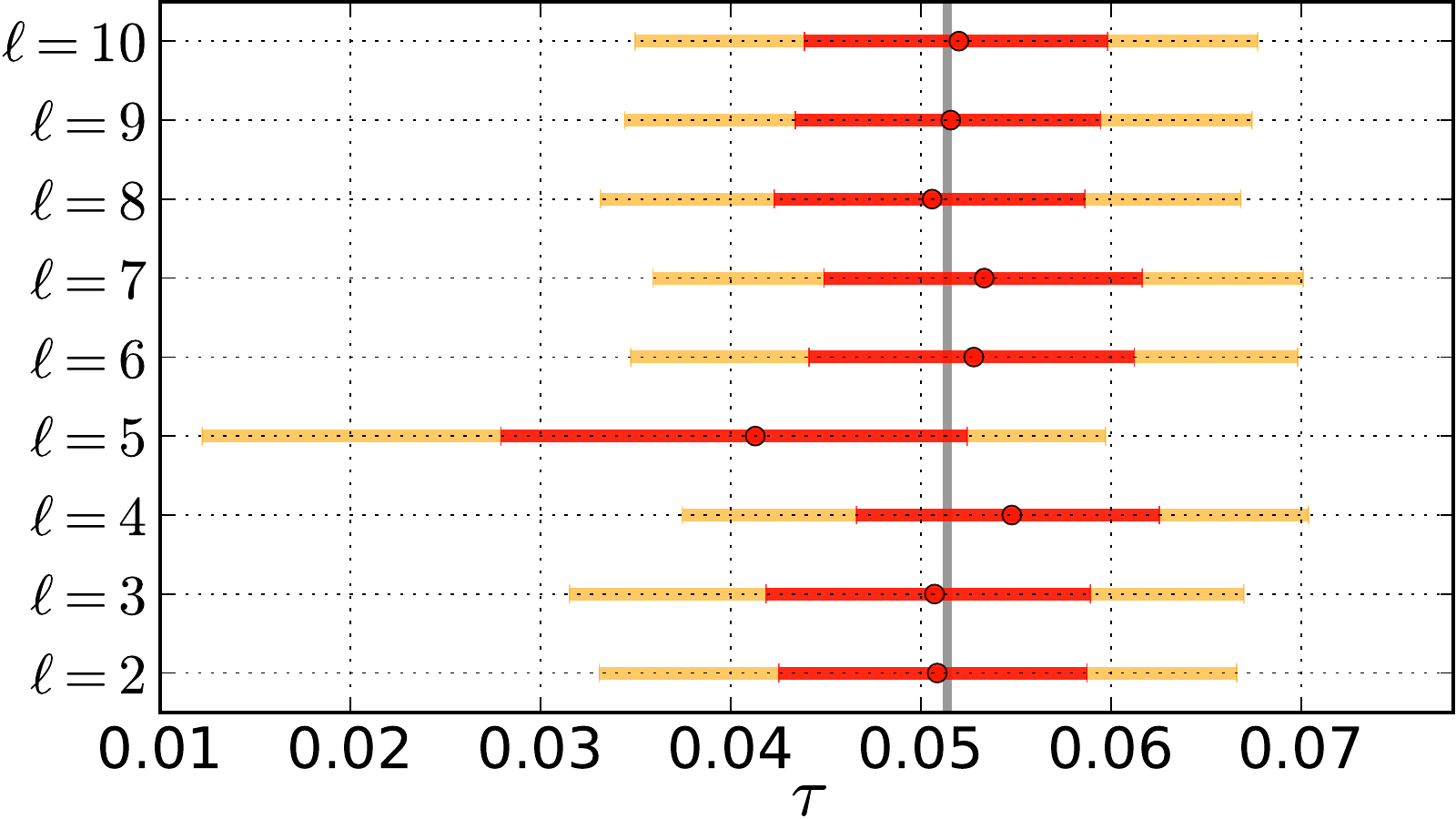}
\caption{Values of $\tau$ obtained when removing one multipole at a time. We show best-fit values (red points), together with 68\,\% (red lines) and 95\,\% (yellow lines) confidence levels. The vertical line shows the best-fit value from the full analysis.}
\label{fig:plot_tau_removedell}
\end{center}
\end{figure}

As was done in \citet{planck2014-a10}, we have tested the stability of the likelihood with respect to the removal of one $EE$ multipole at a time. Figure~\ref{fig:plot_tau_removedell} shows the values of $\tau$ obtained following this procedure. The posteriors are rather stable, except for a $\Delta \tau \approx 0.01$ shift when $\ell\,{=}\,5$ is removed. In order to characterize the significance of this feature we compare the width of the empirical distribution for $\tau$ obtained analysing 300 simulations using all the multipoles between $\ell\,{=}\,2$ and $\ell\,{=}\,29$, to the width obtained with $\ell\,{=}\,5$ removed. In the first case the width is $\sigma_{\tau}({\rm all}\ \ell)=0.0082$, in the latter $\sigma_{\tau}({\rm no}\ \ell=5)=0.010$. Subtracting in quadrature the former error from the latter gives us a measure on the expected parameter shifts between the two analyses \citep[see][]{gc2019} of $\sigma_{\tau}(\ell=5)=0.0057$.  The observed shift in $\tau$ of approximately $0.009$ is thus consistent with shifts expected from the simulations.

\paragraph{Stability for different data cuts:}

\begin{figure}[htbp!]
\begin{center}
\includegraphics[width=0.5\textwidth]{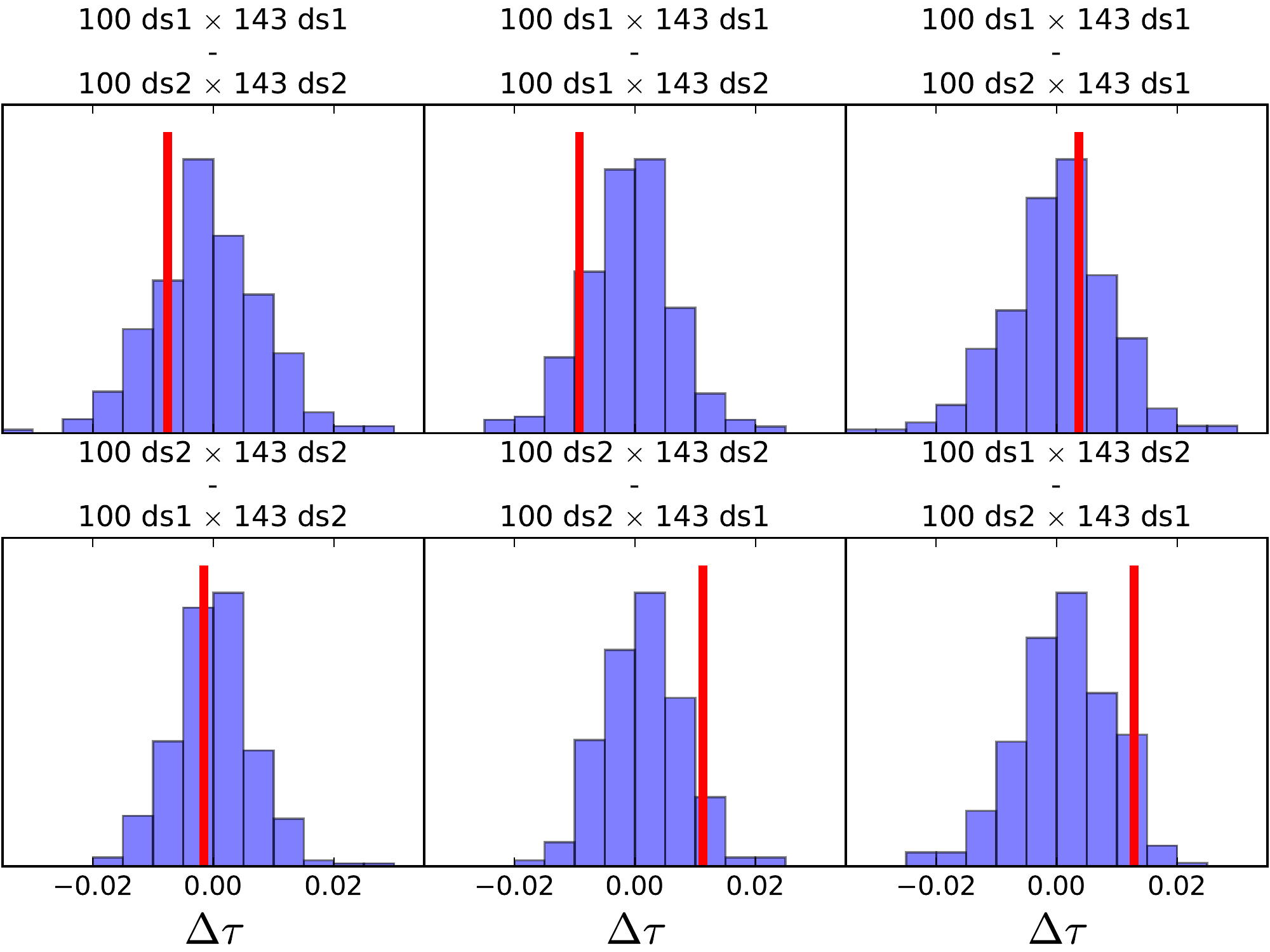}
\includegraphics[width=0.5\textwidth]{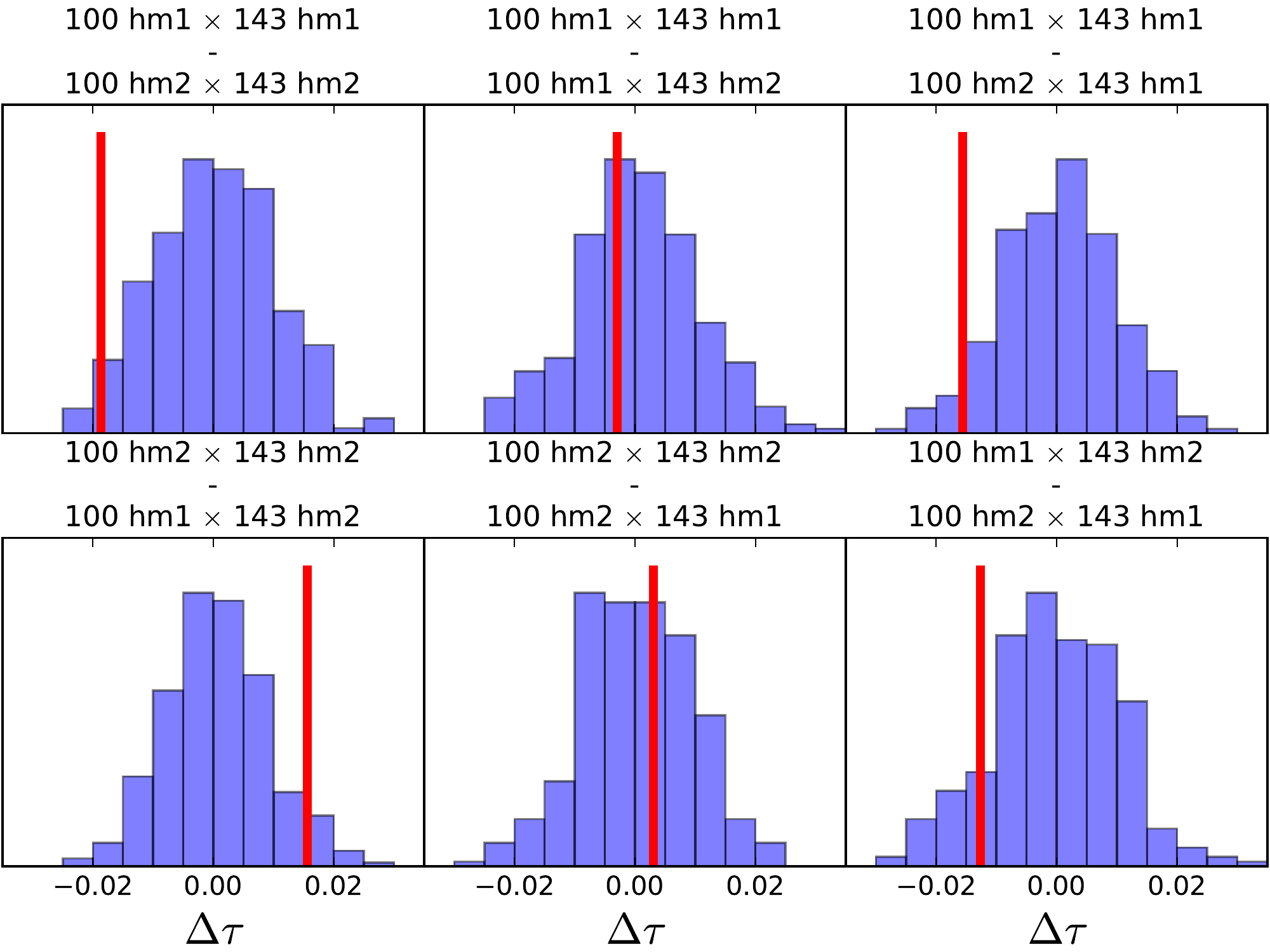}
\includegraphics[width=0.5\textwidth]{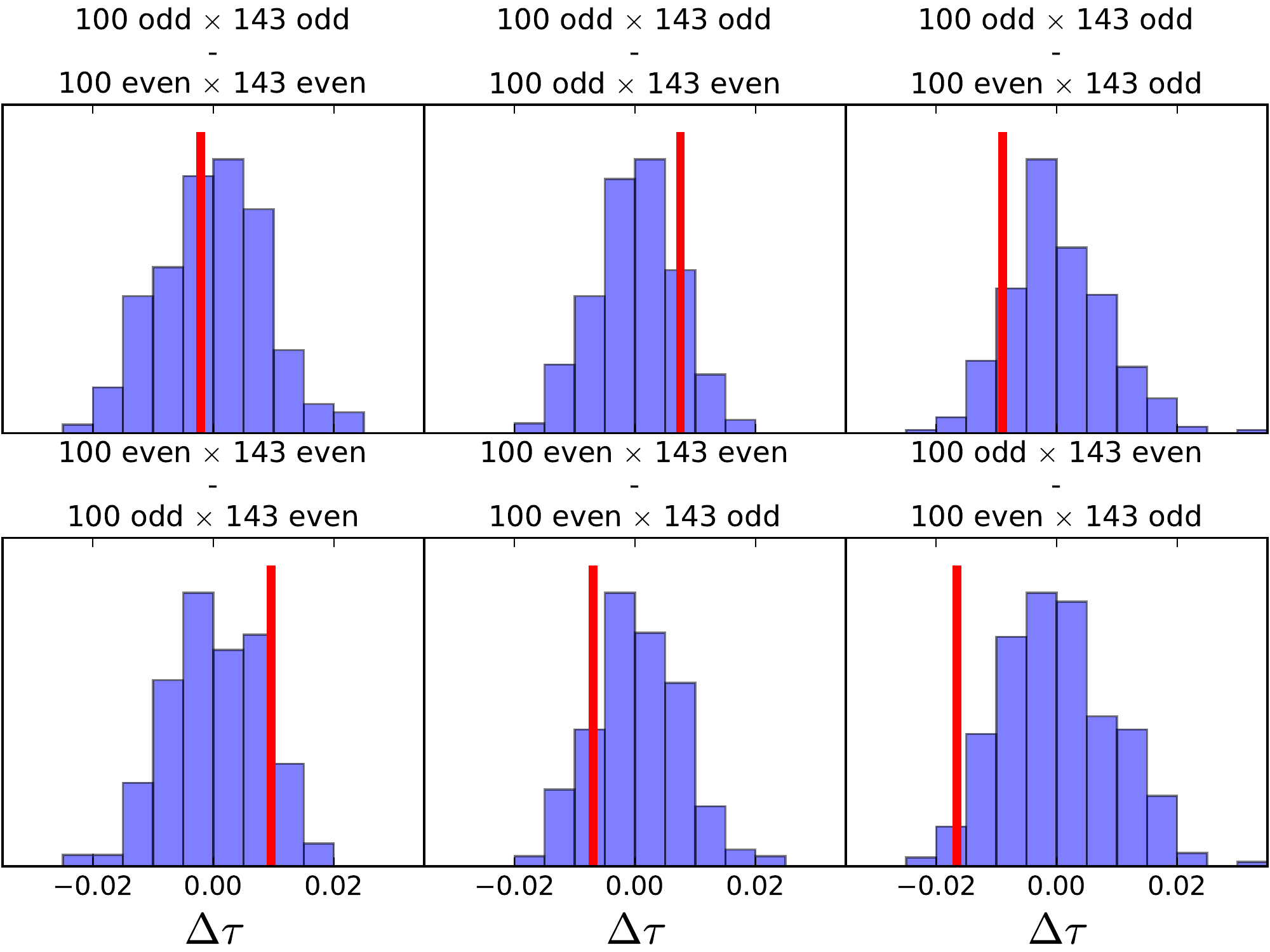}
\caption{Shifts in $\tau$ values between different data splits (red vertical lines) compared with histograms of the corresponding differences measured on simulations.}
\label{fig:hist_delta_tau_splits}
\end{center}
\end{figure}

Continuing the discussion started in Sect.~\ref{subsec:spectra}, we compute $\tau$ constraints from the $EE$ spectra for all of the data splits analysed (i.e., DSs, HMs, and OE). Following the same procedure as for the mask consistency test, we verify that the different $\tau$ values we obtain on different data splits are consistent with the statistics of the FFP10 simulations. We clean all the single data splits independently using 30\GHz\ and 353\GHz, following the procedure described in Sect.~\ref{subsec:cleaning}. Then we estimate $\tau$ from all the data splits using the 50\,\% mask to compute the cross-QML spectra. 
Figure~\ref{fig:hist_delta_tau_splits} shows the differences in $\tau$ found between data splits, as calculated with the data, compared to histograms of the corresponding quantities computed on simulations. For all the data cuts analysed the $\tau$ values obtained are consistent with the distribution of the FFP10 noise and systematics simulations.

\paragraph{Synchrotron-tracer stability:}\label{subsubsec:sync_cleaning_test}

In the power spectra and likelihood analysis we adopt the 30-GHz LFI channel as our polarized synchrotron tracer. Furthermore we remove synchrotron only from the 100-GHz channel, ignoring its contribution at 143\GHz.

\begin{figure}[htbp!]
\begin{center}
\includegraphics[width=0.475\textwidth]{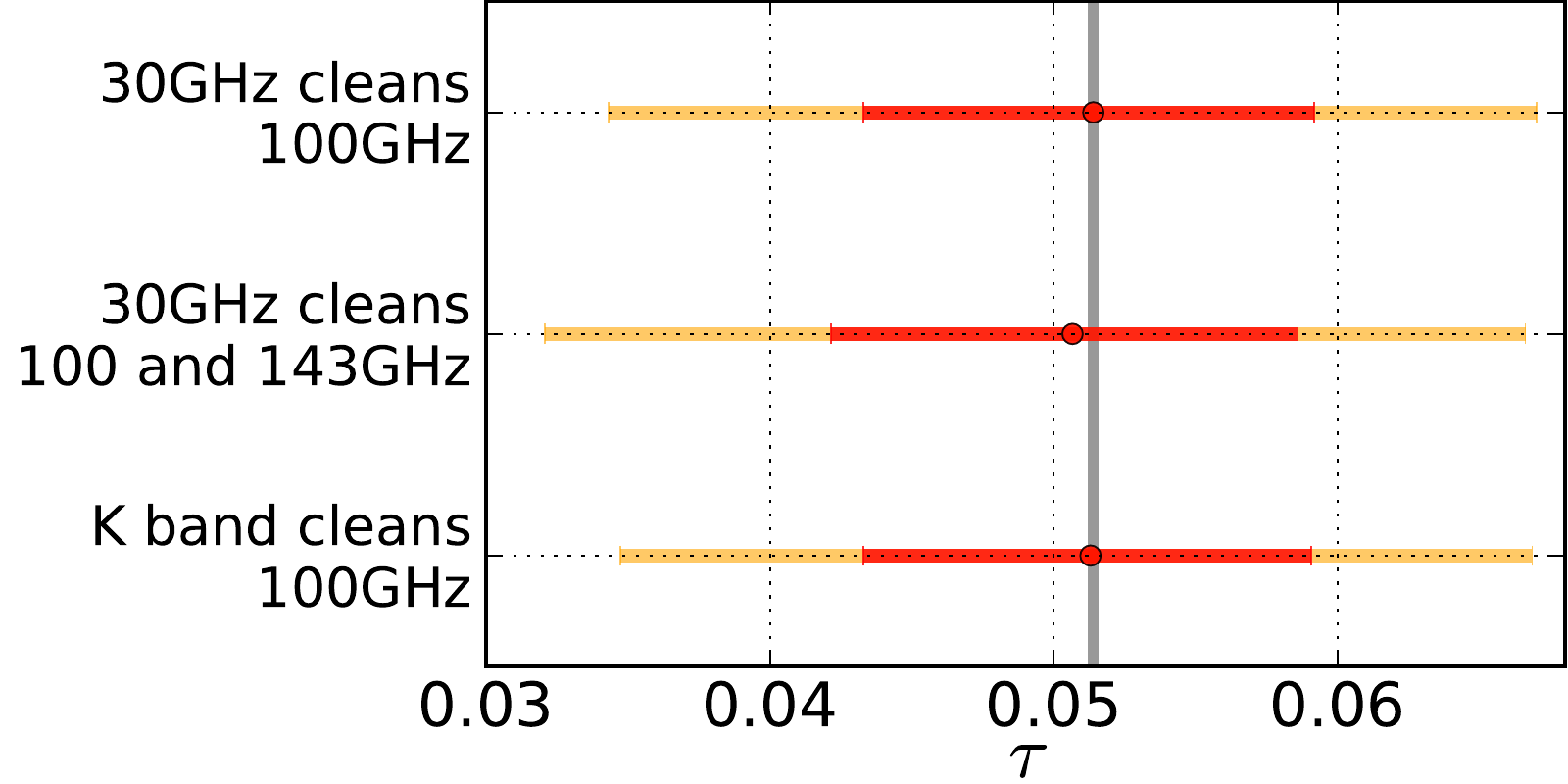}
\caption{Posteriors on $\tau$ obtained from cross-spectra between the 100- and 143-GHz channels; only the polarized synchrotron-tracer handling is changed among the three cases. From top to bottom we use: 30\GHz\ to clean 100\GHz\ only; 30\GHz\ to clean both 100 and 143\GHz; and WMAP K band to clean 100\GHz\ only. Intense and fainter bars show 68\,\% and 95\,\% confidence levels, respectively. The vertical line shows the best-fit value from the default analysis.}
\label{fig:plot_tau_different_sync_cleaning}
\end{center}
\end{figure}

We can test the validity of these two assumptions. First we process with the usual smoothing procedure the map and covariance of the WMAP K-band data. We substitute WMAP K band for \Planck\ 30\GHz\ in the cleaning of the 100-GHz map, finding for synchrotron a scaling of $\alpha=0.0119\pm0.0006$.  We then calculate power spectra, construct a likelihood, and compute $\tau$ for $100\times143$.
Secondly we clean both 100 and 143\GHz\ with 30\GHz, finding for synchrotron a scaling of $\alpha= 0.0073\pm0.0011$ for the 143-GHz map. Again we then calculate power spectra, construct a likelihood, and compute $\tau$ for $100\times143$.
The results of these tests are shown in Fig.~\ref{fig:plot_tau_different_sync_cleaning}; the posterior on $\tau$ is not substantially affected by these variations in the handling of synchrotron contamination.

\paragraph{LFI-HFI cross-spectra:}\label{subsubsec:lfi_cross_hfi}

We may perform a similar analysis to that done for the HFI channels, but instead using the 70-GHz LFI full channel for one of the maps. We filter and degrade the 70-GHz band-pass-mismatch-corrected map with the procedure described above.  We also apply the same smoothing to the gain template described in \citetalias{planck2016-l02}, which is meant to correct for dipole and foreground leakage to polarization from calibration mismatch.

The foreground-cleaning procedure is performed using the same foreground tracers as used for the HFI channels, adopting the 70-GHz R2.2x mask (see Table~\ref{table:mask_fsky} for more details), which retains about 67\,\% of the sky. In the cleaning procedure, along with the foreground templates, we also fit the gain template (see Sect.~\ref{sec:lo-ell:lfi} and Eq.~\ref{eq_cleaning} of this paper, and section~3.3 of \citetalias{planck2016-l02}). The amplitudes found for the templates are: $\alpha=0.0580\pm0.0044$ for synchrotron; $\beta=0.00923\pm 0.00036$ for the dust; and $\gamma=0.960\pm0.083$ for the gain template.  These are consistent with the values shown in Fig.~\ref{fig_alpha_beta_gamma_chi2}.

The procedure followed for $\tau$ estimation is the same as that described in Sect.~\ref{subsubsec:likelihood} above. For the 70-GHz channel, we use the LFI FFP10 simulations, containing noise and systematic effects related to calibration, namely temperature-to-polarization leakage of dipole and intensity foregrounds due to calibration mismatch \citepalias[see appendix~B of][for details]{planck2016-l02}. As we do for data, we fit a template of gain-related systematics also for the simulations. We build a leakage template from simulations using the following procedure: we first remove, from each of the 300 simulations, the fiducial FFP10 CMB map, the input foreground map, and the noise realization, obtaining a realization of dipole and foreground leakage; then we average all of the 300 simulations.

In the likelihood construction we use the 47\,\% of the sky that is in common between the 50\,\% mask used for the $100\times143$ cross-spectrum analysis and the 70-GHz R1.4 mask (which retains 58\,\% of the sky).

\begin{figure}[htbp!]
\begin{center}
\includegraphics[width=0.475\textwidth]{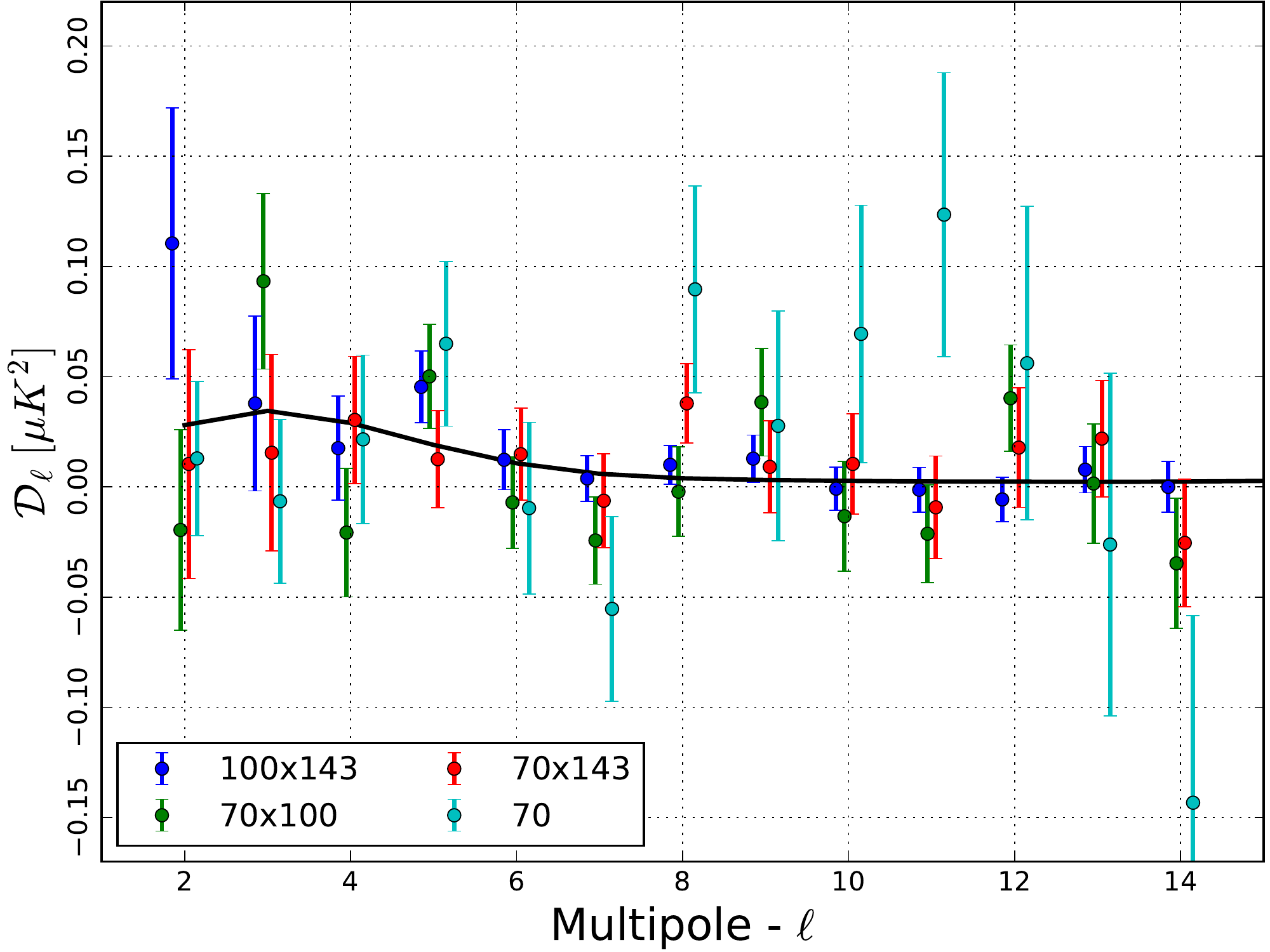}
\caption{$EE$ cross-spectra for $100\times143$, $70\times100$, and $70\times143$, together with the $EE$ 70-GHz auto-spectrum. The error bars are based on MC simulations for the cross-spectra and on the Fisher matrix for the auto-spectrum. The sky fractions are 50\,\% for $100\times143$, 47\,\% for $70\times100$ and $70\times143$, and 62\,\% for 70\GHz. The black line shows a theoretical power spectrum with $\tau=0.05$.}
\label{fig:plot_EE_lfi_hfi}
\end{center}
\end{figure}
\begin{figure}[htbp!]
\begin{center}
\includegraphics[width=0.475\textwidth]{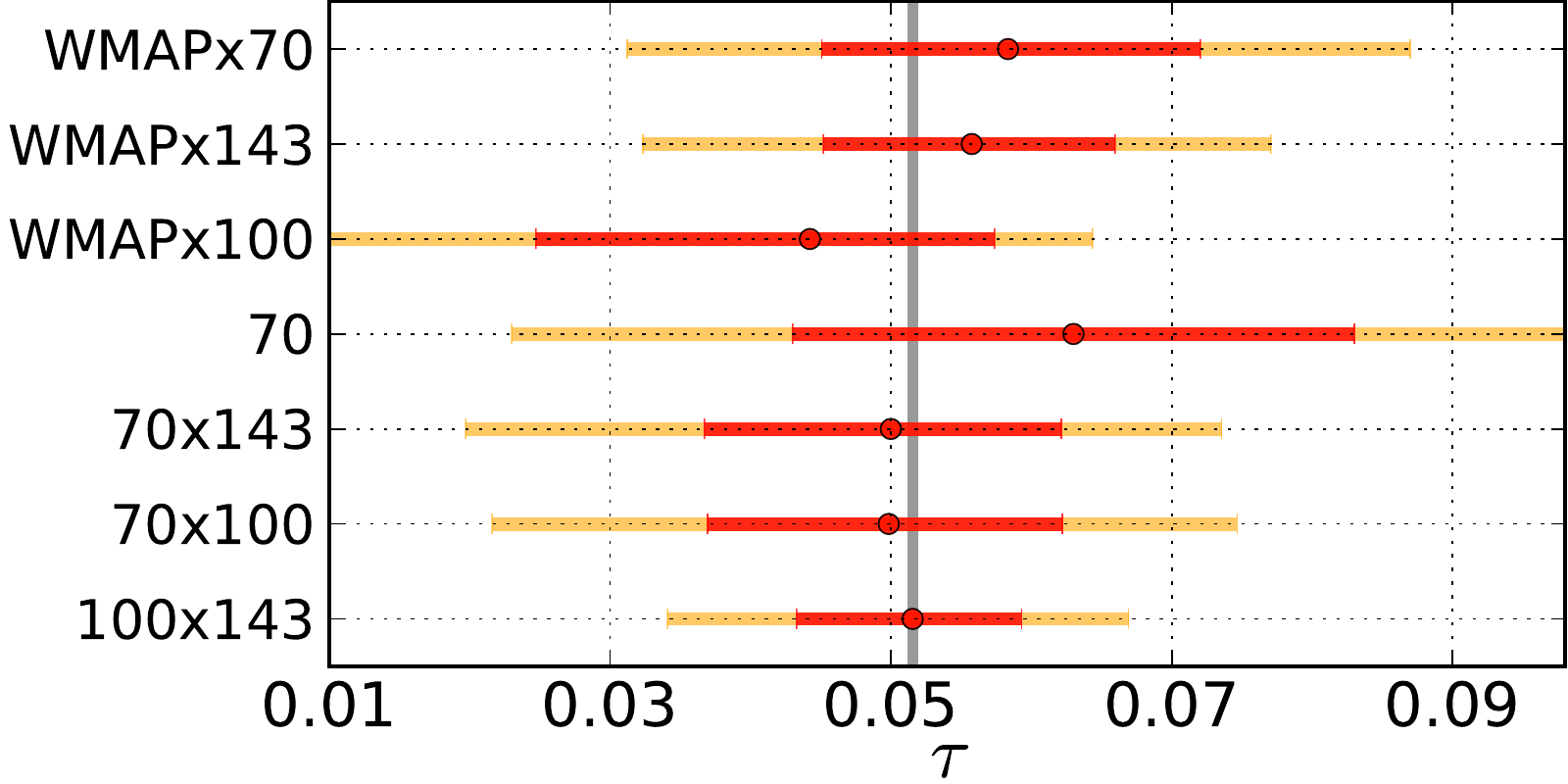}
\caption{Whisker plot showing best-fit values and 68\,\% and 95\,\% confidence levels for $\tau$ when combining various \Planck\ and WMAP channels. The vertical line shows the best-fit value from $100\times143$.}
\label{fig:plot_tau_crosses_all}
\end{center}
\end{figure}

In Fig.~\ref{fig:plot_EE_lfi_hfi}{,} we show $EE$ cross-spectra between $70$, $100$, and 143-GHz channels and the auto-spectrum of the 70-GHz channel. The errors shown are computed through MC simulations for the cross-spectra and from the Fisher matrix for the auto-spectrum.  We estimate $\tau$ from the cross-spectra of 70\GHz\ with both 100\GHz\ and 143\GHz, showing the results in Fig.~\ref{fig:plot_tau_crosses_all}.

\paragraph{\Planck-WMAP cross-spectra:}

We also process with the same smoothing and cleaning procedures the Ka, Q, and V bands of WMAP \citep{bennett2012}, using the K band and 353-GHz channels as tracers of polarized synchrotron and dust, respectively. In the cleaning we use the WMAP processing masks \citep{page2007} retaining 93\,\% of the sky for the Ka and Q bands and 96\,\% for V. As covariance matrices we adopt the official WMAP covariance matrices and smooth them in harmonic space with the cosine window function given in Eq.~\eqref{eq:cosdegrade}. After the cleaning we inverse-noise weight the three WMAP bands into a single map. 
Using the same procedure as described in Sect.~\ref{subsubsec:likelihood} we estimate $\tau$ from each of the cross-spectra between WMAP and the \Planck\ 70-, 100-, and 143-GHz channels. In order to produce WMAP noise simulations we generate, for each band, 300 Gaussian realizations from the covariance matrices. As is done for the \Planck\ channels, we process all of the simulations through the same cleaning pipeline, combining the resulting maps in the same manner as done for the data.

For power-spectrum estimation and for the likelihood, for the cross-spectrum between 70\GHz\ and WMAP we use the intersection between the WMAP P06 and the LFI R1.4 masks, retaining 57\,\% of the sky. For the cross-spectra of 100\GHz\ and 143\GHz\ with WMAP we use the intersection between the P06 and the HFI 50\,\% masks, retaining 48\,\% of the sky.

\begin{figure}[htbp!]
\begin{center}
\includegraphics[width=0.475\textwidth]{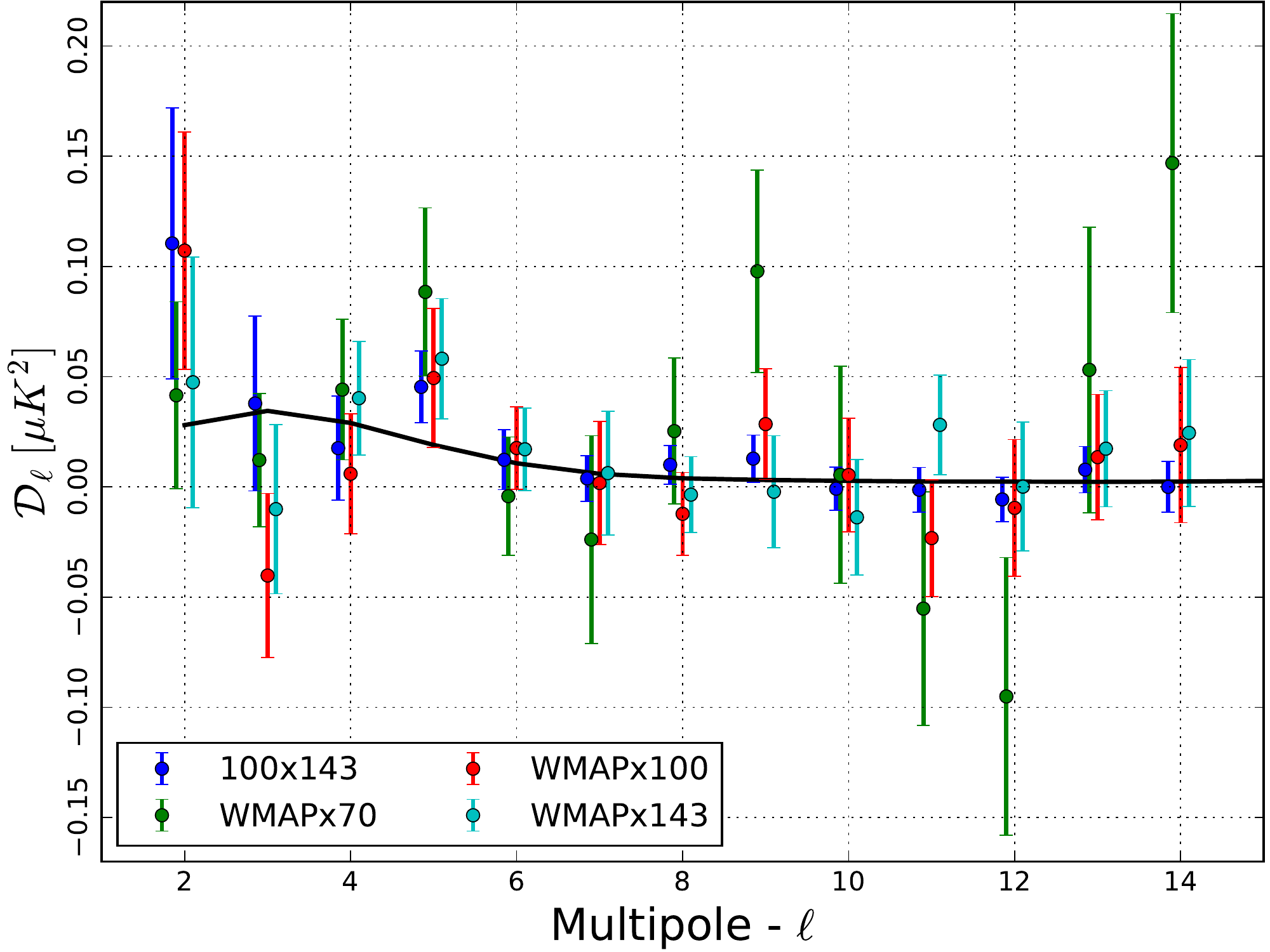}
\caption{$EE$ cross-spectra for ${\rm WMAP}\times70$, ${\rm WMAP}\times100$, and ${\rm WMAP}\times143$. The error bars shown are based on MC simulations. The sky fractions used are 57\,\% for ${\rm WMAP}\times70$ and 48\,\% for both ${\rm WMAP}\times100$ and ${\rm WMAP}\times143$. For reference we also plot the $100\times143$ cross-spectrum. The black line shows a theoretical power spectrum for a model with $\tau=0.05$.}
\label{fig:plot_EE_wmap_planck}
\end{center}
\end{figure}

In Fig.~\ref{fig:plot_EE_wmap_planck}{,} we show the $EE$ cross-spectra between \Planck\ channels and WMAP, with errors computed through MC simulations. The posteriors from all the cross-spectra considered between HFI, LFI, and WMAP channels are plotted in Fig.~\ref{fig:plot_tau_crosses_all}. 

\begin{figure}[htbp!]
\begin{center}
\includegraphics[width=0.5\textwidth]{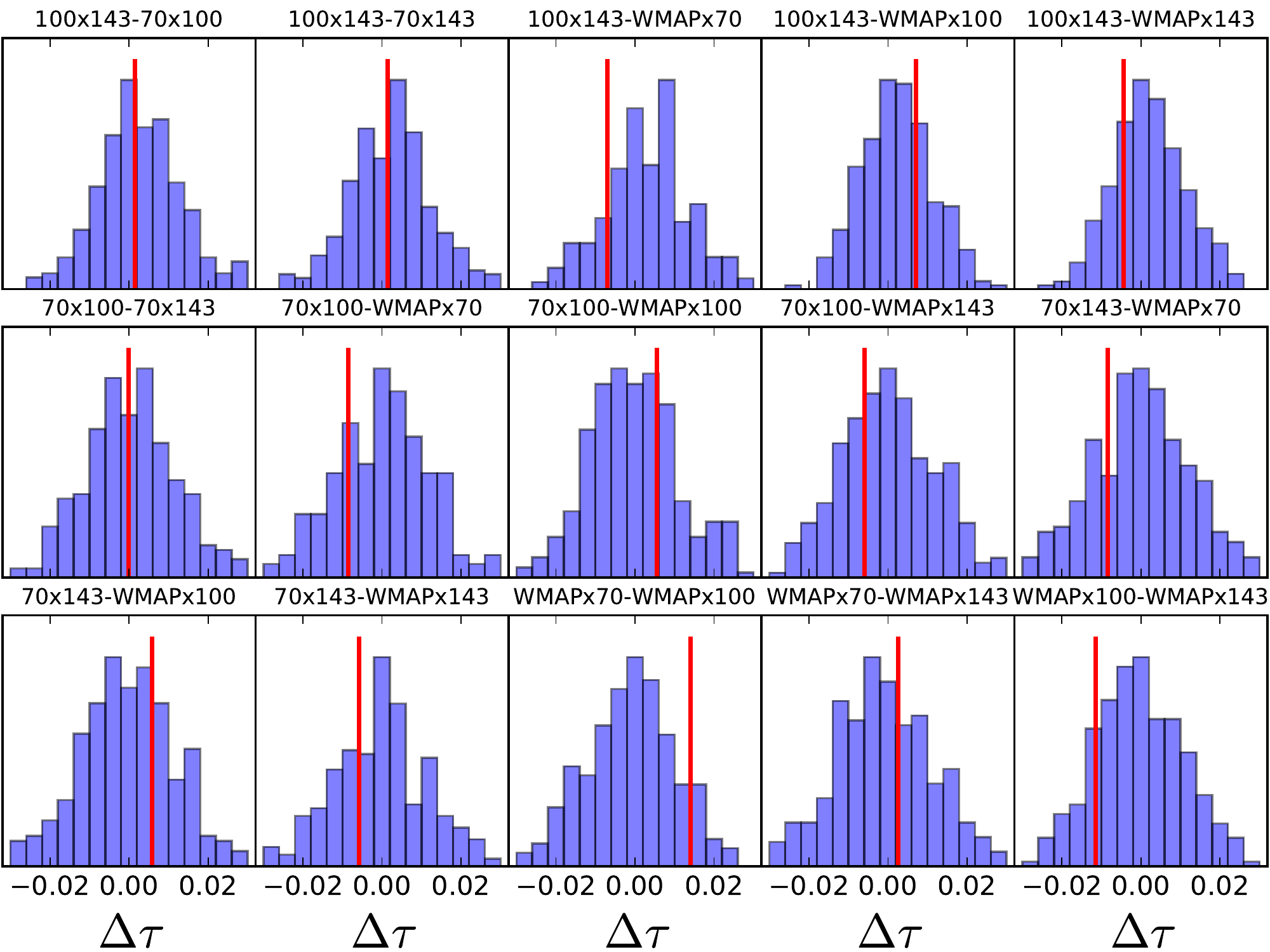}
\caption{Shifts in $\tau$ values obtained from different cross-spectra (red vertical lines), compared with histograms of the corresponding quantities measured on simulations.}
\label{fig:plot_tau_differences_validation_all}
\end{center}
\end{figure}

As was done for the data-split and mask tests, we perform analogous combined MC validation for all the cross-spectra analyses considered involving the \Planck\ 70-, 100-, and 143-GHz channels and the WMAP co-added maps. The results of this validation are presented in Fig.~\ref{fig:plot_tau_differences_validation_all}. The scatter observed for the data is highly compatible with the empirical distribution of the simulations.

\subsubsection{Final considerations on the low-$\ell$ HFI likelihood}\label{subsubsec:final_considerations}

{In this section, we have presented} the low-$\ell$ likelihood analysis based on HFI data. In contrast to what is done in the LFI analysis, we have opted for a simulation-based likelihood, avoiding analytical approaches, in order to overcome the difficulty of:
\begin{itemize}
\item {determining analytically 
the probability distribution of the residual systematics;}
\item {de-biasing maps from noise and residual systematics;}
\item {building reliable noise-plus-systematic-effect covariance matrices.}
\end{itemize}

In addition we have chosen an estimator based on the cross-spectrum between the 100- and 143-GHz maps, as opposed to any auto-spectrum, in order both to be less vulnerable to inaccurate noise-bias determination and to avoid the auto-correlation of systematics within a frequency channel.  We have successfully demonstrated that this method provides an unbiased estimate of cosmological parameters and we have verified the stability of our results with respect to mask, data splits, and channel selection.
Nevertheless{,} this analysis comes with caveats that we emphasize below in detail.
\begin{itemize}
\item The accuracy of the likelihood approximation relies on the capability of the simulations to reproduce the residual systematics present in the data. The simulation pipeline is described and validated extensively in \citetalias{planck2016-l03}. The analyses performed on data splits and null maps, shown in the present paper, confirm those results.
\item The simulations overestimate the ADCNL effect at 100 and 143\GHz\ \citepalias[see section~5.13 of][in particular figures~48 and 49 and the relevant discussion in the text]{planck2016-l03}. This causes a slight overestimation of the variance in the $100\times143$ spectrum.
\item As explained in detail in \citetalias{planck2016-l03}, the gain-variation fit introduced in the 2018 mapmaking procedure corrects for the first-order approximation of the ADCNL systematic effect. However, large signals, such as dipoles and foreground emission close to the Galactic plane, are distorted by the second-order ADCNL effect, leaving a signature at very low multipoles in polarization, which is not corrected by the 2018 \Planck-HFI analysis, {for more details about this effect see section~3.2 of \citet{delouis2019} and figures 4 and 5 of \citet{Pagano:2019tci}.}
\item As a consequence of the preceding point, our constraining power on $\tau$ comes almost entirely from $\ell\,{=}\,4$ and $\ell\,{=}\,5$. The variance of residual systematics at $\ell\,{=}\,2$ and $\ell\,{=}\,3$ is large enough to weaken the constraint on $\tau$ coming the very largest scales.
\item The substantial impact of $\ell\,{=}\,5$ on the $\tau$ estimation, already observed by \citet{planck2014-a10}, is perfectly compatible with a statistical fluke, as described by FFP10 simulations. Furthermore we could not find any direct link between this effect and the shift at the same multipole in the temperature power spectrum, since it is mostly related to sky regions masked in the polarization analysis. The $TE$ null tests for $\ell\,{=}\,5$ support this interpretation. 

\item In the likelihood approximation we sample the theoretical $C_\ell$ space, extracting cosmological models assuming $\Lambda$CDM+$r$ cosmology. This does not represent a limitation for more complicated models, as long as they do not depart too dramatically from $\Lambda$CDM. 
\end{itemize}

In addition to these issues, we also 
discard the $TE$ spectrum in the low-$\ell$ likelihood analysis.
The main reason for this choice is linked to the restricted number of simulations and their limitations. As we discussed in Sect.~\ref{subsec:spectra} and show in Table~\ref{tab:pte_null_test}, the $\ell>10$ modes of the $TE$ spectra fail our null test and we believe that a plausible reason for this is the lack of fidelity of our simulation suite, which assumes perfect dust cleaning in temperature, while some residual can still be present in the data. Furthermore, the limited number of simulations makes it difficult to correctly explore the correlations between $TT$ and $TE$, as well as $EE$ and $TE$. Building an uncorrelated likelihood, as we did for $EE$, would certainly result in an overestimated and possibly biased constraint. While it is possible to overcome this issue with an analytic (or semi-analytic) approach \citep[such as described in e.g.,][]{2017arXiv170808479G}, it has been decided not to pursue such possibilities, given the null-test failure of the $TE$ spectrum. With this limitation in mind, it is still worth noting that restricting the spectrum to $\ell_{\rm max}\,{=}\,10$, and measuring $\tau$ only from $TE$ gives a value $0.051\pm0.015$. From simulations, we can evaluate that the contribution of the $10\,{<}\,\ell\,{<}\,30$ $TE$ modes gives only a $7\,\%$ improvement in the $\tau$ constraint.
It is thus likely that, even employing a much larger number of simulations 
and improving the foreground model, one could only slightly increase 
\Planck’s constraining power on $\tau$ by using the low-$\ell$ $TE$ spectra in addition to
the $EE$ spectrum, compared to using the latter alone. 

As extensively discussed in \citetalias{planck2016-l03}, the determination of polarization efficiency remains one of the main limitations of the \Planck\ HFI 2018 products. Nevertheless, regarding the low-$\ell$ polarization likelihood, we have verified that possible residual systematic effects, related to the accuracy of the polarization efficiency determination, have a negligible impact on our $\tau$ determination\footnote{For the impact on the high-$\ell$ likelihood see Sec.~\ref{sec:hi-ell:datamodel:inst}.}.

\subsection{LFI-based low-$\ell$ likelihood}
\label{sec:lo-ell:lfi}

The LFI low-$\ell$ polarization likelihood follows the methodology of the 2015 release \citepalias{planck2014-a13}, with several improvements discussed below. We employ a pixel-based approach to compute the likelihood $\mathcal{L}(C_{\ell})$, defined by the conditional probability $\mathcal{P}(\vec{m}|C_{\ell})$:
\begin{equation}
\mathcal{L}(C_{\ell}) =
\mathcal{P}(\vec{m}|C_{\ell})=\frac{1}{(2\pi)^{N/2}|\tens{M}|^{1/2}}
\exp\left(-\frac{1}{2}\vec{m}^{\tens{T}}\,\tens{M}^{-1}\vec{m}\right)\, .
\label{eq:pbLike}
\end{equation} 
Here, $\vec{m}$ is a foreground-mitigated temperature and linear polarization map array of total length $N$ pixels, whose signal-plus-noise covariance matrix is $\tens{M}$. For the temperature map we employ the \texttt{Commander} solution described in the previous section. As in the 2015 \Planck\ release, linear polarization CMB maps are estimated from the LFI 70-GHz channel, using the 353- and 30-GHz channels as tracers to minimize the polarized dust and synchrotron emission, respectively. {The main improvements with respect to the 2015 analysis, mostly due to revisiting the calibration \citepalias[][]{planck2016-l02}, are}: 
\begin{itemize}
\item {the inclusion in the data set of Surveys~2 and 4, which were left out in the 2015 analysis because of null-test issues that are now under control} \citepalias[see][]{planck2016-l02};
\item {the use of a cosine window function to better suppress high-resolution noise;}
\item {a larger sky fraction used in the baseline likelihood;}
\item {an improved comparison with \wmap\ Ka, Q, and V data.}
\end{itemize}
We also improved the analysis of consistency with respect to the 2015 release, and applied new consistency tests. {As in the 2015 \Planck\ release, we do not include 44\,GHz as a cosmological channel, since preliminary analysis showed poor component-separation $\chi^2$ values even for relatively small sky fractions.}

\subsubsection{LFI masks}\label{sec:masks}

\begin{figure}[htbp!]
\includegraphics[width=0.5\textwidth]{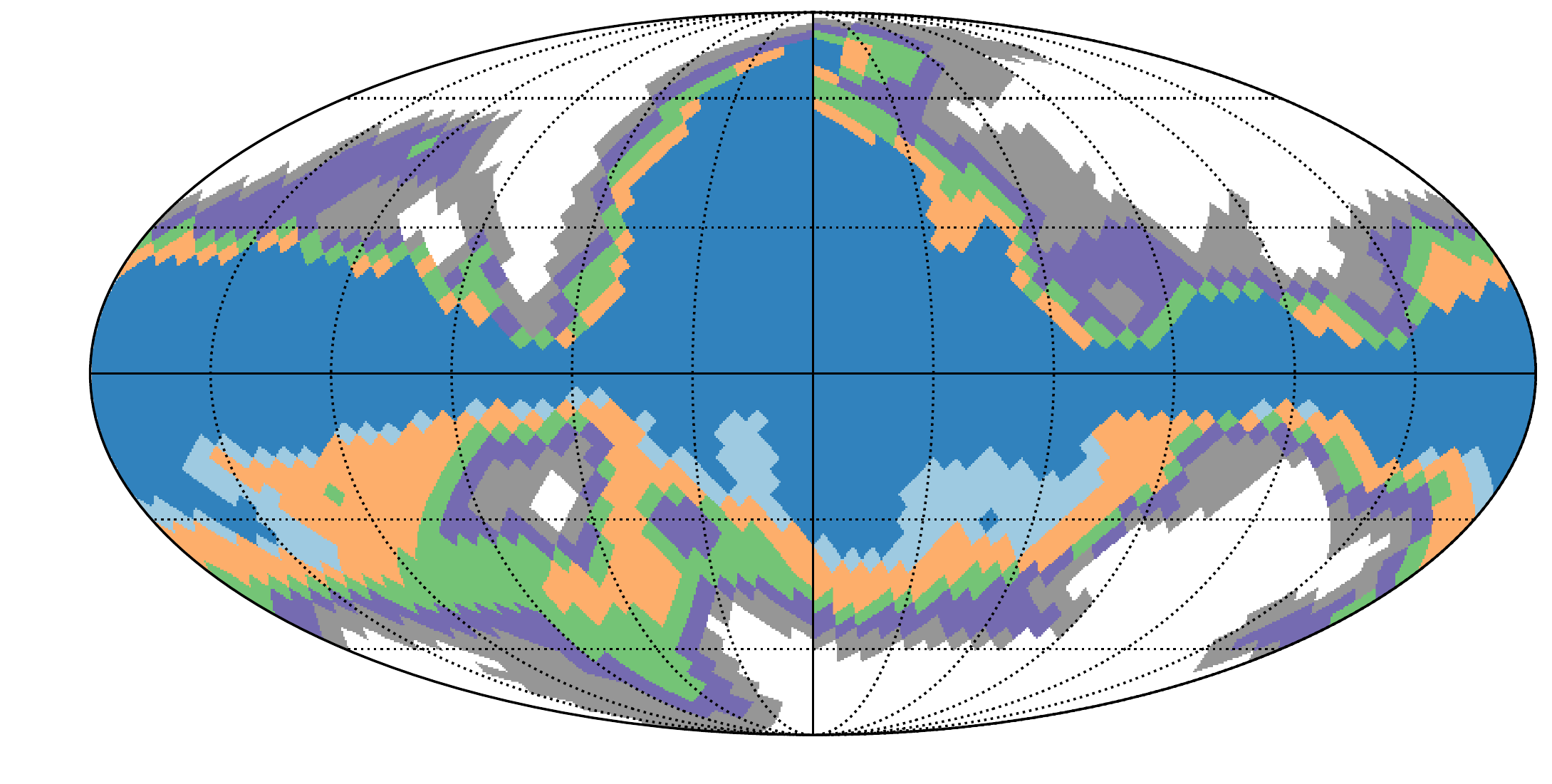}
\caption[]{For illustration purposes, we here show the polarization masks R0.6, R0.8, R1.0, R1.2, R1.8x, and R2.2x; see text for the definition of the $R$ parameter.}
\label{fig_masks}
\end{figure}

We now describe the procedure we follow to build masks of the polarized Galactic diffuse emission for the 70-GHz-based low-$\ell$ likelihood. Both synchrotron and dust emission need to be considered. Following the scheme employed in the 2015 release, we use the (band-pass corrected) 30-GHz and the 353-GHz maps as tracers of the synchrotron and dust emission, respectively. We smooth both maps to $1^\circ$ angular resolution, working at $N_\mathrm{side}\,{=}\,1024$ (the 353-GHz data are downsampled from the native $N_\mathrm{side}\,{=}\,2048$ resolution). The synchrotron contribution at 70\GHz\ is estimated by extrapolating in frequency the 30-GHz map, assuming a power-law energy distribution with a fixed spectral index $\alpha\,{=}\,-3.0$ \citep{planck2014-XXII}. For the 353-GHz map, we instead assume that the dust emits like a modified blackbody with $\beta\,{=}\,1.59$ and a temperature $T_{\rm d}\,{=}\,19.6\,{\rm K}$ \citep{planck2014-XXII}. 
We build the polarization amplitude maps, $P=\sqrt{Q^2+U^2}$, separately for dust and synchrotron and generate one mask for each signal. In particular, these masks are obtained by masking out on the extrapolated maps all the pixels where $\sqrt{{\rm var}(Q)+{\rm var}(U)} > 0.73\,R\,\mu\mathrm{K}$. The value $0.73\,\mu{\rm K}$ comes from the square root of the sum of the variance in $Q$ and $U$ for the CMB (generated from the best-fit $\Lambda$CDM model) computed over the whole sky. On the other hand, $R$ represents a parameter that we change to consider different thresholds, and thus to build our set of masks (which are labelled as ``R$X.Y$,'' where ``$X.Y$'' is the value of $R$). While this procedure is in principle susceptible to some bias because of the CMB polarization emission in the maps, we find in practice that this contribution impacts the mask selection in a negligible manner. The dust and synchrotron masks are then combined and the resulting mask is smoothed with a Gaussian window of ${\rm FWHM}\,{=}\,10^\circ$. In the resulting mask, all pixels with values lower than 0.5 are set to 0, while those with values greater than 0.5 are set to 1. The masks obtained for each value of the threshold are finally downsampled to a resolution of $N_\mathrm{side}\,{=}\,16$, appropriate for the low-$\ell$ studies.  In Fig.~\ref{fig_masks} we show a subset of the masks that we consider; 
for the R1.8 and R2.2 masks we also consider an ``extended'' version, labelled ``x,'' made by the union of the northern portion of the R1.4 mask and the southern part of the corresponding R mask. These ``extended'' versions remove a few extra pixels close to the North Galactic Spur. The sky fraction retained by each mask is reported in Table~\ref{table:mask_fsky}.

\begin{table}[htbp!]
\begingroup
\caption{Masks used for polarization analysis at 70\,GHz. Mask R2.2x is used for the foreground cleaning procedure, while R1.8x is used for cosmological parameter estimation.}
\label{table:mask_fsky}
\nointerlineskip
\vskip -3mm
\setbox\tablebox=\vbox{
   \newdimen\digitwidth 
   \setbox0=\hbox{\rm 0} 
   \digitwidth=\wd0 
   \catcode`*=\active 
   \def*{\kern\digitwidth}
   \newdimen\signwidth 
   \setbox0=\hbox{+} 
   \signwidth=\wd0 
   \catcode`!=\active 
   \def!{\kern\signwidth}
\halign{\hbox to 0.8in{#\leaderfil}\tabskip=2em& 
  \hfil#\hfil\tabskip=0pt\cr
\noalign{\doubleline}
\omit\hfil Mask\hfil& $f_{\rm sky}$ [\%]\cr
\noalign{\vskip 3pt\hrule\vskip 5pt}
 R0.6 & 21.4 \cr
 R0.7 & 27.1 \cr
 R0.8 & 32.8 \cr
 R0.9 & 37.9 \cr
 R1.0 & 43.5 \cr
 R1.1 & 48.2\cr
 R1.2 & 52.2 \cr
 R1.4 & 58.1 \cr
 R1.8x & 62.4 \cr
 R2.2x & 66.6 \cr
\noalign{\vskip 5pt\hrule\vskip 3pt}}}
\endPlancktable
\endgroup
\end{table}

\subsubsection{Data set at 70\,GHz}\label{sec:dataset_lfi}

As mentioned above, we use for temperature the \texttt{Commander} map at a resolution of $N_\mathrm{side}\,{=}\,16$. This map is smoothed with a $440^\prime$ Gaussian beam. In order to numerically regularize the inversion of the corresponding covariance matrix, we add to this map a white noise realization with $\sigma_{\rm N}\,{=}\,2\,\mu\mathrm{K}$ in each pixel.\footnote{We add noise to make the covariances invertible {\it after\/} the smoothing.}  When building the pixel-based likelihood, the regularization noise is consistently accounted for in the temperature part of the noise-covariance matrix. 

As in the 2015 analysis, the low-ell pixel-based likelihood in polarization is built using data from the 70-GHz LFI channel.  For the current release, Surveys~2 and 4 that were removed in 2015 can now be included, as suggested by null-test analysis \citepalias[see][for details]{planck2016-l02}. Thus for this legacy release, we use data from all of the eight surveys carried out by \Planck\ LFI.

Linear polarization maps have been downsampled from their native resolution to $N_\mathrm{side}\,{=}\,16$, applying cosine window smoothing (see Eq.~\ref{eq:cosdegrade}).
To clean the 70-GHz $Q$ and $U$ maps we follow a template-fitting procedure, analogous to the one employed in 2015. The cleaned map $\vec{m} \equiv [Q, U]$ is derived as
\begin{equation}\label{eq_cleaning}
\vec{m} = \frac{1}{1-\alpha-\beta}\left ( \vec{m}_{70} - \alpha \vec{m}_{30} - \beta \vec{m}_{353} - \gamma \vec{m}_\mathrm{gain}\right)\,,
\end{equation}
where $\vec{m}_{70}$, $\vec{m}_{30}$, and $\vec{m}_{353}$ are bandpass-corrected versions of the 70-, 30-, and 353-GHz maps. This last equation and the following ones are consistent with Eqs.~\eqref{eq:cleaned_dataset} and~\eqref{eq:cov_cleaned_dataset} of Sect.~\ref{sec:lo-ell:hfi}; however, we repeat them here for convenience.  
  The 353-GHz map employs only PSB data \citepalias{planck2016-l03}.  In Eq.~(\ref{eq_cleaning}), $\alpha$ and $\beta$ are the scaling coefficients for synchrotron and dust emission, respectively, and $\vec{m}_\mathrm{gain}$ is a gain template with its own scaling coefficient $\gamma$ that accounts for the residual systematic leakage occurring during the timeline calibration process (while LFI calibration is photometric, any biases in the process would generate mismatches in the polarization maps; see \citetalias{planck2016-l02} for more details). This last template can be either subtracted with fiducial value (i.e., $\gamma = 1$) or its amplitude can be fitted by letting the coefficients $\alpha$, $\beta$, and $\gamma$ vary jointly. In either case, the fitted coefficients in Eq.~(\ref{eq_cleaning}) are estimated by minimizing the figure of merit 
\begin{equation}
\chi^2 = (1-\alpha-\beta)^2 \vec{m}^\tens{T} \tens{C}_{\tens{S}+\tens{N}}^{-1} \,\vec{m}\,, \label{chi2_alphabeta}
\end{equation}
where
\begin{align}
\tens{C}_{\tens{S}+\tens{N}} & \equiv (1-\alpha-\beta)^2\, \langle \vec{m} \vec{m}^\tens{T} \rangle \\ 
 & = (1-\alpha-\beta)^2\, \tens{S}(C_{\ell}^\mathrm{th})+ \tens{N}_{70} +\alpha^2 \tens{N}_{30}+\beta^2 \tens{N}_{353}\,.\label{CSN_alphabeta}
\end{align}
Here $C_{\ell}^\mathrm{th}$ is the fiducial model given in \citet{planck2014-a10},\footnote{To be explicit, we set $\ln(10^{10} A_{\rm s})=3.0343$ and $\mathrm{\tau}=0.05$, corresponding to $10^9A_{\rm s} e^{-2\tau}\approx1.8808$.} while $\tens{N}_{70}$, $\tens{N}_{30}$, and $\tens{N}_{353}$ are the pure polarization parts of the 70-, 30- and 353-GHz noise-covariance matrices.\footnote{We assume, and have checked explicitly, that the noise-induced $TQ$ and $TU$ correlations are negligible.} The $\tens{N}_{70}$ matrix has been rescaled to match the noise level estimated from the half-ring half-difference (HRHD) map at 70\GHz. In more detail, we have verified that both the 70-GHz HRHD map and its timeline-to-map simulated counterpart show statistically significant excess noise with respect to that predicted by the noise-covariance matrix, obtained as a by-product of the mapmaking process \citep{keskitalo2010}. By decomposing the matrix into spherical harmonics, we find that this bias is not constant, but increases at low multipoles.  Thus an overall multiplicative factor is not appropriate to address this issue. In order to reduce the discrepancy between the noise level described by the original $\tens{N}_{70}$ with respect to what is observed in the data, we apply to its eigenvectors a harmonic filter
\[
f(\ell)=F_0 \left[1 + c \left (\frac{\ell}{16}\right)^\kappa\right],
\] 
where the parameters $F_0$, $c$, and $\kappa$ are determined by maximizing the full-sky log-likelihood of the filtered noise matrix 
with respect to the HRHD map. The values found in the minimization and used in the filter are $F_0 = 0.992$, $c= 0.040$, and $k = -0.966$. This correction also improves in a consistent way the full-sky log-likelihood of the FFP10 simulations HRHD. The origin of this mismatch is not understood.

For the 353-GHz data we use a downsampled version of the high-resolution mapmaking covariance matrix, which only provides $IQU$ correlations within a pixel, while pixel-pixel correlations are assumed to be zero. All noise-covariance matrices and the signal matrix have been smoothed with the same cosine window apodization applied to the $Q$ and $U$ maps. To ensure that the smoothed covariance matrix is numerically well conditioned, we add to the map $\vec{m}_{70}$ a regularization white-noise realization with $20\,\mathrm{nK}$ standard deviation,\footnote{This regularization noise has the same purpose as the temperature regularization described above, yet the value is smaller to ensure that it does not perturb the cosmological parameter estimate in the lower signal-to-noise polarization case.} which is also accounted for in the covariance matrix. 

The final polarization noise-covariance matrix used in the pixel-based likelihood and associated to the {\it foreground-cleaned\/} map in Eq.~(\ref{eq_cleaning}) is given by
\begin{align}
\tens{N} =\,& \frac{1}{(1-\alpha-\beta)^2}\left( \tens{N}_{70} + \alpha^2 \tens{N}_{30}+\beta^2 \tens{N}_{353} \right. \nonumber\\
&\qquad\qquad\qquad \left. +\,\sigma_\alpha^2 \vec{m}_{30} \vec{m}^\tens{T}_{30} + \sigma_\beta^2 \vec{m}_{353} \vec{m}^\tens{T}_{353}\right), 
\label{lowl_covmat}
\end{align}
where $\sigma_\alpha^2$ and $\sigma_\beta^2$ are the uncertainties associated with the foreground scaling-coefficients we estimate. When fitting for the gain template, the effective noise-covariance matrix should properly include a contribution from the uncertainty on the scaling factor $\gamma$, as is done for the foreground scaling-coefficients. Since for the final results we fix $\gamma=1$, we do not include such a term in Eq.~(\ref{lowl_covmat}).

\begin{figure}[htbp!]
\includegraphics[width=0.475\textwidth]{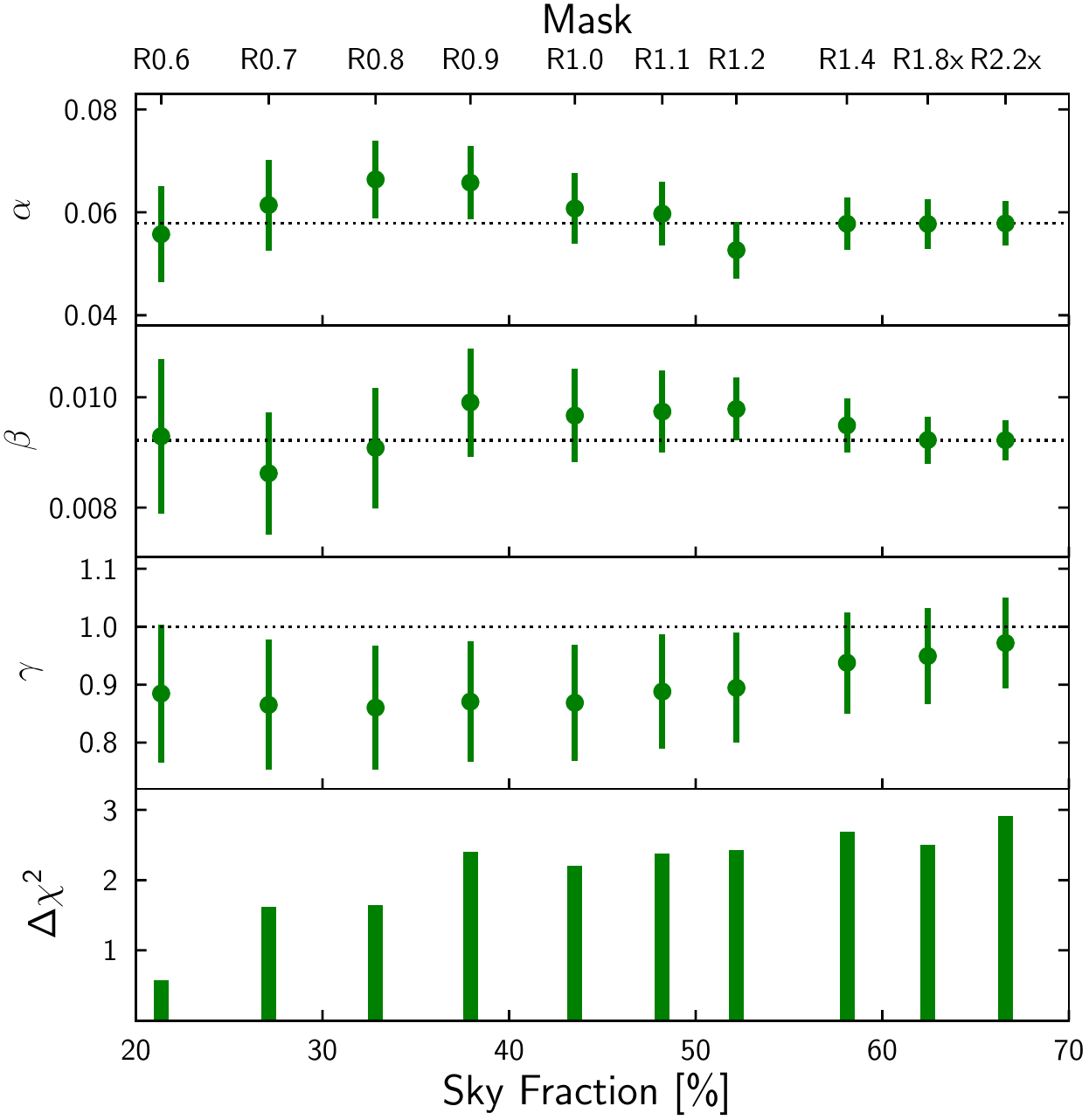}
\caption[]{Scaling coefficients for synchrotron ($\alpha$), dust ($\beta$), and the amplitude of the gain template ($\gamma$) estimated on several masks. 
Dotted lines mark the values of the coefficients for mask R2.2x, which are the ones we use to build the cleaned data set for the pixel-based likelihood.
The bottom panel shows the $\chi^2$ excess of the fit, defined as $\Delta \chi^2 = (\chi^2 - N_{\rm dof})/\sqrt{2 N_{\rm dof}}$.
}
\label{fig_alpha_beta_gamma_chi2}
\end{figure}

In Fig.~\ref{fig_alpha_beta_gamma_chi2} we show $\alpha$, $\beta$, and $\gamma$ estimated on the different masks described in the previous sub-section. For each mask, we also report the $\chi^2$ excess of the fit, defined as $\Delta \chi^2 \equiv (\chi^2 - N_{\rm dof})/\sqrt{2 N_{\rm dof}}$, $N_{\rm dof}$ being the number of active pixels in the $Q$ and $U$ masks. We consider only masks for which $\Delta \chi^2 < 3$. We find good consistency among the coefficients estimated on different sky fractions and therefore choose as the processing mask the one retaining the largest sky fraction, {that is}, mask R2.2x, with $f_{\rm sky}=66.6\,\%$. 

\begin{figure}[htbp!]
\includegraphics[width=0.5\textwidth]{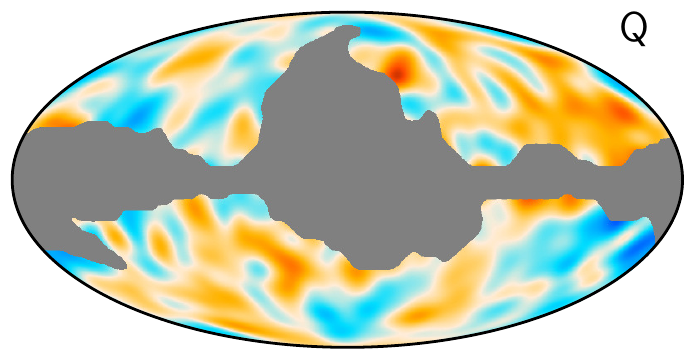}
\includegraphics[width=0.5\textwidth]{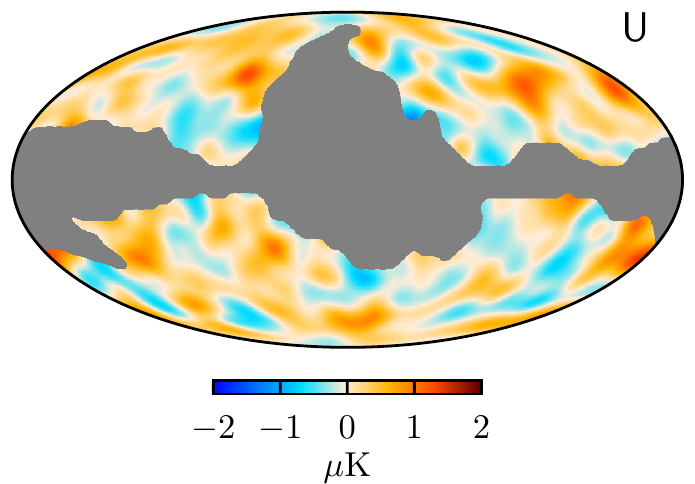}
\caption[]{Foreground-cleaned 70-GHz maps of the Stokes parameters $Q$ (top) and $U$ (bottom) 
on 62.4\,\% of the sky (mask R1.8x). In order to highlight the largest angular scales relevant for the low-$\ell$ likelihood, these 
maps have been smoothed with a Gaussian filter of ${\rm FWHM}\,{=}\,10^\circ$. As a reference, for $\tau\approx0.06$ the expected range for the signal is $\pm 0.4\,\mu {\rm K}$.}
\label{mollview_cleanmaps}
\end{figure}

For the pixel-based likelihood baseline, we build the 70-GHz cleaned maps of Eq.~(\ref{eq_cleaning}) with the coefficients $\alpha=0.058 \pm 0.004$, $\beta=0.0092 \pm 0.0004$, and $\gamma =1$, estimated on mask R2.2x. These maps are shown in Fig.~\ref{mollview_cleanmaps}. We have verified that if we keep the amplitude of the gain template fixed at $\gamma =1$, the foreground coefficients we estimate are indistinguishable in practice from the case where we let $\gamma$ vary. We have also verified that this choice has negligible impact on the parameter $\tau$.  If we compare these estimates to the foreground coefficients of the 2015 analysis, we find that, thanks to the larger sky fraction retained by the processing mask (66.6\,\% compared to 43\,\% in 2015), the uncertainties have been almost halved. The scaling coefficients for the synchrotron emission are compatible between the two analyses at the level of $0.6\,\sigma$, where $\sigma$ refers to the largest of the two uncertainties. The foreground coefficient for dust, on the other hand, has shifted higher by about $2\,\sigma$ with respect to 2015. This shift is likely due to the improved quality of the Planck 2018 353-GHz map compared with the 2015 release \citepalias[see figures~11 and 14 of][]{planck2016-l03}.

\subsubsection{Power spectra}\label{sec:lfi_powerspec}

We use the quadratic maximum-likelihood (QML) code {\tt Bolpol} \citep{Gruppuso:2009,2012JCAP...02..023G} to estimate the polarized angular power spectra of the data set used in the low-$\ell$ pixel-based likelihood. Specifically, we use the \texttt{Commander} map in temperature and the 70-GHz cleaned $Q$ and $U$ maps given in Eq.~(\ref{eq_cleaning}) and shown in Fig.~\ref{mollview_cleanmaps}. For the covariance matrix, we use the noise-covariance matrix of Eq.~(\ref{lowl_covmat}) for the polarization part, while for temperature we assume only a $4\,\mu \mathrm{K}^2$ diagonal regularization noise. Consistently, a white noise realization of the corresponding variance has been added to the \texttt{Commander} map. 

\begin{figure}[htbp!]
\includegraphics[width=0.475\textwidth]{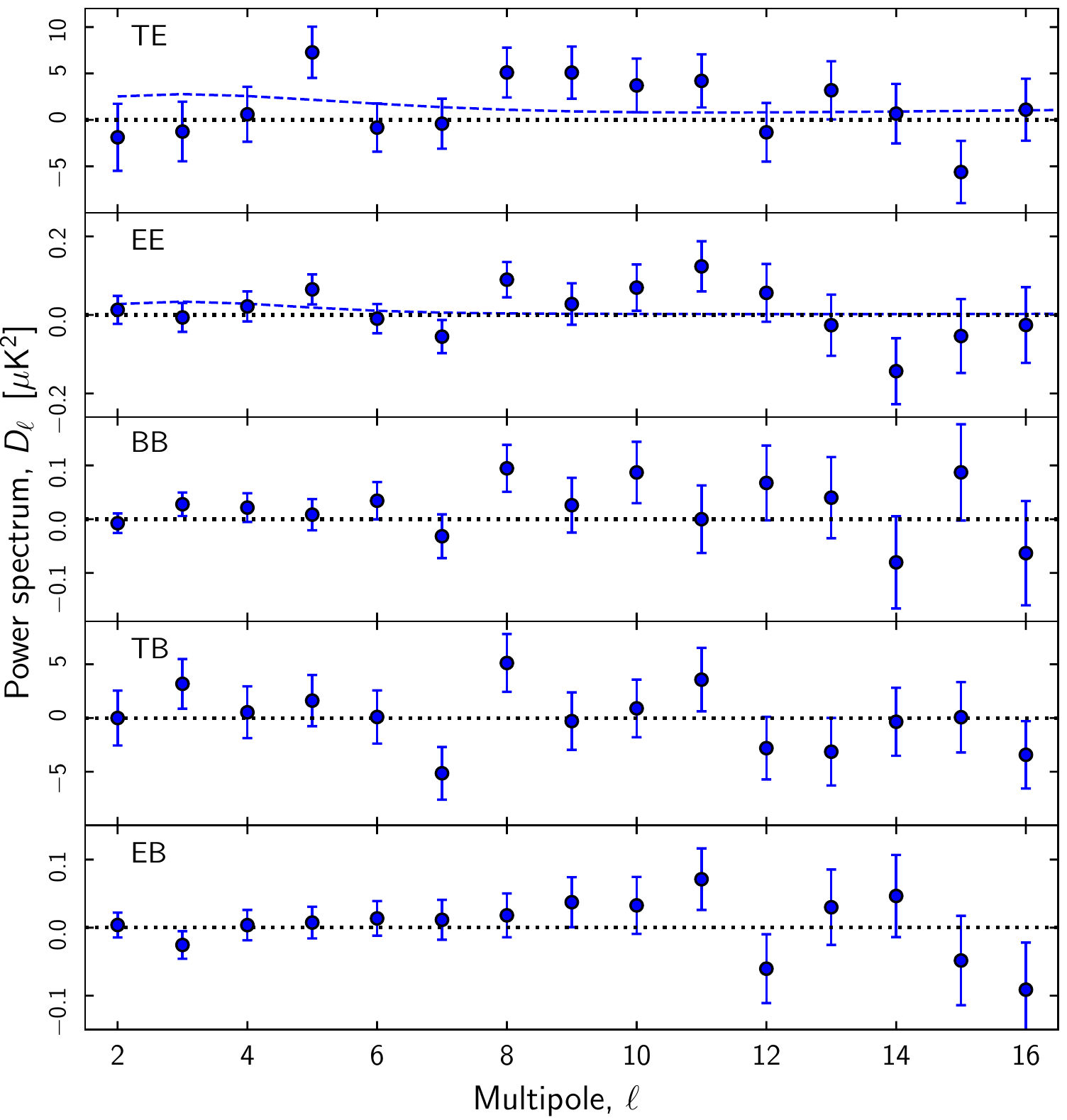}
\caption[]{Polarization power spectra of the cleaned LFI 70-GHz maps, estimated on the \texttt{Commander} mask in temperature and the cosmological analysis mask R1.8x ($f_{\rm sky}=62.4\,\%$) in polarization. Blue dashed lines represent the model with $\tau = 0.05$.}
\label{fig_spectra}
\end{figure}

Figure~\ref{fig_spectra} shows the spectra estimated employing the \texttt{Commander} mask in temperature.  For polarization, we use the mask employed for cosmological analysis, R1.8x ($f_{\rm sky}=62.4\,\%$).  The choice of this last mask as the reference for polarization will be explained in Sect.~\ref{sec:lfi_valsim}.  Error bars on the individual entries of the power spectrum have been derived from the Fisher information matrix of the QML estimator.

In order to test the consistency of the estimated power spectra $C_{\ell}$ with the model $C_{\ell}^\textrm{th}$ introduced in the previous section, we compute the harmonic $\chi^2_{\rm h}$:
\begin{equation}\label{chi2hred}
\chi_\textrm{h}^2 = \sum_{\ell,\ell^{\prime}=2}^{\ell_\textrm{max}} (C_{\ell} - C_{\ell}^\textrm{th}) \, \tens{M}_{\ell \ell^{\prime}}^{-1} \, (C_{\ell^{\prime}} - C_{\ell^{\prime}}^\textrm{th})\, ,
\end{equation} 
where $\tens{M}_{\ell \ell^{\prime}} = \langle (C_{\ell} - C_{\ell}^\textrm{th}) (C_{\ell^{\prime}} - C_{\ell^{\prime}}^\textrm{th}) \rangle $, and the average is taken over 1000 signal and noise simulations. Signal simulations are drawn from the $C_{\ell}^\textrm{th}$ model, while noise simulations are generated using the noise-covariance matrix given in Eq.~(\ref{lowl_covmat}). We also use the simulations to sample the empirical distribution for $\chi_\textrm{h}^2$, considering $\ell_\textrm{max}=6$, 12, and 30, for each of the six CMB polarization spectra ($TT$, $TE$, $TB$, etc.).  We report in Table~\ref{table:lowl_harmPTE} the empirical probability of observing a value of $\chi_\textrm{h}^2$ greater than that for the data (i.e., the probability to exceed or PTE). This test supports the hypothesis that the observed spectra are consistent with the best-fit model from \citet{planck2014-a10}, {that is}, $\ln(10^{10} A_{\rm s})=3.0343$ and $\mathrm{\tau}=0.05$, and the propagated instrumental uncertainties. {We note} that for spectra involving $B$ modes, the fiducial model is null, making this, in fact, a null test, probing instrumental characteristics and data processing independently of any cosmological assumption. 


\begin{table}[htbp!]
\begingroup
\newdimen\tblskip \tblskip=5pt
\caption{Empirical probability of observing a value of $\chi_{\rm h}^2$ greater than that calculated from the data.
}
\label{table:lowl_harmPTE}
\nointerlineskip
\vskip -3mm
\setbox\tablebox=\vbox{
   \newdimen\digitwidth 
   \setbox0=\hbox{\rm 0} 
   \digitwidth=\wd0 
   \catcode`*=\active 
   \def*{\kern\digitwidth}
   \newdimen\signwidth 
   \setbox0=\hbox{+} 
   \signwidth=\wd0 
   \catcode`!=\active 
   \def!{\kern\signwidth}
\halign{\hbox to 0.8in{#\leaderfil}\tabskip=2em&
   \hfil#\hfil&
   \hfil#\hfil&
   \hfil#\hfil\tabskip=0pt\cr
\noalign{\doubleline}
\omit&\multispan3\hfil PTE [\%]\hfil\cr
\noalign{\vskip -5pt}
\omit&\multispan3\hrulefill\cr
\omit\hfil Spectrum\hfil& $\ell_\textrm{max}=6$ & $\ell_\textrm{max}=12$& $\ell_\textrm{max}=30$\cr
\noalign{\vskip 3pt\hrule\vskip 5pt}
$TT$& 34.9& 78.3& 90.2\cr
$TE$& 17.2& 15.8& 29.9\cr
$EE$& 60.9& 17.7& 25.0\cr
$BB$& 51.2& 24.2& *4.5\cr
$TB$& 72.3& 26.6& 63.8\cr
$EB$& 76.0& 61.6& 28.2\cr
\noalign{\vskip 5pt\hrule\vskip 3pt}}}
\endPlancktable 
\endgroup
\end{table}


\subsubsection{Cosmological parameters}\label{sec:lfi_cosmopar}

We combine the \commander\ solution in temperature with the 70-GHz data set (with foreground mitigated as discussed in Sect.~\ref{sec:dataset_lfi} above) to estimate cosmological parameters via the pixel-based likelihood approach of Eq.~(\ref{eq:pbLike}). As done for Table~\ref{tab:results_lowell_hfi}, we fix all parameters except $\ln(10^{10} A_{\rm s})$, $\tau$, and $r$ to the following fiducial values: $\{\Omega_{\rm b}h^2\,{=}\,0.0221, \Omega_{\rm c}h^2\,{=}\,0.12, \theta_\ast\,{=}\,1.0411, n_{\rm s}\,{=}\,0.96\}$. We also assume the inflationary consistency relation for $n_{\rm t}$ \citep{planck2016-l10}.

\begin{figure}[htbp!]
\centering
\includegraphics[width=0.5\textwidth]{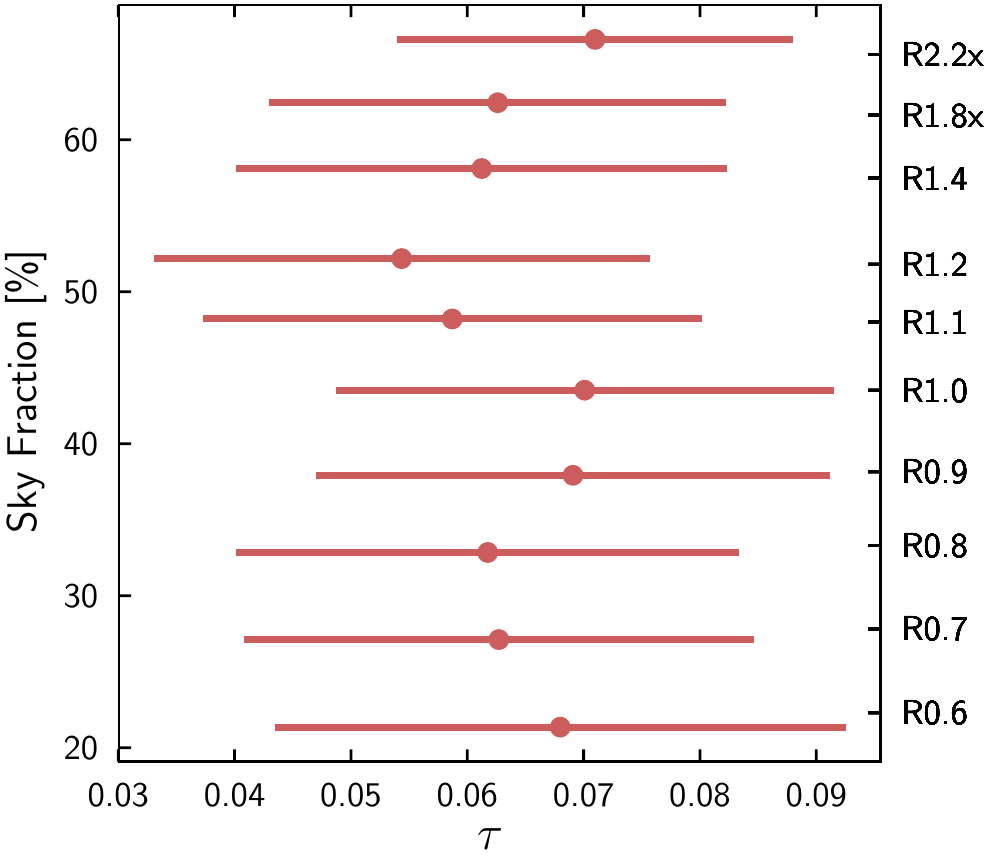}
\caption[]{Optical depth to reionization, $\tau$, and 68\,\% CL uncertainties estimated with the LFI polarization pixel-based likelihood on several masks, with sky fractions ranging from about 20\,\% to 70\,\%.}
\label{fig_cosmopar_fsky}
\end{figure}

We first investigate the dependence of the parameter $\tau$, the optical depth to reionization, on the amount of sky left unmasked. 
To this end, we only fit for $\tau$ and $A_{\rm s}$ on a grid of models, while keeping $r=0$. Figure~\ref{fig_cosmopar_fsky} shows the estimated values of $\tau$, 
and the associated $68\,\%$ confidence limit (hereafter ``CL'') uncertainties, for the ten masks of Table~\ref{table:mask_fsky}.

As will be explained in the following section, we choose mask R1.8x as the baseline for the polarization analysis. The corresponding constraint on the reionization optical depth (shown in Table~\ref{table:params_ext}) is $\tau = 0.063 \pm 0.020$
, to be compared to the value obtained in the 2015 analysis, which was $\tau = 0.067 \pm 0.023$. The values of $\tau$ are compatible at the level of $0.2\,\sigma$, where $\sigma$ refers to the uncertainty on the parameter derived in the 2015 analysis. The uncertainty itself has decreased with respect to 2015, mainly for the following reasons: the larger sky fraction used in the analysis (62.4\,\% for the cosmology mask compared to 46\,\%\footnote{The conservative choice for the mask in the 2015 analysis was driven by the $BB$ residual present for masks using larger sky fractions.}); the inclusion of the full set of data collected by the LFI, including Surveys~2 and 4, which were not considered in the 2015 analysis; and, finally, the decision to use a cosine-apodized pixel window in place of the square window used in 2015. If, however, we compare the uncertainty on $\tau$ from the 2018 data to what we see in the simulations, we find that the expected error should be lower, at $0.014$.  This is a nearly $3\,\sigma$ outlier, as discussed in more detail in Sect.~\ref{sec:lfi_valsim} below. On the other hand, the expected error for $\ln(10^{10}A_{\rm s})$ is in line with expectations from the simulations.

For all the estimates discussed above, we have assumed the tensor-to-scalar ratio $r=0$. If instead we also fit for $r$, we obtain a 95\,\% CL upper limit of $\mathrm{r}_{0.002} \leq 1.7$ (or, equivalently, $r_{0.05} \leq 1.2$, where the subscript gives the $k$ scale in units of ${\rm Mpc}^{-1}$). This value is slightly weaker than the corresponding \Planck\ 2015 constraint. The reason is likely linked to the $B$-mode excess visible in Fig.~\ref{fig_spectra}, which drives the posterior on $r$ towards higher values, without however producing any significant detection. In any case, when using the LFI low-$\ell$ likelihood in conjunction with the high-$\ell$ results, the corresponding constraint on $r$ is not impacted, since it is dominated by the statistically more powerful temperature constraint.

\begin{table}[htbp!]
\begingroup
\caption{Constraints on $\ln(10^{10}A_{\rm s})$, $\tau$, and $r$ from the low-$\ell$ LFI likelihood. We show mean and 68\,\% confidence levels. For $r$, the 95\,\% upper limit is shown.}
\label{table:params_ext}
\nointerlineskip
\vskip -3mm
\setbox\tablebox=\vbox{
   \newdimen\digitwidth 
   \setbox0=\hbox{\rm 0} 
   \digitwidth=\wd0 
   \catcode`*=\active 
   \def*{\kern\digitwidth}
   \newdimen\signwidth 
   \setbox0=\hbox{+} 
   \signwidth=\wd0 
   \catcode`!=\active 
   \def!{\kern\signwidth}
\halign{\hbox to 0.9in{#\leaderfil}\tabskip=1em&
  \hfil#\hfil\tabskip=10pt& 
  \hfil#\hfil\tabskip=0pt\cr
\noalign{\doubleline}
\omit\hfil Parameter \hfil& $\Lambda$CDM & $\Lambda$CDM + $r$ \cr
\noalign{\vskip 5pt\hrule\vskip 5pt}
$\ln(10^{10} A_{\rm s})$& $2.965\pm0.055$& $2.70^{+0.23}_{-0.13}$\cr
\noalign{\vskip 2pt}
$\tau$& $0.063 \pm 0.020$& $0.064\pm0.019$\cr
$r_{0.002}$& \dots& $\leq1.7$\cr
\noalign{\vskip 2pt}
$10^9A_{\rm s}e^{-2\tau}$& $1.710^{+0.086}_{-0.087}$& $1.33^{+0.21}_{-0.26}$\cr
\noalign{\vskip 4pt\hrule\vskip 3pt}}}
\endPlancktable
\endgroup
\end{table}

\subsubsection{Validation}\label{sec:lfi_valsim}

We now describe the Monte Carlo validation of the LFI 70-GHz low-$\ell$ polarization likelihood. This validation is based on realistic simulations to test the entire pipeline, including both the likelihood template-fitting procedure to model and remove dust and synchrotron contamination in 70-GHz $Q$ and $U$ maps, and the pixel-based Gaussian likelihood. We build a set of 1000 simulations combining CMB, foregrounds, and noise at 30, 70 and 353\GHz. To be consistent with the data, all the simulated maps and the noise-covariance matrices are smoothed using a Gaussian beam of $440'$ in temperature and a cosine window in polarization. {For} temperature{,} we assume perfect foreground cleaning{,} and{,} therefore{,} we use pure CMB maps to which we add regularization white noise with $2\,\mu\mathrm{K}$ standard deviation. The CMB realizations are generated from the fiducial model of \citet{planck2014-a10}, where we set $\ln(10^{10} A_{\rm s})=3.0343$ and $\mathrm{\tau}=0.05$. In polarization, foregrounds are taken from the Full Focal Plane 10 (FFP10) suite of simulations, whereas the noise simulations have been randomly generated from the noise-covariance matrices for each of the different frequency channels. 
Analogously to what is done to the data, we add regularization noise with standard deviation $\sigma_{\rm N}=20\,{\rm nK}$ also to the simulated maps and associated noise-covariance matrices at 70\GHz. We estimate parameters using the \commander\ mask for temperature ($f_{\rm sky} = 86\,\%$), and for polarization the ten masks given in Table~\ref{table:mask_fsky}, with $f_{\rm sky}$ ranging from 21\,\% to 67\,\%. However, in order to replicate what is done to the data, we perform the foreground cleaning of each simulation on mask R2.2x, as described in Sect.~\ref{sec:dataset_lfi}.

When estimating the cosmological parameters of the simulations, we only focus on $\ln(10^{10} A_{\rm s})$ and $\tau$, keeping the other cosmological parameters fixed to their fiducial values. We verify that the mean values of $\ln(10^{10} A_{\rm s})$ and $\tau$ measured from the 1000 simulations properly recover the input values for all the masks considered in this analysis. In particular, for the mask that we select as the baseline for cosmological studies, R1.8x, we find a bias from the input, expressed in units of the standard deviation of the mean, at the level of $0.56$ for $\tau$ and $0.30$ for $\ln(10^{10} A_{\rm s})$.

The main motivation behind the validation test, however, is to assess how likely are the variations we see in the $\tau$ values measured on the data for different masks (see Fig.~\ref{fig_cosmopar_fsky}). For each simulation and each mask given in Sect.~\ref{sec:masks} we estimate the cosmological parameters, $\ln(10^{10} A_{\rm s})$ and $\tau$. We then compute the variation of parameter estimates between all 45 pairs of different masks, across the 1000 simulations described above. We thus effectively build empirical distributions of the parameter variations between masks, properly taking into account the correlations between parameters from different masks, which in several cases exhibit large overlaps. When comparing the $\Delta \tau$ and $\Delta \ln(10^{10} A_{\rm s})$ measured on the data to the empirical distributions, we find that the observed variations in the data are compatible with expectations from simulations. 

\begin{figure}[htbp!]
\centering
\includegraphics[width=0.49\textwidth]{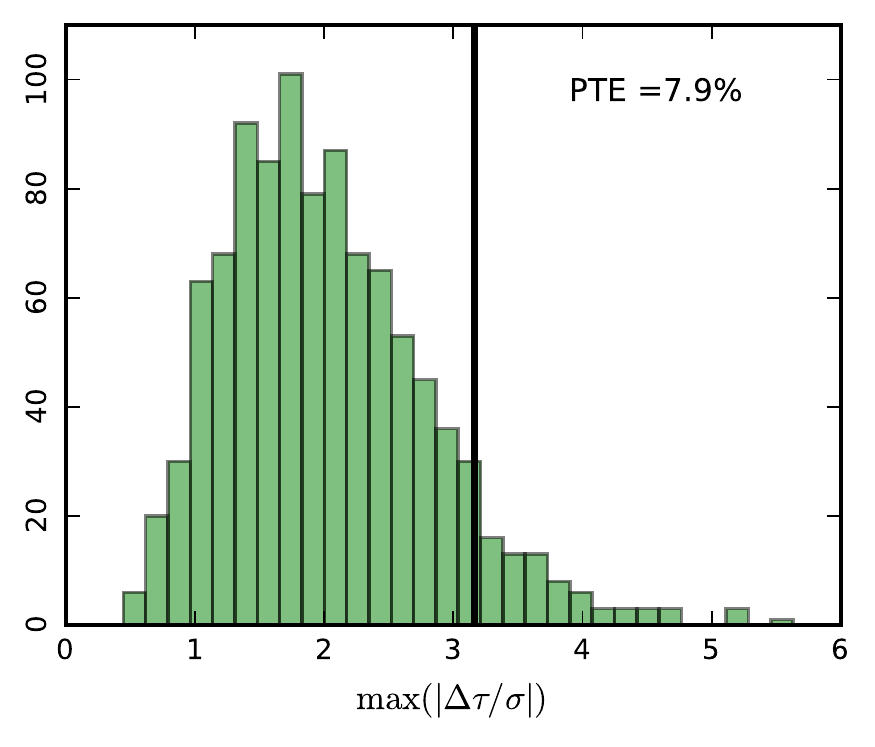}
\caption[]{Histogram of the the most discrepant shift for $\Delta \tau$ measured on 1000 simulations compared with the data (vertical line), as explained in Sect.~\ref{sec:lfi_valsim}.}

\label{fig:pte_global}
\end{figure}

Based on the distribution of the shifts from simulations, we provide in Table~\ref{table:pte_maskvalidation1}, for each pair of masks, the PTE for the absolute value of the shifts in $\tau$ and $\ln(10^{10} A_{\rm s})$ observed in the data. Since we find low probabilities for mask differences involving mask R2.2x, we choose to conservatively use R1.8x as the baseline for cosmological results. However, we also find low probabilities for some mask differences at intermediate sky fractions (e.g., R1.0 $-$ R1.1, and R1.0 $-$ R1.2), and we need to assess the actual global significance of these outliers in order to establish the overall consistency of the data. We focus on $\tau$, because $\ln(10^{10} A_{\rm s})$ behaves similarly, given that the two parameters are highly correlated. For each simulated sky, we search for the most discrepant shift across all the available mask differences, and we express it in units of the standard deviation estimated from simulations, $\mathrm{max(|\Delta \tau/\sigma|)}$. We then compare the parameter shift of the data for the maximally discrepant mask difference against the distribution of the maximally discrepant shifts from simulations, and we compute the associated PTE, as shown in Fig.~\ref{fig:pte_global}.  We conclude that the data are actually consistent across a broad range of sky fractions.

If we now focus on the R1.8x mask that we choose as the baseline for cosmological results and we look at the distribution of the uncertainties on the parameters from the 1000 simulations, we find an average uncertainty of $0.014\pm 0.002$ on $\tau$ and of $0.0522\pm0.0016$ on $\ln(10^{10} A_{\rm s})$.  The quoted error on these uncertainties is computed from the dispersion in the outcomes of the simulations.  {In} the case of $\tau$ the expected uncertainty from simulations is significantly smaller than the one derived from the data, which is $0.020$ at 1$\,\sigma$. In fact, only 0.9\,\% of the simulations exhibit an error on $\tau$ larger than the data, so our results are a nearly 3$\,\sigma$ outlier. These results, while not statistically significant, may well hint at the possibility that the final covariance matrix in Eq.~(\ref{lowl_covmat}) is not capturing the full error budget in the data. On the other hand, the uncertainty derived for $\ln(10^{10} A_{\rm s})$ is well within 2$\,\sigma$ of the expected value.

\begin{table*}[htbp!]
\begingroup
\newdimen\tblskip \tblskip=5pt
\caption{Consistency of the parameters $\tau$ and $\ln(10^{10} A_{\rm s})$ estimated on different masks. For each pair of masks defined in Sect.~\ref{sec:masks}, the probability to exceed (PTE) gives the percentage of simulations with absolute parameter shifts greater than the ones measured on the data.
}
\label{table:pte_maskvalidation1}
\nointerlineskip
\vskip -3mm
\setbox\tablebox=\vbox{
   \newdimen\digitwidth 
   \setbox0=\hbox{\rm 0} 
   \digitwidth=\wd0 
   \catcode`*=\active 
   \def*{\kern\digitwidth}
   \newdimen\signwidth 
   \setbox0=\hbox{+} 
   \signwidth=\wd0 
   \catcode`!=\active 
   \def!{\kern\signwidth}
\halign{\hbox to 1.0in{#\leaderfil}\tabskip=0.5em& 
  \hfil#\hfil\tabskip=1em&\hfil#\hfil\tabskip=4em&
  \hbox to 1.0in{#\leaderfil}\tabskip=0.5em& 
  \hfil#\hfil\tabskip=1em&\hfil#\hfil\tabskip=4em&
  \hbox to 1.0in{#\leaderfil}\tabskip=0.5em& 
  \hfil#\hfil\tabskip=1em&\hfil#\hfil\tabskip=0pt\cr
\noalign{\doubleline}
\omit&\multispan2\hfil PTE [\%]\hfil&\omit&\multispan2\hfil PTE [\%]\hfil&\omit&\multispan2\hfil PTE [\%]\hfil\cr
\noalign{\vskip -4pt}
\omit&\multispan2\hrulefill&\omit&\multispan2\hrulefill&\omit&\multispan2\hrulefill\cr
\omit\hfil Masks\hfil & $\lvert\Delta \tau\rvert$ & $\lvert\Delta \ln(10^{10} A_{\rm s})\rvert$&\omit\hfil Masks\hfil & $\lvert\Delta \tau\rvert$ & $\lvert\Delta \ln(10^{10} A_{\rm s})\rvert$&\omit\hfil Masks\hfil & $\lvert\Delta \tau\rvert$ & $\lvert\Delta \ln(10^{10} A_{\rm s})\rvert$\cr
\noalign{\vskip 3pt\hrule\vskip 5pt}
R0.6 $-$ R0.7& 46.0& 54.0&R0.7 $-$ R1.8x& 99.6& 97.6&R1.0 $-$ R1.1&  0.7&  0.8\cr
R0.6 $-$ R0.8& 49.5& 57.9&R0.7 $-$ R2.2x& 44.5& 48.0&R1.0 $-$ R1.2&  0.6&  0.7\cr
R0.6 $-$ R0.9& 91.9& 89.9&R0.8 $-$ R0.9& 16.7& 21.9&R1.0 $-$ R1.4& 15.6& 21.2\cr
R0.6 $-$ R1.0& 85.7& 82.0&R0.8 $-$ R1.0& 24.1& 28.6&R1.0 $-$ R1.8x& 27.4& 33.3\cr
R0.6 $-$ R1.1& 45.0& 49.5&R0.8 $-$ R1.1& 65.1& 63.7&R1.0 $-$ R2.2x& 89.1& 86.2\cr
R0.6 $-$ R1.2& 28.9& 33.5&R0.8 $-$ R1.2& 36.4& 37.3&R1.1 $-$ R1.2& 17.5& 21.8\cr
R0.6 $-$ R1.4& 61.1& 70.2&R0.8 $-$ R1.4& 94.1& 97.9&R1.1 $-$ R1.4& 60.2& 53.9\cr
R0.6 $-$ R1.8x& 68.8& 76.8&R0.8 $-$ R1.8x& 91.9& 90.5&R1.1 $-$ R1.8x& 50.3& 45.7\cr
R0.6 $-$ R2.2x& 84.7& 80.7&R0.8 $-$ R2.2x& 36.0& 37.8&R1.1 $-$ R2.2x&  8.2&  9.0\cr
R0.7 $-$ R0.8& 85.8& 88.7&R0.9 $-$ R1.0& 83.0& 76.8&R1.2 $-$ R1.4&  9.3&  8.7\cr
R0.7 $-$ R0.9& 39.7& 44.9&R0.9 $-$ R1.1&  9.4& 12.6&R1.2 $-$ R1.8x& 11.8& 11.4\cr
R0.7 $-$ R1.0& 41.5& 44.8&R0.9 $-$ R1.2&  4.3&  5.8&R1.2 $-$ R2.2x&  1.5&  1.6\cr
R0.7 $-$ R1.1& 67.3& 65.8&R0.9 $-$ R1.4& 30.1& 40.0&R1.4 $-$ R1.8x& 66.2& 68.6\cr
R0.7 $-$ R1.2& 41.3& 42.6&R0.9 $-$ R1.8x& 42.4& 52.1&R1.4 $-$ R2.2x&  4.8&  6.0\cr
R0.7 $-$ R1.4& 89.3& 92.5&R0.9 $-$ R2.2x& 81.5& 77.7&R1.8x $-$ R2.2x&  1.7&  2.4\cr
\noalign{\vskip 5pt\hrule\vskip 3pt}}}
\endPlancktablewide
\endgroup
\end{table*}


\subsubsection{Comparison with WMAP}
We now turn to a comparison of the results we obtain from LFI with those obtained from the WMAP-9 data set \citep{bennett2012} when it is processed in a similar way to the LFI data. In polarization, the two data sets have been foreground-cleaned separately, while for temperature we keep the \Planck\ \texttt{Commander} solution for both. For LFI, we follow the procedure described in Sect.~\ref{sec:dataset_lfi}, relying on mask R2.2x. For WMAP, instead, we follow a procedure similar to the above, but we use the K-band map as a synchrotron tracer, while keeping the 353-GHz map for dust. The Ka, Q, and V channel maps have first been smoothed with the same cosine window used for LFI, then foreground-cleaned individually on the WMAP analysis mask P06. The estimated foreground coefficients and associated $\chi^2$ values are given in Table~\ref{table:lowl_WMAPscalings_p06}. We then combine the different channels into a single noise-weighted average map. Noise levels used in the weighting are based on measurements taken outside of the analysis mask. The result is the foreground-cleaned WMAP map used for the analyses below. 

\begin{figure}[htbp!]
\centering
\includegraphics[width=0.475\textwidth]{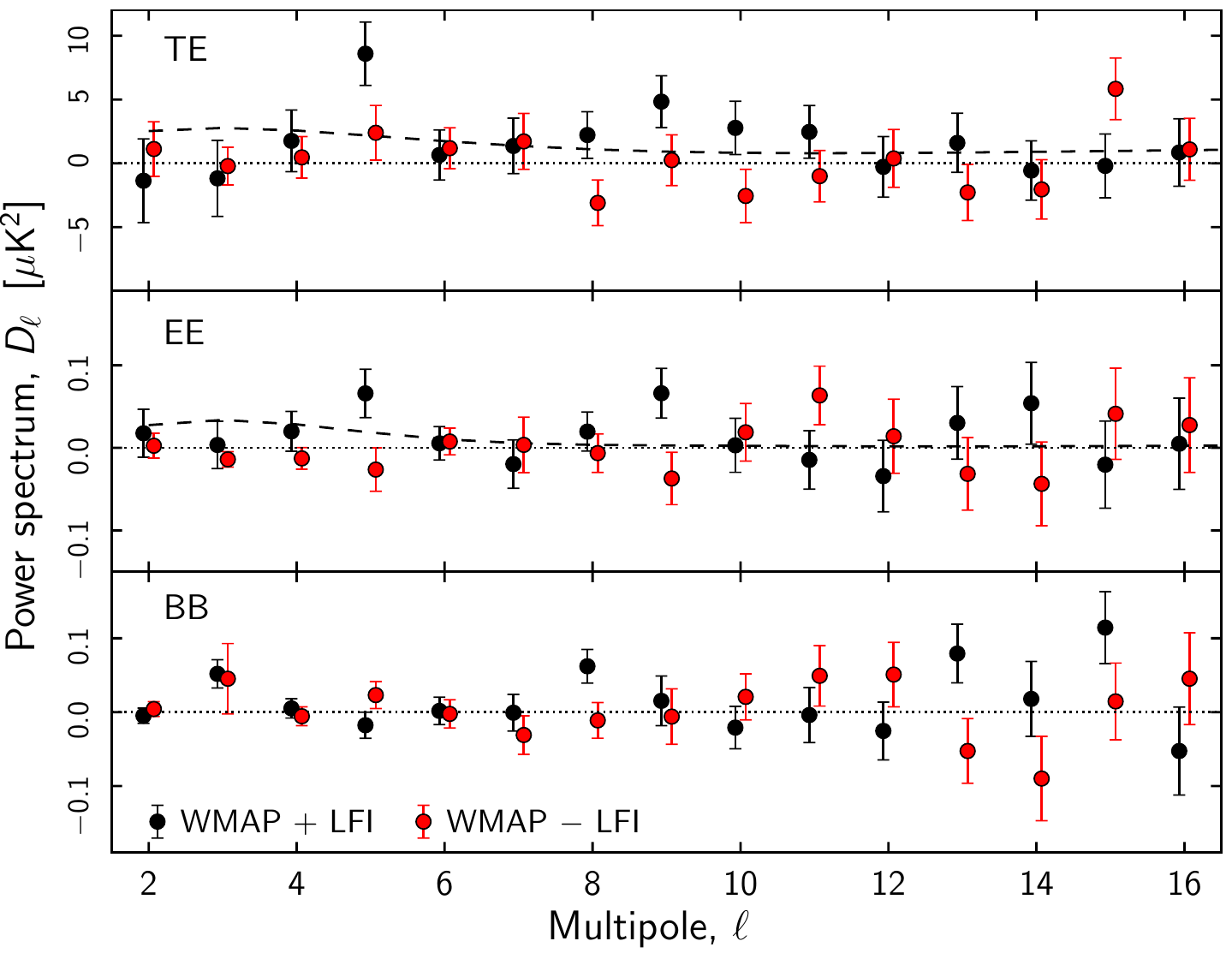}
\caption[]{ 
Angular power spectra, computed using the QML approach, of the noise-weighted sum of the WMAP and LFI data sets (see text for details), compared to spectra from their half-difference. The model predictions shown with the dashed lines are the same as shown in Fig.~\ref{fig_spectra}, with $\tau = 0.05$.
}
\label{plot_ps_lfi_wmap_fgcP06}
\end{figure}

To ease the comparison of the CMB polarization signal measured by the two experiments, we also consider a noise-weighted combination of the LFI and WMAP foreground-cleaned maps (WMAP\,+\,LFI), together with their half-difference (WMAP\,$-$\,LFI). Both have been conservatively obtained employing a mask given by the union of the mask R2.2x with the WMAP P06 analysis mask. The QML power spectra of these maps are shown in Fig.~\ref{plot_ps_lfi_wmap_fgcP06}.

In order to assess the consistency of the polarization signal measured by LFI and WMAP in $TE$ and $EE$ over the multipole range of the reionization bump, we evaluate the harmonic $\chi_\textrm{h}^2$ for $C_{\ell}^{TE}$, $C_{\ell}^{EE}$, and the two spectra jointly $[C_{\ell}^{TE}, C_{\ell}^{EE}]$, including data up to $\ell_{\mathrm{max}}=12$. We first compare $\chi_\textrm{h}^2$ for the noise-weighted sum (WMAP\,+\,LFI) to an empirical distribution sampled from 1000 Monte Carlo simulations containing only noise and residual foregrounds. We find that none of the simulations has a $\chi_\textrm{h}^2$ value higher than the data in $TE$, and an empirical PTE of 2.1\,\% in $EE$. As a second test, we compare the same data $\chi_\textrm{h}^2$ to an empirical distribution, again sampled from 1000 Monte Carlo simulations, which this time, in addition to noise and residual foregrounds, also contains a CMB signal realization, generated assuming the cosmological model given in the previous section (again with $\tau = 0.05$). In this case, we find empirical PTEs of $17.0\,\%$, $22.1\,\%$, and $33.5\,\%$ for $C_{\ell}^{TE}$, $C_{\ell}^{EE}$, and $[C_{\ell}^{TE}, C_{\ell}^{EE}]$, respectively. If we repeat the test for the half-difference data set (WMAP\,$-$\,LFI) and compare the $\chi_\textrm{h}^2$ of the data to a set of 1000 Monte Carlo simulations of noise and residual foregrounds, we find empirical PTEs of $67.7\,\%$, $59.5\,\%$, and $73.1\,\%$. These results strongly support the hypothesis that the spectra for the weighted combination WMAP\,+\,LFI does contain a significant cosmological signal, while the half-difference WMAP\,$-$\,LFI is compatible with noise and residual foregrounds. Furthermore, the $BB$ spectrum is consistent with the null-hypothesis for both the noise-weighted sum (PTE = 13.2\,\%) and half-difference (PTE = 59.6\,\%).

\begin{table}[htbp!]
\begingroup
\caption{Scaling coefficients for synchrotron ($\alpha$) and dust ($\beta$) obtained for the WMAP data set described in the text, using WMAP K band and \Planck\ 353-GHz maps as templates. We also report the associated reduced $\chi^2$ values.}
\label{table:lowl_WMAPscalings_p06}
\nointerlineskip
\vskip -3mm
\setbox\tablebox=\vbox{
   \newdimen\digitwidth 
   \setbox0=\hbox{\rm 0} 
   \digitwidth=\wd0 
   \catcode`*=\active 
   \def*{\kern\digitwidth}
   \newdimen\signwidth 
   \setbox0=\hbox{+} 
   \signwidth=\wd0 
   \catcode`!=\active 
   \def!{\kern\signwidth}
\halign{\hbox to 0.7in{#\leaderfil}\tabskip=1em&
  \hfil#\hfil& 
  \hfil#\hfil\tabskip=8pt&
  \hfil#\hfil\tabskip=6pt&
  \hfil#\hfil\tabskip=0pt\cr
\noalign{\doubleline}
\noalign{\vskip -2pt}
\omit\hfil Band\hfil& $f_{\rm sky}$&$\alpha$& $\beta$ & $\chi^2_{\mathrm{red}}$\cr
\noalign{\vskip 3pt\hrule\vskip 5pt}
Ka& 73\,\%& $0.318\pm0.003$& $0.0036\pm0.0005$& $1.004$\cr
Q& 73\,\%&  $0.161\pm0.003$& $0.0040\pm0.0005$& $1.048$\cr
V& 73\,\%&  $0.048\pm0.003$& $0.0079\pm0.0005$& $0.970$\cr
\noalign{\vskip 5pt\hrule\vskip 3pt}}}
\endPlancktable
\endgroup
\end{table}

\begin{figure}[htbp!]
\centering
\includegraphics[width=0.475\textwidth]{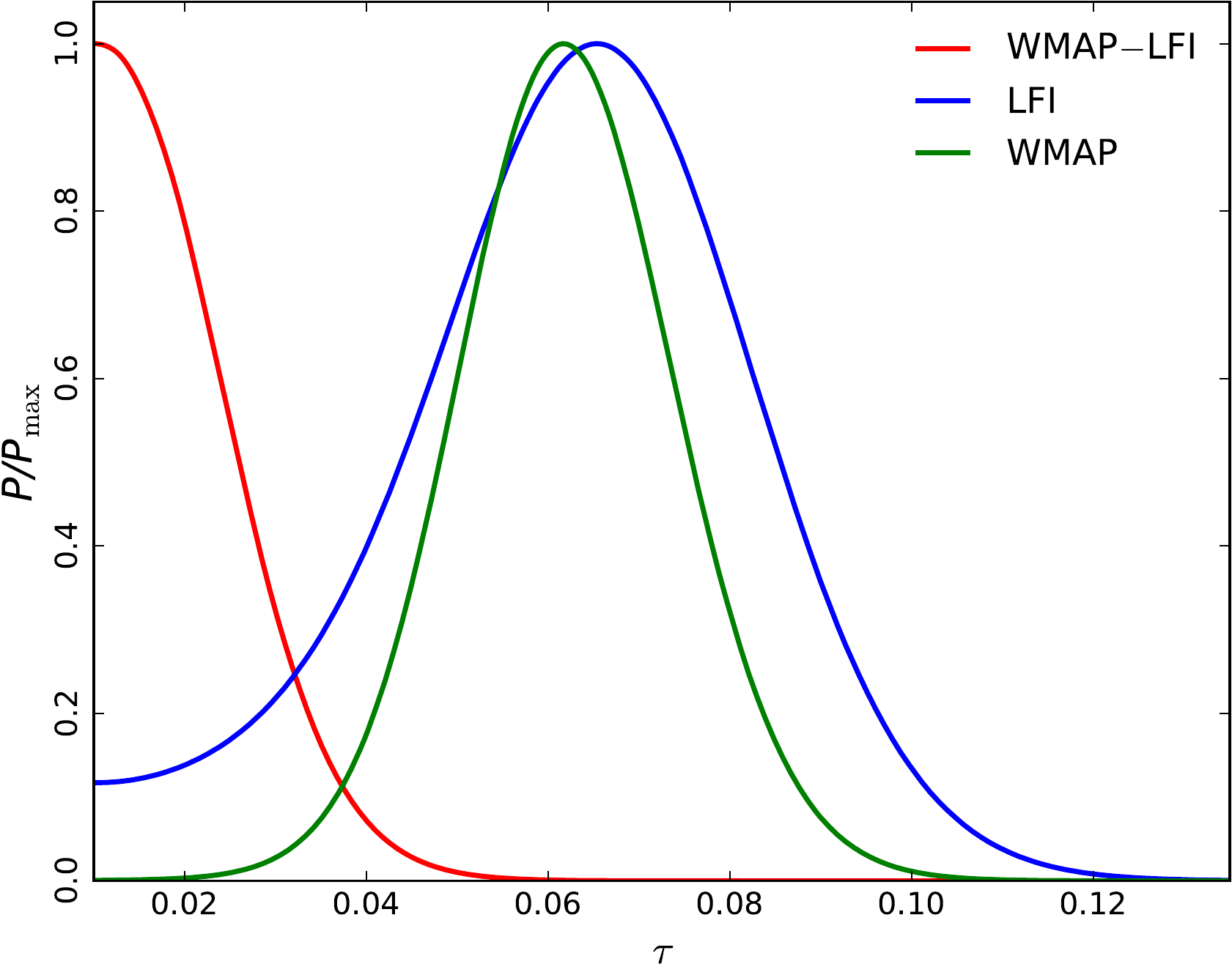}
\caption[]{Posterior distributions for $\tau$ estimated from: the LFI baseline likelihood ($\tau=0.063\pm0.020$); the WMAP data set smoothed with a cosine window, cleaned as given in Table~\ref{table:lowl_WMAPscalings_p06}, and masked with the analysis mask P06 ($\tau=0.062\pm0.012$); and the difference between these two data sets, WMAP\,$-$\,LFI, to which we apply a mask given by the intersection of the R1.8x and P06 masks ($\tau\leq 0.033$ at 95\,\% CL).}
\label{plot_tau_lfi_wmap_defaultmask}
\end{figure}

We further extend this comparison to the estimates of $\tau$. In Fig.~\ref{plot_tau_lfi_wmap_defaultmask} we show the posteriors for LFI, WMAP, and the half-difference of LFI and WMAP.

The constraints from LFI (with mean value $\tau=0.063\pm0.020$) and our re-analysis of WMAP data using the \Planck\ 353-GHz map as a dust template ($\tau=0.062\pm0.012$) are in very good agreement. Despite the re-introduction of Surveys~2 and 4, LFI 70-GHz data alone are still a factor of approximately 1.7 less sensitive than the combination of WMAP's Ka, Q, and V bands. This result can only in part be explained by the larger sky fraction employed in the WMAP data sets. As remarked above (see last paragraph of Sect.~\ref{sec:lfi_valsim}), the error on $\tau$ for the LFI data set is about $40\,\%$ larger than that expected from simulations (assuming its final noise-covariance matrix). 
Cleaning the WMAP maps of polarized dust contamination using the 353-GHz, rather than the WMAP dust template assumed in \citet{page2007}, lowers $\tau$ by roughly 2$\,\sigma$, confirming the results already derived in \citetalias{planck2014-a13}, where the \Planck\ 2015 products were used. A new analysis based on the cross-spectrum between \Planck\ 70\GHz\ and WMAP is presented in the previous section (Sect.~\ref{subsubsec:consistency_tests}, in the paragraph headed ``\Planck-WMAP cross-spectra''), giving consistent results on $\tau$. 

Finally, the good consistency between the two data sets is also confirmed by the estimate of $\tau$ for WMAP\,$-$\,LFI. This has been obtained in the part of the sky shared by the LFI R1.8x and WMAP P06 masks, and it provides a value of $\tau$ that is consistent with the null hypothesis, giving an upper limit of $\tau\leq 0.033$ (95\,\% CL).

%

\section{High multipoles}
\label{sec:hi-ell}

The 2018 high-$\ell$ likelihood approach is very similar to the one used for the 2015 release \citepalias{planck2014-a13}: 
from multipoles of $\ell\,{=}\,30$ and higher, we use a Gaussian likelihood
approximation, based on pseudo-$C_\ell$s calculated from a selection of data from the \Planck-HFI channels. It is still based exclusively on the temperature and $E$-mode polarization data.

Several important improvements have been implemented over the 2015 release. A new beam-leakage correction (described
in Sect.~\ref{sec:hi-ell:datamodel:beamleak}), a better Galactic contamination correction (Sect.~\ref{sec:hi-ell:datamodel:gal}), the use of the reprocessed HFI data described in \citetalias{planck2016-l03}, and a better estimation of the
polarization efficiencies (Sect.~\ref{sec:hi-ell:datamodel:inst}), allow us to validate the use of the polarized part of the data (specifically $TE$ and $EE$
spectra) for cosmological applications. The knowledge of the effective polarization efficiencies of the HFI channels is the largest source of uncertainty for the \Planck\ 2018 legacy release,
and we will investigate how it affects our determination of the cosmological model in Sect.~\ref{sec:valandro:PE}.

The following sections will quickly remind the reader of the methodology adopted (Sect.~\ref{sec:hi-ell:method}),
describe and validate our data cut choices (Sect.~\ref{sec:hi-ell:datamodel:data}), and highlight the differences with the
2015 data. The foreground model, similar to the 2015 one, is explained
in Sect.~\ref{sec:hi-ell:datamodel:fg}, while the
improved Galactic model is described in its own section, Sect.~\ref{sec:hi-ell:datamodel:gal}. The 
instrument model is described in
Sect.~\ref{sec:hi-ell:datamodel:inst}, and
Sect.~\ref{sec:hi-ell:datamodel:beamleak} will be devoted to the
improvements in the beam-leakage model, using the results from \citet{quickpolHivon}. 
A summary of the methodology and a description of the reference \plik\ likelihood is given in Sect.~\ref{sec:hi-ell:datamodel:alltogether}.

As for the 2013 and 2015 releases, we have investigated different implementations of the likelihood, allowing us to explore
alternative modelling choices. In 2018, the \plik\ likelihood that was the baseline for the 2015 results remains the reference implementation.
The \camspec\ likelihood allows us to test different choices and cross-validate the reference results, while 
the \pliklite\ likelihood is a foreground and nuisance marginalized version of the \plik\ likelihood.
Both the \plik\ and \camspec\ likelihoods have improved and evolved since 2015 and while they are, to a large extent, very
similar, they differ in a few important points. In the following description, while we will mainly focus on the \plik\ description, 
we will note whenever the \camspec\ likelihood differs in implementation or in modelling choices. 
As we did in Sect.~\ref{sec:hi-ell:datamodel:alltogether} for \plik, we will recap the description of the \camspec\ likelihood in
Sect.~\ref{sec:hi-ell:prod}, and also describe the changes implemented
since 2015 in the \pliklite\ likelihood.

Validation of the high-$\ell$ likelihood, including tests on end-to-end
simulations, is described  in Sect.~\ref{sec:valandro} and following sections. The consistency between the data and our model is discussed in Sect.~\ref{sec:valandro}, which investigates the
goodness-of-fit determinations, inter-frequency agreement,
cross-spectra residuals, temperature and polarization conditional
predictions, and consistency of cosmological parameters from $TT$, $TE$,
and $EE$.  In Sect.~\ref{sec:valandro:PE} we look at the impact of
polarization-efficiency corrections, and in Sect.~\ref{sec:valandro:priors}
we discuss priors. Different data cuts are investigated in
Sect.~\ref{sec:valandro:cuts}, including
inter-frequency agreement and comparisons of parameters derived from
$\ell\,{<}\,800$ and $\ell\,{>}\,800$.  This continues investigations that were
described in a paper focused specifically on parameter shifts in the 2015 data
release, \citet{planck2016-LI}.  We also here revisit the curious situation
with the $\Alens$ consistency parameter in Sect.~\ref{sec:valandro:alens}.
The effects of aberration are discussed in Sect.~\ref{sec:valandro:aberration}
and we finish this extensive section with a
description of the simulations used in Sect.~\ref{sec:valandro:sims}.

\subsection{Methodology}
\label{sec:hi-ell:method}

We retain the general approach of the 2013 and 2015 analyses
\citepalias{planck2013-p08,planck2014-a13}.
The 2018 high-$\ell$ likelihood approximation
is based on the three cleanest \Planck-HFI channels, namely the 100-GHz,
143-GHz, and 217-GHz ones. These channels represent the best optimization between resolution, sensitivity, and low foreground contamination.
In the ideal case (a full sky
free of foregrounds, with isotropic noise) and ignoring secondary effects (such as lensing; \citet{planck2016-l08}),
the CMB temperature anisotropies and polarization distribution being
Gaussian, all of the cosmological information is extractable from the
auto- and cross-spectra. In this case, the statistical properties of these 
spectra are well known, with each multipole following an independent
Wishart distribution. However, the diffuse Galactic foreground contamination,
along with that from point sources, requires certain areas of the sky
to be ignored in the analysis.  This masking couples the power
spectrum multipoles and moves their distribution away from the
analytic form.  Along with the anisotropic behaviour of the noise,
this forces one to adopt an approximate likelihood. 
As discussed in \citetalias{planck2013-p08}, \citetalias{planck2014-a13}, and following studies performed in \citet{HL08} and \citet{2012A&A...542A..60E}, at high enough
multipoles a correlated Gaussian distribution provides a sufficient approximation:
\begin{equation}
-\ln{\cal L}(\vec{\hat C} | \vec{C}(\theta)) = \frac{1}{2} \left[\vec{\hat C} - \vec{C}(\theta)\right]^{\tens{T}} \tens{\Sigma}^{-1}
 \left[\vec{\hat C} - \vec{C}(\theta)\right] + {\rm const.}\;,
 \label{eq:basic-likelihood}
 \end{equation}
 where $\vec{\hat C}$ is the data vector, $\vec{C}(\theta)$ is the
 prediction for the model with parameter values $\theta$, and
 $\tens{\Sigma}$ is the covariance matrix (computed for a fiducial cosmology). 
The Gaussian shape, suggested by the central limit theorem, performs reasonably well, even for $30\le\ell<100$, 
where the symmetry of the adopted distribution shape causes a small level of bias,\footnote{Estimated in 2015 at the level of $0.1\,\sigma$ on $\ns$. We will ignore this, as we did in 2015.} as discussed 
in \citetalias{planck2014-a13}.

To take account of the differing components of foreground emission
 and residual systematics (including temperature-to-polarization leakage) at different frequencies, the data vector
 (and corresponding data model) consists of
 all relevant cross-frequency spectra (as opposed to a reduced number
 of co-added ones).  This also allows us to assess the inter-frequency coherence of our data set directly at the likelihood level.

The covariance matrix $\tens{\Sigma}$ encodes the expected
correlations between the various spectra, as computed from different
pairs of maps, coming from the properties of the CMB, instrument noise, and 
attempts at capturing the effects of the different non-idealities listed above. For small excursions around the cosmological model that the strong 
constraining power of the \Planck\ data set permits, the theoretical correlation between the $TT$, $TE$, and $EE$ spectra 
can be computed using a fiducial model (containing the CMB and the
different frequency-dependent foregrounds and systematic biases; \citealt{HL08}). 
Corrections to account for masks and anisotropic noise are described in great detail in \citetalias{planck2014-a13}. In particular, 
large-scale masks (mainly covering the Galactic emission) will be treated following the {\tt MASTER} approach \citep{Hietal02}, while we correct
the covariance matrix to account for the effect of point-source masks using a Monte Carlo-based estimation \citepalias{planck2014-a13}. 
The effect of anisotropy in the noise, induced by the pointing strategy, is treated by a heuristic approximation described in \citetalias{planck2013-p08} and extended to
polarization in \citetalias{planck2014-a13}. 
Following the tests performed on realistic simulations in \citetalias{planck2014-a13}, we will ignore
the effects of the anisotropy of the Galactic contamination, owing to its low level in our masked maps, and treat all of our foreground sources
as Gaussian. We also ignore (at the level of the covariance) the extra
correlations between temperature and polarization induced by the beam
leakage (to be described in Sect.~\ref{sec:hi-ell:datamodel:beamleak}).

\subsection{Data}
\label{sec:hi-ell:datamodel:data}

We now discuss the content of the data vector part of Eq.~\eqref{eq:basic-likelihood}. In particular, we describe in
detail the data cuts that were implemented to avoid either noise biases and correlated residuals (Sect.~\ref{sec:hi-ell:datamodel:datasel}) 
or strong foreground contamination
(Sect.~\ref{sec:hi-ell:datamodel:mask}).  While beams are
corrected for in the data vector (Sect.~\ref{sec:datamodel:beam}), calibration residuals and beam-induced leakage
will be taken into account in the model vector as described later in Sect.~\ref{sec:hi-ell:datamodel:intro}.

\subsubsection{Cut selection}
\label{sec:hi-ell:datamodel:datasel}

We estimate the spectra of the CMB at different frequencies using
cross-spectra formed with pairs of different maps. 
{We discuss later in this section why we will eventually use the ``half-mission'' cross-spectra as we already did in \citetalias{planck2014-a13}.}
To the extent that the maps in a pair are uncorrelated, cross-spectra are not
biased by any noise contribution.  Thus we use
such cross-spectra in the likelihood, and do not include any auto-spectra
formed from a single map. 
The noise power spectrum must still be modelled, however, because of its
contribution to the scatter of the cross-spectra in the covariance. 
Nevertheless, the requirement on the fidelity of this model is lower that what would be needed to 
debias the auto-spectra directly \citepalias{planck2013-p08}.
We describe
the noise model in Sect.~\ref{sec:hi-ell:datamodel:noise}. 

We based the results of the two previous releases on sets of maps made
with two differing data cuts. In 2013, we measured the cross-spectra
between different bolometers and ``detsets'' (i.e., ``detector sets,'' which are two independent assemblies of polarized detectors) and combined them to estimate
unbiased auto- and cross-frequency-channel power spectra. 
However, we showed in 2015 \citepalias{planck2014-a13}
that the data processing had left
correlated terms between the bolometer and detset maps, which
translated into (a low level of) bias.  While this bias could be mitigated to some extent using an empirical, data-inspired,
correction, we based the 2015 likelihood on a different data cut. Taking advantage of the redundant sky coverage
accessible in 2015, we used the so-called ``half-mission'' (HM)
maps.  These are full-sky maps constructed from data from all
detectors at a given 
frequency collected during either the first (HM1) or second (HM2) half of the
full mission. While this cut discards more information
than the previous one (producing fewer maps and so having a smaller
ratio of cross-spectra to auto-spectra), it was shown to be immune to the cross-detector noise correlation at the level of
sensitivity of our tests. 

\begin{figure*}[htbp!]
\begin{centering}
\includegraphics[width=60.0mm,angle=0]{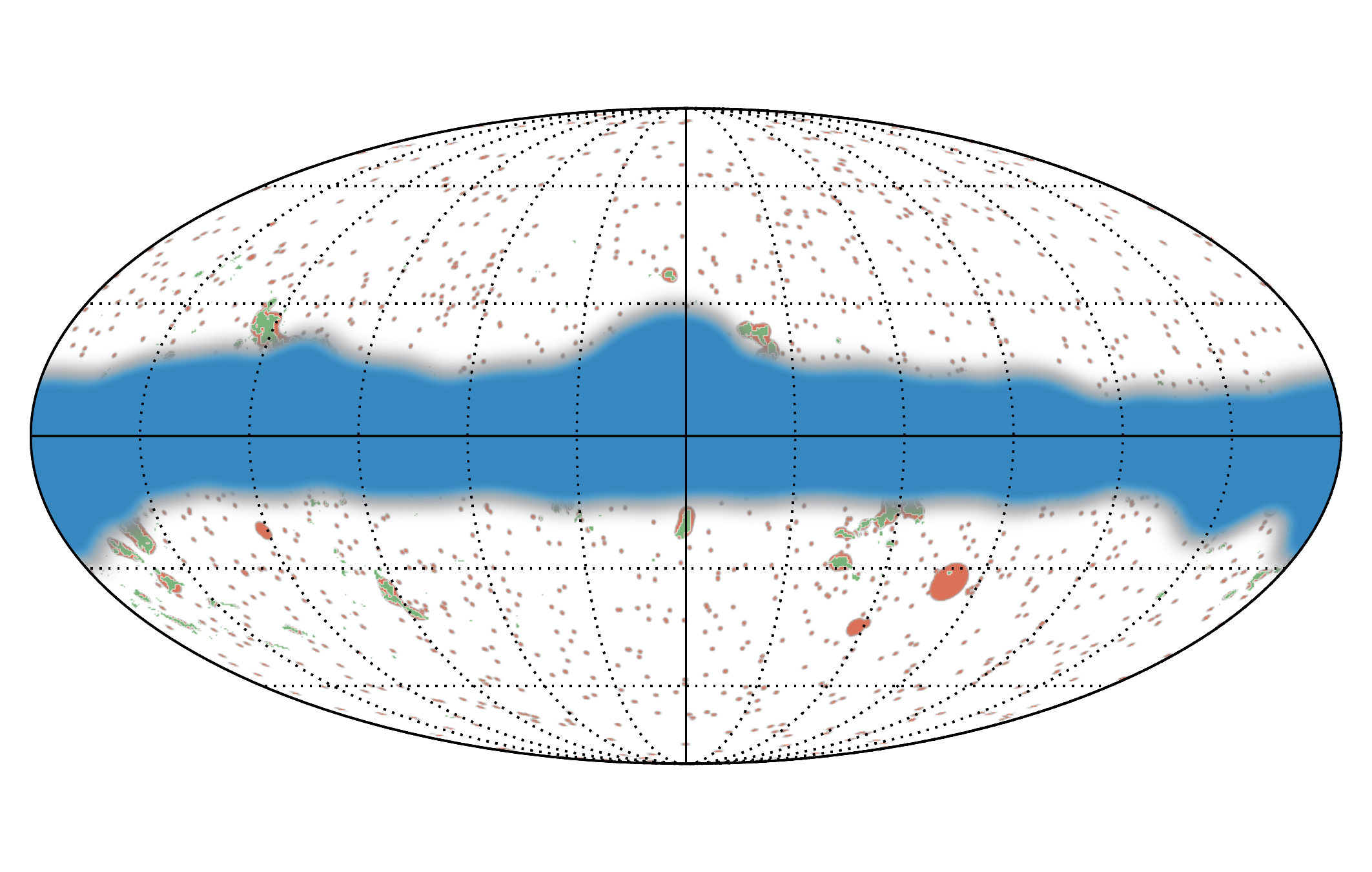}
\includegraphics[width=60.0mm,angle=0]{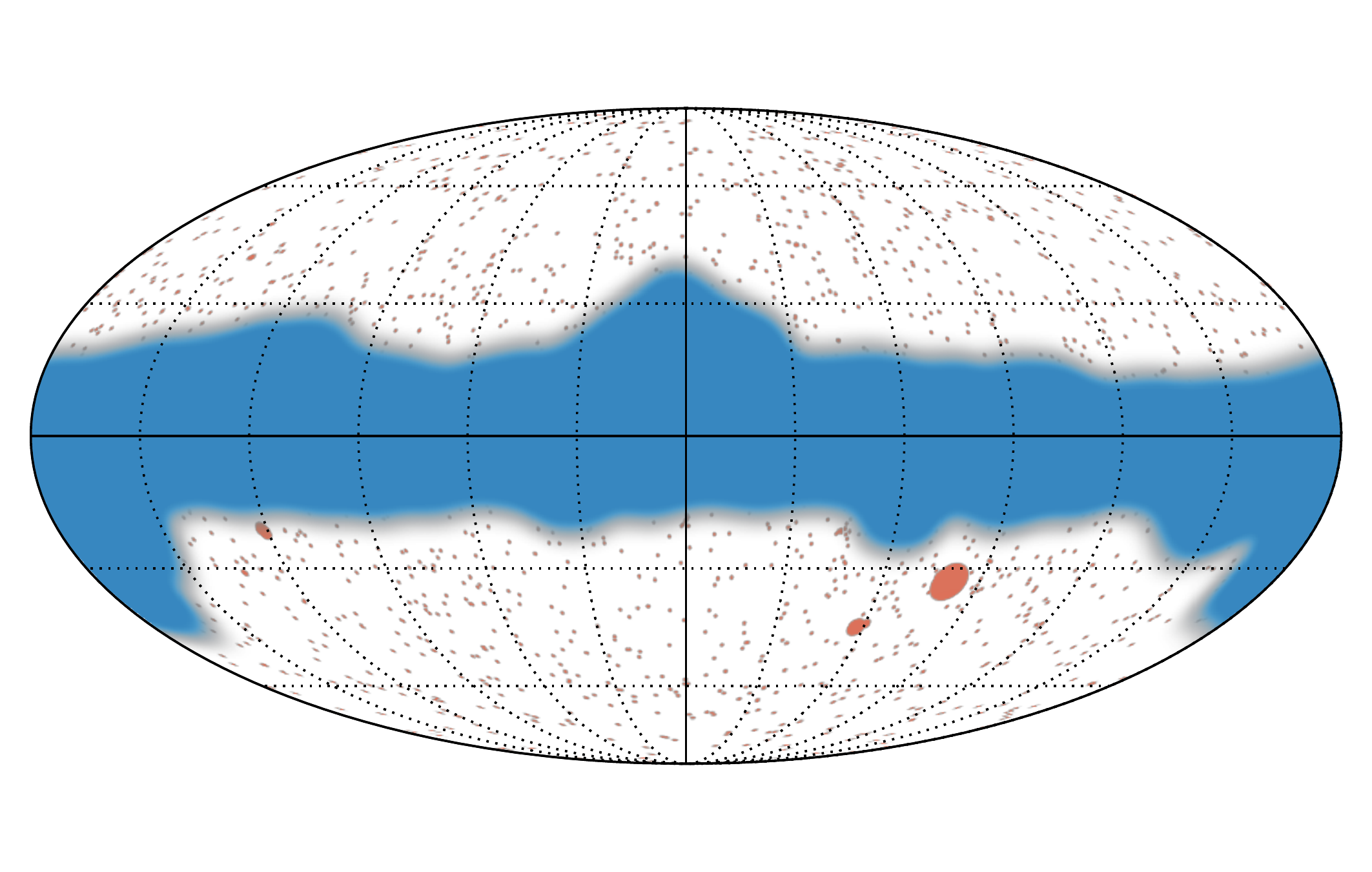}
\includegraphics[width=60.0mm,angle=0]{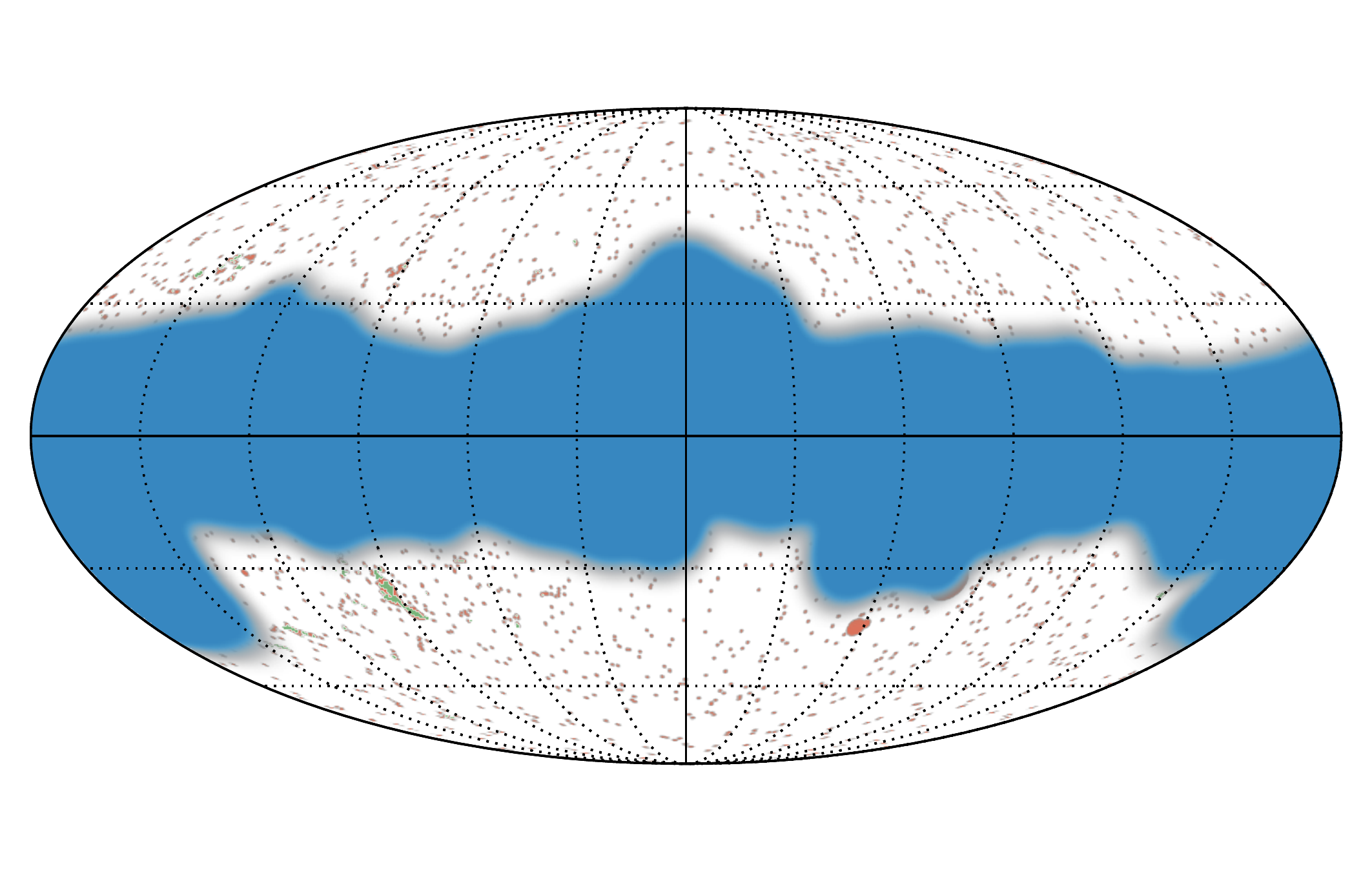}\\
\includegraphics[width=60.0mm,angle=0]{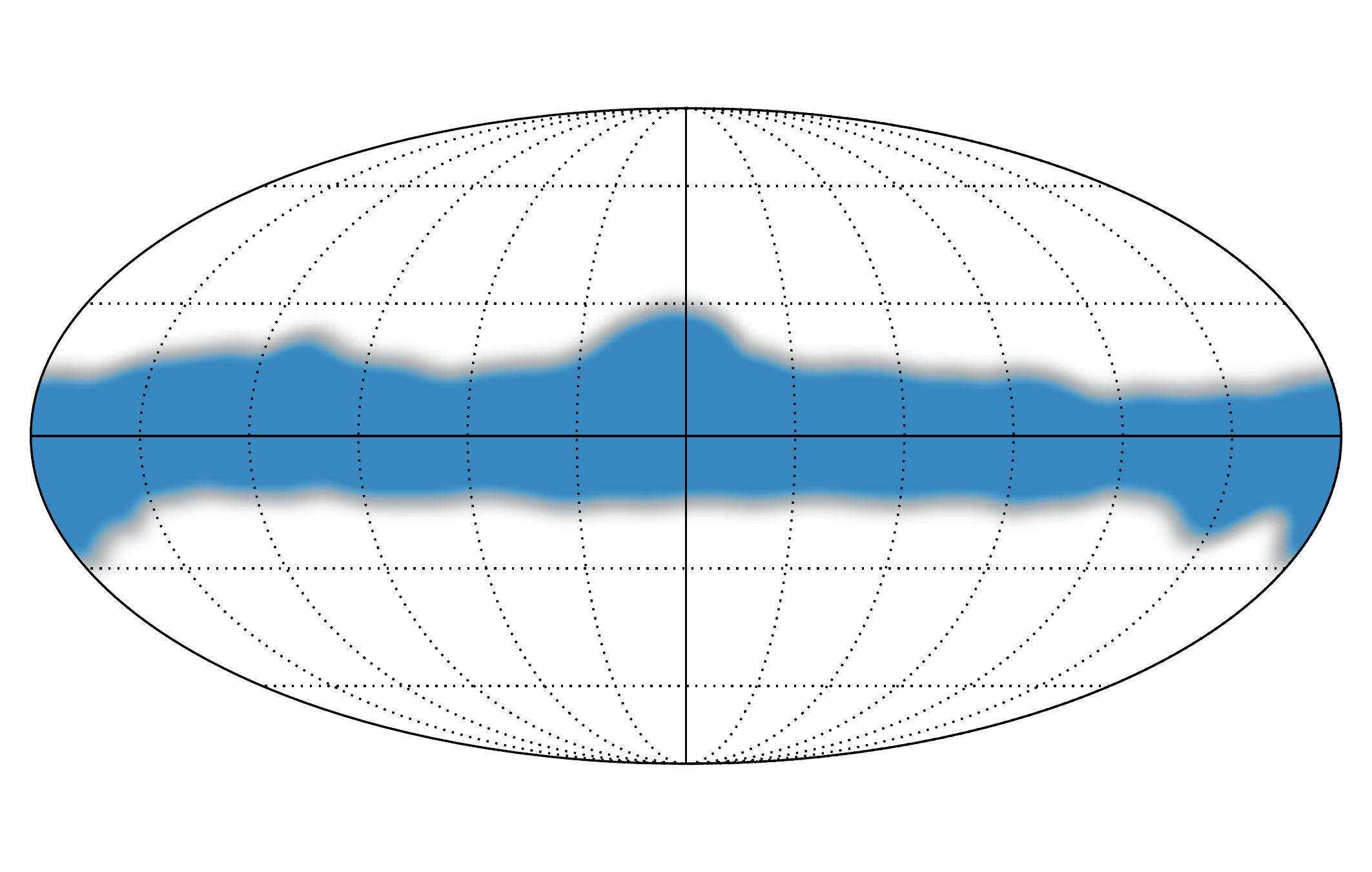}
\includegraphics[width=60.0mm,angle=0]{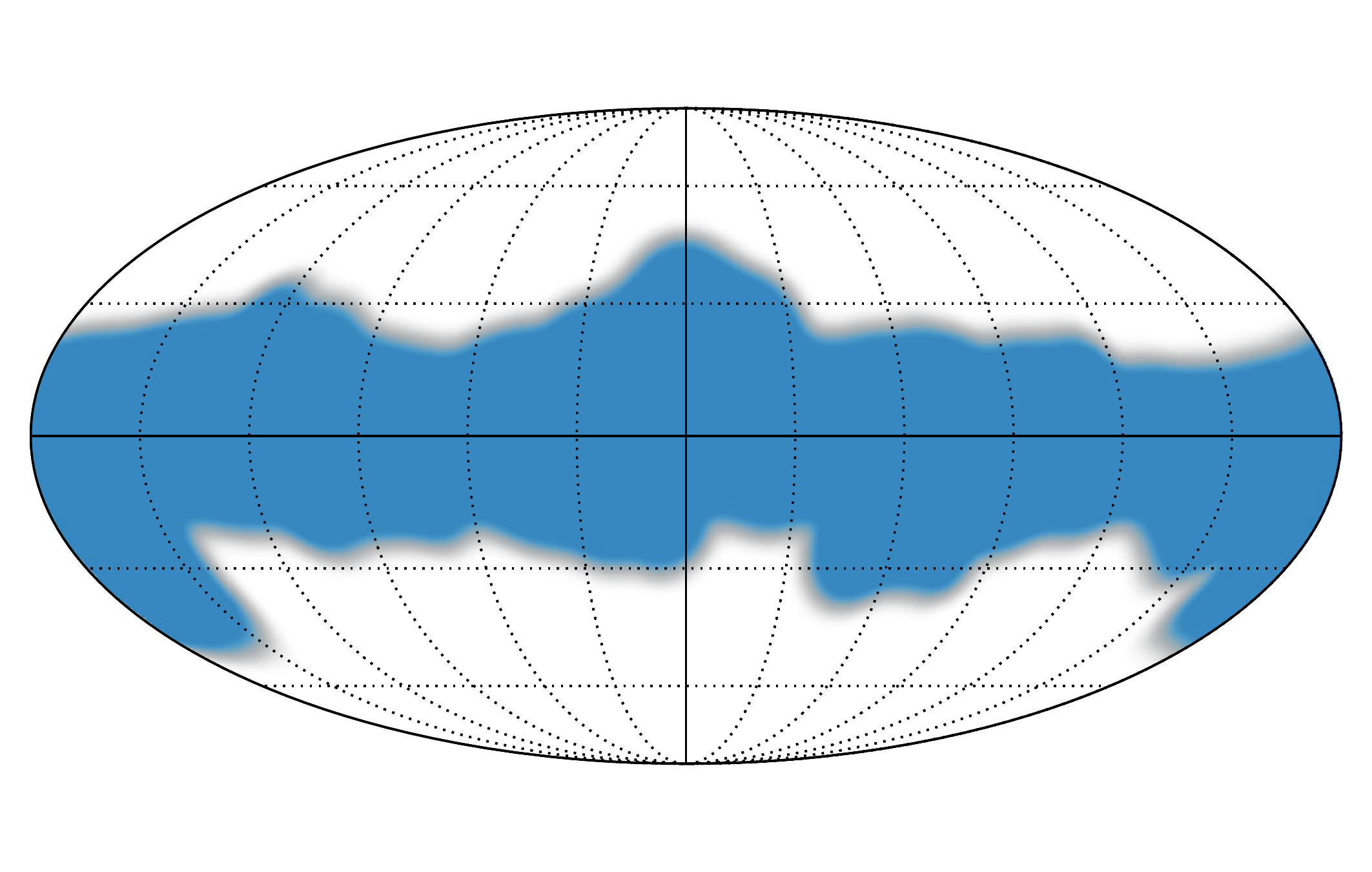}
\includegraphics[width=60.0mm,angle=0]{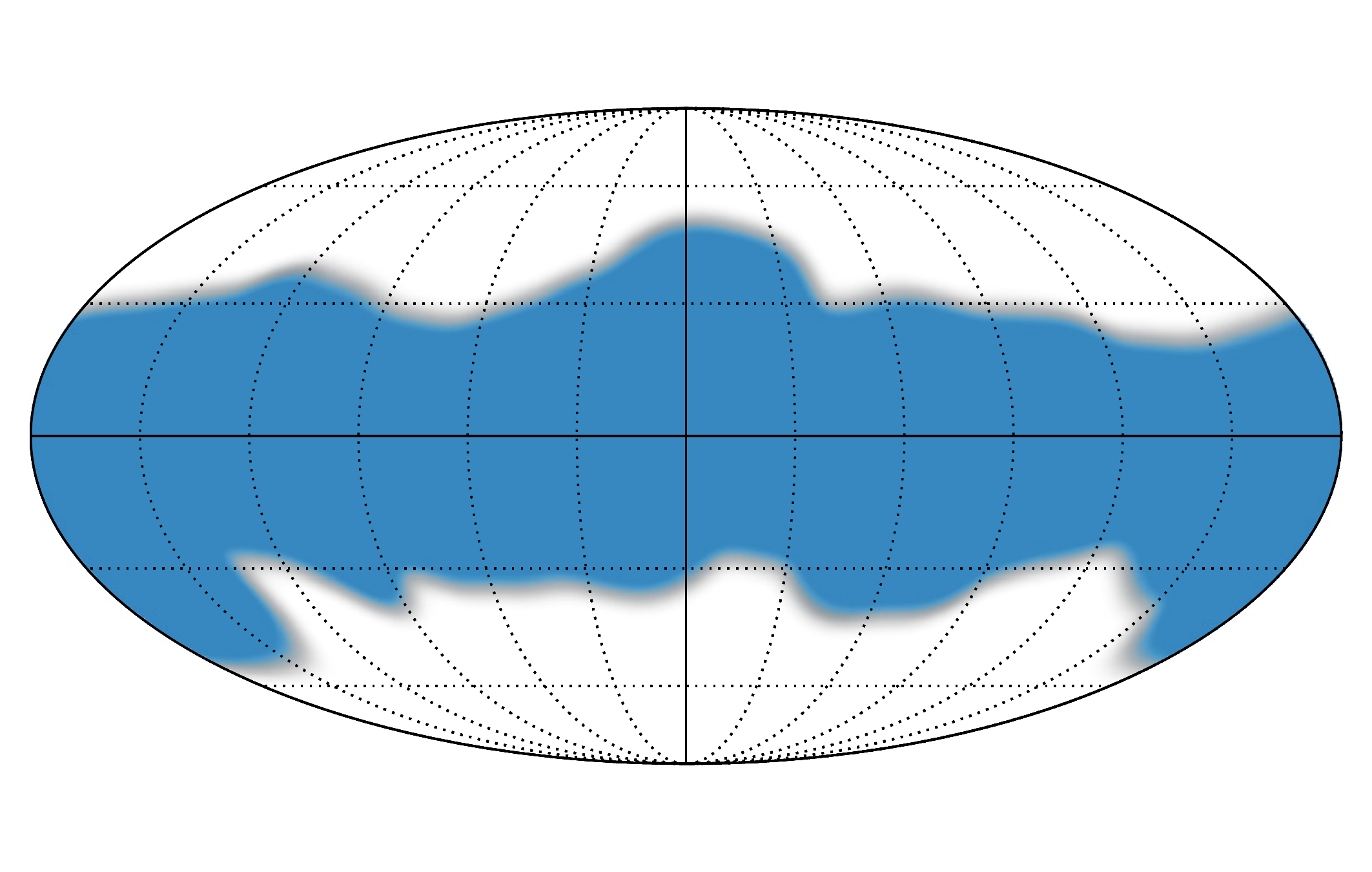}
\caption{Masks used in the \plik\ likelihood in temperature (top) and polarization (bottom).  From left to right we show masks for 100\,GHz (T66 top, P70 bottom), 143\,GHz (T57 top, P50 bottom), and 217\,GHz (T47 top, P41 bottom). Blue correspond to the Galactic (dust) mask, green is CO, and orange is the point-source and extended-source mask.}
\label{fig:hi-ell:data:all_mask}
\end{centering}
\end{figure*}

We coadd the $TE$ and $ET$ spectra into a single ``$TE$'' spectrum for each frequency pair, taking into account the different levels of noise in the different frequencies and half-mission maps {(i.e., using the individual $TE$ and $ET$ covariance matrices, ignoring their $\ell$ correlations, following equation~19 of \citetalias{planck2014-a13}). In the range of multipoles that we will eventually use, the weighting of the $TE$ and $ET$ spectra deviates from 0.5 by at most 20\,\%.} Beam  {and leakage corrections are computed for those coadded $TE$ spectra using the same weighting.} Polarization efficiency corrections will accordingly be made at the level of  coadded  spectra, {and we discuss in Sect.~\ref{sec:hi-ell:datamodel:inst} the maximum amplitude of the residual polarization efficiency correction error due to the coaddition.} {Coadding the $ET$ and $TE$ spectra allows for a reduction of the size of the covariance matrix, which, for an unbinned $TE+ET$ case will be around 8\,GB for the full TTTEEE likelihood. The $TE$ and $ET$ spectra have large correlations; contrary to the $TT$ case, the foreground contamination is smaller and of less help in regularizing the matrix. We discuss in Sect.~\ref{sec:valandro:gof} how we already see hints of limitations in our statistical model in the coadded $TE$ case. This fact might argue for combining all the different $TE$ cross-frequency spectra into a single spectrum, as proposed by the alternative \camspec\ likelihood. However, this choice prevents us from exploring the foreground and nuisance parameters along with the cosmological model parameters. Keeping the $TE$ cross-frequency spectra separate (but coadding the individual $ET$ and $TE$ ones) provides a reasonable compromise.}

Other data cuts were also investigated, in particular the so-called ``half-ring'' (HR) data cut. This is similar to the 
HM one in using data from all detectors at a given frequency, except
now using data from either the first half or the second half of
each pointing period (or ring). However, HR maps were shown to be 
correlated at a low level (\citet{planck2013-p03},\citetalias{planck2016-l03}), as a result of the glitch correction step in the processing pipeline. 
The HR cross-spectra (and specifically the difference between HR maps of each HM)
are nevertheless used in 
estimating the noise level in the HM spectra, as needed to build the
covariance matrix for the latter (see Sect.~\ref{sec:hi-ell:datamodel:noise}, where we also discuss how the correlations in HR maps are corrected for). 

While the HM cut effectively suppresses correlated time-dependent
systematics (up to a $\Delta T\,{\approx}\,1$-yr baseline), it does
not eliminate scanning-dependent systematic effects.
The \Planck\ scanning is nearly identical in each 
half mission, {in other words}, a majority of the rings that constitute each HM map are identical (pointing wise) in each half mission. 
For this reason, any signal-dependent systematic will be correlated
between maps from either half. 

To investigate this issue, \citetalias{planck2016-l03} introduced the
so-called ``odd-even'' (OE) cut, where
frequency maps are produced using every other pointing period (or
ring), {that is}, restricting to either the odd- or the even-numbered rings.
This choice renders the odd and even maps uncorrelated, up to time
scales of order of the pointing period (approximately 1\,hour), and
partially avoids scanning-related systematic effects. At high
multipoles, however, we find traces of residual correlation between the
odd and even maps (see Appendix~\ref{app:oe_correlation} and Sect.~\ref{sec:hi-ell:datamodel:noise}.)
For this reason the 2018 data selection at high multipoles is chosen
to be the same as in 2015, based on HM maps.


\subsubsection{Masks}
\label{sec:hi-ell:datamodel:mask}

We mask the temperature and polarization maps to avoid areas of the sky
that are strongly contaminated by non-cosmological foregrounds,
principally diffuse Galactic emission and, in temperature, point-source emission.  We use the same masks as we did for the 2015 release
\citepalias{planck2014-a13} and here only summarize the content of those masks.

Masks are constructed at each frequency based on the Galactic contamination, point-source contamination, and
instrumental resolution (for the point-source mask). The Galactic contamination 
being dominated by the dust emission (see discussion in Sect.\
\ref{sec:hi-ell:datamodel:gal}), we only aim to mask the main dust-contaminated regions. In temperature, Galactic masks
are built by thresholding high-frequency HFI maps.  As discussed in the
2015 papers, the dust temperature maps may be taken as a proxy for the
polarized emission, so the polarization masks are also built by thresholding
the same high-frequency maps. The 100-GHz and 217-GHz detectors are sensitive
to carbon monoxide (CO) emission, so we also mask strong CO-emitting regions at
these two frequencies. Point-source polarization being negligible at
the resolution and sensitivity of \Planck, we do not include any point
source contributions in the polarized masks.  

We investigated, at the power spectrum and at the cosmological parameter levels, the
effect of masking point sources in polarization as well. We found no impact on $TE$, which is expected since point sources 
are masked in temperature. In $EE$ a systematic offset is observed in the $100\times100$ 
spectrum when comparing the spectra obtained with and without the point-source mask. 
The offset is smaller than $3\times10^-6\,\mu{\rm K}^2$ (this is
in $C_\ell$ not ${\cal D}_\ell$ units), much below the other sources of 
contamination in the multipole region we eventually use for cosmology. 
The impact on cosmological parameters is of order $0.1\,\sigma$ on $n_\mathrm{s}$ and $A_\mathrm{s}$ 
when using the $EE$ spectrum only, and no effect is detected when using the joint likelihood.
{In} this test, contrary to what we eventually do in temperature, we do not attempt 
to correct the covariance matrix for the effect of the point-source mask on the polarized maps.
We ignore the bias, which is at the limit of the precision of our cosmological parameter estimations.

In order to mitigate correlations between the $C_\ell$ estimators at
different multipoles, we apodize the masks. The large
scale Galactic masks are apodized with a 4\pdeg71 FWHM ($\sigma\,{=}\,2\degree$) Gaussian window function, while the point-source and CO masks have a ${\rm FWHM}\,{=}\,30\arcm$ window applied to them.\footnote{In detail, the apodization is computed by applying the apodization kernel to the mask distance map (i.e., the map of the distance of each pixel to the closest mask border),
which ensures that the apodized mask indeed goes to zero in the masked region. The mask distance map is easily computed using {\tt HEALPix} routines. To avoid spiky features of the distance map in the convex regions of the mask, the distance map is first smoothed using a $1\farcm5$ FWHM Gaussian window.}  As a rule of thumb, the large apodization of the Galactic
masks reduces the effective sky fraction by about $10\,\%$. Final temperature masks are the product of the
apodized Galactic and apodized point-source (and CO at 100\,GHz and 217\,GHz) masks, while the polarization ones are
simply the apodized Galactic masks. In the following, we refer to the masks by the fraction of sky they retain and by
their use. We call ``Gxx,'' ``Txx,'' and ``Pxx'' the apodized Galactic masks, the final temperature masks, and the
final polarized masks, respectively, with the suffix indicating the percentage of
sky retained.

As in 2015, we use the T66, T57, and T47 masks (based on the G70, G60,
and G50 Galactic masks) in temperature 
and the P70, P50, and P41 masks (corresponding to the G70,
G50, and G41 Galactic masks) in polarization, at 100\,GHz, 143\,GHz, and
217\,GHz, respectively.  These masks are illustrated in Fig.~\ref{fig:hi-ell:data:all_mask}.

\begin{figure}[htbp!]
\begin{centering}
\includegraphics[width=60.0mm,angle=0]{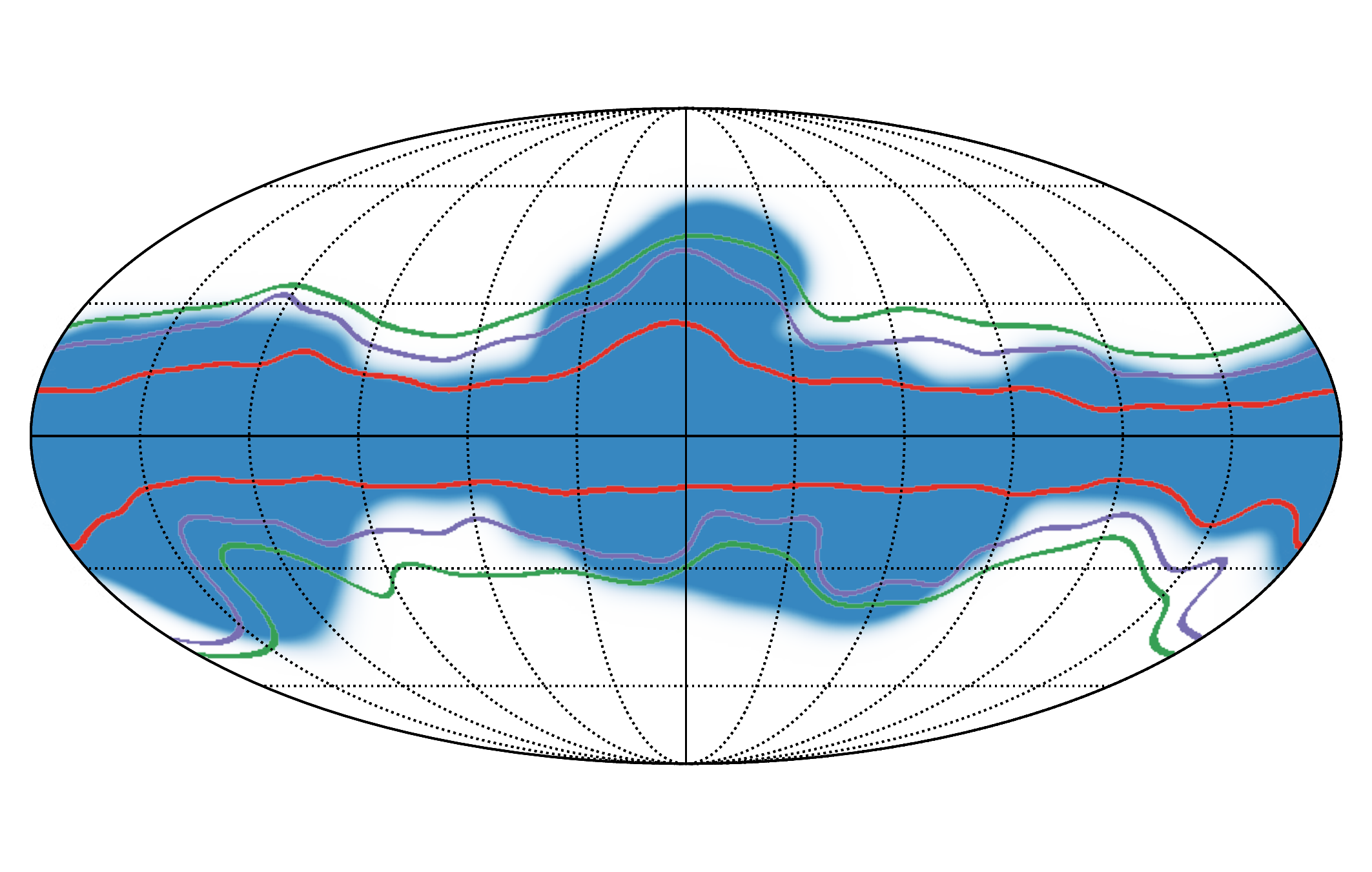}
\caption{Blue is the C50 mask, while the outline of the P70, P50, and P41 masks are shown in red, purple, and green, respectively.}
\label{fig:hi-ell:data:cammsk}
\end{centering}
\end{figure}

The alternative \camspec\ likelihood uses a different polarization
mask, based on the thresholding of 353-GHz polarization maps.
This C50 mask is built as follows.  First, 143-GHz $Q$ and $U$ maps are used as proxies of the CMB and
are subtracted out of the 353-GHz ones. The resulting maps are smoothed with a Gaussian of ${\rm FWHM}\,{=}\,10\deg$
and then added in quadrature to make a $P$ map.  This 
$P$ map is thresholded, and the result is smoothed with a Gaussian of
${\rm FWHM}\,{=}\,5\deg$.  The thresholding and smoothing is repeated four
times, so as to avoid disconnected regions in the mask. The thresholding
level was adjusted to retain about 60\,\% of the sky (before smoothing) on
the final iteration. 
A comparison between the C50 and Pxx masks is shown in Fig.~\ref{fig:hi-ell:data:cammsk}.

\subsubsection{Beams}
\label{sec:datamodel:beam}

Effective beam models are built by fitting spline functions to the planet-transit observations \citep{planck2014-a09}. 
Those models are then convolved with the complex scanning 
strategy of \Planck, and combined according to map and cross-spectra weightings to estimate the beam matrices $W^{XY,\; X'Y'}_\ell$ that relate 
the observed map $C^{XY}_\ell$ spectra to the sky spectra $C^{\mathrm{sky},X'Y'}_{\ell}$, with $X,Y,X',Y' \in \{T,E,B\}$ \citep[hereafter \citetalias{quickpolHivon}]{quickpolHivon}:
\begin{equation}
       C^{XY}_{\ell} = \sum_{X'Y'} W^{XY,\; X'Y'}_\ell C^{\mathrm{sky},X'Y'}_{\ell}.
        \label{eq:beam_matrix}
\end{equation}
{The} equation is only exact for the ensemble average over sky realizations of the observed and sky spectra.
Retaining the terminology used in 2013 and 2015, we call the diagonal
part of those matrices $W^{XY,\; XY}_\ell$ the ``effective beam,'' and correct
the pseudo-spectra with this transfer function to form the mask-deconvolved pseudo-spectra that will constitute the data vector of the likelihood. 
The non-diagonal terms
of the matrix are treated as leakage, and are corrected for in the
model vector. This correction and a related sub-pixel effect are discussed in Sect.~\ref{sec:hi-ell:datamodel:beamleak}. As explained
in \citetalias{planck2014-a13} and \citetalias{quickpolHivon}, because of the complex
scanning strategy, beam matrices and the effective beams depend on the details 
of the masks.  The effective beam for a cross-spectrum is a
 priori different from the geometrical mean of the effective beams
for the auto-spectra of each of the maps entering, and a different beam matrix needs to be computed for each frequency cross-spectrum combination.

While for the 2013 release we ignored mask effects in the beams and for the 2015 release we
targeted our computation for an average sky fraction, for this release effective beams are 
calculated for the actual masks used in the likelihood. 
In the range of scales we retain for cosmology, this affects mainly 
the $143\times143$ spectrum, with a correction at $\ell\,{=}\,2000$ of amplitude $<0.7\,\%$ in $EE$, $<0.5\,\%$ 
in $TE$, and $<0.2\,\%$ in $TT$. 

\subsubsection{Multipole range}

\begin{table}[htbp!] 
\begingroup 
\newdimen\tblskip \tblskip=5pt
 \caption{Multipole cuts for the temperature and polarization spectra at high $\ell$.}
  \label{tab:hi-ell:data:lrange}
\vskip -3mm
\footnotesize
\setbox\tablebox=\vbox{
\newdimen\digitwidth
\setbox0=\hbox{\rm 0}
\digitwidth=\wd0
\catcode`*=\active
\def*{\kern\digitwidth}
\newdimen\signwidth
\setbox0=\hbox{+}
\signwidth=\wd0
\catcode`!=\active
\def!{\kern\signwidth}
\newdimen\decimalwidth
\setbox0=\hbox{.}
\decimalwidth=\wd0
\catcode`@=\active
\def@{\kern\decimalwidth}
\openup 3pt
\halign{
\hbox to 1.2in{#\leaderfil}\tabskip=2em&
    \hfil#\hfil\tabskip=0pt\cr
\noalign{\doubleline}
\omit\hfil Frequency [GHz]\hfil& Multipole range\cr
\noalign{\vskip 3pt\hrule\vskip 5pt}
\multispan2\hfil $TT$\hfil\cr
$100\times100$&*30--1197\cr
$143\times143$&*30--1996\cr
$143\times217$&*30--2508\cr
$217\times217$&*30--2508\cr
\noalign{\vskip 5pt\hrule\vskip 5pt}
\multispan2\hfil $TE$ and $TE$\hfil\cr
$100\times100$&*30--*999\cr
$100\times143$&*30--*999\cr
$100\times217$&505--*999\cr
$143\times143$&*30--1996\cr
$143\times217$&505--1996\cr
$217\times217$&505--1996\cr
\noalign{\vskip 5pt\hrule\vskip 3pt}
}}
\endPlancktable
\endgroup
\end{table}

We retain the approach followed for the 2015 release and use similar multipole range cuts, discarding scales where either the signal-to-noise (S/N) ratio is too low, 
the Galactic contamination is too high, or where we identify possible systematics. Table~\ref{tab:hi-ell:data:lrange} 
summarizes our choices. 

It is of interest to note the low value of the maximum multipole
cutoff that we use for
$100\times100$ $TT$, namely $\ell\,{=}\,1197$. In 2015, we used this value,
which is conservative, given the 
expected S/N of this spectrum. The rationale for this was to
allow for easier comparison with the detset-based likelihood, which
required a 
correlated noise correction that was difficult to
establish at higher multipoles. 
As discussed in Appendix~\ref{app:oe_correlation}, there are signs of
failure of the odd-even/half-mission null test at similar angular
scales, and so we 
retain the 2015 value.

As in 2013 and 2015, we do not include the $100\times143$ $TT$ and $100\times217$ $TT$ spectra. {Their inclusion} bring{s} little extra S/N and require{s} more complexity in the foreground model.

We also do not change the choices made in 2015 for polarization.  We cut out all multipoles higher than $\ell\,{=}\,1000$ 
for cross-spectra involving the 100-GHz data given its high noise
level, and ignore all multipoles below $\ell\,{=}\,500$
for cross-spectra involving the 217-GHz map, owing to difficulties in modelling its dust contamination.

As discussed in Sect.~\ref{sec:hi-ell:prod:camspec}, the \camspec\ likelihood uses slightly different 
multipole cuts, being a bit more conservative in $TT$ at low multipoles
for 217\,GHz, a bit more aggressive at 100\,GHz at high multipoles in
temperature and polarization, and a bit more aggressive at 217\,GHz at low multipoles in polarization.
The impact of the different multipole range in temperature has already been
discussed in \cite{planck2014-a09}. The change of range in
polarization is found to have a very low impact on
parameters; differences between the \plik\ and \camspec\ results are
dominated by other differences between the likelihoods.

\begin{figure}[h!]
\centering
\includegraphics[width=0.495\textwidth,trim=0 .5cm 0 .6cm, clip]{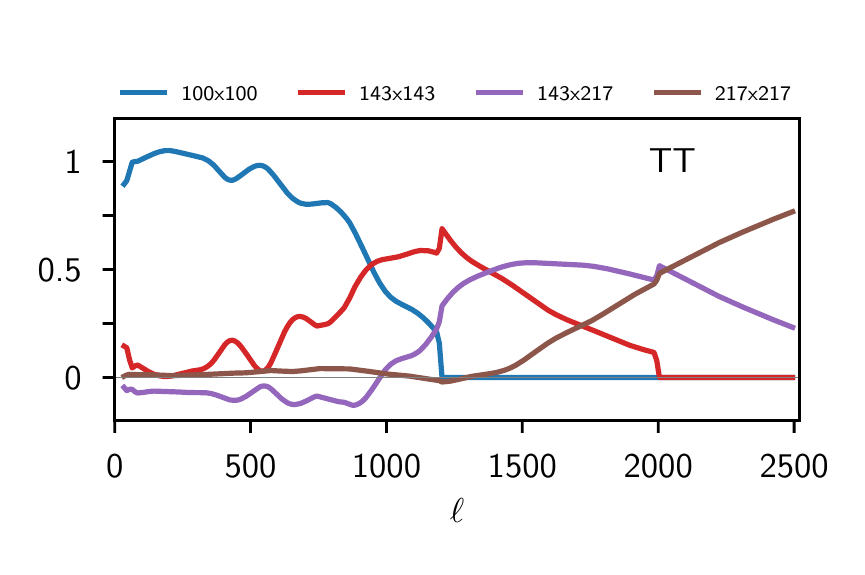}\\
\includegraphics[width=0.495\textwidth,trim=0  .5cm 0 .4cm, clip]{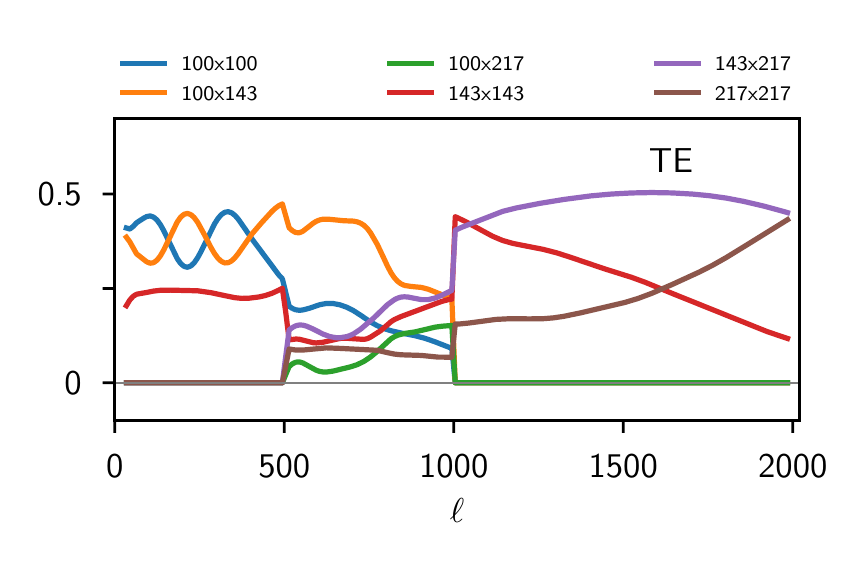}\\
\includegraphics[width=0.495\textwidth,trim=0  .5cm 0 .4cm, clip]{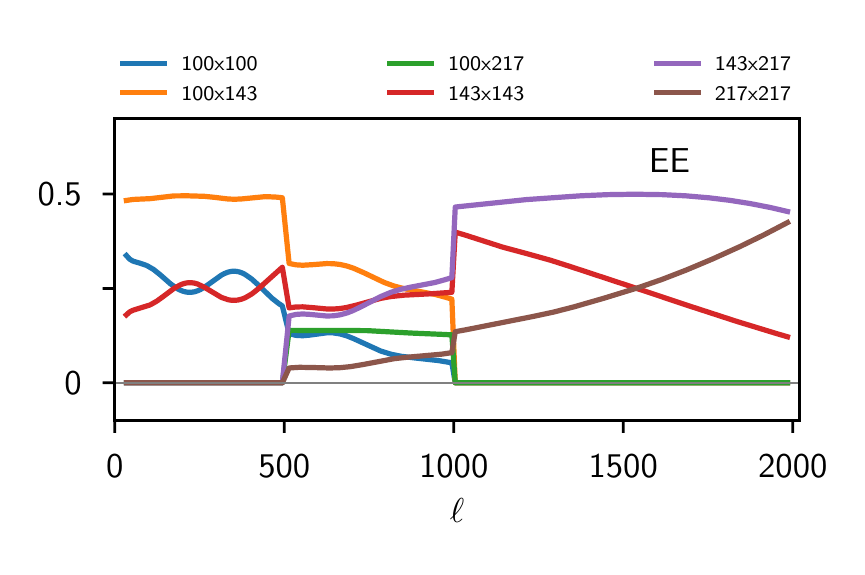}\\

\caption{Relative weights of each frequency cross-spectrum in the $TT$ (top), $TE$ (middle), and $EE$ (bottom) coadded spectra (see Sect.~\ref{sec:valandro} for definitions). Weighting is obtained using the optimal 
mixing matrix for each spectrum individually. For each multipole $\ell$, the elements of the matrix are rescaled so as to sum to unity, ignoring the $\ell\ell'$ contributions. Each colour corresponds to the weighting of a given cross-spectrum.
Sharp jumps in the plots are due to the multipole selection. The figure is very similar to the 2015 one, with small differences being the result of refinements in the foreground and systematics models.}
\label{fig:hi_ell:data:mixing}
\end{figure}

Given the masks, level of foreground contamination, beams, and multipole
ranges, it is not obvious which are the 
dominant frequencies at each multipole bin.
Figure~\ref{fig:hi_ell:data:mixing} presents the effective weights of each
cross-spectrum, as determined by the likelihood.
It can be seen that $143\times143$ has a high contribution, in
particular for $TE$ and $EE$, where it is in both cases the only spectrum
retained over the full $\ell$ range, and thus it has a particular
importance for cosmological parameters coming from from $TE$ and $EE$.


\subsubsection{Binning}

We reduce the size of the covariance matrix by binning, using the same
scheme as for the 2015 release. The spectra are summed into bins of width 
$\Delta\ell\,{=}\,5$ for $30\,{\le}\,\ell\,{\le}\,99$, 
$\Delta\ell\,{=}\,9$ for $100\,{\le}\,\ell\,{\le}\,1503$,
$\Delta\ell\,{=}\,17$ for $1504\,{\le}\,\ell\,{\le}\,2013$, and 
$\Delta\ell\,{=}\,33$ for $2014\,{\le}\,\ell\,{\le}\,2508$, 
with a weighting of the $C_\ell$ proportional to
$\ell(\ell+1)$ over the bin widths, {that is},
\begin{equation}
C_b = \sum_{\ell=\ell_b^{\mathrm{min}}}^{\ell_b^{\mathrm{max}}} w_b^\ell C_\ell, \ \mathrm{with}\ w_b^\ell = \frac{\ell(\ell+1)}{\sum_{\ell=\ell_b^{\mathrm{min}}}^{\ell_b^{\mathrm{max}}} \ell(\ell+1) } .
\end{equation}

This binning does not significantly affect the determination of
cosmological parameters in \LCDM-type models (including extensions)
that have smooth power spectra \citepalias{planck2014-a13}. Non-binned
versions of the likelihood have also been produced for use in
cosmological applications where the power spectra are not so smooth,
such as when searching for features in the primordial power spectrum.

\subsubsection{2015-to-2018 release data differences}

The HFI data-processing paper \citepalias{planck2016-l03} describes in detail the changes
made to the data processing for 2018.  Most of the developments have
focused on improving the polarization maps on large angular scales.  We summarize here the
differences relevant to the high-$\ell$ likelihood. 

In temperature, the 2015 and 2018 power spectra are remarkably similar, with differences smaller than $0.3\,\%$, corresponding to:
\begin{enumerate}[(i)]
\item the calibration corrections described in section~5.1 of \citetalias{planck2016-l03}; and
\item the beam changes described in detail in Sect.~\ref{sec:datamodel:beam}. 
\end{enumerate}

In polarization, the main changes are the correction of the analogue-to-digital convertor nonlinearity (ADCNL) effects
 and the bandpass-mismatch corrections. 

The largest difference between the 2018 and 2015 $EE$ spectra is for the $100\times100$ $EE$ spectrum at large scales, with an amplitude of the 
order of $3\times 10^{-5}\,\muK^2$ in $C_\ell$ at $\ell\,{=}\,30$ and decreasing sharply down to $\ell\,{=}\,250$. 
This correction is of order $20\,\%$ of the dust contribution,
which dominates the spectrum at low multipoles in the P70 mask that we use for $100\times100$. 
At 143\,GHz the difference is 3 times smaller, with the opposite sign. 
In both cases, the change between 2015 and 2018 is found to be
dominated by the bandpass-leakage correction; 2018 data without the
bandpass-leakage corrections reproduce, to a large extent, the 2015 spectra at $30\,{<}\,\ell\,{<}\,250$. 
This modification of the behaviour of the multipoles at the low end of
our range of interest affects the dust contamination estimations. With
the bandpass-leakage correction having a different slope to the dust contribution at low multipoles, the determination of the spectral index parameter $n_{\rm s}$
from the $EE$ spectra differs between 2015 and 2018, as will be discussed in Sect.~\ref{sec:hi-ell:datamodel:alltogether}.

The $TE$ cross-spectra changes between 2015 and 2018 do not seem to exhibit any clear trends or sharp features. 
Below $\ell\,{=}\,1000$, in the signal-dominated regime, the difference
between the 2015 and 2018 spectra scatters at a level below $1\,\%$ of
the CMB spectrum itself (in bins of $\Delta\ell\,{=}\,30$). 
This level corresponds to about a quarter of the peak amplitude of the
beam-leakage correction for the $143\times143$ $TE$ spectrum (one of the
main contributors to the coadded CMB spectrum, as can be seen for
example in Fig.~\ref{fig:hi_ell:data:mixing}). 

Comparisons between the 2015 and 2018 foreground- and
nuisance-parameter-corrected spectra are presented in
Figs.~\ref{fig:hi_ell:valid:residualTT},
\ref{fig:hi_ell:valid:residualTE}, and
\ref{fig:hi_ell:valid:residualEE}.  {All} figures are produced using
the parameter values of the 2018 \LCDM\ best-fit model. 

Thanks to improvements in the mapmaking algorithm, the effective noise levels in the polarization maps have decreased between 2015 and 2018. The amplitude of this change and its consequences are discussed in the noise estimation section, Sect.~\ref{sec:hi-ell:datamodel:noise}.

\subsection{Model}
\label{sec:hi-ell:datamodel:intro}
\begin{figure*}[htbp!]
\begin{centering}
  \includegraphics[angle=0,width=0.99\textwidth]{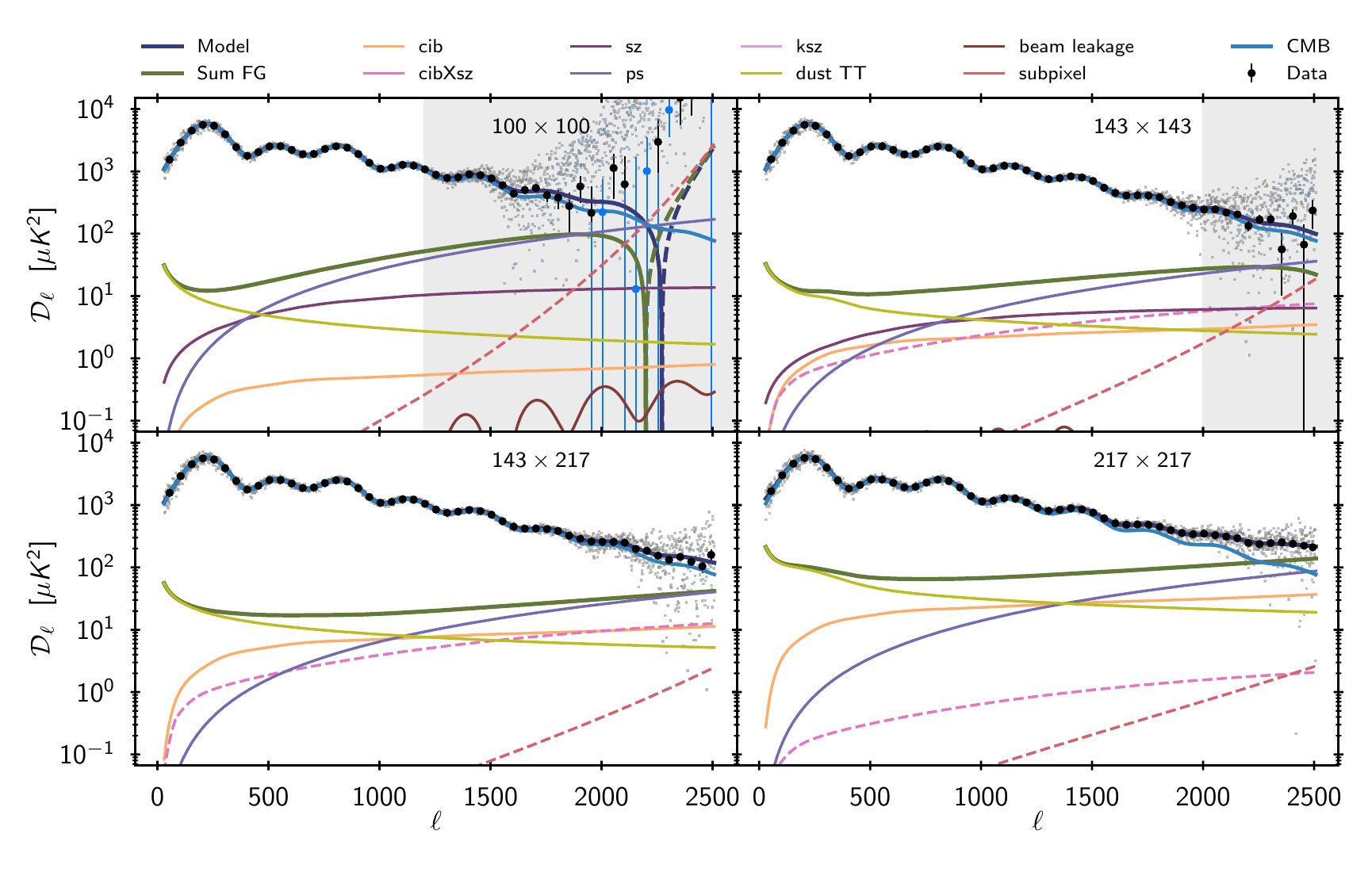}
  \end{centering}
  \caption{Best-fit (\planckalllensing) cross-spectra used for the temperature high-$\ell$ likelihood. Black circles represent the data, binned at $\Delta\ell=50$, while thin grey points show each multipole to give a sense of the scatter. Blue points and dashed lines correspond to negative values. The thick dark blue lines show the full model at each multipole (sum of all components), and the dark green one is the sum of foreground and nuisance contributions. Grey shaded areas are not used for cosmology.
 }
 \label{fig:hi-ell:fgcmbTT}
\end{figure*}

\begin{figure*}[htbp!]
\begin{centering}
 \includegraphics[angle=0,width=0.99\textwidth]{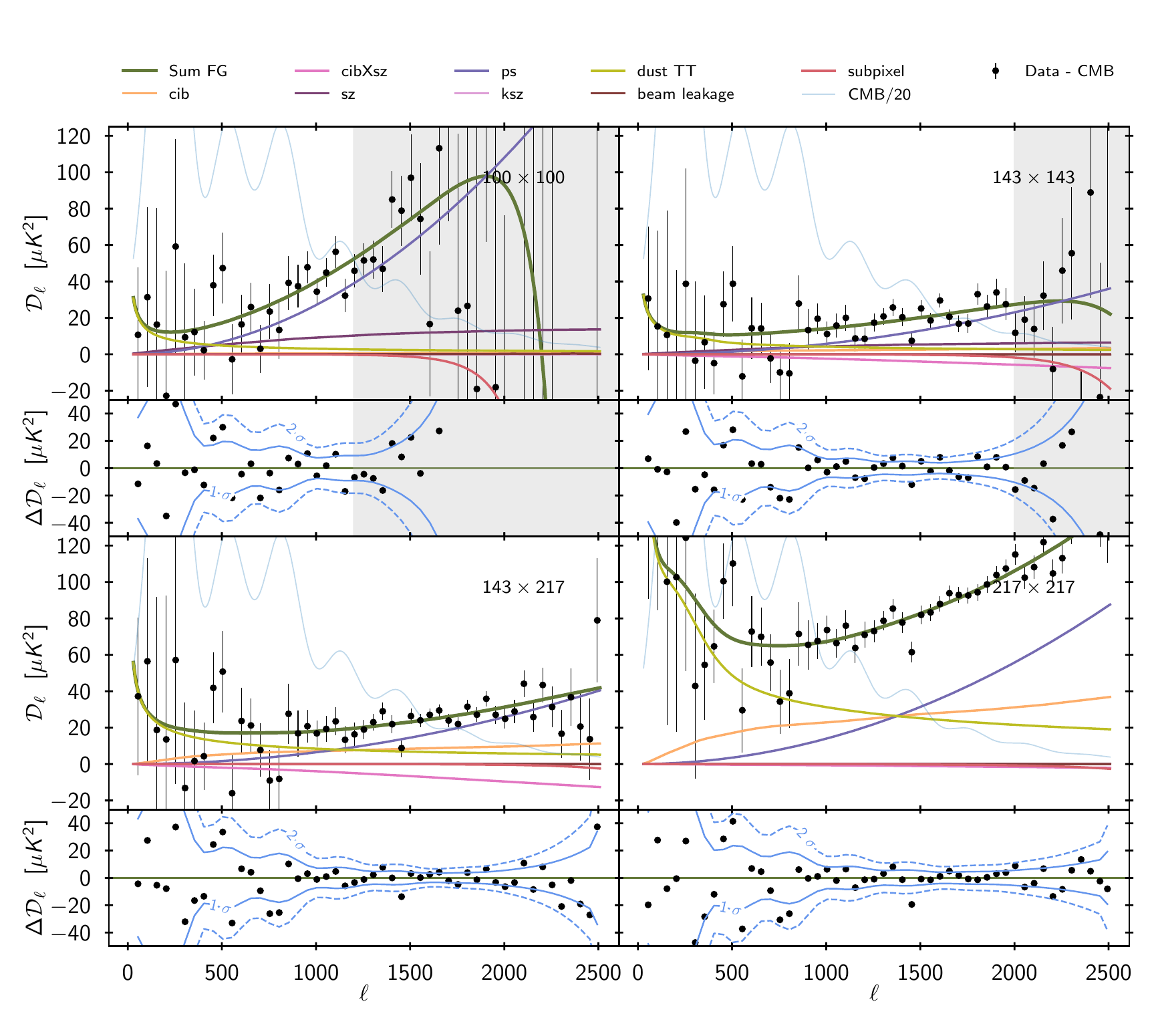}
\end{centering}
 \caption{Best-fit foreground and nuisance models for all of the
    cross-spectra used in the TT high-$\ell$ likelihood. The data,
    with the best-fit theoretical CMB $C_\ell$ spectrum subtracted,
    are shown in black with a binning of $\Delta\ell=50$. The $TT$ CMB
    spectra divided by 20 are shown in the top parts of each panel. The bottom
    parts of each panel show the residual after foreground and nuisance
    correction. Grey shaded areas indicate regions of multipole space
    not used for cosmology. Nuisance contributions (i.e., beam leakage and sub-pixel effects) are always much smaller than astrophysical foregrounds in the $\ell$ ranges we keep.
}
  \label{fig:hi-ell:fgTT}
\end{figure*}

\begin{figure*}[htbp!]
\begin{centering}
  \includegraphics[angle=0,width=0.99\textwidth]{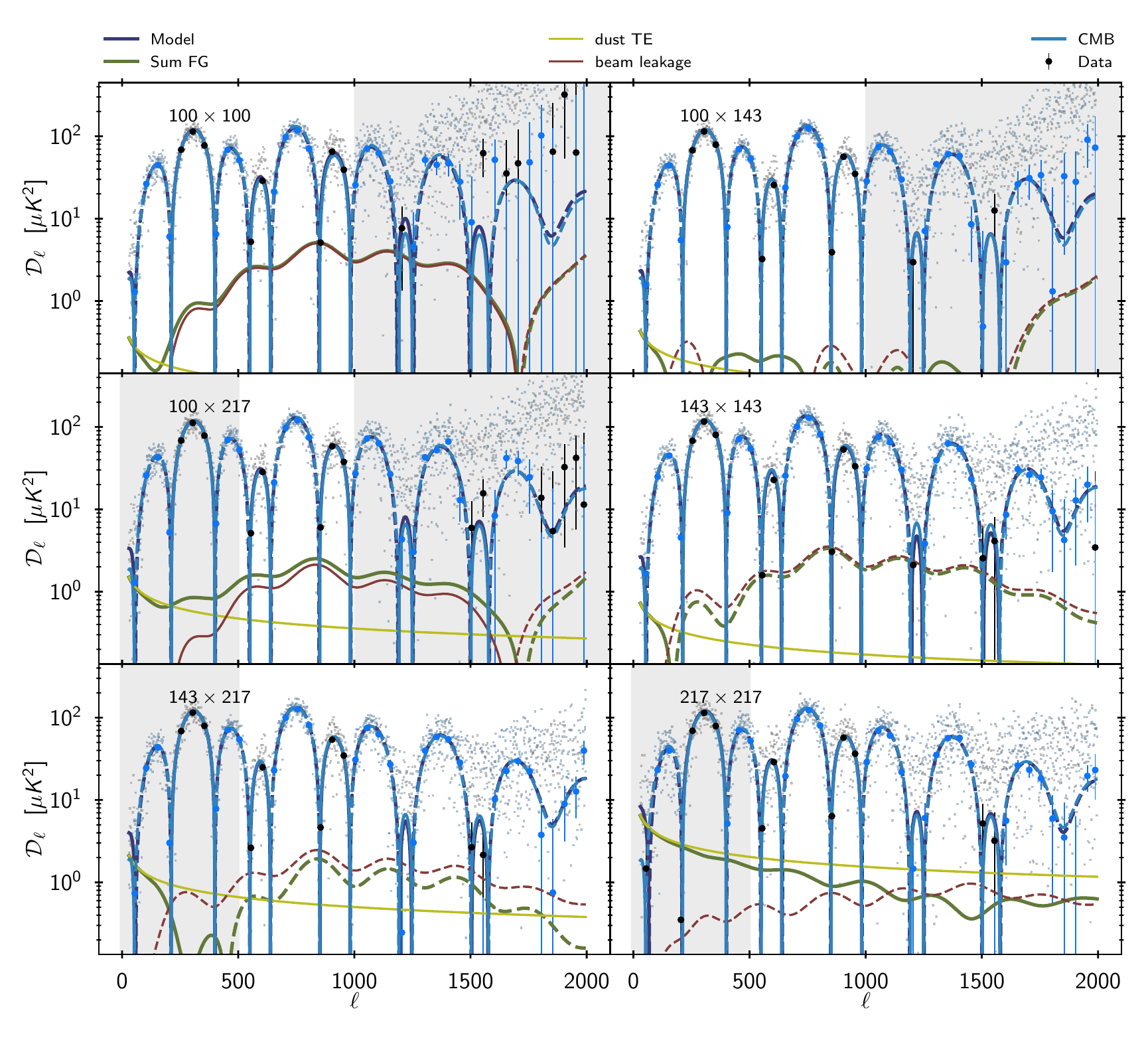}
  \end{centering}
  \caption{Best-fit model cross-spectra used for the TE high-$\ell$ likelihood. Black circles represent the data, binned at $\Delta\ell=50$, while thin grey points show each multipole, to give a sense of the scatter. Blue points and dashed lines correspond to negative values. The thick dark blue lines show the full model (sum of all components), while the dark green one is the sum of foreground and nuisance signals. Grey shaded area are not used for cosmology. }
  \label{fig:hi-ell:fgcmbTE}
\end{figure*}

\begin{figure*}[htbp!]
\begin{centering}
  \includegraphics[angle=0,width=0.99\textwidth]{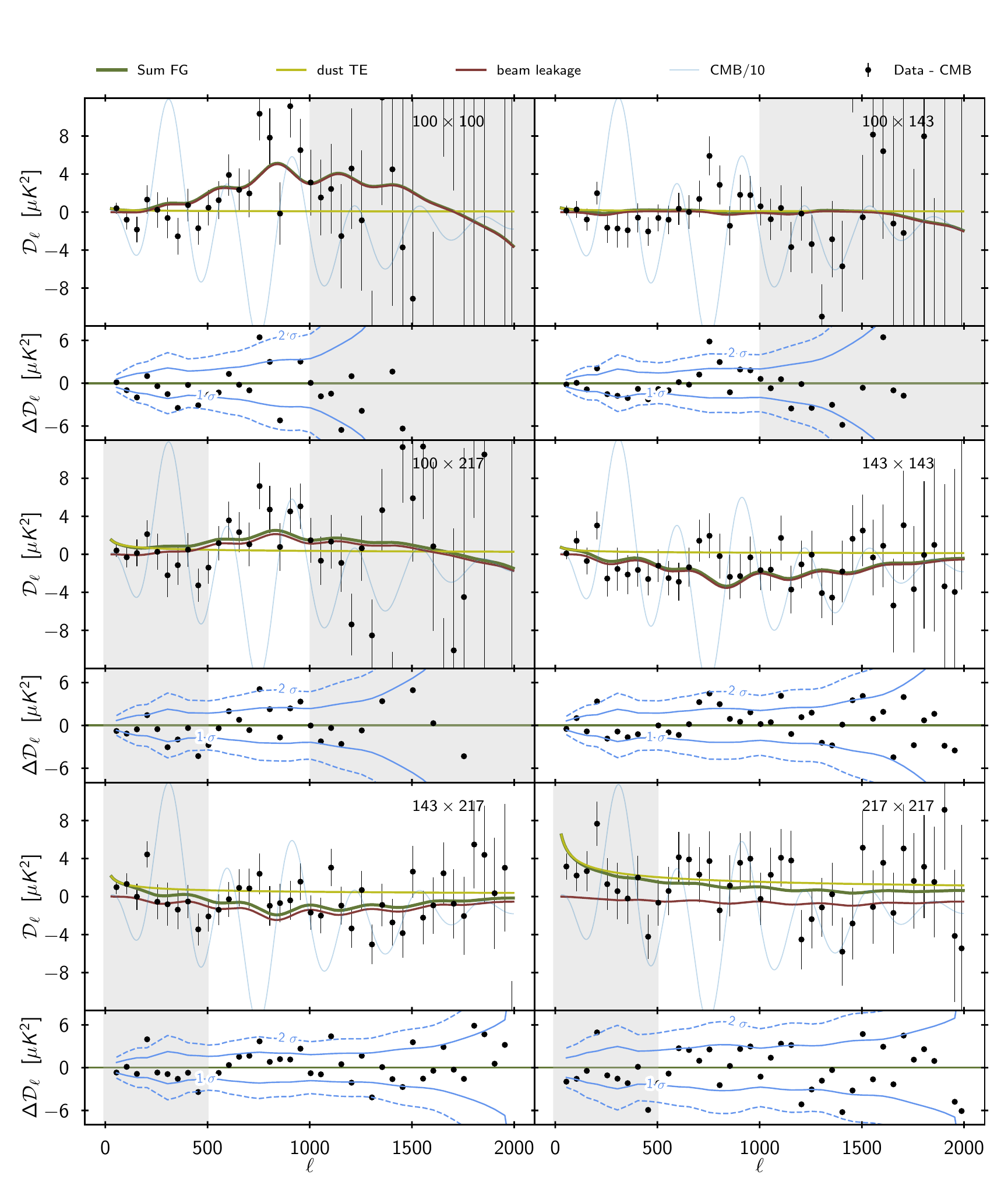}
\end{centering}
\vspace{-0.3cm}
  \caption{Best-fit foreground and nuisance models for all of the cross-spectra used in the TE high-$\ell$ likelihood. The data,
    with the best-fit theoretical CMB $C_\ell$ spectrum subtracted, are shown in black with a binning of $\Delta\ell=50$. The $TE$ CMB spectra divided by 10 are shown in the top parts of each panel. The bottom parts of each panel show the residual after foreground and nuisance correction. Grey shaded areas indicate regions of multipole space
    not used for cosmology. Except at $217\times217$, beam leakage
    largely dominates the dust contributions in the $\ell$ ranges we keep.}
  \label{fig:hi-ell:fgTE}
\end{figure*}

\begin{figure*}[htbp!]
\begin{centering}
  \includegraphics[angle=0,width=0.99\textwidth]{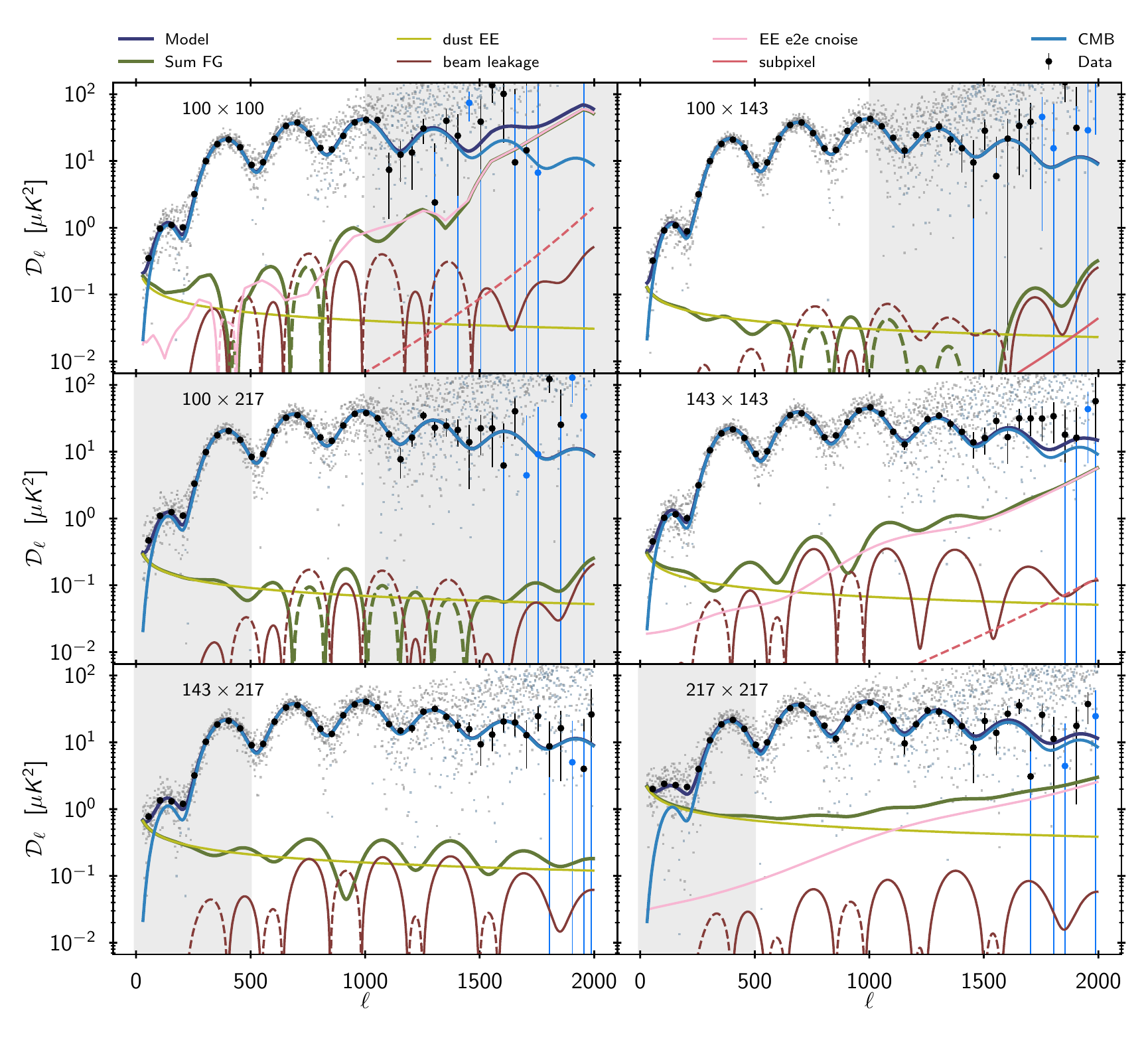}
  \end{centering}
  \caption{Best-fit model cross-spectra used for the EE high-$\ell$ likelihood. Black circles represent the data, binned at $\Delta\ell=50$, while pale grey points show each multipole to give a sense of the scatter. Blue points and dashed lines correspond to negative values. The thick dark blue lines show the full model (sum of all components), and the dark green one is the sum of foreground and nuisance parts. Grey shaded areas are not used for cosmology.}
  \label{fig:hi-ell:fgcmbEE}
\end{figure*}

\begin{figure*}[htbp!]
  \begin{centering}
\includegraphics[angle=0,width=0.99\textwidth]{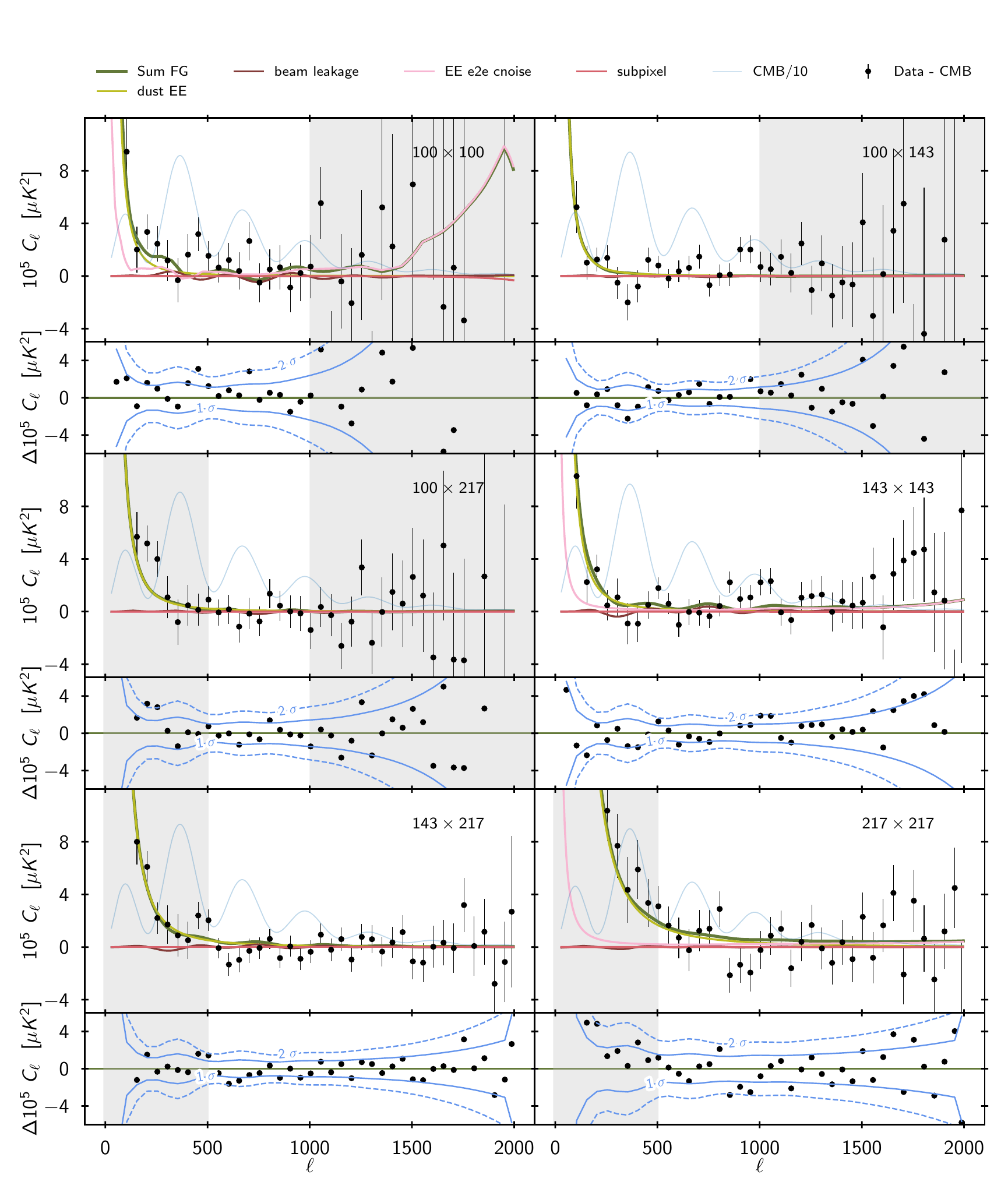}
  \end{centering}
  \caption{Best-fit foreground and nuisance models for all of the
    cross-spectra used in the EE high-$\ell$ likelihood. The data,
    with the best-fit theoretical CMB $C_\ell$ spectrum subtracted,
    are shown in black with a binning of $\Delta\ell=50$. The $EE$ CMB
    spectra divided by 10 are shown in the top parts of each panel. The bottom
    parts of each panel show the residual after foreground and nuisance
    correction. Grey shaded areas indicate regions of multipole space not used for cosmology.}
  \label{fig:hi-ell:fgEE}
\end{figure*}

We now turn to the model vector description. Sections~\ref{sec:hi-ell:datamodel:fg} and \ref{sec:hi-ell:datamodel:gal} focus on the astrophysical foregrounds part of the model, while the other sections 
describe the instrumental model. A comparison of the data and the
model is presented in Figs.~\ref{fig:hi-ell:fgcmbTT},
\ref{fig:hi-ell:fgcmbTE}, and \ref{fig:hi-ell:fgcmbEE} (for $TT$, $TE$, and
$EE$, respectively), 
and Figs.~\ref{fig:hi-ell:fgTT}, \ref{fig:hi-ell:fgTE}, and \ref{fig:hi-ell:fgEE} display the residuals after removing the CMB contribution.

\subsubsection{Extragalactic foregrounds }
\label{sec:hi-ell:datamodel:fg}
We ignore any extragalactic foreground contamination in the
polarization segments of the likelihood (i.e., in the $TE$ and $EE$ spectra). 
Observations from
ground-based CMB experiments have shown that, in the regions surveyed
by those experiments, polarized point-source contamination is smaller
than the level of sensitivity of \Planck\
\citep[e.g.,][]{2017arXiv170709353H}. 
As a further test of this hypothesis, we also tried to add our point-source mask to the polarized masks and 
observed negligible shifts in the cosmological parameters recovered from this new EE likelihood. Since the point sources are masked
in temperature, masking them in polarization is only relevant for EE (see also Sect.~\ref{sec:hi-ell:datamodel:mask}).

For temperature, we employ the same extragalactic foreground model as used in 2015. It includes templates to take into account the contribution of the following components.
\begin{description}
\item [\emph{Cosmic infrared background (CIB)}]\ We use a
  template power spectrum and emission law based on the one-plus-two
  halo model described in \citet{planck2013-pip56}. The template is
  close to a power law at high multipoles. As for the 2015 release, we assume
  perfect correlation between the emission at 100, 143, and
  217\,GHz. The template is rigidly rescaled by a single amplitude
  corresponding to the CIB contamination in the 217-GHz channel in
  ${\cal D}_\ell$ at $\ell=3000$ (without any colour correction). The
  amplitude of the observed CIB contribution is highly dependent on
  the dust model. We discuss in Sect.~\ref{sec:hi-ell:datamodel:gal}
  how this amplitude has changed between 2015 and 2018 as a result of
  the modification of the dust model. {We also note} that the \camspec\ 
  likelihood utilizes a slightly different model in which the
  correlation between frequencies is allowed to vary; this does not
  lead to significant differences in the recovered cosmological parameters.
\item [\emph{Point sources (PS)}]\ At the likelihood level, infrared
  and radio point sources cannot be separated and are accounted for by
  a single flat power spectrum for each auto- and cross-frequency
  spectrum.  The amplitude for each cross-spectrum is given in ${\cal D}_\ell$ at $\ell=3000$. 
\item [\emph{Kinetic SZ (kSZ)}]\ The kSZ emission is parameterized by a single amplitude in ${\cal D}_\ell$ at $\ell=3000$, scaling a fixed template from \citet{TBO11}.
\item [\emph{Thermal SZ (tSZ)}]\ The tSZ emission is also
  parameterized by a single amplitude in ${\cal D}_\ell$ corresponding to the
  emission at $\ell=3000$ at 143\,GHz (colour-corrected), which scales a
  fixed template given by the $\epsilon=0.5$ model from \citet{EM012}.
\item [\emph{Thermal SZ $\times$ CIB correlation}]\ Given the CIB
  and tSZ levels, the cross-correlation between the thermal SZ and the
  CIB, or tSZ $\times$ CIB, is parameterized by a single
  correlation parameter, $\xi$, which scales a fixed template from \citet{Aetal12b}.
\end{description}

The \Planck\ temperature S/N at high-$\ell$ has not changed
significantly between 2015 and 2018, and we continue to recommend the
use of a prior\footnote{{This prior, together with all the others discussed in the rest of the text (and summarized Table~\ref{tab:fg-params})
are used in all cosmological parameter estimations and tests presented in this paper, in \citetalias{planck2016-l06}, or in \citet{planck2016-l10}. These priors are not automatically included in the \plik\ or \camspec\ likelihoods, but are externally enforced when estimating parameters. Furthermore, these priors are used when marginalizing over nuisance parameters to produce the \pliklite\ likelihood.}
} to constrain the SZ amplitudes according to 
\begin{equation}
{\cal D}^{\rm kSZ} + 1.6 {\cal D}^{\rm tSZ} = (9.5 \pm 3)\,\mu\textrm{K}^2,
\end{equation}
in good agreement with the estimates of \citet{reichardt12}. 
As can be seen in Fig.~\ref{fig:hi-ell:fgcmbTT}, the kSZ, tSZ, and
tSZ$\times$CIB contributions are always small compared to those of the
dust, CIB, and point sources. The $100\times100$ and $143\times143$
spectra provide us with most of our constraining power on the tSZ,
with the tSZ contribution reaching a similar amplitude to that of the dust or PS between $\ell=500$ and $\ell = 800$.

{Given the relative amplitude of the different extragalactic components, the largest source of modelling uncertainty  is
the CIB. We discussed in \citetalias{planck2014-a13} why we  replaced the power-law model used in \citetalias{planck2013-p08} by the
halo model from \citet{planck2013-pip56}. The agreement of the halo model with a likelihood-based measurement of the CIB at higher frequency was investigated in 
\citet{Mak:2017}. The result of this work was to show that, in the range of scales we are exploring here, the halo template is in reasonable agreement with those improved CIB measurements, and also the amplitude of the CIB is highly correlated between at least the high frequencies. As discussed above, we assume full correlation for the baseline \plik\ likelihood CIB model. The alternative \camspec\ likelihood has investigated 
the effect of this last assumption, allowing for independent CIB amplitudes for $143\times 143$, $143\times 217$, and $217\times 217$ cross-spectra, and found no impact on the cosmological parameters even in extended models (as discussed in section~2.2.2 of \citetalias{planck2016-l06}). We present in Appendix~\ref{app:CIBextra} an evaluation of different CIB models in the \LCDM\ and \LCDM+$\Alens$ cases. As expected, we find no significant change in the cosmological parameters, but correlated changes on the CIB and PS parameters.}

{We show in Sect.~\ref{sec:valandro:priors} that ignoring the SZ prior translates into small shifts in $n_{\rm s}$. This is not unsurprising, since we showed in figure~44 of \citetalias{planck2014-a13} that the kSZ amplitude is mildly correlated with the slope of the primordial scalar fluctuations.
}

\subsubsection{Galactic foregrounds }
\label{sec:hi-ell:datamodel:gal}

\begin{table*}[htbp!] 
\begingroup 
\newdimen\tblskip \tblskip=5pt
\caption{Dust-plus-CIB contamination level (in $\mu{\rm K}^2$) at each
frequency, ${\cal D}_{\ell=200}^{\rm a}$ for different Galactic masks.}
\label{table:hil:dustTT}
\vskip -3mm
\footnotesize 
\setbox\tablebox=\vbox{
\newdimen\digitwidth
\setbox0=\hbox{\rm 0}
\digitwidth=\wd0
\catcode`*=\active
\def*{\kern\digitwidth}
\newdimen\signwidth
\setbox0=\hbox{+}
\signwidth=\wd0
\catcode`!=\active
\def!{\kern\signwidth}
\newdimen\decimalwidth
\setbox0=\hbox{.}
\decimalwidth=\wd0
\catcode`@=\active
\def@{\kern\decimalwidth}
\halign{ 
\hbox to 1.25in{#\leaderfil}\tabskip=1em& 
    \hfil#\hfil\tabskip=2em& 
    \hfil#\hfil&  
    \hfil#\hfil&  
    \hfil#\hfil&  
    \hfil#\hfil\tabskip=0pt\cr 
\noalign{\doubleline}
\omit\hfil Frequencies [GHz]\hfil& *G70& *G60& G50& G41& CIB*\cr 
\noalign{\vskip 3pt\hrule\vskip 5pt}
$100\times100$& $**8.8\pm*1.7$& $**4.3\pm*1.1$& $*2.56\pm*0.88$& $*1.61\pm*0.83$& $0.17\pm0.023$\cr
$143\times143$& $*25.7\pm*3.9$& $*11.3\pm*2.2$& $*6.5*\pm*1.6*$& $*4.7*\pm*1.6*$& $0.75\pm0.10*$\cr
$217\times217$& $309.4\pm39.9$& $151.7\pm21.7$& $99.8*\pm16.0*$& $73.9*\pm13.9*$& $7.9*\pm1.1**$\cr
\noalign{\vskip 5pt\hrule\vskip 3pt}
}}
\endPlancktablewide 
\tablenote {{\rm a}} The levels reported in this table correspond to the amplitude of the dust-plus-CIB emission, $\mathcal{D}_\ell$, at $\ell=200$ in $\muK^2$. They are obtained at each frequency by fitting the 545-GHz cross-half-mission spectra against the CMB-corrected $545\times 100$, $545\times 143$, and $545\times 217$ spectra over the range of multipoles $30\leq\ell\le500$. The CMB correction is obtained using the 2015 CMB best-fit. This contamination is dominated by dust, with a small CIB contribution. The columns labelled with a Galactic mask name (G41, G50, G60, and G70) correspond to the results when combining those masks with the CO, extended-source, and frequency point-source mask specific to each frequency. The CIB contribution is shown in the last column. This is calculated in an iterative way, estimating the CIB amplitude at $\ell=3000$ in the $217\times 217$ spectrum from a full Markov chain Monte Carlo (MCMC) estimation of cosmological parameters in an \LCDM\ model; the CIB contributions at different frequencies at $\ell=200$ used here are then inferred using the spectral energy distribution and multipole templates described in Sect.~\ref{sec:hi-ell:datamodel:fg}.\par
\endgroup
\end{table*} 

\begin{figure}[htbp!]
\includegraphics[angle=0,width=0.5\textwidth]{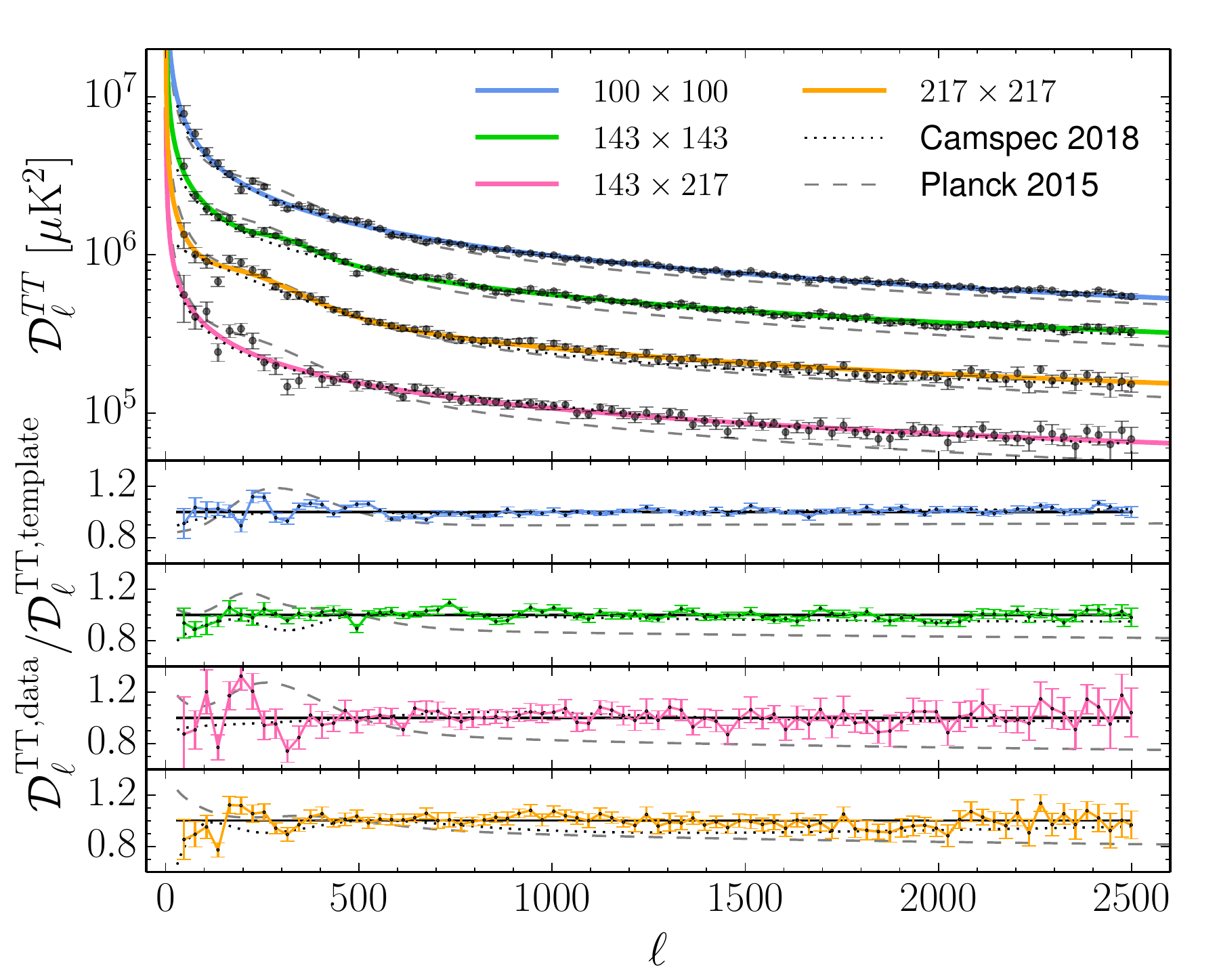}
\caption{Dust $TT$ templates. In the top panel the black data points are the 545-GHz, mask-differenced, half-mission spectra, calculated using the same masks as the ones used for the $100\times 100$, $143\times 143$, $217\times 217$, and $143\times 217$ power spectra (from top to bottom, respectively) in the likelihood. The solid lines show the dust templates estimated by fitting the data to the model of Eq.~\eqref{eq:dust:TTtemplate}. The grey dashed line shows the dust template used in 2015, while the black dotted line shows the templates used in the 2018 \camspec\ likelihood, both normalized at $\ell=500$ to the same amplitude as the baseline template.
The other (smaller) panels show the ratio between the data and the baseline dust templates (the solid lines of the first panel). Features that are not well modelled by our smooth template are at the level of 15\,\%. We also show that the baseline templates differ from the \camspec\ ones at the level of approximately 10\,\%. Finally, the 2015 template (which was the same at all frequencies) differs from the ones used in 2018 by up to 25\,\%, due to a different fit of the bump at $\ell\approx300$, the high-$\ell$ slope remaining essentially the same. This difference is compensated by a change in extragalactic foregrounds such as point sources and the CIB amplitude, as shown in Fig.~\ref{fig:hil:dustpars}.}
\label{fig:hil:dust545model}
\end{figure}
\begin{figure}[h!]
  \includegraphics[angle=0,width=0.49\textwidth]{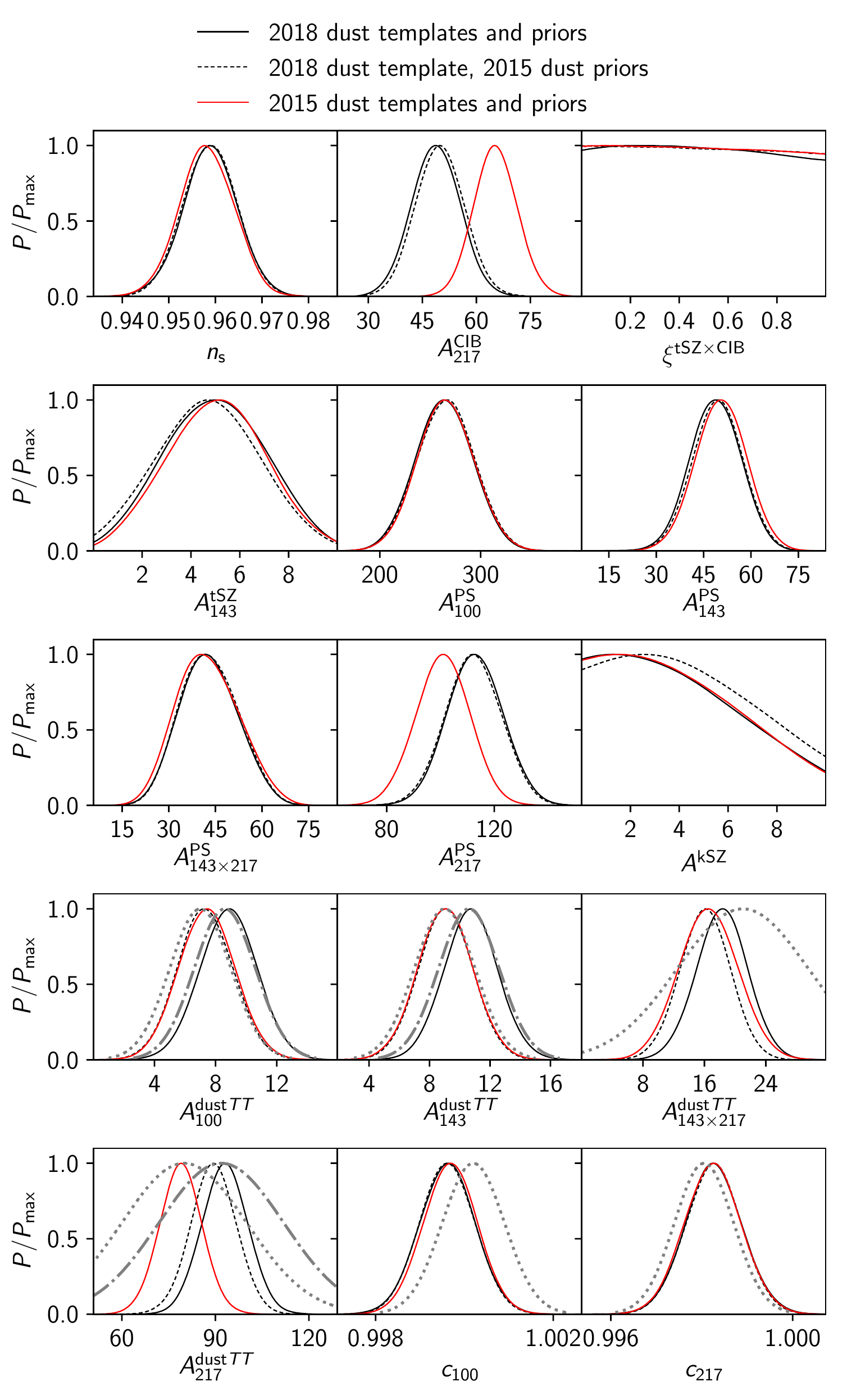}
  \caption{One-dimensional posterior distributions for cosmological and nuisance parameters using the TT\ high-$\ell$ likelihood with a $\tau$-prior of $0.055\pm0.09$ \citep[we used the \PICO\ code for this test,][]{Fendt:2006uh}. We show the results using the dust templates and priors from the 2018 release (black solid line), 2018 templates with the 2015 priors (black dashed), and the 2015 dust templates with the 2015 priors (red solid), assuming a \LCDM\ model. The dot-dashed grey lines show the Gaussian priors applied in 2018, while the dotted grey lines show the ones applied in 2015.  We only show the most significantly affected parameters, with $\ns$ being the cosmological parameter most altered, at the level of about $0.1\,\sigma$. On the other hand, the change in the shape of the template has a large impact on foreground parameters. This is due to the fact that the 2015 templates had a more pronounced bump at $\ell\approx300$, but similar high-$\ell$ slopes and similar amplitude priors at $\ell\approx200$, which induced a correlated shift in the point-source and CIB amplitudes at 217\,GHz.}
\label{fig:hil:dustpars}
\end{figure}

\begin{figure}[h!]
  \includegraphics[angle=0,width=0.495\textwidth]{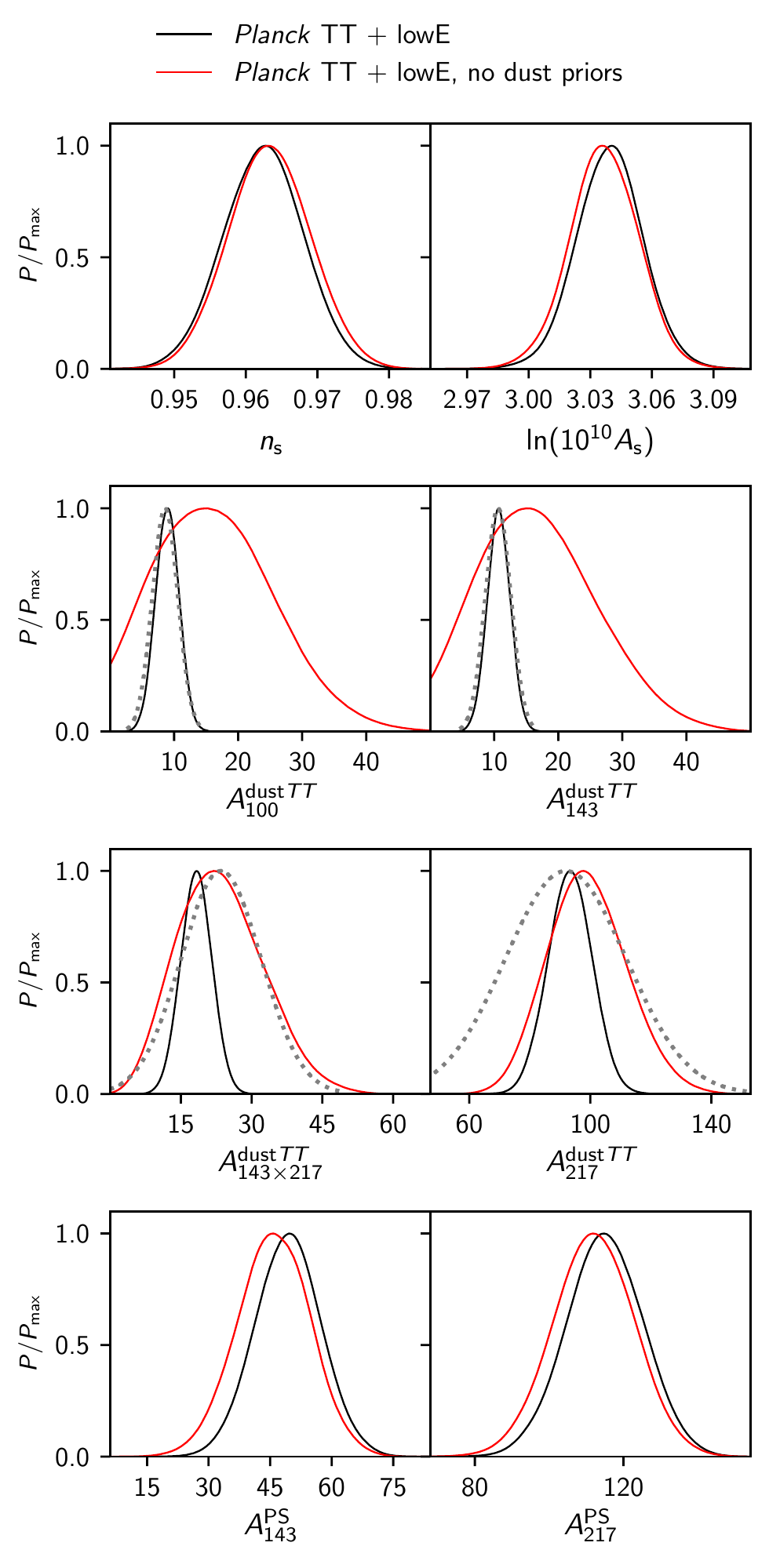}
  \caption{One-dimensional posterior distributions for cosmological and nuisance parameters using TT+\simall. The black line shows the baseline set of results, which adopts priors on the dust amplitudes. The red lines show the case when we leave the amplitudes free to vary; these are in good agreement with the priors set in the baseline, which are indicated with grey dotted lines. As a consequence, leaving the dust amplitudes free to vary without priors has small to negligible impact on other parameters.}
\label{fig:hil:TTdustfree}
\end{figure}
 
 We model the dust contribution to the power spectra following the 2015
results paper on cosmological parameters \citepalias{planck2014-a15} as
 \begin{equation}
\left(
C^{XY,\rm dust}_{\nu\times\nu'}\right)_{\ell} = A^{XY,\rm dust}_{\nu\times\nu'}\times C_{\ell}^{XY,\nu\nu'\rm dust}\;.
\label{eq:galactic-dust-intensity} 
\end{equation} 
Here $XY=TT,EE,TE$, while $\nu$ and $\nu'$ take values from 100, 143, and 217\,GHz, denoting
frequencies of the maps used in the cross-spectra of the high-$\ell$ likelihood, $C_{\ell}^{XY,\nu\nu'\rm dust}$ is the dust template power spectrum normalized to unity at $\ell=200$ in $TT$ and $\ell=500$ in $TE$ and $EE$, and with corresponding amplitude $A^{XY,\rm dust}_{\nu\times\nu'}$. While in 2015 we used the same dust
template at all frequencies, we now use channel-specific templates, dependent on the frequency and sky fraction for each
$TT$ cross-spectrum. In particular, different point-source masks have different effects on the shape of the dust contamination residual and this must
be taken into account.
In polarization, since we ignore point-source contributions and do not mask them, we use an identical template at each frequency and we
verify that it is a fair approximation to the dust contamination behaviour in $TE$ and $EE$ across frequencies and in our range of Galactic masks.
As in 2015, a relation between dust amplitudes at different frequencies using the dust spectral energy distribution as calculated in e.g., \cite{planck2016-l11A} cannot be easily derived, since different Galactic masks are employed at different frequencies and thus
the dust amplitudes also vary simply because of the different parts of the Galaxy being observed.

As for the 2015 release, for the TT and TE likelihoods all Galactic contributions are sampled together with the cosmological parameters, using priors obtained from higher frequency channels. However, in contrast to what was done in 2015, we now fix the amplitude of the dust contribution in EE. We will detail the reasons of this choice later in this section.

In the following, we will describe how we build our template dust power spectrum from high-frequency data and evaluate the amplitude of the dust contamination at each frequency and for each mask. {We} do not apply colour corrections in any of the results presented below.\footnote{We do not apply colour corrections on the dust amplitudes, consistently with our choice for other foreground parameters, such as the CIB or the point-source amplitudes, since these corrections have uncertainties linked to the knowledge of the spectral energy distribution of the foregrounds themselves and of the \Planck\ detector bandpasses.} Therefore, the dust amplitudes reported are those as measured by the full bandpass of the \Planck\ detectors, contrary to the choice adopted in \citet{planck2016-l11A}, where colour corrections are applied. Note also that we work with maps in thermodynamic units, so we convert the 545-GHz and the 857-GHz maps (which are in units of flux density per unit area) from ${\rm MJy}\,{\rm sr}^{-1}$ to $\mathrm{K_{CMB}}$ by dividing the maps by the conversion coefficients $U_{\rm c}\,[{\rm MJy}\,{\rm sr}^{-1}/{\rm K}_{\rm CMB}]=58.062295$ and $U_{\rm c}\,[{\rm MJy}\,{\rm sr}^{-1}/{\rm K}_{\rm CMB}]=2.2703657$, respectively. These are slightly updated with respect to the ones provided in table~3 of \citet{planck2014-a12}.

\paragraph{\textbf{Galactic \textit{TT} dust emission}:}\label{sub:galactic-dust:TT}

Let us first describe the situation for $TT$, and discuss $TE$ and $EE$ afterwards.
As in 2015, we calculate dust templates using the 545-GHz maps, where the dust contribution to the power spectrum is large with respect to the CMB power spectrum (and thus also with respect to the related CMB cosmic variance and chance correlation contributions). We cross-check that we obtain broadly compatible results using the 353-GHz or 857-GHz maps as well.
Half-mission cross-spectra at 545\,GHz are then calculated for each of the mask combinations used in the likelihood, in order to provide us with a good estimate of the large-scale behaviour of the dust. 

The masks applied to the various maps used in the likelihood differ not only because of the different Galactic sky fractions retained, but more importantly also
because of the frequency-dependent point-source masks applied (together with masks for CO and extended sources). These point-source masks have been produced as described in \citetalias{planck2014-a15}, and they remove some of the brightest Galactic areas that lie in regions not covered by our Galactic masks, altering the simple power-law modelling advocated in \citet{planck2013-p06b}. This is the reason why a specific dust-residual template must be calculated for each specific combination of masks employed in the cross-spectra used in the likelihood.

For each of our masks in intensity, we estimate the shape of the dust template using differences of the 545-GHz HM spectra measured with that mask and with a second one whose covered sky fraction is smaller by 10\,\% or 20\,\%. Assuming that the shape of the dust contamination does not vary substantially between this pair of masks, this procedure allows us to eliminate the contribution of isotropic extragalactic foregrounds such as CIB and point sources. We used a similar procedure in 2015, but instead of estimating specific templates for each frequency mask, as done here, we calculated only one template to be used at all frequencies; we did this using a joint point-source mask (together with masks for CO and extended sources) and calculating differences of spectra calculated with pairs of Galactic masks with only one set of bracketing sky fraction,
retaining 60\,\% and 40\,\% of the sky. Thus, in 2015 we calculated only one template for all frequencies, while for this release we use a specific template for each frequency cross-spectrum.

The power-spectrum difference is then fitted with a simple analytic model so as to smooth out noise. As in 2015, we use
\begin{equation}
  C_{\ell}^{TT,\rm dust} \propto (1 + h \, \ell^{\,k} \, e^{-\ell/t}) \times (\ell/\ell_{\rm p})^n,
  \label{eq:dust:TTtemplate}
\end{equation}
with $h$, $k$, $t$, and $n$ being free parameters that will differ for each mask, and $\ell_{\rm p}=200$. The fit is performed for $30 \leq\ell\leq 2000$. Figure~\ref{fig:hil:dust545model} shows the 2018 templates, and compares them to the 2015 one. The \camspec\ likelihood uses a similar procedure, with a slightly different analytic shape, also illustrated in Fig.~\ref{fig:hil:dust545model}. In this figure, all templates are normalized to have the same amplitude at $\ell=500$. The differences between the \plik\ and \camspec\ templates are small and of the order of 10\,\%, while the difference with the 2015 one is larger, up to 25\,\%. As mentioned above, this is mostly due to the fact that in 2015 we used a fixed set of bracketing masks and the combination of all point-source masks at all frequencies. As a consequence, the 2015 template featured a more prominent bump at $\ell\approx 300$ with respect to the templates we find here, for all frequencies, except for the template at $217\times217$, and featured a slope at high $\ell$ with $n=-2.63$, while here we find $n=-2.65$, $-2.57$, and $-2.55$ for the 100-, 143-, and 217-GHz masks, respectively. 
These changes induce a correlated change in the foreground solution, as shown in Fig.~\ref{fig:hil:dustpars}. The main impact is an increase in the point-source amplitude and a decrease in the CIB at 217\,GHz. However, we verified that the best-fit sum of the foreground contribution at $\ell=2000$ for $143\times143$, $143\times217$ and $217\times217$ changed by less than ${\cal D}_{\ell=2000}=2\,\mu {\rm K}^2$, which corresponds to 0.4, 0.6, and $1.1\,\sigma$ shifts, respectively.

After calculating the dust templates, we determined an estimate of the amplitude of the dust contamination in each of the frequency maps used for CMB analysis by computing their cross-spectra with the 545-GHz HM maps, similarly to what was done in 2015. Estimates are also checked using correlation with the 353-GHz and 857-GHz HM maps. 
These values are then used to set Gaussian priors on the amplitude of the dust emission, which are then sampled together with cosmological parameters.

The procedure for estimating such dust amplitude priors is the following.
We assume that at large scales all our maps $\vec{m}_{\nu}$ have in common only the CMB, CIB, and a variable amount of Galactic dust, so that 
\begin{equation}
\vec{m}_{\nu} = \vec{m}^{\mathrm{CMB}} + a_{\nu} \vec{m}^{\mathrm{dust+CIB}},
\label{eq:hil:dustmap}
\end{equation}
where $\vec{m}^{\mathrm{CMB}}$ is the CMB map and $a_{{\nu}}$ is the rescaling factor at frequency $\nu$ of the dust-plus-CIB map $\vec{m}^{\mathrm{dust+CIB}}$ measured at a reference frequency, which in our case is the 545-GHz one (so that $a_{{545}}=1$).  Here we assumed that the CIB and dust amplitudes scale in frequency with the same rescaling factor $a_{{\nu}}$, since the CIB emission law is similar enough to the dust one that differences in emission laws can be ignored. However, since the CIB and dust multipole templates are different (and therefore treated separately in the likelihood), it is necessary to disentangle the level of dust contamination alone from the CIB one. In order to do so, in the following, we will limit our fits to large scales ($\ell<500$), where the CIB contribution is small compared to Galactic dust, and we will subtract at the end the expected CIB residual to estimate the dust-only amplitudes, as we did in 2015.  Other extragalactic components at these scales are small and thus ignored. 

From Eq.~\eqref{eq:hil:dustmap}, the cross-spectrum between frequencies $\nu_1$ and $\nu_2$ is
\begin{align}
 \left ( C^{TT}_{{\nu_1}\times {\nu_2}}\right)_{\ell} = C^{TT,\rm CMB}_{\ell} +\,& a_{{\nu_1}}a_{{\nu_2}}\, C^{TT,\dustcib}_{\ell}\nonumber\\ +\,& (a_{{\nu_1}}+a_{{\nu_2}})\, C^{\rm chance}_{\ell},\label{eq:hil:chancedust}
\end{align}
where $C^{\rm chance}_{\ell}$ is the contribution from the chance correlation between the CMB and the dust (which would vanish on average over many sky realizations) and $C^{TT,\dustcib}_{\ell}$ is the dust spectrum at the reference frequency. 
Eq.~\eqref{eq:hil:chancedust} implies
\begin{align}
 \left ( C^{TT}_{{\nu}\times {\nu}}\right)_{\ell} =\,& C^{TT,\rm CMB}_{\ell} + a_{{\nu}}^2\, C^{TT,\dustcib}_{\ell} + 2a_{{\nu}}\, C^{\rm chance}_{\ell}, \nonumber\\
\left ( C^{TT}_{{\nu}\times {545}}\right)_{\ell} =\,& C^{TT,\rm CMB}_{\ell} + a_{{\nu}}\, C^{TT,\dustcib}_{\ell} + (a_{{\nu}}+1)\, C^{\rm chance}_{\ell}, \nonumber\\
\left ( C^{TT}_{{545}\times {545}}\right)_{\ell} =\,& C^{TT,\rm CMB}_{\ell} + C^{TT,\dustcib}_{\ell} + 2\, C^{\rm chance}_{\ell}.\label{eq:hil:dustcl}
\end{align}
To evaluate the dust contribution in the $\nu\times\nu$ power spectrum, $a^2_{\nu}C^{TT,\dustcib}_{\ell}$, we use the $C^{TT}_{545\times \nu}$ and $C^{TT}_{545\times 545}$ spectra, after subtracting an estimate of the CMB contribution. This is analogous to the procedure adopted in \cite{planck2016-l11A}, where estimates focused on scales larger than the ones considered here. We use the best fit of the 2015 \Planck\ data release and verify that the dust estimates are robust against small changes in these best fits (e.g., by using the 2018 best fit instead of the 2015 one). This is more important when using the 353-GHz map as a dust tracer, but we only do that as a cross-check of our 545-GHz-based estimations. Assuming that the chance correlations average out over the multipole range considered, the dust rescaling factor can be measured from Eq.~\eqref{eq:hil:dustcl} as
\begin{equation}
a_{\nu}= \left< C^{TT}_{545\times \nu-{\rm CMB}}/C^{TT}_{545\times 545-{\rm CMB}} \right>_{\ell=30-500},
\label{eq:hil:alphadust}
\end{equation} 
where $\left< \right>_{\ell=30-500}$ indicates the weighted mean over the multipole range $30 \leq \ell \leq 500$ (uncertainties are dominated by the scatter in the $545\times \nu$ cross-spectrum). The estimate of the dust contribution at $\ell=200$ is then calculated as $a_{\nu}^2 \left(C^{TT}_{545\times 545-{\rm CMB}}\right)_{\ell=200}$, after smoothing the $545\times 545$ spectrum around the multipole of interest. This approach has the advantage of estimating the dust amplitude in the spectrum $\nu\times\nu$ without using that spectrum directly, albeit assuming knowledge of the CMB power spectrum. 

Table~\ref{table:hil:dustTT} reports the results of the fits at each frequency, for each Galactic mask. The error range quoted corresponds to the uncertainty in the fits. The values reported correspond to the sum of the CIB and the dust contamination at $\ell=200$. The last column of Table~\ref{table:hil:dustTT} gives the estimate of the CIB contamination at the same multipole from the joint cosmology and foreground fit, which is removed to determine the dust prior we use in the likelihood.

We derive our priors on the foreground amplitudes from this table, combining the 545-GHz fit with the estimated residual CIB contamination, to obtain the following values:
$(8.6\pm2)\,\muK^2$ for the $100\times100$ spectrum (G70); $(10.6\pm2)\,\muK^2$ for $143\times143$ (G60); and $(91.9\pm20)\,\muK^2$ for $217\times217$ (G50). Finally the $143\times217$ value is obtained by computing the geometrical average between the two auto-spectra for the larger of the two masks (G60), yielding $(23.5\pm8.5)\,\muK^2$.
These are comparable with the values obtained in 2015 to within 10\,\%.
Figure~\ref{fig:hil:dustpars} shows the impact of using the new dust templates and the new priors in place of the old ones. We also checked that when leaving the dust amplitudes free to vary without priors, the recovered values are within 0.5$\,\sigma$ of the estimated priors, with no impact on the \LCDM\ parameters and only small shifts in the point-source amplitudes, as shown in Fig.~\ref{fig:hil:TTdustfree}. As expected, the posterior distributions of the dust amplitudes are much wider than the priors, since the dust-amplitude estimates inferred using the cross-correlation with the 545-GHz maps are much more precise than what is estimated directly from the CMB frequencies used in the likelihood.

We further verify these results using a ``dust-cleaning'' procedure (also sometimes referred to as ``undusting''), already described in section~3.2 of \citetalias{planck2014-a15} (where the rescaling factor $a_\nu$ used here is related to the $\alpha$ one used in that paper through $a_\nu=\alpha/(1+\alpha)$). From Eq.~\eqref{eq:hil:dustmap}, a CMB map cleaned of the dust contamination can be obtained as $\vec{m}_\nu^{\rm clean}=(\vec{m}_\nu-a_{\nu} \vec{m}_{545})/(1-a_{\nu})$. Therefore, the $C^{TT,\rm clean}_{\nu_1\times\nu_2}$ CMB-only spectrum at $\nu_1\times\nu_2$ cleaned from the dust contribution can be obtained from
\begin{align}
\left(C^{TT,\rm clean}_{\nu_1\times\nu_2}\right)_\ell=\,& \frac{1}{(1-a_{\nu_1})(1-a_{\nu_2})}\times
\bigg[\left(C^{TT}_{\nu_1\times\nu_2}\right)_\ell
 +a_{\nu_1}a_{\nu_2} \left(C^{TT}_{545\times545}\right)_\ell \bigg.
 \nonumber\\
\bigg. &\qquad\qquad -a_{\nu_1} \left(C^{TT}_{\nu2\times545}\right)_\ell -a_{\nu_2} \left(C^{TT}_{\nu1\times545}\right)_\ell\bigg] .
\label{eq:hil:undusting}
\end{align}
The dust rescaling factor $a_\nu$ is calculated by minimizing

\begin{equation}
\sum_\ell \left(C^{TT,\rm clean}_{\nu_1\times\nu_2}\right)_\ell \tens{C}^{-1}_{\ell\ell'} \left(C^{TT,\rm clean}_{\nu_1\times\nu_2}\right)_{\ell'}\,,
\label{eq:hil:undustingminimize}
\end{equation}
where the covariance matrix $\tens{C}$ of the $C^{TT,\rm clean}_{\nu_1\times\nu_2}$ spectrum does not contain the variance related to the dust (plus CIB) contribution (in practice, this covariance matrix is calculated using a fiducial model that does not contain foregrounds). 

This procedure is analogous to the previous one, with the difference that it correctly takes the rescaling of the CMB-dust chance correlations into account and it does not use a best-fit CMB to subtract for that contribution, but directly uses the $\nu\times\nu$ spectrum.

We find good agreement, as expected, in the rescaling factors found with the two procedures, with estimates agreeing at better than the 10\,\% level at 100\,GHz and 2\,\% at 217\,GHz. 

Lastly, the dust (plus CIB) contribution can be estimated as
\begin{equation}
\left(C^{TT,\dustcib}_{\nu_1\times\nu_2}\right)_\ell=\left(C^{TT}_{\nu_1\times\nu_2}\right)_\ell- \left(C^{TT,\rm clean}_{\nu_1\times\nu_2}\right)_\ell\,. \label{eq:hil:dustundust}
\end{equation}

\begin{figure}[htbp!]
\includegraphics[angle=0,width=0.5\textwidth]{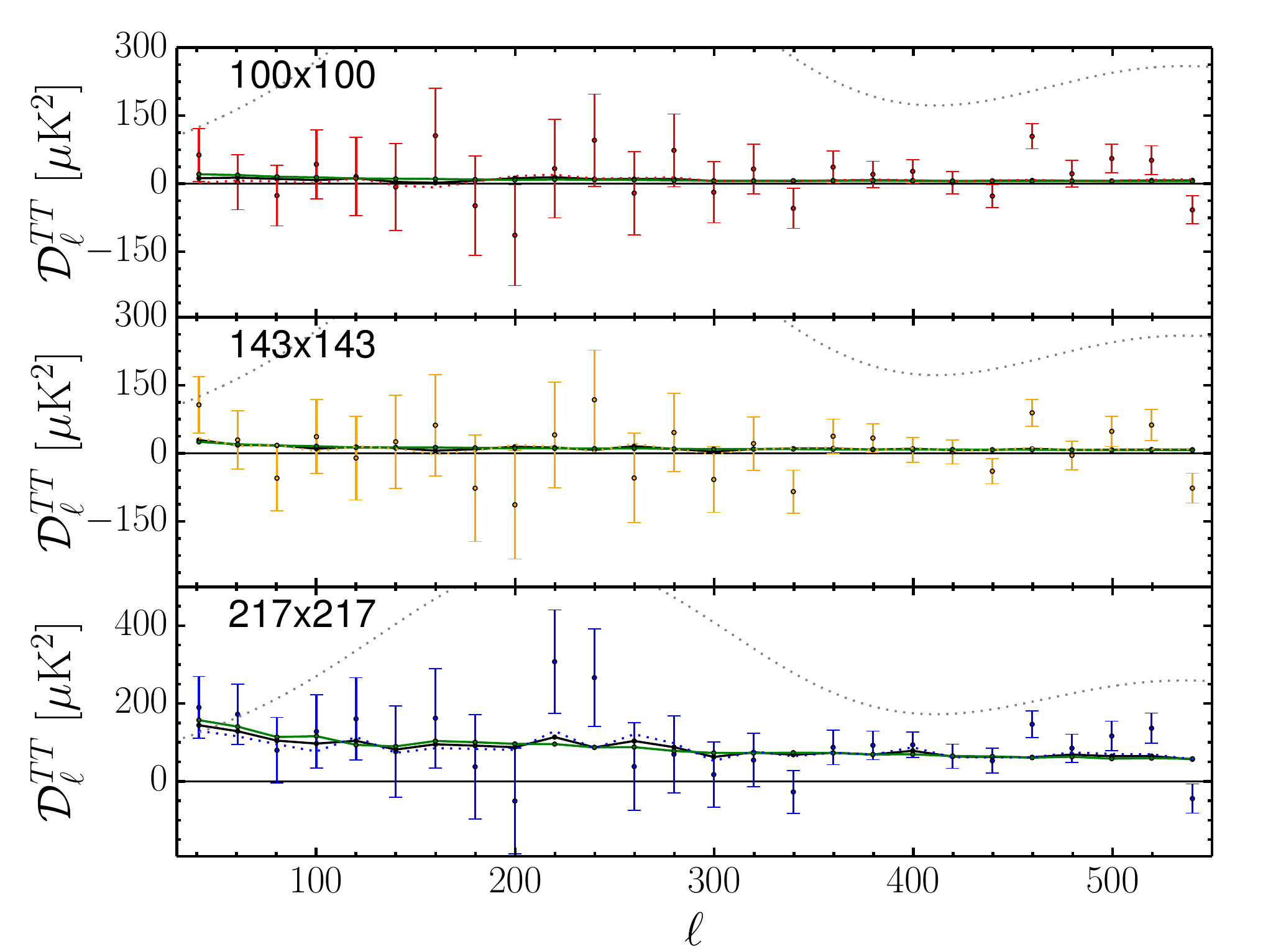}
\caption{Dust-plus-CIB contamination estimates in temperature. The $100\times100$, $143\times143$, and $217\times217$ half-mission cross-spectra with the 2018 \Planck\ CMB best-fit subtracted are shown from top to bottom. The green line shows $a_\nu^2 C^{TT}_{545\times 545-{\rm CMB}}$, with the $a_\nu$ rescaling factor estimated from Eq.~\eqref{eq:hil:alphadust}, the black line shows $a_\nu C^{TT}_{\nu\times 545-{\rm CMB}}$, while the dotted coloured line shows the dust contribution as estimated from the cleaning procedure in Eq.~\eqref{eq:hil:dustundust}. The grey dotted line shows the \TT{}\ CMB best-fit.}
\label{fig:hil:comparecross545}
\end{figure}

Figure~\ref{fig:hil:comparecross545} shows the $C^{TT}_{\nu\times \nu-{\rm CMB}}$ power spectrum minus the best-fit CMB contribution, compared to the dust-plus-CIB contribution at $\nu\times\nu$, estimated as $a_\nu^2 C^{TT}_{545\times 545-{\rm CMB}}$ and $a_\nu C^{TT}_{\nu\times 545-{\rm CMB}}$, with the rescaling coefficients measured from Eq.~\eqref{eq:hil:alphadust} and the dust estimate from Eq.~\eqref{eq:hil:dustundust}. Once again, we find good agreement between these different estimates.

As illustrated by the \camspec\ ''cleaned'' likelihood described in appendix~A of \citetalias{planck2016-l06} and section~3.2 of \citetalias{planck2014-a15}, and the \texttt{mspec} likelihood described in appendix~D.1 of \citetalias{planck2014-a13} one could have used the cleaned CMB spectrum $C^{TT,\rm clean}_{\nu_1\times\nu_2}$ directly in the likelihood, avoiding the need to fit for dust amplitudes with priors, as is done in the baseline approach. However, the cleaning procedure removes not only dust, but also extragalactic components, whose residuals are harder to model in a robust fashion. Furthermore, while the cleaning procedure minimizes (by construction) the dust sample variance and the CMB-dust chance correlations, it also adds the noise of the 545-GHz maps, correlated across the frequencies used in the likelihood. While we tried this procedure as a cross-check, we consider the baseline approach to be more robust.

 
\paragraph{\textbf{Galactic \textit{TE} and \textit{EE} dust emission}:}
\label{sub:galactic-dustPol}
\begin{figure}[t!]
\includegraphics[angle=0,width=0.49\textwidth]{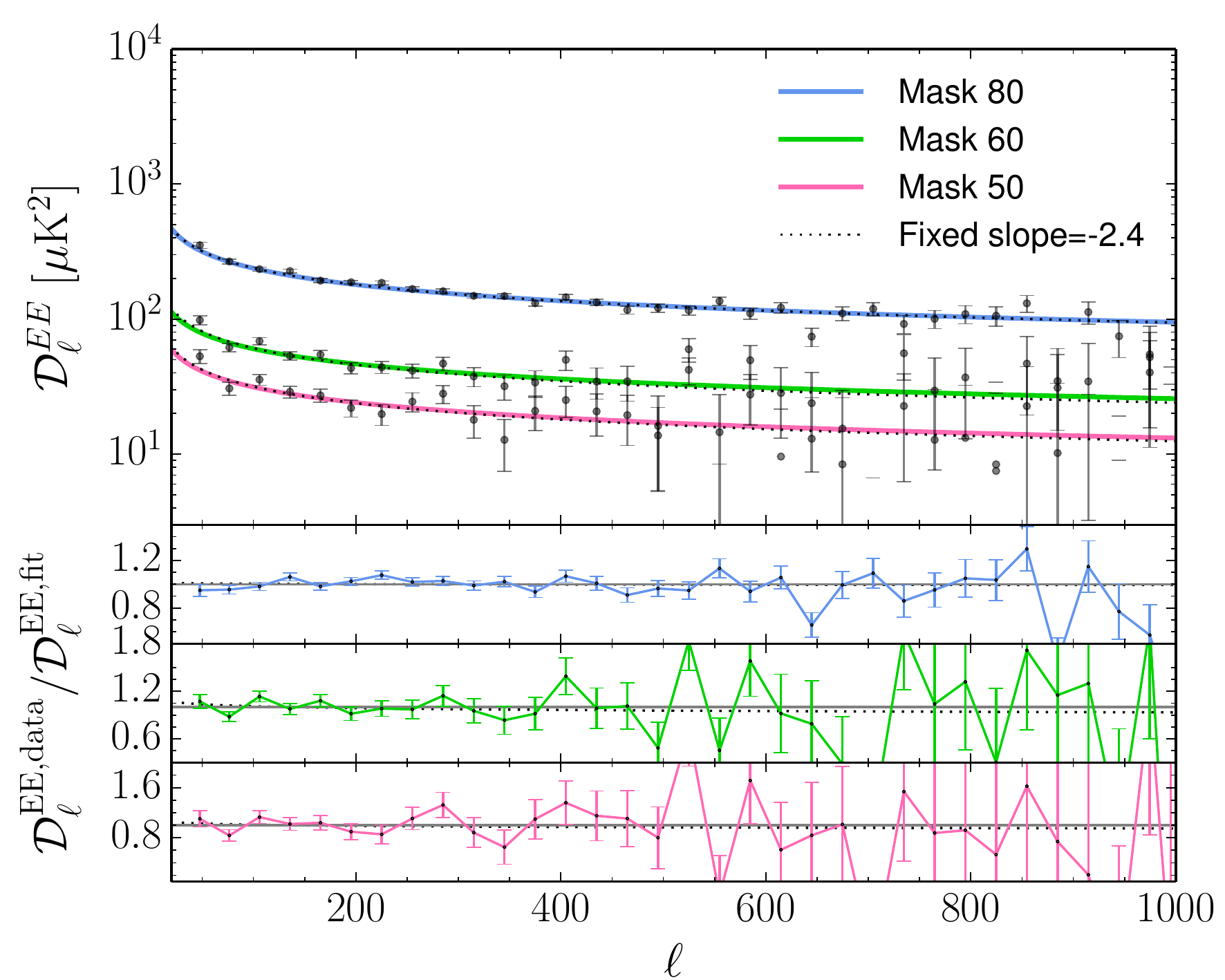}
\caption{Dust $EE$ templates. In the upper panel, the black data points are the 353-GHz, mask-differenced, half-mission spectra, calculated using the same masks as the ones used for the $100\times 100$, $143\times 143$, and $217\times 217$ power spectra (from top to bottom, respectively) in the likelihood. The solid lines show the dust templates estimated fitting the data with a power law. The black dotted line shows templates with fixed slope $C_\ell\propto \ell^{\,-2.4}$, fit to the data. The fits with free slopes are in good agreement with the ones obtained with slope fixed at $-2.4$. The dust templates used in the likelihood thus assume the same slope of $-2.4$ at all frequencies.
The lower panels show the ratio between the data and the dust templates. The templates with fixed slope, whose ratios with the ones with free slopes are shown by the black dotted lines, differ from the ones with free slopes by less than 10\,\%.}
\label{fig:hil:dustEEmodel}
\end{figure}

We check that a simple power law with a slope of $n=-2.4$ is a suitable template for the dust contamination in polarization, similar to what was used in 2015.  This was done using the 353-GHz maps, which is \Planck's highest frequency with polarization information. The procedure is similar to the one used for temperature, using the fact that at 353\,GHz the dust contribution is large with respect to the CMB one. We calculate half-mission cross-spectra at this frequency for the masks used in the likelihood. Unlike for the temperature masks, those for polarization only include the Galactic and missing pixel masks (which do change at each frequency, but which have a small impact on the final results). The dust templates are thus expected to be similar for different masks and frequencies, since they do not contain frequency-dependent point-source masks.

For each of our polarization masks, we estimate the shape of the dust template using the difference of 353-GHz half-mission spectra computed on two masks covering a sky fraction bracketing the target mask, thus getting rid of isotropic components such as the CMB. The bracketing is obtained by varying the size of the Galactic mask by 20\,\% (10\,\% for the G41 mask). 

\begin{figure}[t!]
\includegraphics[angle=0,width=0.49\textwidth]{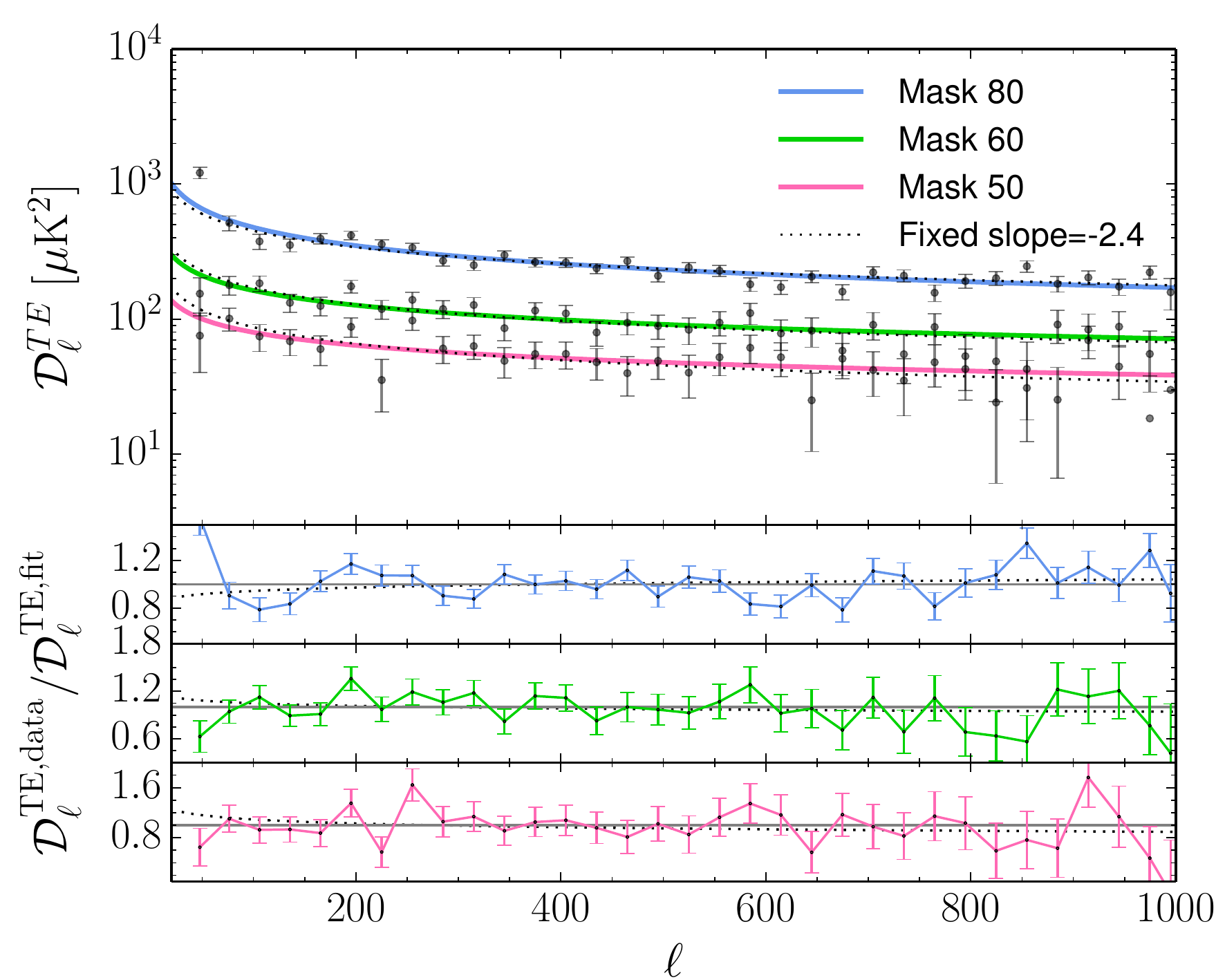}
\caption{Same as for Fig.~\ref{fig:hil:dustEEmodel}, except now for $TE$.}
\label{fig:hil:dustTEmodel}
\end{figure}

We fit the power spectrum difference as a simple power law, $ C_{\ell}^{TE,EE,\rm dust} \propto \ell^{\,n} $,
either fixing the slope to $n=-2.4$ or letting it vary freely, over the range $30 \leq \ell \leq 1000$. We find that for all our masks 
assuming a slope of $n=-2.4$ is in reasonably good agreement with the results obtained when the slope is free to vary, with slopes that can change by less than 5\,\% between different masks, confirming the findings of \citet{planck2016-l11A}. This is also shown in Fig.~\ref{fig:hil:dustEEmodel} for $EE$ and in Fig.~\ref{fig:hil:dustTEmodel} for $TE$.
In detail, we obtain: $n=-2.44\pm0.03$, $n=-2.35\pm0.08$, and $n=-2.31\pm0.12$ for the masks used in $TE$ for $100\times100$, $143\times143$, and $217\times217$; and $n=-2.40\pm0.02$, $n=-2.37\pm0.07$, and $n=-2.37\pm0.10$ for the masks used in $EE$, with similar results for the cross-frequency spectra. We verified that subtracting the best-fit CMB instead of a power spectrum calculated from a bracketing mask, as done in \citet{planck2016-l11A}, yields similar results with smaller uncertainties.

\begin{figure}[htbp!]
\includegraphics[angle=0,width=0.5\textwidth]{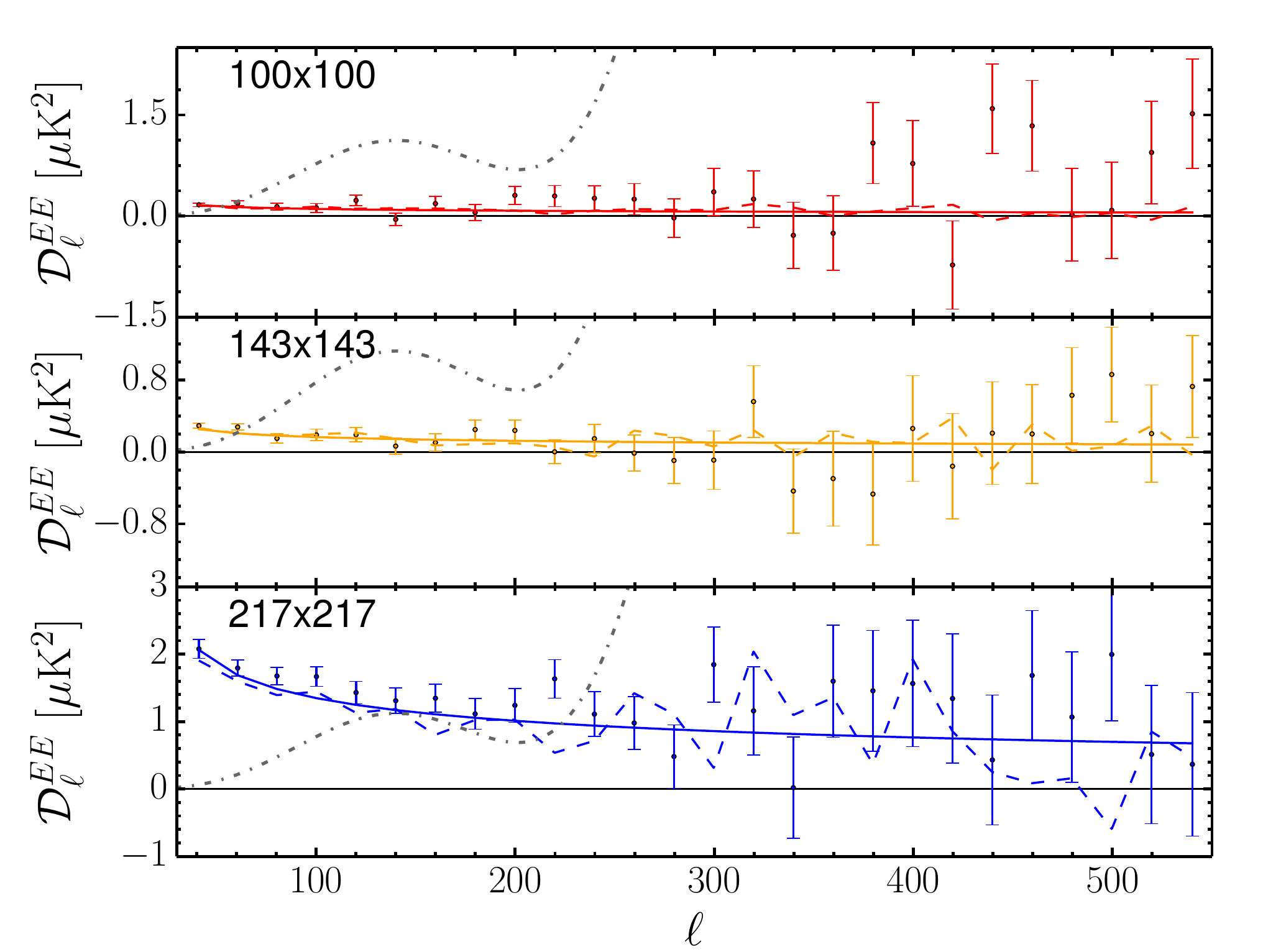}
\caption{Comparison between the dust contribution to the $EE$ power spectra, as estimated from the ``dust cleaning'' procedure in Eq.~\eqref{eq:hil:dustundust} (dashed lines), and the smooth power-law model used in the likelihood (solid lines). The data points are the $100\times 100$, $143\times 143$, and $217\times 217$ spectra (from top to bottom) minus the best-fit \Planck\ baseline 2018 results (shown in dot-dashed grey).
}
\label{fig:hil:EEdust}
\end{figure}

As far as fitting the dust amplitudes at the frequencies and masks used in the likelihood are concerned, for the $TE$ and $EE$ power spectra we use similar procedures to the ones used in temperature, with the simplification that in polarization there is no CIB contribution or extragalactic component, and now we use 353\,GHz instead of the 545\,GHz as a dust monitor. 
In particular, for $TE$ we calculate the dust rescaling factors as in Eq.~\eqref{eq:hil:alphadust}, where we assume that the dust rescaling factors in $TT$ and in $EE$ are the same, $a^{T}_\nu=a^{E}_\nu$, following section~5.3 of \cite{planck2016-l11A}. The fit is performed in the range $30 \leq \ell \leq 300$. Contrary to the temperature case, the $TE$ estimates are very noisy, due to the small dust amplitude, which translates into a low S/N even at $353\times353$. We thus set priors with large uncertainties of about 30\,\%.
For $EE$, we find more stable results by fitting a smooth template to the dust component evaluated in Eq.~\eqref{eq:hil:dustundust}. 
{W}hen evaluating dust amplitudes we do take into account the polar efficiency and the correlated noise corrections described in Sects.~\ref{sec:hi-ell:datamodel:noise} and \ref{sec:hi-ell:datamodel:inst}. However, we do not have estimates of the polar efficiency corrections at 353\,GHz (at this frequency it was only possible to calculate relative ones between detectors, as discussed in Sect.~\ref{sec:hi-ell:datamodel:inst} of this paper and in section~5.10.3 of \citetalias{planck2016-l03}). Since these corrections are expected to be as large as a few percent, they project into additional uncertainties in the dust estimates. Finally, the estimates performed here used the 353-GHz maps produced using both the PSBs and SWBs. We verified that using the 353-GHz maps produced without the PSBs, as suggested in \citetalias{planck2016-l03}, changes results by less than 1\,\% in \EE\ and 10\,\% in \TE, within estimates of statistical fluctuations.

\begin{figure}[htbp!]
\includegraphics[angle=0,width=0.5\textwidth]{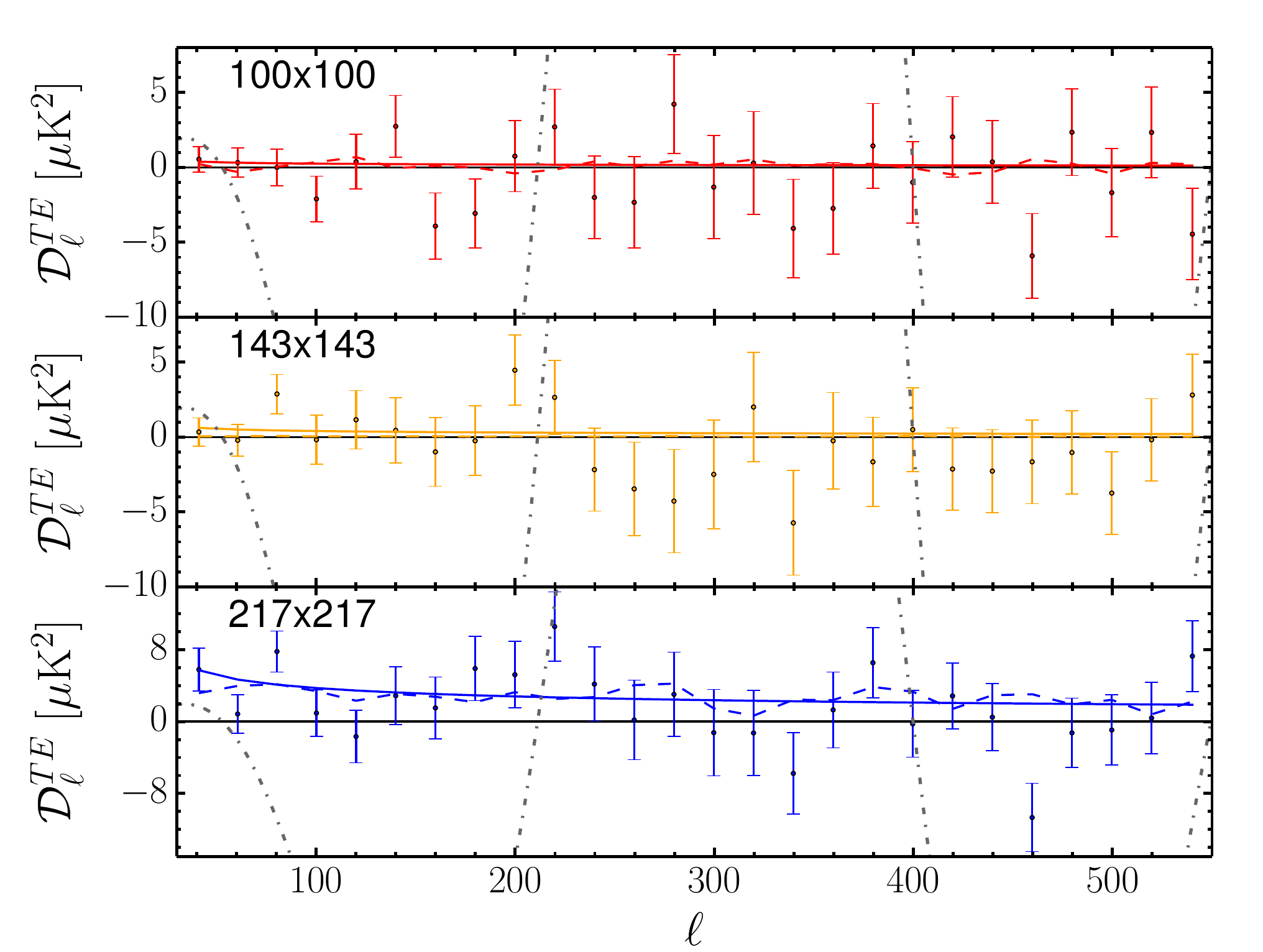}
\caption{Same as Fig.~\ref{fig:hil:EEdust}, but for the $TE$ power spectra.
}
\label{fig:hil:TEdust}
\end{figure}

\begin{figure}[htbp!]
\includegraphics[angle=0,width=0.5\textwidth]{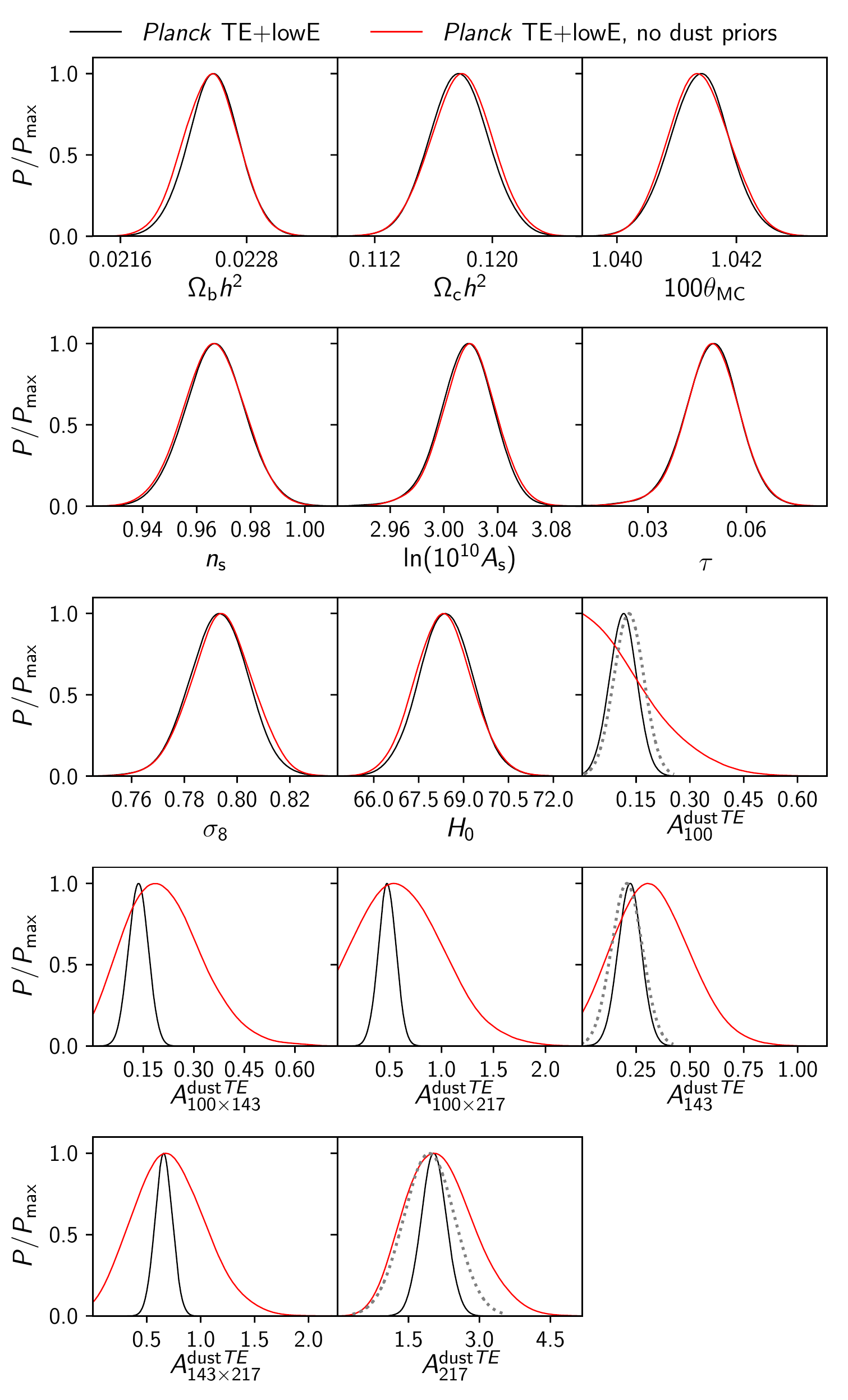}
\caption{One-dimensional posterior distributions of cosmological parameters and dust amplitudes for \TE{}+\simall. The black line shows the baseline results, which have priors on the dust amplitude, while the red one shows results when letting the dust amplitudes be completely free to vary. The grey dotted line shows the priors we set on the parameters. There is good agreement between the recovered dust amplitudes and the priors.
}
\label{fig:hil:1dTEdustprior}
\end{figure}
Table~\ref{table:hil:dustPol} gives the resulting values. Cross-frequency, cross-mask estimates are obtained by computing the 
geometric averages of the auto-frequency contaminations for the largest mask of each pair.

As done for $TT$, we compare the Galactic dust contribution calculated using a smooth power-law template, with amplitudes given in Table~\ref{table:hil:dustPol}, to the dust contribution obtained using the cleaning procedure in Eq.~\eqref{eq:hil:dustundust}.
This is shown in Fig.~\ref{fig:hil:EEdust} for $EE$ and Fig.~\ref{fig:hil:TEdust} for $TE$ and shows good agreement.
Additionally, we verified in $EE$ that using the dust-cleaned spectrum $C^{TT,{\rm clean}}_{\nu_1\times\nu_2}$ from Eq.~\eqref{eq:hil:undusting} directly in the likelihood, instead of including the dust contribution in the model vector with a fixed amplitude (as done in the baseline approach), does not change cosmological parameters.

\begin{table*}[htbp!] 
\begingroup 
\newdimen\tblskip \tblskip=5pt
\caption{$TE$ and $EE$ dust contamination levels, ${\cal D}_{\ell}$ at
$\ell=500$ in $\mu{\rm K}^2$.  Values reported in the table correspond to
the dust priors set on the contamination level at each frequency.}
\label{table:hil:dustPol}
\vskip -3mm
\footnotesize 
\setbox\tablebox=\vbox{
\newdimen\digitwidth
\setbox0=\hbox{\rm 0}
\digitwidth=\wd0
\catcode`*=\active
\def*{\kern\digitwidth}
\newdimen\signwidth
\setbox0=\hbox{+}
\signwidth=\wd0
\catcode`!=\active
\def!{\kern\signwidth}
\newdimen\decimalwidth
\setbox0=\hbox{.}
\decimalwidth=\wd0
\catcode`@=\active
\def@{\kern\decimalwidth}
\halign{ 
\hbox to 1.5in{#\leaderfil}\tabskip=0.5em& 
    \hfil#\hfil\tabskip=2em&
    \hfil#\hfil&
    \hfil#\hfil\tabskip=0pt\cr
\noalign{\doubleline}
\omit\hfil Spectrum\hfil& 100\,GHz (G70)& 143\,GHz (G50)& 217\,GHz (G41)\cr 
\noalign{\vskip 3pt\hrule\vskip 5pt}
\multispan4\hfil ${\cal D}^{TE}_{\ell=500}$\hfil\cr
\noalign{\vskip 3pt}
100\,GHz (G70)& $0.13\pm0.04$& $0.13\pm0.04$& $0.46\pm0.09$\cr
143\,GHz (G50)&              & $0.21\pm0.07$& $0.69\pm0.09$\cr
217\,GHz (G41)&              &              & $1.94\pm0.54$\cr
\noalign{\vskip 3pt\hrule\vskip 5pt}
\multispan4\hfil ${\cal D}^{EE}_{\ell=500}$\hfil\cr
\noalign{\vskip 3pt}
100\,GHz (G70)& $0.055\pm0.014$& $0.040\pm0.010$& $0.094\pm0.023$\cr
143\,GHz (G50)&                & $0.086\pm0.022$& $0.206\pm0.051$\cr
217\,GHz (G41)&                &                & $0.70*\pm0.18*$\cr
\noalign{\vskip 3pt\hrule\vskip 3pt}
}}
\endPlancktablewide
\endgroup
\end{table*} 
%

\begin{figure}[h!]
\includegraphics[angle=0,width=0.5\textwidth]{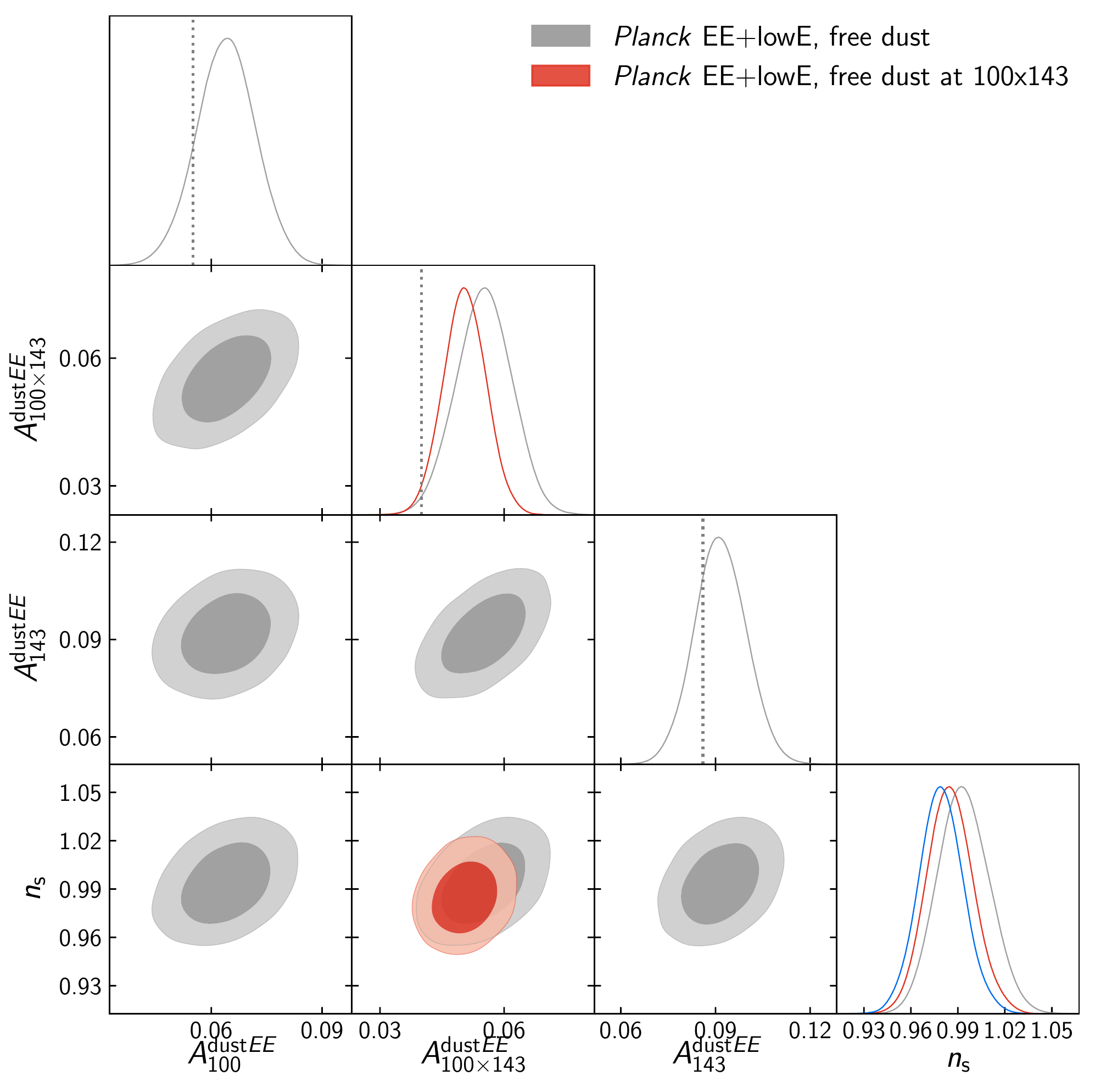}
\caption{Posterior distributions in one and two dimensions for the dust amplitudes and $\ns$ for EE+\simall. The grey lines and contours show results when letting the dust amplitudes be completely free to vary, the red line and contours show results when letting only the dust amplitude of the $100\times143$ spectrum be free to vary, while the blue line shows the baseline results, fixing the amplitudes to the values in Table~\ref{table:hil:dustPol}. The grey dashed lines show the baseline values of the dust amplitudes. When the dust amplitudes are allowed to be free to vary, the amplitude of the $100\times143$ cross-spectrum increases to values that are incompatible with those obtained using the 353-GHz map.
 }
\label{fig:hil:2dEEdustprior}
\end{figure}

Figure~\ref{fig:hil:EEdust} suggests that in $EE$ the dust contribution estimated with the baseline procedure at $217\times217$ might be underestimated by about 10\,\%. We remind the reader here, however, that we do not use this spectrum at low multipoles, $\ell<500$. Furthermore, we have verified that using a higher dust amplitude at this frequency does not have any impact on cosmological results from $EE$ alone in the \LCDM\ case.

Finally, we have investigated the impact of leaving the amplitude of the dust priors free to vary, as opposed to setting the dust amplitudes to the values shown in Table~\ref{table:hil:dustPol}. 
In $TE$ we recover results that are in agreement with the priors, with no impact on cosmology, as shown in Fig.~\ref{fig:hil:1dTEdustprior}. 
However, in $EE$
the posterior distribution of the dust amplitudes for $100\times 100$, $100\times 143$, and $143\times 143$ are higher than the value of the prior by $1\,\sigma$, $2.2\,\sigma$, and $0.6\,\sigma$, respectively, as shown in Fig.~\ref{fig:hil:2dEEdustprior}. For the spectra involving the 217-GHz map, the dust amplitudes are weakly constrained (due to the use only of multipoles larger than $\ell=500$), so letting them freely vary has no impact on cosmological results. The increase in the dust amplitudes is correlated with an increase of the value of $\ns$ of $0.9\,\sigma$, of which about half is due to the increase in the $100\times143$ dust amplitude. 
Such large dust amplitudes, especially the almost 40\,\% increase at $100\times 143$, are in slight disagreement with the dust estimates obtained using the 353-GHz maps. While this might be due to a statistical fluctuation, in Sect.~\ref{sec:valandro:cuts} we also check that eliminating this cross-spectrum from the likelihood shifts cosmological results by an amount that is within expectations. {Furthermore, assuming that the data set without priors can be considered as a subset of the data set that includes them, we calculated the expected standard deviation ($\sigmaexp$) of the difference between parameters obtained in the two cases using the formalism described in \citet{gc2019} and also described further in Sect.~\ref{sec:valandro:cuts}. We find that this shift in $\ns$ is only deviant by $1.7\,\sigmaexp$.}
We also verified that this cannot be accounted for by synchrotron contamination. This was done using a procedure similar to the one leading to Eq.~\eqref{eq:hil:alphadust}, using as a synchrotron monitor the 30-GHz LFI maps (in place of the 353-GHz HFI maps used for dust) and cross-correlating them with the 100-GHz maps using the G70 mask (see also Sect.~\ref{subsec:cleaning} for a similar procedure at low-$\ell$). This leads to an estimate of the synchrotron amplitude of less than 5\,\% of that of the dust at $\ell=80$.
For the baseline runs we thus decided simply to fix the dust amplitudes to the values in Table~\ref{table:hil:dustPol}, retaining the possibility of letting them freely vary for cross-checks.

\subsubsection{Noise model }
\label{sec:hi-ell:datamodel:noise}

Noise modelling enters into the likelihood construction pipeline in
two places:
\begin{enumerate}[(i)]
\item An estimate of statistical properties of the noise is needed to compute 
the noise contribution to the $C_{\ell} C_{\ell'}$ covariance
matrix. As in previous \Planck\ likelihood releases, we assume that the
noise contributions may be adequately computed from noise angular power
spectra and pixel-variance maps.
\item While we use cross-correlations between data cuts to avoid a
  direct noise contribution to a power spectrum, there might remain small correlations between 
the different maps.  These would induce ``correlated noise residuals'' in the
cross-spectra and need to be checked for.
\end{enumerate}

\paragraph{Noise statistical properties:}
\begin{figure}[htbp!]
\begin{centering}
  
  \includegraphics[angle=0,width=0.495\textwidth]{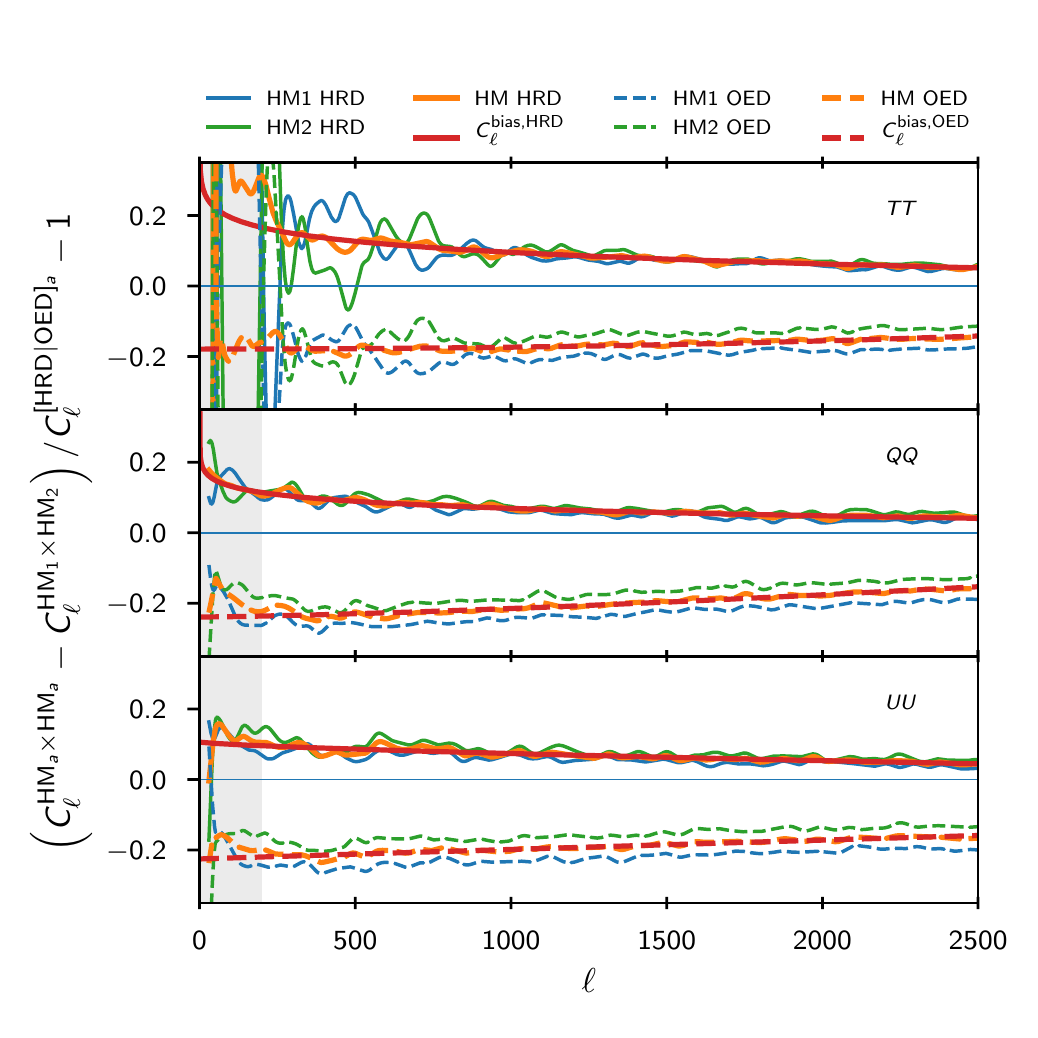}
  \end{centering}
  \caption{Correction ($C^{\mathrm{bias}}_{\ell}$) for the noise estimates based on HRDs (solid lines) and OEDs (dashed lines), as described in Eqs.~\eqref{eq:biasnoise} and \eqref{eq:plik_hrd_noise_bias}. Blue and green lines shows the two contribution to the bias estimates (i.e., for HM1 and HM2), showing that the bias is similar enough in each so that we can use the mean for the correction (orange line). The model fit (red) is obtained by restraining the fit to the non-shaded region ($\ell>200$). As we found in \citetalias{planck2014-a13}, HRD noise estimates are found to be biased low (due to correlations between half-ring maps, as explained in \citealt{planck2013-p03}). OED noise estimates, on the other hand, are biased high (in particular in polarization). Once the corresponding correction is applied to either case (HRD or OED), the overall noise models are similar. We only present here the 100-GHz channel results where the effect is the most striking. OEDs have very low anti-correlation for $TT$ at 143\,GHz and a slight correlation at 217\,GHz, while the $QQ$ and $UU$ equivalents are still anticorrelated at all frequencies.
 }
  \label{fig:hi-ell:noise_bias}
\end{figure}

We follow the method described in \citetalias{planck2014-a13} to build the
noise model that will enter into the computation of the covariance.
We remind the reader that this model is estimated in two steps. First, for each half-mission map, a rough estimate of the noise power 
spectrum shape in $T$, $Q$, and $U$ is obtained using ``half-ring
 difference'' (HRD) maps, and fit using a low-order polynomial-like
approximation of the form
\begin{equation}
  \log(C^{\mathrm{HRD}^{(i)}}_{\ell}) = \sum_{j = 0}^{4} \alpha^{(i)}_j \,
  \ell^{j} + \alpha^{(i)}_5 \log(\ell + \alpha^{(i)}_6) \, ,
  \label{eq:plik_hrd_noise_fit}
\end{equation}
where $(i)$ stands for $1$ or $2$, denoting the {first or last} HM. 
{A set of $\alpha^{(i)}_j$ is needed for each $T$, $Q$, and $U$ and each HM map at each frequency channel.}
 {Cotemporal half-ring} maps are known to be
correlated to one another by the deglitching and mapmaking
steps.  {Hence HRD} spectra underestimate the noise level and need to be 
corrected for a slowly varying bias that we assume to be identical in each half-mission and which, following \citetalias{planck2014-a13} notation, we will denote $C_\ell^{\rm bias}$:
\begin{equation}
C_\ell^{{\mathrm{noise}^{(i)}}} = C^{\mathrm{HRD}^{(i)}}_{\ell} \times C_\ell^{\rm bias}.
\end{equation}

For each frequency and each $T$, $Q$, and $U$ map, for a given half-mission $(i)$, we can estimate an empirical
multiplicative bias ${\tilde C_\ell^{\rm bias}}$ by computing the ratio between 
the difference of the HM auto-spectrum of the half-mission $(i)$ and the
cross-HM spectrum, and the HRD noise
estimate $C^{\mathrm{HRD^{(i)}}}_{\ell}$. We show in Fig.~\ref{fig:hi-ell:noise_bias} that the empirical bias estimate 
is very close for each half mission, and thus we form our overall empirical estimate by taking the average of the two {and assume that the same debiasing can be applied to Eq.~\eqref{eq:plik_hrd_noise_fit} for both HMs}:
\begin{align}
{\tilde C_\ell^{\rm bias}} &\,= {\left(\frac{C_{\ell}^{\mathrm{HM}^{(1)}\times\mathrm{HM}^{(1)}}-C_{\ell}^{\mathrm{HM}^{(1)}\times\mathrm{HM}^{(2)}}}{2C_\ell^{\mathrm{HRD}^{(1)}}}\right.} \nonumber\\
&\qquad\qquad{\left.+\frac{C_{\ell}^{\mathrm{HM}^{(2)}\times\mathrm{HM}^{(2)}}-C_{\ell}^{\mathrm{HM}^{(1)}\times\mathrm{HM}^{(2)}}}{2C_\ell^{\mathrm{HRD}^{(2)}}} \right )}.
\label{eq:biasnoise}
\end{align}
A smooth multiplicative bias correction ${C_\ell^{\rm bias}}$
is obtained by fitting a simple power-law model to the empirical ${\tilde C_\ell^{\rm bias}}$:
\begin{equation}
  C^{\mathrm{bias}}_{\ell} = \beta_0 \, \ell^{\,\beta_1} + \beta_2 \, .
  \label{eq:plik_hrd_noise_bias}
\end{equation}
{A different set of $\beta_k$ parameters is determined for each frequency and
map-type ($T$, $Q$, or $U$) combination, but are common between the two HMs.}
Examples of the bias model are shown in Fig.~\ref{fig:hi-ell:noise_bias}. 

{In} Sect.~\ref{sec:hi-ell:datamodel:datasel}{, we describe} the odd-even
data selection.  This can also be used for noise estimation by using
odd-even difference (OED) maps instead of HRD maps.
Unfortunately, OEDs do not solve the correlation issues found in the
HRDs. We show in Fig.~\ref{fig:hi-ell:noise_bias} that the OE maps
are found to be slightly anticorrelated, so that, if used in the simplest
way, the OED
results would overestimate the noise level.  Just as in the HRD case,
the OED results can be corrected using a corresponding procedure,
yielding essentially equivalent noise spectra estimates. {We} {n}ote that the
2018 \camspec\ likelihood uses uncorrected OED results for its noise
estimate, lowering its $\chi^2$ values slightly.

{The two-step procedure described above uses the HRD (or OED) spectra as an intermediate step, but after debiasing with Eq.~\eqref{eq:biasnoise} the baseline likelihood noise model is entirely based
on the difference between the HM maps auto- and cross-spectra $C_{\ell}^{\mathrm{HM}^{(i)}\times\mathrm{HM}^{(i)}}-C_{\ell}^{\mathrm{HM}^{(1)}\times\mathrm{HM}^{(2)}}$. 
The intermediate step can be bypassed and the noise model can be reproduced without the help of the HDR spectra. 
Indeed, combining Eqs.~\eqref{eq:plik_hrd_noise_fit}, ~\eqref{eq:biasnoise}, and ~\eqref{eq:plik_hrd_noise_bias}, the final noise 
approximation for each HM map reads
\begin{equation}
C_\ell^{\mathrm{noise}^{(i)}} = \exp \left(\sum_{j = 0}^{4}
\alpha^{(i)}_j \, \ell^{j} + \alpha^{(i)}_5 \log(\ell + \alpha^{(i)}_6) \right)
\times (\beta_0 \, \ell^{\,\beta_1} + \beta_2 ),
\end{equation}
with the $\beta_k$ parameters common to both HM maps of a given frequency channel, and the $\alpha^{(i)}_j$ ones being different for each one (here the index $i$ is 1 or 2, corresponding to the first or last HM map). 
The noise approximation of each frequency channel and $T$, $Q$, or $U$ HM maps can thus be equivalently obtained by 
fitting for the $\alpha^{(i)}_j$ and $\beta_k$ parameters jointly on both differences between the HM maps' spectra and their cross-spectrum.}

Overall, the noise spectra have a very similar shape to those of 2015
(i.e., slightly non-white, as shown in figure~24 of \citetalias{planck2014-a13}). 
The levels are similar, with a slight increase of the order of 2\,\%, except at 217\,GHz in temperature, which is 4\,\% higher, and 143\,GHz in polarization, which is about 12\,\% lower. 
As discussed in section~3.2 of \citetalias{planck2016-l03}, the large difference for the 143-GHz polarization is mainly due to the change in the destriper smoothing scale. 
The other smaller differences can be traced back principally to the
discarding of the last 1000 rings of the mission and to a lesser
extent to the change in the numbers of bad pixels that are discarded
from the maps (which have increased between 2015 and 2018, after a
change in the condition number threshold at the mapmaking
stage). Indeed, if we apply the 2018 ring-selection and
condition-number mask to the 2015 data, 
we recover noise levels similar to those of 2018.

Variations of the pixel noise variance over the sky are handled as in
2015, using a heuristic correction to the analytic white-noise approximation for
the covariance matrix. 
As described in appendix~C of \citetalias{planck2014-a13}, we rescale the
expressions that are valid for white anisotropic noise to be correct
in the case of non-white isotropic noise. We expect that this
correction is sufficient in our case where departure from white noise
is small and noise anisotropy is highly localized in space. This
assumption was checked 
numerically in 2015 with the help of Monte Carlo simulations.

The 2018 pixel variance maps are very similar to those of 2015, except for the removal of the last 1000 rings, and an increase in the number of bad pixels, as described above.
We already discussed the effect of the removed rings, which slightly changes the overall noise level. We checked using simulations that the new masked pixels do not create significant extra variance in the spectra.

\paragraph{Correlated noise residuals:}
\begin{figure}[htbp!]
\begin{centering}
  
  \includegraphics[angle=0,width=0.495\textwidth]{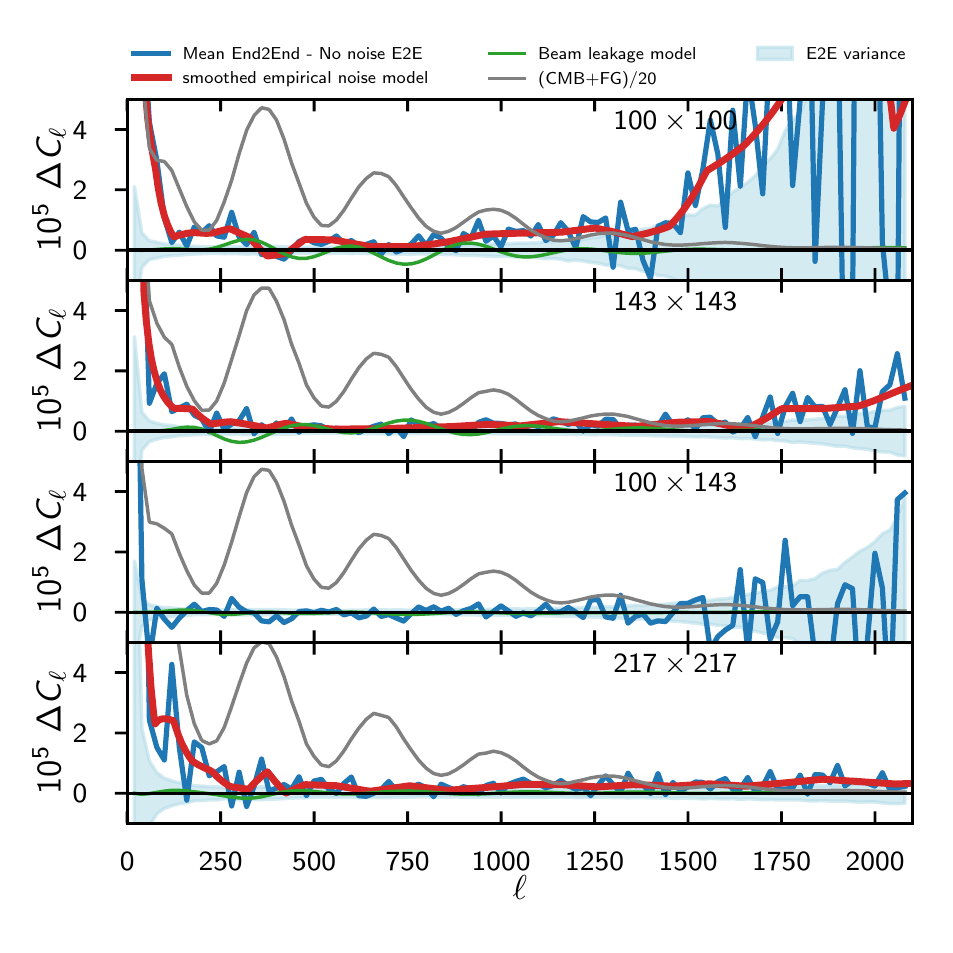}
  \end{centering}
  \vspace{-.5cm}
  \caption{Difference between the mean $EE$ spectra of the 300 end-to-end simulations and the {\it noise-free\/} simulation; both share the same input CMB and foreground maps. The difference between the two spectra, shown in blue, is smoothed in $\Delta\ell=20$ flat bins. The blue shaded area gives the variance between the 300 simulations. The green line is the beam-leakage correction model shown for scale; it is cancelled in the difference between the mean end-to-end simulation and the noise-free spectra. For scale as well, the grey line presents $5\,\%$ of the input simulation power (e.g., CMB plus foreground from the Planck Sky Model). From top to bottom the panels are the $100\times100$,$143\times143$, $100\times143$, and $217\times217$ spectra. Apart from the lowest bin ($\ell=20$, which is not used for the likelihood analysis), the $100\times143$ spectra do not show any significant residual, while all of the others show signs of possible correlated noise. We do not show the other cross-spectra that behave similarly. In the auto-frequency spectra plot, we show (with a red line) the empirical end-to-end $EE$ correlated noise model, produced by fitting smoothing splines to the blue lines, after removing strongly varying trends at low $\ell$ and high $\ell$. 
}
  \label{fig:hi_ell:data:e2e_EE_noise}
\end{figure}

In 2015, we reported important correlated noise residuals between the
detset maps, but could not find any significant signature of
correlated noise between the half-mission maps. Whilst we could have built a data-inspired correction for the detset spectra, which would have 
improved the agreement between cosmological parameters obtained from detsets and half-mission maps, we decided to use only the latter in 2015 and are doing likewise for the 2018 release. 

Correlated noise was identified in 2015 by measuring the cross-spectra between different detset HRD maps. The cross-spectra between half-mission HRD maps, however, was compatible with zero in the multipole ranges used in our likelihood.
We have reproduced those tests for the 2018 half-mission maps, checking that changes in the processing have not introduced any new correlations. As for the 2015 release, we do not find any significant sign of correlated noise for the likelihood's $\ell$ range. 
We also performed similar tests, discussed in detail in Appendix~\ref{app:oe_correlation}, also using odd-even differences. While we did not find any significant sign of noise correlation between the HM maps, in the likelihood's $\ell$ range, we do, however, find signs of noise correlations between the odd and even maps.

All of these tests, being based on data-cut differences, cannot detect
any residual present in all of our data selections.
However, the end-to-end simulations (described in more detail in
Sect.~\ref{sec:valandro:sims}) can be used to constrain any residual correlations. 
We form the means of the HM spectra of the 300 end-to-end simulations and compare them to the spectra 
of the noise-free simulation (which includes foreground and projection effects). 
In the case of EE, we find excesses of power at large and
small scales in (only) the single-frequency spectra 
(i.e., for $100\times100$ $EE$, $143\times143$ $EE$, and $217\times217$ $EE$),
which exceed the end-to-end simulation dispersions (Fig.~\ref{fig:hi_ell:data:e2e_EE_noise}).
The fitted empirical correlated noise model that we produce using the
end-to-end simulations is added to the model vector of the $EE$ likelihood. It can be compared to
the contribution of the other nuisance and foreground contributions in Figs.~\ref{fig:hi-ell:fgcmbEE} and~\ref{fig:hi-ell:fgEE}. 
At large scales, its amplitude is within one order of magnitude of the
dust emission. The effect of this correlated noise on the cosmological
parameters is discussed in Sect.~\ref{sec:valandro:sims} (see in particular Figs.~\ref{fig:cnoise_e2e_EE_sim} and \ref{fig:cnoise_e2e_EE_data}).
Switching the correction on and off, we find a shift in $\omb$ of
about $0.25\,\sigma$ for both the simulations and the data.
In simulations, we show that, as expected given its shape and amplitude at large scales, the correlated noise is mainly degenerate with the dust contamination.  
Allowing the amplitudes of the empirical corrections to vary along with the cosmological parameters, 
we find marginal detections of the templates at 100\,GHz and 143\,GHz
(and no detection at 217\,GHz, which, however, is not surprising given that we cut all multipoles for $\ell<500$ at this frequency),
showing that the simulations indeed marginally prefer the correction. 
When doing likewise with the data, the recovered amplitude is only
compatible with unity in the case of the 143\,GHz data.
The correction brings an overall $\Delta\chi^2\,{=}\,9$ improvement (over
762 degrees of freedom for the binned case), entirely
due to its effect on the $100\times100$ and $143\times143$ spectra and dominated by the low-$\ell$ parts of the empirical templates. 

\subsubsection{Calibration and polarization-efficiency uncertainties }
\label{sec:hi-ell:datamodel:inst}

As we did in 2013 and 2015, we perform a minor recalibration of the different cross-spectra to account for 
\begin{enumerate}[(i)]
  \item small residual errors after the dipole modulation calibration in temperature,
  \item uncertainties in the determination of the effective polarization efficiencies of the acquisition pipeline in polarization.
\end{enumerate}
We observe in \citetalias{planck2016-l03} that while temperature calibration is now extremely accurate, the determination of the polarization efficiencies (PEs) is much more difficult and remains one of the main limitations of the \Planck\ 2018 products. 
In the following, we will use ``calibration'' to refer to both the correction to the temperature calibration and to the PE.

\paragraph{Methodology:}

As in 2015, we do not correct the calibration errors by rescaling the
data, but rather apply them to the model vector, multiplying each part
 $C^{XY}_{\ell,\nu\times\nu'}$ (the $XY$ cross-spectrum
 between the $\nu$ and $\nu'$ maps, where $X$ and $Y$ stand for
 either $T$ or $P$, for temperature or polarization), by a
 recalibration factor $G^{XY}_{\nu\times\nu'}$.

In 2015, we used a {\it map-based\/} calibration model, where the
calibration of any cross-spectrum is given by the product of the
calibration factor for each map.  So the calibration matrix is
entirely defined by the knowledge of the calibration correction to be
applied to the temperature maps and the PE correction required for the
polarization ones. The alternative would be a {\it spectrum-based\/}
model, where each individual cross-spectra can have a specific
calibration correction that can differ from the geometrical average
of the calibration correction of the 
auto-spectra.
Defining our calibration parameters at the spectrum level (i.e., the square of the value to be applied on the maps), we call $\calibC^{TT}_{\nu}$ the calibration parameter for temperature at frequency $\nu$ and $\calibC^{PP}_{\nu}$ the same thing for polarization. With this notation
the calibration matrix $G^{XY}_{\nu\times\nu'}$ is given by 
\begin{align}
G^{XY}_{\nu\times\nu'} = \frac{1}{\calibM_{\rm P}^2\calibM_{\rm Pol}^{(X)}\calibM_{\rm Pol}^{(Y)}} \left( \frac{1}{2\sqrt{\calibC^{XX}_{\nu}\calibC^{YY}_{\nu'}}} + \frac{1}{2\sqrt{\calibC^{XX}_{\nu'}\calibC^{YY}_{\nu}}} \right) \,, \label{eq:caldef}
\end{align}
where we further introduced $\calibM_{\rm P}$, an overall \planck\
calibration uncertainty factor (defined at the map level), and $\calibM_{\rm Pol}^{(X)}$,
a degree of freedom allowing us to explore a global temperature-to-polarization calibration factor (also defined at the map level).
The \planck\ calibration uncertainty has not been changed from 2015 to
2018 and is still taken to be a Gaussian with 
\begin{equation}
\calibM_{\rm P} = 1 \pm 0.0025.
\label{eq:calplanck}
\end{equation} 
This is redundant with the temperature calibration parameters, so we {chose} 
\begin{equation}
\calibC^{TT}_{143} = 1, 
\end{equation}
taking the 143-GHz temperature map as the calibration reference.
The $\calibM_{\rm Pol}^{(X)}$ parameter is used to explore a possible
global T-to-P calibration, so we set $\calibM_{\rm Pol}^{(T)}=1$ and
only allow $\calibM_{\rm Pol}^{(P)}$ to vary. This is redundant with the
$\calibC^{PP}_{\nu}$ parameters and so will only be used when those are fixed.

In using Eq.~\eqref{eq:caldef}, we ignore the fact that our effective
spectra combine different half-mission cross-spectra, equivalent to
assuming that different half-mission maps
have much better inter-calibration within a frequency channel than
between frequencies. We have also ignored the variation of the
weighting used in the different cross-spectra combinations. 

{The impact of this approximation can be easily estimated using a Taylor 
expansion. At dominant order, in the case of a $TE$ cross-frequency spectrum, $\delta G^{TP}_{\nu\times\nu'}$,
the residual calibration error induced by the approximation made in Eq.~\eqref{eq:caldef}, 
goes as the half difference between the $\calibC^{PP}$ of the two frequencies times the 
half difference between the $TE$ and $ET$ weights:
\begin{equation}
\delta G^{TP}_{\nu\times\nu'} = \frac{\calibC^{PP}_{\nu}-\calibC^{PP}_{\nu'}}{2}\times\frac{w^{TE}_{\nu\nu'}-w^{ET}_{\nu\nu'}}{2}. 
\end{equation}
We will see in Eq.~\eqref{eq:relcalEE}
that the maximum half difference between $\calibC^{PP}_{\nu}$ is below $0.04$, $ {\left |\calibC^{PP}_{143}-\calibC^{PP}_{217}\right |}/{2} = 0.037$.
The half difference between the $TE$ and $ET$ weights is at most $ {\left |w^{TE}_{\nu\nu'}-w^{ET}_{\nu\nu'}\right |}/{2}  \leq 0.1$ in the 
range of multipoles retained for cosmology, translating into a residual error that is always smaller
than the half-percent level.}

\paragraph{Temperature inter-calibration:}

We determine the 100-GHz and 217-GHz temperature-calibration
parameters ($\calibC^{TT}_{100}$ and $\calibC^{TT}_{217}$) by
measuring the inter-calibration of the $100\times100$ and
$217\times217$ cross-half-mission spectra with the $143\times143$ one, minimizing
\begin{equation}
\chi^2(\calibC^{TT}_{\nu}) = \Big(C_\ell^{\nu\times\nu}\times\calibC^{TT}_{\nu}-C_\ell^{143\times143}\Big)^{\sf T}\!\tens{\Sigma}^{-1}_{\ell\ell'}\Big(C_{\ell'}^{\nu\times\nu}\times\calibC^{TT}_{\nu}-C_{\ell'}^{143\times143}\Big). \label{eq:mincal}
\end{equation} 
Since the $C_\ell$ above are being measured on the data, foreground contamination must be accounted for or mitigated.
For the 2013 release \citepalias{planck2013-p08}, we did so by computing the spectra only at high Galactic latitude (i.e., small sky fraction), 
and we further reduced the multipole range to relatively low $\ell$ to avoid extragalactic contamination. 
This approach can be improved, as we do in 2018, by using dust-cleaned
maps (as described in Eq.~\ref{eq:hil:dustundust} of Sect.~\ref{sub:galactic-dust:TT}), which reduces the Galactic contamination. Another option is to jointly estimate the foreground contribution and the intercalibration, as is done in the {\tt SMICA} foreground-cleaning algorithm \citep{planck2016-l04}. This was the basis of the intercalibration procedure for the 2015 release. 
We cross-check between the calibration estimates obtained using dust-cleaned maps and the {\tt SMICA} approach in different Galactic masks and multipole ranges ($30<\ell_{\rm min}<100$ and $400<\ell_{\rm max}<600$).
In all cases, we obtain remarkably stable determinations of the temperature calibration parameters and see excellent compatibility between the two approaches.  We therefore use as our calibration correction priors the values
\begin{align}
\label{eq:relcal}
\calibC^{TT}_{100} =\,& 1.0002 \pm 0.0007 \,, \nonumber \\
\calibC^{TT}_{217} =\,& 0.99805 \pm 0.00065 \,.
\end{align}
These are a significant improvement over 2015, and in good agreement
with the values quoted in \citetalias{planck2016-l03}, which were obtained with different multipole ranges and masks.
We investigate the effect of leaving the calibration parameters free in Sect.~\ref{sec:valandro:priors}. 
Leaving the calibration parameter free is found to have very little effect on the cosmological parameters in the \lcdm{} model.

\paragraph{Polarization efficiencies:}
The determination of PE corrections is more difficult,
as emphasized in \citetalias{planck2016-l03}.
The PEs were {measured} on the ground \citep[see table~B.1 in][]{rosset2010}. The{y}  were done detector by detector, before the assembly of the focal plane, while the polarization
angle measurements were performed on the assembled focal plane. Values
$90\,\%$ and $96\,\%$ at 100\,GHz, $83\,\%$ and $93\,\%$ at 143\,GHz,
of the PE are of the order of $90\,\%$ (at map level), ranging between
and $94\,\%$ and $95\,\%$ at 217\,GHz. Statistical uncertainties of
the ground measurements were found to be approximately $0.3\,\%$ at
100\,GHz and 143\,GHz, and $0.2\,\%$ at 217\,GHz. 
The ground values are incorporated into the data analysis pipeline,
specifically in the mapmaking and beam-estimation steps. We should
thus expect the $EE$ and $TE$ power spectra to be intercalibrated to
within a fraction of a percent. Given the size of the errors on the polarized
power spectra and the modelling uncertainties for the temperature-to-polarization leakage, it seemed reasonable in 2015 to ignore any residual
error in the PEs, and to set all
$\calibC^{PP}_{\nu}$ parameters to unity.

As described in section~5.10.3 of \citetalias{planck2016-l03}, progress in the data analysis and description of the polarized dust contamination has allowed us to reassess 
the determination of the PEs of the 353-GHz detectors using observations of highly polarized dust regions. Figure~35 of \citetalias{planck2016-l03}
shows that {in order to ensure compatibility between the different detectors, their PEs needed to be corrected at the level} of a
few percent, about five times larger than the estimated uncertainty of
the ground measurement{, which was deemed acceptable by the HFI instrument team}. {{We note} that this only provides the relative PE corrections for each detector within the 353-GHz channel and cannot provide us with the absolute
PE determination that we need for cosmology.}

Corrections of this {magnitude} should be measurable at lower frequencies, thanks to the sensitivity of the detectors{,}  provided that the beam-leakage corrections are well under control (as will be discussed in the next section).
 Unlike the 353-GHz channel, for which dust emission has a high
S/N, the CMB at 100, 143, and 217\,GHz does not
have a high enough S/N to allow for a direct estimate of
the PE corrections {(which would in any case still be limited to relative corrections)}.  Hence we shall rely on a
different method and estimate the $\calibC^{PP}_{\nu}$ parameters by
trying to improve the agreement between $C_\ell^{EE,\nu\times\nu}$
(the $EE$ power spectrum at a frequency $\nu$), and
$C_\ell^{EE,\mathrm{best}}$ (the theoretical $EE$ power spectrum
predicted using the best-fit \LCDM\ parameters based on the $TT$
data alone).  This will be done by minimizing functions of the form
\begin{align}
\chi^2(\calibC^{PP}_{\nu}) &= \Big(C_\ell^{EE,\nu\times\nu}\times\calibC^{PP}_{\nu}-C_\ell^{EE,\mathrm{best}}\Big)^{\sf T}\nonumber\\
&\qquad\qquad\tens{\Sigma}^{-1}_{\ell\ell'}\,\Big(C_\ell^{EE,\nu\times\nu}\times\calibC^{PP}_{\nu}-C_\ell^{EE,\mathrm{best}}\Big).
\label{eq:relcalEE_FIT}
\end{align} 
This scheme will provide better constraints than those that would have been
obtained had we followed a $TT$-like procedure, cross-calibrating {directly} between different observed power 
spectra, the noise in $EE$ being much larger than in $TT$. 
Furthermore this method will effectively provide the {absolute} PEs
of each frequency channel (i.e., including the temperature-to-polarization calibration ratio), 
which cannot be obtained by an intercalibration between the different
$E$ maps.  {This procedure makes use of the excellent agreement between temperature and polarization, {\it even before PE correction}. 
A fact that we already demonstrated in 2015.} However, this procedure is dependent on the cosmological
model, and  {could} have the tendency to push the polarization-based
parameters towards the temperature ones. {We note, however, that the agreement between the $TT$, $TE$, and $EE$ parameters discussed in Sect.~\ref{sec:valandro:parcompTTTEEE}, while good, does not seem to corroborate this concern. If anything this agreement has decreased since 2015, as a result of the PE corrections and leakage corrections that we will discuss in the next section. We also note later in this paragraph that marginalizing over the absolute temperature-to-polarization calibration makes no difference to the cosmological parameters, reducing further the concern of artificially enforcing the temperature and polarization agreement.} One might also worry that the 
apparent extra smoothing of the $TT$ peaks at high-$\ell$ could affect the PE estimation (see Sect.~\ref{sec:valandro:alens}). As we will describe later, we checked that on the relatively low multipole range 
we retain for the PE correction estimation, using either the $TT$
best-fit \LCDM\ model or the \LCDM+$\Alens$ model provides similar results. 
{We discuss later in this section how the limitations of our PE estimations could be overcome in the future, if new calibration data become available.}

As before, we estimate the PE correction using as
input the $EE$ power spectra obtained using multiple combinations of
masks, either paralleling those used for the likelihood (with different
masks at different frequencies), or using the same C50 or P50 mask at
all frequencies. We mitigate the effect of the dust contamination
either by first correcting the observed spectra by a power-law
template, as described in Sect.~\ref{sec:hi-ell:datamodel:gal}, or by
using dust-cleaned $E$ maps, as described in
Sect.~\ref{sub:galactic-dust:TT}. In all cases, we discard the low
($\ell<200$) multipoles where the Galactic contamination residuals can
affect our PE determinations. Similarly, we
discard the high ($\ell\,{>}\,1000$) multipoles, where the S/N is
low. We also correct the observed power spectra by subtracting the
temperature-to-polarization leakage templates that we describe in
Sect.~\ref{sec:hi-ell:datamodel:beamleak}. These are small in $EE$ and
have little effect on the PE
estimations. Similarly, the auto-spectra-correlated noise templates
discussed in Sect.~\ref{sec:hi-ell:datamodel:noise} are subtracted
from the observed spectra. These also have a low impact on the final
results, since they mainly affect the low and high multipoles, which are
already cut out. Finally, we also explored a case where we modified
Eq.~\eqref{eq:relcalEE_FIT} to fit more than one frequency at a time
and include the cross-frequency spectra. This allowed us to
cross-check our results by estimating the PEs
while discarding some or all of the auto-spectra, verifying the
coherence of the calibration model represented by the calibration
matrix defined in Eq.~\eqref{eq:caldef}.

All of these different procedures yield very similar results, which
are compatible within uncertainties. We retain the following values for the likelihood:
\begin{align}
\label{eq:relcalEE}
\calibC^{PP}_{100} =\,& 1.021 \pm 0.010 \,; \nonumber \\
\calibC^{PP}_{143} =\,& 0.966 \pm 0.010 \,; \nonumber \\
\calibC^{PP}_{217} =\,& 1.040 \pm 0.010 \,.
\end{align}
We  see below that we will eventually not let the PE correction vary in the priors, but fix them to their central values.
For $EE$, the \camspec\ likelihood follows a similar procedure, comparing the $EE$ power spectra obtained using the C50 mask with a theoretical CMB spectrum computed using the $TT$ best-fit cosmology. Maps are dust-cleaned and temperature-to-polarization leakage is corrected at the spectrum level. However, unlike \plik, which
enforces through Eq.~\eqref{eq:caldef} a {\it map-based\/} calibration
model, allowing a single PE correction per
channel, \camspec\ uses a {\it spectrum-based\/} one, so that a unique
correction is determined for each cross-half-mission spectrum. In
other words, \camspec\ will allow the correction of the
100-HM1$\times$143-HM2 to be different from the
100-HM2$\times$143-HM1, as well as being
different from the geometrical average of the
100-HM1$\times100$-HM2 and the
143-HM1$\times143$-HM2 ones.
At this stage, it is reassuring to note that the \camspec\
PE corrections for $EE$ quoted in the first
column of Table~\ref{tab:camcal} are in excellent agreement with those
of Eq.~\eqref{eq:relcalEE} above (numbers quoted in the table
correspond to $1/\calibC^{PP}_\nu$) and that in this case the
map-based and spectrum-based approaches give very similar results.

The {\tt SMICA} algorithm can also be applied to the $E$ maps. This provides an intercalibration of the different $E$ maps, but not an absolute PE determination. Up to a global rescaling, the {\tt SMICA} values are in good agreement with the map-based and spectrum-based calibrations, as reported in table~9 of \citetalias{planck2016-l03}. 

In each of these schemes (map-based, spectrum-based, and {\tt SMICA}), we
note that the 217-GHz calibration at lower ($200<\ell<500$) and higher
($500<\ell<1000$) multipoles are in tension, with the higher part of
the multipole range pulling $\calibC^{PP}_{217}$ to high 
values ($\approx 1.07$). Such a large  {correction} {is somewhat} surprising, compared to the typical correction measured on the 353-GHz polarized detectors. 
A possible explanation for this could be a sign of a residual transfer
function in the 217-GHz channel. {We also note that in the $217\times 217$ residual plot shown in the bottom panel of Fig.~\ref{fig:hi_ell:valid:residualEE} the multipole region $700\,{<}\,\ell\,{<}\,1000$ is systematically low and can explain the preference for higher PE correction. We note, however, that this excursion is not sufficient to degrade the agreement between the data and the model or between the different cross-frequencies, as discussed in Sect.~\ref{sec:valandro:cond}.} The other frequencies, however, are much more
stable to changes to the multipole range used in the fit.
Owing to the low weight of the 217-GHz data
in the EE and TE parts of the likelihood (see
Fig.~\ref{fig:hi_ell:data:mixing}), cosmological parameters are not
significantly affected by shifts of the PE for
217\,GHz. For this reason, we decide to fix the polarization
calibration parameters to the central values of Eq.~\eqref{eq:relcalEE}
and set $\calibM_{\rm Pol}^{(P)}=1$. As discussed Sect.~\ref{sec:valandro:PE}, if we were to
allow the calibration parameters to vary with priors given by Eq.~\eqref{eq:relcalEE}, we would end up with an almost $3\,\sigma$ pull at
217\,GHz, since the likelihood does not use the $\ell<500$ multipoles
at this frequency. 
{We also investigated the effect of fixing the PE correction, but marginalizing over the overall temperature-to-polarization calibration. This procedure corresponds to inter-calibrating the different PE correction and alleviates the possible 
dependency on the cosmological model that we highlighted above. In this case, we recover the expected temperature-to-polarization calibration to better than $0.5\,\sigma$ and see no cosmological parameter modification, in both \LCDM\ and \LCDM+$\Alens$ cases.}
The \camspec\ likelihood also uses fixed
PE corrections during the construction of the coadded $EE$ and $TE$ spectra.

Correcting the PEs with the values given in Eq.~\eqref{eq:relcalEE}
 affects mainly the $EE$ inter-frequency agreement,
as can be seen in Fig.~\ref{fig:hi_ell:data:EE_triangle}. The effect
on the cosmological parameters is shown in Fig.~\ref{fig:hi_ell:2015_vs_2018_systeffects}. 
Quantitatively, the $\chi^2$ improves by about 50 for
$EE$ (over 762 degrees of freedom for the binned case) and remains almost identical for $TE$. 

\paragraph{\textit{TE}-based polarization efficiency estimation:}

In the discussion above, we used only the $EE$ spectra to estimate the
PE corrections. We can also use the $TE$ spectra,
now minimizing the function
\begin{align}
\chi^2(\calibC^{TE}_{\nu}) =\,& \Big(C_\ell^{TE,\nu\times\nu}\times\sqrt{\calibC^{TE}_{\nu}\calibC^{TT}_{\nu}}-C_\ell^{TE,\mathrm{best}}\Big)^{\sf T}\tens{\Sigma}^{-1}_{\ell\ell'}\nonumber\\
&\qquad\qquad \Big(C_\ell^{TE,\nu\times\nu}\times\sqrt{\calibC^{TE}_{\nu}\calibC^{TT}_{\nu}}-C_\ell^{TE,\mathrm{best}}\Big)\,,
\end{align} 
where the temperature calibration parameters $\calibC^{TT}_{\nu}$ are fixed to the values given earlier in this section.
If the map-based calibration model is accurate, this provides a nice
cross-check,since the $\calibC^{TE}_{\nu}$ should then be equal to $\calibC^{PP}_{\nu}$.
Results of this test show marginal agreement between the two
procedures, with the $\calibC^{TE}_{100}$ and $\calibC^{TE}_{217}$ values
about $1\,\sigma$ away from the $EE$ results, and a larger
disagreement at 143\,GHz: 
\begin{align}
\label{eq:relcalTE}
\calibC^{TE}_{100} =\,& 1.04 \pm 0.02 \,; \nonumber \\
\calibC^{TE}_{143} =\,& 1.00 \pm 0.02 \,; \nonumber \\
\calibC^{TE}_{217} =\,& 1.02 \pm 0.02 \,.
\end{align}
Owing to the relatively low weight of spectra involving 100\,GHz and
217\,GHz in the polarized part of the likelihood, changing
between using the $\calibC^{EE}$ and the $\calibC^{TE}$ values would only have a small impact on the cosmological fits. 
However, the difference at 143\,GHz can translate into a larger effect.

As was the case when using $EE$ to estimate the PE corrections, the results using
$TE$ are stable (within error bars) to variations of masks, multipole
ranges, and methodology (i.e., either fitting for one calibration at a
time or performing a multifrequency fit taking cross-spectra into
account). The \camspec\ calibration procedure, which provides an
effective spectrum-based PE correction for each
half-mission and frequency pair, finds a similar behaviour, as shown in
Table~\ref{tab:camcal}, with a similar calibration discrepancy between
the $143\times143$ $EE$ and $TE$ spectra. Up to $2\,\%$ level differences can be observed in the $ET$ and $TE$ auto-spectra (in particular at 217\,GHz), which could be due to noise scatter between HM1 and HM2.

The discrepancy between the $143\times143$ $EE$ and $TE$ apparent
PE corrections is also found to be robust under
variation of the reference CMB model, for example if instead of using
the best-fit \LCDM\ cosmology we instead use the best-fit
\LCDM+$\Alens$ parameters. {We note} that in the range of multipoles we use for PE-correction estimation, 
the impact on the $EE$ and $TE$ spectra of opening the $\Alens$ parameter is small.
The outlier in the
143-GHz spectrum around $\ell=750$ (see
Fig.~\ref{fig:hi_ell:valid:residualTE}) also does not appear to be
responsible for the larger
discrepancy at 143\,GHz, as tested for by restraining the multipole range to $\ell<700$.

We found, following the exact same procedures, that the end-to-end simulations (described
Sect.~\ref{sec:valandro:sims}) do not display such a difference in the
estimated corrections of the PEs. Rather, in the
simulations we measure the $\calibC^{TE}_{143}$ and
$\calibC^{PP}_{143}$ parameters and find them all to be perfectly
compatible with the input value (in that case, $1$). 

We can also rule out a frequency-specific correlated noise component
at 143\,GHz (either in $TE$ or $EE$) as a possible explanation of the
discrepancy. As described above, we can perform the PE
estimate without the frequency auto-spectra and instead use
only cross-frequency ones (i.e., $100\times143$ and $143\times217$). For both $TE$ and $EE$ we obtain similar central values. 

A residual transfer function seems unlikely to explain the difference
either. We tried to fit both for a correction of the PE
and an additive correction of the form $\Delta
C^{EE,\mathrm{trans}}_\ell = \alpha \ell(\ell+1) C^{EE}_\ell$ for $EE$
and $\Delta C^{TE,\mathrm{trans}}_\ell = \alpha
\ell(\ell+1) C^{TE}_\ell / 2$ for $TE$. The slopes obtained at 143\,GHz
using $TE$ and $EE$ were found to be incompatible with one another,
with a value within $1\,\sigma$ of zero for $EE$, but non-zero for
$TE$ at the $1.5\,\sigma$ level.  The latter value would correspond to
an arcminute-scale beam-shape error, 5 times our expected beam uncertainty.

Unlike for $EE$, in which the temperature-to-polarization leakage corrections are small, the temperature-to-polarization leakage cannot be ignored in $TE$. Error modes of the uncertainty on the leakage template at large scales are partially degenerate with an effective recalibration of the $TE$ spectrum. 
As discussed in Sect.~\ref{sec:hi-ell:datamodel:beamleak}, we
assess the leakage
uncertainty budget by propagating the uncertainties on the beam
determinations, polarization angles, and PEs
 through to power spectra in simulations. The leakage
template uncertainty is found to be too small by a factor of around 5 to account for the difference between $\calibC^{TE}_{143}$ and $\calibC^{PP}_{143}$.

At this stage, two different choices can be made: either we can assume that
the discrepancy is the sign of an unknown systematic effect, not reproduced in our 
end-to-end analysis, which projects partially onto the PE correction; or we can assume that the discrepancy is a statistical accident, and enforce the map-based calibrations, coherent with our instrument model and our simulations. 

The \plik\ likelihood was implemented choosing to keep the polarization
cross-spectra separate (at the level of channel cross-spectra), and hence can
use either
a map-based or spectrum based calibration, and can explore the calibration parameters jointly with the cosmology. However, for the baseline, we choose to restrict the calibration model to a map-based one. 
The alternative (i.e., spectrum based calibration) can be tested to its full extent (with different PE correction for $EE$, $ET$, and $TE$ cross-half-mission spectra) using the \camspec\ likelihood, as explained in Sect.~\ref{sec:hi-ell:prod:camspec}. \camspec\ co-adds all $EE$ and $TE$ cross-spectra and only allows for the joint determination of an overall $EE$ and $TE$ residual correction along with the cosmological parameters. 

Future ground-based surveys will allow us to test the map-based calibration using both bright polarization sources (such as the Crab Nebula) and possibly cross-calibration on the CMB sky. It is unclear at this stage whether the \Planck\ observations of bright polarized sources will have enough S/N to allow for a strong improvement of the PE determinations. {Cross-calibration on observations of the CMB sky from future ground-based surveys will also be limited by the sky coverage of such surveys and by the accuracy of the \Planck\ polarized data.}

We present in Sect.~\ref{sec:valandro:PE} the detailed
consequences of the two different calibration models, implemented in the \plik\ likelihood. 
Quantitatively, in \plik, the map-based PE correction improves
the joint $\chi^2$ by about 50, mainly gained in the EE part
of the likelihood. Also in \plik, implementing the spectrum-based
model (i.e., using both Eq.~\ref{eq:relcalEE} in EE and
Eq.~\ref{eq:relcalTE} in TE) further improves the $\chi^2$ by about 10
(all in the TE part of the likelihood). The different
corrections, however, translate into shifts of below $0.6\,\sigma$ on
parameters in extended models, and little change on the 6-parameter \LCDM.

\subsubsection{Secondary beam effects}
\label{sec:hi-ell:datamodel:beamleak}
We describe in Sect.~\ref{sec:datamodel:beam} how the new \quickpol\ method \citepalias{quickpolHivon} is used to compute the full matrix
that links the observed spectra with the underlying sky ones (see Eq.~\ref{eq:beam_matrix}). 
{We do not reproduce here the technical derivation presented in \citetalias{quickpolHivon}, but will briefly describe the most relevant parts.}

{In general, using the signal equation given by equation~12 of \citetalias{quickpolHivon}, one can relate the signal measured by a given detector with the sky temperature and polarization, as a function of its optical efficiency (which we call ``calibration'' in Sect.~\ref{sec:hi-ell:datamodel:inst}), polarization efficiency, optical beam, and polarization orientation. The \Planck \  frequency maps are built by optimally combining all the different valid observations of the sky at a given frequency that fall in the same pixel \citepalias{planck2016-l03}. In doing so, the mapmaking algorithm ignores beam shapes and orientations and assumes identical circular beams. This is encoded formally in equation~A.2 of \citetalias{quickpolHivon}, which shows how the temperature and polarization maps  
are a convolution of the sky with the beams of the different detectors, modulated by a position-dependent weighting. For a given pixel and a given detector, this weighting is expressed as a function of the coverage in orientation, the hit map (including the masked pixels and discarded data), and the polarization efficiencies of the detector. From this result, equation~38 of \citetalias{quickpolHivon} gives the relation between the power spectra measured from those maps and the underlying sky ones. In particular, equation~38 shows how the circular beam assumption made at the mapmaking stage mixes all the different temperature and polarization spectra, as well as the power at different multipoles. Mixing can also arise from a mismatch between the parameters of the real detectors (optical efficiencies, polarization efficiencies, and orientation angles) and the ones assumed at the mapmaking stage.}

\begin{figure*}[htbp!]
\begin{center}
\includegraphics[width=0.33\textwidth]{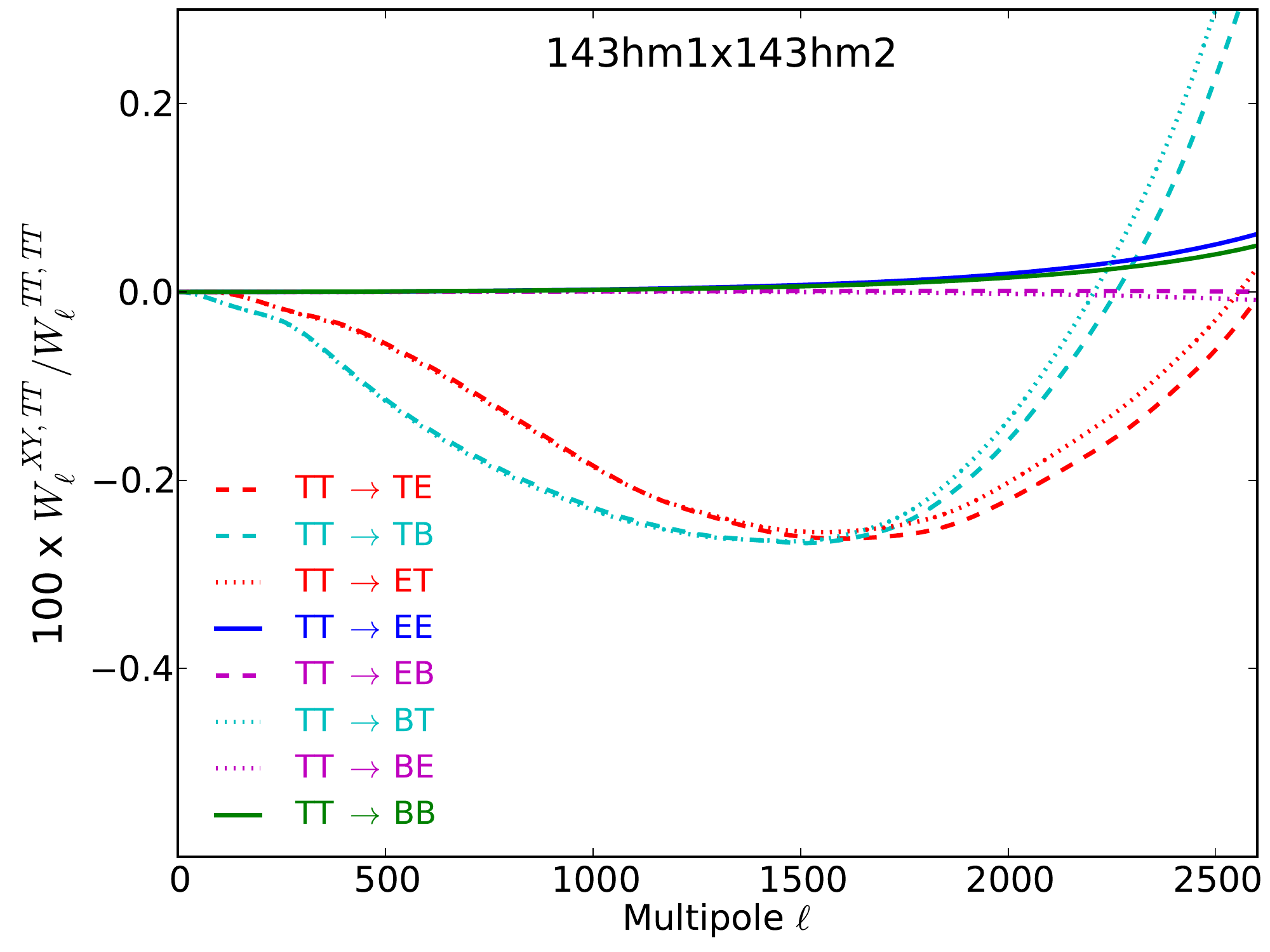}
\includegraphics[width=0.33\textwidth]{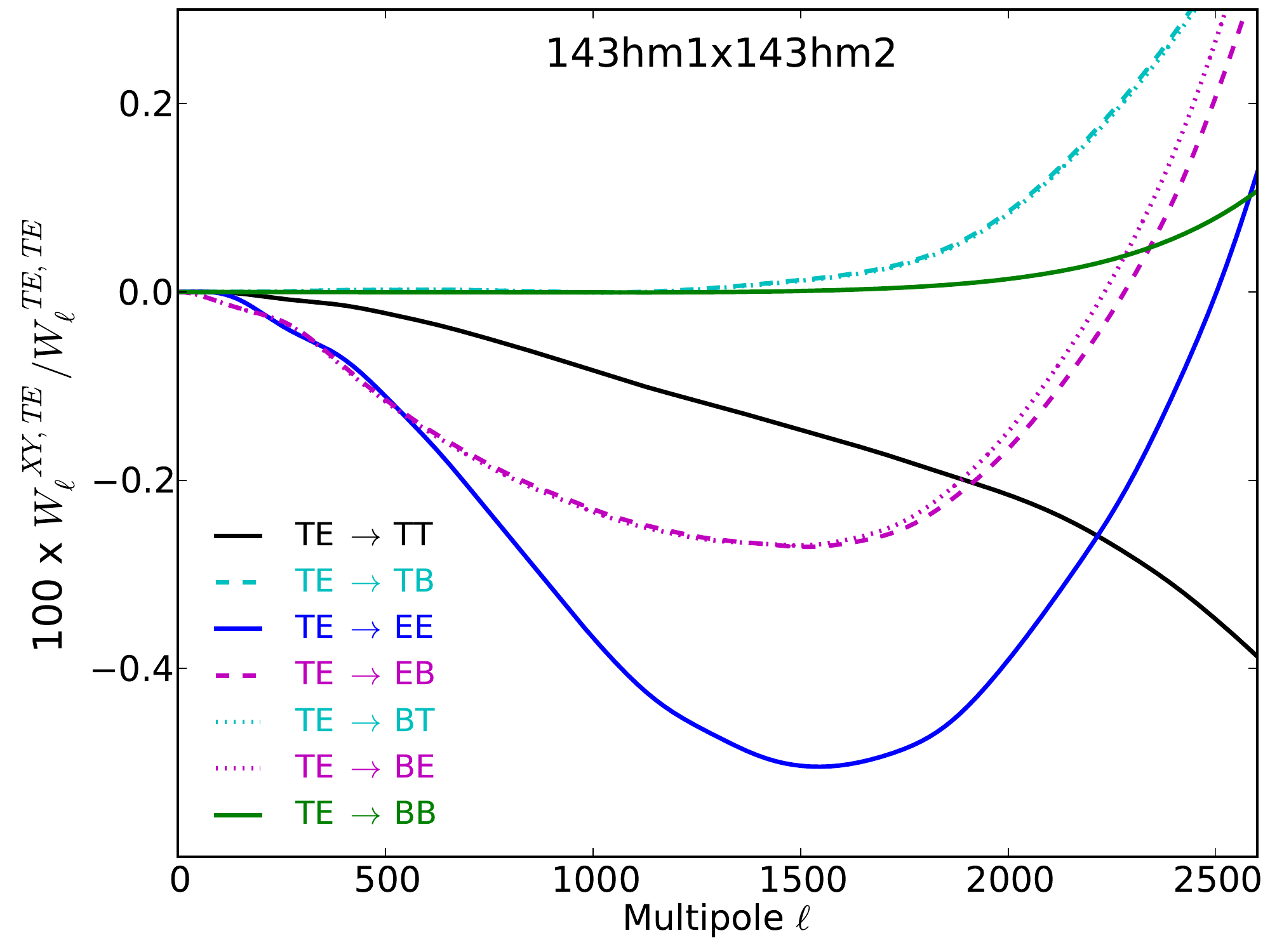}
\includegraphics[width=0.33\textwidth]{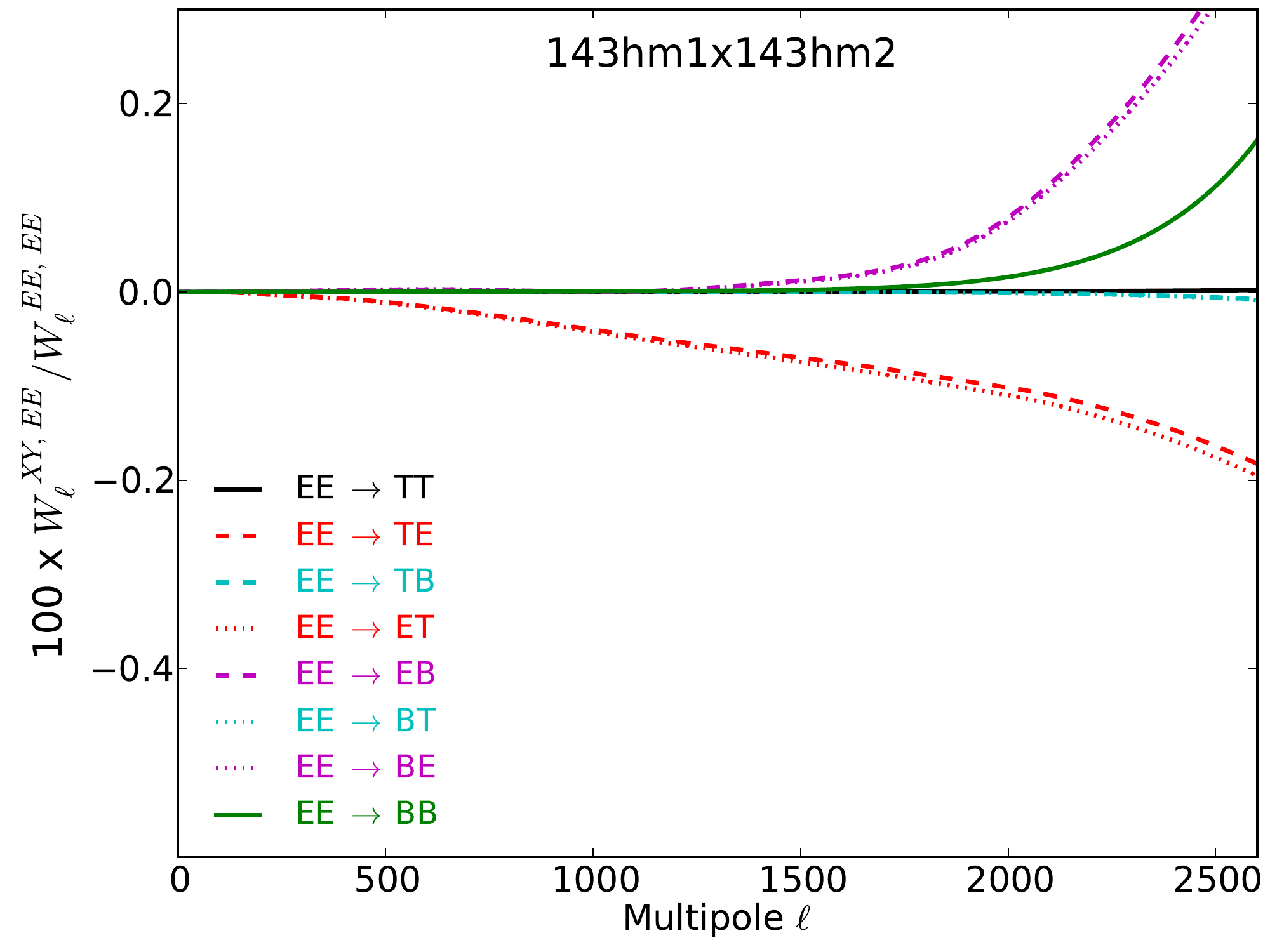}
\includegraphics[width=0.33\textwidth]{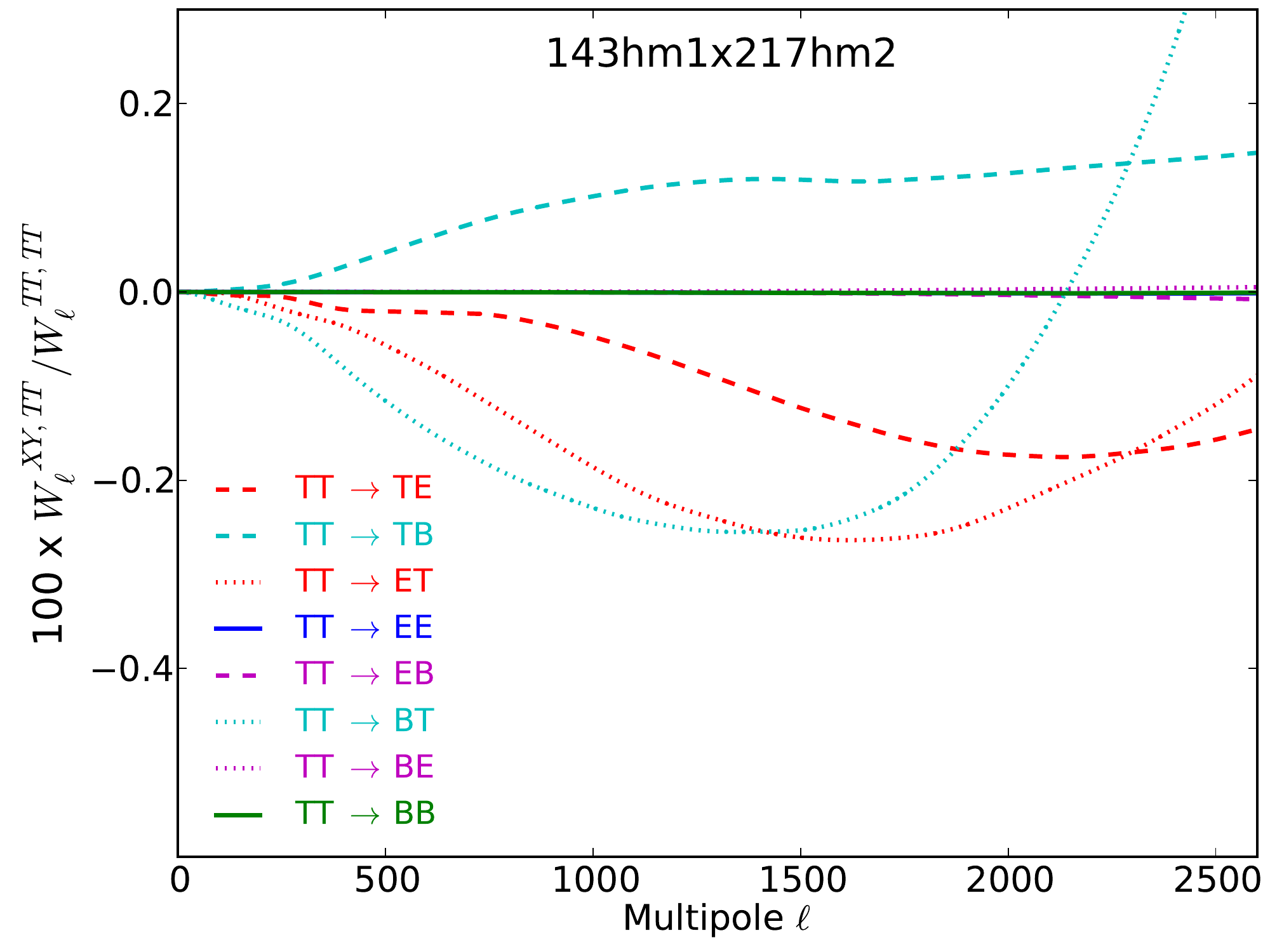}
\includegraphics[width=0.33\textwidth]{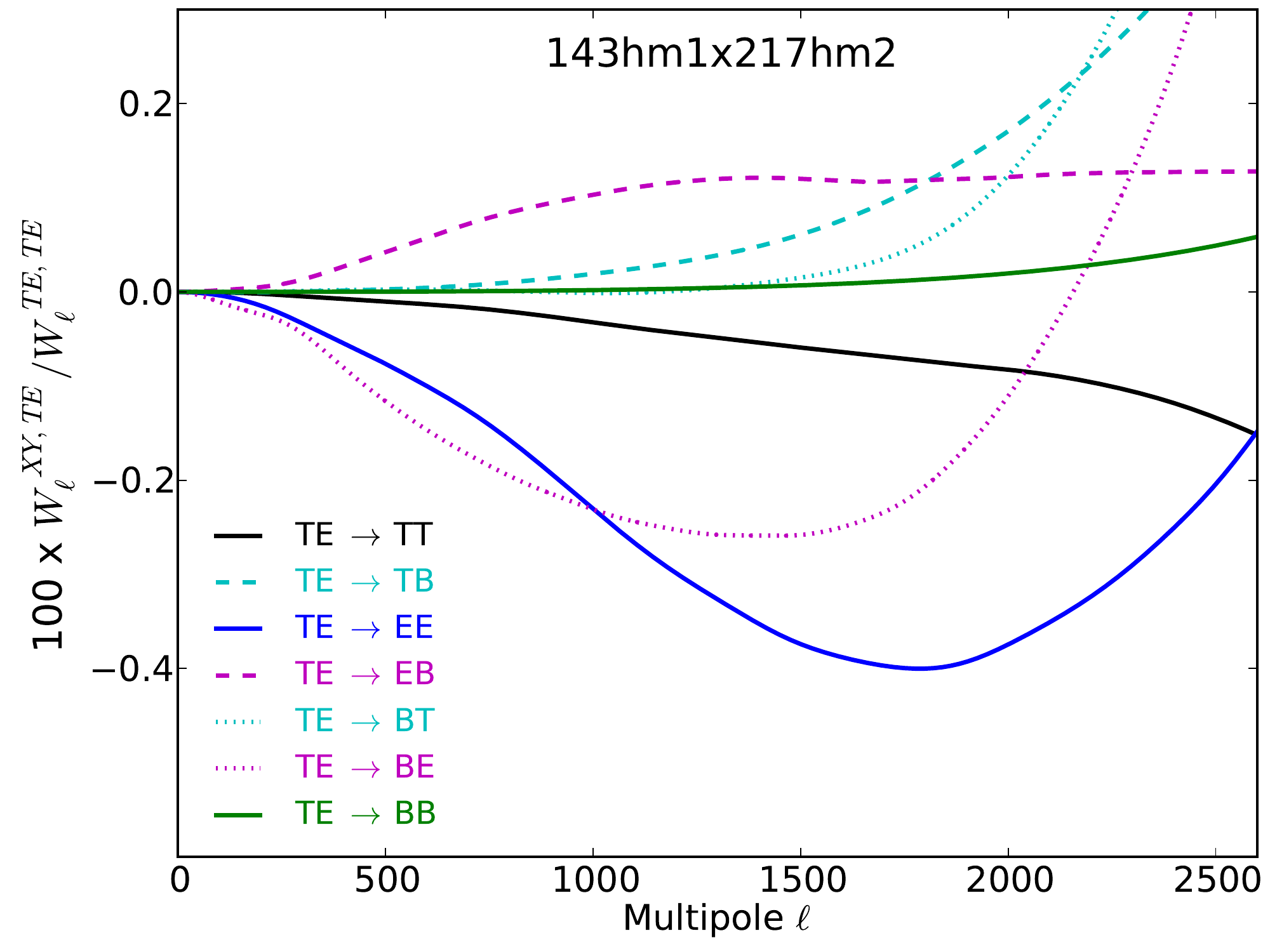}
\includegraphics[width=0.33\textwidth]{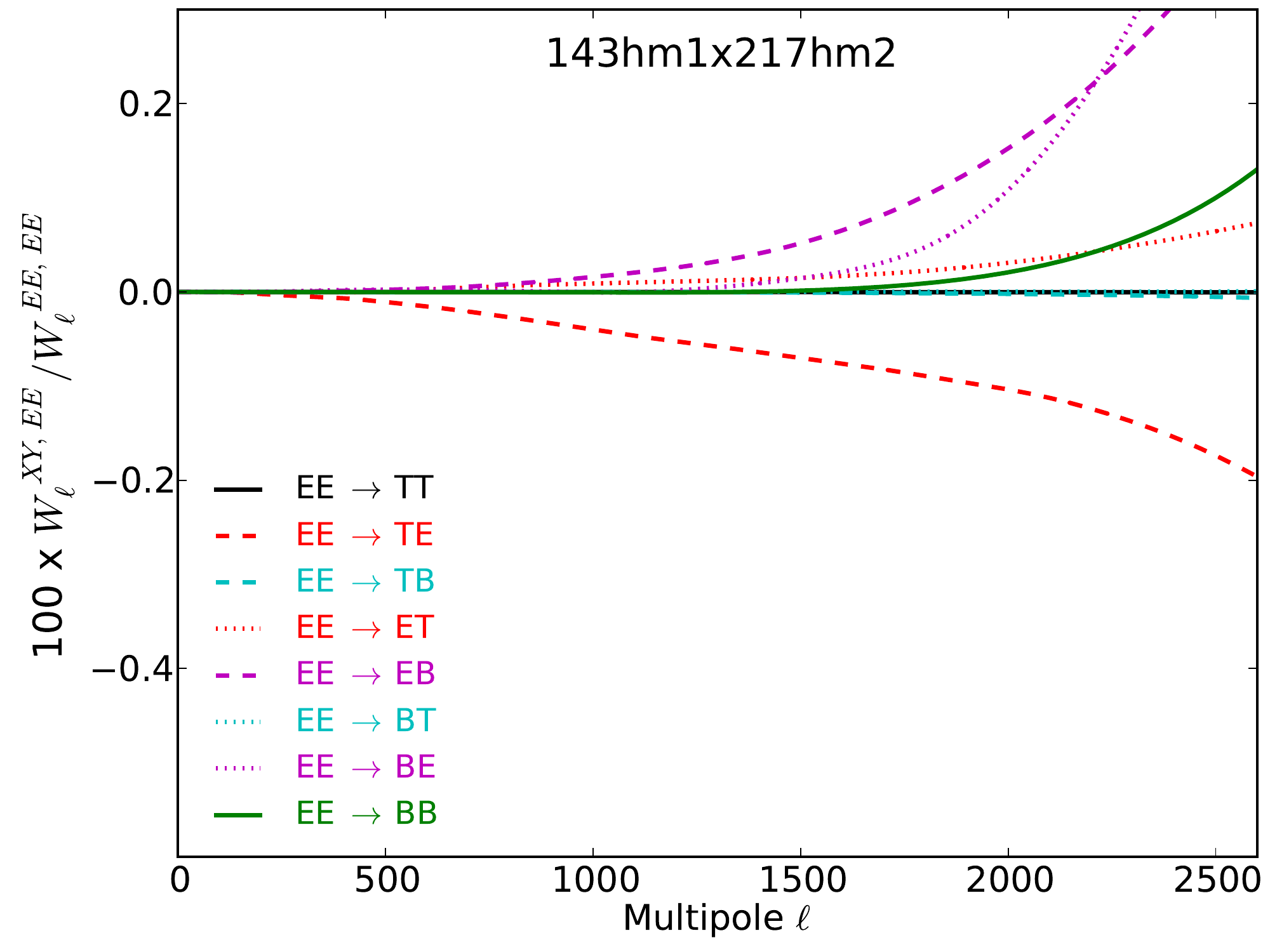}
\caption{Elements of the beam matrices $\widehat{W}^{XY,\;TT} \equiv W^{XY,\;TT}/W^{TT,\;TT}$ (left panels),
$\widehat{W}^{XY,\;TE}$ (middle panels),
and $\widehat{W}^{XY,\;EE}$ (right panels), for the cross-analysis of 143-HM1 and 143-HM2 (upper panels)
and of 143-HM1 and 217-HM2 (lower panels).
}
\label{fig:Wmatrix}
\end{center}
\end{figure*}

{In the case of a so-called ``smooth scanning'' of the sky (i.e., when the weighting functions that encode the orientation coverage of the pixels and the hit map vary slowly across the sky), as is the case for \Planck, \citetalias{quickpolHivon} showed that the mixing simplifies to an $\ell$-dependent matrix linking each temperature and polarization cross-spectrum, that we reproduce Eq.~\eqref{eq:beam_matrix}. As discussed in Sect~\ref{sec:datamodel:beam}, the diagonal part of this matrix is the effective beam and is corrected for in the data vector. In this section, we will be interested in the (small) off-diagonal part that represents the temperature-to-polarization leakage and will treat the effect as additive corrections that we include 
as templates in the model vector of Eq.~\eqref{eq:basic-likelihood}.}

{Following {the} previous work  presented in \citet{planck2013-p03c}, \citetalias{quickpolHivon} also shows how to evaluate the effects of the finite size of the pixels, usually called ``sub-pixel effects,'' which are found to cause a small extra smoothing (integrated in our mixing matrices) and an extra noise term. We discuss these later in this section.}

{The mixing matrices and sub-pixel effects we use here are computed following the 
methodology higlighted in {Sect.}~4 of \citetalias{quickpolHivon}, using the detector parameters, scanning strategy, and data flags assumed for mapmaking in \citetalias{planck2016-l03}, and the masks we retained described in Sect.~\ref{sec:hi-ell:datamodel:mask}. {This} ignores the leakage due to the mismatch between the PE assumed during mapmaking \citep[from][]{rosset2010} and the one we evaluate in Sect.~\ref{sec:hi-ell:datamodel:inst}. We discuss later in the section how this extra leakage is actually taken care of by our PE correction.}

\paragraph{Beam leakage:}

\begin{figure}[h!]
\centering
\includegraphics[width=0.5\textwidth]{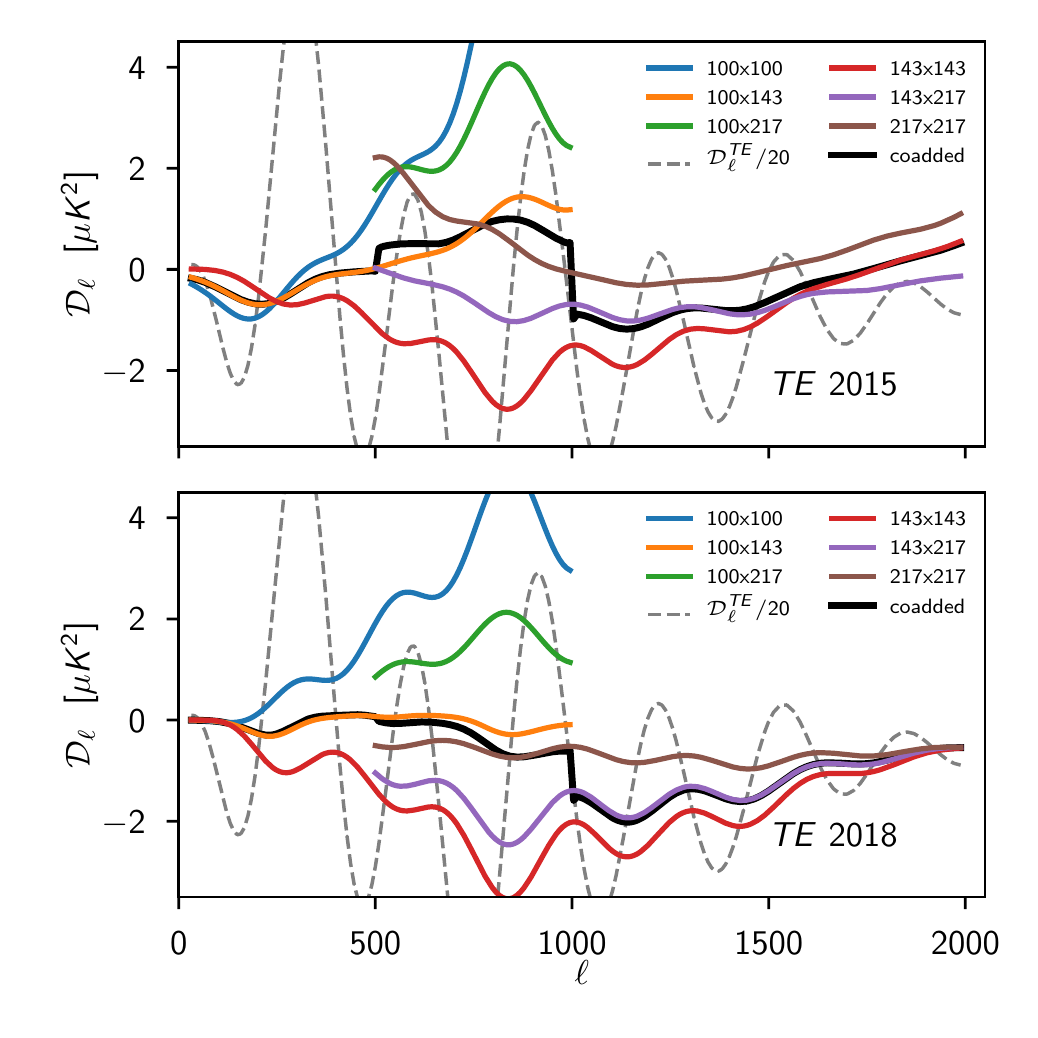}\\
\vspace{-0.4cm}
\includegraphics[width=0.5\textwidth]{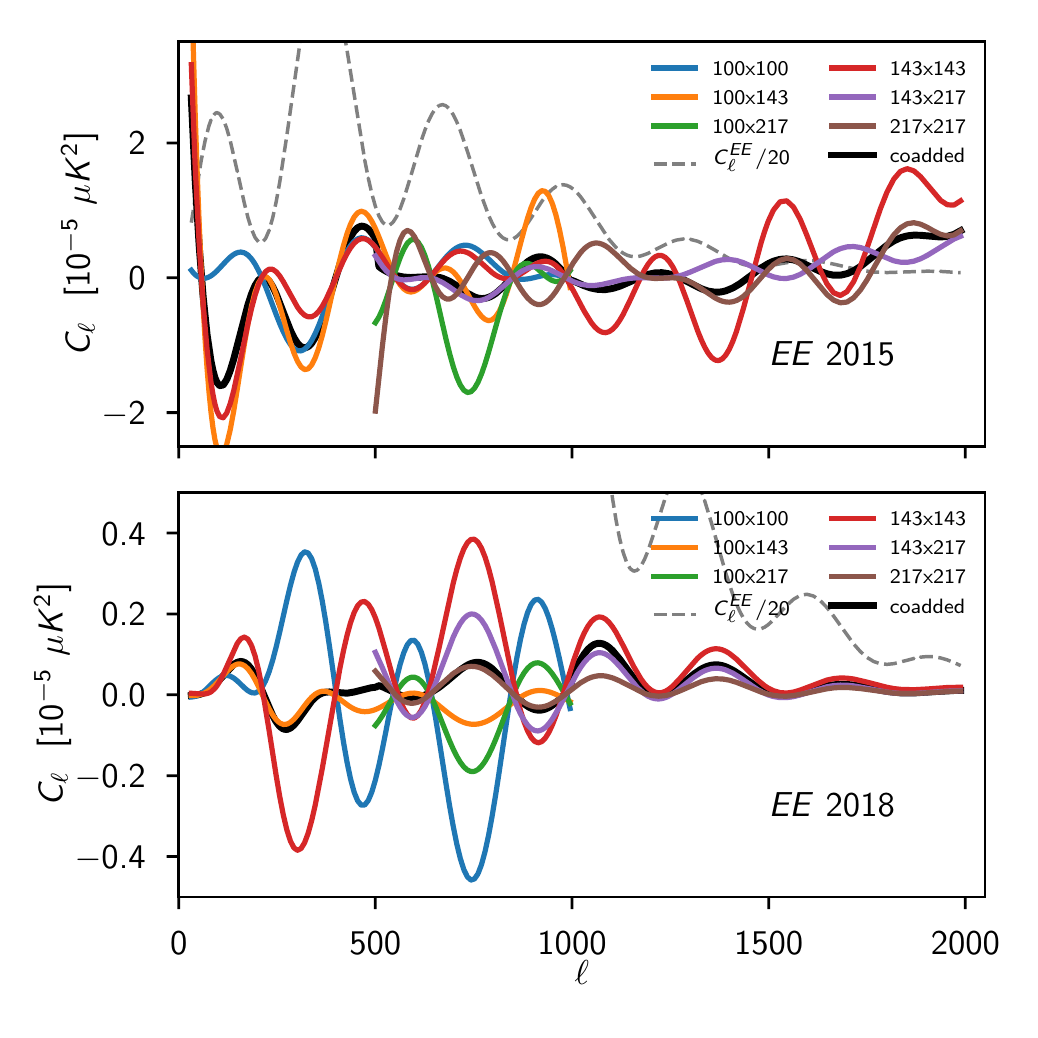} 
\vspace{-0.6cm}
\caption{Models of the beam leakage for $TE$ and $EE$ in 2015 and
 2018. Differences are striking in $EE$, where the 2015 model was
 overfitting, affected by PE errors. {I}n the bottom panel, the 2015 and 2018 $EE$ models are presented with different vertical scales.
The $TE$ leakage models are similar, except for the 
$500<\ell<1000$ multipole range, where the correction changes sign, 
pushing the $TE$ best cosmology further away 
from the $TT$ one.}
\label{fig:hi_ell:data:leakages}
\end{figure}

Going back to Eq.~\eqref{eq:beam_matrix} and dividing both left- and
right-hand sides with the diagonal part of the matrix, we can relate how each of the different data spectra
are affected by leakage:\footnote{As for Eq.~\eqref{eq:beam_matrix}, the formula is exact only for the ensemble average over sky spectra.}
\begin{subequations}
\begin{align}
  \widehat{C}^{XY}_{\ell} &=\, \widetilde{C}^{XY}_{\ell} / W^{XY,\;XY}_\ell, \\
        &=\, {C}^{XY}_{\ell} + \sum_{X'Y'\ne XY} \widehat{W}^{XY,\; X'Y'}_\ell\ C^{X'Y'}_{\ell},\\
        &=\, {C}^{XY}_{\ell} + \delta C^{XY}_{\ell},
\end{align}
\end{subequations}
where $\widehat{C}^{XY}_{\ell}$ is a beam-corrected data spectrum and $\widetilde{C}^{XY}_{\ell}$ the $XY$ data spectrum, $X$ and $Y$ standing for either $T$ or $E$. 
Some elements of the mixing matrix $\widehat{W}^{XY,\; X'Y'}_\ell
\equiv W^{XY,\; X'Y'}_\ell/W^{XY,\; XY}_\ell $are shown
in Fig.~\ref{fig:Wmatrix} for different combinations of frequencies.
The $\delta C^{XY}_{\ell}$, computed for a fiducial cosmological model that we will use as a template in our likelihood theory vector is displayed in Fig.~\ref{fig:hi_ell:data:leakages}. 
As can be seen in Fig.~\ref{fig:hi_ell:data:leakages}, as well as in Figs.~\ref{fig:hi-ell:fgcmbTE} and \ref{fig:hi-ell:fgcmbEE}, the net effect of the beam leakage is to increase the disagreement between frequencies. It is also clear that the leakage is to some extent degenerate with a rescaling of the power spectra (i.e., PE errors).

In 2015, we identified large incompatibilities between the different $TE$ and $EE$ cross-spectra, and we correctly identified beam leakage as the dominant source of discrepancy, at least for $TE$.
To tackle this problem, and lacking then the full beam-matrix 
computation brought by \quickpol, we used an effective model to estimate the beam-leakage templates 
{describing each element of the matrix as a fourth-order polynomial in $\ell$. 
This approximation is justified in \citetalias{planck2014-a13}, noting that symmmetries in the \Planck\
scanning strategy strongly suggested this form in the case of elliptical beams}. In detail, the parameters of the model
were determined by trying to enforce agreement between the foreground-corrected empirical 
$TE$ or $EE$ spectra with a theoretical model predicted from the $TT$ best-fit cosmological parameters. 
Our best 2015 
model is displayed and compared with the new model in Fig.~\ref{fig:hi_ell:data:leakages}.
While the 2015 model is close to the new one, it had serious limitations that were discussed in \citetalias{planck2014-a13}. 
In particular, lacking strong enough priors on the effective beam-leakage parameters, the approach was largely degenerate with 
the cosmological parameters (when exploring jointly the model and
cosmological parameters). Using instead the $TT$-based cosmology and
optimizing the model parameters 
pushes the polarization-based cosmology towards the $TT$ one. 

For the 2018 release, the beam matrix is fixed and computed from our best knowledge of the \Planck\ mission and the new model is less sensitive to this issue; 
the amplitude of the template is below $10\,\%$ of the
CMB (see Fig.~\ref{fig:hi-ell:fgTE} for example), so an $O(1\,\%)$ error in the shape of the input cosmology only translates
into an $O(0.1\,\%)$ error on the observed spectra, much below the uncertainty of the PE corrections.

The largest effect of the leakage correction is on the $TE$ spectra (which was also the case in 2015, with its effective model). 
The correction dramatically improves the inter-frequency agreement,
as shown in Fig.~\ref{fig:hi_ell:data:TE_triangle}, and resolves in large part the 
inter-frequency disagreement that was the reason not to include polarization data in the baseline 2015 results. Quantitatively, the leakage correction improves the $\chi^2$ 
by about $37$ (over 762 degrees of freedom for the binned TE likelihood). 
While the individual cross-spectra corrections co-add into a small
correction for the combined spectrum, they still have a non-negligible
effect on the cosmological parameters, as discussed in Sect.~\ref{sec:hi-ell:datamodel:alltogether} and shown in Fig.~\ref{fig:hi_ell:2015_vs_2018_systeffects}.

{The $\chi^2$ improvement brought by the leakage correction clearly shows that the $TE$ cross-spectra inter-agreement is improved by the correction. 
This can be further characterized by allowing the amplitude of the six $TE$ leakage templates to freely vary and fit them along with cosmological parameters. 
When doing so, we recover the same cosmological parameters that we obtained when fixing the template amplitudes. Furthermore, the six amplitudes are recovered to within $1.5\,\sigma$ (in the worst case) of the model-predicted ones. Opening up those six new parameters only improves the fit by $\Delta\chi^2=6$, so we see no reason not to fix them to their predicted values.}

The effect is much smaller in $EE$.  That this is to be expected can be
confirmed by evaluating the ratio of the amplitude of the leakage
effects in $TE$ and $EE$ in both the signal- and the noise-dominated
regimes, as we discuss in Appendix~\ref{app:ee_leakage_small}.

\paragraph{Sub-pixel effects:}
The coupling between the pixel-scale signal gradient and the sampling strategy creates a source of extra correlated noise \citepalias[for detailed discussion and derivations we refer the reader to][]{quickpolHivon}.
This is estimated along with the beam transfer function and leakage, and depends on the details of the masks and data cuts.
The effect is illustrated in Fig.~\ref{fig:hi-ell:fgTT} for $TT$ and in
Fig.~\ref{fig:hi-ell:fgEE} for $EE$, where it is always subdominant.
Sub-pixel effects are not included in $TE$, where they are much
smaller still. As can be seen in Figs.~\ref{fig:hi-ell:fgcmbTT}
and~\ref{fig:hi-ell:fgTT}, the correction is mainly relevant for the
$100\times100$ and $143\times143$ $TT$ spectra. Inclusion of the
sub-pixel corrections leaves the cosmological parameters essentially
unchanged, and has only a small effect on the foreground parameters (in
particular on the point-source contribution at 143\,GHz).

\begin{figure*}[htbp!]
\begin{centering}
\includegraphics[angle=0,width=0.95\textwidth]{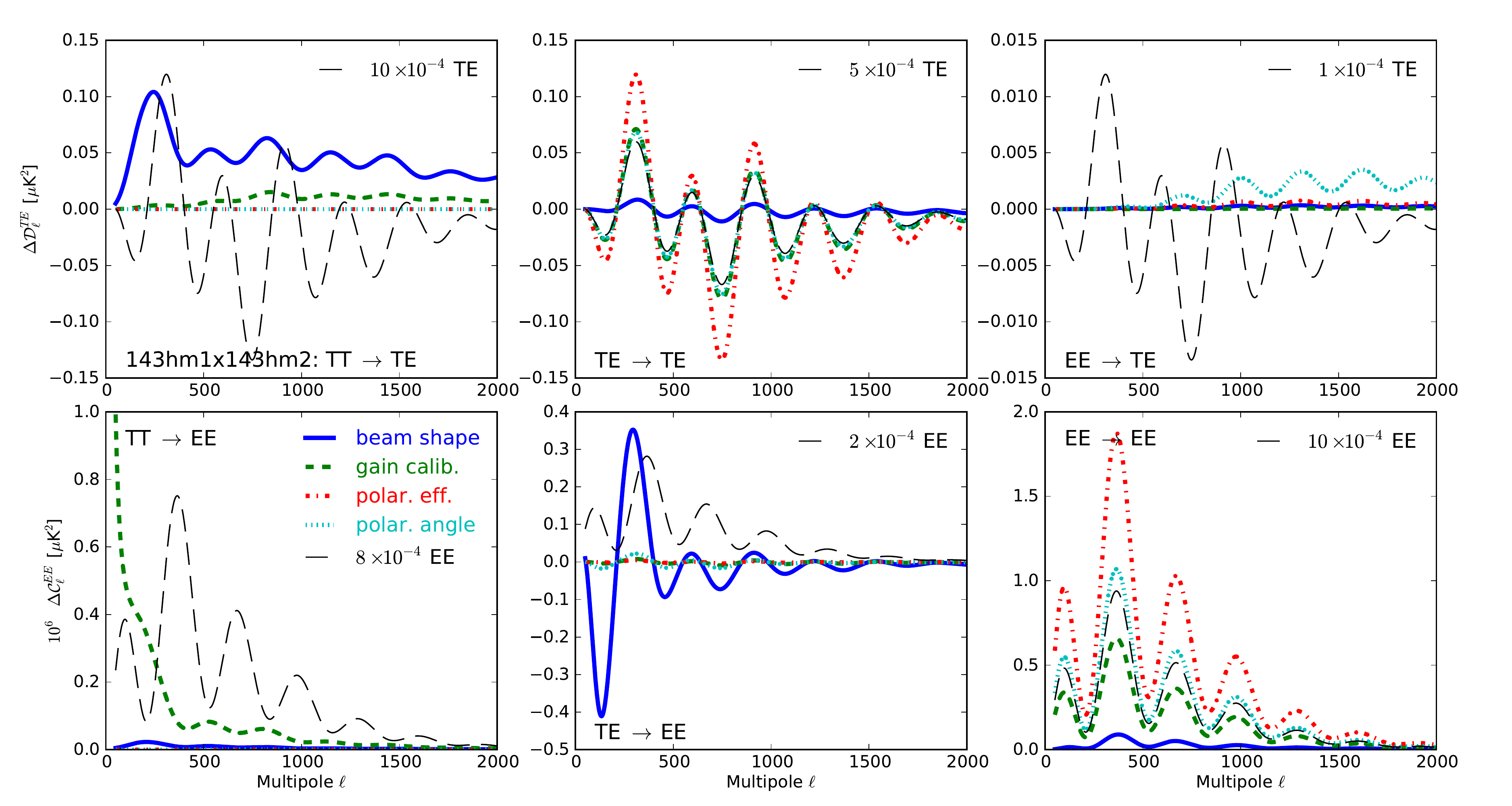}
\includegraphics[angle=0,width=0.95\textwidth]{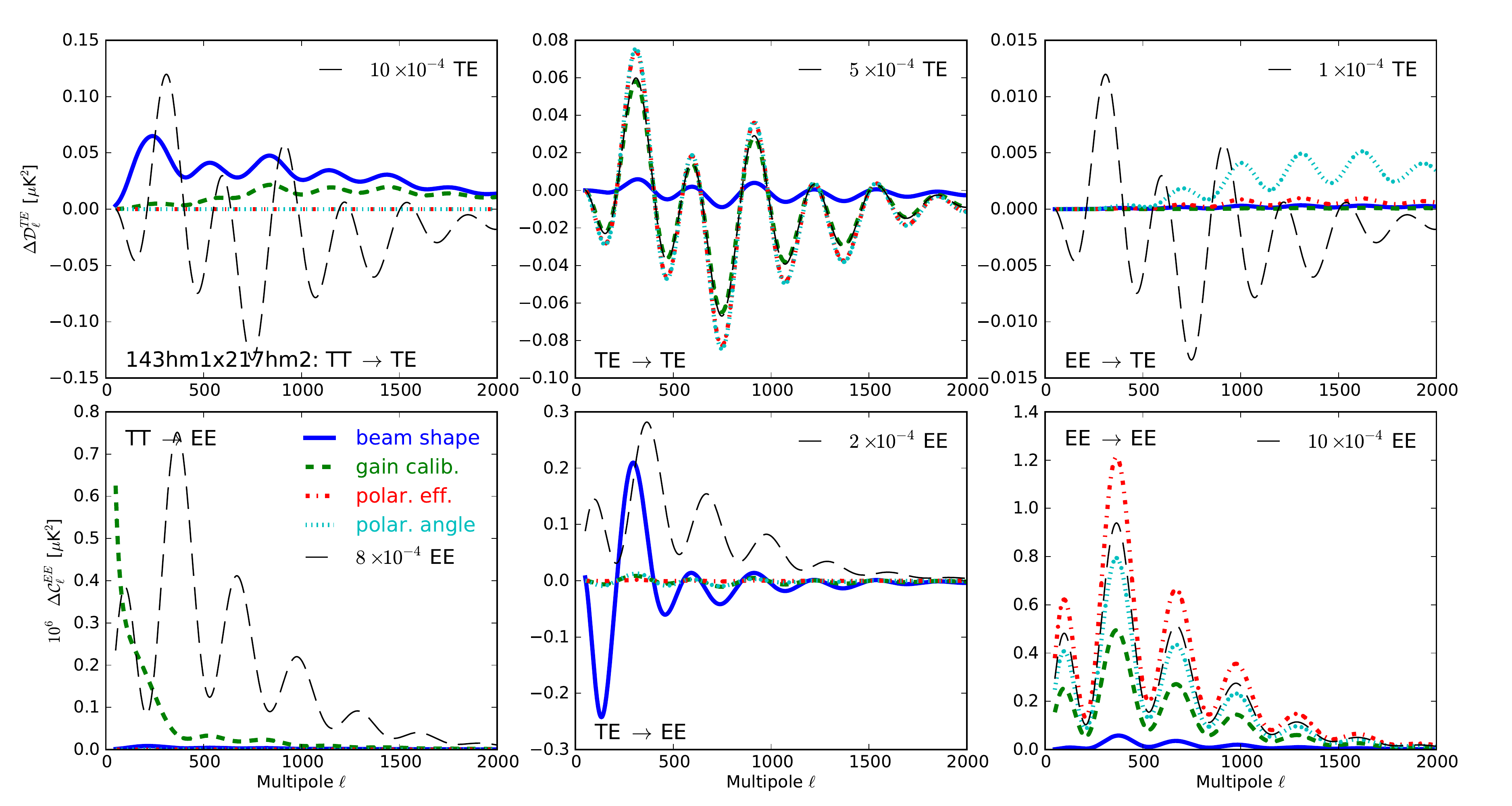}
\caption{Contribution of various sources of uncertainties to the
  cross-spectra leakage, for 143-HM1$\times$143-HM2 (upper panels) and
  143-HM1$\times$217-HM2 (lower panels). The beam errors are based on
  Monte Carlo simulations of planet observations, mimicking the actual
  beam measurements. The gain calibration error assumes an independent relative uncertainty of $0.1\,\%$ for each detector. 
The PE (between $0$ and $1$ by definition) is
assumed to be known with an absolute error of a few tenths of a percent, as
determined on pre-flight measurements \citep{rosset2010}, while the polarization angles are assumed to be known with a precision of $1\degr$ for PSBs and $5\degr$ for SWBs.
}
\label{fig:hi-ell:leakage_error}
\end{centering}
\end{figure*}

\paragraph{Beams and beam-leakage uncertainty:}
In 2013 and 2015, we used Monte Carlo simulations to propagate
uncertainties in the beam estimations through to uncertainties in the effective beams.
The first few eigenvectors of the effective-beam uncertainties
were then retained, and these could act as
multiplicative templates in the theory vectors.  In 2013 the
coefficients were sampled over with a suitable prior, whereas in 2015
they were analytically marginalized over, adding a contribution to
the likelihood covariance matrix. In 2015, the improvements in the beam 
estimation pipeline rendered this contribution effectively negligible. For this reason, 
we ignore entirely the effect of beam determination uncertainties in 2018.

We have also investigated other sources of uncertainties that were not evaluated in 2015.
First, in all of our analysis, and in particular in our computation of effective beams, we use beam measurements on non-polarized sources (i.e., planets) and 
assume perfect co-polarization (i.e., the polarization of the incoming
radiation is preserved by the beam).
\texttt{GRASP} simulations of the beams\footnote{Performed using the so-named software from TICRA, \url{www.ticra.com}\,.} (described in detail in \citealt{planck2013-p03c} and \citealt{maffei2010}) allow us to assess the validity of this assumption.
Using these simulations, one can show that the slight departures from
co-polarization of the \planck\ beams translate into a small
systematic error in the beam estimations, the amplitude of which depends mainly on the location of each detector in the 
focal plane (the closer to the centre, the smaller the effect). 
These errors project onto both a small effective recalibration of the polarized data, which we estimate to be below $0.3\,\%$ on the spectrum amplitudes, 
and an effective increase of the beam FWHM, which is smaller than $0\farcm4$ (using the \citetalias{quickpolHivon} definition of the beam as the diagonal of the beam matrix). 
The calibration difference is absorbed in the PE correction described in
Sect.~\ref{sec:hi-ell:datamodel:inst}. It is about half of the PE estimation uncertainty (in the case of $EE$) and cannot explain the PE-estimation 
differences at 143\,GHz when comparing $EE$ and $TE$. 
The beam FWHM rescaling corresponds to effects always less than
$0.5\,\%$ for $\ell<2000$. This is much smaller than our PE
errors.  Hence we ignore the non-co-polarization of the
beams in our 2018 analysis.

 As a second investigation, similarly to what was done for the beam
uncertainties, we also explore the beam-leakage uncertainties.  We
propagate beam, calibration, PE, and
angle uncertainties through a Monte Carlo analysis into uncertainties in the beam-leakage corrections. 
Results are shown
in Fig.~\ref{fig:hi-ell:leakage_error} for a subset of cross-spectra. 

Simulations were produced for the 2015 release and the levels of the uncertainty propagated correspond to the best available estimates at the time. In the case of PEs, where the ground-based estimated uncertainty was found to be about a tenth of the error reported in Sect.~\ref{sec:hi-ell:datamodel:inst}, we are thus underestimating the effect. 
This is only an issue for the $TE$-to-$TE$ and $EE$-to-$EE$ leakage, which are (as expected) dominated by the PE errors. Of course, residual errors in those leakage terms are by construction absorbed into our PE corrections; thus we only have to worry about the next sources of uncertainty, namely the gain errors and the polarization angles, whose associated uncertainties are a fraction of that of the PE errors.
Note also that the amplitude of the gain uncertainty ($0.1\,\%$) and polarization angle errors ($1\degr$) used in the simulations are larger than the current estimates given in Table~7 (better than $3\times10^{-4}$ precision on absolute gain calibration for CMB channels) and section~5.10.2 (ground-measured angles used in the data analysis are coherent with the IRAM measurement of the Crab Nebula within $0\pdeg3$) of \citetalias{planck2016-l03}, respectively.

In the case of the two dominant leakage terms, $TT$-to-$TE$ and $TE$-to-$EE$, the leakage errors are found to be dominated by the beam
 uncertainties.
  For the $143\times143$ spectrum, the largest error in
  the $TT$-to-$TE$ leakage is always less than $0.1\,\mu{\rm K}^2$ in $\mathcal{D}_\ell$ at 
$\ell\approx200$, near the first $TT$ acoustic peak. This corresponds
to a tenth of the beam-leakage template for this spectrum{. As can be seen in Fig.~\ref{fig:hi-ell:leakage_error}, the error is found to be of the order of 0.1\,\% of the $TE$ spectrum, well below our main source of uncertainty, namely the PE correction.}

In the case of $EE$, the uncertainty in the gain determination translates into low-$\ell$ biases arising from the subdominant $TT$-to-$EE$ term, below a tenth of the expected dust contamination at $\ell<500$. We noted above that this is a generous upper limit, since the gain uncertainty is smaller than what was used in the simulations.

This Monte Carlo estimation shows that uncertainties in the beam leakage
cannot account for the differences in the PE
estimates obtained using either $EE$ or $TE$, discussed in Sect.~\ref{sec:hi-ell:datamodel:inst}. 
Given the very low level of these uncertainties when compared to the $TE$ and $EE$ CMB spectra, we neglect them in the rest of the analysis.

{Owing to the small error budget we discussed above, we did not allow for any free parameter in the model.
To assess whether a  modified leakage model could improve the agreement between the different $TE$ cross-spectra, we explore corrections to the templates in the form of the polynomial model used in 2015. We showed in \citetalias{planck2014-a13} that the parametric model is degenerate with cosmological parameters. 
For this reason, the parameters of the polynomial correction will be measured assuming a reference cosmological model. 
Exploring the 18 new parameters, we find that the overall $\chi^2$ can only be improved by $\Delta\chi^2=17.5$.
Unsurprisingly, the correction modifies the template to make it much more similar to the 2015 model shown in the top panel of Fig.~\ref{fig:hi_ell:data:leakages}. 
This corresponds to $\muK^2$ level corrections, in particular in the third trough region, around $\ell\,{=}\,750$. We discuss in Sect.~\ref{sec:valandro:cond} that this region is where we see the largest residual disagreement between frequencies. However, this level of correction is much larger that what our Monte Carlo propagation of uncertainties allows. Given the shape of the error modes of the dominant sources of uncertainties (displayed in Fig.~\ref{fig:hi-ell:leakage_error}), it is difficult to localize the peak of the template correction in the third trough region. Given the small improvement in $\chi^2$ (comparable to the number of parameters) and the shape of the correction, we see no compelling reason to modify the \quickpol-based template and use it in the likelihood.}

\subsection{Summary description of \plik}
\label{sec:hi-ell:datamodel:alltogether}

In summary, the high-$\ell$ likelihood is formed using a
Gaussian approximation. The data consist of cross-spectra constructed
from half-mission maps using masks of varying sizes, 
corrected for beams that take into account the detailed shape of the
masks and scanning strategy. The model vector contains the CMB contribution, 
as well as astrophysical foregrounds and systematics templates
(temperature-to-polarization leakage is the dominant
 systematic in $TE$). The model vector also includes the corrections for residual calibration and PE errors.
A code, \plik\ is distributed along with this paper and implements this likelihood approximation. 
It is the baseline high-$\ell$ likelihood product for the \Planck\ 2018 legacy release.

In detail, for temperature, the foreground model contains:
\begin{itemize}
\item the dust contribution, described in Sect.~\ref{sec:hi-ell:datamodel:gal}, with a fixed template adapted to each frequency and sky fraction, and parameterized by a free amplitude, constrained by a prior; and
\item the extragalactic foreground model, described in
 Sect.~\ref{sec:hi-ell:datamodel:fg}, consisting of point-source
 contributions (grouping radio and infrared sources), parameterized by free amplitudes at each frequency, and the clustered-CIB, tSZ, kSZ, and the CIB-SZ correlation, each parameterized by a single amplitude rescaling templates at the different frequencies.
\end{itemize}
The systematic effects residual model contains:
\begin{itemize}
\item a sub-pixel noise correction; and
\item a beam-leakage template.
\end{itemize}
The latter two are very small corrections, and the templates (described Sect.~\ref{sec:hi-ell:datamodel:beamleak}) are fixed.
The calibration parameters are described in
Sect.~\ref{sec:hi-ell:datamodel:inst}, and are allowed to vary, subject
to tight priors. A comparison of the different foreground and nuisance contributions is presented in Figs.~\ref{fig:hi-ell:fgcmbTT} and~\ref{fig:hi-ell:fgTT}.

The EE polarization likelihood has a very simple foreground model,
consisting only of the dust correction described in
Sect.~\ref{sec:hi-ell:datamodel:gal}, ignoring the small synchrotron
contribution at low frequencies.  Given the degeneracy of the dust
contribution with residual systematics, the dust template amplitudes
have been fixed. The residual systematics model consists of:
\begin{itemize}
  \item empirical correlated noise corrections for the
    $100\times100$, $143\times143$, and $217\times217$ spectra,
    determined by the end-to-end simulations (see Sect.~\ref{sec:hi-ell:datamodel:noise});
  \item sub-pixel noise corrections (Sect.~\ref{sec:hi-ell:datamodel:beamleak}); and
\item beam-leakage templates (also Sect.~\ref{sec:hi-ell:datamodel:beamleak}).
\end{itemize}
Templates of these three types have fixed amplitude. Both the sub-pixel noise and the beam-leakage templates are small. 
The PE corrections, however, are extremely important for
the $EE$ spectrum, as described in Sect.~\ref{sec:hi-ell:datamodel:inst}. 
Percent-level corrections are applied to the model vector, and ensure
a good level of inter-frequency agreement, as shown in Fig.~\ref{fig:hi_ell:data:EE_triangle}. Comparisons of the different foreground and nuisance contributions are presented in Figs.~\ref{fig:hi-ell:fgcmbEE} and~\ref{fig:hi-ell:fgEE}.

Finally, the $TE$ cross-spectrum likelihood is very similar to the $EE$
one, with a single component for the foreground model, corresponding
to the dust contribution. As in the $TT$ case, the dust model is
parameterized by a free amplitude at each frequency, subject to a prior computed using the 353-GHz maps. 
The S/N is low enough that using the same power-law template for each frequency is sufficient. The systematic effects residual model reduces to the beam-leakage template, the sub-pixel
noise being even more negligible for the $TE$ cross-spectra (see
Sect.~\ref{sec:hi-ell:datamodel:beamleak}) than for the $TT$
 ones. The beam-leakage template is important and is the cause of
the improvement of the inter-frequency agreement in $TE$ shown in
Fig.~\ref{fig:hi_ell:data:TE_triangle} (we remind the reader that
temperature-to-polarization leakage was our main source of residual uncertainty in 2015). 

As described in Sect.~\ref{sec:hi-ell:datamodel:inst}, we propagate
the $TT$ calibration and $EE$ PE correction to $TE$,
using a map-based model. Using an alternative spectrum-based
calibration model, shows some tension with
 the map-based one and leads to fraction-of-$\sigma$-level changes for the cosmological parameters in extended models (see Sect.~\ref{sec:valandro:PE} for full discussion). 
A comparison of the different foreground and nuisance contributions is presented in Figs.~\ref{fig:hi-ell:fgcmbTE} and~\ref{fig:hi-ell:fgTE}.

The full list of foreground and nuisance parameters is given in Table~\ref{tab:fg-params}. 
Apart from small changes in the dust template shapes, the main change
compared to the 2015 likelihood model consists in the refined
systematic effects residual model, and in particular with the introduction of
the beam-leakage model (particularly important in TE) 
and the PE corrections (particularly important in EE).

{We decided to recommend fixing some of the nuisance parameters. We investigated the effect of exploring those jointly with the cosmological parameters, either with a prior
or letting them be entirely free to vary. Results of those tests are reported in Sect.~\ref{sec:valandro:PE} for the PE and Sect.~\ref{sec:valandro:priors} for the foreground priors, while we discuss in Sects.~\ref{sec:hi-ell:datamodel:beamleak} and \ref{sec:hi-ell:datamodel:noise} the beam leakage template and $EE$ correlated noise, respectively. In those last two cases, we show, by opening up the amplitudes of the templates, that the data prefer the corrections with amplitudes compatible with the model-predicted ones. 
We discussed in Sect~\ref{sub:galactic-dustPol} why we decided to eventually fix the \textit{EE} dust amplitudes, since opening them up will push the $100\times 143$ recovered dust
contribution toward values that are in disagreement with the dust estimates from the 353-GHz maps. The opening up of the \textit{EE} dust amplitudes is shown to have only a small effect on cosmological parameters for the joint \TTTEEE\ likelihood, and statistically acceptable when considering EE only. We also fix the PE corrections. We discuss in Sect~\ref{sec:hi-ell:datamodel:inst} how the $\ell\,{>}\,500$ modes at 217\,GHz, which are the only multipoles retained in the EE likelihood, are pushing the PE correction toward a large value that is in disagreement at the $3\,\sigma$ level with the value we determine using only the larger scales. We show in Sect.~\ref{sec:valandro:PE} that opening up the PE correction parameters or changing the correction model from the map-based one to the spectrum-based one has very little impact on the cosmological parameters for the joint TTTEEE likelihood, and we discuss how extensions such as the \LCDM+$\Alens$ model are affected by the change to a spectrum-based correction model at the $0.5\,\sigma$ level. 
The likelihood package distributed allows other users to reproduce the different tests performed in this paper, and in particular to open up the dust \textit{EE} amplitude and PE correction parameters and to change the PE correction model from the baseline map-based one to the spectrum-based one.}

\begin{figure}[htbp!]
\centering
\includegraphics[width=0.495\textwidth]{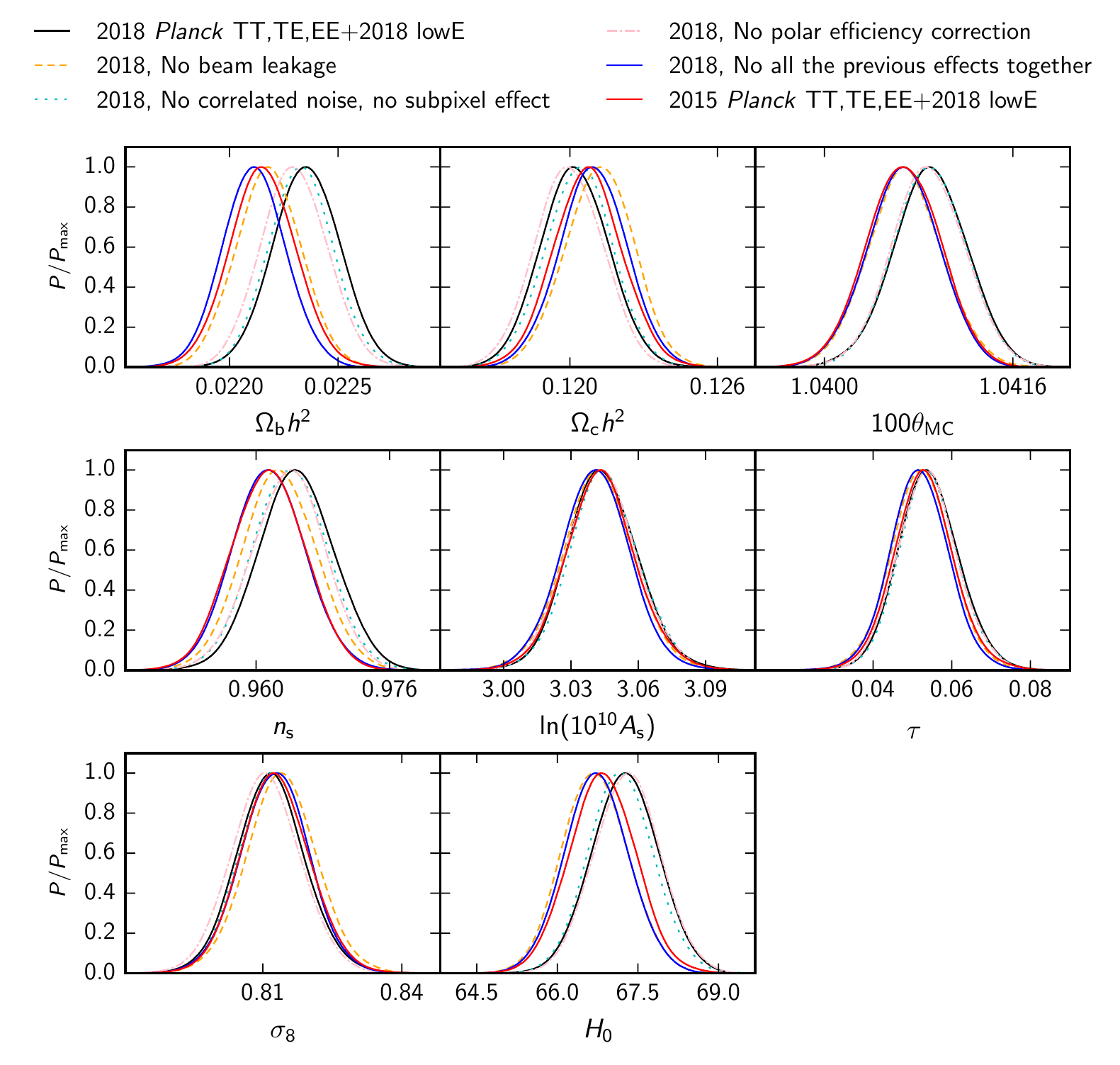}

\caption{Impact of corrections of systematic effects on cosmological
  parameters. The figure show 1-dimensional posterior distributions
  for the \LCDM\ models. We show
  results for the baseline 2018 \planckall (solid black
  line) likelihood (which includes the 2018 \plikTTTEEE\ and the 2018 \lowl\ likelihoods), as well as the results obtained with the old 2015
  TT,TE,EE (solid red line) likelihood (which includes the 2015 \plikTTTEEE\ the 2015 \lowl\ likelihoods). In both cases we also used the
  2018 lowE likelihood at large scales in polarization, in order to
  compare the 2015 versus 2018 results using the same
  constraint on the optical depth to reionization. All the other lines
  show the impact of neglecting various systematic effects, whose
  corrections have been introduced after the second, 2015, data release. In particular, we show the impact on the 2018 results of neglecting the beam leakage (dashed orange), the correlated noise and sub-pixel effects (dotted cyan), the PE corrections (dot-dashed pink), and all of the above (solid blue). Most of the changes in parameters between 2015 and 2018 can be attributed to these systematic effects (mostly the beam-leakage correction in TE). Other remaining modifications, such as those in the mapmaking or dust modelling, have a smaller impact.
}
\label{fig:hi_ell:2015_vs_2018_systeffects}
\end{figure}

Figure~\ref{fig:hi_ell:2015_vs_2018_systeffects} shows a comparison between 2015 and the 2018 results on cosmological parameters. It demonstrates how the correction for systematic effects in polarization (in particular beam leakage in $TE$ but also correlated noise, sub-pixel effects, and PE corrections) can mostly account for the changes between the 2015 and 2018 likelihoods. 

\begin{table*}[htbp!] 
\caption{Parameters and priors
  used for astrophysical foregrounds and instrumental modelling for the baseline likelihood. Uniform priors are given as ranges in square brackets, while Gaussian priors are given by their mean and standard deviation in parenthesis. We also give the fixed values of parameters that are not allowed to vary in the baseline likelihood.
} 
\label{tab:fg-params}
\begingroup 
\newdimen\tblskip \tblskip=5pt
\nointerlineskip
\vskip -3mm
\footnotesize
\setbox\tablebox=\vbox{
\newdimen\digitwidth
\setbox0=\hbox{\rm 0}
\digitwidth=\wd0
\catcode`*=\active
\def*{\kern\digitwidth}
\newdimen\signwidth
\setbox0=\hbox{+}
\signwidth=\wd0
\catcode`!=\active
\def!{\kern\signwidth}
\halign{\hbox to 2cm{$#$\leaderfil}\tabskip=2em& 
   $#$\hfil\tabskip=2em& 
   #\hfil\tabskip=0pt\cr
\noalign{\doubleline}
\omit\hfil Parameter\hfil&\omit\hfil Prior range\hfil&\omit\hfil Definition\hfil\cr
\noalign{\vskip 3pt\hrule\vskip 5pt}
A^{\rm PS}_{100}&                [0,400]&             Contribution of Poisson point-source power to $\mathcal{D}^{100\times 100}_{3000}$ for \planck\ (in $\mu\mathrm{K}^2$)\cr
A^{\rm PS}_{143}&                [0,400]&             As for $A^{\rm PS}_{100}$, but at $143\times143\,$GHz\cr
A^{\rm PS}_{217}&                [0,400]&             As for $A^{\rm PS}_{100}$, but at $217\times217\,$GHz\cr
A^{\rm PS}_{143\times217}&       [0,400]&             As for $A^{\rm PS}_{100}$, but at $143\times217\,$GHz\cr
A^{\mathrm{CIB}}_{217}&          [0,200]&             Contribution of CIB power to $\mathcal{D}^{217}_{3000}$ at the \planck\ CMB frequency for $217\,$GHz (in $\mu\mathrm{K}^2$)\cr 
A^{\mathrm{tSZ}}&                [0,10]&              Contribution of tSZ to $\mathcal{D}_{3000}^{143\times 143}$ at $143\,$GHz (in $\mu\mathrm{K}^2$)\cr
A^{\mathrm{kSZ}}&                [0,10]&              Contribution of kSZ to $\mathcal{D}_{3000}$ (in $\mu\mathrm{K}^2$) \cr
\omit&&[We apply a joint tSZ-kSZ prior with ${\cal D}^{\rm kSZ} + 1.6 {\cal D}^{\rm tSZ} = (9.5 \pm 3)\,\mu\textrm{K}^2$]\cr
\xi^{{\rm tSZ}\times {\rm CIB}}& [0,1]&               Correlation coefficient between the CIB and tSZ\cr
A^{{\rm dust}TT}_{100}&          [0,50]&              Amplitude of Galactic dust power at $\ell=200$ at $100\,$GHz (in $\mu\mathrm{K}^2$)\cr
\omit&                           (8.6\pm 2)\cr
A^{{\rm dust}TT}_{143}&          [0,50]&              As for $A^{{\rm dust}TT}_{100}$, but at $143\times143\,$GHz\cr
\omit&                           (10.6 \pm 2)\cr
A^{{\rm dust}TT}_{143\times 217}&[0,100]&             As for $A^{{\rm dust}TT}_{100}$, but at $143\times 217\,$GHz\cr
\omit&                           (23.5\pm 8.5)\cr
A^{{\rm dust}TT}_{217}&          [0,400]&             As for $A^{{\rm dust}TT}_{100}$, but at $217\times217\,$GHz\cr
\omit&                           (91.9\pm 20)\cr
\noalign{\vskip 5pt\hrule\vskip 5pt}
c_{100}&                         [0,3]&               Power spectrum calibration at $100\,$GHz\cr
\omit&                           (1.0002\pm 0.0007)\cr
c_{217}&                         [0,3]&               Power spectrum calibration at $217\,$GHz\cr
\omit&                           (0.99805\pm 0.00065)\cr
y_{\rm cal}&                     [0.9,1.1]&           Absolute map calibration for \planck\ \cr
\omit&                           (1\pm 0.0025)\cr
\noalign{\vskip 5pt\hrule\vskip 5pt}
A^{{\rm dust}EE}_{100}&          0.055&              Amplitude of Galactic dust power at $\ell=500$ at $100\,$GHz (in $\mu\mathrm{K}^2$)\cr
A^{{\rm dust}EE}_{100\times 143}&0.040&              As for $A^{{\rm dust}EE}_{100}$, but at $100\times 143\,$GHz\cr
A^{{\rm dust}EE}_{100\times 217}&0.094&              As for $A^{{\rm dust}EE}_{100}$, but at $100\times 217\,$GHz\cr
A^{{\rm dust}EE}_{143}&          0.086&               As for $A^{{\rm dust}EE}_{100}$, but at $143\times143\,$GHz\cr
A^{{\rm dust}EE}_{143\times 217}&0.21&              As for $A^{{\rm dust}EE}_{100}$, but at $143\times 217\,$GHz\cr
A^{{\rm dust}EE}_{217}&          0.70&               As for $A^{{\rm dust}EE}_{100}$, but at $217\times217\,$GHz\cr
\noalign{\vskip 5pt\hrule\vskip 5pt}
A^{{\rm dust}TE}_{100}&          [0,10]&              Amplitude of Galactic dust power at $\ell=500$ at $100\,$GHz (in $\mu\mathrm{K}^2$)\cr
\omit&                           (0.13\pm 0.042)\cr
A^{{\rm dust}TE}_{100\times 143}&[0,10]&              As for $A^{{\rm dust}TE}_{100}$, but at $100\times 143\,$GHz\cr
\omit&                           (0.13\pm 0.036)\cr
A^{{\rm dust}TE}_{100\times 217}&[0,10]&              As for $A^{{\rm dust}TE}_{100}$, but at $100\times 217\,$GHz\cr
\omit&                           (0.46\pm 0.09)\cr
A^{{\rm dust}TE}_{143}&          [0,10]&              As for $A^{{\rm dust}TE}_{100}$, but at $143\times143\,$GHz\cr
\omit&                           (0.207 \pm 0.072)\cr
A^{{\rm dust}TE}_{143\times 217}&[0,10]&              As for $A^{{\rm dust}TE}_{100}$, but at $143\times 217\,$GHz\cr
\omit&                           (0.69\pm 0.09)\cr
A^{{\rm dust}TE}_{217}&          [0,10]&               As for $A^{{\rm dust}TE}_{100}$, but at $217\times217\,$GHz\cr
\omit&                           (1.938\pm 0.54)\cr
\noalign{\vskip 5pt\hrule\vskip 5pt}
c_{EE100}&			1.021 &						Polarization efficiency correction at $100\times 100\,$GHz (called $\calibC^{PP}_{100}$ in Eq.~\ref{eq:relcalEE})\cr
c_{EE143}  &			0.966 &						As for $c_{EE100}$, but at $143\times143\,$GHz\cr
c_{EE217} &			1.04  &						As for $c_{EE100}$, but at $217\times217\,$GHz\cr
\noalign{\vskip 5pt\hrule\vskip 3pt}
}}
\endPlancktablewide
\endgroup
\end{table*}
\subsection{ Other high-$\ell$ likelihood products}
\label{sec:hi-ell:prod}
Alongside this paper, the Planck Collaboration is releasing multiple
high-$\ell$ likelihood products:
\begin{itemize}
\item the \plik\ likelihood, corresponding to the data choices and
 approximations described in the preceding sections;
\item the \camspec\ likelihood, which correspond to a variation on some key data and model choices 
(e.g., polarization mask and PE), which have been flagged in the previous sections; and
\item \pliklite, a nuisance-marginalized
version of \plik.
\end{itemize}
\camspec\ and \pliklite\ will be described in more detail in
the following sections.

\subsubsection{\camspec}
\label{sec:hi-ell:prod:camspec}

The \camspec\ likelihood was the baseline for the 2013 release and described in detail in \citetalias{planck2013-p08}.  It used cross-spectra formed from 
detector-set temperature maps made out of data from the nominal mission period.  For the 2015 release \citepalias{planck2014-a13}, \camspec\ was extended 
to include both polarization and temperature-polarization cross-spectra and used data from the full-mission period.  To mitigate the effects of
correlated noise between detectors, the cross-spectra were formed from frequency maps constructed from separate halves of the
full-mission data.  The foreground modelling was also adjusted, and the sky fraction retained at each frequency was increased.  

For 2018, further changes of the foreground model have been made, in particular for the CIB.
The noise-modelling procedure has also been modified, going from using differences of maps constructed from the first and second halves of
each pointing period (HRD maps) to differences of maps constructed from alternating pointing periods (OED maps, see Sect.~\ref{sec:hi-ell:datamodel:datasel} for definition and Sect.~\ref{sec:hi-ell:datamodel:noise} for discussion of different noise models).  This change was made because of the impact of correlated glitch residuals in the half missions
led to an underestimate of the noise, particularly in the polarization spectra at multipoles $\ell\la500$.   The 
masks have also been refined, principally in polarization.  Individual polarization spectra have had temperature-to-polarization leakage-correction templates removed before their addition for inclusion in the likelihood (see Sect.~\ref{sec:hi-ell:datamodel:beamleak}).  The spectra have also been recalibrated before
being combined, to correct for polarization efficiency errors and small transfer-function effects (see Sect.~\ref{sec:hi-ell:datamodel:inst}).

{We go on to} elaborate on differences between \camspec\ and \plik\ in polarization handling, in particular covering:
\begin{itemize}{}
\item the use of a single mask;
\item Galactic dust subtraction in polarization;
\item effective calibration handling for TE and EE;
\item the coaddition process; and
\item the absence of polarized dust nuisance parameters.
\end{itemize}

\camspec\ uses a single mask for polarization to reduce the amount of computation required to calculate covariances.  The \camspec\ polarization mask is illustrated in blue in Fig.~\ref{fig:hi-ell:data:cammsk}.

Component separation is performed for \camspec\ in polarization for
$\ell\,{\leq}\,300$, using 353\,GHz as a dust tracer.  To avoid introducing correlated noise, each ``half'' of a cross-spectrum is cleaned with a ``half'' 353-GHz map.  For a cross-half-mission likelihood, for example, whenever a first-half-mission map is used, it is cleaned using a first-half-mission 353-GHz map, and similarly for the second-half maps.  Cleaning coefficients are determined by minimizing a quadratic function of the spectra (the cleaned spectra are very insensitive to the precise values of the coefficients used, once they are in approximately the right range).  Values of the cleaning coefficients $\alpha$ applied
are presented in Table~\ref{tab:camclcoeff}, with a cleaned spectrum $C_\ell^{\mathrm{clean}\, ij}$ between maps $i$ and $j$ then given in terms of beam- and mask-deconvolved spectra $C^{p q }_\ell$ by
\begin{align}
C^{\mathrm{clean} \, i j}_\ell &= (1+\alpha^i)(1+\alpha^j)\,C^{i  j}_\ell -  \alpha^i (1+\alpha^j)\,C^{353 j}_\ell \nonumber\\
&\qquad\qquad-(1+\alpha^i) \alpha^j C^{i \,353}_\ell
+ \alpha^i \alpha^j C^{353\, 353}_\ell .
\end{align}
The 353-GHz maps become noisy at high multipoles.  So at $\ell\,{>}\,300$
smooth power laws are subtracted from the spectra instead of using
the cleaning procedure we just described.  The power
laws are determined by fitting to estimates of the dust at the lower
multipoles (see also Sect.~\ref{sub:galactic-dustPol} for a comparison of this approach with the template-based one.).

\begin{table}[htbp!]
\begingroup
\newdimen\tblskip \tblskip=5pt
\caption{Cleaning coefficients applied in the construction of the polarization segment of the \camspec\ likelihood, specifying
the amount of 353-GHz subtraction required at each frequency.}
\label{tab:camclcoeff}
\vskip -3mm
\footnotesize
\setbox\tablebox=\vbox{
\newdimen\digitwidth
\setbox0=\hbox{\rm 0}
\digitwidth=\wd0
\catcode`*=\active
\def*{\kern\digitwidth}
\newdimen\signwidth
\setbox0=\hbox{+}
\signwidth=\wd0
\catcode`!=\active
\def!{\kern\signwidth}
\newdimen\decimalwidth
\setbox0=\hbox{.}
\decimalwidth=\wd0
\catcode`@=\active
\def@{\kern\decimalwidth}
\openup 3pt
\halign{
\hbox to 1.2in{#\leaderfil}\tabskip=1em&
    \hfil#\hfil\tabskip=2em&
    \hfil#\hfil\tabskip=2em&
    \hfil#\hfil\tabskip=0pt\cr
\noalign{\doubleline}
\omit\hfil Map type\hfil& 100\,GHz& 143\,GHz& 217\,GHz\cr
\noalign{\vskip 3pt\hrule\vskip 5pt}
Temperature & 0.0208& 0.0341& 0.143\cr
Polarization& 0.0192& 0.0392& 0.141\cr
\noalign{\vskip 5pt\hrule\vskip 3pt}
}}
\endPlancktable
\endgroup
\end{table}

A relatively conservative Galactic mask is used, sufficient to render post-cleaning dust residuals negligible, even at the most contaminated frequency (217\,GHz).  We consider any remaining synchrotron contamination to be negligible, even at 100\,GHz (we found in Sect.~\ref{sub:galactic-dustPol} the synchrotron residual to be negligible on a larger sky fraction).  Given the low level of point-like emission detectable above the noise in polarization outside of the Galactic region, we do not apply a point-source mask.

Although instrumental noise should not enter on the average into cross-spectra (noise between the two halves being assumed to be independent), it does contribute to the scatter.  Thus a noise estimate is required to form suitable covariance matrices.  For \camspec\ this requires the estimation of noise power spectra, which are now obtained using odd-even difference maps, as discussed above.  This procedure is not perfect and may in particular overestimate the noise contribution for 100-GHz \EE\ (see Sect.~\ref{sec:hi-ell:datamodel:noise}).  Noise contributions from the 353-GHz cleaning prescription are also currently neglected.

The individual cross-spectra are then relatively calibrated against a
fiducial model ($\Lambda$CDM $TT$ best fit; see Sect.~\ref{sec:hi-ell:datamodel:inst} for the estimation of calibration and polarization efficiency corrections).  Calibration factors are given in Table~\ref{tab:camcal}.  With appropriate multipole cuts applied for each spectrum, given in Table~\ref{tab:camclcut}, the reweighted spectra are then co-added with an $\ell$-dependent weight, each one contributing proportionally to the inverse of the diagonal of its covariance matrix (calculated for a fiducial model).  {I}ndependent relative calibrations are applied for $EE$, $TE$, and $ET$.
The covariance matrices for the pairs of individual spectra are then
correspondingly weighted and combined to yield covariance matrix blocks for and between the
$TT$ spectra, the coadded $TE$ spectrum, and the coadded $EE$ one.  The spectra and covariances are then used to make a ``fiducial Gaussian'' \citep{HL08,planck2013-p08} likelihood.    

\begin{table}[htbp!]
\begingroup
\newdimen\tblskip \tblskip=5pt
\caption{Spectral calibration factors applied in the construction of the \camspec\ likelihood.  For each spectrum in 
question, the factor is defined as that
by which the appropriate $TT$ best-fit \lcdm\ spectrum has to be multiplied in order to minimize a quadratic measure
of its difference with that spectrum.}
\label{tab:camcal}
\vskip -3mm
\footnotesize
\setbox\tablebox=\vbox{
\newdimen\digitwidth
\setbox0=\hbox{\rm 0}
\digitwidth=\wd0
\catcode`*=\active
\def*{\kern\digitwidth}
\newdimen\signwidth
\setbox0=\hbox{+}
\signwidth=\wd0
\catcode`!=\active
\def!{\kern\signwidth}
\newdimen\decimalwidth
\setbox0=\hbox{.}
\decimalwidth=\wd0
\catcode`@=\active
\def@{\kern\decimalwidth}
\openup 3pt
\halign{
    \hfil#\hfil\tabskip=2em&
    \hfil#\hfil\tabskip=2em&
    \hfil#\hfil\tabskip=2em&
    \hfil#\hfil\tabskip=2em&
    \hfil#\hfil\tabskip=0pt\cr
\noalign{\doubleline}
\multispan2\hfil Data splits\hfil& \multispan3\hfil Calibration factors\hfil\cr
\noalign{\vskip -8pt}
\multispan2\hrulefill& \multispan3\hrulefill\cr
\noalign{\vskip -2pt}
HM1& HM2& $EE$ factor& $TE$ factor& $ET$ factor\cr
\noalign{\vskip 5pt\hrule\vskip 5pt}
100& 100& 0.975& 0.982& 0.989\cr
100& 143& 1.005& 1.000& 0.980\cr
100& 217& 0.944& 0.973& 0.982\cr
143& 100& 0.995& 0.975& 1.007\cr
143& 143& 1.032& 0.998& 1.011\cr
143& 217& 0.977& 0.979& 1.005\cr
217& 100& 0.951& 0.976& 0.964\cr
217& 143& 0.982& 1.005& 0.966\cr
217& 217& 0.951& 0.986& 0.966\cr
\noalign{\vskip 5pt\hrule\vskip 3pt}
}}
\endPlancktable
\endgroup
\end{table}

\begin{table}[htbp!]
\begingroup
\newdimen\tblskip \tblskip=5pt
\caption{Multipole cuts applied in the construction of the \camspec\ likelihood.}
\label{tab:camclcut}
\vskip -3mm
\footnotesize
\setbox\tablebox=\vbox{
\newdimen\digitwidth
\setbox0=\hbox{\rm 0}
\digitwidth=\wd0
\catcode`*=\active
\def*{\kern\digitwidth}
\newdimen\signwidth
\setbox0=\hbox{+}
\signwidth=\wd0
\catcode`!=\active
\def!{\kern\signwidth}
\newdimen\decimalwidth
\setbox0=\hbox{.}
\decimalwidth=\wd0
\catcode`@=\active
\def@{\kern\decimalwidth}
\openup 3pt
\halign{
\hbox to 1.0in{#\leaderfil}\tabskip=1em&
    \hfil#\hfil\tabskip=1em&  \hfil#\hfil\tabskip=2em&
    \hfil#\hfil\tabskip=1em&  \hfil#\hfil\tabskip=2em&
    \hfil#\hfil\tabskip=1em&  \hfil#\hfil\tabskip=0pt\cr
\noalign{\doubleline}
\omit& \multispan2\hfil TT\hfil& \multispan2\hfil TE\hfil&
\multispan2\hfil EE\hfil\cr
\noalign{\vskip -8pt}
\omit& \multispan2\hrulefill& \multispan2\hrulefill& \multispan2\hrulefill\cr
\noalign{\vskip -2pt}
\omit\hfil Spectrum\hfil& $\ell_{\rm min}$& $\ell_{\rm max}$&
 $\ell_{\rm min}$& $\ell_{\rm max}$& $\ell_{\rm min}$& $\ell_{\rm max}$\cr
\noalign{\vskip 5pt\hrule\vskip 5pt}
$100\times100$&   *30&  1200& *30& 1200& *30& 1000\cr
$100\times143$& \dots& \dots& *30& 1200& *30& 1200\cr
$100\times217$& \dots& \dots& *30& 1200& 200& 1200\cr
$143\times143$&   *30&  2000& *30& 2000& *30& 1500\cr
$143\times217$&   500&  2500& *30& 2000& 300& 2000\cr
$217\times217$&   500&  2500& 500& 2500& 500& 2000\cr
\noalign{\vskip 5pt\hrule\vskip 3pt}
}}
\endPlancktable
\endgroup
\end{table}

As a result of the masking and the dust-cleaning procedure, there is no need for modelling of any further components in the polarization part of the likelihood; up to overall relative calibration factors, the \camspec\ polarization likelihood is free of nuisance parameters.

The 2018 release cosmological parameters paper \citepalias{planck2016-l06} gives
 a complementary description of \camspec\ to that given here,
 including a discussion of the changes to the CIB and dust modelling
 for TT from 2015.  A comprehensive presentation of \camspec\ is given
 in \citet{eg2019}. We show in Appendix \ref{app:camvspliPS} a comparison between the \plik\ and \camspec\ $TE$ and $EE$ power spectra, which are found to be in good agreement. A comparison between the two likelihoods at the level of cosmological parameters is presented in \citetalias{planck2016-l06}.

\subsubsection{\pliklite}
\label{sec:hi-ell:prod:pliklite}
As in 2015, we follow the methodology developed in \citet{dunkley2013} and \citet{calabrese2013} and extended to \Planck\ in \citetalias{planck2014-a13} to generate and release a more compact CMB-only \planck\ likelihood, \pliklite.
This is a compressed and faster high-$\ell$ likelihood, which already includes marginalization over foregrounds and residual systematics. 
\begin{figure}[htbp!]
\centering
 \includegraphics[width=\columnwidth]{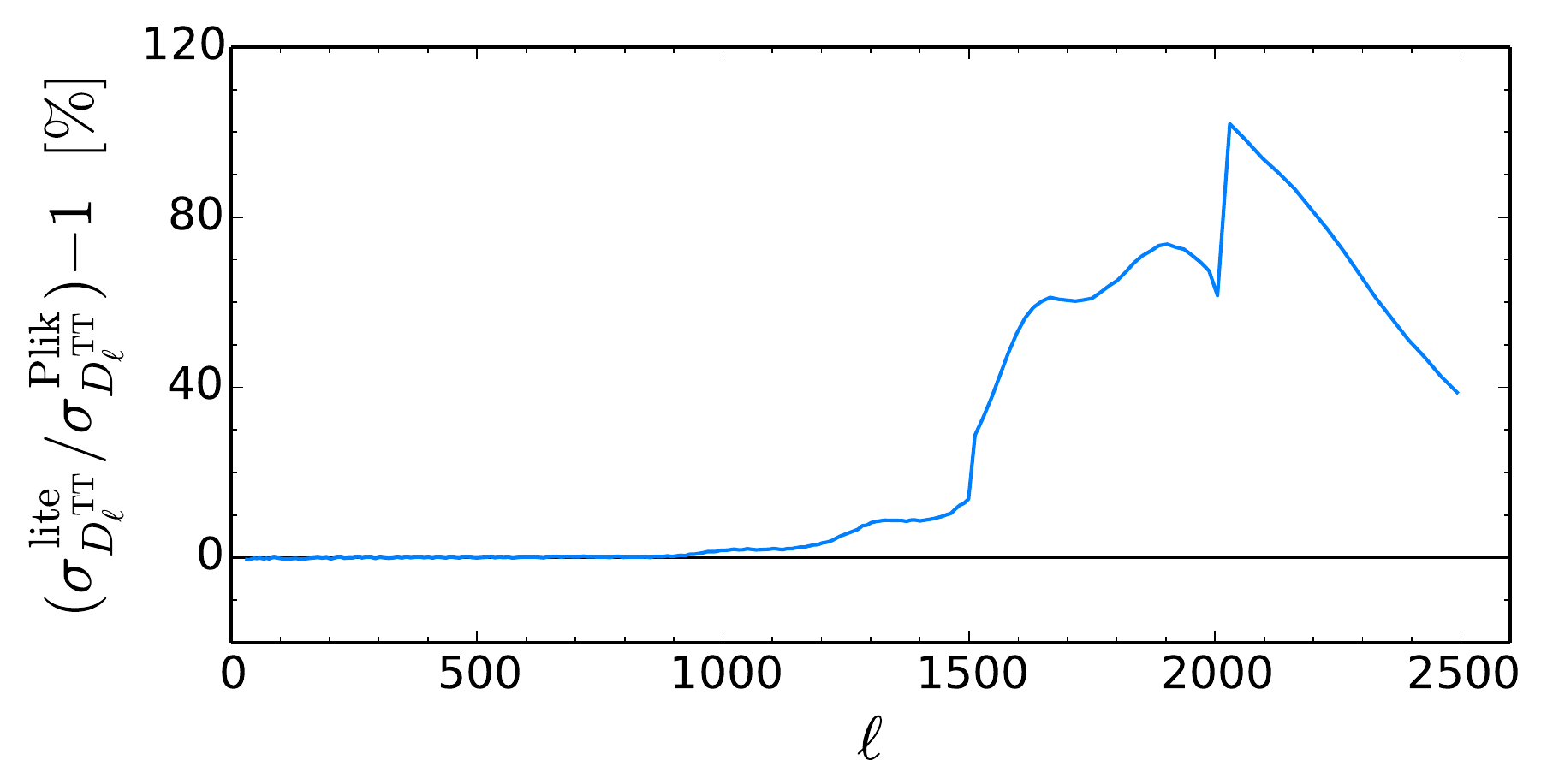}
\caption{Comparison of the uncertainties, i.e., the square root of the diagonals of the covariance matrices, of the \plik\ coadded power spectrum (calculated as described in appendix~C.4 of \citetalias{planck2014-a13}) and of the {\tt Plik\_lite} CMB bandpowers in TT. The plot shows the ratio between the {\tt Plik\_lite} and the \plik\ errors. As expected, the nuisance parameter marginalization (in particular over the point-source amplitudes), which is performed to produce the {\tt Plik\_lite} bandpowers, increases the errors on the CMB-only power spectrum by up to a factor of 2. We do not show the same results for EE and TE, since in those cases the errors are the same to within fractions of a percent. } 
\label{fig:plikvslite}
\end{figure}
\paragraph{Temperature and polarization CMB-only spectra:}
\begin{figure*}[htbp!]
\centering
 \includegraphics[width=\textwidth]{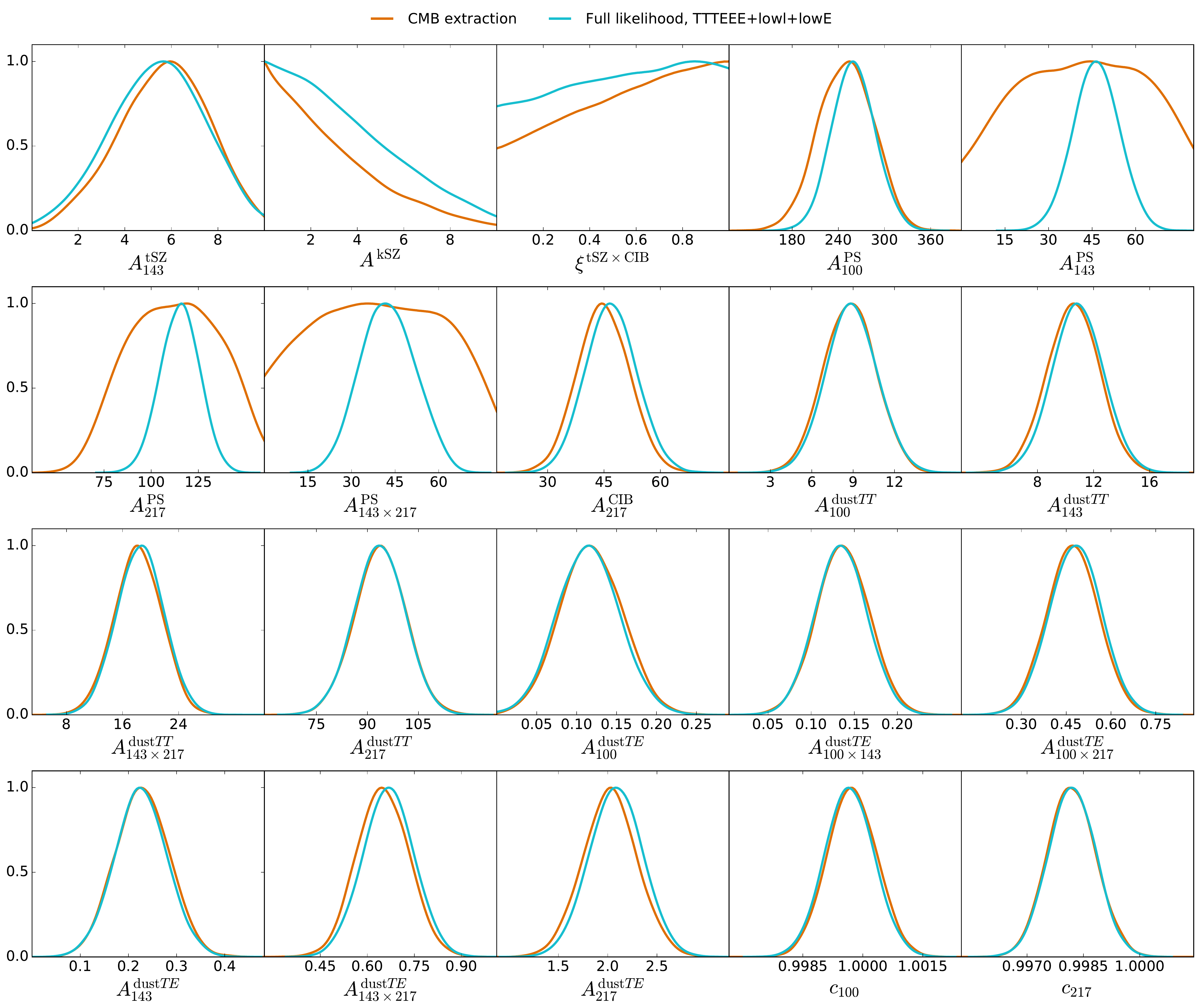}
\caption{Comparison of the nuisance parameters estimated simultaneously with the CMB bandpowers (red lines) and the results from the \plik\  likelihood (blue lines).
} 
\label{fig:fgcomp}
\end{figure*}
\begin{figure*}[htbp!] 
\centering
 \includegraphics[width=0.7\textwidth]{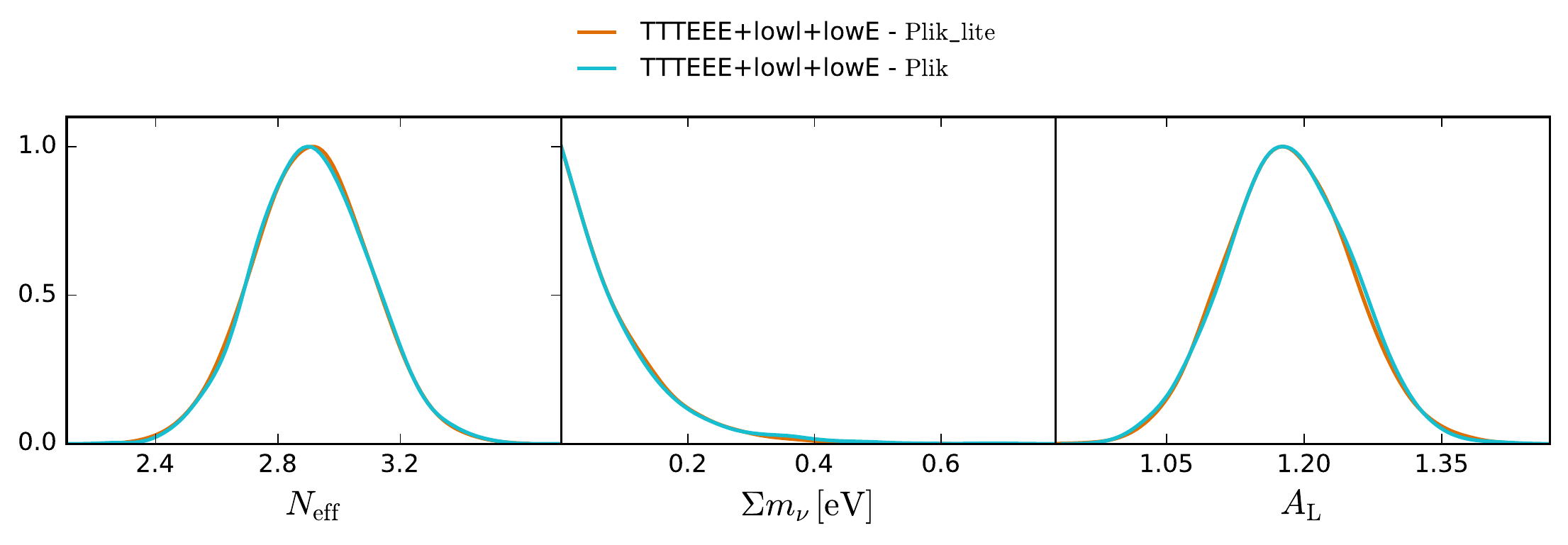}
\caption{Comparison of extensions to the \lcdm\ model from the CMB-only likelihood \pliklite\ (red) and the baseline likelihood \plik\  (blue). The curves almost completely overlap.} 
\label{fig:cmbonly_ext}
\end{figure*}
For this data release, the model for the theoretical power for a single cross-frequency binned spectrum (between frequencies $i$ and $j$) in temperature or polarization, $C_b^{{\rm th},ij}$, is written as
\be
C_b^{{\rm th},ij}= C_b^{\rm CMB} + C_b^{{\rm sec},ij}(\theta) + C_b^{{\rm sys},ij} , \label{eq:model}
\ee
where $C_b^{{\rm sec},ij}(\theta)$ is the secondary signal from Galactic and extragalactic foregrounds and is a function of 20 secondary nuisance parameters $\theta$, while $C_b^{{\rm sys},ij}$ is the residual level of beam leakage, sub-pixel effects, and correlated noise added with fixed amplitudes at different frequencies. The model is calibrated as in the full multi-frequency likelihood (see Sect.~\ref{sec:hi-ell:datamodel:inst}), fixing the 143-GHz temperature calibration factor to 1, the polarization efficiencies to 1.021, 0.966, and 1.04 for 100, 143, and 217\,GHz, respectively, and sampling the 100 and 217-GHz temperature calibration factors as nuisance parameters (i.e., as part of the $\theta$ vector). $C_b^{\rm CMB}$ is the vector of CMB-only bandpowers, that we want to extract from the data. For the \Planck\ high-$\ell$ likelihood this vector is composed of $N_{\rm b}\,{=}\,613$ CMB bandpowers, consisting of 215 data points describing $TT$, followed by 199 elements for the $EE$ spectrum, and 199 for $TE$. 

In vector form Eq.~\eqref{eq:model} reads
\be
C_b^{\rm th}= {\tens A} C_b^{\rm CMB} + C_b^{\rm sec}(\theta)+C_b^{\rm sys},
\ee
where $C_b^{\rm th}$, $C_b^{\rm sec}$ and $C_b^{\rm sys}$ are multi-frequency spectra, and the mapping matrix ${\tens A}$, with elements that are either 1 or 0, maps the $C_b^{\rm CMB}$ vector (of length $N_{\rm b}$), which is the same at all frequencies, onto the multi-frequency data. With this formalism, the multi-frequency \Planck\ likelihood can be rewritten as a multivariate Gaussian distribution:
\begin{align}
-2 \ln\mathscr{L} & = \big({\tens A}C_b^{\rm CMB}+C_b^{\rm sec} +C_b^{\rm sys} - C_b\big)^{\sf T} \tens{\Sigma}^{-1}\nonumber\\
&\qquad \big({\tens A}C_b^{\rm CMB}+C_b^{\rm sec}+C_b^{\rm sys} -C_b\big) + \mathrm{constant} ,
\label{eqn:likemf}
\end{align}
where $C_b$ is the full multi-frequency data vector. 

To extract from this the $C_b^{\rm CMB}$ vector we need to map the posterior distribution
\be
p(C_b^{\rm CMB}\,|\,C_b) = \int p(C_b^{\rm CMB},\theta\,|\,C_b)\, p(\theta)\, d\theta\,;
\ee 
however, given the size of the parameter space that we want to cover, sampling the full distribution $p(C_b^{\rm CMB},\theta\,|\,C_b)$ is a computational challenge. 

Noticing that we know the form of the distribution of the $C_b^{\rm CMB}$ vector (i.e., the components are Gaussian variables), we simplify the problem by using a Gibbs sampling technique.  We divide the process of generating a sample from the full joint distribution into two steps in the conditional distributions $p(C_b^{\rm CMB}\,|\,\theta,C_b)$ and $p(\theta\,|\,C_b^{\rm CMB},C_b)$. First, we consider a sample in the $\theta$ vector: to initialize the process we take for this some fiducial input values and, after the first iteration, we draw a trial sample according to the usual Metropolis-Hastings scheme in a standard MCMC sampling. Second, keeping the $\theta$ vector fixed, we form a $C_b^{\rm CMB}$ vector according to the conditional distribution $p(C_b^{\rm CMB}\,|\,\theta,C_b)$. This is straightforward because the conditional slice through $p(C_b^{\rm CMB},\theta\,|\,C_b)$ is a Gaussian and therefore the mean and covariance of this conditional distribution is immediately obtained by taking derivatives of the likelihood in Eq.~\eqref{eqn:likemf}, using the new $\theta$ to calculate the $C_b^{\rm sec}$ contribution to the mean \citep[see][for details]{dunkley2013}. We alternate a step in $\theta$ and one in $C_b^{\rm CMB}$ until reaching convergence for the full joint distribution; for this about 500\,000 steps are required. At this end of this process, the chain containing the samples of $\theta$ and $C_b^{\rm CMB}$ is stored for future analyses, e.g., to obtain posterior distributions for the foreground power contributions or to compute statistics of the $C_b^{\rm CMB}$ vector. 

Figure~\ref{fig:fgcomp} compares the nuisance parameters $\theta$ recovered in this model-independent sampling and the distributions obtained with the full likelihood in the baseline $\Lambda$CDM case. The parameters are consistent, although we see a significant broadening of the distribution for the \Planck\ Poisson sources. This degradation in measuring Poisson power is observed because the sources can mimic blackbody emission and are therefore degenerate with the freely-varying $C_b^{\rm CMB}$ parameters.

\paragraph{The {\tt Plik\_lite} CMB-only likelihood:}\label{app:plik_lite}
The marginalized mean and covariance matrix for the $C_b^{\rm CMB}$ vector, which we call ${\tilde C}^{\rm CMB}_b$ and $\tens{\tilde \Sigma}$, are estimated from the samples using a standard MCMC prescription, and then used to construct a compressed CMB-only Gaussian likelihood: 
\be
 -2 \ln\mathscr{L}({\tilde C}_b^{\rm CMB}\,|\,C_b^{\rm th}) = \vec{x}^{\sf T} \tens{\tilde \Sigma}^{-1}\vec{x} \,,
\label{eqn:cmblike}
\ee
where $\vec{x} = {\tilde C}^{\rm CMB}_b/y_{\rm P}^2 - C_b^{\rm th}$ and $C_b^{\rm th}$ is the binned CMB theory spectrum (including the lensing).
The overall \Planck\ calibration $y_{\rm P}$ is the only nuisance parameter left in this compressed likelihood. The Gaussianity assumption is a good approximation in the \plik\ $\ell$ range; the extracted C$_b$s are well described by Gaussian distributions over the whole multiple range.  The covariance matrix, $\tens{\tilde \Sigma}$, now incorporates the uncertainty due to foregrounds and systematics (as propagated by the marginalization over the nuisance parameters $\theta$ during the $C_b^{\rm CMB}$ extraction). To illustrate this point, in Fig.~\ref{fig:plikvslite} we plot the ratio between the square root of the diagonal of the {\tt Plik\_lite} and of the \plik\ coadded covariance matrices (the latter calculated as in appendix~C.4 of \citetalias{planck2014-a13}). As expected, the plot shows that marginalizing over the nuisance parameters, in particular over Poisson source amplitudes, in a model-independent way as performed in the {\tt Plik\_lite} algorithm, increases the error bars on the CMB-compressed power spectrum by up to a factor of 2 at small scales. On the other hand, for $EE$ and $TE$ (which have lower foreground contributions), we find that the error bars match to within fractions of a percent.

To test the performance of this compressed likelihood, we compare results using both the full multi-frequency likelihood and the CMB-only version. We report below examples for the baseline {\plik}+\lowl+\lowE\  case. 

We first estimate cosmological parameters with {\tt Plik\_lite} for the restricted $\Lambda$CDM 6-parameter model 
and compare them with the full-likelihood results. The agreement between the two methods is excellent, showing consistency to better than 0.1$\,\sigma$ for all parameters.

We then extend the comparison to a set of three \lcdm\ extensions, adding one parameter at a time to the base-\lcdm\ model: the effective number of neutrino species $\neff$; the neutrino mass $\mnu$; and the lensing amplitude consistency parameter $\Alens$. These parameters enable more extensive variations of the damping tail than with the base ones alone, and so are more correlated with the foreground parameters. Distributions for the added parameter in each of the three extensions are shown in Fig.~\ref{fig:cmbonly_ext}. Also in these cases we note that the agreement between the two methods is excellent, with all parameters differing by less than $0.1\,\sigma$.

We find similar consistency when using only a subset of the data, e.g., with $EE$ spectra alone. 

\subsection{Data and model consistency}
\label{sec:valandro}

\begin{figure*}[htbp!]
\begin{center}
\includegraphics[width=1.0\textwidth]{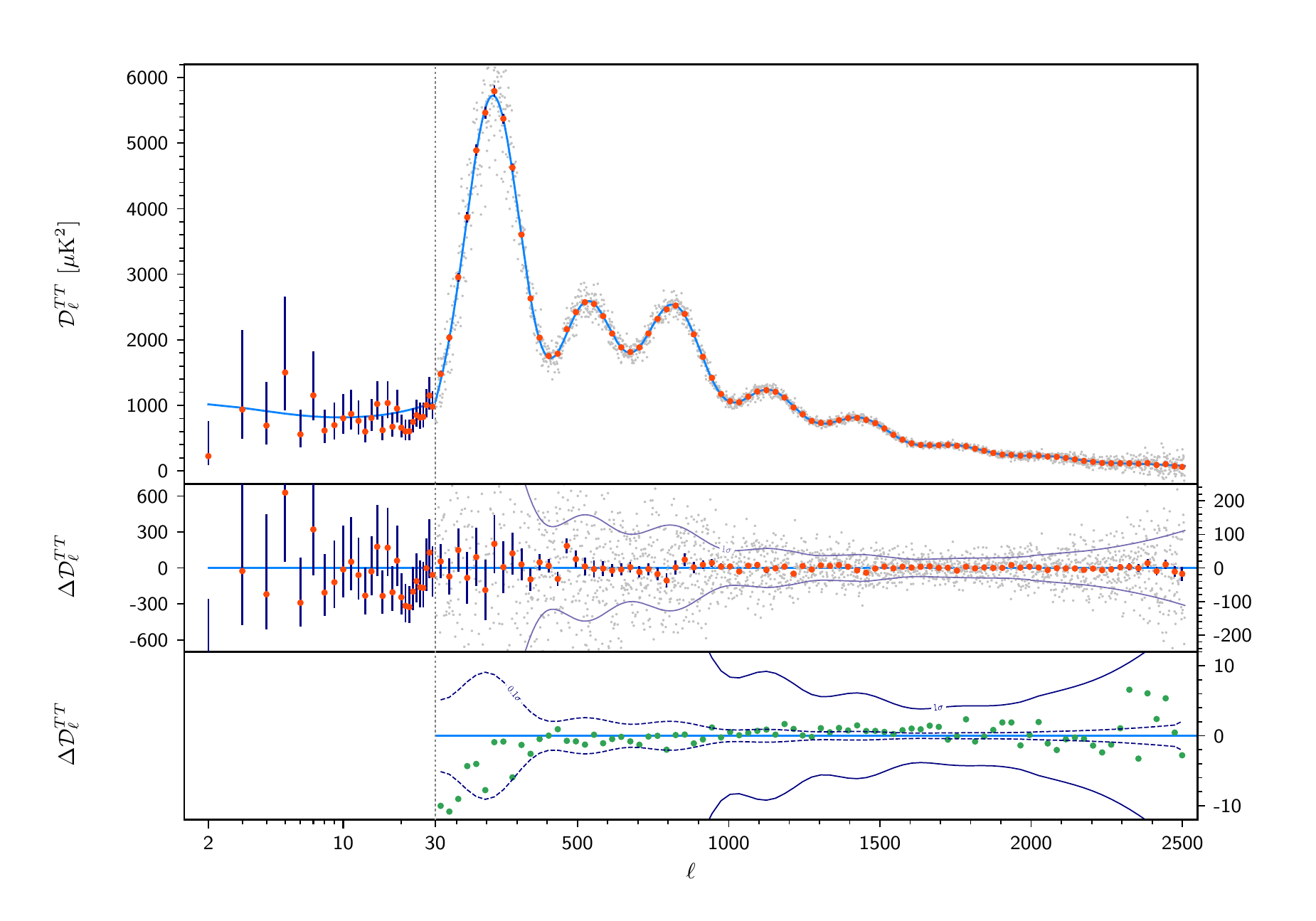}
\vspace{-1.0cm}
\end{center}
\caption{
\Planck\ 2018 temperature power spectrum. At multipoles $\ell\,{\geq}\,30$ we show the frequency-coadded temperature spectrum computed from the \plik\ cross-half-mission likelihood, with foreground and other nuisance parameters fixed to a best fit assuming the base-\lcdm\ cosmology. In the multipole range $2\,{\leq}\,\ell\,{\leq}\,29$, we plot the power-spectrum estimates from the \commander\  component-separation algorithm, computed over 86\,\% of the sky (see Sect.~\ref{ValidationTT}). The base-\lcdm\ theoretical spectrum best fit to the likelihoods is plotted in light blue in the upper panel. Residuals with respect to this model are shown in the middle panel. {The} vertical scale changes at $\ell\,{=}\,30$, where the horizontal axis switches from logarithmic to linear. The error bars show $\pm 1\,\sigma$ diagonal uncertainties, including cosmic variance (approximated as Gaussian) and not including uncertainties in the foreground model at $\ell\,{\ge}\,30$.
The $1\,\sigma$ region in the middle panel corresponds to the errors of the unbinned data points (which are in grey). The bottom panel displays the difference between the 2015 and 2018 coadded high-multipole spectra (green points). The $1\,\sigma$ region corresponds to the binned data errors. {The} vertical scale differs from the one of the middle panel. 
The trend seen for $\ell\,{<}\,300$ corresponds to the change in the dust correction model described in Sect.~\ref{sec:hi-ell:datamodel:gal}. }
\label{fig:coaddedTT}
\end{figure*}

\begin{figure*}[htbp!]
\begin{center}
\includegraphics[width=1.0\textwidth]{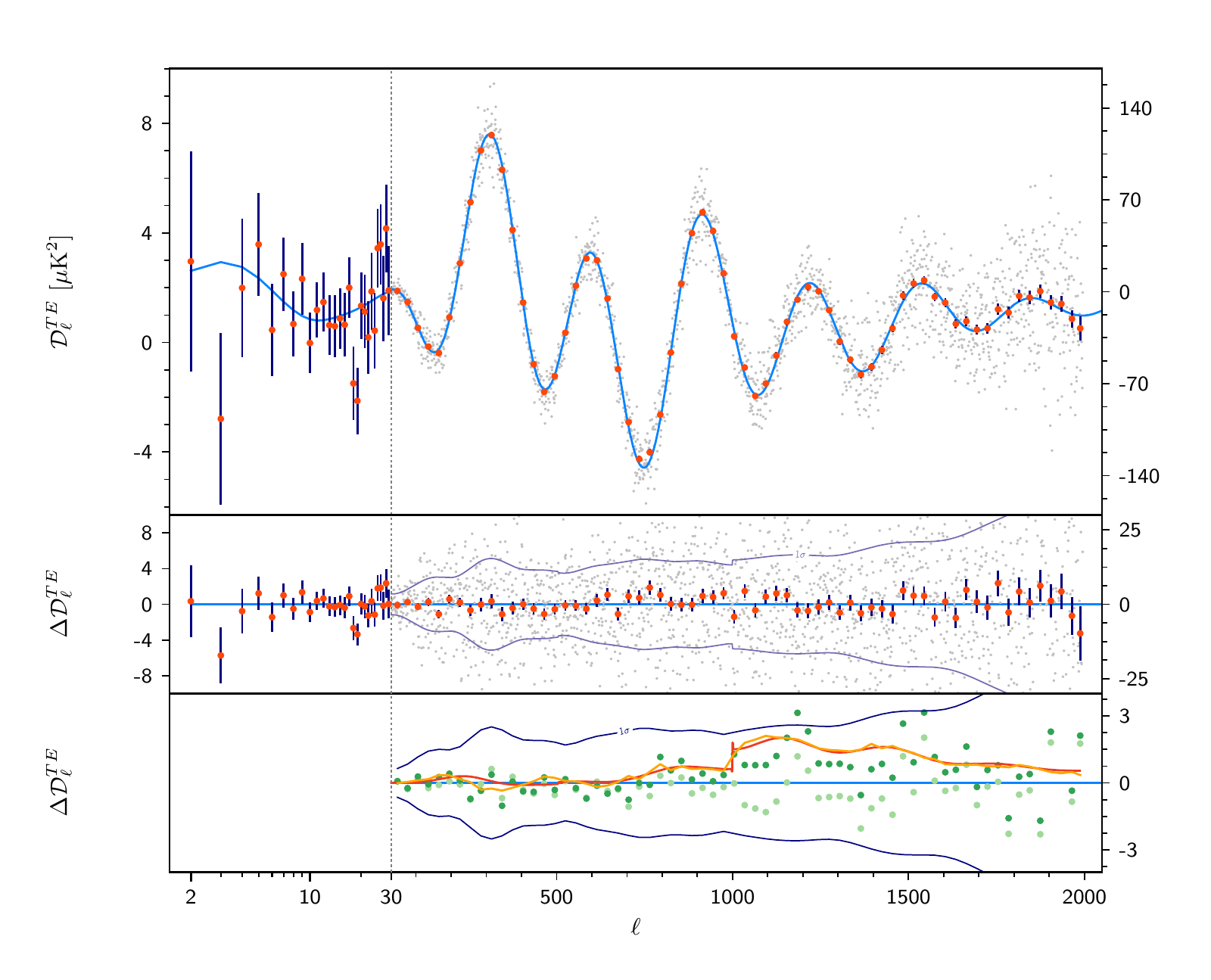}
\vspace{-1.0cm}
\end{center}
\caption{
\Planck\ 2018 $TE$ power spectrum. Figure conventions are similar to those of Fig.~\ref{fig:coaddedTT}. In the multipole range $2\leq\ell\leq 29$, we plot the power spectrum estimates from the \simall\ likelihood (although this is not used in the baseline parameter analysis for $\ell\,{\le}\,29$; see the discussion of Sect.~\ref{subsubsec:final_considerations}). The bottom panels display the difference between the 2015 and 2018 coadded high-multipole spectra (green points).  The red and orange lines correspond to the effect of the beam-leakage correction and the addition of the beam-leakage and the polarization-efficiency corrections, respectively. Both corrections were absent in the 2015 data. The light green points show the difference between the 2015 and 2018 coadded spectrum, after correction of the 2015 data by these two effects. The difference here is dominated by the leakage correction. \label{fig:coaddedTE}
}
\end{figure*}

\begin{figure*}[htbp!]
\begin{center}
\includegraphics[width=1.0\textwidth]{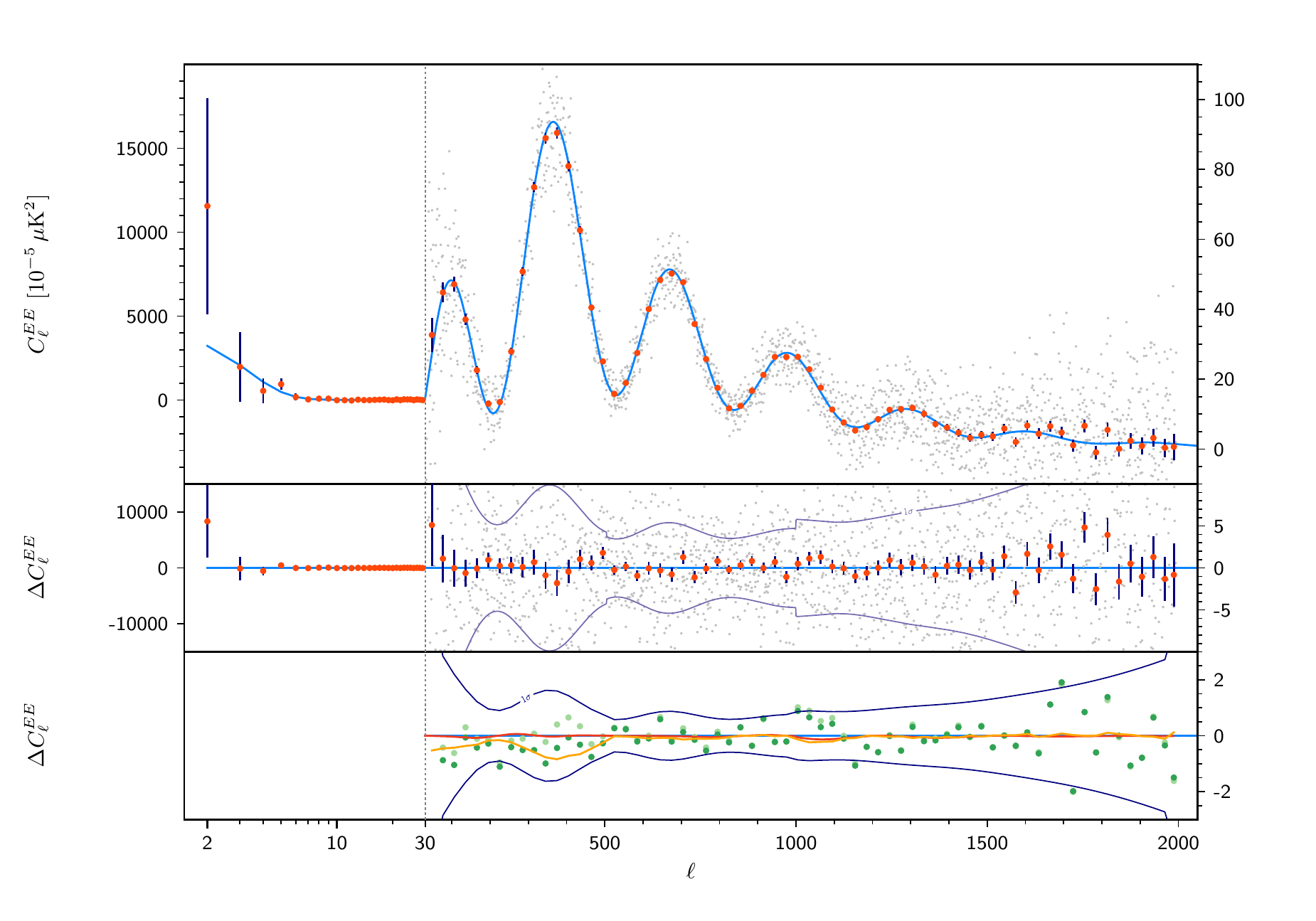}
\vspace{-1.0cm}
\end{center}
\caption{
\Planck\ 2018 $EE$ power spectrum. Figure conventions are similar to those of Fig.~\ref{fig:coaddedTT}. In the multipole range $2\leq\ell\leq 29$, we plot the power spectra estimates from the \simall\ likelihood.
The bottom panels display the difference between the 2015 and 2018 coadded high-multipole spectra (green points).  The red and orange lines correspond to the effect of the beam-leakage correction and the addition of the beam-leakage and the polarization-efficiency corrections, respectively. Both corrections were absent in the 2015 data. The light green points show the difference between the 2015 and 2018 coadded spectra, after correction of the 2015 data by the two effects. The difference in $EE$ is dominated by the polarization-efficiency correction. \label{fig:coaddedEE}
}
\end{figure*}

The overall agreement between the high-$\ell$ \plik\ likelihood and the best-fit \LCDM\ model is illustrated in part by Figs.~\ref{fig:coaddedTT}, \ref{fig:coaddedTE}, and \ref{fig:coaddedEE}, which show coadded power spectra and their residuals. The following section will present more quantitative tests of this agreement. In the figures, the best-fit cosmology is the baseline
\Planck\ result that uses the \planckalllensing\ likelihood combination. 
Differences with the 2015 result are highlighted at $\ell\,{\ge}\,30$ in the bottom panel of each figure. Differences at larger scales in $TT$ have already been discussed in Sect.~\ref{sec:lo-ell:TT}. 
The large-scale polarization results have evolved significantly since 2015. As discussed in Sect.~\ref{sec:lo-ell:hfi}, while those were based exclusively on the LFI data in 2015, the remarkable improvements in data processing have enabled us to now rely entirely on the HFI 100- and 143-GHz data and to use LFI as validation.

For the high-$\ell$ likelihood, the main difference in the temperature results between 2015 and 2018 is on the largest scales, $30\,{\le}\,\ell\,{\le}\,400$. The difference is always smaller than $10 \muK^2$ and corresponds to a change in the 
dust correction template, as discussed in Sect.~\ref{sec:hi-ell:datamodel:gal}. Comparison with the data model in Fig.~\ref{fig:hi-ell:fgTT} shows that the amplitude of the change corresponds to about a third of the $100\times100$ or the $143\times143$ dust correction, in line with the template shape change shown in Fig.~\ref{fig:hil:dust545model}.

In polarization, the main differences originate from the beam-leakage correction (for $TE$) and from the polarization-efficiency (PE) correction (for $EE$). These corrections were discussed in Sects.~\ref{sec:hi-ell:datamodel:beamleak} and \ref{sec:hi-ell:datamodel:inst}, respectively. 

\subsubsection{Goodness of fit}
\label{sec:valandro:gof}
\begin{table*}[htbp!]
\caption{
Goodness-of-fit tests for the 2018 \Planck\ temperature and polarization spectra. The reference model is our best-fit base-\LCDM\
cosmology from the \planckalllensing\ likelihood combination. The first set of rows show the results for the full
binned \plik\ high-$\ell$ likelihood, while the second set of rows
show the results from the coadded, unbinned CMB high-$\ell$
spectra. The last two rows show the results for the low-$\ell$ TT
and low-$\ell$ EE likelihoods. The effective $\chi^2$ is defined
as $\chi_\mathrm{eff}=-2\log(\mathcal{L})$. The number of degrees of
freedom ($N_{\rm dof}$) is set equal to the number of multipoles. The
last column lists the probability to exceed (PTE).  {T}he $\log(\mathcal{L})$ value for
the low-$\ell$ EE likelihood is normalized up to an arbitrary
additive constant and so further entries are not shown in this case.
Similarly, the non-Gaussian nature of the low-$\ell$ TT likelihood
precludes us from displaying a PTE in this case. 
\label{table:chi2}
}
\vskip -4mm
\newdimen\tblskip \tblskip=5pt
\setbox\tablebox=\vbox{
 \newdimen\digitwidth
 \setbox0=\hbox{\rm 0}
 \digitwidth=\wd0
 \catcode`*=\active
 \def*{\kern\digitwidth}
 \newdimen\signwidth
 \setbox0=\hbox{+}
 \signwidth=\wd0
 \catcode`!=\active
 \def!{\kern\signwidth}
 \newdimen\dpwidth
 \setbox0=\hbox{.}
 \dpwidth=\wd0
 \catcode`?=\active
 \def?{\kern\dpwidth}
\halign{\hbox to 1.7in{#\leaderfil}\tabskip 2em&
\hfil#\hfil \tabskip 2em& \hfil#\hfil& \hfil#\hfil& \hfil#\hfil&
\hfil#\hfil\tabskip 0pt\cr
\noalign{\doubleline}
\noalign{\vskip -2pt}
\omit\hfil Likelihood\hfil& Multipoles& *$\log(\mathcal{L})$& $*\chi_{\mathrm{eff}}^2$& $N_{\rm dof}$& PTE\cr
\noalign{\vskip 3pt\hrule\vskip 3pt}
TT, full, binned& 30--2508&      $*-380.34$& *760.68& *765& 0.54**\cr
TE, full, binned& 30--1996&      $*-428.68$& *857.36& *762& 0.0090\cr
EE, full, binned& 30--1996&      $*-371.48$& *742.96& *762& 0.68**\cr
TTTEEE, full, binned& 30--2508&  $-1172.47$& 2344.94& 2289& 0.20**\cr
\noalign{\vskip 3pt\hrule\vskip 3pt}
TT, coadded, unbinned& 30--2508& $-1274.57$& 2549.14& 2479& 0.16**\cr
TE, coadded, unbinned& 30--1996& $-1035.77$& 2071.54& 1967& 0.050*\cr
EE, coadded, unbinned& 30--1996& $-1028.55$& 2057.10& 1967& 0.077*\cr
TTTEEE, coadded, unbinned& 30--2508& $-3328.51$& 6657.02& 6413& 0.016*\cr
\noalign{\vskip 3pt\hrule\vskip 3pt}
Low-$\ell$ TT (\commander)& *2--29**&$**-11.63$&$**23.25$&**27&\dots \cr
Low-$\ell$ EE (\simall)& *2--29**&$*-198.02$&**\dots&**27&\dots \cr
\noalign{\vskip 3pt\hrule\vskip 3pt}}}
\endPlancktablewide
\end{table*}

We start the discussion of the data and model consistency by looking at the goodness of fit of the \lcdm\ model. In all cases, the reference model will be the best \lcdm\ fit from the \planckalllensing\ likelihoods.
We find that the general agreement of the data with the cosmological model is reasonable. Values of $\chi^2$ and PTEs are presented in Table~\ref{table:chi2}. The statistics are computed using the same best-fit reference model (\planckalllensing). We report both the values obtained from the full-frequency binned likelihood, which will encapsulate possible small disagreements between the different cross-spectra, and the coadded unbinned likelihood, where those inter-frequency disagreements average out, but which will be more sensitive to our statistical description of the multipole-by-multipole scatter. The coadded spectra (and associated covariances) are computed following the same procedure as described in section~C.4 of \citetalias{planck2014-a13}, with the difference that we are now including the calibration and PE corrections in the $\tens{J}$ matrix that links the CMB spectrum to the CMB part of the model of the frequency cross-spectra (see equation~C.40 of \citetalias{planck2014-a13} for the definition). Such corrections were ignored in 2015, owing to the small amplitude of the TT calibration corrections. {We} computed the coadded spectra independently for TT, TE, and EE, and only took into account their correlation when dealing with the \TTTEEE\ case.

We first discuss the case of the full-frequency binned likelihoods. The $\chi^2$ and PTEs of TT, EE, and \TTTEEE\ for this case are good, with PTEs of $0.54$, $0.68$, and $0.20$, respectively. 
The \TTTEEE\ PTE is slightly lower compared to the TT and EE ones because of the low value of the TE PTE of $0.009$.
In fact, while the $\chi^2$ of TE improved very significantly since 2015 (with $\chi^2_{(2018)}-\chi^2_{(2015)}=-73.72$ from $\chi^2_{(2015)}=931.08$ for $762$ degrees of freedom for the full-frequency TE likelihood), thanks in particular to the beam-leakage correction (which accounts for half of the improvement), there are still hints of disagreements between the different cross-spectra. 

We check whether modifying the nuisance parameters or computing the $\chi^2$ relative to the $TE$ best-fit spectra could substantially improve the fit to the $TE$ data.
Using the $TE$-specific PE corrections from Eq.~\eqref{eq:relcalTE} only marginally improves the situation, 
reducing the $\chi^2$ by about $\Delta\chi^2=-9$, which is not enough to significantly improve the PTE. 
Similarly, computing the statistics against the TE-only best-fit \lcdm\ cosmology instead of the \planckalllensing\ one only improves the $\chi^2$ by $\Delta\chi^2=-4.6$. 
In the covariance matrix, we ignore the extra variance induced by the leakage between temperature and polarization. A rough estimate of the correction can be obtained by rescaling of the terms of the covariance matrix by the elements of the beam matrix, $W^{XY,\; XY}_\ell$, displayed Fig.~\ref{fig:Wmatrix}. Those are found to be $O(10^{-3})$. Including the extra variance terms in the covariance would modify the $\chi^2$ by a few more points, which would still not solve our issue. 
{We remind the reader that we discussed in Sect.~\ref{sec:hi-ell:datamodel:beamleak} the fact that trying to improve the leakage model, either by allowing the amplitude of the templates for each cross-spectrum to freely vary, or by exploring the parameters of a fourth-order polynomial correction, yielded insignificant improvements on the $\chi^2$ ($\Delta\chi^2=-6$ for 6 new parameters in the first case, and $\Delta\chi^2=-17.5$ for 18 new parameters in the second one) and could not improve the PTE. }
{We note} that one would need a $\Delta\chi^2$ of approximately $-70$ to reach a PTE around $0.25$. 

Of note is the fact that the PTE (not given in the table) for the coadded \TE{} binned spectrum (computed under the standard binning scheme) is in fact much better, with a value of $0.71$ for 199 degrees of freedom. This might be an indication that the scatter between the frequency cross-spectra is not entirely captured by our model and statistical description.  We note, however, that the cross-spectrum comparison of Fig.~\ref{fig:hi_ell:valid:triangleTE} (and further discussed in Sect.~\ref{sec:valandro:triangle}) does not show a particularly large disagreement between any given pair of cross-spectra (at the coarse binning scale used in the figure). Figure~\ref{fig:hi_ell:valid:conditionalTE} (which will be discussed in detail in Sect.~\ref{sec:valandro:cond}) nevertheless shows a relatively low PTE for the conditional residuals, with features localized around the first and third troughs of the $TE$ spectrum{, especially for the $100\times143$ and $100\times217$ spectra. Removing any of the six cross-spectra from the likelihood brings the PTE up to between $0.02$ and $0.14$, with the largest gain being obtained by the removal of the $100\times217$. We will discuss in Sect.~\ref{sec:valandro:cuts} that removing any cross-frequency spectrum has little effect on the cosmological constraints (i.e., their changes are compatible with the modification of the constraining power). This gives us further reasons to conclude that the low PTE is due to limitations in our numerical ability to model the scatter between the different and correlated cross-spectra entering the likelihood.}

We now turn to the unbinned coadded case. PTEs are lower for almost all of the different cases. For TE, however, the PTE improves slightly, in line with our expectation that the low statistic is due to inconsistencies between the cross-spectra (under our modelling of the covariance). While the TT coadded unbinned $\chi^2$ and PTE values are very similar to 2015 (as can be seen in table~16 of \citetalias{planck2014-a13}, with a value of PTE of $0.172$), TE and EE have lower PTEs. Two reasons can explain the slight decrease. First, because of the leakage corrections, the polarization best fits are now further away from the TT one than they were in 2015. As  discussed in Sect.~\ref{sec:valandro:parcompTTTEEE}, we believe that the agreement between TT (and especially TE) was{,} by chance{,} too good in 2015 because of the absence of the leakage corrections. Second, after the changes in mapmaking, we end up with lower estimates of the noise level in polarization (see Sect.~\ref{sec:hi-ell:datamodel:noise} for discussion). This translates into a reduction of the variance, in particular for EE, which then makes the statistics more sensitive to the scatter of the data.

As a consequence, the joint \TTTEEE\ unbinned coadded PTE has a low value. As we remarked above, binning the spectra significantly increases the PTE values, pointing towards a possible limitation in the statistical description of the behaviour of our cross-spectra at the multipole-by-multipole level.


\begin{figure}[htbp!]
\centering
\includegraphics[width=\columnwidth]{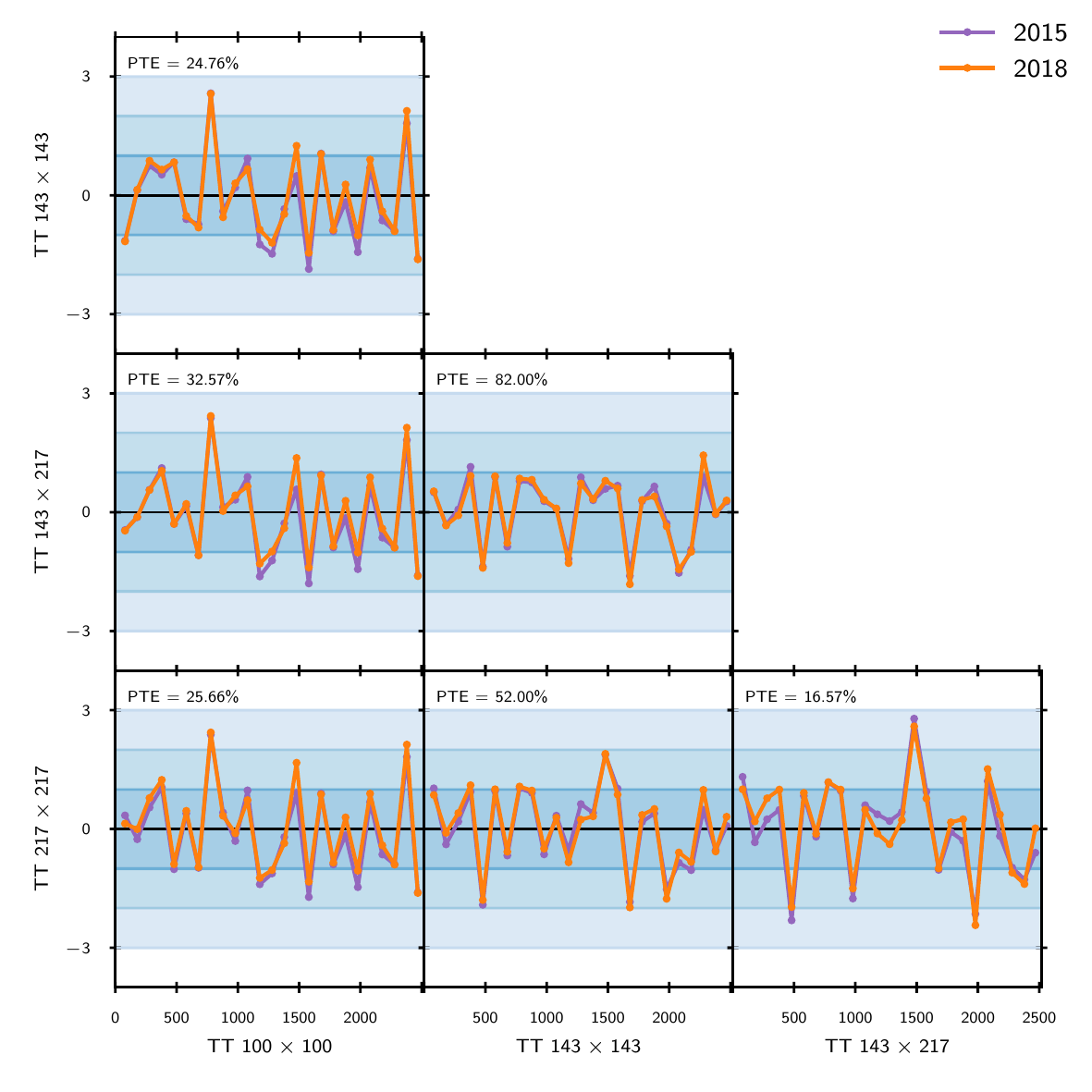}
\caption{Inter-frequency foreground-cleaned $TT$ power-spectra statistical deviations (in units of standard deviations). Each sub-panel shows differences between the foreground-cleaned cross-spectra, indicated by the row and column labels (row minus column). We show here each spectral comparison over the full multipole range, even though the likelihood discards certain portions of the data. The two lines correspond to the 2015 data and nuisance model (purple) and the 2018 one (orange). In each case, foreground and nuisance cleaning is performed at the spectrum level, as is done in the likelihood, and using the best-fit nuisance parameters from the baseline likelihood for each release (i.e., \planckall\ for 2018). The PTE value quoted in each sub-panel corresponds to the 2018 data (and model) for the full range presented in the plot and with the same $\Delta\ell\,{=}\,100$ flat binning that is used for the figure. 
}
\label{fig:hi_ell:valid:triangleTT}
\end{figure}

\begin{figure}[h!]
\centering
\includegraphics[width=\columnwidth]{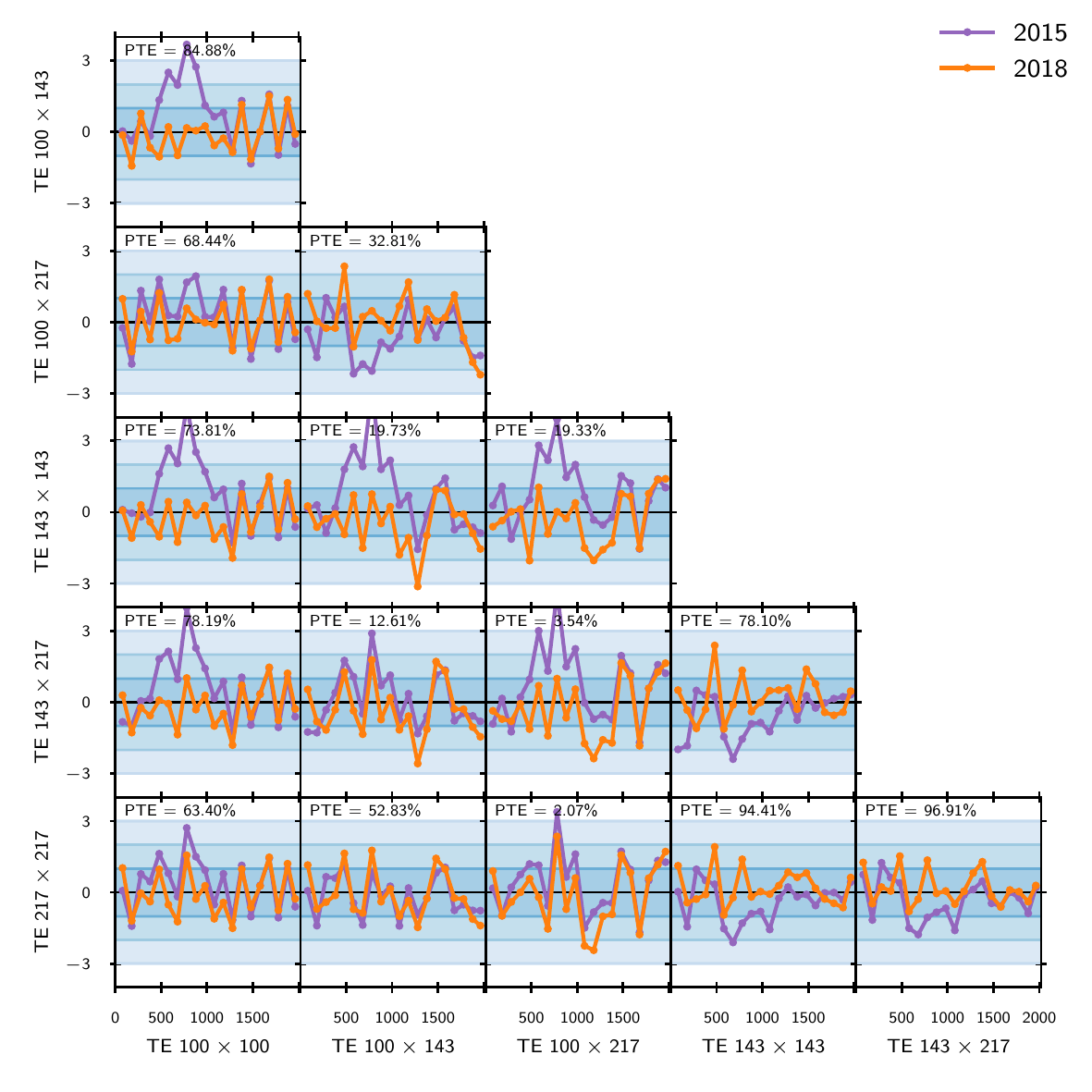}
\caption{
Same as for Fig.~\ref{fig:hi_ell:valid:triangleTT}, but now for $TE$ instead of $TT$.
}

\label{fig:hi_ell:valid:triangleTE}
\label{fig:hi_ell:data:TE_triangle}
\end{figure}

\begin{figure}[h!]
\centering
\includegraphics[width=\columnwidth]{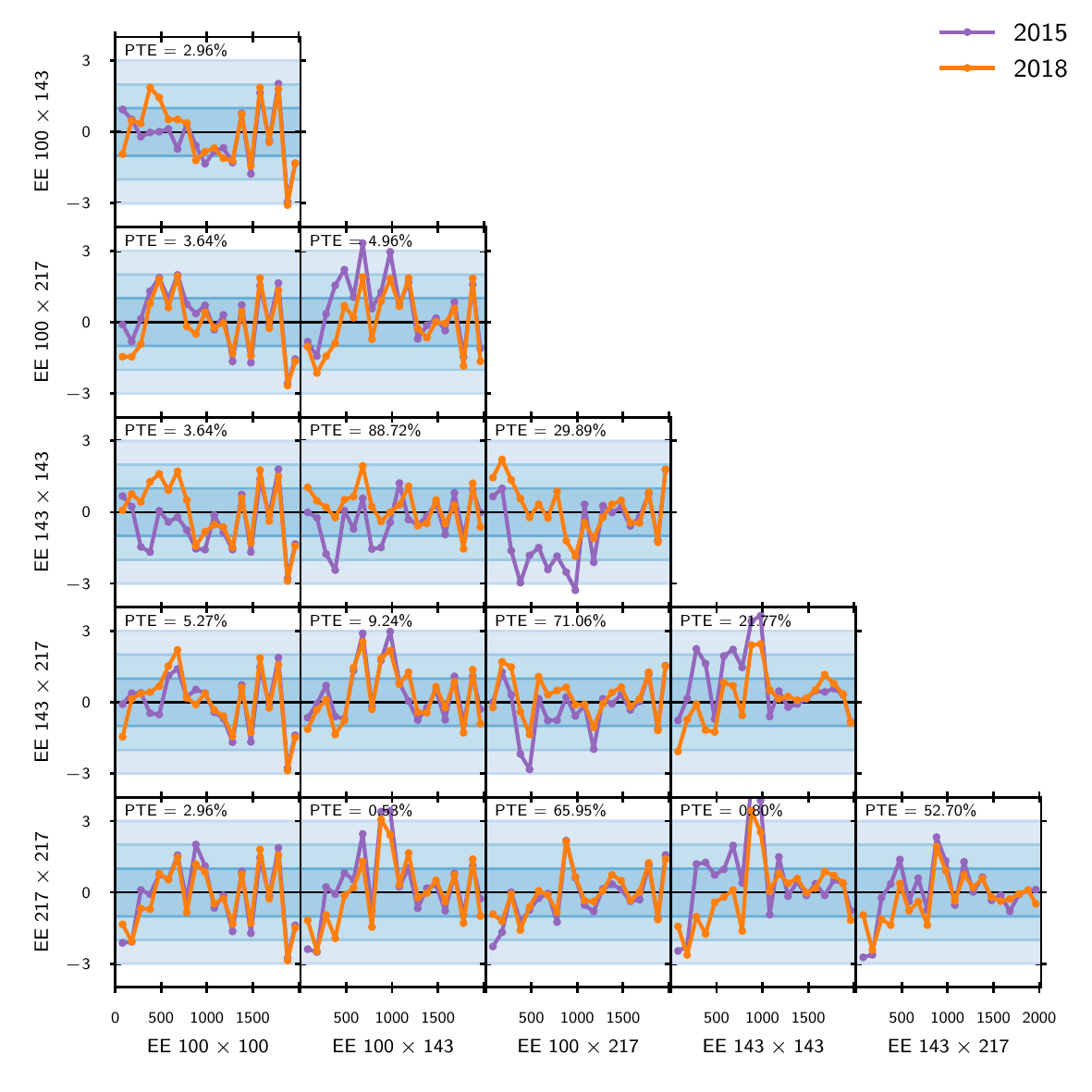}
\caption{
Same as for Fig.~\ref{fig:hi_ell:valid:triangleTT}, but now for $EE$ instead of $TT$. 
}
\label{fig:hi_ell:valid:triangleEE}
\label{fig:hi_ell:data:EE_triangle}
\end{figure}

\subsubsection{Inter-frequency agreement}
\label{sec:valandro:triangle}
We now turn our attention to the agreement between the individual cross-spectra that enter into the TT, TE, and EE likelihoods.
We will assess the inter-frequency agreement using two different kinds of tests that have different levels of dependence on the cosmological model. In this section, we will use differences between foreground-and-nuisance-corrected cross-spectra, while the next section will compare each individual cross-spectrum to the best-fit cosmological model.

The triangle plots displayed in Figs.~\ref{fig:hi_ell:valid:triangleTT}, \ref{fig:hi_ell:valid:triangleTE}, and \ref{fig:hi_ell:valid:triangleEE} present the cross-spectra differences.
These differences have only a weak dependence on the cosmological model (in this case, the best-fit \lcdm\ model using the \planckalllensing\ likelihood), which is used only to determine the values of the foreground and nuisance parameters. We show the differences between cross-spectra in units of $\sigma$, using the square roots of the diagonals of the appropriate covariance matrices for the spectral differences (see appendix~C.3.2 of \citetalias{planck2014-a13}). 
{W}e display the inter-frequency agreement even in multipole regions that are not retained for the likelihood.

The inter-frequency agreement in temperature is essentially unchanged between 2015 and 2018, remaining good. However, the situation is very different in polarization, where the inter-frequency agreement has 
improved spectacularly. We remind the reader that the inter-frequency agreement was mediocre in polarization in 2015. A plausible explanation for this lack of agreement was temperature-to-polarization leakage and our model of the leakage was not sufficient in 2015 to correct for its effect on the polarization power spectra. This was the main reason why we decided not to recommend the use of polarization at high $\ell$ for cosmology in 2015. 

In $TE$, shown in Fig.~\ref{fig:hi_ell:valid:triangleTE}, the larger than $3\,\sigma$ excursion around $\ell\,{=}\,700$ seen in 2015 has indeed been removed by the beam-leakage correction. 
The inter-frequency agreement is now much better, even in the multipole ranges that we are discarding from the likelihood. We also see that the $143\times143$ $TE$ cross-spectrum would prefer a different 
PE calibration than the map-based one that we retain, as would be expected from the discussion in Sect.~\ref{sec:hi-ell:datamodel:inst}. Nevertheless, the agreement is good.

In $EE$, shown in Fig.~\ref{fig:hi_ell:valid:triangleEE}, the improvement is not as spectacular. Changes from 2015 are dominated by the PE corrections, which we know significantly improve the $\chi^2$.  Overall, we judge the agreement to be satisfactory.

\begin{figure*}[htbp!]
\centering
\includegraphics[width=\textwidth]{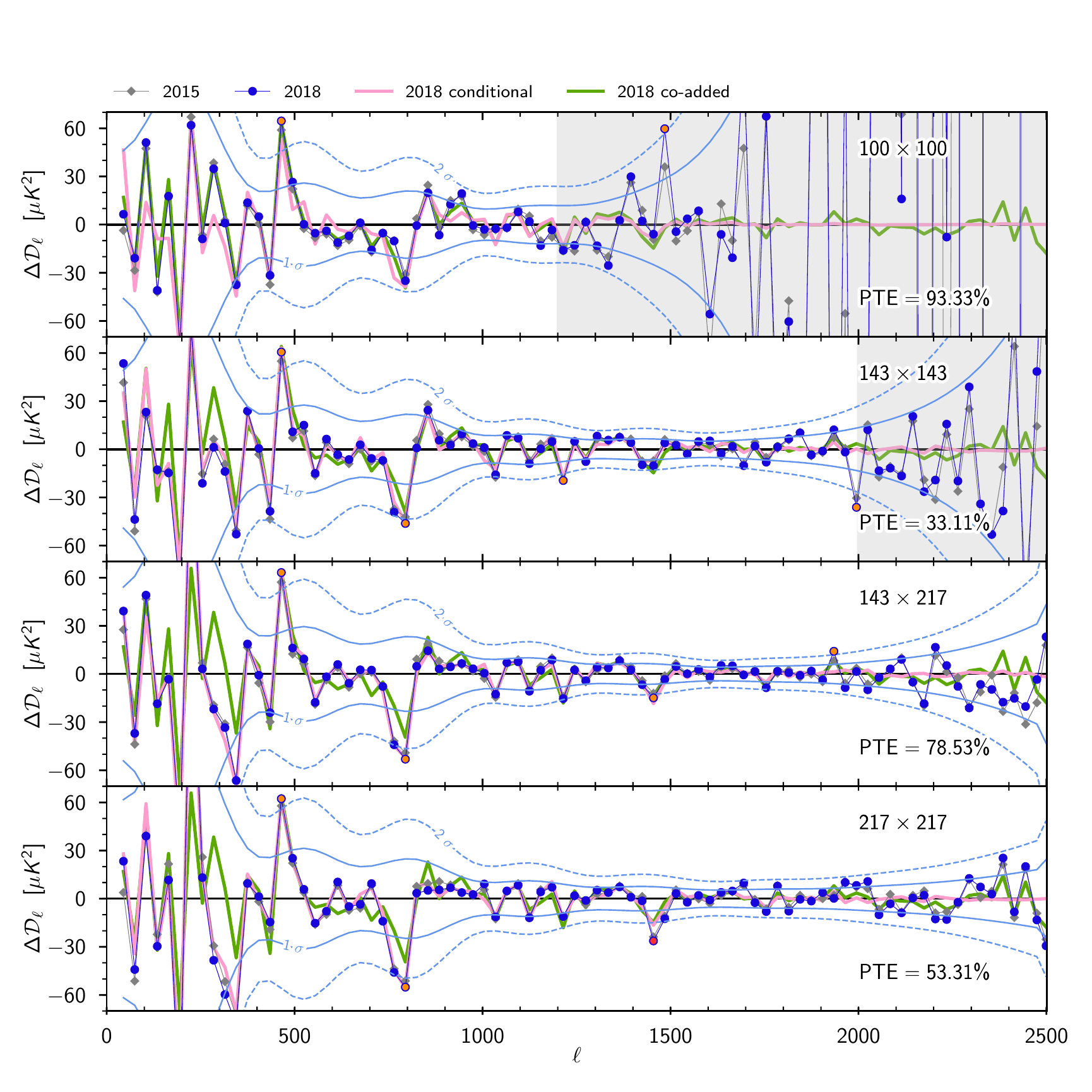}
\caption{Residuals for each cross-spectrum entering into the TT high-$\ell$ likelihood. The blue points show the residuals compared to the baseline \LCDM\ model for the 2018 release, while the grey points show the 2015 residuals (compared to their own best-fit model). The green lines show the foreground-corrected co-added CMB spectra, computed using the multipole cuts described in Table~\ref{tab:hi-ell:data:lrange}, i.e., excluding the shaded regions in each panel, while the pink lines are the conditional prediction, obtained from all of the other spectra that enter the likelihood. The blue lines show the $\pm1\,\sigma$ and $\pm2\,\sigma$ contours corresponding to the diagonals of the blocks of the 2018 covariance matrix appropriate to the cross-spectrum shown. Orange and red points correspond to $2$ and $3\,\sigma$ outliers. PTEs are computed for the non-shaded regions.}
\label{fig:hi_ell:valid:residualTT}
\end{figure*}

\begin{figure*}[htbp!]
\centering
\includegraphics[width=\textwidth]{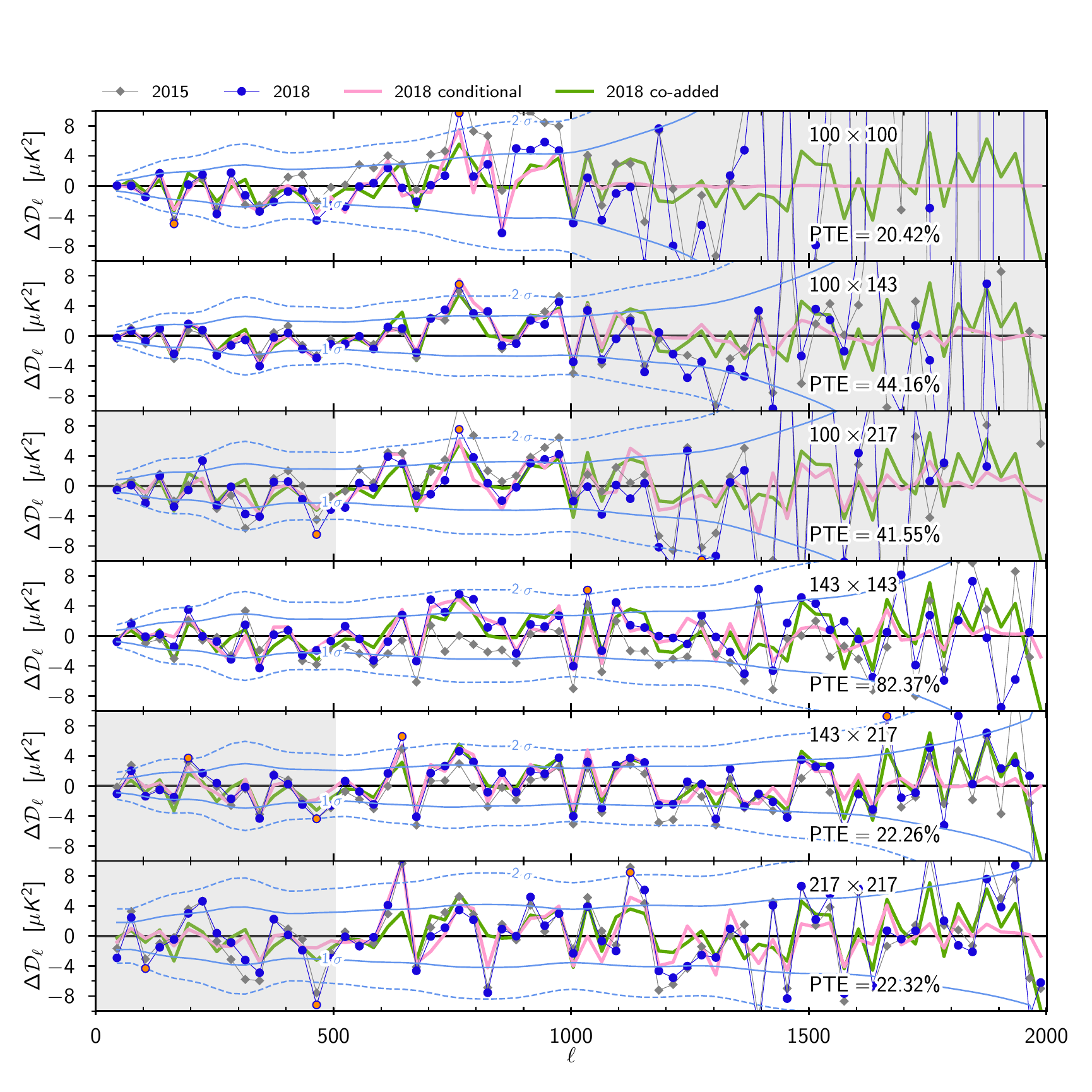}
\caption{Same as for Fig.~\ref{fig:hi_ell:valid:residualTT} but for TE instead of \TT.}
\label{fig:hi_ell:valid:residualTE}
\end{figure*}

\begin{figure*}[htbp!]
\centering
\includegraphics[width=\textwidth]{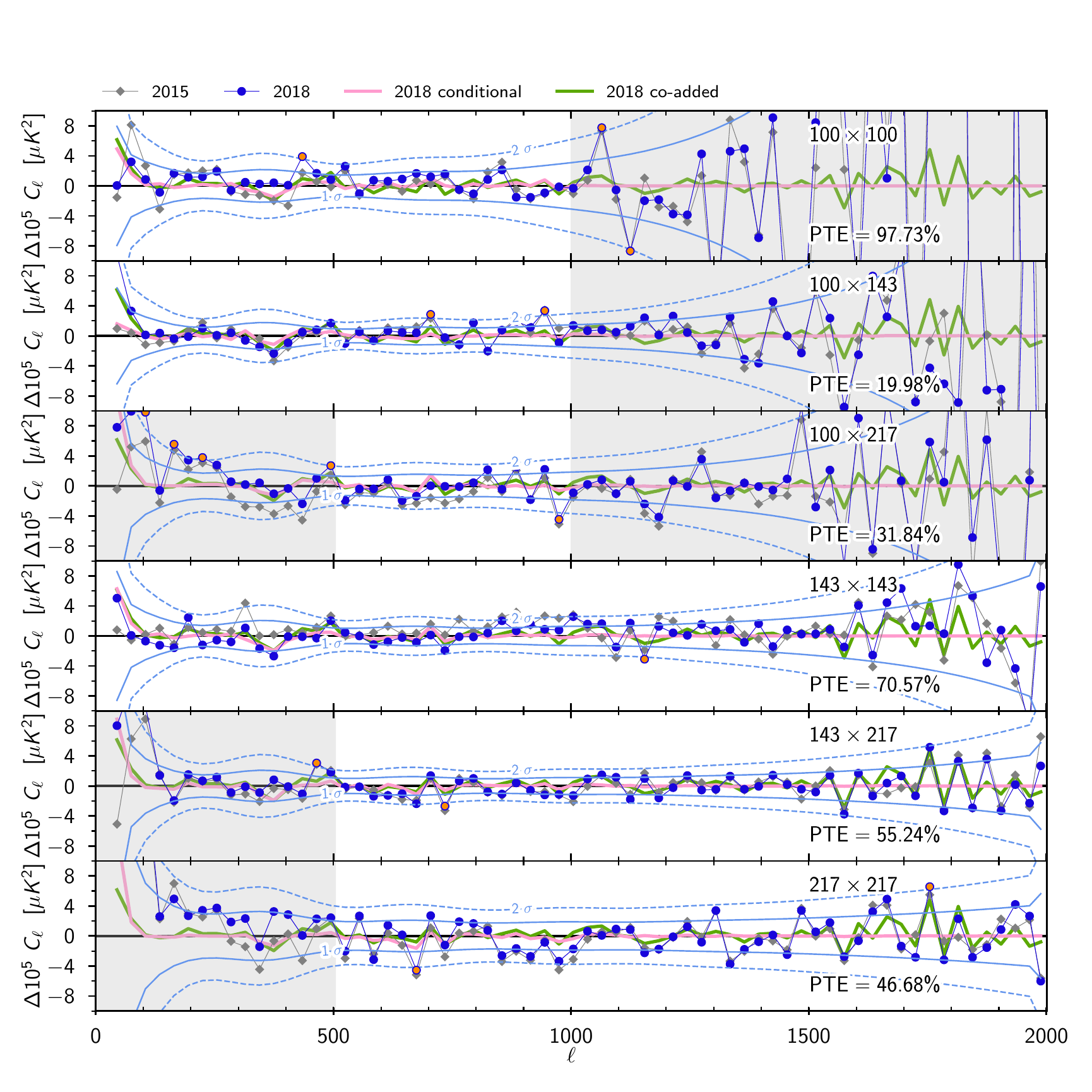}
\caption{Same as for Fig.~\ref{fig:hi_ell:valid:residualTT} but for EE instead place of TT.}
\label{fig:hi_ell:valid:residualEE}
\end{figure*}

\subsubsection{Cross-spectra residuals and conditional residuals}
\label{sec:valandro:cond}
Another way to assess the inter-frequency agreement is to look at the cross-spectrum-by-cross-spectrum best-fit residuals. These are shown in Fig.~\ref{fig:hi_ell:valid:residualTT} for $TT$, with regions of multipoles not used for the likelihood shaded in grey.\footnote{The PTEs displayed in the figure are computed only for the retained multipole region.} Similarly to what was done for the triangle plots, foreground and nuisance contaminations are corrected using the best-fit foreground model obtained by jointly fitting for their parameters along with the cosmology. The residual plots have a stronger dependence on cosmology than the triangle plots, since they require the use of a best-fit cosmological model, in this case the \planckalllensing\ best-fit model. In the figure, the cross-spectra residuals for the 2018 data can be compared to the 2015 residuals (using the same model). They can also be compared to the coadded CMB spectrum obtained with all the cross-spectra (including the one considered in each panel) and with the conditional prediction obtained from all of the other spectra that enter the likelihood. This conditional prediction is obtained by noting that the likelihood data and model vectors as well as the covariance can be decomposed into blocks. The data vector is decomposed into 
\begin{equation}
\vec{\hat C}=\left [ \vec{\hat C}^{\nu\nu'}, \vec{\hat C}^{\rm others}\right ]^{\sf T},
\end{equation}
with $\vec{\hat C}^{\nu\nu'}$ the part of the data vector corresponding to the cross-spectra currently considered. The theory vector is decomposed in a similar fashion, $\vec{C}=\left [ \vec{C}^{\nu\nu'}, \vec{C}^{\mathrm{others}}\right ]^{\sf T}$, and correspondingly the covariance is decomposed into
\begin{equation}
\tens{\Sigma} = \begin{pmatrix} 
\tens{\Sigma}^{\nu\nu'\times\nu\nu'} & \tens{\Sigma}^{\nu\nu'\times{\rm others}} \\
\tens{\Sigma}^{{\rm others}\times\nu\nu'} & \tens{\Sigma}^{{\rm others}\times{\rm others}} 
\end{pmatrix}.
\end{equation}
With this decomposition, and assuming a Gaussian shape for the likelihood, the conditional prediction is
\begin{equation}
\vec{C}^{\nu\nu',{\rm cond}} = \vec{C}^{\nu\nu'} + \tens{\Sigma}^{\nu\nu'\times{\rm others}}\left (\tens{\Sigma}^{{\rm others}\times{\rm others}}\right)^{-1} \left (\vec{\hat C^{\rm others}} - \vec{C^{\rm others}} \right), 
\label{eq:condcoadd}
\end{equation}
with a covariance of 
\begin{equation}
\tens{\Sigma}^{\nu\nu'\times\nu\nu',{\rm cond}} = \tens{\Sigma}^{\nu\nu'\times\nu\nu'} - \tens{\Sigma}^{\nu\nu'\times{\rm others}}\left (\tens{\Sigma}^{{\rm others}\times{\rm others}}\right)^{-1} \tens{\Sigma}^{{\rm others}\times\nu\nu'}.
\label{eq:condcov}
\end{equation}
{T}he computation of the conditional prediction requires a cosmological best-fit model to form the $\vec{C}^{\nu\nu'} $ and $\vec{C}^{\rm others}$ vectors.

\begin{figure}[htbp!]
\centering
\includegraphics[width=\columnwidth]{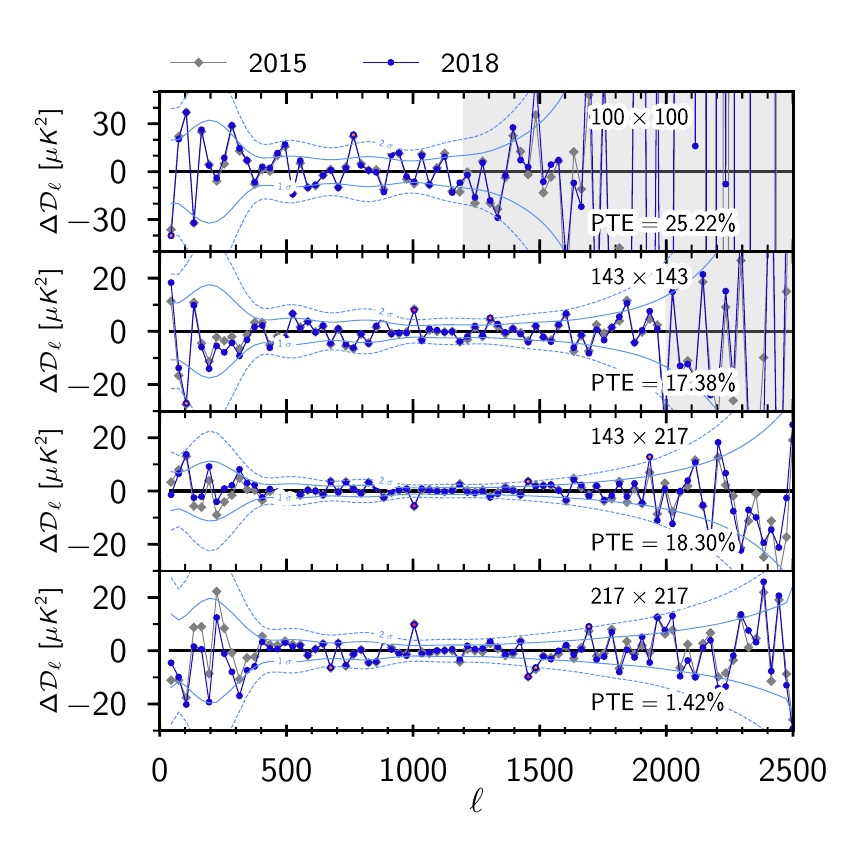}
\caption{Conditional residuals for each cross-spectrum entering into the TT high-$\ell$ likelihood. The blue points show the difference between the data of the displayed cross-spectrum and the conditional prediction for it based on the other cross-spectra (within the \LCDM\  baseline model) for the 2018 release, while the grey ones correspond to the 2015 analysis (with its own best-fit model). The conditional predictions are computed taking into account the multipole cuts described in Table~\ref{tab:hi-ell:data:lrange}, i.e., excluding the shaded regions in each panel. The blue lines shows the $\pm1\,\sigma$ and $\pm2\,\sigma$ contours corresponding to the diagonal of the block of the conditional covariance based on the 2018 data and covariance matrix corresponding to the cross-spectrum shown. Orange and red points correspond to $2$ and $3\,\sigma$ outliers. PTEs are computed for the non-shaded regions, using the $\Delta\ell\,{=}\,30$ binning scale as used in the plots.}
\label{fig:hi_ell:valid:conditionalTT}
\end{figure}

To better assess the agreement between the conditional prediction and the data, the difference between the two is also displayed in the conditional residual of Fig.~\ref{fig:hi_ell:valid:conditionalTT} for $TT$. The errors displayed in the figure correspond to the diagonal of the covariance obtained from Eq.~\eqref{eq:condcov}, and are thus much tighter than the ones displayed in the residual plot (Fig.~\ref{fig:hi_ell:valid:residualTT}), which correspond to the diagonal of $\Sigma^{\nu\nu'\times\nu\nu'}$.

We recover (in the residual plot of Fig.~\ref{fig:hi_ell:valid:residualTT}) the excellent agreement between the data and the cosmological model for each cross-spectrum that was already discussed in \citetalias{planck2014-a13}. Probabilities to exceed are perhaps a little large, especially in the case of the $100\times100$ cross-spectrum. In general, the scatter at large scales, where the Galactic contamination is important, seems to be small compared to the expected dispersion. That could be a sign of a slight overestimation of the variance of the dust contamination.
The residual outliers are identical to those found in 2015 and are usually common to all frequencies. We also confirm the $3\,\sigma$ outlier at $\ell\,{\approx}\,1460$, which is much more pronounced at high frequencies. We discussed in section~3.8 of \citetalias{planck2014-a13} how this outlier seemed to be partially of Galactic origin.  As can be seen in figure~A.2 of \citetalias{planck2016-l06}, the power in this bin moves up by about $1\,\sigma$ in the map-level dust-cleaned version of the \camspec\ likelihood.

The conditional residuals shown in Fig.~\ref{fig:hi_ell:valid:conditionalTT} provide a deeper assessment of the agreement between frequencies.  We recover here the fact that the $\ell\,{\approx}\,1460$ excursion is more pronounced in $143\times217$ and particularly in $217\times217$. We remark that $143\times143$, $143\times217$, and $217\times217$ all have a strong outlier compared to the conditional prediction at $\ell\,{=}\,1000$. 
In the residuals of Fig.~\ref{fig:hi_ell:valid:residualTT}, it seems that this disagreement in the conditional originates from a small outlier in $217\times217$ at this multipole, which is not particularly worrying. 

\begin{figure}[htbp!]
\centering
\includegraphics[width=\columnwidth]{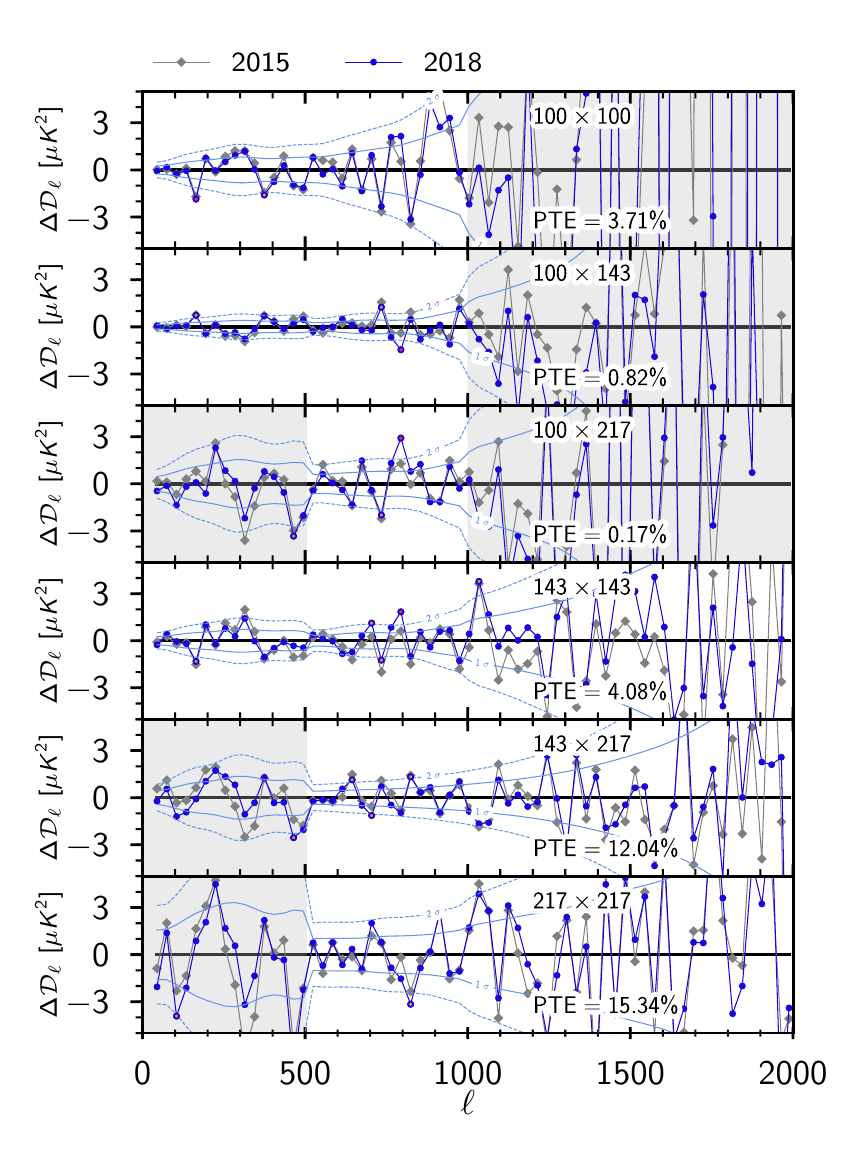}
\caption{Same as for Fig.~\ref{fig:hi_ell:valid:conditionalTT} but now for TE instead of TT.}
\label{fig:hi_ell:valid:conditionalTE}
\end{figure}

\begin{figure}[htbp!]
\centering
\includegraphics[width=\columnwidth]{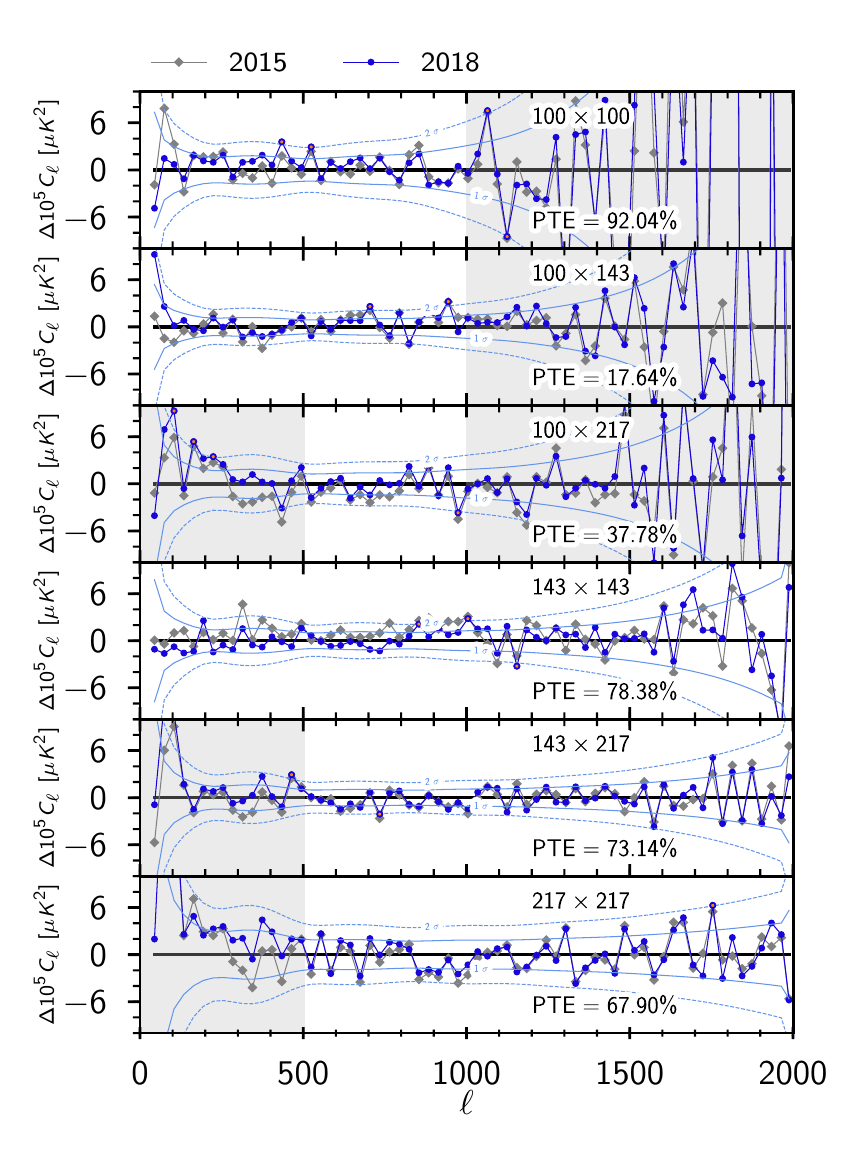}
\caption{Same as for Fig.~\ref{fig:hi_ell:valid:conditionalTT} but now for EE instead of TT.}
\label{fig:hi_ell:valid:conditionalEE}
\end{figure}

The $TE$ cross-spectrum residuals are presented in Fig.~\ref{fig:hi_ell:valid:residualTE}. Each individual cross-spectrum is in excellent agreement with the cosmological model, as shown by the PTE values displayed in the different panels. As can be seen in the figure, with the beam-leakage correction being smooth, the outliers of the $TE$ residuals are similar to the 2015 ones. However, their significance is in most cases reduced. This is particularly striking for the outliers present in 2015 in the $100\times100$ and $100\times217$ cross-spectra around $\ell\,{=}\,750$ and between $\ell\,{=}\,850$ and $\ell\,{=}\,1000$ for $100\times100$. Their amplitudes have been significantly reduced when comparing to 2015, but the feature at $\ell\,{\approx}\,750$ remains present at the $2\,\sigma$\ level. This feature was absent in 2015 in the $143\times143$ spectrum. After leakage correction, the feature now appears in all the cross-spectra at a similar level, and thus seems to be in the sky. {We note} that the agreement between cross-spectra in this multipole range after leakage correction is still not perfect, since the conditional residual plot, Fig.~\ref{fig:hi_ell:valid:conditionalTE}, shows a $2\,\sigma$-level difference between $143\times143$ and the conditional prediction for it from the other cross-spectra. As can be seen in Fig.~\ref{fig:hi_ell:data:leakages}, the multipole region around $\ell\,{=}\,750$ is the peak of the beam-leakage correction model and the residual disagreement could be a sign of limitations of our correction template. {We note}, however, that the amplitude of the conditional residual outliers around $\ell\,{=}\,750$ are of the order of $2\muK^2$, similar to the amplitude of the leakage correction for the $100\times143$ or $143\times143$\ cross-spectra {while the error budget of the leakage correction is below $0.1\muK^2$, as we discussed in Sect.~\ref{sec:hi-ell:datamodel:beamleak}}. It would thus require a large modification to the template for it to fully explain the observed outlier, and the extra variance induced by the leakage correction that we ignore in the likelihood implementation can at most modify the errors by a fraction of a percent. {We also noted in Sect.~\ref{sec:hi-ell:datamodel:beamleak} that a parametric correction to the leakage template, similar to the 2015 leakage model, does not significantly improve the agreement between cross-spectra.}
 
 A $2\,\sigma$ outlier, also noted by \cite{2017PhRvD..96h3526O}, remains at $\ell=165$ at the first trough of the $TE$ spectrum. It seems to be particularly driven by the $100\times100$ spectrum, as can also be seen in the conditional residual plot, where $100\times100$ disagrees with the other cross-spectra. 

In general the conditional residual $TE$ plots shown Fig.~\ref{fig:hi_ell:valid:conditionalTE} exhibit an excess of 2 and $3\,\sigma$ outliers, resulting in relatively low PTEs. The worst case is $100\times217$, where the two outliers around $\ell\,{=}\,750$ drive the PTE of the conditional residual to the low value of $0.0017$; removing the two outlying bins brings the PTE up to $0.23$. The situation is similar for the second worst case, $100\times143$, where the removal of the same two bins improves the PTE to a similar value. The relatively bad agreement between the cross-spectra is in line with the poor PTE of the full-frequency TE likelihood, and an excess of scatter between cross-spectra. Visually, the outliers of the conditional residuals seem to be particularly located around the first and third troughs of the $TE$ spectrum (i.e., around $\ell\,{=}\,165$ and $\ell\,{=}\,750$).  Removing the multipole regions $120\,{<}\,\ell\,{<}\,200$ and $700\,{<}\,\ell\,{<}\,800$ (for all frequencies in the full-frequency binned likelihood)  improves the PTE of the TE likelihood to $0.085$. {Removing the $100\times217$ or the $100\times143$ spectra that exhibit the worst outliers (with opposite signs between the two cross-spectra) brings up the PTE of conditional residual predictions (not shown), with the lowest PTEs being for the $100\times143$ or $100\times217$ spectra, at around $0.05$}. {Those results are in line with the general conclusion of the discussion of Sect.~\ref{sec:valandro:gof}. There is no particular cross-spectrum to blame for the overall low PTE. Each of them, as well as the coadded $TE$ spectra, are in excellent agreement with the model. However, the numerical limitation of our modelling of the statistical properties of the six different and highly correlated cross-spectra starts to show in our PTE-based tests.}

Finally, there does not seem to be any notable outliers in the $EE$ inter-frequency comparisons, in either the residual plot of Fig.~\ref{fig:hi_ell:valid:residualEE} or the conditional plot of Fig.~\ref{fig:hi_ell:valid:conditionalEE}. As can be seen in the low-multipole part of the residual plot, an upward trend is visible, compatible with some residual dust contamination (albeit, below the $1\,\sigma$ level). It is clear from Figs.~\ref{fig:hi_ell:valid:residualEE} and ~\ref{fig:hi_ell:valid:conditionalTE} that the trend is dominated by residuals in the $100\times143$ spectrum. As was discussed in Sect.~\ref{sec:hi-ell:datamodel:gal}, whenever jointly fitting the amplitude of dust residuals in $EE$ with the cosmological parameters, we indeed see a preference for a higher level
of contamination than what our cross-correlations with the 353-GHz dust polarized tracer map would lead us to expect. We discussed in the same section that opening the dust amplitude in $EE$
at $100\times143$ causes a $0.9\,\sigma$ upward shift in $\ns$ (see also Fig.~\ref{fig:hil:2dEEdustprior}). As we discussed in Sect.~\ref{sec:hi-ell:datamodel:noise}, while we do find hints of (and correct for) correlated noise 
residuals in the $100\times100$, $143\times143$, and $217\times217$ spectra, we do not find any sign of a residual systematic issues at large scales in the FFP10 simulations that could explain the behaviour of $100\times143$.

\begin{figure}[htbp!]
\centering
\includegraphics[width=\columnwidth]{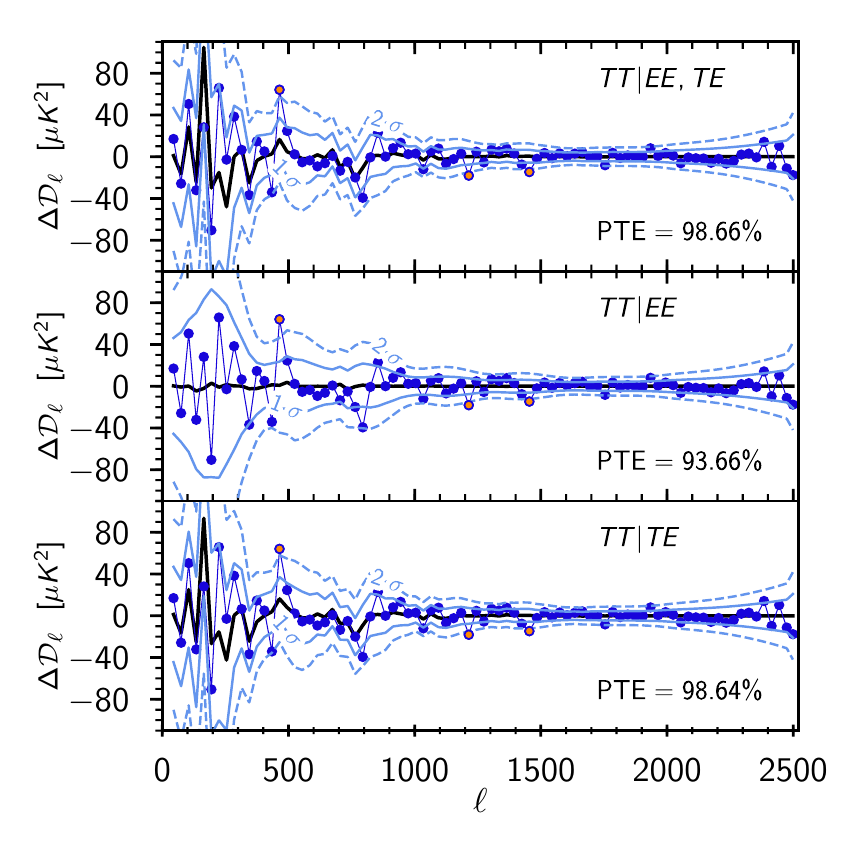}
\caption{Conditional residuals for the co-added CMB spectrum in the TT likelihood. The blue points show the difference between the co-added $TT$ spectrum and the 2018 \LCDM\ baseline results, while the black lines show the differences between conditional predictions of the spectrum and the baseline model. Predictions are made either conditional on both the $TE$ and $EE$ data together (top), the $EE$ data only (middle), or the $TE$ data only (bottom). The blue lines show the $\pm1\,\sigma$ and $\pm2\,\sigma$ contours (around the black line) corresponding to the diagonals of the blocks of the conditional covariances computed from the 2018 covariance matrix and data. Orange and red points correspond to $2$ and $3\,\sigma$ outliers. For each panel the PTE is computed for the difference between the data and its conditional prediction using the conditional covariance.}
\label{fig:hi_ell:valid:conditional_coaddTT}
\end{figure}

\begin{figure}[htbp!]
\centering
\includegraphics[width=\columnwidth]{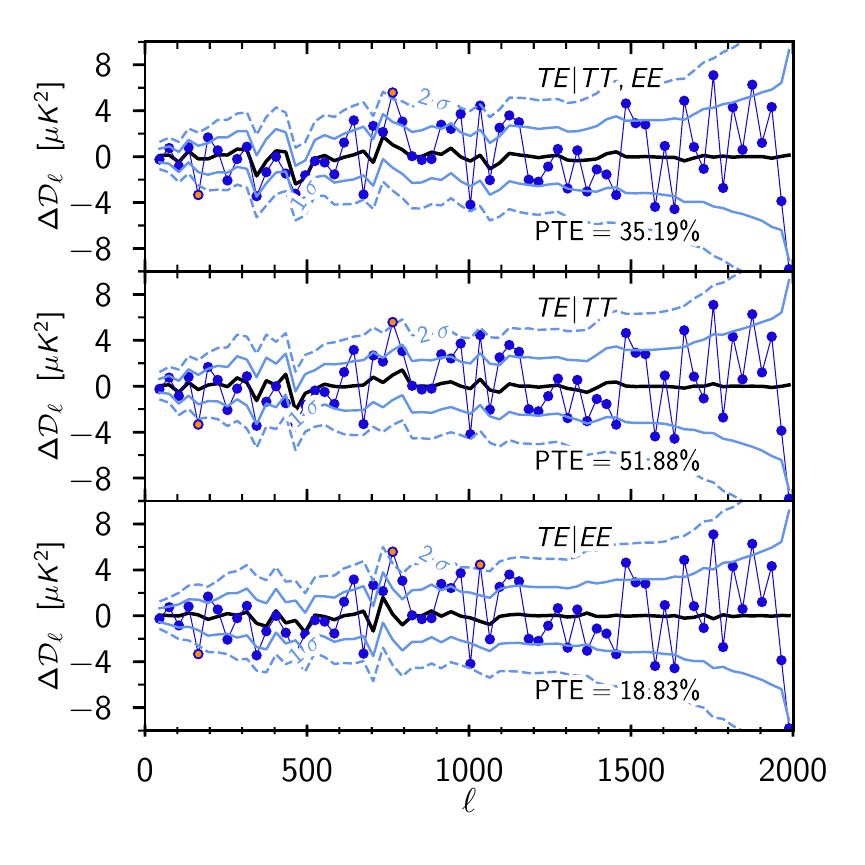}
\caption{Same as for Fig.~\ref{fig:hi_ell:valid:conditional_coaddTT} but now for TE (conditioned on TT and/or EE) instead of TT.}
\label{fig:hi_ell:valid:conditional_coaddTE}
\end{figure}

\subsubsection{Temperature and polarization conditional predictions}
\label{sec:hi_ell:valid:condcoadd}
\begin{figure}[!t!]
\centering
\includegraphics[width=\columnwidth]{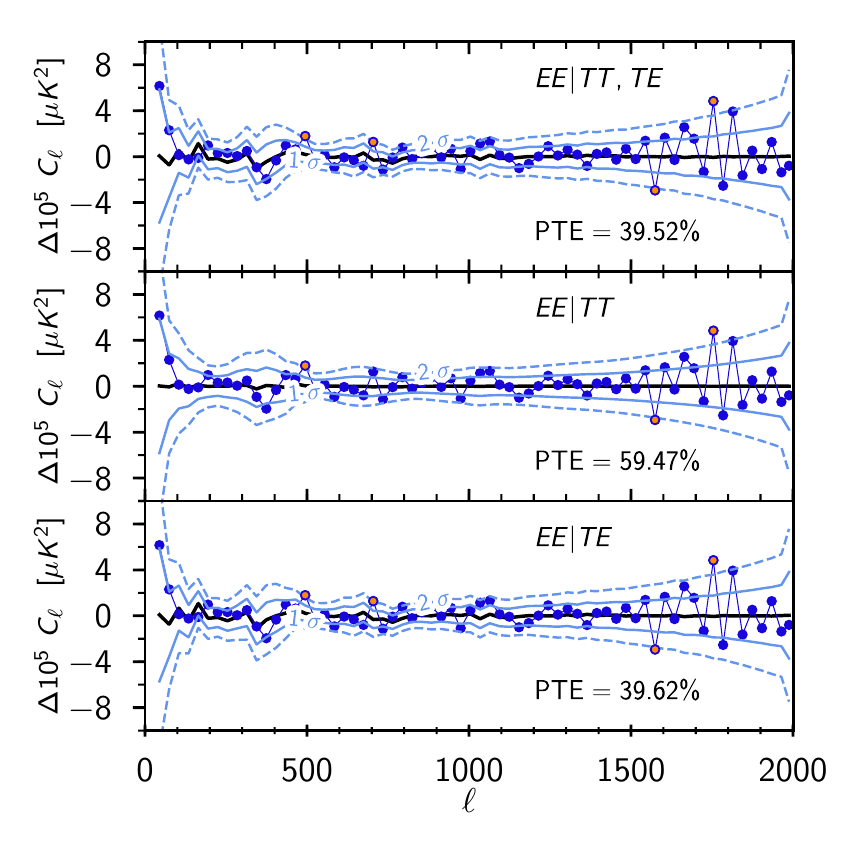}
\caption{Same as for Fig.~\ref{fig:hi_ell:valid:conditional_coaddTT} but now for EE (conditioned on TT and/or TE) instead of TT.}
\label{fig:hi_ell:valid:conditional_coaddEE}
\end{figure}

The coadded conditional plots for $TT$, $TE$, and $EE$ (Figs.~\ref{fig:hi_ell:valid:conditional_coaddTT}, \ref{fig:hi_ell:valid:conditional_coaddTE} and \ref{fig:hi_ell:valid:conditional_coaddEE}) allow us to assess the overall agreement between the different CMB power spectra, assuming a best-fit \LCDM\ model. The conditional predictions were obtained first by computing the coadded $TT$, $TE$, and $EE$ spectra and joint covariance, as described earlier in Sect.~\ref{sec:valandro:gof}, and then using Eq.~\eqref{eq:condcoadd}.
The figures show good compatibility between the different spectra and indicate the relative correlation between $TT$, $TE$, and $EE$. This is particularly striking in the case of $TT$,
where $TE$ predicts some of the scatter of the temperature spectrum. It is also true for $TE$, where $TT$ and $EE$ contribute to explain some of its outliers. The PTEs for the $TT$ predictions actually seems too good. 
The low-multipole range in particular appears to have too low a scatter compared to that predicted from $EE$, for example. It is also true (albeit more difficult to assess from the figure) for $TE$. This could be a sign of the limitations of our model of the statistical properties of the dust, and in particular the correlations between the temperature and polarization contributions.

\subsubsection{Consistency of cosmological parameters from TT, TE, and EE}
\label{sec:valandro:parcompTTTEEE}

We {conclude} this set of discussions of the consistency between our data and the model by exploring the consistency between the best-fit cosmological parameters obtained from each of the TT, TE, and EE likelihoods.
Similarly to what was done in \citetalias{planck2014-a13}, we simulate 100 sets of $\plik\TE{}$ or $\plik\EE{}$ frequency power spectra from the power spectrum covariance matrix conditioned on the $\plik\TT{}$ power spectrum. As a fiducial model, we use the best-fit solution of the $\Lambda$CDM \TT{}+\simall\ data combination. For all the other polarization-related nuisance parameters varied in the runs (i.e., the \TE{}\ Galactic dust amplitudes) we use the mean values of the priors we set on them as fiducial values. From each of these simulations we estimate the best-fit cosmological parameters using the same assumptions as those adopted for the real data, and then evaluate their means and standard deviations. 
For all runs (simulations and data) we run the \plikEE\ or \plikTE\ likelihoods in combination with a Gaussian prior on $\tau$ centred on the fiducial cosmology and with an uncertainty similar to the one obtained from \simall, {that is}, $\tau=0.052\pm0.009$. 
This is because \plikEE\ and \plikTE\ are mostly sensitive to the combination $A_{\rm s} e^{-2 \tau}$ and can barely disentangle between the $\As$ and $\tau$ values. Thus, the $\tau$ and $\As$ values recovered from the simulations almost completely depend on the prior set on $\tau$. Therefore, to avoid an artificial bias, we centre the prior for $\tau$ on the fiducial value used to produce the simulations, and then for this test we treat the data (to which we compare the simulations) with the same prior.
{We note} that since we only simulate here the \plik\ likelihoods and not \simall, our simulations do not capture the additional scatter expected due to uncertainties on $\tau$.  Hence, since \plikEE\ and \plikTE\ are mostly sensitive to the combination $A_{\rm s} \exp{-2 \tau}$, we expect the spread of the best-fit values from the simulations of $\tau$ and $\As$ to be much smaller than the standard deviations of the corresponding posterior distributions obtained from the MCMC runs on the data \citep[further details can be found in][]{planck2016-LI}.

 We then evaluate for the data and for the simulations the deviation statistic ${\cal P}$, which quantifies how much  {\it all\/} the parameters differ from their means, taking into account the correlations among them. It is defined as
\begin{equation}
{\cal P} = \big(\vec{P} - \left<\vec{P}\right>\big)^{\sf T} \tens{P}^{-1} \big(\vec{P} - \left<\vec{P}\right>\big) ,
\label{app:DEV}
\end{equation}
Here, $\vec{P}$ is the vector of the best-fit parameters varied in the run, $\tens{P}$ is the covariance matrix of the parameters evaluated from the simulations, and $\langle\vec{P}\rangle$ is the mean of the best-fit parameters over the 100 simulations. We evaluate ${\cal P}$ both for the six \LCDM\ parameters alone and for all of the parameters varied in the runs, {in other words}, the \LCDM\ parameters plus the calibration parameter $y_{\rm P}$ for \EE{}\ (seven parameters in total), and additionally the dust amplitudes and the inter-frequency intensity calibration parameters $\calibC^{TT}_{100}$ and $\calibC^{TT}_{217}$ for \TE{}\ (15 parameters in total).

For $\EE{}$, we find there are 32 simulations (or 23 when considering cosmological parameters only) with a deviation ${\cal P}$ larger than that found for the data, suggesting that the shifts in the best-fit parameters between \EE\ and \TT\ are in good statistical agreement. This confirms the results we found in \citetalias{planck2014-a13}.
Similarly, for $\TE{}$ there are 10 simulations (or 23 when considering cosmological parameters only) with a deviation ${\cal P}$ larger than that for the data, again pointing towards good statistical agreement between the TE and TT cosmologies. Note that this is an improvement with respect to what was found in \citetalias{planck2014-a13}, where the TE cosmology was too close to the TT one. This change is mainly due to the beam-leakage correction in TE, described in Sect.~\ref{sec:hi-ell:datamodel:beamleak}, which pushes the TE best-fit solution slightly away from the TT one. However, as already discussed in Sect.~\ref{sec:valandro:gof}, the best-fit $\chi^2$ for TE using the data is much higher than what might be expected given the simulations. 

\begin{figure*}[htbp!] 
\centering
\includegraphics[width=\columnwidth]{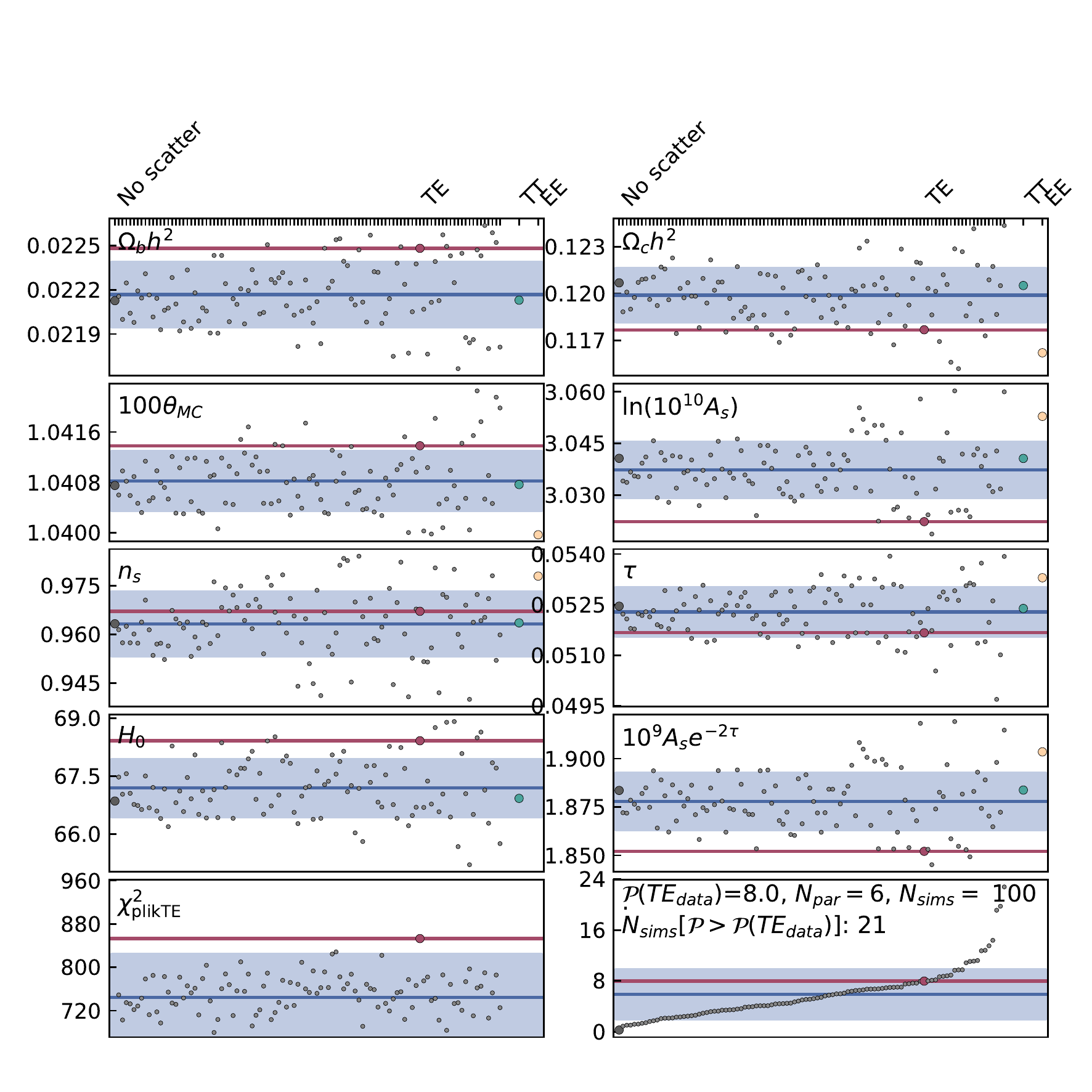}
\includegraphics[width=\columnwidth]{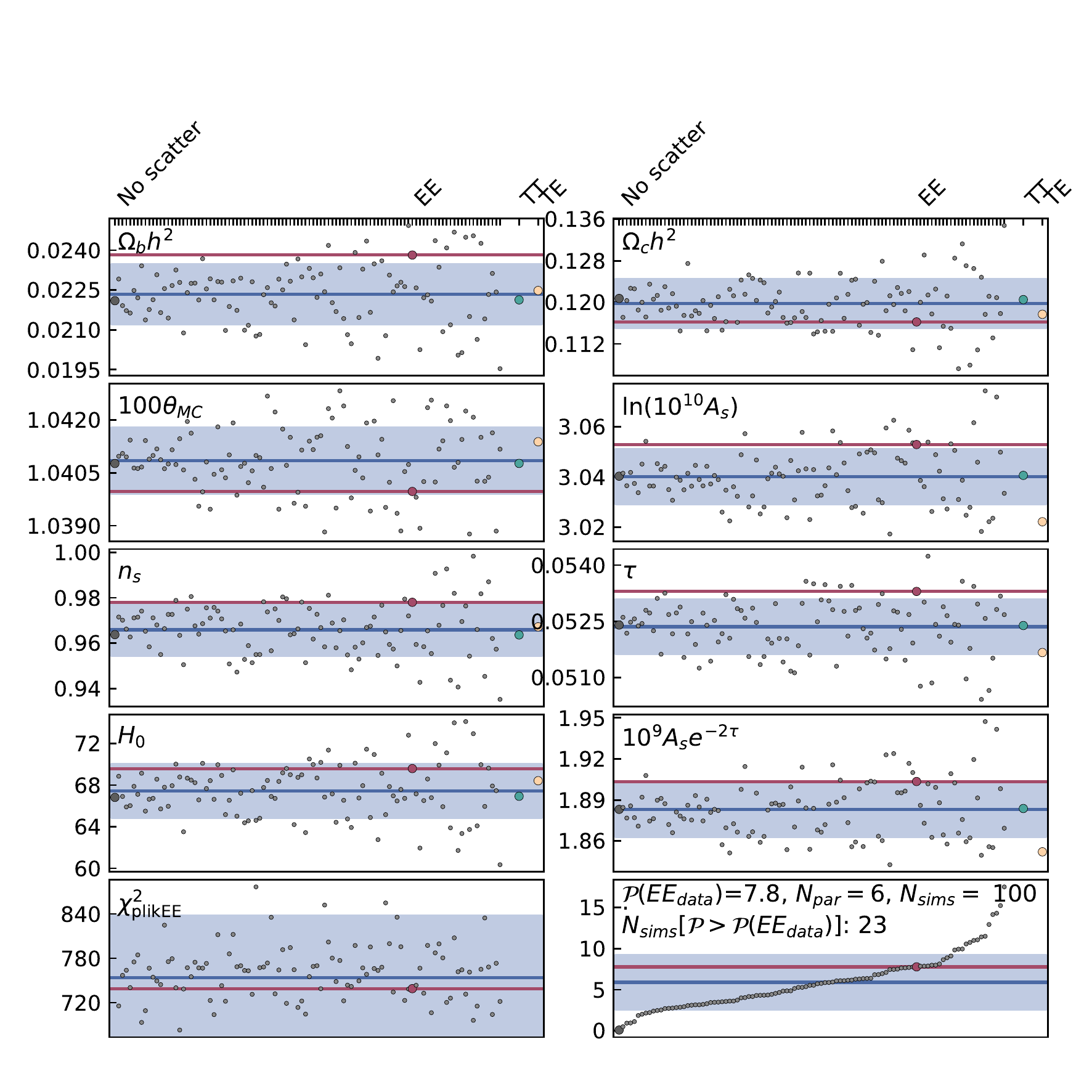}
\caption{Best-fit cosmological parameters from 100 simulations (grey circles) of the \plikTE\ (left) and \plikEE\ (right) power-spectra, conditioned on the \plikTT\ power spectrum, assuming as a fiducial cosmology the best-fit of the $\Lambda$CDM {\TT{}+\simall} results (green circle). The blue lines show the means of the simulations, while the blue bands show the standard deviations of the simulations, centred on the means. The points are ordered from lowest to highest values of $\cal P$ parameter, defined in Eq.~\eqref{app:DEV}, and shown in the bottom-right panels. The best fits from the real data combinations \plikEE+$\tau$ and \plikTE+$\tau$ are shown with red circles and lines and yellow circles, respectively. Both data and simulations are run in combination with a Gaussian prior on $\tau$ centred on the fiducial value from the TT best-fit and with an uncertainty comparable to the one provided when using \simall, i.e., with $\tau=0.052\pm0.009$. The best fits for the fiducial $EE$ or $TE$ power spectra, without added scatter, are shown with dark grey circles. All of the results shown here were obtained using the \CAMB\ code \citep{Lewis:1999bs} and are maximum-likelihood best fits. These plots demonstrate that the best-fit cosmology from either TE or EE is consistent with that obtained from TT.}
\label{fig:hi_ell:valid:consistencypars}
\end{figure*}

Figure~\ref{fig:hi_ell:valid:consistencypars} shows the best-fit cosmological parameters obtained from the simulations (grey points), together with their means (blue lines). The light-blue bands show the standard deviations of the simulations centred on their means. The points shown in the figure are ordered from left to right by their ${\cal P}$ parameter values.

As expected, the mean parameters of the simulations, both for EE and TE, are very close to those of the fiducial cosmology obtained using the TT likelihood.
Also, the standard deviations from the simulations are close to those obtained from the posterior distributions of the MCMC chains (except for $\tau$ and $\lnAs$ for the reason discussed above). 

Overall, we conclude that consistency between TT, TE, and EE, already found at the power spectrum level in Sect.~\ref{sec:valandro:gof}, is confirmed at the cosmological parameters level.

\subsection{Impact of polarization-efficiency corrections}
\label{sec:valandro:PE}
\begin{figure}[htp!]
\centering
\includegraphics[width=\columnwidth]{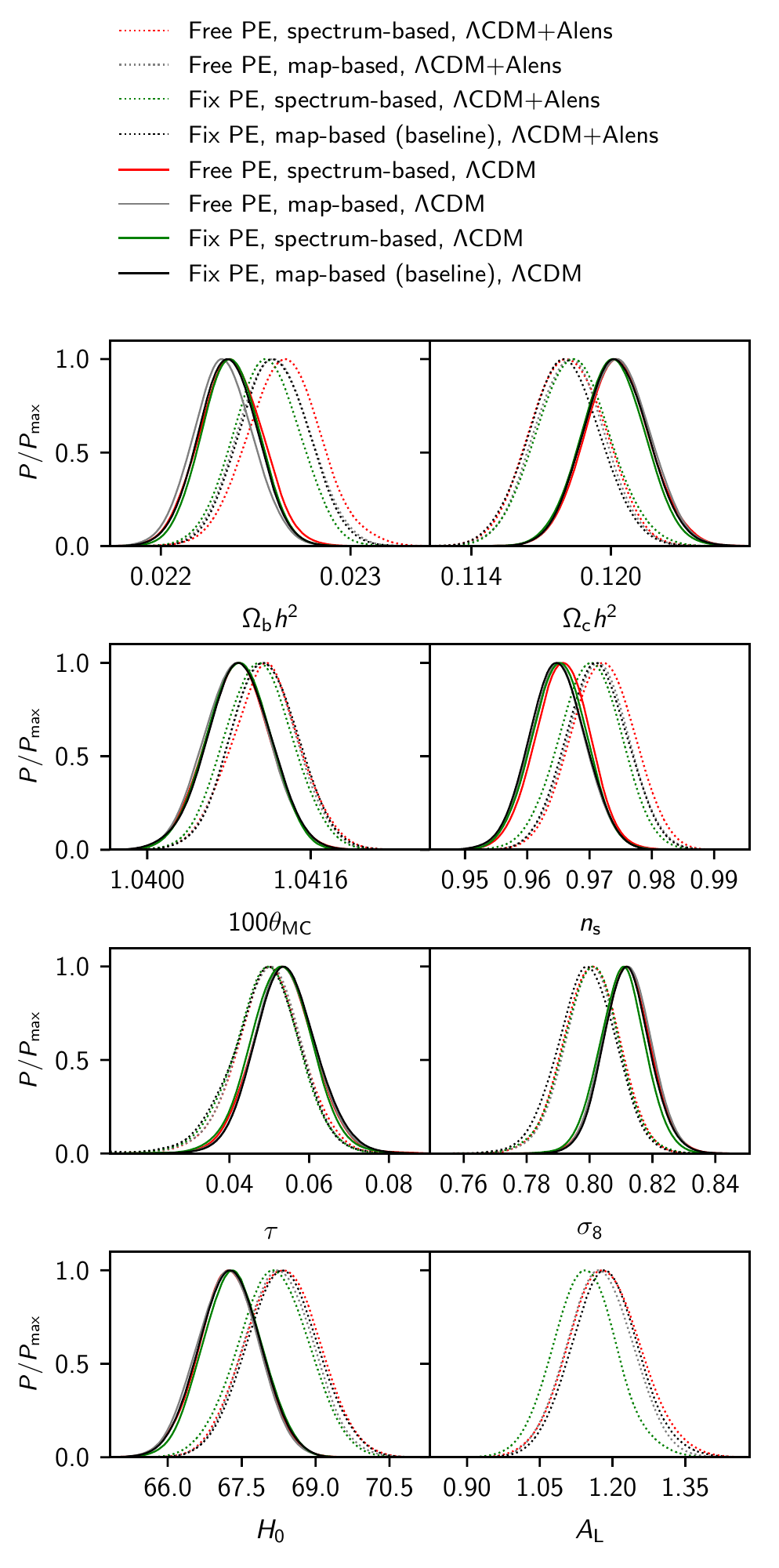}
\caption{{Impact of different choices for the PE corrections on 1-dimensional posterior distributions of cosmological parameters from \planckall. The red lines show results when leaving the PE parameters free to vary independently for EE and TE, while the grey lines assume coupled parameters for the EE and TE spectra. The green lines show results obtained when fixing the polarization efficiencies in the spectrum-based approach to the values given by Eq.~\eqref{eq:relcalTE}. Finally the black lines show results obtained when fixing the polar efficiencies in the map-based, baseline case (Eq.~\ref{eq:relcalEE}). The solid lines show results for the \LCDM\ case, while the dotted ones show results for the \LCDM+$\Alens$ model. This plot demonstrates that the cosmological parameters in the \LCDM\ case are robust to changes in the PE-correction model.} 
}
\label{fig:hi_ell:valid:PEcosmo}
\end{figure}

\begin{figure}[htp!]
\centering
\includegraphics[width=\columnwidth]{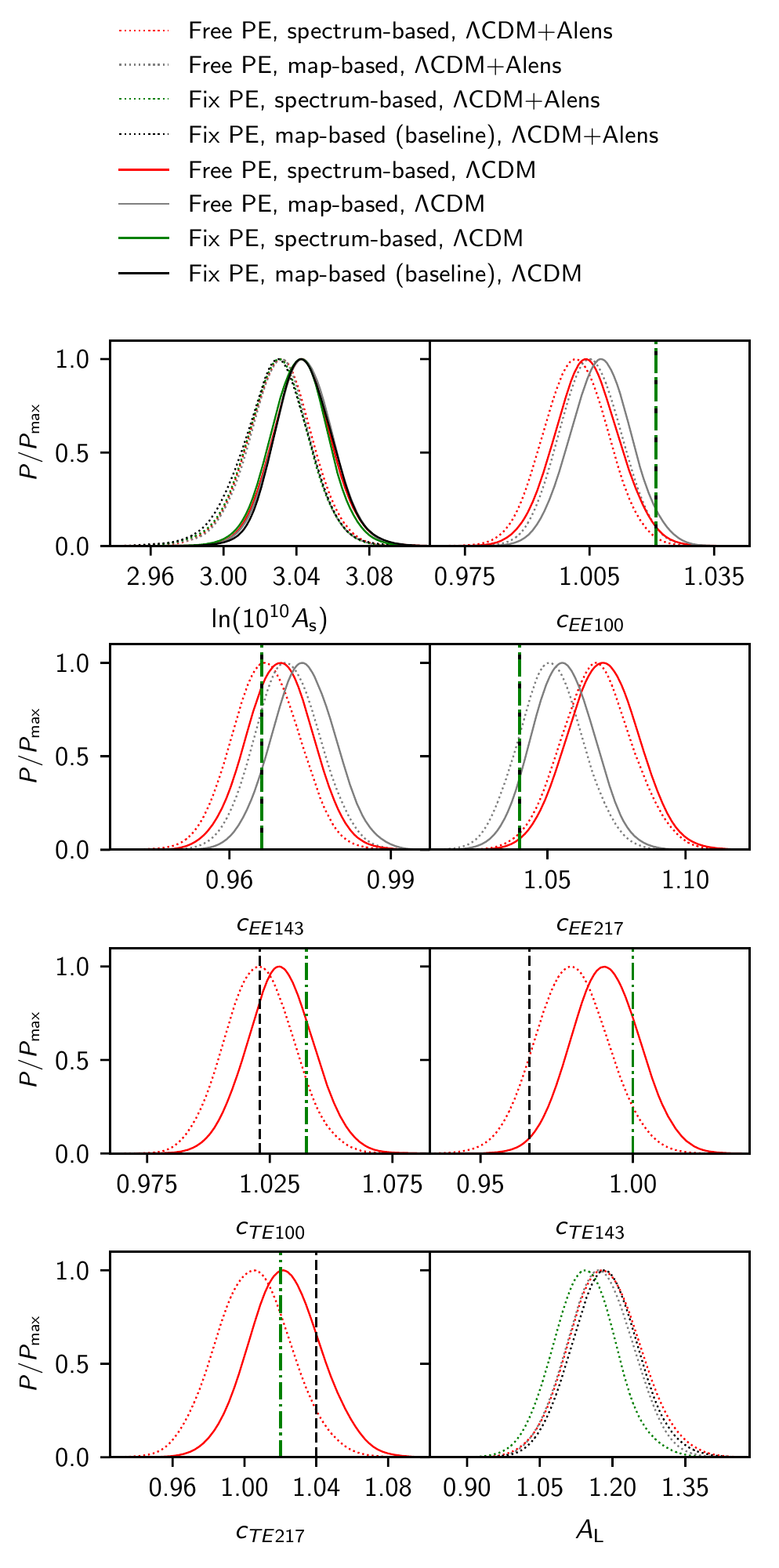}
\caption{Impact of different choices for the PE corrections on 1-dimensional posterior distributions of cosmological parameters and PE amplitudes from \planckall. The red lines show results when leaving the PE parameters free to vary independently for EE and TE, while the grey lines assume coupled parameters for the EE and TE spectra. The green lines show results obtained when fixing the polarization efficiencies in the spectrum-based approach to the values indicated by the vertical green dotted-dashed lines. Finally the black lines show results obtained when fixing the polar efficiencies in the map-based, baseline case, to the values indicated by the vertical black dashed lines. The solid lines show results for the \LCDM\ case, while the dotted ones show results for the \LCDM+$\Alens$ model. This plot shows that the recovered values for PE change according to the cosmological model. 
}
\label{fig:hi_ell:valid:PE}
\end{figure}

We tested the impact of different effective PE corrections $\calibC^{PP}_{\nu}$ on cosmological results from \plik\TT{}TEEE+\commander+\simall, as described in Sect.~\ref{sec:hi-ell:datamodel:inst}. The baseline map-based approach uses the same corrections, given in Eq.~\eqref{eq:relcalEE}, for both the $EE$ and $TE$ power spectra and applied using Eq.~\eqref{eq:caldef}. These corrections are inferred by comparing the $EE$ frequency power spectra, in the multipole range $200\,{\leq}\,\ell\,{\leq}\,1000$, to the best-fit \LCDM\ model obtained from \plik\TT{}+\simall\ .

We show in this section the impact of using instead a spectrum-based approach, which applies to $EE$ the same corrections as in the baseline, but applies to $TE$ the corrections reported in Eq.~\eqref{eq:relcalTE}. The latter are calculated by comparing the $TE$ frequency power-spectra, in the multipole range $200\,{\leq}\,\ell\,{\leq}\,1000$, to the best-fit \LCDM\ model obtained from \plik\TT{}+\simall\ .

We find that using the spectrum-based approach does not introduce shifts in the \LCDM\ parameters compared to the baseline map-based one  {(Fig.~\ref{fig:hi_ell:valid:PEcosmo})}. However, it impacts some extensions of \LCDM, notably the best-fit value of the lensing-consistency parameter $\Alens$, lowering it by $0.6\,\sigma$ with respect to the baseline case.
This is shown in Fig.~\ref{fig:hi_ell:valid:PE}, which presents both the \lcdm\ and the $\Alens$ case and only displays the parameters affected by the PE correction parameters. We checked that other extensions are less affected, e.g., $\nnu$ changes by approximately $0.3\,\sigma$, and the upper limit on $\mnu$ relaxes by 7\,\%.

We investigate this further by letting the PE parameters vary freely using a spectrum-based approach (i.e., with independent efficiencies in EE and TE).
This is shown in Fig.~\ref{fig:hi_ell:valid:PE}.

\begin{figure*}[htbp!]
\centering
\includegraphics[width=0.7\textwidth]{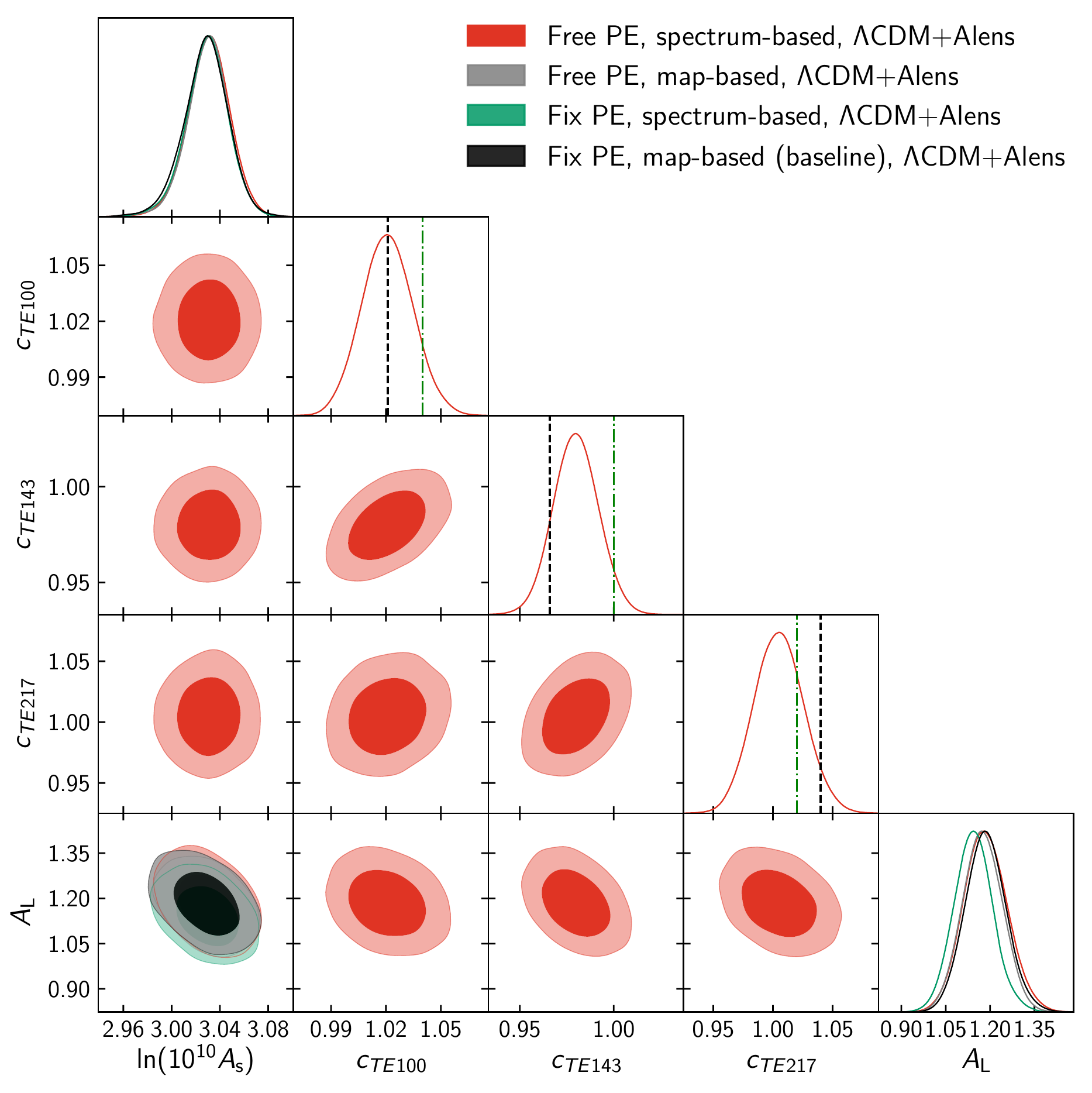}
\caption{Degeneracy between the $\Alens$ parameter and the PE corrections. We show results for \planckall, leaving PE corrections free to vary independently for EE and TE (red contours). 
For the cosmological parameters shown, i.e., $\lnAs$ and $\Alens$, we also plot the cases already described in Fig.~\ref{fig:hi_ell:valid:PE}, with the same colours as used there. }
\label{fig:hi_ell:valid:PE2d}
\end{figure*}

First, we find that the recovered values of $\calibC^{PP}$ for EE and $\calibC^{TE}$ for TE are somewhat different from those calculated in Sect.~\ref{sec:hi-ell:datamodel:inst}, typically by less than $2\,\sigma$. This is not surprising, since in that section the efficiencies were calculated in a restricted multipole range where foreground, noise modelling, and cosmological model uncertainties are reduced. We find $\calibC^{PP}_{100}=1.0043\pm0.0077$, $\calibC^{PP}_{143}=0.9693\pm0.0062$, and $\calibC^{PP}_{217}=1.070\pm0.013$ for EE and $\calibC^{TE}_{100}=1.029\pm0.014$, $\calibC^{TE}_{143}=0.991\pm0.011$, and $\calibC^{TE}_{217}=1.023\pm0.020$ for TE. The recovered PE at 217\,GHz in EE, $\calibC^{PP}_{217}$, is the one most discrepant from the values calculated in Eq.~\eqref{eq:relcalEE}, being higher by about $3\,\sigma$. As already discussed in Sect.~\ref{sec:hi-ell:datamodel:inst}, this is due to the fact that including the multipoles larger than $\ell\,{=}\,1000$ at 217\,GHz, as done in the likelihood, pushes $\calibC^{PP}_{217}$ to have values around $1.07$, which might be a symptom of an additional, unmodelled systematic affecting the high $\ell$ data at this frequency. 
We checked that the recovered values of the PEs do not depend on the specific treatment used for the dust amplitudes, by seeing that we obtain the same results when allowing the dust amplitudes to freely vary as we obtain when fixing them in the baseline case.

Second, while we recover similar PEs in the \LCDM\ and \LCDM+$\Alens$ cases in \EE{}, we find a dependence on the cosmological model for \TE{}, contrary to what was found in Sect.~\ref{sec:hi-ell:datamodel:inst}. Once again, this is due to the inclusion of higher multipoles in the likelihood, which induces a degeneracy between $\Alens$ and the PE values (also shown in Fig.~\ref{fig:hi_ell:valid:PE2d}). 
The recovered value of $\calibC^{TE}_{143}$ in particular is closer to the spectrum-based values of Eq.~\eqref{eq:relcalTE} in the \LCDM\ case, but then it shifts closer to the map-based values of Eq.~\eqref{eq:relcalEE} in the \LCDM+$\Alens$ case. This is the reason why, when letting the efficiencies freely vary, we recover an $\Alens$ that is almost identical to the baseline one obtained with the map-based values, whereas fixing the polarization efficiencies to the spectrum-based values decreases $\Alens$, pushing it closer to unity.

Overall, we conclude that the uncertainties in the different modellings of the PE corrections induce systematic uncertainties in some extensions of the \LCDM\ model, which are, however, smaller than about $0.6\,\sigma$.
We choose to quote our baseline results with the map-based approach, but we warn the reader that around $0.5\,\sigma$ shifts can be introduced by using a different modelling of the PE corrections. Given our current knowledge of instrument performance, systematics, and astrophysical foregrounds, this is an effect that cannot be further reduced at present in the \Planck\ data.

\subsection{Impact of priors}
\label{sec:valandro:priors}

\begin{figure}[htbp!]
\includegraphics[width=\columnwidth]{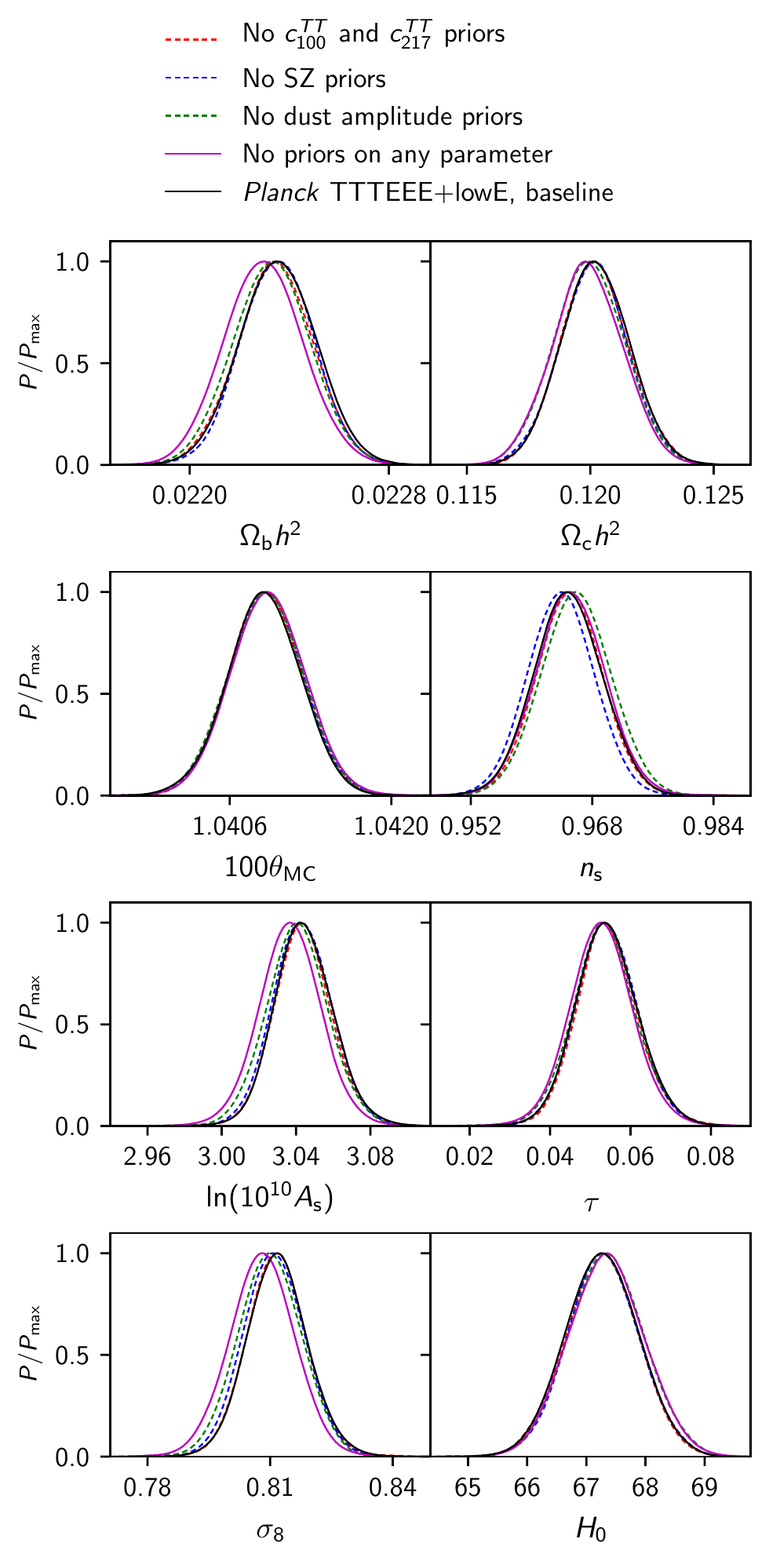}
\caption{Impact of lifting priors on \LCDM\ cosmological parameters. We show results obtained when lifting the TT inter-frequency parameters $\calibC^{TT}_{100}$ and $\calibC^{TT}_{217}$ described in Sect.~\ref{sec:hi-ell:datamodel:inst} (dashed red), when lifting the SZ amplitude priors described in Sect.~\ref{sec:hi-ell:datamodel:fg} (dashed blue), and when lifting the dust amplitude priors in TT and TE, while allowing the dust-amplitude values in EE to vary freely (dashed green). We also show the impact of letting all of the nuisance parameters vary freely (keeping only a prior on the overall calibration $\calibM_{\rm P}$, as given in Eq.~\ref{eq:calplanck}, solid magenta). The baseline \TT{}TEEE+\simall\ case is in solid black. In all these cases, shifts with respect to the baseline are smaller than $0.5\,\sigma$.}
\label{fig:hi_ell:valid:priors}
\end{figure}

\begin{figure}[htbp!]
\includegraphics[width=\columnwidth]{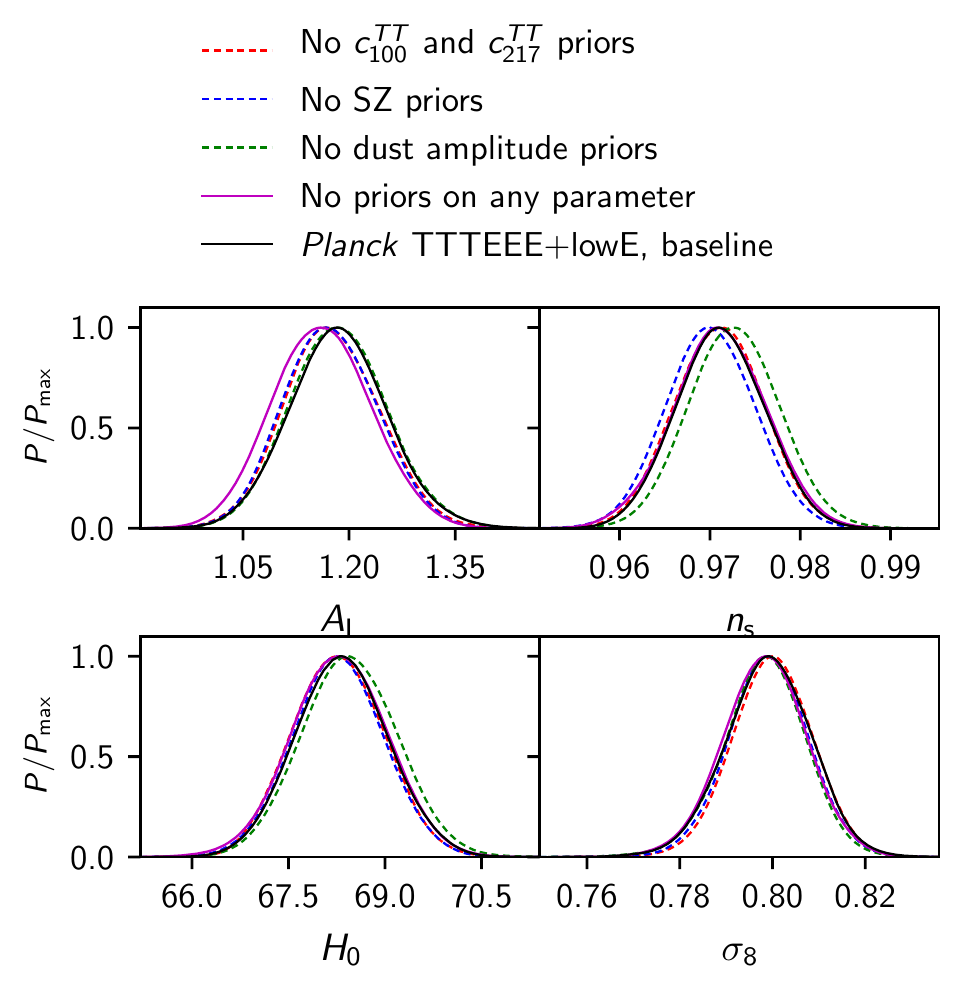}
\includegraphics[width=\columnwidth]{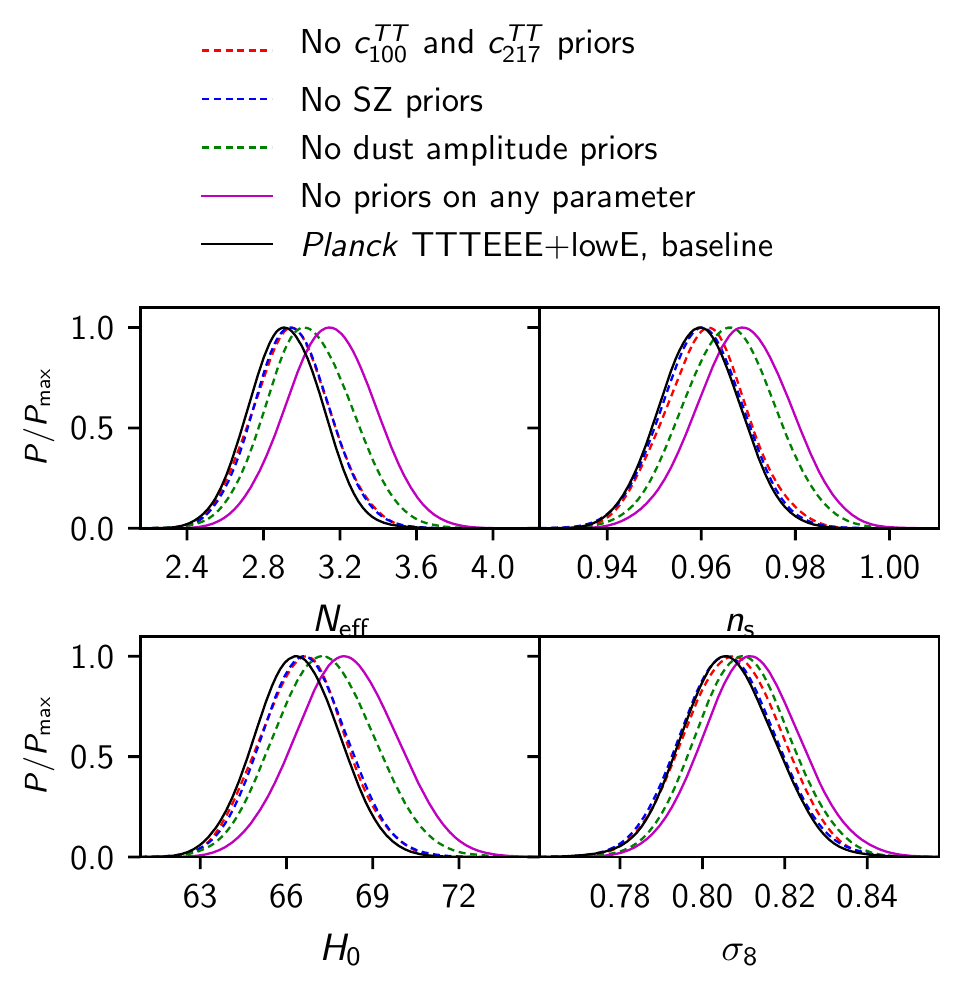}
\caption{Impact of lifting priors on the \LCDM+$\nnu$ and \LCDM+$\Alens$ cosmological parameters (of which we only show a subset). The combinations of priors lifted are as in Fig.~\ref{fig:hi_ell:valid:priors}. The largest shift seen is on $\nnu$, by $1.1\,\sigma$, when lifting all of the priors, due to large degeneracies with $\ns$, dust amplitudes, and inter-frequency calibration parameters. }
\label{fig:hi_ell:valid:priorsAlensNeff}
\end{figure}

We also test the impact of lifting the priors we set on the nuisance parameters.
Figure~\ref{fig:hi_ell:valid:priors} shows the impact on cosmological parameters for the \LCDM\ model when lifting one set of priors at time and when lifting them all (except for the overall calibration parameter $y_{\rm P}$, which always has the prior of Eq.~\ref{eq:calplanck} applied). Specifically, we test the impact of lifting the priors on the TT inter-frequency parameters $\calibC^{TT}_{100}$ and $\calibC^{TT}_{217}$ (described in Sect.~\ref{sec:hi-ell:datamodel:inst}), on the SZ amplitudes (described in Sect.~\ref{sec:hi-ell:datamodel:fg}), and on the dust amplitudes in TT, TE, and EE (i.e., letting the EE dust amplitude freely vary rather than being fixed as in the baseline). 
We find that lifting all the priors induces shifts in \LCDM\ parameters that are less than around $0.5\,\sigma$, with the largest shift being on $\lnAs$. 

When considering extensions to the base cosmology, for the \LCDM+$\Alens$ case we find that shifts are smaller than $0.4\,\sigma$, while in the \LCDM+$\nnu$ model we find a larger impact, up to $1.1\,\sigma$ for $\nnu$, as shown in Fig.~\ref{fig:hi_ell:valid:priorsAlensNeff}. {However, errors on $\nnu$ also increase by about 20\,\%. Since the data set without priors can be considered as a subset of the data set that includes them, we calculated the expected standard deviation ($\sigmaexp$) of the difference between parameters obtained in the two cases using the formalism described in \citet{gc2019} and Sect.~\ref{sec:valandro:cuts}. We find that such a shift is only deviant by $1.8\,\sigmaexp$.
Thus, this shift is compatible with statistical fluctuations, and we also observe that it is correlated with shifts in the dust amplitudes, inter-frequency calibration amplitudes, and $\ns$.} Specifically, due to degeneracies with other nuisance parameters without priors, the uncertainties on dust amplitudes become much larger, allowing values of parameters that are incompatible with those obtained in Sect.~\ref{sec:hi-ell:datamodel:gal} with the help of the 353- and 545-GHz channels.

We thus conclude that the cosmological results we obtain are  relatively insensitive to the priors that we set.

\subsection{Data cuts}
\label{sec:valandro:cuts}
In \citetalias{planck2014-a13}, we explored the robustness of the high-$\ell$ likelihood to numerous choices of data cuts, varying the types of input maps used (either HM or DS, see Sect.~\ref{sec:hi-ell:datamodel:datasel} for definition), changing masks, and exploring the effect of removing cross-spectra or multipole ranges from the likelihood. 
We will only reproduce here the most interesting examples, namely the removal of certain cross-spectra and the comparison between the high- and low-multipole ranges of the likelihood. The outcome of the first such test in polarization has changed compared to 2015, thanks to the leakage and PE corrections, which improve the inter-frequency agreement in TE and EE. The second test was not presented for polarization in 2015, given the uncertainty on the polarization systematic corrections.

Owing to the difficulties in correcting for correlated noise in the DS likelihood (that we discussed in 2015), we do not reproduce such a test for this paper. We do, however, explore the behaviour of the Odd-Even data cut (OE, see Sect.~\ref{sec:hi-ell:datamodel:datasel} for definition). Unfortunately, as discussed in Sect.~\ref{app:oe_correlation}, we also detect hints of correlated noise in the OE spectra.   We nevertheless build an OE likelihood, without trying to correct for this correlated noise, and see what the consequences are for parameters. We observe $\sigma$-level shifts in the foreground parameters for the $143\times143$ and $143\times217$ cross-spectra in TT, and up to $1\,\sigma$ shifts on $n_{\rm s}$ in TE and EE. Similarly, the OE $\Alens$ value is reduced by $1\,\sigma$ compared to the HM one. We can link some of these changes to the shape of the correlated noise described in Sect.~\ref{app:oe_correlation}. In the TT case this is also confirmed by the fact that when we reduce the value of $\ell_{\rm max}$ used in the TT likelihood to 1600, where the correlated noise seen in Fig.~\ref{fig:hi-ell:data:oe_hm_comparisons} is small in OE, we retrieve a similar $\Alens$ value for the HM and OE splits. We do not perform further investigations of the OE-based likelihood.

\subsubsection{Inter-frequency agreement}

\begin{figure*}[htbp!]
\includegraphics[width=18cm]{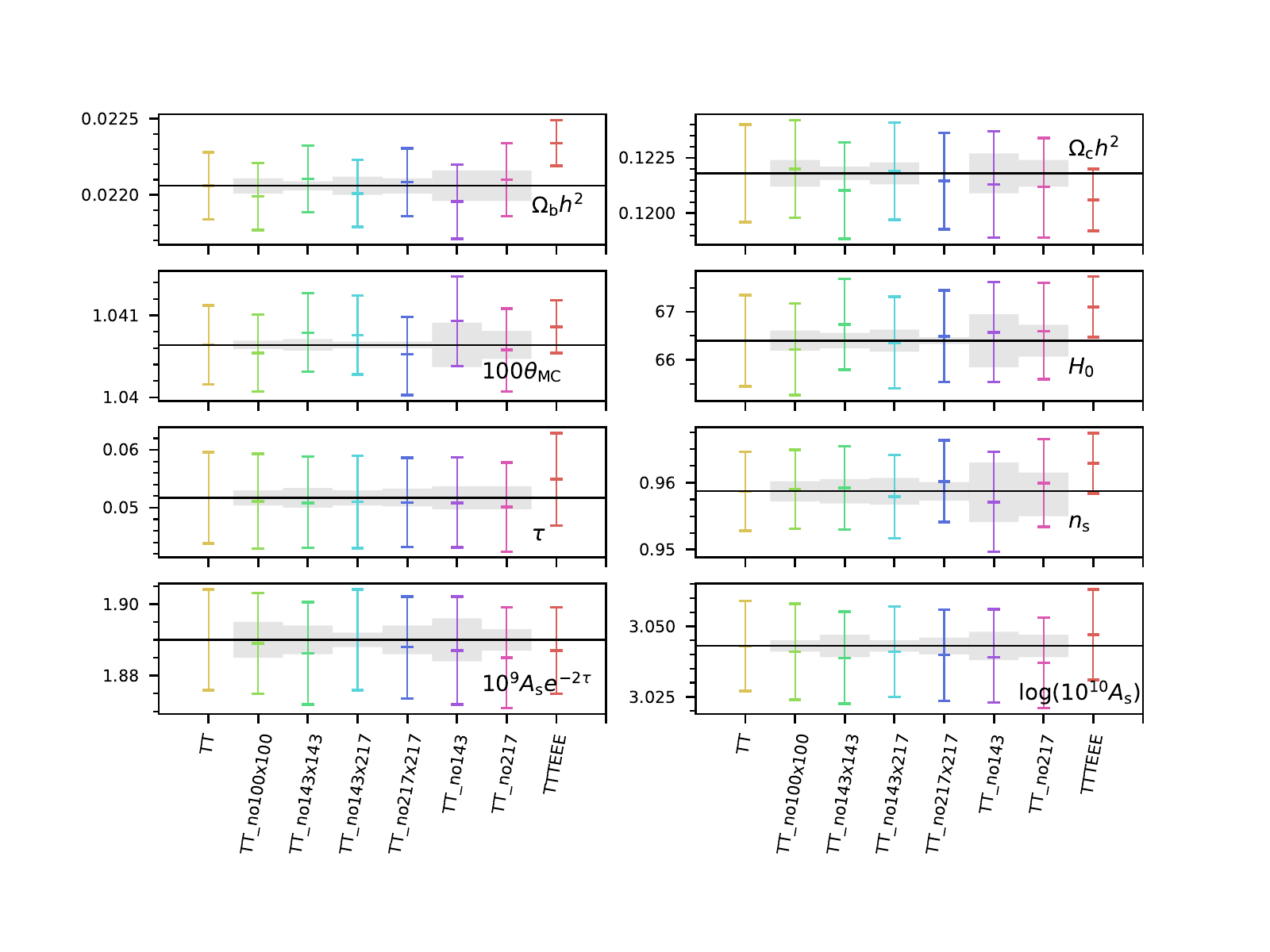}
\vspace{-1.0cm}
\caption{Marginal means and 68\,\% error bars on the cosmological parameters for \plik\TT\ combined with \simall, when removing either a single cross-spectrum, or all cross-spectra that include a given frequency channel. Shifts in cosmological parameters are in agreement with expectations. Grey bands represent the expected deviation with respect to the full-frequency case, computed from Eq.~\eqref{greybands}. For comparison, we also show the case \plikTTTEEE+\simall.}
\label{fig:valandro:TT_byfreq}
\end{figure*}

We check the impact on parameter estimates of excluding one cross-frequency spectrum or one full-frequency channel at a time from the \plik\ likelihoods. For this particular test, we analyse the \plik\ TT, TE, and EE likelihoods in combination with the \lowE\ one (i.e., the \simall\ likelihood, but not the \commander\ one.). All our tests are performed within the $\Lambda$CDM cosmological framework, but we also check the impact of letting the $\Alens$ parameter vary freely. 
In Fig.~\ref{fig:valandro:TT_byfreq} we report the constraints for \plik\TT{}+\simall. Here, the ``no$-\nu\times\nu'$'' labels indicate the constraints obtained by excluding the $\nu\times\nu'$ frequency spectrum, while the ``no$-\nu$'' labels show the case where we remove all the spectra calculated using frequency $\nu$ from the likelihood. It should be remembered that the baseline \plik\TT{}\ likelihood uses the $100\times100$, $143\times143$, $143\times217$, and $217\times217$ half-mission cross-spectra.

\begin{figure*}[htbp!]
\includegraphics[width=18cm]{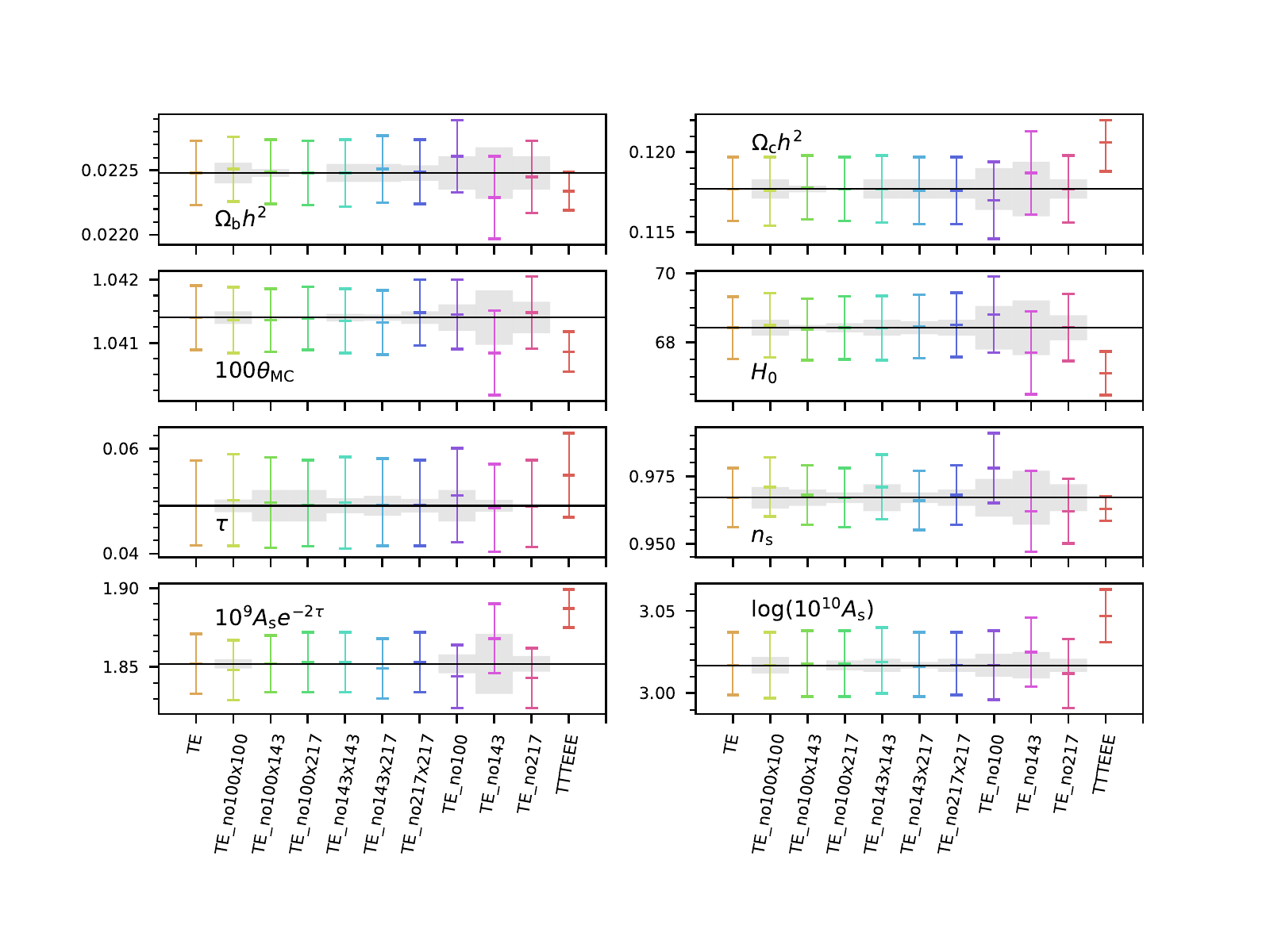}
\vspace{-1.0cm}
\caption{Same as Fig.~\ref{fig:valandro:TT_byfreq}, but for \plik\TE{}. Shifts in cosmological parameters are in agreement with expectations.}
\label{fig:valandro:TE_byfreq}
\end{figure*}

\begin{figure*}[htbp!]
\includegraphics[width=18cm]{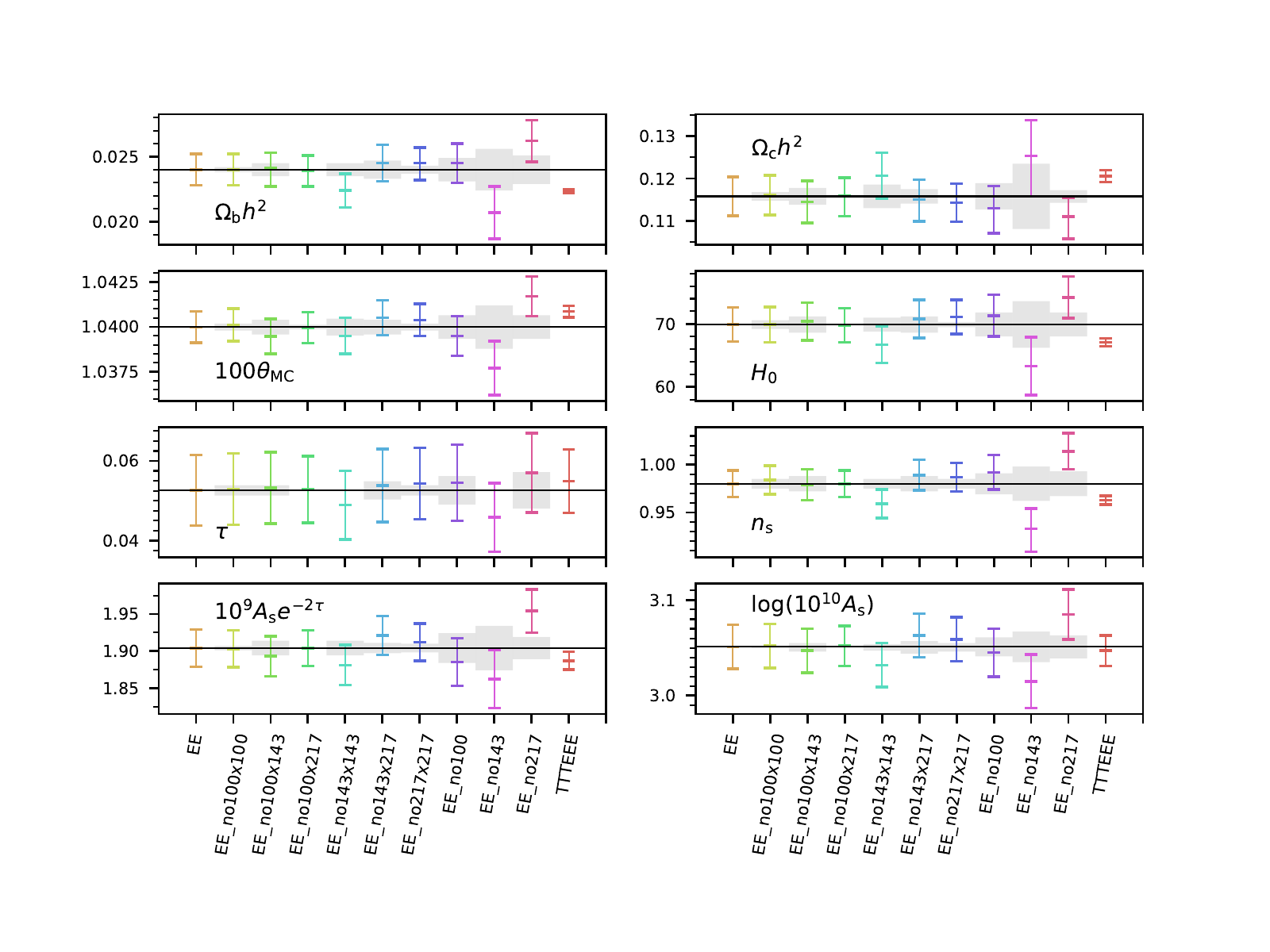}
\vspace{-1.0cm}
\caption{Same as Fig.~\ref{fig:valandro:TT_byfreq}, but for \plik\EE. Shifts in cosmological parameters are in agreement with expectations within $2.6\,\sigmaexp$, and showing some mild disagreement between frequencies.}
\label{fig:valandro:EE_byfreq}
\end{figure*}
The grey bands in Fig.~\ref{fig:valandro:TT_byfreq} show the expected standard deviation ($\sigmaexp$) of the difference between parameters obtained from the full-likelihood case and those obtained with a subset of the data (in this case, from a likelihood formed by excluding certain spectra). As in \citet{gc2019} and \citet{planck2014-a11}, it can be shown that the covariance of  the parameter difference is equal to the difference of the covariances of the subset case and of the full-likelihood one:

\begin{align}
{\rm cov}(\vec{P}_\mathrm{subset}-\vec{P}_\mathrm{full})&=\overline{(\vec{P}_\mathrm{subset}-\vec{P}_\mathrm{full})(\vec{P}_\mathrm{subset}-\vec{P}_\mathrm{full})^{\sf T}}\\ &= {\rm cov}(\vec{P}_\mathrm{subset}) -{\rm cov}(\vec{P}_\mathrm{full})\, ,
\label{greybands}
\end{align}
where \noindent $\vec{P}_\mathrm{full}$ and $\vec{P}_\mathrm{subset}$ are vectors of the maximum-likelihood parameter values\footnote{In practice, we use the posterior means instead of the maximum-likelihood parameters. This is a fair approximation for posterior distributions that are close to Gaussian.} for the two data sets and $\overline{\cdots}$ denotes the mean over the sampling distribution. The standard deviations $\sigmaexp$ for single parameters are thus calculated as $\sigmaexp^2=\sigma_\mathrm{subset}^2-\sigma_\mathrm{full}^2$.
The $\sigmaexp$ calculation is affected by the MCMC numerical errors on $\sigma_\mathrm{full}$ and $\sigma_\mathrm{subset}$, which we estimate to be at least at the level of a few percent for a Gelman-Rubin convergence diagnostic of $R\leq0.01$. Furthermore, the differences between the best-fit parameters of less than $0.1\,\sigma$ are also subject to numerical errors. We thus ignore the cases where the differences between parameters are of this order.

As shown in Fig.~\ref{fig:valandro:TT_byfreq}, the constraints obtained with the \plik\ \TT{}\ likelihood are robust against exclusions of frequency spectra. Parameters shift by less than $0.6\,\sigma_\mathrm{full}$\footnote{To avoid confusion, we will specify whether we quantify the shifts in terms of the uncertainty of the full likelihood case, $\sigma_\mathrm{full}$, or in terms of the expected shifts $\sigmaexp$. The latter is the one that quantifies statistical consistency.} with respect to the baseline case, with the largest change being in $\theta_{\rm MC}$ when considering the ``no143'' case (by $0.6\,\sigma_\mathrm{full}$ or $1.0\,\sigmaexp$). We find that all of the shifts with respect to the baseline case are in agreement with the expected ones from Eq.~\eqref{greybands}, within $2\,\sigmaexp$, once we take into account numerical uncertainties.

In Figs.~\ref{fig:valandro:TE_byfreq} and~\ref{fig:valandro:EE_byfreq} we also report the tests for \plik\ TE and EE, respectively. Additionally to the spectra also used in TT, recall that the baseline EE and TE likelihoods also include the $100\times143$ and the $100\times217$ spectra, so for these cases we check the robustness of the results when excluding these frequency spectra as well. 
For \plik\ TE we find that constraints obtained excluding frequency spectra are in agreement with the baseline results within $1\,\sigma_{\rm full}$ , with the largest shift being in $\theta_{\rm MC}$ for the ``no143'' case by about $1.1\,\sigma_{\rm full}$ ($1.2\,\sigmaexp$).\footnote{The shifts expressed in terms of $\sigma_{\rm full}$ and $\sigmaexp$ can have similar values when $\sigma_\mathrm{subset}^2\approx 2\,\sigma_{\rm full}^2$, so that $\sigmaexp^2=\sigma_\mathrm{subset}^2-\sigma_\mathrm{full}^2\approx \sigma_\mathrm{full}^2$, as in the case considered in the text.} All cases are in agreement with expected shifts to within $2\,\sigmaexp$.

Finally, the effect of removing frequency spectra from the \plik\EE{}\ likelihood is larger than for the previous cases, as shown in Fig.~\ref{fig:valandro:EE_byfreq}. In particular, completely excluding the 143-GHz frequency channel (the ``no143'' case) induces shifts in most of the cosmological parameters, including changes up to $3.2\,\sigma_\mathrm{full}$ ($2.7\,\sigmaexp$)  on $\ns$. Smaller but still significant deviations are present when excluding only the $143\times143$ spectrum. Similarly, shifts in the direction opposite to the ``no$143$'' ones (and slightly less significant) are induced when completely removing the 217-GHz frequency channel (the ``no217'' case), with the maximum shift on $\ns$, which increases by $2.4\,\sigma_\mathrm{full}$ ($2.6\,\sigmaexp$). Therefore, deviations agree with expectations only within $2.6\,\sigmaexp$, which might point towards a remaining disagreement between the 143- and 217-GHz channels in EE. We have checked that these differences remain even when letting the EE\ dust amplitude parameters freely vary with priors, instead of fixing them as in the baseline case, suggesting that incorrect modelling of the Galactic foregrounds cannot account for these differences. 

We thus conclude that the cosmological parameters for the $\Lambda$CDM model for the \plik\TT\ and TE parameters are stable against exclusions of frequency spectra. This is important since almost all the constraining power of the joint TTTEEE likelihood comes from TT and TE \citep{2014PhRvD..90f3504G}. For EE, we find that a few parameters shifts are in mild disagreement with expectations at the level of about $2.5\,\sigmaexp$.

\subsubsection{$\ell<800$ versus $\ell>800$ comparisons}
\label{sec:valandro:cuts:800}

We have tested the stability of our results by splitting the multipole range of \plik\ in two different regions, $\ell\,{<}\,800$ and $\ell\,{>}\,800$, for each likelihood combination, following similar tests performed in \citet{planck2014-a11}, \citet{2016ApJ...818..132A}, and \citet{planck2016-LI}, as well as in section~6.1 of \citetalias{planck2016-l06}.

\begin{figure*}[htbp!]
\includegraphics[width=18cm]{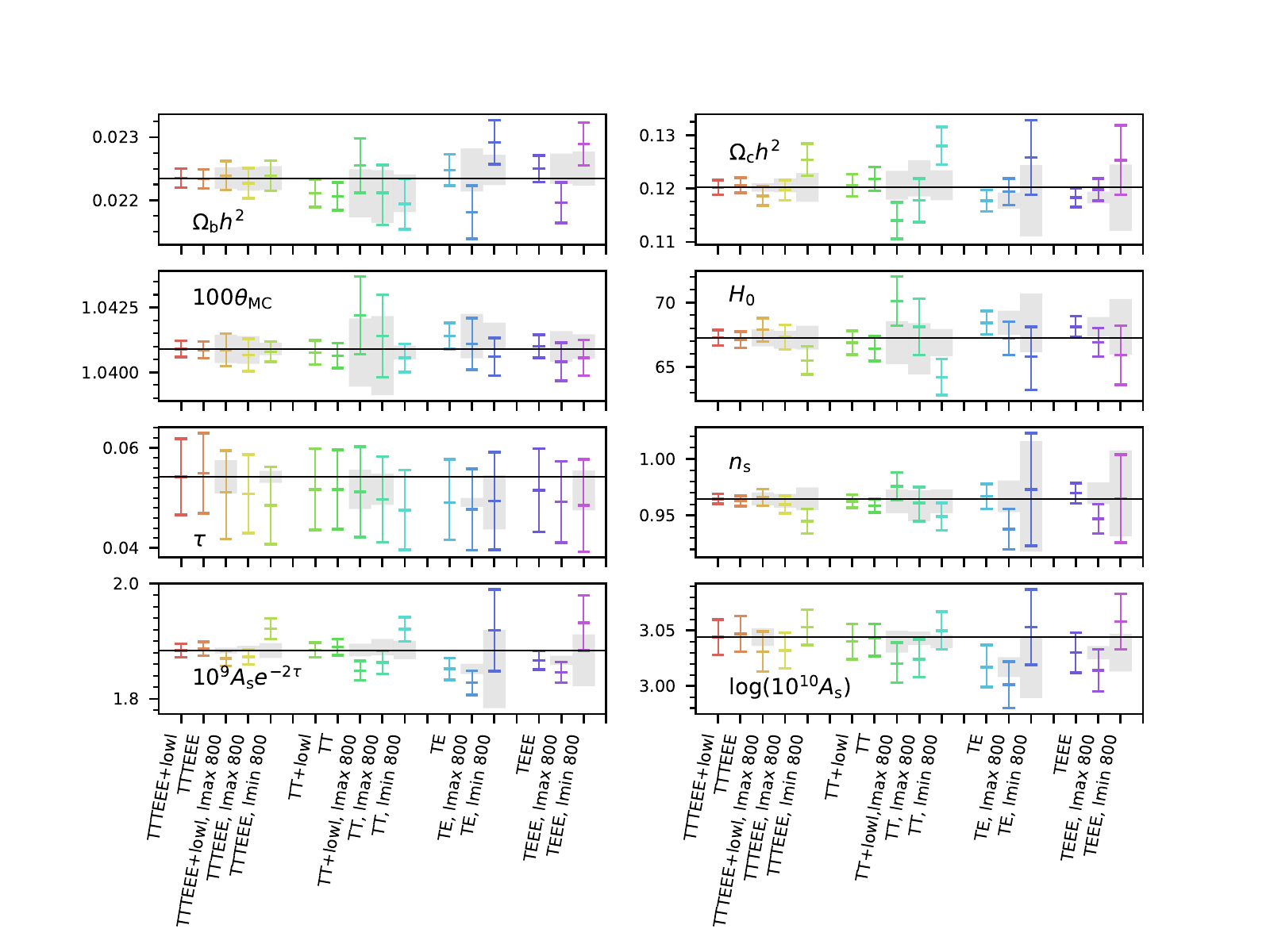}
\caption{Marginal means and 68\,\% error bars on the cosmological parameters for \plik{}TT, TE, TEEE, and TTTEEE, combined with \simall and in some cases \lowl\ (as indicated in the labels), in the full multipole range and for two different multipole cuts, namely $\ell\,{<}\,800$ and $\ell\,{>}\,800$. }
\label{fig:valandro:lmin-lmax800}
\end{figure*}

We perform these tests not only for the \plikTT\ likelihood, but also for the TTTEEE, TE, TEEE, and EE likelihoods. The choice of splitting at $\ell\,{=}\,800$ was made to give equal constraining power on the \LCDM\ parameter volume to the two multipole ranges in TT\footnote{To avoid unnecessary detail, we write $\ell\,{<}\,800$ and $\ell\,{>}\,800$, although the binned \plik\ likelihoods are effectively split at $\ell\,{\leq}\,801$ and $\ell\,{>}\,801$.} For TE, EE, TTTEEE, and TEEE, the appropriate choices of multipole split to satisfy this criterion might vary from the value needed for TT, but we keep the same value in all cases to facilitate comparison of the results from the different likelihoods. The $\ell\,{<}\,800$ \TT\ and \TTTEEE\ cases include the \plik\ likelihood and the \lowl\ one. Furthermore, all cases considered here ($\ell\,{<}\,800$ and $\ell\,{>}\,800$ in temperature or polarization) include the \simall\ likelihood to constrain the optical depth to reionization.
All the tests are performed for the $\Lambda$CDM model and we checked that our conclusions are robust even when the foreground parameters are held fixed to the best fit of the baseline case.

As shown in Fig.~\ref{fig:valandro:lmin-lmax800}, we confirm that parameters estimated from $\ell\,{<}\,800$ (i.e., $\ell_{\rm max}\,{=}\,800$) and $\ell\,{>}\,800$ (i.e., $\ell_{\rm min}\,{=}\,800$) from the TT likelihood feature moderately significant shifts. 
In order to quantify the magnitude of these differences, we calculate the statistics $\chi_\mathrm{diff}^2$ of the parameter shifts, following the simplified procedure already used in \cite{2016ApJ...818..132A} and \cite{planck2016-LI}. We assume that the parameters inferred from the two multipole ranges are independent; this assumption holds only if one excludes $\tau$ from the set of cosmological parameters considered, since in both ranges this parameter is constrained by the use of the \simall\ likelihood. For the same reason, we include $A_{\rm s}e^{-2\tau}$ instead of $\lnAs$ in the parameter set considered, since the former is less degenerate with the constraint on $\tau$. Thus, we consider $\Omch,\Ombh,\theta,\ns,A_{\rm s}e^{-2\tau}$ as the reference parameter set (excluding $\tau$ for the reasons given above). Assuming that the parameters are uncorrelated between the two multipole ranges allows us to calculate the covariance of the parameter differences between the two subsamples as the sum of the parameter covariances from each set:
\begin{equation}
\tens{\Sigma}_\mathrm{diff}=\tens{\Sigma}_\mathrm{\ell<800}+\tens{\Sigma}_\mathrm{\ell>800}.
\end{equation}

 As a consequence, the standard deviation on the shift of a single parameter is calculated from $\sigmadiff^2=\sigma_\mathrm{\ell<800}^2+\sigma_\mathrm{\ell>800}^2$. 
Finally, we define a statistic based on the parameter shifts as
\begin{equation}
\chi_\mathrm{diff}^2=\Delta \vec{p}^{\sf T} \, \tens{\Sigma_\mathrm{diff}}^{-1} \, \Delta \vec{p},
\label{chi2shifts}
\end{equation}
where $\Delta \vec{p}$ is the vector of shifts in parameters between the two
data sets. We then quantify the consistency between two data sets by assuming that the statistic in Eq.~\eqref{chi2shifts} has a $\chi^2$ distribution. This allows us to calculate the PTE of the parameter differences, and thus the equivalent number of Gaussian $\sigmagauss$ for the deviation, where $\sigmagauss$ is defined such that a 1-dimensional Gaussian would have the same two-tailed PTE.
As shown in \cite{planck2016-LI}, this is an approximate procedure that does not take into account remaining correlations between the subsets of data and assumes that the $\chi_\mathrm{diff}^2$ statistics follows a $\chi^2$ distribution. However, in that paper it was also shown using simulations that this procedure provides estimates of the deviation that are accurate at the level of a few tenths
of $\sigma$.
Following this procedure, we calculate that for the five cosmological parameters listed above, the difference between the two subsamples is $1.8\,\sigmagauss$. This is similar to what was obtained with the baseline data from the 2015 \Planck\ release, for which we found $1.7\,\sigmagauss$.

Considering single parameters, the largest differences are found for $\Omega_{\rm c} h^2$ and $A_{\rm s}e^{-2\tau}$, which shift by $2.9$ and $2.7\,\sigmadiff$, respectively. As a consequence, derived parameters such as $H_0$ and $\sigma_8$ are different by $2.5\,\sigmadiff$ and $2.9\,\sigmadiff$, respectively. Once again, these shifts are degenerate with the other cosmological parameters, so when considering all of them together the shifts are less significant. 

As already discussed in \cite{planck2016-LI} and \cite{planck2014-a11}, these parameter differences are related to the low-$\ell$ power deficit and to the $\Alens$ deviation.
In fact, as in 2015, we find that when excising the $\ell\,{<}\,30$ data from the $\ell\,{<}\,800$ case, the the shifts are reduced to $1.2\,\sigmagauss$ for the five cosmological parameters, and to $1.9\,\sigmadiff$ and $2.0\,\sigmadiff$ for $\Omch$ and $A_{\rm s}e^{-2\tau}$, respectively. This shows that part of the differences in parameters is due to the deficit in power at low multipoles. This does not mean that one should excise those multipoles, this is just a test that allows us to identify which fluctuations of the power spectrum are responsible for for the differences in parameters between the two ranges.
Similarly, when letting the $\Alens$ parameter vary freely (either for both multipole ranges or for the $\ell\,{>}\,800$ case alone), these shifts are reduced to $1.2\,\sigmagauss$ for the five cosmological parameters, and to $0.2\,\sigmadiff$ and $1.6\,\sigmadiff$ for $\Omch$ and $A_{\rm s}e^{-2\tau}$, respectively, with $\Alens$ determined from $\TT{}(\ell\,{>}\,800)+\lowE$ at the level of $\Alens({\ell\,{>}\,800})=1.6\pm0.3\,\sigma$.

We also perform the same test using the \plik\ TE likelihood. As shown in Fig~\ref{fig:valandro:lmin-lmax800}, also in this case we find shifts between the $\ell\,{<}\,800$ and the $\ell\,{>}\,800$ parameters, although with less statistical significance with respect to TT, also due to the fact that the $\ell\,{>}\,800$ TE likelihood has much less constraining power with respect to the $\ell\,{<}\,800$ one.
The largest difference is for the baryon density $\Omega_{\rm b}h^2$, at the level of $2\,\sigmadiff$, with a total difference on the five cosmological parameters of $1.5\,\sigmagauss$.

For the Plik EE case (not shown in Fig.~\ref{fig:valandro:lmin-lmax800} for readibility), parameters inferred from the two multipole regions, $\ell\,{<}\,800$ and $\ell\,{>}\,800$, are completely in agreement within $1\,\sigmadiff$. When we then combine the likelihoods into the \plik\ TEEE case, we can see roughly the same behaviour found for the \plik\TE\ case.

Finally, we find that for the full combination \plik{}TTTEEE, shifts become less significant compared to the TT case, due to the fact that in some cases the differences in TT are in opposite directions with respect to the TEEE ones. Once again, the two most deviant parameters are $\Omch$ and $A_{\rm s}e^{-2\tau}$, with shifts of $2.0$ and $2.2\,\sigmadiff$, respectively; while considering the full set of five cosmological parameters, we obtain $1.2\,\sigmagauss$.

To conclude, we confirm that there is a shift in cosmological parameters between $\ell\,{<}\,800$ and $\ell\,{>}\,800$ in TT, which is not statistically very significant, at the level of $1.8\,\sigmagauss$ once degeneracies between cosmological parameters are taken into account. At a smaller statistical level, similar trends are also observed in $\Omch$ and $A_{\rm s}e^{-2\tau}$ for the two multipole ranges of the TE likelihood , while shifts in other parameter directions are also observed.
In EE, the high-multipole range has too low a statistical power to enable a valuable cross-check of these changes. Overall, the weak statistical significance of these differences do not provide sufficient information to infer that these shifts might have a physical or systematic origin; they are consistent with being merely statistical fluctuations.  The excellent consistency of the results obtained from different subsets of frequencies (and therefore, of detectors) in TT and TE shown in the previous subsection suggest that a systematic source for these shifts would require a common mode impacting many different detectors \citep[see also Figs.~11 and 12 of][]{planck2016-LI}.

\subsection{The $\Alens$ consistency parameter}
\label{sec:valandro:alens}

\begin{figure*}[htbp!]
\includegraphics[width=\textwidth]{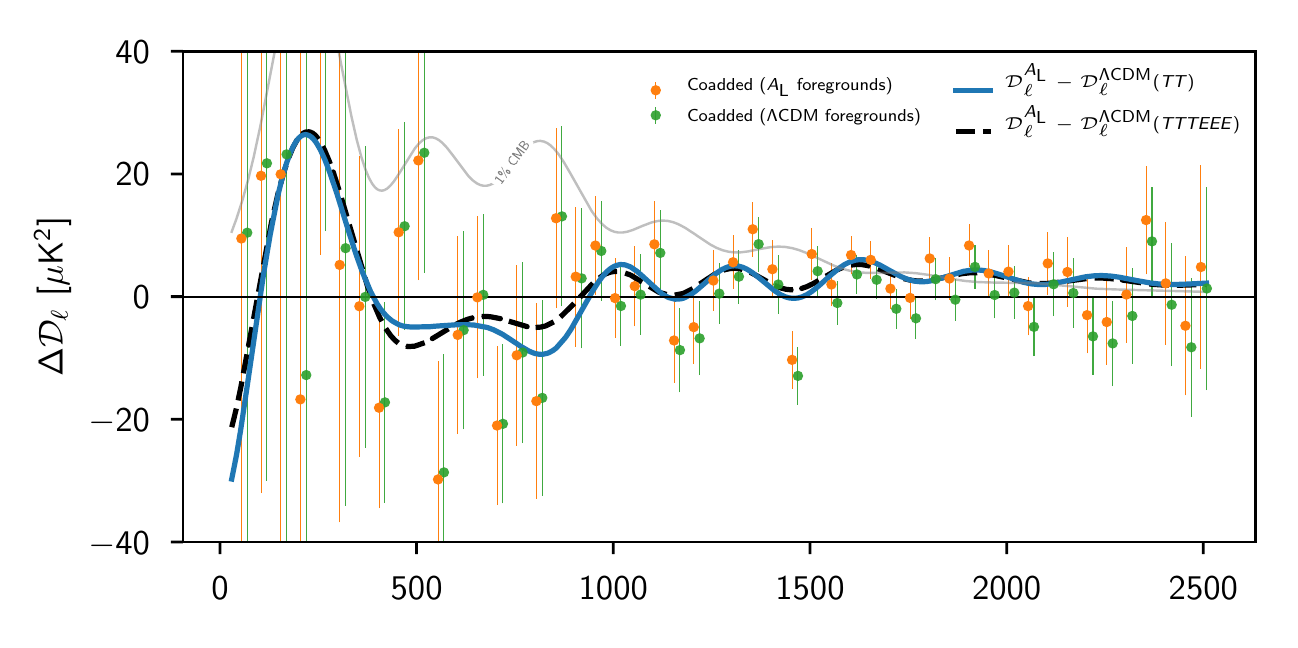}
\vspace{-0.25cm}
\caption{Comparison of the residuals of the coadded $TT$ spectrum with the difference between the \lcdm\ and \lcdm$+\Alens$ best-fit models. The green points show the coadded $TT$ data when using the foreground parameters obtained with a \lcdm\ best fit, while the orange points are obtained using the foreground parameters from a \lcdm$+\Alens$ one. In both cases, we display the residual to the best-fit \lcdm\ cosmology (using the \planckTT+\lowE\ likelihood) in bins of size $\Delta\ell\,{=}\,50$ and we shift the bin locations slightly to improve legibility. The blue line is the difference between the $\Alens$ and \lcdm\ best-fit models obtained using the \planckTT+\lowE\ likelihood, while the black dashed line is obtained with the \planckall\ likelihood. The thin grey line gives a sense of the scale of the residuals by displaying the $TT$ power spectrum divided by 100. The residuals display an approximately oscillatory behaviour, which does not seem to be statistically anomalous, but which appears to roughly match the particular oscillatory feature of the $\Alens$ effect. The difference between the two residuals shows the degeneracy between $\Alens$ and the foreground parameters. We discuss in the text how fixing the foreground parameters to their \lcdm\ values decreases the $\Alens$ deviation from unity by about $1\,\sigma$.}
\label{fig:valandro:alensresdiuals} 
\end{figure*}

The $\Alens$ issue is discussed at length in Sect.~6.2 of \citetalias{planck2016-l06}. Here, we briefly remind the reader of the most relevant facts of this ``curious'' feature of the \planck\ data, and we mention some of the tests we performed to investigate if its origin might be systematic or not.

$\Alens$ is a phenomenological parameter multiplying the amplitude of the lensing potential power, which produces the smoothing of the acoustic peaks and the transfer of power to the damping tail in the CMB angular power spectra. It enables us to check whether or not the lensing effect in the power spectrum is consistent with expectations (for which $\Alens=1$) and with the amplitude measured by the lensing reconstruction.  

Since the first \Planck\ data release in 2013 \citep{planck2013-p11}, the $TT$ data have shown a preference for a value of $\Alens$ larger than unity by about $2\,\sigma$. This result has not substantially changed across releases, despite the many changes in the likelihoods. These include increasing the data from nominal- to full-mission, improving the constraint on the optical depth to reionization, changing the data cuts for the maps used in the likelihood (from detector sets to half-mission maps), improving the treatment of systematics (in particular calibration and beams) and of foregrounds, among others.

For the 2018 \Planck\ release, we find $\Alens = 1.243\pm 0.096 \quad\onesig{\planckTTonly+\lowE}$,\footnote{We remind the reader that the \planckTTonly\ likelihood includes the low-$\ell$ TT part. Unless stated otherwise, it will always been used in the results of this section.} a $2.5\,\sigma$ deviation from unity, with the best-fit $\chi^2$ for the \LCDM+$\Alens$ model improved by $8.7$ points over the one for the \LCDM\ model. Most of this improvement comes from the high-$\ell$ likelihood (mostly in the range $600< \ell<1500$); however, the low-$\ell$ temperature \commander\  likelihood fit is also improved if $\Alens$ is allowed to float freely, with $\Delta \chi^2=-2.3$. This shows that the $\Alens$ deviation is mainly caused by oscillatory residuals at high $\ell$, but that it also allows the model to better fit the low power at $\ell\,{<}\,30$, as was discussed at length in \cite{planck2016-LI}. The oscillatory residual is shown in Fig.~\ref{fig:valandro:alensresdiuals}, which illustrates how $\Alens$ improves the fit to the data, both at low and high multipoles. Although the fit seems to improve also for $\ell\,{<}\,1000$, it has already been shown that the lower-multipole part of the high-$\ell$ likelihood by itself does not particularly prefer a high value of $\Alens$;  the preference is rather dominated by the features at $\ell\,{>}\,1000$ (\citetalias{planck2014-a13}; \citealt{planck2016-LI}).

As already stated, the $\Alens$ parameter is not a physical one, and hence it may be simply providing a degree of freedom to fit to a particular kind of statistical fluctuation. However, we have used $\Alens$ as a proxy for all of those extensions of the \LCDM\ model that can effectively change the amplitude of lensing and thus produce effects similar to $\Alens$, such as the dark energy equation of state $w$ or the curvature of the Universe $\omegak$. These parameters also deviate from their \LCDM\ values when evaluated from the CMB power spectra alone \citepalias[see][]{planck2016-l06}. None of these mild deviations are really interpreted as signs of a breaking of the vanilla \LCDM\ model, since their statistical significance is relatively low, and since the amplitude of the lensing potential power spectrum measured from the lensing reconstruction (coming independently from the 4-point function, giving a result which is three to four times more precise than that evaluated from the CMB power spectra), does not show such an excess \citep{planck2016-l08}. Therefore, adding the lensing reconstruction or other external data sets such as BAO, brings the fits of these extensions of the \LCDM\ model back to consistency with their \LCDM\ values.

In order to design a physical model that can successfully fit for the power-spectrum residuals that cause the $\Alens$ deviation, it is thus necessary to evade constraints from CMB lensing and external data sets. Apparently promising models, such as compensated isocurvature \citep{Grin:2011tf,Grin:2011nk} ones were recently highly constrained by the inclusion of data on the largest scales of the lensing reconstruction power spectrum (the lowest multipole bin at $L\,{=}\,8$--40 was introduced in the lensing baseline in this release), as described in \citet{Valiviita:2017fbx} and \citet{planck2016-l10}.

Polarization is also sensitive to $\Alens$. The $TE$ polarization data alone slightly prefer a value lower than unity, $\Alens=0.76\pm 0.22$ (\TE+\lowE), while the $EE$ data slightly favour a value higher than one, $\Alens=1.32^{+0.24}_{-0.27}$ (\EE+\lowE). Both are consistent with $\Alens=1$ within $2\,\sigma$, while the joint constraint gives $\Alens = 1.180\pm 0.065\ \onesig {{\rm Planck}\TTTEEE+\lowE}$, a $2.8\,\sigma$ deviation from unity. However, as already discussed in Sect.~\ref{sec:valandro:PE}, the results on $\Alens$ from 
polarization are not robust against changes of treatment of the PE corrections, since using a spectrum-based approach instead of the map-based approach can lower the value of $\Alens$ by about $0.6\,\sigma$ (see Fig.~\ref{fig:hi_ell:valid:PE2d}). Polarization thus does not provide a robust confirmation of this deviation.

We also perform a number of tests to verify the robustness of the $\Alens$ deviation and to check whether it could have its origin in a systematic effect. We briefly describe these tests below.

First of all, the same $\Alens$ deviation is measured both with the \plik\ (baseline) and \camspec\ likelihoods in temperature, when combining with \lowl\ and \lowE. The two likelihoods share many common features (Gaussian approximation, similar foreground model, same Galactic-, extended-object-, and point-source- masks in intensity), but also have some differences, such as a slightly different treatment of foregrounds (e.g., Galactic dust and the CIB in particular), independent evaluations of the power spectra and covariance matrices, and the way that missing pixels are treated.
On the other hand, the $\Alens$ values recovered in polarization by the two likelihoods differ, due mainly to the different treatment of PE corrections (map-based for \plik, spectrum-based for \camspec). As already discussed, it is not possible at this point to determine if one of the two PE correction approaches is more correct than the other.

\begin{figure}[htbp!]
\includegraphics[width=\columnwidth]{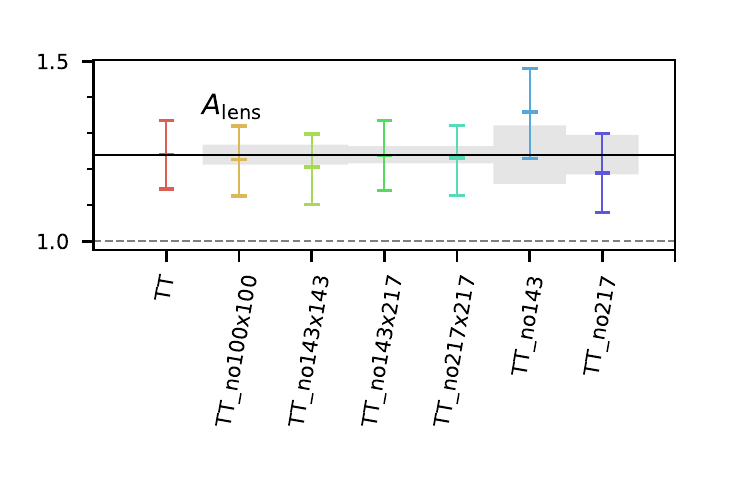}
\vspace{-1.0cm}
\caption{Marginal means and 68\,\% error bars for the $\Alens$ consistency parameter for \plik\TT{}\ combined with \commander\  and \simall\  EE, when removing either a single cross-spectrum, or all cross-spectra within a given frequency channel from the high-$\ell$ part of the likelihood (i.e., \plik). Grey bars represent the variance with respect to the full-frequency case, computed using Eq.~\eqref{greybands}. }
\label{fig:valandro:Alens_TT_byfreq}
\end{figure}

We also verify that the $\Alens$ deviation in temperature cannot be explained by systematic effects operating on one particular frequency. Figure~\ref{fig:valandro:Alens_TT_byfreq} shows the results on $\Alens$ when eliminating one cross-spectrum at a time from the \plik\ likelihood, or all spectra involving one particular frequency. This is a very interesting test, since it indicates that if the $\Alens$ deviation has a systematic origin, it must be a coherent effect across frequencies, and thus across different detectors in the focal plane. Furthermore, this test also shows that unmodelled residual Galactic dust contamination cannot fully explain the $\Alens$ deviation, since these would cause frequency-dependent $\Alens$ results. This was also investigated in \citetalias{planck2014-a13} (see figure~36 there), where we showed that the value of $\Alens$ is unchanged when using more conservative galactic masks.

Furthermore, as already shown in \citetalias{planck2014-a13} and \citet{planck2016-LI}, we know that the $\Alens$ deviation is partially degenerate with extragalactic foreground parameters in TT. In particular for \plik\TT{}+\lowl+\lowE, $\Alens$ is degenerate with the point-source amplitude at 143\,GHz (by $-34$\,\%) and at 100\,GHz by ($-18$\,\%), with the CIB amplitude (by $-16$\,\%), and with the kinetic SZ amplitude (by $-16$\,\%). A consequence of this is that if we fix foreground parameters to the best-fit \LCDM\ values, the value of $\Alens$ decreases from $\Alens=1.24\pm 0.09$ in the baseline case to $\Alens=1.16\pm 0.07$. The reason for this degeneracy is that $\Alens$ provides the extra peak-smoothing required at high $\ell$. This allows $\omc$ and $\lnAs$ to decrease with respect to their \LCDM\ values (in the \LCDM\ case, higher values of these parameters improve the fit by increasing smoothing). Following degeneracy directions, $\ns$ then increases to better fit the low-$\ell$ data. This however produces an increase in power at high multipoles, which can be compensated by reducing the amplitude of foreground parameters. As shown in Sect.~\ref{sec:valandro:priors}, the priors that we set on nuisance parameters (specially on the SZ parameters) cannot account for this deviation.  The degeneracy with foreground parameters is clear in Fig.~\ref{fig:valandro:alensresdiuals}, where we show a comparison between the coadded \TT{} data when assuming either the \lcdm\ foreground parameters or the $\Alens$ ones. 

As discussed in \cite{planck2016-LI}, the $\Alens$ deviation is somewhat dependent on the low power in the multipoles below $\ell\,{\approx}\,30$. Cutting those multipoles yields slightly lower values of $\Alens=1.21\pm0.10$. Also, the $\Alens$ deviation is related to the difference between the \LCDM\ parameters measured from \TT{}\ at $\ell\,{<}\,800$ and $\ell\,{>}\,800$, as already discussed in the previous subsection. 

We also check that the $\Alens$ deviation cannot be explained by a mistreatment of the aberration due to the motion of the Solar System with respect to the CMB frame, which, as discussed in the following section, might also create an oscillatory residual in the power spectrum (but of too small an amplitude).

Finally we check whether an error in the modelling of the transfer function could account for this deviation. For a small transfer function error, the additive correction is an oscillatory template. We therefore add to the theory model a free polynomial transfer function for each $\mu\times\nu$ frequency spectrum of the form
\begin{equation}
\Delta C_\ell^{(\mu\times\nu)}(\theta) = (b^\mu b^\nu -1)\, C_\ell^{(\mu\times\nu),\mathrm{fid}}(\theta),
\end{equation}
where $b^\nu = 1+\sum_{i=1}^N \epsilon_i^\nu \ell^i$ and $C_\ell^{(\mu\times\nu),\mathrm{fid}}(\theta)$ is a fiducial CMB-plus-foreground theory model.
We fit for the $\epsilon^\nu$ parameters both in the case of a quadratic ($N\,{=}\,2$) or cubic ($N\,{=}\,3$) polynomial expansion. In the \LCDM+$\Alens$ case, we find that marginalizing over the $\epsilon^\nu$ parameters substantially increases the errors on cosmological parameters such as $\Ombh$, $\Alens$, $\ns$ and $\lnAs$, but does not change their central values. We thus conclude that an error in the transfer function cannot account for the $\Alens$ deviation.

In conclusion, the $\Alens$ deviation is a feature of the \planck\ data mainly driven by the $TT$ power spectrum. The polarization data cannot provide a robust test of whether this has a physical origin, due to the degeneracy with the PE corrections. A systematics explanation of this feature would require us to introduce a common mode to different frequencies and hence detectors in the focal plane, which is not trivial to achieve. At present, we have not found any conclusive evidence that this deviation is due to a systematic effect. A physical explanation of this deviation is also not easy to find, since many extensions of the \LCDM\ model that we tested, which can provide effects similar to $\Alens$, are constrained by the addition of external data sets (and in particular the lensing reconstruction). A modification of the foreground model could potentially do the trick, but would probably require some structure in the smooth templates that we are using. We did not explore this particular route any further. We note that the reduction of the $\ell\,{\approx}\,1460$ outlier in the 217-GHz spectra shown in the dust-cleaned version of the \camspec\ likelihood \citepalias[discussed in][]{planck2016-l06} slightly reduces the significance of the deviation of $\Alens$ from 1, but does not change the general $\Alens>1$ trend. The most viable explanation for the $\Alens$ behaviour at this point seems to be that there is a set of statistical excursions in the $TT$ power spectrum that happens to be fit by varying this particular consistency parameter.  

\subsection{Aberration}
\label{sec:valandro:aberration}

We also examine the possibility that mistreatment of the effect of frame-boosting, particularly through aberration of the CMB sky, could significantly impact the results on cosmological parameters. While the orbital motion of the satellite with respect to the Solar System barycentre is corrected in the \planck\ maps (i.e., the ``orbital dipole,'' as well as associated stellar aberration), the motion of the Solar System barycentre with respect to the CMB frame is not, since this would alter the properties of foregrounds, which are not in the CMB frame.
Measuring in the Solar System frame, rather than the CMB frame, causes Doppler boosting not only of the CMB monopole into the dipole, but also of the anisotropies on smaller scales, producing a frequency-dependent dipolar modulation that increases the amplitude of the temperature power spectrum by about 0.25\,\% in the velocity direction (with the opposite effect in the anti-velocity direction). Furthermore, the frame change causes an aberration effect, in which the apparent arrival direction of CMB photons is pushed toward the velocity direction, with peak deflection of $\beta = 4\farcm20$ and a root-mean-squared (rms) deflection over the sky of 3\farcm0\ \citep{planck2013-pipaberration}.
Both the modulation and aberration effects are strongly suppressed when averaged over the full sky, and are expected to be small for the sky fractions used in the \plik\ likelihoods.

In fact{,} the aberration effect turns out to be the largest of the two, and so we now focus on that.  To directly test its amplitude in \Planck\ data, we follow the approach of \citet{2014PhRvD..89b3003J} and \citet{2017JCAP...06..031L}, modelling the effect at first order on the power spectrum as proportional to the logarithmic derivative of the power spectrum \citep[e.g.,][]{moss2011}:
\begin{equation}
\Delta C_\ell= -C_\ell \frac{d\ln C_\ell}{d\ln\ell} \beta \left\langle\cos{\theta}\right\rangle,
\end{equation}
where $\beta=v/c=1.23\times 10^{-3}$, $v$ is the speed of the Sun with respect to the CMB rest-frame, $\theta$ is the angle between the pixel position and the velocity direction $\vec{d}=(l,b)=264\pdeg0, 48\pdeg25$  \citep{planck2016-l01} and the average is over all the observed pixels. For each cross-spectrum used in the likelihood, we build a template $T_\ell=-C_\ell\,\beta\,d\ln C_\ell/d\ln\ell$, using a fiducial best-fit model to calculate $C_\ell$ and its derivative.\footnote{Owing to the smoothness of the CIB template, we ignore its contribution to the aberration effect.} We then evaluate the amplitude of this template either as $\left\langle\cos{\theta}\right\rangle$ for each of the masks used in the likelihood ($\left\langle\cos{\theta}\right\rangle=0.012$,
 $0.016$, and $0.0326$ for the masks at 100, 143, and 217\,GHz, respectively) or leaving it free to vary independently for each cross-spectrum. In the first case we find no impact on cosmological parameters. In the second case, we find that, when leaving the amplitudes free to vary over a large range ($-2.5$, $2.5$), these amplitudes are unconstrained and the uncertainty on the angular scale of the sound horizon is increased. We thus conclude that residual effects due to aberration, or indeed any other effect proportional to the derivative of the power spectrum, have negligible impact on our cosmological results.

\subsection{Simulations}
\label{sec:valandro:sims}
\begin{figure*}[htbp!]
\begin{center}
\includegraphics[width=18cm]{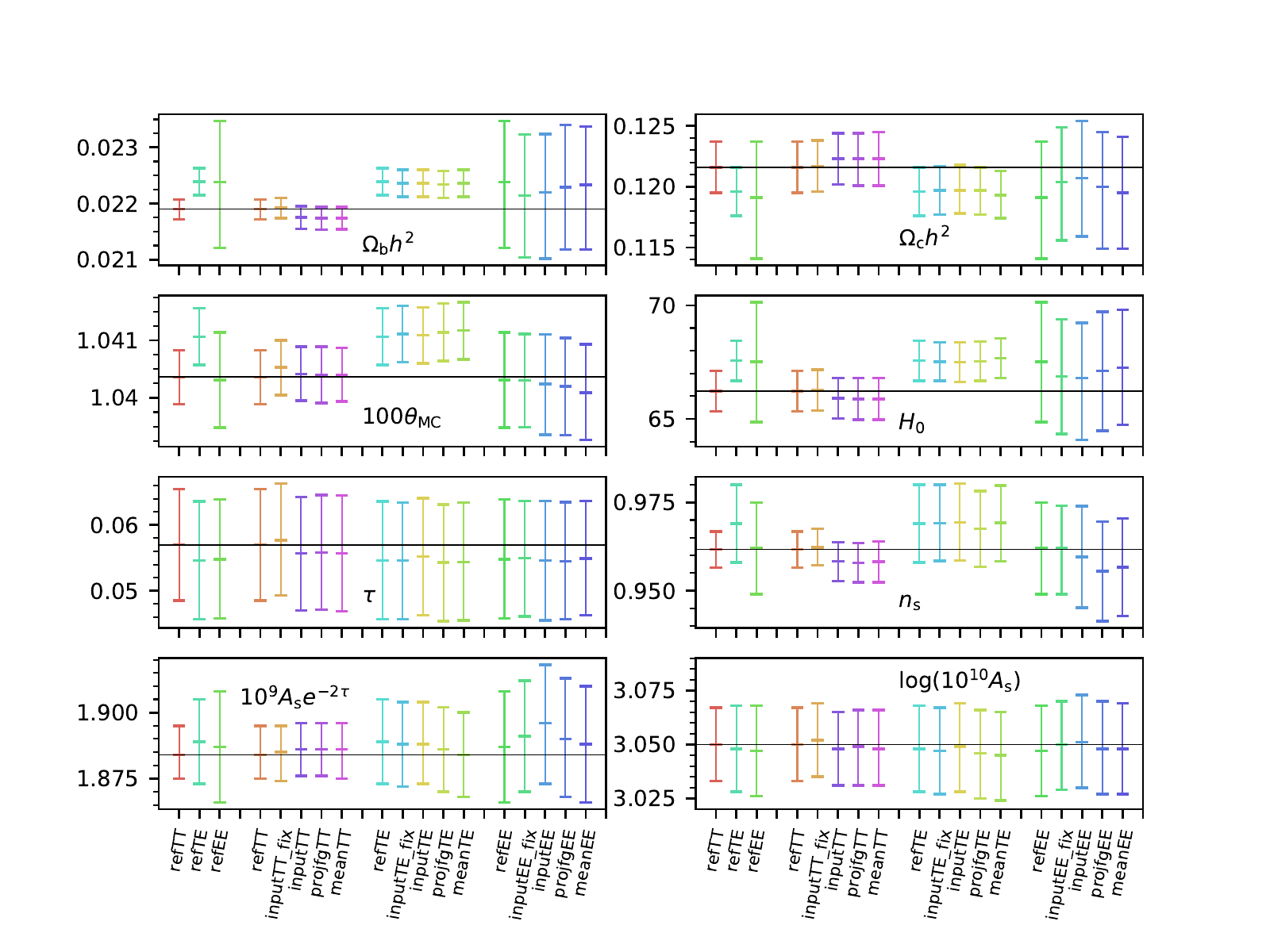}
\caption{Cosmological parameters for the set of end-to-end simulations cases discussed in Sect.~\ref{sec:valandro:simsresults}. In all cases the input for the simulations is the same input CMB (and in some cases foreground) map(s). Error bars correspond to the uncertainty for a single realization (including noise variance, even in cases where the simulations contains no noise). The horizontal lines corresponds to the best-fit values of the \texttt{refTT} case.}
\label{fig:TTEETE}
\end{center}
\end{figure*}

\begin{figure}[htbp!]
\includegraphics[width=\columnwidth]{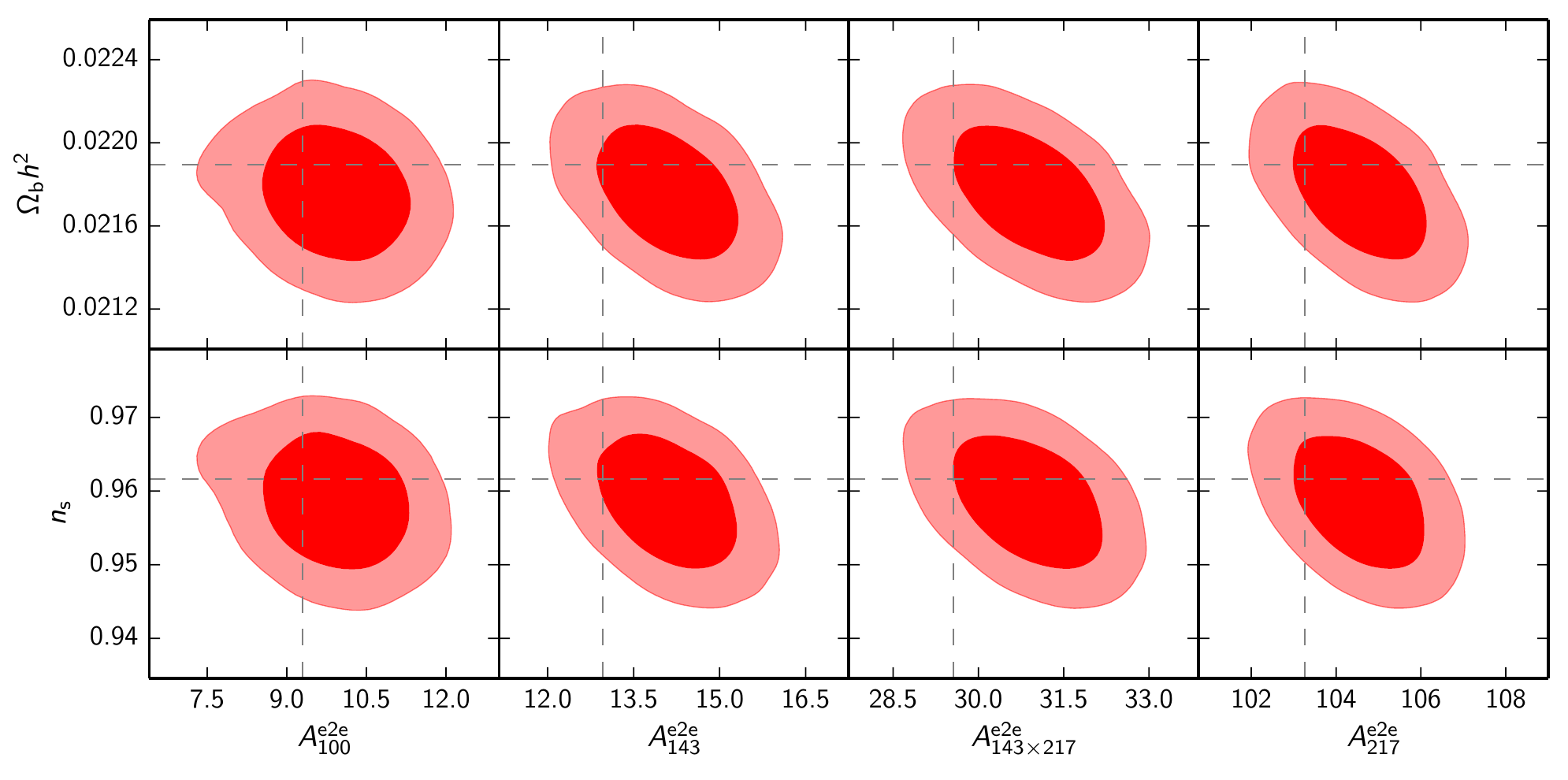}
\caption{Degeneracy between $n_{\rm s}$, $\Omega_{\rm b} h^2$, and foreground amplitudes in \TT\ for the single CMB and foreground realization used in the E2E simulations.  Horizontal and vertical dashed lines indicate the cosmological parameters obtained in the \texttt{refTT} case and the known foreground amplitudes of the simulation, while the red contours indicate the values obtained in the \texttt{inputTT} case. Upward shifts of up to $1.5\,\sigma$ in foregrounds are associated with downward shifts in cosmological parameters of close to $1\,\sigma$ and are due to chance correlations in the particular CMB and foreground realization that we are using for the simulations. This accounts for the shifts in foreground amplitudes found in Fig.~\ref{fig:TTTEEE_fg_whisker}}.
\label{fig:TT_fg_whisker}
\end{figure}

\subsubsection{Description}
The \Planck\ high-$\ell$ likelihood was already validated using simulations in 2015. The covariance-matrix analytical approximations were validated using 10\,000 dedicated CMB realizations \citep[see appendix~C.1.3 of][]{planck2014-a13}, and the likelihood implementation was shown to be able to recover the cosmological parameters on 300 FFP8 simulations (section~3.6 and appendix~C.2 of the same paper). The FFP8 simulations explored the CMB and noise realizations, but used the same fixed foreground signals \citep{planck2014-a14}.
For those validation tests, some modifications had to be implemented to the likelihood in order to correctly account for the FFP8 foreground and noise model, which were both slightly different from the real sky. 
With these modifications, we showed that both the TT and joint TT, TE, and EE likelihood were able to recover the input cosmological parameters on average. 
A small bias was exhibited in $\ns$ for the TT likelihood (around $0.1\,\sigma$). This was attributed to limitations of the Gaussian approximation in the lower multipole range of the likelihood ($\ell\,{<}\,100$), since 
a similar effect was found when changing the hybridization scale between the low- and high-multipole regimes. As discussed in \citetalias{planck2014-a13}, this bias was reduced when including the TE and EE data in the joint likelihood.

While these results on FFP8 were found to be very reassuring, those simulations only incorporated a simplified model of the noise and systematic residuals.
This limitation was overcome by the FFP10 simulations, which have a higher fidelity modelling of noise and systematics. 
In \citetalias{planck2014-a13} (Sect.~3.7), only the first few FFP10 simulations available at the time were analysed.
In this work, we expand this analysis to the 300 FFP10 simulations, which were also used to build and validate the HFI-based low-$\ell$ polarization likelihood presented Sect.~\ref{sec:lo-ell:hfi}. 

The FFP10 simulations are described in great detail in \citetalias{planck2016-l03}, and we repeat here only their most important features. 
Ideally, in order to build a simulation, one should generate a sky realization (CMB plus foregrounds in different frequency bands), apply the instrumental beam, and produce synthetic TOIs using the \Planck\ scanning strategy. Those TOIs then need to be convolved with the instrumental effects, which include noise and systematics, before applying the \Planck\ data-processing and mapmaking pipeline. 
All these steps are numerically very costly, and to simplify them it was decided to divide the process in two parts. Firstly, 1000 sky realizations (including convolution by the beams and scanning strategy at the map level) are produced. Secondly, 300 end-to-end (E2E) simulations of the HFI data are made, implementing the full pipeline described above (going through all the steps of the full simulation, including TOI production and processing), but for a single-sky realization (CMB plus foregrounds). Then, approximate E2E realizations can be produced by ``CMB-swapping,'' which means that the input CMB sky (convolved by the beam and scanning strategy) is removed from the E2E realizations and replaced by one of the 999 remaining CMB simulations.
This approach was found to be satisfactory for the low-$\ell$ data analysis (where we can ignore beam effects), and good enough for the exploration of Monte Carlo based uncertainties for lensing and non-Gaussianity tests (which for example ignore the temperature to polarization leakages). 
However, our investigations showed that it was not suitable for the validation of the high-$\ell$ likelihood.
A part of the systematic effects is a function of the sky realization; thus removing the input CMB map from the E2E realizations leaves an additional small systematic residual, which depends on that particular realization and that is common to all our approximate E2E simulations. We find that this residual is particularly important at small scales, where beam effects, temperature-to-polarization leakage, and subpixel effects have to be taken into account. While the effect on cosmological parameters can be ignored for a single realization, when averaging our results over numerous CMB realizations the bias becomes significant. 
For this reason, we limit our exploration of the noise and systematic residuals to the set of the 300 single-sky-realization end-to-end simulations. 

Restricting the simulations to a single-sky realization makes the evaluation of small biases difficult, since we are not, in our simulations, exploring the CMB cosmic variance. For the same reason we will first compare the cosmological parameters recovered from the E2E simulations to the ones obtained on the input CMB maps (i.e., the CMB before applying beam and scanning strategy effects) and to the ones obtained at different levels of complexity of the simulations, which will separately explore the chance correlations with foregrounds and with beams and scanning-strategy effects. We already validated most of the likelihood using the FFP8 simulations in \citetalias{planck2014-a13}, so we will only focus here on the effect of the higher fidelity modelling of noise and systematics. Similarly, we
 will separately present the TT, TE, and EE results, while we will not discuss joint results that have already been validated using FFP8. 

\begin{figure*}[htbp!]
\includegraphics[width=18cm]{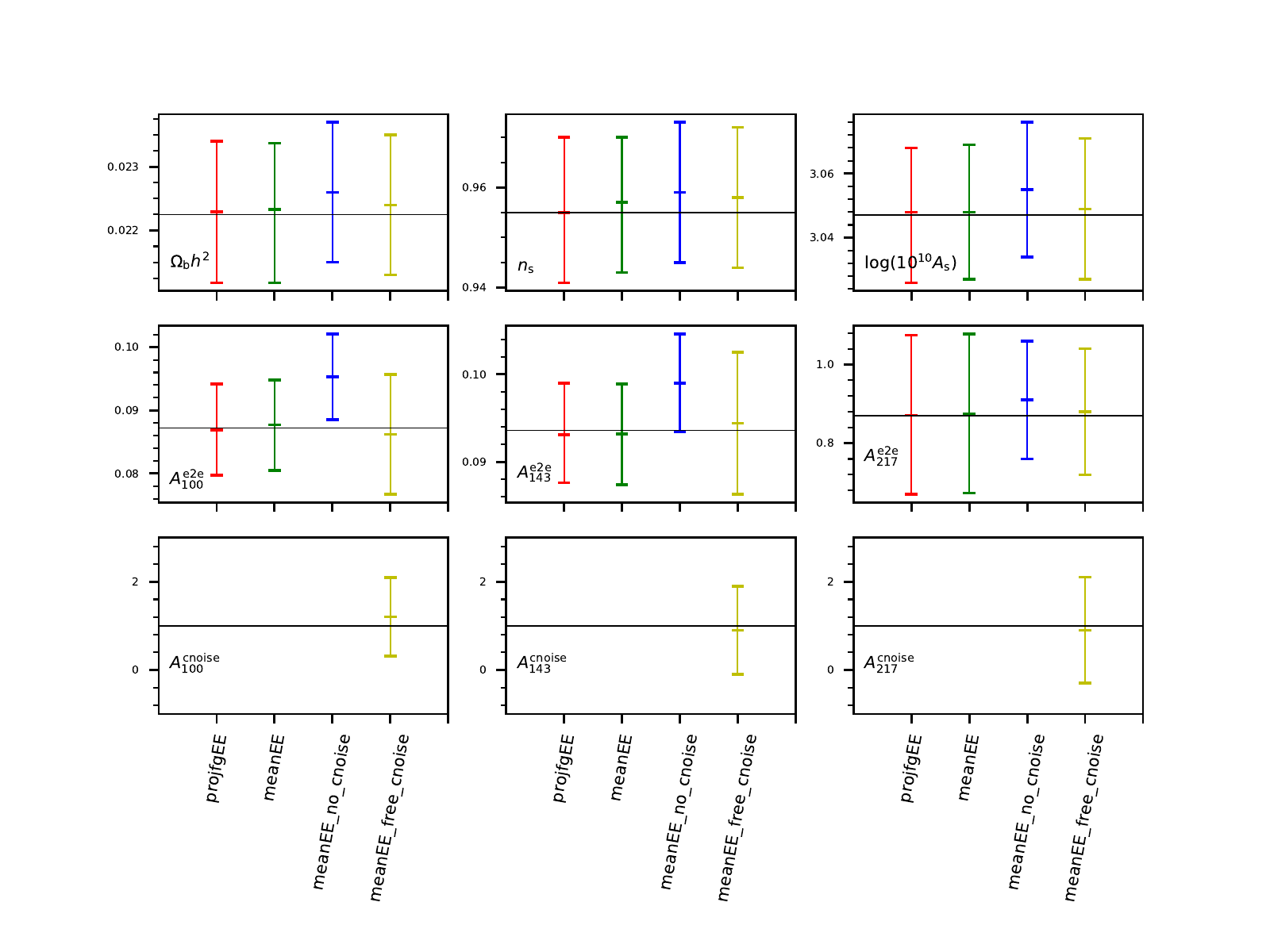}
\vspace{-0.5cm}
\caption{Marginal means and 68\,\% error bars on the cosmological parameters for \plik{}EE for the end-to-end simulations. We compare here the noise-free case ({\tt projfgEE}) with the mean of the 300 end-to-end simulations (to reduce the effect of the noise scatter), including the empirical correction (\texttt{meanEE}), ignoring it (\texttt{meanEE\_no\_cnoise}), or fitting for its amplitude but leaving the template constant (\texttt{meanEE\_free\_cnoise}). We only show here the parameters affected by template. Ignoring the effect results in a $1\,\sigma$ bias on the recovery of the dust contamination level at 100 and 143\,GHz, and smaller shift in $\omb$ and $\ns$.   When fitting for the amplitude of the effect, we obtain a marginal detection of the correlated noise.}
\label{fig:cnoise_e2e_EE_sim}
\end{figure*}

\begin{figure*}[htbp!]
\includegraphics[width=18cm]{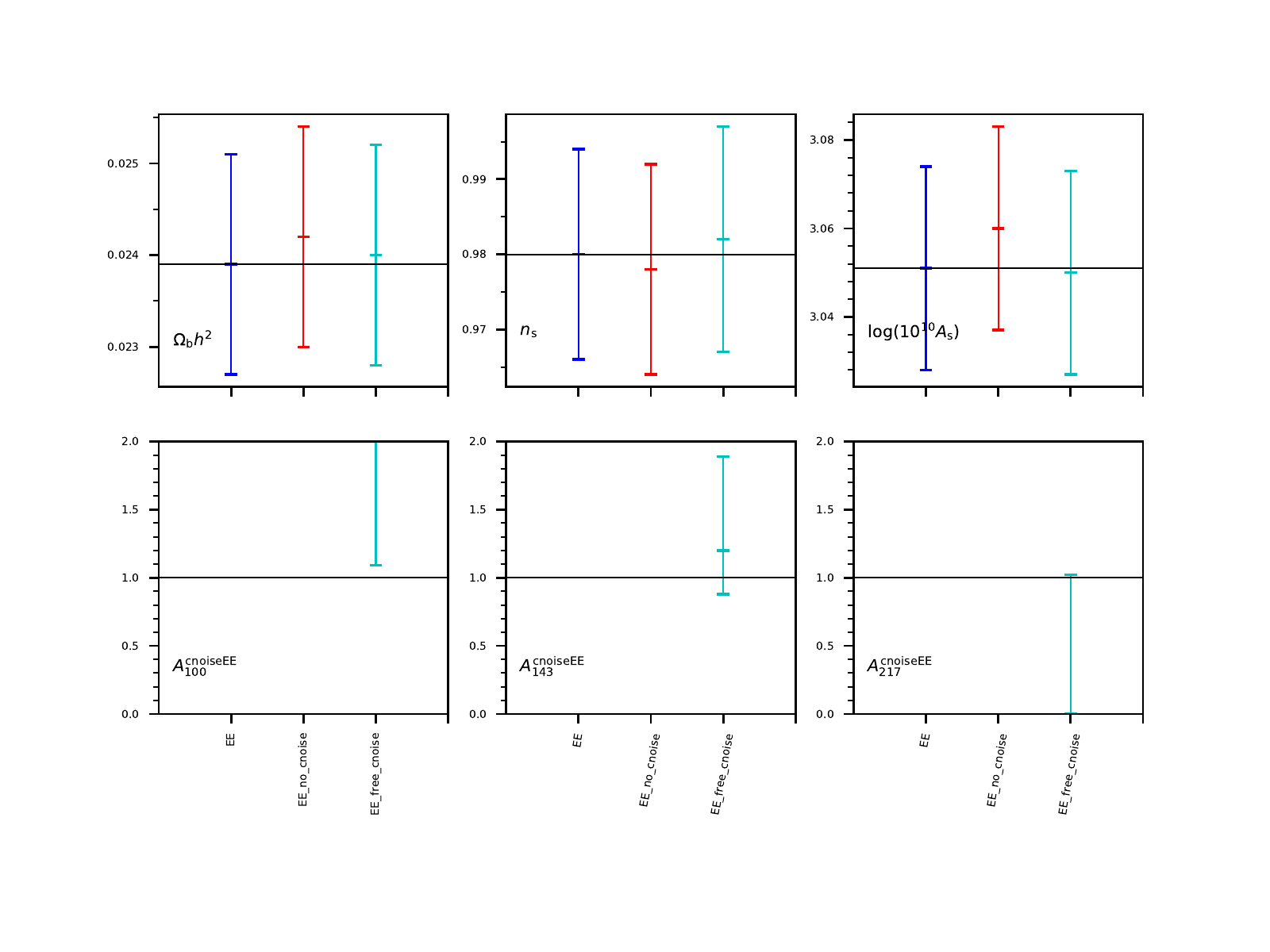}
\vspace{-1.0cm}
\caption{Similar to Fig.~\ref{fig:cnoise_e2e_EE_sim} (but now for the \plik{}EE likelihood on the data), we explore the impact on the cosmological parameter posteriors of the empirical correlated-noise correction. We compare here the baseline analysis (correction included) with the cases where the correction is ignored or its amplitude is left to vary.}
\label{fig:cnoise_e2e_EE_data}
\end{figure*}

We adapt the likelihood approximation we used for the data to the E2E specific case. Power spectra are still computed on the simulated half-mission E2E maps using the same masks as used for the data (even though the E2E simulations do not include point sources). The power spectra are beam-corrected using the same window functions used in the data (including the convolution by the scanning strategy), since the E2E simulations are designed to have the same effective beams.\footnote{The input CMB-plus-foreground maps from which we simulate the TOIs for the E2E realizations are convolved by the effective beams that includes residuals from the time-response corrections in the data processing. To avoid applying the time-response correction twice in the E2E maps, the processing is only applied to the noise and systematics parts of the TOIs.} 
The noise evaluation, which is needed to build the covariance matrices, is performed differently than for the data, since the HR maps are not available in the simulations (see Sect.~\ref{sec:hi-ell:datamodel:noise}); we measure the noise directly from the variance of the E2E simulations (since we have the same sky realization in each). 
Moreover, we simplify the foreground model using the approach we validated for the analysis of the FFP8 simulations in 2015. While for the data analysis, the foreground model is built from analytical calculations and empirical templates, the foreground model used in the simulations is directly measured on the input foreground maps. We use a single amplitude parameter for each frequency cross-spectrum (which will be called $A^{\rm e2e}$ in the following figures and tables).
As in the data case, we include in the nuisance model a correction for the temperature-to-polarization leakage and we account for the subpixel effects. The simulations do not include any polarization efficiency errors and in all the cases described below, we assume that we know it perfectly and set the polarization-efficiency correction parameters to one.We discussed in Sect.~\ref{sec:hi-ell:datamodel:inst} how when measuring the polarization efficiencies on the E2E simulations we do not reproduce the difference between the $EE$ and $TE$ based estimations observed for the data. As noted above for the beam transfer functions, the way the E2E simulations are produced ensures that we can use the same templates we use when analysing the data.
Finally, we need to account for a correlated noise component in the $100\times100$, $143\times143$, and $217\times217$ \EE{} power spectra. This correlated noise is measured directly in the simulations, and is also corrected for in the data (see Sect.~\ref{sec:hi-ell:datamodel:noise}). 
The covariance matrix is computed from those ingredients in the same way that we compute the data covariance. 

\subsubsection{Results on parameters from simulations}
\label{sec:valandro:simsresults}

\begin{figure}[h!]
\begin{center}
\includegraphics[trim={0.7cm 0.3cm 1.2cm 1.3cm},clip,width=\columnwidth]{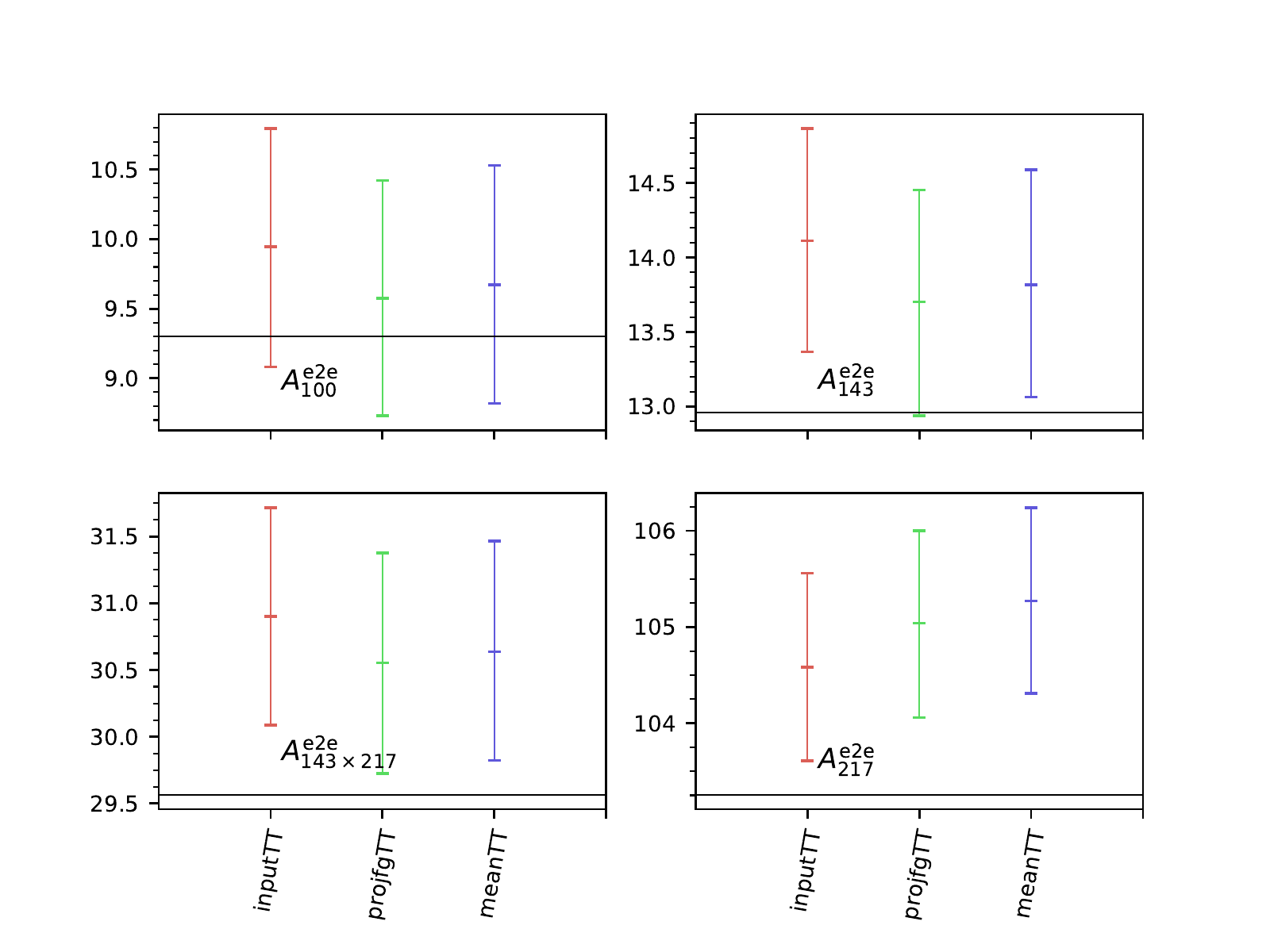}\\
\includegraphics[trim={0.7cm 0.3cm 1.2cm 1.3cm},clip,width=\columnwidth]{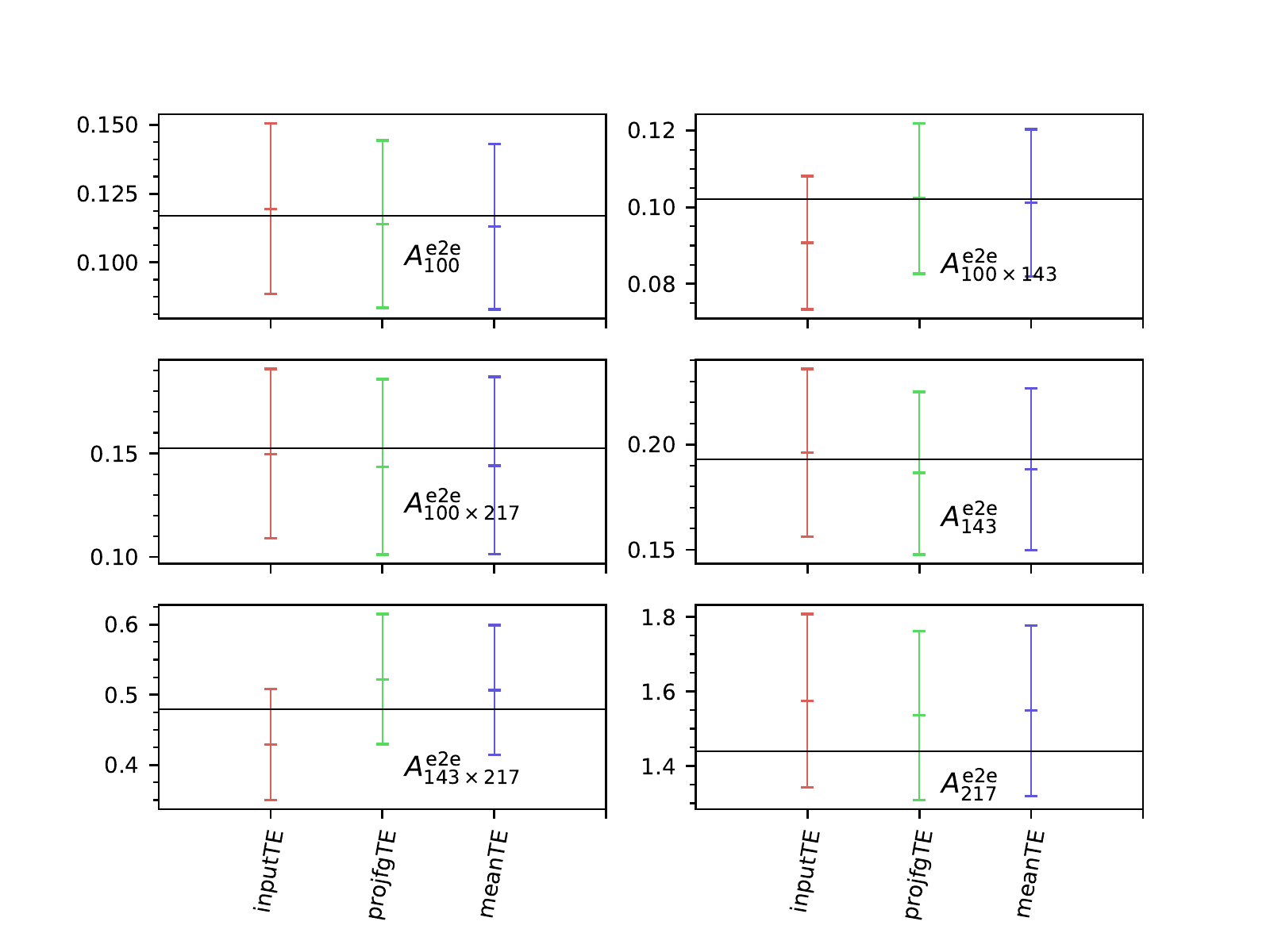}\\
\includegraphics[trim={0.7cm 0.3cm 1.2cm 1.3cm},clip,width=\columnwidth]{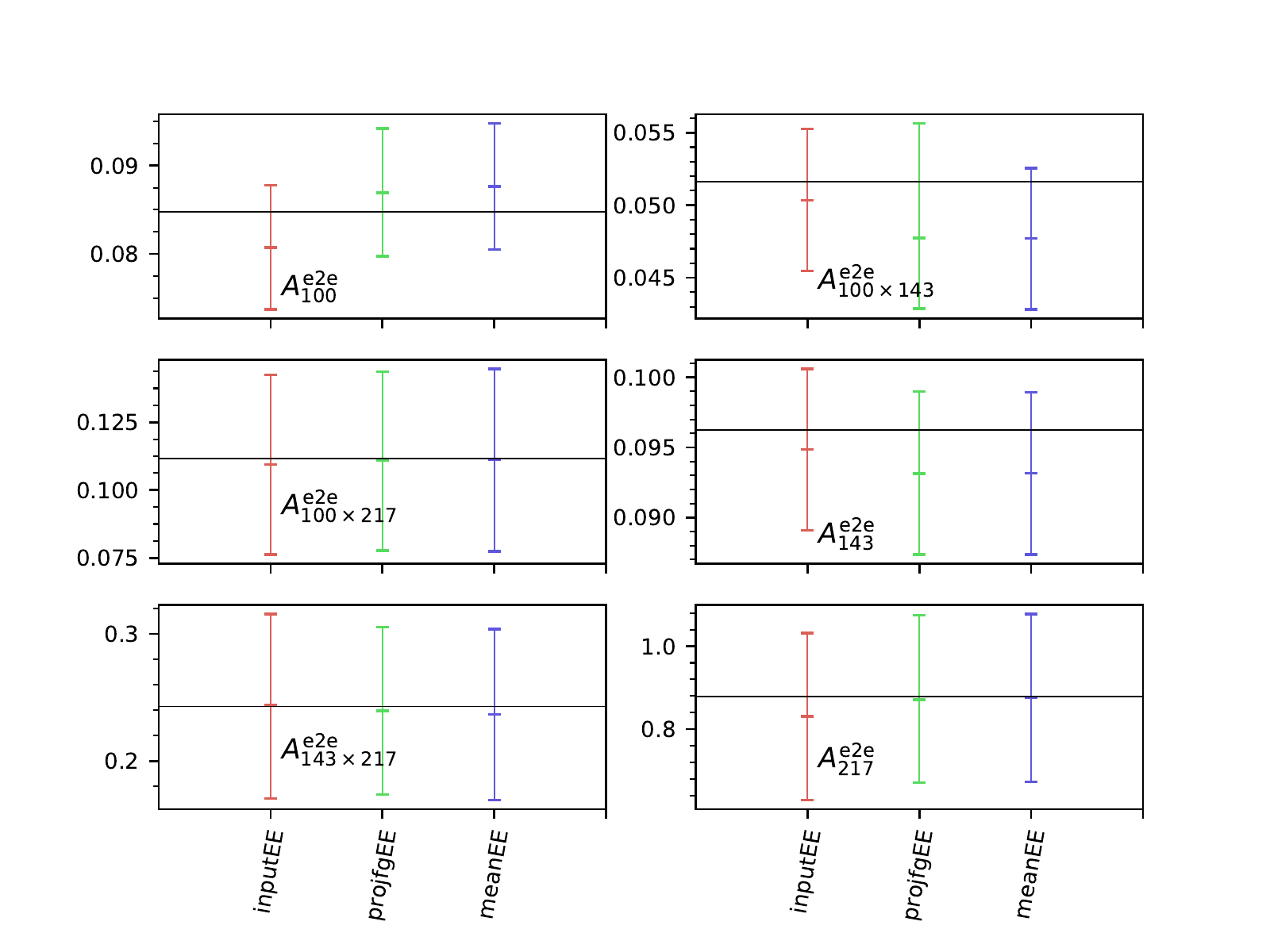}
\caption{\TT{} (top), \TE{} (middle), and \EE{} (bottom) foreground amplitudes obtained for the different simulation cases discussed Sect.~\ref{sec:valandro:simsresults}. The horizontal reference lines correspond to the central value of the foreground amplitude prior.}
\label{fig:TTTEEE_fg_whisker}
\end{center}
\end{figure}

\begin{figure*}[htbp!]
\begin{center}
\includegraphics[width=18cm]{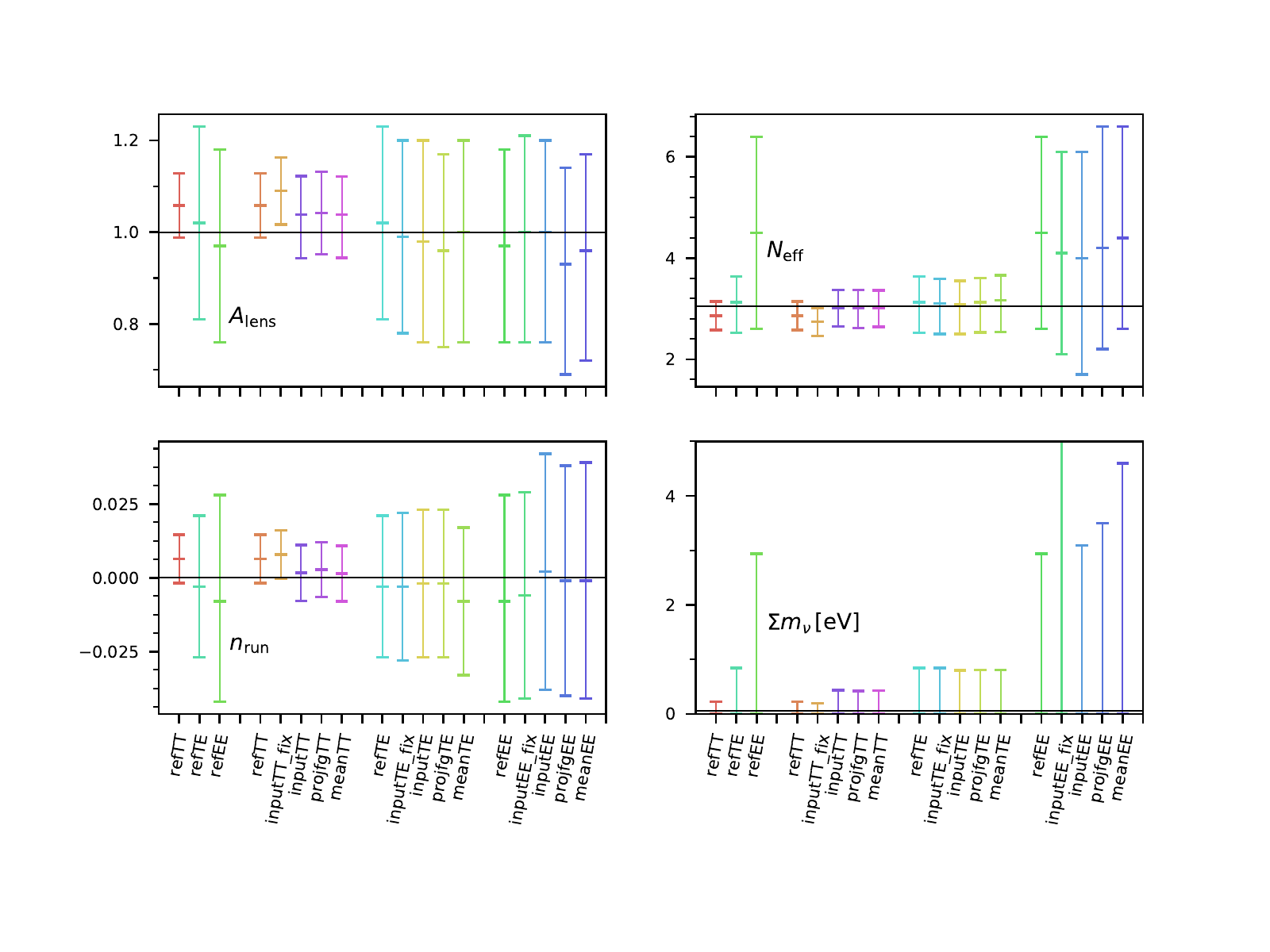}
\vspace{-1.0cm}
\caption{Extended parameters for the set of end-to-end simulation cases discussed in Sect.~\ref{sec:valandro:simsresults}. The horizontal reference line corresponds to the expected value for each extended parameter.}
\label{fig:ext}
\end{center}
\end{figure*}

Figure~\ref{fig:TTEETE} presents the cosmological results obtained on simulations when varying assumptions are made, as indicated with the following labelling convention:
\begin{enumerate}
\item \texttt{ref} is for the case when parameters are recovered from the input CMB maps without beams and scanning strategy convolution, and without foregrounds, noise, or systematics; 
\item \texttt{input} is for a similar case to \texttt{ref}, but when the CMB-plus-foregrounds input maps are used;
\item \texttt{projfg} is for the case when beams and scanning strategy effects are added to the input maps of \texttt{input};
\item \texttt{mean} is for when the means of the spectra over the 300 simulations are used in the likelihood.
\end{enumerate}
In each case, the likelihood described above is adapted to take into account the effects included in each case. For example, beam-transfer functions, beam-leakage effects, subpixel effects, as well as foregrounds or correlated-noise corrections, are absent from the likelihood used for the \texttt{ref} cases. {We, } however, keep the same covariance for all cases. In the example of \texttt{input} we run two different tests, one where the foreground parameters are explored along with the cosmological parameters (\texttt{input}) and one where we fix these to the actual level of foregrounds in the simulations (\texttt{input\_fix}). {We note} that in the latter case, the chance correlations between CMB and foregrounds are still in the spectra. Contrary to the data case, where we fix the dust amplitudes in EE, we will let them freely vary in the \texttt{inputEE} exploration.

In this subsection, we only discuss the values of the cosmological parameters, while in Sect.~\ref{sec:sim:var} we will present the statistical significance of parameter value differences over the 300 E2E realizations. 

We discuss first the \texttt{ref} group of results. As can be seen in Fig.~\ref{fig:TTEETE}, in the case of the particular sky realization used in the E2E, up to $2\,\sigma$ shifts can be observed in cosmological parameter constraints between the TT and TE likelihoods, which is to be expected. In the 2015 paper
\citepalias{planck2014-a13} we already checked that such shifts vanish on average when exploring the CMB cosmic variance in a similar setting.
Comparing the \texttt{input} explorations with the \texttt{ref} ones, we see small deviations (up to $1\,\sigma$) of the cosmological parameters, which are signs of the degeneracy between the cosmological parameters and foregrounds, as also shown by the joint contours in Fig.~\ref{fig:TT_fg_whisker}. This is true for EE and in particular for TT, for which the effect is stronger. This is due to the chance correlations between the CMB and the foregrounds, which are not removed even in the case where the foreground-amplitude parameters are fixed (as in the \texttt{input\_fix} case). We know from the FFP8 test performed in 2015, that when marginalizing over 300 CMB realizations, the bias due to the chance correlations vanishes, as expected \citepalias[section~3.6 of][]{planck2014-a13}.

When additionally considering the scanning strategy and beam effects, the \texttt{projfg} explorations are in good agreement with the \texttt{input} ones. The TT and TE cases, where the beams and leakage effects (especially for TE) are important, are in excellent agreement with the \texttt{input} ones. 
We observe $\approx0.3\,\sigma$ shifts of $\ns$ in EE.\footnote{Throughout this section all of the shifts that are reported in units of $\sigma$ refer to possible biases estimated in terms of the expected error on the cosmological parameters recovered from the corresponding ($TT$, $TE$, or $EE$) data.} Since we have only a single realization, we have not explored this shift any further.

Finally, we compare those results with the \texttt{mean} cases, which would contain any uncorrected residual systematic captured in the E2E simulations. Here the cosmological parameters are in excellent agreement with the \texttt{projfg} ones and we do not observe any significant residual bias in $TT$ and $TE$. In $EE$ also, we do not observe a significant bias, but this is expected, since we remind the reader that we correct the $EE$ power spectra at this stage for a bias observed in the simulations for the $100\times100$, $143\times143$, and $217\times217$ $EE$ spectra. A similar correction is applied to the data, as discussed in Sect.~\ref{sec:hi-ell:datamodel:noise}. 
The effect of the EE bias on cosmological parameters is explored in Fig.~\ref{fig:cnoise_e2e_EE_sim}, where the \texttt{projfgEE} constraints are compared with those of the \texttt{meanEE} (with the correction) and to the case where the correction is ignored (\texttt{EE no cnoise}), or left free to vary in amplitude (\texttt{EE free cnoise}).
When ignoring the correction, small systematic shifts are found in the cosmological parameters, in particular a bias of $0.25\,\sigma$ in the case of $\omb$, a smaller one on $\ns$ and a $1\,\sigma$ bias is observed in the foreground parameters. The latter is not surprising and we discussed in Sect.~\ref{sec:hi-ell:datamodel:noise} how the bias template was similar in shape to the dust one. When fitting jointly the correlated-noise amplitude and the other cosmological and nuisance parameters, we recover a non-zero amplitude for the correlated noise, with the right order of magnitude, but with a low significance (around $1\,\sigma$). Figure~\ref{fig:cnoise_e2e_EE_data} shows the same test for the data, where ignoring the correction causes a similar $0.25\,\sigma$ shift in $\omb$. When jointly fitting the correlated-noise amplitude and the other cosmological and nuisance parameters in the data, the amplitude of the correlated-noise template prefers values larger than unity. As noted before, the correlated-noise template has a shape similar to that of the dust and we discussed in Sect.~\ref{sec:hi-ell:datamodel:gal} (as shown in Fig.~\ref{fig:hil:2dEEdustprior}) that the $100 \times 100$\ $EE$ spectrum prefers a higher amplitude of dust than the one deduced from correlations with the 353-GHz map.  As in the simulations, uncertainties on the amplitude are large and we cannot conclude that there is a preference in the data for the correction. We noted in Sect.~\ref{sec:hi-ell:datamodel:noise} that the correction improves the $\chi^2$ in the data by $\Delta\chi^2=9$.

Figure~\ref{fig:TTTEEE_fg_whisker} is the equivalent of Fig.~\ref{fig:TTEETE} for the foreground parameters. Recall that, contrary to the data case, here we use a single foreground template, accounting for all sources of astrophysical contaminants, and fit for a single amplitude for each. As discussed above, in TT, the CMB-foreground chance correlations have an almost $2\,\sigma$ impact on the recovered foreground parameters compared to the input values (the black lines in Fig.~\ref{fig:TTTEEE_fg_whisker}). These input values are obtained by measuring the amplitude of the foreground contamination in the unique foreground realization that is added to the CMB maps to build the reference sky. In TE and EE, we report small shifts in the foreground recovered amplitude when including the beam and scanning effects and possible residual systematics.

Finally, Fig.~\ref{fig:ext} shows the results when exploring classical \LCDM{} extensions (specifically $A_{\mathrm{L}}$, $N_{\mathrm{eff}}$, $d\ln n_{\rm s}/d\ln k$, and $\Sigma m_\nu$). As before, the extended parameters have a similar level of agreement with the input as in the \LCDM\ case. 

\subsubsection{Statistical quantification of biases from simulations}
\label{sec:sim:var}
Table~\ref{tab:e2e} summarizes the shifts and spreads of the individual end-to-end realizations with respect to the cosmological parameter values and foreground amplitudes obtained using \texttt{projfg}.   The shifts in Table~\ref{tab:e2e} are computed using
\begin{equation}
\Delta = \sqrt{N_\mathrm{sim}}\left(\frac{1}{N_\mathrm{sim}}\sum_{i=1}^{N_\mathrm{sim}}\frac{X_{\texttt{e2e}} - X_{\texttt{projfg}}}{\sigma_i}\right),\ \ \ \mathrm{with\  }N_\mathrm{sim}=300,
\label{eq:e2e_shifts}
\end{equation}
where $X$ is a cosmological or foreground parameter. Measuring shifts compared to the \texttt{projfg} parameters nulls the effects of the cosmic variance of the CMB and foregrounds, as well as chance correlations. However, the chance correlations between CMB, foregrounds, and noise and systematic biases are not fully explored, since we are only varying the noise. Using \texttt{projfg} as a reference also ensures that we are only exploring the noise and systematic-induced biases, and not residual beam effects (e.g., residual beam errors, leakage, and subpixel template inaccuracies).
The shifts are overall satisfactory, apart from a $2.8\,\Delta$ one in $\ns$ for TE. This would correspond to a $0.16\,\sigma$ bias for a single realization, which can actually be seen in Fig.~\ref{fig:TTEETE} when comparing the 
\texttt{projfgTE} and \texttt{meanTE} results. Given that we cannot properly explore the CMB cosmic variance, it is difficult to better assess this possible bias. Contrary to EE, where we found correlated noise in the simulations, we do not find a similar effect in TE and have no template to correct for this bias at the spectrum level. 
For the single E2E CMB sky, we do not reproduce the $\ns$ bias on TT that was reported in 2015. 

Table~\ref{tab:e2e_fg} displays the shifts for foreground parameters. These are much larger and perhaps less reassuring than the cosmological-parameter ones, in particular in TT. For 217 GHz, the measured bias reaches almost $5\,\Delta$ (corresponding to a systematic $0.3\,\sigma$ shift for a single realization). It seems clear that there is an important degeneracy between the foreground parameters and the noise realizations. This is not entirely surprising, since the foreground contamination corresponds to a small contribution to the spectra compared to the CMB. We do see variations of the foreground parameters in the data whenever modifying the cosmological model. {We note}, however, that when exploring the CMB variance (but not the foreground one), as we did with the FFP8 simulations in 2015, and using a foreground model with an extra degree of freedom accounting for the CIB instead of using a single template, we did not find such  biases on the foreground parameters \citepalias[see section~3.6 of][]{planck2014-a13}.

To summarize, we report a $0.16\,\sigma$ potential systematic bias on $\ns$ in TE. We observed a $0.3\,\sigma$ shift for $\ns$ for EE, seen when including projection effects in the simulations.  Finally, we have potentially introduced a $0.25\,\sigma$ bias in $\omb$ for EE if the correlated noise that is observed in the simulations and then used to form a correction is not actually present in the data. We remind the reader that the correction improves the data by $\Delta\chi^2=9$, but is known to be degenerate with the dust correction.

\begin{table}[htbp!] 
\begingroup 
\newdimen\tblskip \tblskip=5pt
\caption{Mean cosmological parameter shifts in units of $\Delta$ (defined in  
Eq.~\ref{eq:e2e_shifts}).}
\label{tab:e2e}
\vskip -3mm
\footnotesize
\setbox\tablebox=\vbox{
\newdimen\digitwidth
\setbox0=\hbox{\rm 0}
\digitwidth=\wd0
\catcode`*=\active
\def*{\kern\digitwidth}
\newdimen\signwidth
\setbox0=\hbox{+}
\signwidth=\wd0
\catcode`!=\active
\def!{\kern\signwidth}
\newdimen\decimalwidth
\setbox0=\hbox{.}
\decimalwidth=\wd0
\catcode`@=\active
\def@{\kern\decimalwidth}
\openup 3pt
\halign{
\hbox to 1.0in{#\leaderfil}\tabskip=1em&
    \hfil#\hfil\tabskip=2em&
    \hfil#\hfil\tabskip=2em&
    \hfil#\hfil\tabskip=0pt\cr
\noalign{\doubleline}
\omit\hfil Parameter \hfil& *TT& *TE& *EE\cr
\noalign{\vskip -2pt}
\omit& *$[\Delta]$& *$[\Delta]$& *$[\Delta]$\cr
\noalign{\vskip 5pt\hrule\vskip 5pt}
$\Omega_{\rm b} h^2$&    $-0.827$& $!0.593$& $-0.140$\cr
$\Omega_{\rm c} h^2$&    $!0.255$& $-1.84*$& $-0.768$\cr
$\theta$&                $-0.068$& $-0.442$& $-1.39*$\cr
$\tau$&                  $-0.858$& $!0.112$& $!0.203$\cr
$\ln(10^{10}A_{\rm s})$& $-0.907$& $-0.724$& $-0.264$\cr
$n_{\rm s}$&             $!0.428$& $!2.82*$& $!1.18*$\cr
\noalign{\vskip 5pt\hrule\vskip 3pt}
}}
\endPlancktable
\endgroup
\end{table}

\begin{table}[htbp!] 
\begingroup 
\newdimen\tblskip \tblskip=5pt
\caption{Mean foreground amplitude shifts in units of $\Delta$ (defined in  
Eq.~\ref{eq:e2e_shifts}).}
\label{tab:e2e_fg}
\vskip -3mm
\footnotesize
\setbox\tablebox=\vbox{
\newdimen\digitwidth
\setbox0=\hbox{\rm 0}
\digitwidth=\wd0
\catcode`*=\active
\def*{\kern\digitwidth}
\newdimen\signwidth
\setbox0=\hbox{+}
\signwidth=\wd0
\catcode`!=\active
\def!{\kern\signwidth}
\newdimen\decimalwidth
\setbox0=\hbox{.}
\decimalwidth=\wd0
\catcode`@=\active
\def@{\kern\decimalwidth}
\openup 3pt
\halign{
\hbox to 1.0in{#\leaderfil}\tabskip=1.0em&
    \hfil#\hfil\tabskip=2em&
    \hfil#\hfil\tabskip=2em&
    \hfil#\hfil\tabskip=0pt\cr
\noalign{\doubleline}
\omit\hfil Amplitude\hfil& TT& TE& EE\cr
\noalign{\vskip -2pt}
\omit& $[\Delta]$& $[\Delta]$& $[\Delta]$\cr
\noalign{\vskip 5pt\hrule\vskip 5pt}
$A^{\rm e2e}_{100}$&           $2.48$& $-0.587$& $!1.69*$\cr
$A^{\rm e2e}_{100\times 143}$& \dots&  $-0.930$& $-0.013$\cr
$A^{\rm e2e}_{100\times 217}$& \dots&  $!0.155$& $!0.202$\cr
$A^{\rm e2e}_{143}$&           $3.36$& $!1.36*$& $!0.104$\cr
$A^{\rm e2e}_{143\times 217}$& $2.57$& $-2.53*$& $-0.902$\cr
$A^{\rm e2e}_{217}$&           $4.95$& $!1.34*$& $!0.558$\cr
\noalign{\vskip 5pt\hrule\vskip 3pt}
}}
\endPlancktable
\endgroup
\end{table}

\section{Joint likelihood}
\label{sec:joint}

As in the previous releases, the final joint \planck\ CMB likelihood is formed using a combination of the different approximations
described in the earlier parts of this paper. This hybrid low-$\ell$/high-$\ell$ approach has been advocated in \cite{Efstathiou:2004} and \citet{Efstathiou:2006} and applied to WMAP data \citep[first
in][]{2007ApJS..170..377S}, as well as being used for previous \planck\ releases \citepalias{planck2013-p08,planck2014-a13}. 

Given the numerous different likelihood approximations described in this paper, we can build different hybridization schemes, depending on specific goals, be it obtaining better constraints on cosmological parameters, exploring all of the available modes in the data, or marginalizing over the numerous nuisance parameters introduced in the high-$\ell$ likelihood to speed up evaluations.
A summary of the low- and high-multipole likelihood options is given in Table~\ref{tab:likelihooddef}, with a reminder of the multipole range, spectrum coverage, and other specifics of each. 

The baseline hybrid likelihood that is used in most of the
2018-release parameters paper \citepalias{planck2016-l06}, labelled
\planckall, consists of a combination of the \commander\ likelihood,
the \simall\ likelihood, and the \plik\ likelihood. This choice yields
the best overall constraints on the cosmological parameters (when comparing with the low-$\ell$ LFI based likelihood, \bflike), owing to the better S/N of the HFI data in polarization at large scales. Note, however, this scheme ignores correlations between the $TT$ and $EE$ large-scale spectra and discards entirely the $TE$ spectra at large scales. We evaluate in Sect.~\ref{subsubsec:final_considerations} the effect of this choice in terms of lost constraining power on $\tau$, which is found to be small.  Retaining the \plik\ likelihood at small scales also allows for the joint exploration of cosmological parameters with foreground and nuisance parameters at small scales (instead of marginalizing over them, as in the \pliklite\ likelihood).
We have particularly focused our efforts on the validation of each of the ingredients of this particular choice.
Our best estimate of the $TT$, $TE$, and $EE$ power spectra using this hybridization scheme are shown in Figs.~\ref{fig:coaddedTT}, \ref{fig:coaddedTE}, and \ref{fig:coaddedEE}. While we present a $TE$ spectrum estimate at low-$\ell$ from the \simall\ approximation, we do not include it in the cosmological results, for the reasons discussed in Sect.~\ref{subsubsec:final_considerations}.

Replacing in this scheme the \commander\ and \simall\ likelihoods by
the \bflike\ one allows us to probe the
temperature-polarization correlations at large scales, at the price of
decreasing the constraining power of the polarization-only data at large scales.

At high multipoles, the \plik\ likelihood can be replaced by the
\camspec\ one to explore alternative model choices, e.g., the
different masks in polarization and the different approaches to
correction of the polarization efficiencies. In the case of
the polarization-efficiency corrections, some extra nuisance
parameters can be explored in the \plik\ likelihood, compared to the baseline setting, in particular
different overall TT-to-EE and TT-to-TE calibration (as allowed by \camspec), as well as switching from the baseline map-based to the spectrum-based polarization-efficiency correction model.
\begin{table}[htbp!] 
\begingroup 
\newdimen\tblskip \tblskip=5pt
 \caption{{Likelihood approximations summary.}}
  \label{tab:likelihooddef}
\vskip -3mm
\footnotesize
\setbox\tablebox=\vbox{
\newdimen\digitwidth
\setbox0=\hbox{\rm 0}
\digitwidth=\wd0
\catcode`*=\active
\def*{\kern\digitwidth}
\newdimen\signwidth
\setbox0=\hbox{+}
\signwidth=\wd0
\catcode`!=\active
\def!{\kern\signwidth}
\newdimen\decimalwidth
\setbox0=\hbox{.}
\decimalwidth=\wd0
\catcode`@=\active
\def@{\kern\decimalwidth}
\openup 3pt
\halign{
\hbox to 1.25in{#\leaderfil}\tabskip=0.5em&
    \hfil#\hfil\tabskip=1em&
    \hfil#\hfil\tabskip=0pt\cr
\noalign{\doubleline}
\omit\hfil Name\hfil& Spectra& $\ell$ range\cr
\noalign{\vskip 5pt\hrule\vskip 5pt}
\multispan3\hfil$\ell<30$\hfil\cr
\noalign{\vskip 3pt}
\commander& \TT{}& *2--29\cr
\simall& \EE{}& *2--29\cr
\bflike& \TT{},\TE{},\EE{},\BB,\TB,\EB& *2--29\cr
\noalign{\vskip 5pt\hrule\vskip 5pt}
\multispan3\hfil$\ell\ge30$\hfil\cr
\noalign{\vskip 3pt}
\plik& \TT& *30--2508\cr
\omit& \TE,\EE& *30--1996\cr
\camspec$^{\rm a}$& \TT& *30--2500\cr
\omit& \TE,\EE& *30--2000\cr
\pliklite$^{\rm b}$& \TT& *30--2508\cr
\omit& \TE,\EE& *30--1996\cr
\noalign{\vskip 5pt\hrule\vskip 3pt}
}}
\endPlancktable
\tablenote {{\rm a}} Alternative polarization masking and PE corrections.\par
\tablenote {{\rm b}} Nuisance-marginalized \plik\ likelihood.\par
\endgroup
\end{table}

Finally, the \plik\ likelihood can also be replaced by the \pliklite\ likelihood if one wants to avoid exploration of the high-$\ell$ nuisance parameters.

In all cases, and following the choice made in 2013 and 2015, all of our possible hybrid schemes implement a sharp transition between the low and high multipole regimes at $\ell\,{=}\,30$, and ignore correlations between the low and high multipoles.
Limitations of this approach in temperature were explored in
\citetalias{planck2014-a13}, where it was shown that pushing the
small-scale approximation down to $\ell_\mathrm{min}\,{=}\,30$ was
potentially biasing the determination of the primordial scalar
perturbation slope $\ns$, but only by around
$0.1\,\sigma$. While this estimate was obtained on simulations, a similar behaviour was observed on the data when
changing the hybridization scale in TT. For this reason, a different
choice of transition multipole might have been better for
temperature. The computational cost of the pixel-based polarization low-$\ell$
likelihood forces us to retain a low transition multipole. The \simall\ likelihood could in theory go to a higher multipole, but only for $EE$, as we discussed in Sect.~\ref{subsubsec:final_considerations}.
We decided to avoid the complexity of having different transition multipoles in temperature and polarization and maintained the transition at $\ell=30$.
Being limited to a low transition multipole, we did not reproduce for polarization the hybridization scheme validation that we implemented for temperature. The polarization data have less constraining power than TT and we expect that biases possibly induced by the hybridization scheme will be even lower in polarization. 
Validation on the FFP8 simulations performed in 2015 showed that the $\ns$ bias was reduced when adding polarization data \citepalias{planck2014-a13}. At the level of the precision of our likelihood approximation in polarization, we did not further explore the effect of the sharp $\ell\,{=}\,30$ transition.

\section{Summary and Conclusions}
\label{sec:conclusion}
The \Planck\ 2018 power spectra are displayed in Figs.~\ref{fig:coaddedTT}, \ref{fig:coaddedTE}, and \ref{fig:coaddedEE}. These CMB power spectra, along with the likelihood approximations we described in this paper and the lensing likelihood described in \citep[][]{planck2016-l08}, form the basis 
of the main cosmological results of \Planck, as presented in \citetalias{planck2016-l06} and \cite{planck2016-l10}, and summarized in \cite{planck2016-l01}. 
The tightening and improved robustness of the cosmological constraints thanks to the full use of the \Planck\ polarized data are the main achievements of the 2018 release and the main topic of this paper.
 
The \Planck\ reference likelihood approximation used in this release is very similar to the approach taken in the previous releases in 2013 \citepalias{planck2013-p08} and 2015 \citepalias{planck2014-a13}. 
Different approaches are used for the low-multipole ($\ell\,{<}\,30$) and high-multipole ($\ell\,{\ge}\,30$) regimes. These have been improved through numerous ameliorations at all stages of the data processing,  both at the time-ordered data (for the LFI 30-GHz channel) and mapmaking stages \citepalias{planck2016-l02,planck2016-l03} as well as at the power-spectrum estimation and likelihood building stages. 
One of the key ingredients to engender these improvements has been the availability of high-fidelity end-to-end simulations, explained in different sections of this paper and summarized below.

\subsection{Low-$\ell$ summary}

At low multipoles, the approximations and data sets have evolved significantly between releases, and the 2018 one is no exception. The 2018 likelihood approximation is a combination of two approximations, one for the large-scale temperature data, and another for large-scale polarization, ignoring correlations between the two.\footnote{We already checked in \citetalias{planck2014-a13} the impact of ignoring the temperature-polarization correlations at low multipoles.}

The temperature likelihood is an evolution of the one we presented in 2013. It uses all of the available \Planck\ data and the \commander\ method to build a Blackwell-Rao approximation of the posterior distribution of the large-scale $TT$ spectrum, marginalized over the large-scale foreground contamination; it is described in Sect.~\ref{sec:lo-ell:TT}.
The polarization likelihood, \simall{}, is an evolution of the method proposed in \cite{planck2014-a10}, which relies on simulations to approximate the likelihood of the large-scale $EE$ cross-power spectrum of two HFI channel maps (100\,GHz and 143\,GHz); this likelihood is discussed in Sect.~\ref{sec:lo-ell:hfi}. The new likelihood at large scales provides the tightest cosmological constraint achieved to date on the reionization optical depth parameter, $\tau$. 

This 2018 approach at large scales differs from the 2013 approximation \citepalias{planck2013-p08}, which used the same \commander\ method for temperature and a pixel-based approximation using WMAP data for polarization, along with \Planck\ HFI 353-GHz polarization data to mitigate dust contamination. It is also different from the 2015 method \citepalias{planck2014-a13}, which used a single pixel-based approximation for both temperature and polarization, adopting a combination of the \commander\ best CMB temperature map and the LFI 70-GHz polarization maps (corrected for dust and synchrotron contamination using, respectively, the HFI 353-GHz and LFI 30-GHz data). The 2015 pixel-based approach allowed us to probe the correlation between temperature and polarization at large scales; this ability is important in order to discriminate between some extended cosmological models involving $B$-mode correlations, as well as non-rotationally invariant cosmologies, provided that the CMB signal-covariance matrix is appropriately modified. {In contrast, a} power-spectrum-based likelihood, such as the 2018 baseline, may not be used directly to test isotropy-violating models, since rotational invariance is assumed in its derivation. For this reason, we also improved in this release the 2015 low-$\ell$ method and have described an updated version of it in Sect.~\ref{sec:lo-ell:lfi}, which includes more data and a more thorough validation than in 2015. Owing to its lower S/N, this \lowTEB\ likelihood is not as constraining as the baseline choice, but its results are entirely consistent with it, providing further validation. We remark that \lowTEB\ is the only data set we provide that accounts for correlations between temperature and polarization at large scales.

In the case of polarization, \citetalias{planck2016-l03} describes the central role of the end-to-end simulations for modelling the large-scale systematics that previously prevented the use of HFI data in this regime \cite[an approach that was already outlined in][]{planck2014-a10}. In Sect.~\ref{sec:lo-ell:hfi}, we have shown how \simall, the low-$\ell$ EE likelihood approximation used in the reference likelihood, relies heavily on the HFI end-to-end part of the FFP10 simulations.

The low-$\ell$ parts of the \Planck\ likelihood have been validated through an extensive suite of tests. We find excellent consistency between the different data selection cuts tested (masks, data splits, etc.). In polarization, we note the pull of the $\ell\,{=}\,5$ multipole of the $EE$ spectrum results, which appears quite high and thereby pushes up the best-fit $\tau$ value. However, using simulations, we have shown that the $\tau$ constraints with and without the $\ell\,{=}\,5$ multipole only disagree at about the $1.6\,\sigma$ level. 

We have listed in Sect.~\ref{subsubsec:final_considerations} the limitations of our low-$\ell$ polarization likelihood approximation. We note that limitations in the systematics corrections (in particular, the analogue-to-digital converter nonlinearities) translate into an overestimation of the variance of the lowest multipoles in our simulations, leading us to suspect that refinements in both the analysis and simulations \citep[such as those carried out in][]{delouis2019} might translate into further tightening of the $\tau$ constraint. 

Given their high computational price, the small number of full-scale simulations that can be run has prevented us from correctly taking into account in our approximation the correlations, especially between the $TE$ and $EE$ spectra, and the $TT$ and $TE$ spectra. This already limits our polarization likelihood to using only the $EE$ power spectrum; adding $TE$ and not taking into account correlations would overestimate the constraining power of the low multipoles. We also reported a null-test failure of the $\ell\,{>}\,10$ part of the low-$\ell$ $TE$ spectrum that might be due to inaccuracies in the foreground model of our simulations, which ignores correlations between the dust contamination in temperature and polarization. Those multipoles have low constraining power on $\tau$.  We note, however, the agreement between the $EE$- and $TE$-based $\tau$ constraints, when restricting $TE$ to the multipole region that passes the problematic null test. Given the current $TE$ constraint on $\tau$, even if the null-test failure and correlation problems were resolved, we would only marginally improve the overall constraining power of the low-$\ell$ polarized data on $\tau$. 

\subsection{High-$\ell$ summary}

The high-multipole likelihood approximation, \plik, was described and extensively investigated in Sect.~\ref{sec:hi-ell}. At these small scales, the large number of modes allows us to assume that the likelihood shape is reasonably well approximated by a Gaussian and the smoothness of the spectra permits the use of a fiducial covariance. The accuracy of the data model is paramount in order to obtain a reliable likelihood approximation, and for this reason the greatest challenge was to build and validate this model, including correction for beam effects, temperature-to-polarization leakage, residual systematics, and foreground contamination.
A great deal of care was also needed to construct an accurate analytical approximation for the covariance between the power spectrum modes, taking into account masks and anisotropic noise effects. 
The 2018 high-$\ell$ approximation follows closely the approach developed in 
2013 and 2015, but most of its specific ingredients have been improved, in particular for the polarized data. 

\paragraph{Temperature-to-polarization leakage}
While the high-$\ell$ likelihood uses the maps produced from the improved data processing of the HFI data, the main progress at these scales has been obtained thanks to improvements in our systematic-effects residuals model. The conclusion of the 2015 power spectrum and likelihood paper \citepalias{planck2014-a13} was that unexplained residuals, at the $\muK^2$ level, were observed in the comparison between individual cross-frequency spectra. Those residuals prevented us from recommending the use of the small-scale $TE$ and $EE$ power spectra for high-precision cosmology, even though we showed that when ignoring the residuals the TE- and EE-based cosmological constraints were still in excellent agreement with the more robust TT ones. 
We showed in 2015 that the temperature-to-polarization leakage was a likely culprit for the source of some of these residuals and was degenerate with cosmological parameters.
We have discussed in detail in Sect.~\ref{sec:hi-ell:datamodel:beamleak} how we built and validated a suitable leakage model for the 2018 release. This model solves most of our residual issues in TE, as we have highlighted in Sect.~\ref{sec:valandro:triangle} and shown in Fig.~\ref{fig:hi_ell:valid:triangleTE}. This leakage model was validated on simulations \citep{quickpolHivon}, and we also used them to estimate the residual uncertainty on our model, which depends on the determination of the beam shapes, gain, and polarization efficiencies.

\paragraph{Polarization efficiency correction}
The temperature-to-polarization leakage is found to be small in the $EE$ power spectra and cannot improve the inter-frequency agreement in this case. For $EE$ the disagreement is caused by small errors on the ground-based polarization efficiency determinations of the HFI polarimeters. 
The estimated errors of the ground-based measurements of the polarization
efficiencies \citep{rosset2010} are found to be at least 5 times smaller that the value we have reported here and in \citetalias{planck2016-l03}. 
New direct measurements of polarization efficiencies on regions of the sky dominated by dust emission have led us to revise the ground-based estimates, first for the 353-GHz channels \citepalias{planck2016-l03}, and then for the cosmological channels, as described in Sect.~\ref{sec:hi-ell:datamodel:inst}. The corrections improve the agreement between the cross-frequency power spectra in $EE$ and result in a significant $\chi^2$ reduction ($\Delta\chi^2\,{=}\,50$). This improvement is shown in Fig.~\ref{fig:hi_ell:valid:triangleEE}, and discussed in Sect.~\ref{sec:valandro:triangle}. 

There are some limitations to the accuracy of our correction of the polarization efficiencies. The determination relies on a cosmological prior (using TT \lcdm\ best-fit cosmology as a reference) and our best robustness test, comparing the polarization-efficiency corrections obtained from the $EE$ and $TE$ power spectra, fails at the $2\,\sigma$ level, and this failure is not reproduced in our simulations. While this is not a substantial shortcoming, we investigated whether the data would prefer so-called ``spectrum-based'' polarization-efficiency corrections, where we would allow for different effective polarization-efficiency corrections in the $TE$ and $EE$ spectra.\footnote{We also refer to the model as ``map-based'' in the case where the polarization-efficiency corrections for $TE$ and $EE$ and related.}  At the level of the \lcdm\ cosmological model, for the joint \TTTEEE\ reference likelihood, we found only a minor impact on the cosmological parameters when switching to this spectrum-based model compared with using the reference map-based one, and we see only a small improvement in $\chi^2$ ($\Delta\chi^2\,{=}\,10$). In the case of extended cosmological models, however, the choice of polarization-efficiency correction model does have an impact, modifying by about $0.5\,\sigma$ some of the cosmological parameters, as we have reported in Sect.~\ref{sec:valandro:PE}. In order to explore the impact of the polarization-efficiency correction model (along with other variations of the choices made in the reference high-$\ell$ likelihood implementation), the \camspec\ likelihood (described in Sect.~\ref{sec:hi-ell:prod:camspec} and summarized below) incorporates the spectrum-based correction, and is used in \citetalias{planck2016-l06} to assess the robustness of the constraints to variations of the likelihoods. In the end, the polarization-efficiency estimation will remain one of the limitations of the \Planck\ results. Comparison of the \Planck\ HFI polarized data with future high-precision CMB experiments could help to mitigate this deficiency, but will be ultimately limited by the \Planck\ noise level.

\paragraph{Robustness tests}
At high $\ell$, we have reproduced some of the most interesting tests we performed in 2015 and have extended them in polarization, as well as introducing new ones. The outcome of these tests is found to be satisfactory. Small high-$\ell$ deviations that were already present in the previous releases, such as the apparent extra smoothing at high-$\ell$ ($\Alens$ enhancement) and the mild disagreement between the $\ell\,{<}\,800$ and $\ell\,{>}\,800$ parts of the spectrum, have been shown to still be present in TT, and no new particularly worrying deviations have been found in polarization. We note that despite a very significant improvement compared to 2015, the PTE of the full TE likelihood remains low. The temperature-to-polarization leakage correction has improved this issue significantly, and further data processing improvements have led to a decrease in the TE $\chi^2$ of about $\Delta\chi^2\,{=}\,74$. Nevertheless, the agreement between frequencies remains rather poor, with a PTE of 0.9\,\%, and using the spectrum-based polarization-efficiency correction model is not sufficient to solve the issue. We note that the coadded binned $TE$ spectrum has a good $\chi^2$ with a PTE of 71\,\%. We have investigated the issue in Sect.~\ref{sec:valandro} and shown that each cross-frequency spectrum is in excellent agreement with the best-fit joint TTTEEE \lcdm\ cosmology. The source of the poor PTE lies in disagreement between cross-frequency spectra located around the first and third troughs of the $TE$ power spectrum, as exemplified in the conditional residual (Fig.~\ref{fig:hi_ell:valid:conditionalTE}). We estimate that known sources of systematic residuals (such as 4-K lines, temperature-to-polarization leakage, or polarization-efficiency errors) cannot explain the small observed features. Apart from this issue, all of our other consistency checks of $TE$ pass, including the comparison of cosmological parameters obtained when removing some of the cross-frequency spectra, the $\ell\,{>}\,800$ versus $\ell\,{<}\,800$ comparisons, or the conditional predictions from $TT$ at the spectrum level (with small outliers within statistical expectation around the first and third troughs) or cosmological parameter level. We have assumed that the low PTE is the result of a possible underestimation of the scatter between frequencies and a consequence of the limitation of the statistical description of our data, rather than any missing correction in our model. At this stage we do prescribe the use of the high-$\ell$ TE within a joint TTTEEE likelihood.

\paragraph{FFP10 simulations}
We have used the FFP10 simulations at high $\ell$ to estimate the robustness of our approximations. The high-$\ell$ likelihood was already tested, especially in 2015, using specifically designed simulations (to check the covariance-matrix analytical computation), as well as through the FFP8 suite of simulations. In both cases, we found that the approximation was in excellent agreement with the simulations. This also allowed us to detect a 
small systematic bias in $\ns$, of about $0.1\,\sigma$, which disappears when changing the starting multipole of the likelihood approximation in TT. A similar behaviour was observed when varying the transition multipole between the high-$\ell$ and the low-$\ell$ regimes in the data. We think that this bias is due to the lack of fidelity of the Gaussian approximation when the number of modes becomes smaller. We have now performed further tests using the FFP10 simulations, which contain a better and more accurate reproduction of the instrumental effects. The FFP10 simulations were designed with the goal of driving the low-$\ell$ data-processing enhancement and providing a simulation suite for the less stringent lensing and non-Gaussianity measurements. We have explained in Sect.~\ref{sec:valandro:sims} how those design goals limit what we can achieve with the simulations at high $\ell$. With this limitation in mind we have tested how the likelihood approximation performs on the end-to-end part of the FFP10 simulations. The behaviour of the approximation appears to be satisfactory. Using the simulations, we have detected a systematic bias in the $100\times100$, $143\times143$, and $217\times217$ cross-half-mission $EE$ spectra, which for lack of better explanation we call ``correlated noise.'' Ignoring this correlated noise in the likelihood model causes a systematic $0.25\,\sigma$ bias in the measured value of $\omb$ when using the EE likelihood alone. Using the FFP10 simulations we have built an empirical correction template that we include in the data model of the likelihood. Looking at the mean of the retrieved cosmological parameters on our 300 simulations, we have found a $2.8\,\Delta$ shift\footnote{In Sect.~\ref{sec:valandro:sims}, we defined $\Delta$ to be the estimate of the bias on our 300 simulations.} on $\ns$ in TE, which corresponds to a possible $0.16\,\sigma$ systematic shift in a single realization (again when using TE alone). Finally, when comparing the cosmological parameters obtained for the different levels of complexity of our simulations, we have found a $0.3\,\sigma$ disagreement on $\ns$ in EE between two versions of the simulations, where we would not have expected such a shift. Given the limitations of the FFP10 simulation suite, it is not possible to determine whether this is due to the particular single sky and foreground realization used in the end-to-end simulations or is an indication of a different problem.

\paragraph{Comparison between likelihoods}
Another way to assess the robustness of the high-$\ell$ likelihood is to compare it with the results from a different implementation, using variations in the model or approximations. We performed a similar comparison in 2013, where the two adopted likelihood approximations used very different data cuts, models for the foregrounds, and approximations of the statistical properties of the $TT$ power spectrum. The reference implementation then was the \camspec\ likelihood, while the \plik\ one, which had a lower constraining power, was used for validation. In 2015, \camspec, as well as several other new high-$\ell$ likelihood implementations, were tested, and compared to the new reference, the \plik\ likelihood. In 2018, we compare only the \camspec\ likelihood with the baseline \plik\ one. The specifications of the \camspec\ likelihood are highlighted in Sect~\ref{sec:hi-ell:prod:camspec}, with the main differences being: a different choice of masks in polarization, using a smaller sky fraction; a difference in the polarization-efficiency correction model, using the spectrum-based model and individual and possibly different corrections for each cross-half-mission cross-spectra; a different choice of multipole selection; a different noise model, based on the odd-even map differences; small variations in the foreground model correction in temperature, using a slightly different dust template in TT, and a slightly different CIB model; and small variations in the foreground model correction in polarization, using a map-cleaning procedure for the lowest multipoles and a power-spectrum template at higher ones, ignoring the dust variance in polarization. Out of those differences, the ones that most affect the cosmological constraints are the mask selection, the polarization-efficiency correction model and possibly the noise model.

The results from the two likelihoods are consistent at the level of cosmological parameters, as discussed extensively in \citetalias{planck2016-l06}. We show a comparison of the two likelihood $TE$ and $EE$ power spectra in Appendix~\ref{app:camvspliPS}, which are found to be in good agreement. The differences in cosmological parameters between the two are below $0.5\,\sigma$ in the base-\lcdm\ model, a result that gives confidence in the main cosmological conclusions.  Similar differences are observed on extended parameters, and in particular those sensitive to the apparent extra smoothing in the highest multipole range of the $TT$ power spectrum, which is the cause of the $\Alens$ curiosity. While both \camspec\ and \plik\ $TT$ exhibit the same high $\Alens$ value, the differences in polarization-efficiency correction models between the two likelihoods shift the value of $\Alens$ in the joint $TTTEEE$ cosmological constraint, as we have discussed in Sect.~\ref{sec:valandro:PE}. 

{\paragraph{Assessment of residual systematics}
Estimating the level of residual systematics and uncertainty is in many ways an ill-posed problem. Whenever a particular systematic effect has been detected and modelled, we correct for it, and we performed a comprehensive series of tests to identify such residual systematics. That process leaves us with some remaining minor failures of our different robustness and coherence tests, which we summarize in this last subsection.}

{Overall, our validation tests of the high-$\ell$ likelihood demonstrate the robustness of the \Planck\ high-$\ell$ temperature and polarization data analysis. Our compatibility tests, at the spectrum level (Sect.~\ref{sec:hi_ell:valid:condcoadd}) or at the parameter level (Sect.~\ref{sec:valandro:parcompTTTEEE}), both testify to the excellent agreement between temperature and polarization in the context of the \LCDM\ model. Furthermore, we showed in many different ways (at the spectra level or 
within the \LCDM\ model) the inter-agreement between our individual temperature and polarization cross-frequency spectra. We also established the compatibility at the level of 
cosmological parameters between the larger and smaller scales of the high-$\ell$ data. All of those results, with the duly noted limitations of a list of related ${>}\,2\,\sigma$ 
curiosities in temperature (including the apparent tension between $\ell>800$ and $\ell<800$ modes and the $\Alens\,{>}\,1$ effect), due to the shape of the scatter of the $TT$ bandpowers, together paint a picture of a solid data set.}

{Throughout this paper we have striven to challenge this picture in many different ways, and we have highlighted the limitations of the data and their statistical description, as well as the impact of the different modelling choices we made. Our main limitation lies in the PE-correction model. We have already recalled above how the PE corrections measured at 143\,GHz in $EE$ and $TE$ differ by about $2\,\sigma$. While this could be the result of statistical fluctuations, it is also possible that it is caused by residual systematics, 
affecting either the $EE$ or $TE$ spectrum calibration. We investigated whether changing the PE-correction model from the baseline map-based one to a spectrum-based one could affect the cosmology. Our findings, described in detail in Sect.~\ref{sec:valandro:PE} show that, with the baseline \plik\ likelihood, the \LCDM\ cosmological parameters are immune to the PE-correction choice and also to the marginalization of the PE-correction parameters. We nevertheless recommend fixing them, since, when jointly exploring them in the likelihood, the 217-GHz PE correction is pulled towards high values by the small-scale data.}

{In contrast, when testing for extended cosmological models, we observed systematic shifts in cosmological parameters, associated with the change in the PE-correction model. Those shifts are more pronounced for the extended parameters that are particularly sensitive to the relative temperature-to-polarization calibration, such as $\Alens$. The shifts are exacerbated when fixing the values of the PE correction as we do, and reach at most about $0.5\,\sigma$ on the extended parameters. Similarly, the alternative high-$\ell$ likelihood, \camspec, which implements the spectrum-based PE correction (but also other variations of some key design choices, and in particular a different mask for polarization) shows some differences in the \LCDM\ cosmological parameters at the $0.5\,\sigma$ level or less. It is for these two reasons that we estimate the robustness of our likelihood validation to be at this level. Note, however, that this does not mean that we are affected by a systematic $0.5\,\sigma$ residual effect. The differences between the different PE-correction models can be entirely explained by statistical fluctuations, and in the particular case of \camspec\ also by the mask differences. Both likelihood products are available, and the \plik\ one allows us to test both PE-correction models and to jointly explore the 
PE corrections along with cosmology and other nuisance parameters, allowing us to test the sensitivity of any cosmological model constraints to different data modelling choices. We followed this approach in both \citetalias{planck2016-l06} and \citet{planck2016-l10}, and highlighted the cases where some small dependency on the modelling choice was observed.}

{In some of our robustness tests, and in particular when relaxing priors, we observed shifts in some of our parameters, both in \LCDM\ and extended models. We highlighted the largest ones, such as for example, a modification of $\nnu$ of about $1.1\,\sigma$. In all of those cases, however, we concluded that the observed shifts could all be caused by statistical fluctuations at below the $2\,\sigma$ level. As such, we do not consider them to be a signature of unknown residuals. 
}

{Our tests on simulations revealed some systematic shifts, when using the TT-, TE-, or EE-only likelihoods. We discussed these in Sect.~\ref{sec:valandro:sims}, and recalled them above; they correspond to a $0.1\,\sigma$ shift in $\ns$ for TT, a $0.16\,\sigma$ $\ns$ shift for TE, and a $0.3\,\sigma$ disagreement in $\ns$ between two different levels of approximation of the E2E simulations in EE. 
As we discussed in Sect.~\ref{sec:valandro:sims}, at this level for each individual likelihood, these possible biases do not translate into worrying issues for the joint likelihood. Indeed, the joint TTTEEE simulation tests performed already in 2015 did not reveal any hint of a bias in $\ns$ at the $0.1\,\sigma$ level.}

\subsection{Future outlook}

Despite the important tightening of the constraints on the cosmological model and its parameters since COBE, the \Planck\ results are still in excellent agreement with the basic 6-parameter \lcdm\ model. However, we do report some ``curiosities,'' consisting of apparent deviations of the data from expectations at between the $2\,\sigma$ and $3\,\sigma$ level, such as: the low-$\ell$ power deficit in temperature; the $\Alens$ deviation; and the mild disagreement between the $\ell\,{<}\,800$ and $\ell\,{>}\,800$ parts of the $TT$ spectrum. These deviations are not independent from each other, but are tightly correlated, and are different ways of projecting the same power-spectrum fluctuations onto cosmological parameters.  Nevertheless, the \Planck\ data by themselves do not give us any compelling reason to extend the cosmological model. 
Those small anomalies were already reported in papers accompanying previous \Planck\ releases.  They are still present in the power spectra we use for this 2018 release and the new effects that we tested here (e.g., aberration or beam residuals) cannot explain them. 
The $TT$ power spectrum carries the bulk of the cosmological information coming from \Planck, and those small deviations have an impact on the joint \TTTEEE\ likelihood. The constraining power of the polarization data is not yet sufficient to confirm or disprove any physical model for those deviations.
The best current explanation for these curiosities is that they are
simply statistical excursions in the $TT$ power spectrum.

There are a few remaining ways of further improving the constraining power of the \Planck\ data. As we noted above, improvements in both the analysis and simulations \citep[such as the ones carried out in][]{delouis2019} might translate into further tightening of the $\tau$ constraint, mainly by reducing the observed extra variance of the $\ell=2$ and $\ell=3$ multipoles of the $EE$ power spectrum, which is found to be large in the current likelihood implementation. Adding the $TE$ power spectrum to the likelihood will require a significant amount of work in order to correctly take into account the $TT$-to-$TE$ and $TE$-to-$EE$ correlations, and will probably bring only small improvements on the $\tau$ constraint (although it might be important for other science objectives).

Regarding LFI data, further low-$\ell$ polarization improvements will likely require a substantial reworking of the calibration methodology at 70\,GHz, which has remained basically the same as for the 2015 analysis \citepalias{planck2016-l02}. While 30- and 44-GHz data did improve significantly in the current analysis, the latter is still not sufficiently consistent to be employed for low-$\ell$ cosmology; future improvements at the level of timeline processing and calibration may change this situation.

There is also room for minor refinements to further tighten the cosmological constraints at high $\ell$. The power spectra themselves, especially the $EE$ and $TE$ ones, can be improved slightly by using variance maps to estimate the optimal pseudo-spectrum, as was done in a slightly different setup for CMB lensing in \citet{planck2016-l08}.  Using correlations between the so-called ``detector-set'' maps (maps from individual bolometers and combination of polarimeters) one can also increase the S/N of the spectra used in the likelihood. This route was followed in 2013, but in 2015 we showed that, at least in temperature, the noise in detector-set maps was correlated. This is the reason why we decided to use correlations between half-mission maps in 2015 and also for this release. A robust model of the noise correlation in detector-set maps would be needed before we could use them for cosmology. The FFP10 simulations were not produced for this specific case, and we cannot rely on them to estimate the correlated noise. In 2015, a data-driven correlation template was built for TT. We observed differences in the cosmological parameters (at the fraction of a $\sigma$ level) and in foreground parameters (at the roughly 1\,$\sigma$ level) when comparing to the half-mission results, highlighting a possible degeneracy with the correction templates. It is unclear whether a similar correlated noise effect is present in the detector-set polarized data. In 2015, we observed approximately $\sigma$-level shifts between the baseline half-mission and detector-set EE-derived parameters. It is unclear whether those differences are due to differences in leakage templates, polarization-efficiency errors, or correlated noise.

Another way to improve the constraining power of the high-$\ell$ likelihood would be to increase the fraction of sky usable for cosmology. This could be achieved by replacing the power-spectra template correction we have used here by a dust-cleaning approach at the map level, using the higher-frequency maps. We have used a similar method in the low-$\ell$ likelihood. This approach was already implemented in 2015 in the \texttt{mspec} likelihood and demonstrated in the \camspec-cleaned likelihood discussed in \citetalias{planck2016-l06}. We are currently retaining about 80\,\% of the sky at 100\,GHz (before apodization) and such cleaning might permit us to reach a similar sky fraction for the other frequencies, thus allowing us to ignore the dust variance in the covariance matrix, which somewhat tightens the covariance at large scales. However, this method requires modelling the noise properties of the dust tracer channels (545\,GHz, and 353\,GHz for polarization) in order to avoid introducing correlations between the power spectra through the cleaning, and important modifications to the temperature foreground model. Further improvements might be achievable by reducing the apodization length of the Galactic masks. All of those steps will require significant work on the covariance estimation and thorough validation.

\Planck\ has shown how much information can be obtained by combining the temperature and polarization anisotropies of the CMB. Despite the limitations of the data and their analysis (and in particular the accuracy of the polarization efficiency of the detectors), both temperature and polarization data have been found to be in excellent agreement and strongly support the \lcdm\ model.
Just as COBE and WMAP did in their time, the \Planck\ legacy sets a solid foundation for future CMB experiments and other probes of the cosmological model.

\begin{acknowledgements}
The Planck Collaboration acknowledges the support of: ESA; CNES and CNRS/INSU-IN2P3-INP (France); ASI, CNR, and INAF (Italy); NASA and DoE (USA); STFC and UKSA (UK); CSIC, MINECO, JA, and RES (Spain); Tekes, AoF and CSC (Finland); DLR and MPG (Germany); CSA (Canada); DTU Space (Denmark); SER/SSO (Switzerland); RCN (Norway); SFI (Ireland); FCT/MCTES (Portugal); and ERC and PRACE (EU). A description of the Planck Collaboration and a list of its members, indicating which technical or scientific activities they have been involved in, can be found at 
\href{url}{http://www.cosmos.esa.int/web/planck/planck-collaboration}.
This research used resources of the IN2P3 Computer Center (\href{url}{http://cc.in2p3.fr}) as well as of the Planck-HFI DPC infrastructure hosted at the Institut d'Astrophysique de Paris (France) and financially supported by CNES.
\end{acknowledgements}

\bibliography{Planck_bib,like_2018}{}
\bibliographystyle{aat}

\appendix

\section{High-multipole supplemental material}
\label{app:hi-ell}

\subsection{Odd-even map correlations}
\label{app:oe_correlation}
To evaluate possible noise correlations between odd and even half maps, and to further check the half-mission map, we split the data in four ways, building the odd and even maps 
of each half mission. For example, we will call $\mathrm{H}_1^{\mathrm{Odd}}$ the map produced using the odd rings from the first half of the mission. 
The two cross-spectra formed between the maps from different half missions and different ring parities ($\mathrm{H}_1^{\mathrm{Odd}}\times\mathrm{H}_2^{\mathrm{Even}}$ and
$\mathrm{H}_2^{\mathrm{Odd}}\times\mathrm{H}_1^{\mathrm{Even}}$) will avoid scanning-related correlations and $\Delta t>1\mathrm{yr}$ ones. 
We will call these double cross-spectra ``super-conservative'' (SC) power spectra.
We can identify in the OE and HM power spectra the contribution of 
each of the $\mathrm{H}_i^{\mathrm{X}}$ maps and note that in both cases, the power spectra are the combination of the two SC power spectra, and in each case, two specific extra terms:
\begin{align}
\mathrm{SC} &= \left( \mathrm{H}_1^{\mathrm{Odd}}\times\mathrm{H}_2^{\mathrm{Even}} + \mathrm{H}_2^{\mathrm{Odd}}\times\mathrm{H}_1^{\mathrm{Even}} \right)/2; \\
\mathrm{HM} &= \left( 2*\mathrm{SC}+ \mathrm{H}_1^{\mathrm{Even}}\times\mathrm{H}_2^{\mathrm{Even}} + \mathrm{H}_1^{\mathrm{Odd}}\times\mathrm{H}_2^{\mathrm{Odd}} \right)/4; \\
\mathrm{OE} &= \left( 2*\mathrm{SC}+ \mathrm{H}_1^{\mathrm{Odd}}\times\mathrm{H}_1^{\mathrm{Even}} + \mathrm{H}_2^{\mathrm{Odd}}\times\mathrm{H}_2^{\mathrm{Even}} \right)/4.
\end{align}

We use the differences between the specific OE and HM cross-spectra and the SC ones to assess the amount of possible spurious correlation. 
Figure~\ref{fig:hi-ell:data:oe_hm_comparisons}, presents this test for the $TT$ and $EE$ $143\times143$ spectra. While the HM specific spectra do not seem to exhibit any particular trend between themselves and with the SC, 
there is a clear residual correlation at $\ell\,{>}\,1600$ in the OE for $TT$, coming specifically from the second half mission. This null-test failure cannot be 
explained by the subpixel effects. A similar failure, with a lower significance, can be observed in the second half-mission contribution of the OE for $143\times217$ (not shown). There are also some sign of correlated noise 
in the $100\times100$ $TT$ spectrum, both in the HM and OE specific terms at $\ell\,{>}\,1200$. We are not using this part of the data for cosmology.

There is also a marginally significant (when looking at the data mode by mode) systematic slope in the difference between the OE specific 
terms and the SC or HM specific terms. The $143\times143$ spectrum gives a large contribution to the cosmological constraints,\footnote{See Fig.~\ref{fig:hi_ell:data:mixing} and note also that, as shown Table~\ref{tab:hi-ell:data:lrange}, it is the only spectrum covering the full range, making it very sensitive to the slope of the model.} so that when building a $EE$ likelihood with OE spectra, this slope translates into a systematic $\ns$ shift of about $1\,\sigma$ (towards a better agreement with $TT$). 

Overall, the other spectra, including $TE$, display much better agreement between the SC, OE, and HM specific terms. 
On this basis, we should at the very least exclude the $143\times143$ $TT$, $143\times217$ $TT$, and $143\times143$ $EE$ OE spectra from any cosmological exploration. Since the OE differences also show sign of correlations (see Sect.~\ref{sec:hi-ell:datamodel:noise}), we exclude them from the high-$\ell$ likelihood.

\subsection{Relative amplitudes of the \textit{TE} and \textit{EE} leakage templates}
\label{app:ee_leakage_small}

\begin{figure}[htbp!]
\begin{center}
\includegraphics[angle=90,width=0.9\columnwidth]{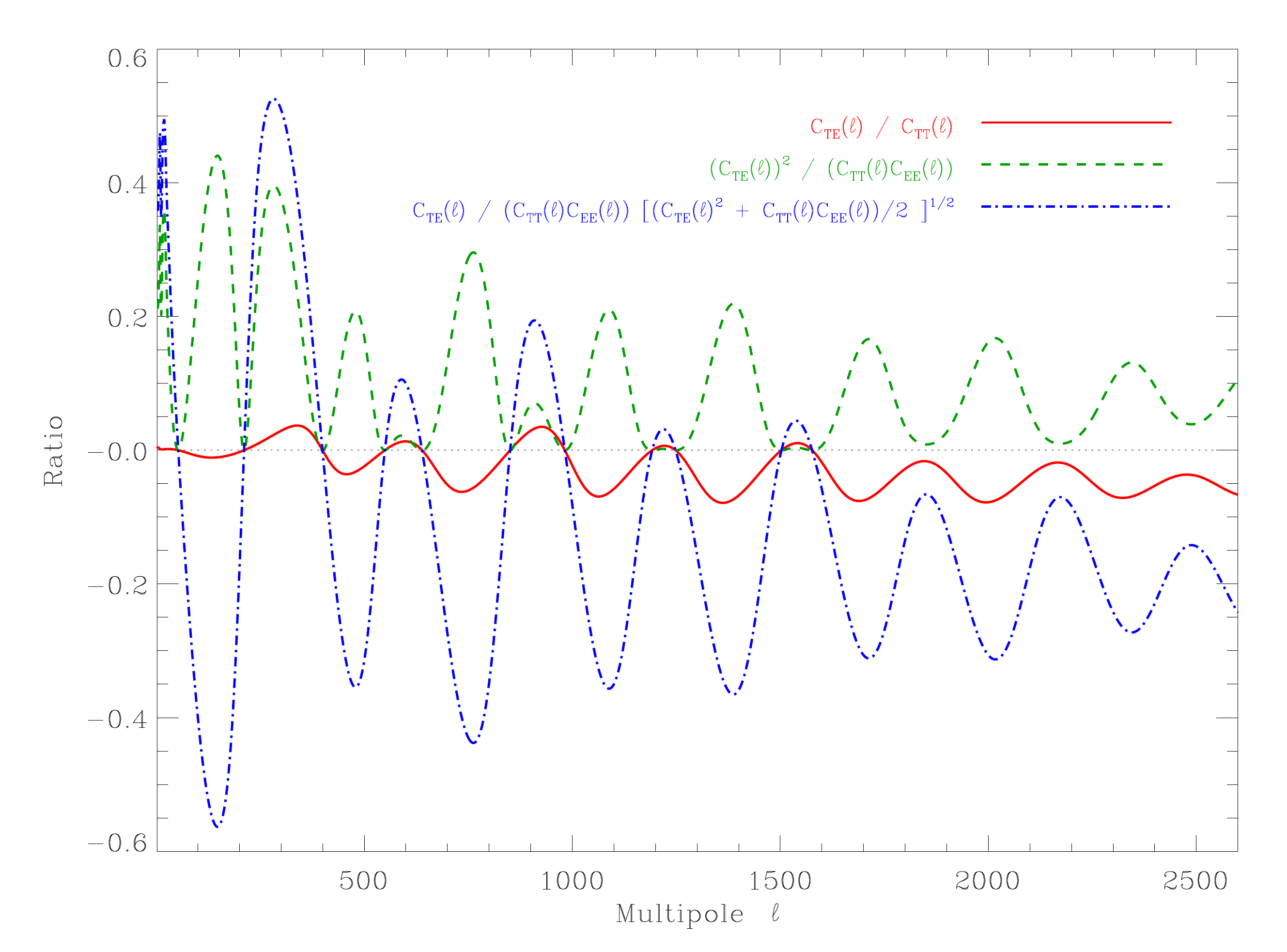}
\caption{Ratios of CMB spectra entering Eq.~\eqref{eq:app:teovtt}, \eqref{eq:app:te2ovttee}, and~\eqref{eq:app:epsilonl}, in red, green, and blue, respectively. All ratios are calculated for the \Planck\ 2015 $\Lambda$CDM best-fit model.
}
\label{fig:CMB_spectra_ratio}
\end{center}
\end{figure}

The estimation of the temperature-to-polarization leakage effect produces qualitatively very different results for the $TE$ and $EE$ spectra. Indeed, as shown Fig.~\ref{fig:hi_ell:data:leakages}, and discussed in Sect.~\ref{sec:hi-ell:datamodel:beamleak}, the leakage effect is predicted to be almost negligible for $EE$, while it is an important bias in $TE$. 
In fact, such a different behaviour is to be expected and can be
confirmed by evaluating the ratio of the amplitude of the leakage
effects in $TE$ and $EE$ in both the signal- and the noise-dominated regimes.

Inspection of Fig.~\ref{fig:Wmatrix} suggests that the leakage contributions to the $TE$ and $EE$
spectra will be dominated respectively by\footnote{This is also clear from equation~(44) in \citetalias{planck2014-a13}.}
\newcommand{\windowsratio}{r_\ell}
\begin{subequations}
\begin{align}
  \delta C^{TE}_{\ell} &= \widehat{W}^{TE,\;TT}_\ell \ C^{TT}_\ell, \\
  \delta C^{EE}_{\ell} &= \widehat{W}^{EE,\;TE}_\ell \ C^{TE}_\ell,
\end{align}
\end{subequations}
while the ratio of the beam window-matrix elements responsible of the leakage is
\begin{align}
  \windowsratio \equiv \left|
  \frac{\widehat{W}^{EE,\;TE}_\ell}{\widehat{W}^{TE,\;TT}_\ell}
  \right| \la 2\, .
\end{align}

\begin{figure*}[htbp!]
\begin{centering}
\includegraphics[width=\textwidth,angle=0]{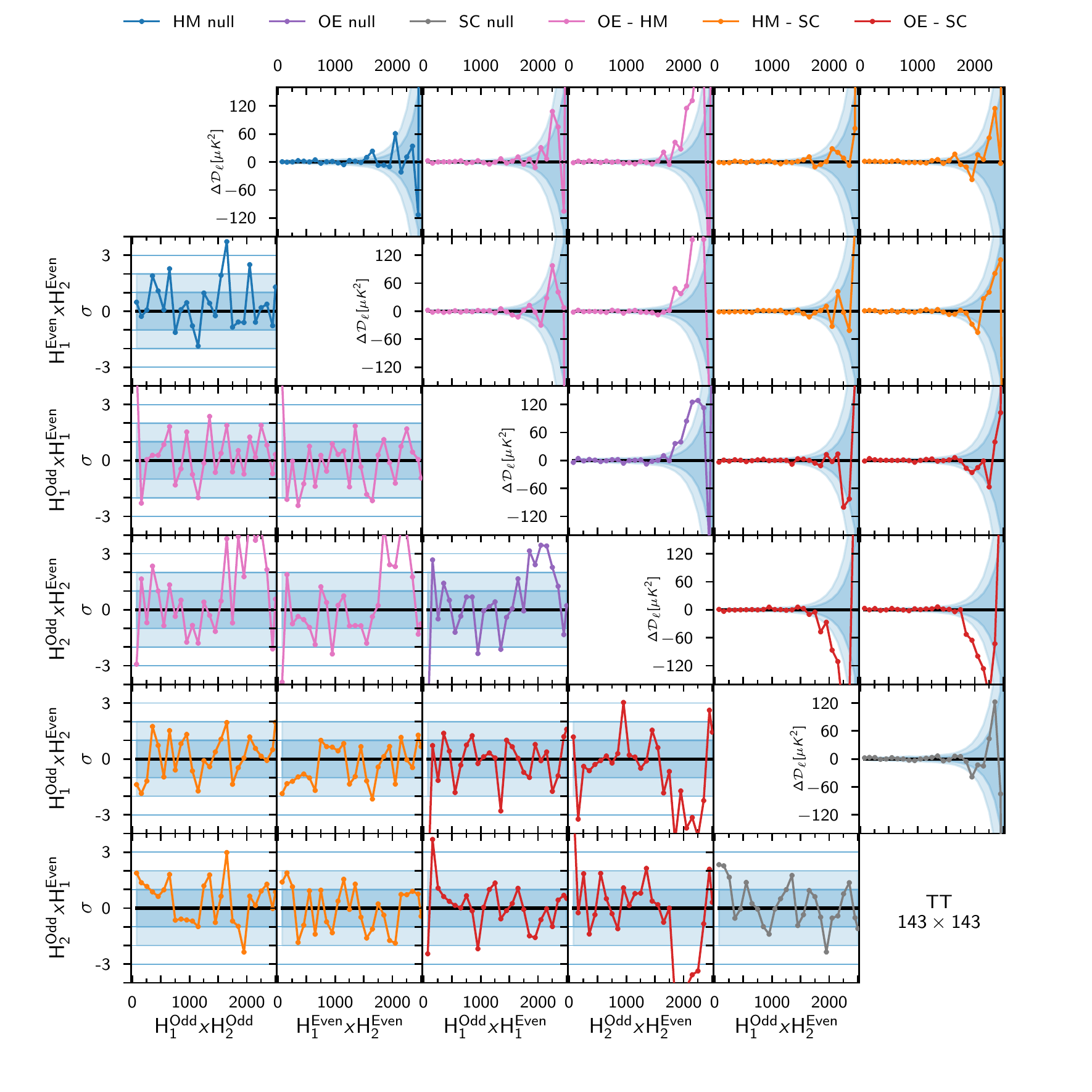}
\caption{Differences between the SC, OE, and HM specific terms entering the $143\times143$ $TT$ power spectrum (see text). The top right triangle shows the various power spectra differences in terms of statistical deviations, while the bottom left shows the spectral difference. Blue, purple, and grey are the null tests for HM, OE, and SC, respectively, obtained by computing the difference between the two terms specific to each case, while orange, red, and pink are HM-specific terms minus SC, OE specific terms minus SC, and HM minus OE, respectively. Error bars are obtained from a simple covariance-matrix approximation, using the empirical spectrum and the sky fractions, and ignoring the mask corrections.}
\label{fig:hi-ell:data:oe_hm_comparisons}
\end{centering}
\end{figure*}

\begin{figure*}[htbp!]
\ContinuedFloat
\begin{centering}
\includegraphics[width=\textwidth,angle=0]{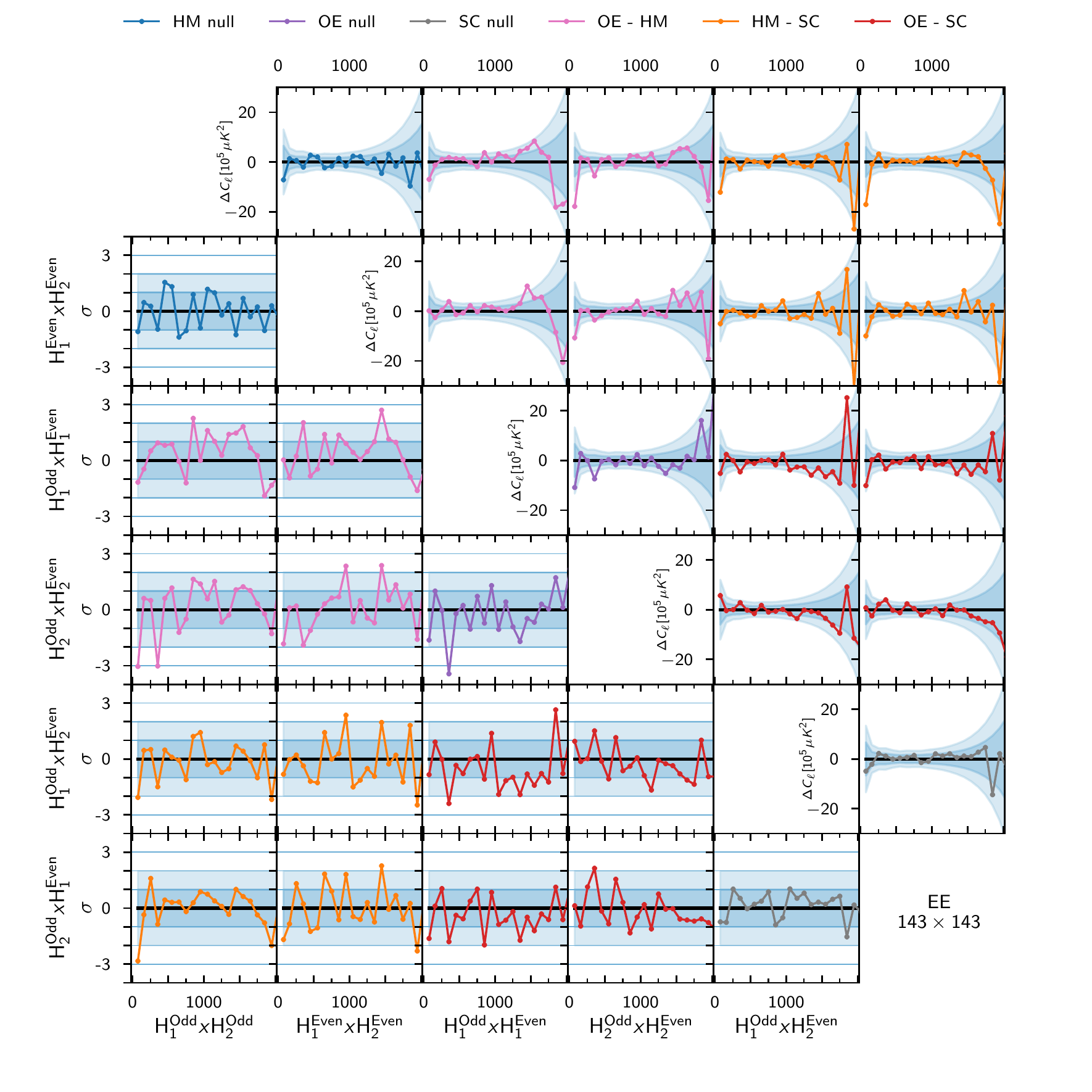}
\caption{Continuation of differences between the SC, OE, and HM specific terms, considering this time the $143\times143$ $EE$ power spectrum.}
\label{fig:hi-ell:data:oe_hm_comparisons}
\end{centering}
\end{figure*}

Noting that the ratio between the $EE$ and $TE$ spectra is always smaller than $0.1$\ (see Fig.~\ref{fig:CMB_spectra_ratio}), we can estimate, in absolute terms, the ratio between the 
$EE$ and $TE$ leakage corrections
\begin{subequations}
\begin{align}
  \left| \frac{\delta C^{EE}_{\ell}}{\delta C^{TE}_{\ell}} \right| &= 
  \windowsratio\ 
  \left|\frac{C^{TE}_\ell}{C^{TT}_\ell}\right|,\\
  & < 0.2\, . \label{eq:app:teovtt}
\end{align}
\end{subequations}

Similarly, we can show that the amplitude of the leakage in $TE$ and
$EE$ relative to $C^{TE}_\ell$ and $C^{EE}_\ell$ is always smaller
in the $EE$ case by evaluating 
\begin{subequations}
\begin{align}
 \left| \frac{\delta C^{EE}_{\ell} / C^{EE}_{\ell} }{\delta C^{TE}_{\ell} / C^{TE}_{\ell}} \right|&= 
 \windowsratio\ 
 \frac{\left(C^{TE}_\ell\right)^2}{C^{TT}_\ell C^{EE}_\ell}, \\  
 & < 1, \label{eq:app:te2ovttee}
\end{align}
\end{subequations}
where the second term is shown in Fig.~\ref{fig:CMB_spectra_ratio} to
be below 0.5 for all $\ell$.

Finally,  one can also compare the leakage power spectra $\delta
C^{XY}_\ell$ to approximations of the scatter of the $XY$ spectra
$\Delta C^{XY}_\ell$.  These are given by
\begin{subequations}
\begin{align}
  \left(\Delta C^{EE}_\ell\right)^2 &= \frac{\left(C^{EE}_{\ell}\right)^2 + 
   \left(C^{EE}_\ell + N^E_{1,\ell}/W^{EE,EE}_\ell\right)
   \left(C^{EE}_\ell + N^E_{2,\ell}/W^{EE,EE}_\ell\right)}{(2\ell+1) \fsky \Delta\ell}, \\
  \left(\Delta C^{TE}_\ell\right)^2 &= \frac{\left(C^{TE}_{\ell}\right)^2 + 
   \left(C^{TT}_\ell + N^T_\ell/W^{TE,TE}_\ell\right)
   \left(C^{EE}_\ell + N^E_\ell/W^{TE,TE}_\ell\right)}{(2\ell+1) \fsky \Delta\ell},
\end{align}
\end{subequations}
where 
$\Delta\ell$ is the bin width, 
$N^E_{1,\ell}$ and $N^E_{2,\ell}$ are the noise power spectra in each of the two polarized maps used to compute $EE$, while
$N^T$ and $N^E$ are the noise spectra of the temperature and polarization maps used to compute $TE$.
One may then consider the ratio
\begin{align}
  \epsilon_\ell \equiv \left| \frac{\delta C^{EE}_{\ell} / \Delta C^{EE}_{\ell} }{\delta C^{TE}_{\ell} / \Delta C^{TE}_{\ell}} \right|&= 
  \windowsratio\ 
  \left|\frac{C^{TE}_\ell}{C^{TT}_\ell}\right|\ 
  \frac{\Delta C^{TE}_\ell}{\Delta C^{EE}_\ell}. 
\end{align}

In the cosmic-variance-dominated regime, where noise can be neglected, one finds
\begin{subequations}
\begin{align}
  \epsilon_\ell &=
  \windowsratio\ 
  \left[ \left| \frac{C^{TE}_\ell}{C^{TT}_\ell}  \right|
    \left(\frac{\left(C^{TE}_\ell\right)^2 + C^{TT}_\ell C^{EE}_\ell}{ 2 \left(C^{EE}_\ell\right)^2}  \right)^{1/2}
  \right], \\
  & \la 1, \label{eq:app:epsilonl}
\end{align}
\end{subequations}
where the term in square brakets is shown in
Fig.~\ref{fig:CMB_spectra_ratio} to be smaller than 0.5 at $\ell\,{>}\,300.$

In the noise-dominated regime
\begin{align}
  \frac{\Delta C^{TE}_\ell}{\Delta C^{EE}_\ell} = \frac{N^T_\ell N^E_\ell}{N^E_{1,\ell}N^E_{w,\ell}}\ 
\frac{W^{TE,TE}_\ell W^{TE,TE}_\ell}{W^{TT,TT}_\ell W^{EE,EE}_\ell}.
\end{align}
This is expected to be less than unity, since for \Planck-HFI all detectors contribute to the measurement
of $T$, while only a subset of them measure $E$ and therefore $\epsilon_\ell < 1$.

Hence in all situations we expect beam leakage to have a lower impact on the
$EE$ spectra than on the $TE$ ones.

\subsection{\plik -\camspec\  polarization power spectrum comparison}
\label{app:camvspliPS}
\begin{figure*}[htbp!]
  \includegraphics[angle=0,width=0.95\textwidth]{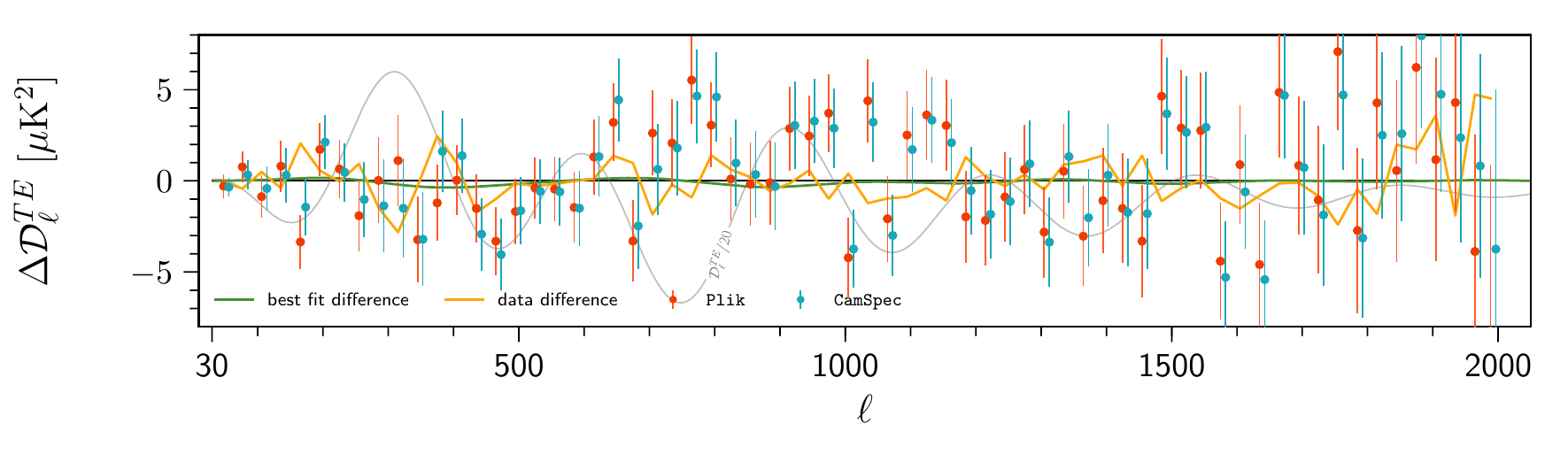}\\
  \includegraphics[angle=0,width=0.95\textwidth]{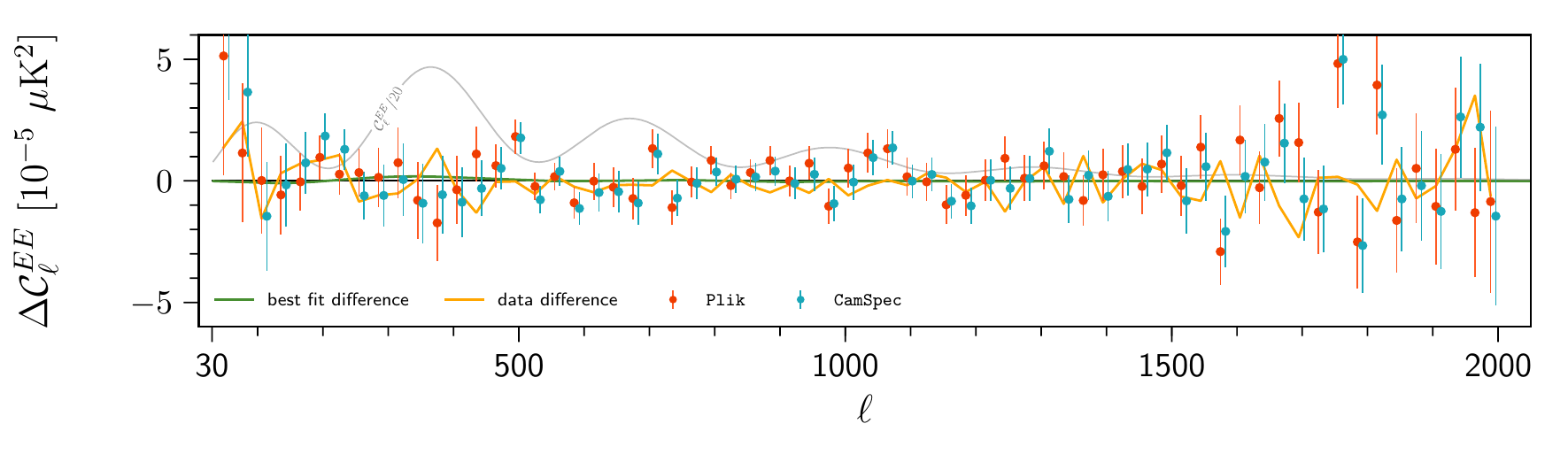}
  \caption{Comparison between the $TE$ (top) and $EE$ (bottom) coadded spectra from \plik\ (red points) and \camspec\ (blue points). We show here the residuals compared to the best-fit \planckalllensing\ \lcdm\ model: \plik\ as the red points; \camspec\ as the blue points; and the difference between the two likelihoods shown as the orange line.  To improve legibility, the \camspec\ data points have been shifted slightly compared to the \plik\ ones. The difference between the best-fit models when replacing \plik\ by \camspec\ in the \planckalllensing\ case is shown in green.  To give the locations of the peaks in $TE$ and $EE$ as a reference, rescaled CMB spectra are displayed in light grey.}
  \label{fig:concl:camspec_vs_plik_pol}
\end{figure*}

Appendix~A of \citetalias{planck2016-l06} gives an overview of the main cosmological parameter differences between the \plik\ and \camspec\  likelihoods, we present here a different way to illustrate the level at which the two likelihoods agree; Fig.~\ref{fig:concl:camspec_vs_plik_pol} shows the comparison between the $EE$ and $TE$ coadded power spectra obtained in the two likelihoods. In both case, we only display the residuals of the coadded power spectra compared to the joint \planckalllensing\ best-fit model obtained using \plik. The figure also shows the differences between the two sets of residuals, as well as the best-fit models when replacing \plik\ by \camspec, highlighting the  good agreement on cosmological parameters.  The \camspec\ points are plotted shifted in $\ell$ to improve legibility. Rescaled $TE$ and $EE$ power spectra are also plotted, to provide a reference for assessing whether the small differences between \plik\ and \camspec\ are correlated with the shape of the power spectrum. We remind the reader that the two likelihoods uses very different masks (shown in Fig.~\ref{fig:hi-ell:data:cammsk}), and we do expect some scatter between the two power spectra. Differences are small, peaking at about $2\muK^2$. There is no clear trend or correlation with the model power spectra. The small scatter between the power spectra of the two likelihoods is the at level at which we estimate that we validate our likelihoods and corresponds to the slight changes in the cosmological parameters discussed in \citetalias{planck2016-l06}.

\subsection{CIB model exploration}
\label{app:CIBextra}
\begin{figure*}
\includegraphics[angle=0,width=0.48\textwidth]{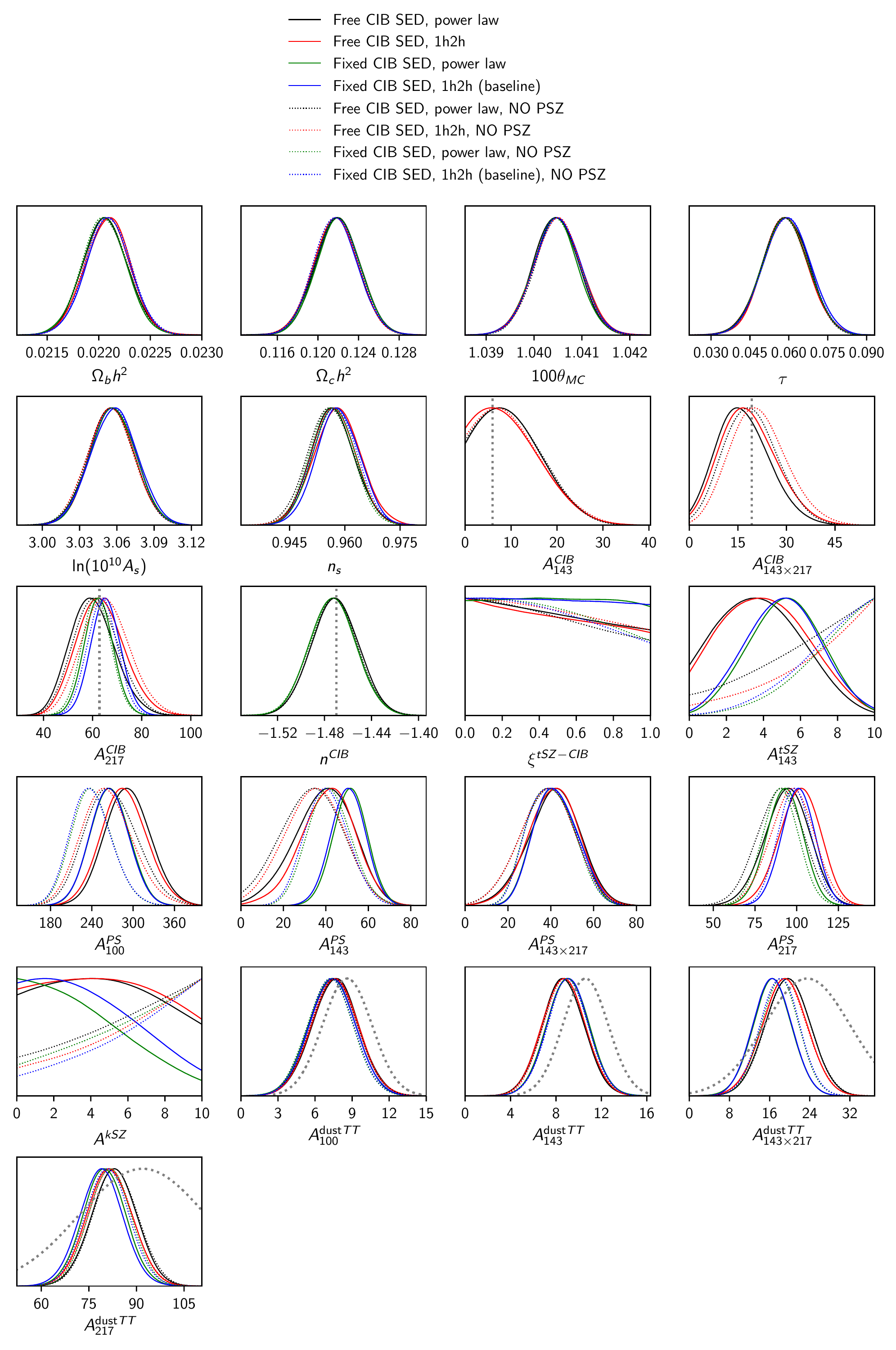}
\includegraphics[angle=0,width=0.48\textwidth]{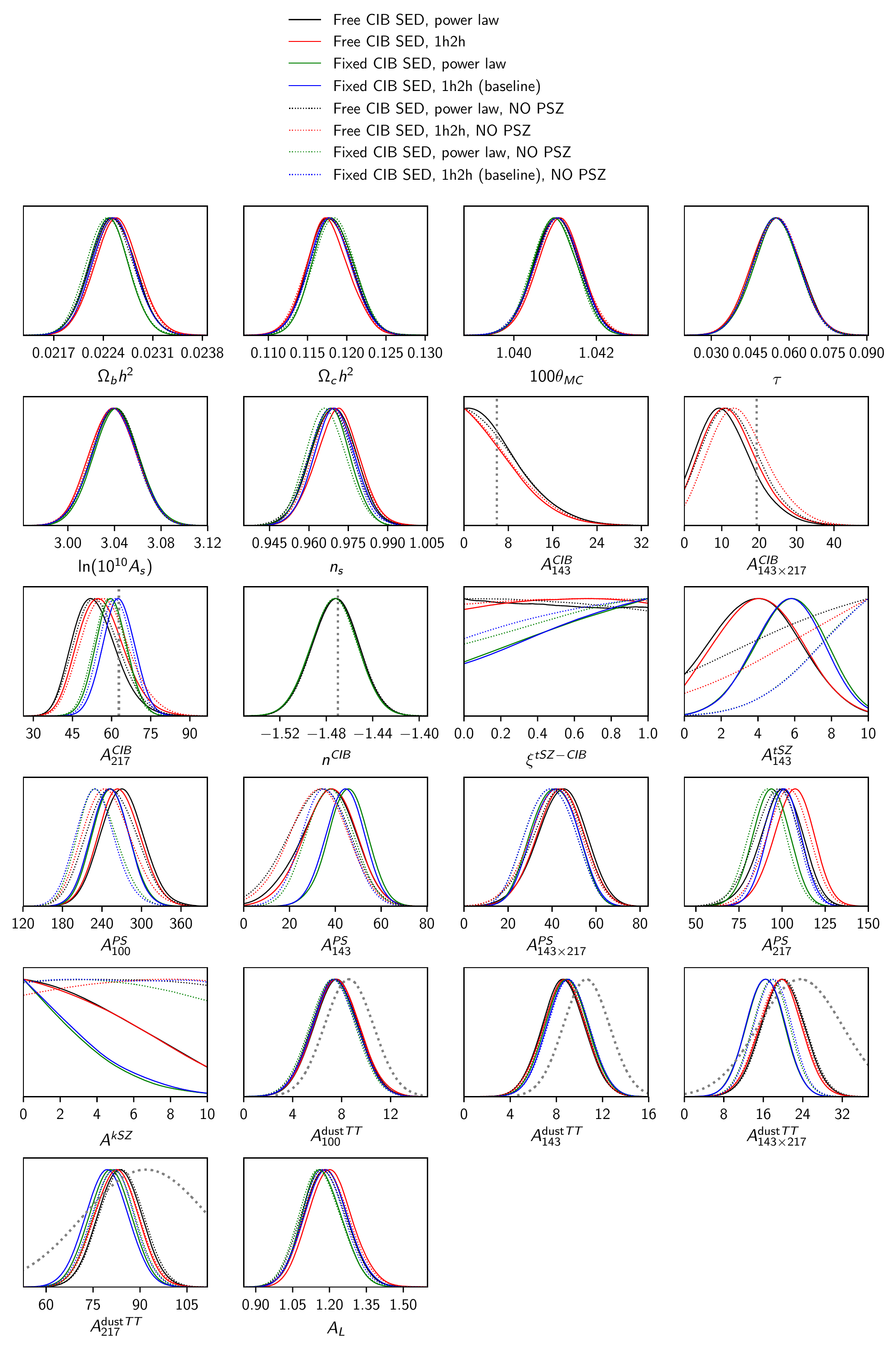}
\caption{Exploration of different CIB models and impact of the SZ prior. In the left panel we present the joint exploration of the \lcdm\ and different foreground parameters and in the right panel, the same when also opening up the $\Alens$\ parameter. In those tests we only use the \planckTTonly+\lowE\ likelihoods. The baseline model correspond to the one-plus-two
  halo model ({\it 1h2h}) described in \citet{planck2013-pip56}, which further predicts the CIB emission law.  We explore cases where the model template is replaced by a power law and where the SED is freely fitted. We further test the impact of relaxing the SZ prior. In all cases, the cosmological parameters are found to be robust to variations of the foreground model, including when opening up the $\Alens$ parameter.}
  \label{fig:app:CIBexpl}
\end{figure*}

In Fig.~\ref{fig:app:CIBexpl} we display the variation of cosmological parameters in \lcdm\ and \lcdm+$\Alens$\ models under modification of the CIB model and removal of the SZ prior. 
We find that the cosmological parameters are robust to variations of the foreground model, and recover the $\ns$\ sensitivity to the SZ prior discussed in Sect.~\ref{sec:hi-ell:datamodel:fg}, which is due to the correlation with the kSZ amplitude. The baseline model corresponds to the one-plus-two
halo model ({\it 1h2h}) described in \citet{planck2013-pip56} and Sect.~\ref{sec:hi-ell:datamodel:fg}, which further predicts the CIB emission law.  
Note that the alternative \camspec\ likelihood uses the same template but independently fits the emission at different frequencies. When using a power-law 
model for the CIB with \plik, we obtain a value of $n_{\rm CIB}=\gamma_{\rm CIB}-2=-1.47$, in perfect agreement with \citet{Mak:2017}. 
The baseline model fixes the CIB SED. When opening it up and independently fitting the CIB contribution for each of the $TT$ cross-spectra, 
we find variations of the CIB and PS contributions that compensate each other. However, the cosmological parameters are left essentially unchanged, both in the \lcdm\ and \lcdm+$\Alens$\ cases.

\end{document}